\documentclass[12pt,10pt]{ucithesis}

\usepackage[utf8]{inputenc}
\usepackage[hebrew,english]{babel}  
\usepackage{cjhebrew}               

\usepackage{amsmath}
\usepackage{amsthm}
\usepackage{array}
\usepackage{graphicx}
\usepackage{csquotes}
\usepackage[style=numeric-comp,sorting=none,giveninits=true,isbn=false,backend=biber]{biblatex}
\DeclareLanguageMapping{hebrew}{english}
\usepackage[backref=true]{atlasbiblatex}
\AtEveryBibitem{%
  \ifboolexpr{ test {\iffieldundef{doi}} and test {\iffieldundef{eprint}} }
    {}
    {\clearfield{url}}%
}

\makeatletter
\newif\if@blx@arabicpages
\@blx@arabicpagesfalse
\AtBeginDocument{%
  \let\blx@orig@pagenumbering\pagenumbering
  \def\pagenumbering#1{%
    \blx@orig@pagenumbering{#1}%
    \def\blx@tmp@style{#1}\def\blx@tmp@arabic{arabic}%
    \ifx\blx@tmp@style\blx@tmp@arabic
      \@blx@arabicpagestrue
    \else
      \@blx@arabicpagesfalse
    \fi}%
  \def\blx@addbackref@i#1{%
    \ifbacktracker
      \blx@leavevmode
      \if@filesw
        \if@blx@arabicpages
          \protected@write\@mainaux{}{\string\abx@aux@backref
            {\the\c@instcount}{#1}{\the\c@refsection}%
            {\number\c@page}{\number\c@page}}%
        \fi
      \fi
    \fi}%
}
\makeatother

\usepackage{relsize}
\usepackage[titletoc]{appendix}

\usepackage{caption}
\usepackage{subcaption}  
\usepackage{multirow}
\usepackage{tabularx}

\usepackage{lipsum}

\usepackage{amsmath}
\usepackage{amssymb}
\usepackage{bm}
\usepackage{slashed}                
\usepackage{booktabs}

\usepackage{graphicx}
\usepackage{subcaption}
\usepackage{hhline}
\usepackage{multirow}
\usepackage{makecell}
\usepackage{siunitx}
\usepackage{float}
\usepackage{lscape}
\usepackage[mathlines]{lineno}
\usepackage[section]{placeins}
\usepackage{fancybox}               
\PassOptionsToPackage{dvipsnames,x11names}{xcolor}
\usepackage[compat=1.1.0]{tikz-feynman}
\usepackage{contour}
\usetikzlibrary{arrows.meta,positioning,patterns,patterns.meta,angles,quotes,intersections,calc,math}
\makeatletter
\newcommand\jetcone[5][blue]{{
    \pgfmathanglebetweenpoints{\pgfpointanchor{#2}{center}}{\pgfpointanchor{#3}{center}}
    \edef\ang{#4/2}
    \edef\e{#5}
    \edef\vang{\pgfmathresult}
    \pgfpointdiff{\pgfpointanchor{#3}{center}}{\pgfpointanchor{#2}{center}}%
    \edef\JCxa{\the\pgf@x}%
    \edef\JCya{\the\pgf@y}%
    \pgfmathparse{\JCxa/1pt}\let\Cx\pgfmathresult
    \pgfmathparse{\JCya/1pt}\let\Cy\pgfmathresult
    \tikzmath{
        \x = veclen(\Cx,\Cy)*\e*sin(\ang)^2;
        \y = tan(\ang)*(veclen(\Cx,\Cy)-\x);
        \a = veclen(\Cx,\Cy)*sqrt(\e)*sin(\ang);
        \b = veclen(\Cx,\Cy)*tan(\ang)*sqrt(1-\e*sin(\ang)^2);
        \angb = acos(sqrt(\e)*sin(\ang));
    }
    \coordinate (tmpL) at ($(#3)-(\vang:\x pt)+(\vang+90:\y pt)$);
    \draw[thin,#1!40!black,fill=#1!50!black!80,rotate=\vang]
        (#3) ellipse({\a pt} and {\b pt});
    \draw[thin,#1!40!black,fill=#1!80!black!40,rotate=\vang]
        (tmpL) arc(180-\angb:180+\angb:{\a pt} and {\b pt})
        -- ($(#2)+(\vang:0.018)$) -- cycle;
}}
\makeatother
\newcommand\Tstrut{\rule{0pt}{2.6ex}}         
\newcommand\Bstrut{\rule[-0.9ex]{0pt}{0pt}}   

\usepackage[plainpages=false,hypertexnames=true,colorlinks=true,linkcolor=blue,citecolor=blue,urlcolor=blue]{hyperref}

\newcommand{\Etmiss}{\ensuremath{E^{\textrm{miss}}_{\textrm{T}}}}
\newcommand{\Zp}{\ensuremath{Z^{\prime}}}
\newcommand{\ttbar}{\ensuremath{t\bar{t}}}
\newcommand{\met}{\ensuremath{E_{\mathrm{T}}^{\mathrm{miss}}}}
\newcommand{\metsig}{\ensuremath{\mathcal{S}(E_{\mathrm{T}}^{\mathrm{miss}})}}
\newcommand{\minmlb}{\ensuremath{\min(\min(m_{\ell b}))}}
\newcommand{\GeV}{\ensuremath{\,\mathrm{GeV}}}
\newcommand{\TeV}{\ensuremath{\,\mathrm{TeV}}}
\newcommand{\pT}{\ensuremath{p_{\mathrm{T}}}}
\newcommand{\ET}{\ensuremath{E_{\mathrm{T}}}}
\newcommand{\MET}{\ensuremath{E_{\mathrm{T}}^{\mathrm{miss}}}}
\newcommand{\dzero}{\ensuremath{d_{0}}}
\newcommand{\Antikt}{anti-$k_{t}$}

\renewcommand*{\hbar}{{\mkern-1mu\mathchar'26\mkern-8mu\mathrm{h}}}

\thesistitle{Searches for new phenomena in final states with leptons
and jets using the ATLAS detector}

\documenttitle{Thesis}

\degreename{Doctor of Philosophy}

\degreefield{Physics}

\authorname{Roy Schimmel Brener}

\committeechair{Professor A}
\othercommitteemembers
{
  Assistant Professor B\\
  Associate Professor C
}

\degreeyear{April 2026}

\copyrightdeclaration
{
  {\copyright} Copyright by \Authorname, \Degreeyear \\
  All Rights Reserved
}

\acknowledgments
{
  The highest praise in physics is often to describe a piece of work as ``clear''. Not ``nice", not ``impressive", not ``great" — simply clear. It has taken me a decade to understand why clarity is, indeed, paramount. 

The concept of truth --- ``\cjRL{'mt}'' in Hebrew --- reflects a continual effort to understand Nature.
By understanding, one means formulating mathematical laws with predictive power over physical phenomena. 
Such theories are ultimately validated only through observation, the essential bridge between ideas and reality. 
When experimental results confirm or refute an idea, one treads incrementally nearer to truth, slowly illuminating a vast unknown. 

It has been my honour to humbly contribute to this collective endeavour, adding a page to a greater story — the ongoing effort to extend the boundaries of our understanding. 
This has involved obtaining several experimental results, thereby clarifying certain ideas. The opportunity to do so using the world's most advanced instruments --- the ATLAS detector and the Large Hadron Collider at CERN --- has been a rare privilege. 

Reaching this point would not have been possible without the support, friendship, collaboration and generosity of many people. 
I am sincerely grateful to all, and would like to personally acknowledge some of them below.

Above all, I extend my deepest appreciation to my mother, Orna Schimmel Brener. 
She has always made me believe I can achieve anything I set my mind on, instilling iron in my spine and putting wind in my sails. 
Thank you for your love and support.

I could not have imagined pursuing my PhD anywhere other than at Weizmann Institute, which has been my home over the past few years. 
Here, I had the privilege of conducting research under the guidance of Noam Tal Hod, and I am honoured to have been his first PhD student. 
His commitment to truth and clarity --- ranging from counting histogram bin entries to overhauling the direction of a research project --- has made him the best advisor I could have hoped for. 
He has brought out the very best in me, and for that I am deeply grateful. 
I also thank Ehud Duchovni, who welcomed me to Weizmann and has been a constant source of support. 

At Weizmann, I wish to thank Lorne J. Levinson, with whom I have been lucky enough to work since my days as a master's student. 
His wisdom, friendship and support throughout my work on the sTGC and beyond are deeply appreciated. 
Furthermore, I should like to express my gratitude to Ranny Budnik, who has chaired my PhD Advisory Committee, for his continued feedback throughout the years. 
In this regard, I also thank Eilam Gross and Liron Barak, who served on my PhD Approval Committee, for a careful read of my thesis and dedicated feedback. 

In our group, let me thank Arka Santra for his friendship and collaboration on the dilepton and Clockwork projects. 
A huge thanks also goes to my friend and office mate, Dvij C. Mankad, whose help and advice will always be cherished. 
Moreover, I thank Giannis M. Maniatis, who has not only been an outstanding colleague on the sTGC project, but also a good friend. 

Still from Weizmann, I thank Etienne Dreyer, a friend whose advice on dilepton matters and beyond is greatly appreciated. 
In the same vein, I am thankful to Anna Ivina --- who even helped me paint my first Rehovot flat --- for her advice on all ATLAS matters. 
I also acknowledge Nilotpal Kakati and Edward B. Shields for their friendship. 
Furthermore, I offer my deepest gratitude to Katharina I. Zittlau for her steadfast support in overcoming many challenges. 
Her generosity and dedication will never be forgotten. 
I am also grateful for the opportunity to become her co-author on a biochemistry paper. 

Within the dilepton project, my recognition goes to Anna Bingham and Frank Ellinghaus, the best colleagues I could have wished for. 
Even S. H\aa land and Ellis Kay also collaborated closely on this analysis, and I thank them both. 
Chairing our Editorial Board, Uta Klein gave a thoughtful review, for which I am grateful. 
I wish to thank Yoav Afik, my Technion office mate, analysis colleague, and LpX Convener --- I was fortunate on all three counts.
I also thank Holly Pacey, whose direction was valuable in furthering the dilepton and lepton+jet searches. 
Likewise, my appreciation goes out Iacopo Longarini and Volker A. Austrup, whose support has played a vital role; with Volker I was privileged to work also as a co-analyser. 

Further, leading the QBH analysis from inception to publication was a remarkable opportunity. 
Here, I thank Sergey Karpov, Yao Fu and Douglas M. Gingrich for their efforts and dedication.  
ATLAS appointed the best Editorial Board one could wish for: I thank the Chair, Christian Sander, for defending our analysis gallantly, and Christos Vergis and Matthias Saimpert, who were exceedingly valuable across all review aspects. 
Let me also seize this opportunity to thank two co-Analysis Contacts: Michael Holzbock and Yanlin Liu, for helping to lead the dilepton/lepton+jet Umbrella. 
I should also like to thank Exotics Conveners whose direction has been profoundly valuable in bringing our analyses forward. 
In particular, to Louie D. Corpe and Giuliano Gustavino, and previously, Daniel Hayden and Tamara Vázquez Schröder --- thank you all. 

I will always be indebted to CERN for laying the ground to meet great people, some of whom have become my good friends. 
To Aaron White: he has been a loyal friend and confidant throughout my years at CERN. 
Thank you Aaron. 
Thanks also to Rongkun Wang, whose generosity in mentoring me over the sTGC trigger chain operation is deeply appreciated. 
Around the same time, I also became acquainted with Alexander N. Tuna, who was another valuable guide on the sTGC. 
In Geneva, I was fortunate enough to meet Ariela Scheinmann, who has been a kind friend. 

In the Clockwork project, I had the good fortune to work with Sean D. Lawlor, who was a valued colleague. Likewise, I want to thank Davide Melini, who mentored me during my first ATLAS analysis, on boosted signatures with photons and fat jets. 

This is also an opportunity to thank a few people who had a positive impact on my path towards PhD. 
From my years at Technion, let me thank one of my best friends, Omer Tsur. 
Likewise, I thank Liat Nemirovsky-Levy, Yuval Abulafia and Ofir Arzi for a wonderful couple of years, which would have been much more difficult and --- of course --- not as fun, without them. 

Lastly, for my time at Imperial College, I express my sincerest gratitude and appreciation to Per Jonsson, who introduced me to particle physics in 2016 through a student project on the Tevatron. 
This led to an internship with T2K and eventually a bachelor's project on COMET. 
I owe a great deal to his mentorship at the beginning of my time in physics as a na\"ive undergraduate student. 
Three more people I wish to thank from my years at Imperial are Federico de Vito, Max Festenstein and David Abraham --- thank you all for your friendship. 

To summarise, I should like to note that this whole journey would not have been possible, nor worth it, without the people whose love, support, friendship, help, advice and dedication I was fortunate enough to receive. 
Coming across so many smart and thoughtful people has been a gift of equal magnitude to the intellectual challenge in pursuing new understandings about Nature. 

Thank you all for bestowing the gift of your acquaintance upon me.

\bigskip
\noindent Roy S. Brener\\
Weizmann Institute, Rehovot\\
April 2026

}

\thesisabstract
{
  The Standard Model is the most precise and consequential theory of science. 
It describes the building blocks of Nature in the form of elementary particles, ranging from matter to force-carrying particles. 
However, its span is not a complete account of all fundamental physics. 
Amongst its limitations, the Standard Model lacks a coherent unification of quantum mechanics and gravity, thus cannot resolve the large hierarchy gap between the weak and Planck scales. 
It does not explain neutrino masses or the evidence implying the existence of Dark Matter. 
Nor does it provide a viable mechanism for baryogenesis sufficient to explain the observed matter-antimatter asymmetry in the Universe. 
Lastly, it is in tension with experimental observations hinting toward charged Lepton Flavour Universality violation. 
This subset of shortcomings indicates that physics beyond the Standard Model must exist. 

Over the last few decades, different experiments have been searching for evidence for New Physics, notably at particle colliders. 
These experiments have accessed rare phenomena at unprecedented precision and energy reach. 
In this thesis, the ATLAS detector at the Large Hadron Collider is used to perform three distinct searches for New Physics. 
One is a study into hypothetical, TeV-scale quantum black holes. 
These objects result from extra-dimensional theories seeking to solve the hierarchy problem. 
The search exploits a unique feature of the model which strongly enhances the production rate with the collider centre-of-mass energy, thus permitting a considerable increase in mass reach. 
Another study tackles anomalies in lepton flavour by searching for heavy resonances that violate flavour conservation, thereby explaining persisting tensions with the Standard Model.   
Finally, a third study presents a search that targets Clockwork, periodic signatures of heavy graviton states appearing in extra dimensional models.

The quantum black holes search uses lepton+jet final states and sets the strongest limits on the model to-date. 
The search for heavy lepton-flavour violating resonances targets signatures consisting of dilepton+$b$-jets and complements previous inclusive results by directly tackling exclusive final states. 
The Clockwork study is performed in dielectron and diphoton final states. 
It establishes a new method to search for periodic signals using wavelet transformations with machine learning, and sets the best limits on the model.

}

\contributionsstatement
{
The analyses presented in this thesis are ordered to reflect the degree of personal contribution, ranging from a project conceived
and executed almost entirely by me to one where I contributed as a part of a larger team. 
A separate detector service project on the New Small Wheel trigger is also discussed, albeit not in the main body of the thesis. 

\medskip
\noindent\textbf{Search for New Physics in $\bm{\ell j}$ at
$\bm{\sqrt{s}=13.6}$~TeV in Run~3}\\[1.5pt]
\noindent Performing this search using a partial Run~3 dataset was proposed to the ATLAS Collaboration by me. 
I initiated the project, and served as its analysis contact and paper editor of the resulting ATLAS publication. 
I carried out all analysis tasks, from beginning to end. 
A small team of collaborators, led by me, provided valued contributions. 
The project is complete; the paper draft is public on arXiv and submitted to Physics Letters B~\cite{ATLAS:2026hzb}. 

\medskip
\noindent\textbf{Search for New Resonances in $\bm{\ell^{+}\ell^{-}+b}$ at $\bm{\sqrt{s}=13}$~TeV in Run~2}\\[1.5pt]
\noindent I was one of two lead analysers on this search, which is part of a broader programme exploring exclusive dilepton final states, $\ell^{+}\ell^{-}+X$, also targeting contact interactions, leptoquarks and a resonance produced in association with missing transverse momentum. 
A preliminary result for the latter is public~\cite{ATLAS-CONF-2023-045}. 
The analysis targetting $b$-jet channels was largely finalised in 2024, when I became gradually more involved in the Run~3 search. 
Since then, the co-analyser took over most tasks. 
The resonant search in $\ell^{+}\ell^{-}+b$ is complete and unblinded since May 2025; the remaining channels are finalised in parallel with the combined ATLAS paper currently in preparation.  
My contributions include co-initiating the project, signal modelling in collaboration with the theorist behind the model, systematic uncertainty estimation, profile-likelihood fits and limit-setting, as well as other tasks shared with the co-analyser.

\medskip
\noindent\textbf{Search for Periodic Signals in $\bm{ee/\gamma\gamma}$ at $\bm{\sqrt{s}=13}$~TeV in Run~2}\\[1.5pt]
\noindent My contribution to this analysis focused on the neural network optimisation: ensuring the classifier and autoencoder networks were optimally configured, as well as contributing to the wavelet transformations. 
This first-of-its-kind analysis is complete, published in JHEP~\cite{ATLAS:2023hbp} and featured in a CERN Courier article~\cite{CERN:Courier2023Clockwork}. 

\medskip
\noindent\textbf{Integration of Strips into the New Small Wheel Trigger}\\[1.5pt]
\noindent This detector service project is separate from the physics analyses above. 
I was one of two main contributors, working under supervision of the Trigger Processor Project Leaders. 

}

\outline
{
  This thesis is arranged in two parts.

\smallskip\noindent\textbf{Part I\quad Background, Chapters 1--5.}

\noindent
Chapter~\ref{chp:sm} introduces the Standard Model: its gauge-symmetry structure, matter
content and the Higgs mechanism of electroweak symmetry breaking.
Chapter~\ref{chp:bsm} motivates physics beyond the Standard Model focusing only on scenarios relevant to this work: quantum black holes in extra-dimensional models, a heavy leptonn-flavour violation resonance produced in association with $b$-quarks and the
Clockwork/Linear Dilaton mechanism.
Chapter~\ref{chp:detector} describes the Large Hadron Collider and the ATLAS detector.
Chapter~\ref{chp:objects} covers object definition (electrons, muons, jets, $b$-tagged
jets, missing transverse energy) common to all analyses presented in the thesis.
Chapter~\ref{chp:stats} presents the statistical tools included in the thesis.

\smallskip\noindent\textbf{Part II\quad Analyses, Chapters 6--8.}

\noindent
The three physics analyses presented in this thesis are described in reverse chronological
order, from most-recent to oldest.
Chapter~\ref{chp:qbh} presents the search for quantum black holes in lepton$+$jet final
states at $\sqrt{s}=13.6$~TeV (Run~3), setting exclusion limits in two benchmark models.
Chapter~\ref{chp:zprime} describes the search for new resonances in $\ell^+\ell^-+b$
final states at $\sqrt{s}=13$~TeV (Run~2).
Chapter~\ref{chp:clockwork} covers the Clockwork/Linear Dilaton search in dielectron and
diphoton final states (Run~2). 
The thesis summarises with Concluding Remarks in Chapter~\ref{chp:conclusions}.

\medskip\noindent The detector service project integrating the strips of the small-strip Thin Gap Chambers into the New Small Wheel trigger system is presented in Appendix~\ref{app:nsw_trigger}.

}

\hypersetup{
	pdftitle={\Thesistitle},
	pdfauthor={\Authorname},
	pdfsubject={\Degreefield},
}

\begin{document}

\pagenumbering{Alph}

\raggedbottom

{%
  \clearpage
  \thispagestyle{empty}%
  \nolinenumbers
  \singlespacing
  \thisfancyput(3.13in,-4.85in){%
    \setlength{\unitlength}{1in}\thicklines\fancyoval(7.8,10.7)}%
  \begin{center}
    \vspace*{1cm}

    \includegraphics[width=7cm]{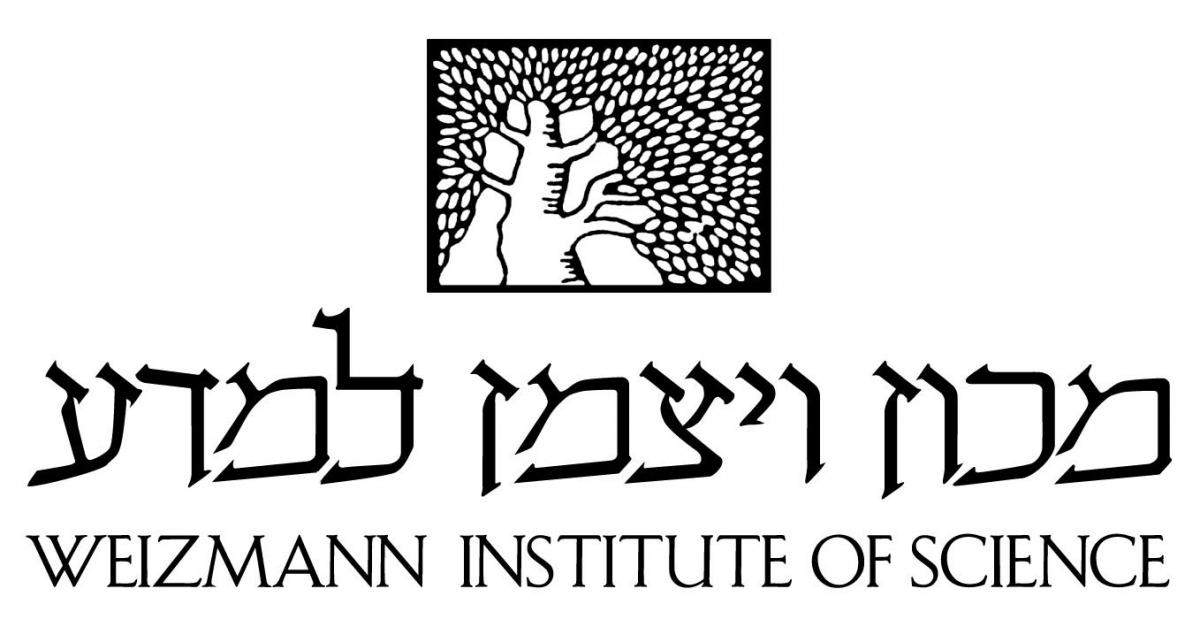}

    \vspace{2cm}

    \large
    \selectlanguage{english}Doctor of Philosophy Thesis \hfill
    \selectlanguage{hebrew}{חיבור גמר לתואר דוקטור לפילוסופיה}

    \vspace{1.5cm}

    \selectlanguage{english}by \hfill
    \selectlanguage{hebrew}{מאת}

    \selectlanguage{english}Roy Schimmel Brener \hfill
    \selectlanguage{hebrew}{רועי שימל ברנר}

    \large
    \vspace{2cm}

    \selectlanguage{hebrew}
    \textbf{חיפושים אחר תופעות חדשות במצבים סופיים עם לפטונים\\[4pt]
            וסילונים באמצעות גלאי אטלס}

    \vspace{1cm}

    \selectlanguage{english}
    \textbf{Searches for new phenomena in final states with leptons\\[4pt]
            and jets using the ATLAS detector}

    \vspace{2.7cm}

    \large

    \selectlanguage{english}Advisers \hfill
    \selectlanguage{hebrew}{מנחים}

    \selectlanguage{english}Noam Tal Hod \hfill
    \selectlanguage{hebrew}{נועם טל הוד}

    \selectlanguage{english}Ehud Duchovni \hfill
    \selectlanguage{hebrew}{אהוד דוכובני}

    \vspace{0.9cm}

    \large
    \selectlanguage{english}April 2026 \hfill
    \selectlanguage{hebrew}{ניסן תשפ"ו}

    \selectlanguage{english}
  \end{center}
  \clearpage
}

\preliminarypages
\setstretch{1.3}

\cleardoublepage
\pagenumbering{arabic}

\clearpage\chapter{The Standard Model}
\label{chp:sm}
Quantum mechanics (QM) and Special Relativity are combined into Quantum Field Theory (QFT), a mathematical framework that describes the behaviour of fundamental particles and their interactions.
The Standard Model (SM) is a QFT that formulates three of the four known forces in the Universe: the electromagnetic, weak and strong interactions.
It has been remarkably successful in providing accurate predictions for a wide range of phenomena in particle physics, which have been confirmed by numerous experimental results. 

This chapter gives an overview of the SM, starting from its underlying symmetries, followed by a description of its fundamental interactions and their matter content.
Running couplings and their role across different SM interactions are then introduced, before the Higgs mechanism responsible for electroweak symmetry breaking and mass generation is presented.

\section{Symmetries and Conservation Laws}

\subsection{Spacetime Symmetries and Noether's Theorem}

Any relativistic QFT must be invariant under the Poincaré group, combining spacetime translations and Lorentz transformations (rotations and boosts), 
\begin{equation}
\mathcal{P} = \mathbb{R}^{1,3} \rtimes \mathrm{SO}(1,3),
\label{eq:poincare_group}
\end{equation}
where $\mathbb{R}^{1,3}$ denotes spacetime translations and $\mathrm{SO}(1,3)$ the Lorentz group. 
The semi-direct product $\rtimes$ reflects that these operations do not commute: a Lorentz transformation followed by a translation differs from the reverse order, since the transformation rotates the direction of translation in spacetime. 
Invariance under spacetime translations means physics is the same everywhere and at all times. 
Lorentz invariance ensures all inertial observers agree on the fundamental laws. 
These symmetries connect directly to conservation laws through Noether's theorem: every continuous symmetry of the action corresponds to a conserved current. 
Consider an action,
\begin{equation}
S = \int \mathrm{d}^{4}x \, \mathcal{L}(\phi, \partial_{\mu}\phi),
\label{eq:action_general}
\end{equation}
where $\mathcal{L}$ is the Lagrangian density and $\phi$ the fields. 
If the action is invariant under a continuous field transformation, 
\begin{equation}
\phi(x) \rightarrow \phi(x) + \delta \phi(x),
\label{eq:field_variation}
\end{equation}
then there exists a conserved current $j^{\mu}$ such that
\begin{equation}
\partial_{\mu} j^{\mu} = 0.
\label{eq:current_conservation}
\end{equation}

This implies the charge
\begin{equation}
Q = \int \mathrm{d}^{3}x \, j^{0}(x),
\label{eq:conserved_charge}
\end{equation}
is, in turn, constant in time. Time translation invariance, spatial translations and rotations yield conservation of energy, momentum and angular momentum, respectively.

\subsection{Internal Symmetries and Gauge Fields}

Beyond spacetime symmetries, fields can have internal symmetries, transformations that act on the fields themselves while leaving coordinates unchanged. 
Consider a complex scalar field with Lagrangian
\begin{equation}
\mathcal{L} = \partial_{\mu}\phi^{\dagger} \partial^{\mu}\phi - m^{2}\phi^{\dagger}\phi.
\label{eq:complex_scalar_lagrangian}
\end{equation}
The field $\phi(x)$ is invariant under the global phase transformation,
\begin{equation}
\phi(x) \rightarrow e^{i\alpha} \phi(x),
\label{eq:global_u1}
\end{equation}
where $\alpha$ is constant. 
This is a global $U(1)$ symmetry, the group of unit complex numbers under multiplication. 
Noether's theorem gives the conserved current as
\begin{equation}
j^{\mu} = i \left( \phi^{\dagger} \partial^{\mu}\phi - \partial^{\mu}\phi^{\dagger} \phi \right),
\label{eq:u1_current}
\end{equation}
with $\partial_{\mu} j^{\mu} = 0$, as in Eq.~\eqref{eq:current_conservation}.
The conserved charge may represent particle number or electric charge. 
This symmetry is Abelian: transformations commute.

A natural question arises: what if we demand that this symmetry holds \emph{locally}, independently at each point in spacetime? 
This means allowing the phase parameter to vary with position---instead of a constant $\alpha$, we now consider a position-dependent $\alpha(x)$:
\begin{equation}
\phi(x) \rightarrow e^{i\alpha(x)} \phi(x).
\label{eq:local_u1}
\end{equation}
Under this local transformation, the Lagrangian in Eq.~\eqref{eq:complex_scalar_lagrangian} is no longer invariant. 
The reason is that when $\alpha$ depends on position, the covariant derivatives of $\phi$ generate terms involving $\partial_{\mu}\alpha(x)$, spoiling the invariance. 
Restoring invariance requires introducing a new vector field $A_{\mu}(x)$, the gauge field, and replacing ordinary derivatives with covariant derivatives,
\begin{equation}
D_{\mu} = \partial_{\mu} + i e A_{\mu},
\label{eq:covariant_derivative}
\end{equation}
where $e$ is a coupling constant. 
Under local transformation, the gauge field must transform as $A_{\mu} \rightarrow A_{\mu} - (1/e)\partial_{\mu}\alpha(x)$ to compensate for the position-dependent phase variation. 
This procedure—demanding local gauge invariance—naturally produces quantum electrodynamics (QED), where $A_{\mu}$ is identified with the photon field and $e$ with the electric charge. 

The profound lesson is that requiring local symmetry inevitably introduces interaction: the gauge field mediates forces between charged particles. 
This principle, applied to additional symmetry groups, generates the complete structure of the SM.

\subsection*{Non-Abelian Gauge Symmetries}

The gauge principle extends beyond the Abelian $U(1)$ case to non-Abelian groups, where generators do not commute with one another. 
These non-Abelian symmetries lead to richer physics: the gauge fields themselves carry the charges associated with the symmetry, leading to self-interactions that are absent in QED. 
In the SM, the full gauge group is
\begin{equation}
SU(3)_{\mathrm{C}} \times SU(2)_{\mathrm{L}} \times U(1)_{\mathrm{Y}},
\label{eq:sm_gauge_group}
\end{equation}
where $SU(3)_{\mathrm{C}}$ describes the strong interactions between quarks (quantum chromodynamics), $SU(2)_{\mathrm{L}}$ governs the weak interactions acting on left-handed fermions and $U(1)_{\mathrm{Y}}$ represents the hypercharge interaction. 
The $SU(3)$ and $SU(2)$ groups are non-Abelian, resulting in the self-interactions of gluons and weak bosons, distinguishing these forces from electromagnetism. 

The complete realisation of this gauge structure requires additional elements, particularly spontaneous symmetry breaking through the Higgs mechanism, which generates masses for the weak gauge bosons while preserving the underlying gauge invariance.
These aspects, along with the detailed structure of fermion representations under the gauge group, loop diagrams and quantum corrections, and the running of couplings across interaction régimes, will be developed in the sections that follow.

\section{Perturbation Theory and Quantum Corrections}
\label{sec:sm:perturbation}

The Lagrangian formulation and gauge symmetries of the preceding section determine the dynamics of SM fields.
Making quantitative predictions from these Lagrangians requires the usage of perturbation theory.

\subsection{Feynman Diagrams and Physical Amplitudes}
\label{sec:sm:feynman}

The probability amplitude for a transition from initial state $|i\rangle$ to final state $|f\rangle$ is encoded in the S-matrix element $\langle f | S | i \rangle$.
The \emph{Feynman amplitude} $\mathcal{M}$ is the reduced matrix element extracted from this, with overall kinematic conventions factored out.
Physical observables follow from $|\mathcal{M}|^2$: the differential cross-section and the differential decay rate of an unstable particle both satisfy
\begin{equation}
    \frac{\mathrm{d}\sigma}{\mathrm{d}\Omega} \propto |\mathcal{M}|^2, \qquad
    \frac{\mathrm{d}\Gamma}{\mathrm{d}\Phi} \propto |\mathcal{M}|^2,
    \label{eq:sm:amplitude_obs}
\end{equation}
where $\Omega$ is the solid angle parameterising the angular directions of the outgoing particles, and $\mathrm{d}\Phi$ is the invariant phase-space element for the final-state momenta.

When the coupling $\alpha = g^2/(4\pi)$ is small, the S-matrix is expanded perturbatively in powers of the interaction Lagrangian $\mathcal{L}_\mathrm{int}$.
As a concrete example, consider the Lagrangian of a real scalar field $\phi$ with a quartic self-interaction,
\begin{equation}
    \mathcal{L}_{\phi^4} = \tfrac{1}{2}(\partial_\mu\phi)^2 - \tfrac{1}{2}m^2\phi^2 - \frac{\lambda}{4!}\phi^4,
    \label{eq:sm:phi4}
\end{equation}
where $\lambda$ is a dimensionless coupling constant and $\mathcal{L}_\mathrm{int} = -\lambda\phi^4/4!$ is the interaction term.
Each term in the perturbative expansion of the S-matrix is evaluated using \emph{Feynman rules}: a systematic correspondence between graphical elements and algebraic factors derived from $\mathcal{L}$.
Lines represent propagating particles and carry the corresponding momentum-space propagator; vertices represent interaction terms and carry the coupling.
The resulting graphs are \emph{Feynman diagrams}, each providing a compact visual representation of a distinct contribution to $\mathcal{M}$.

\subsection{Tree Level and Loop Corrections}
\label{sec:sm:tree_loop}

Tree-level diagrams contain no closed loops and represent the leading contributions from interaction vertices; their predictions are equivalent to those of classical field theory.
Beyond tree level, diagrams with one or more \emph{loops} introduce quantum corrections to the classical result: virtual particles circulate in closed paths, with their four-momenta summed over by integration.
Each loop carries a relative suppression of $\alpha/(2\pi)$ with respect to the tree-level contribution.
For QED, where $\alpha_\mathrm{em} \approx 1/137$, this factor is $\approx 10^{-3}$, so loop corrections are small but non-negligible.

\begin{figure}[htbp]
  \centering
  \begin{subfigure}[b]{0.44\textwidth}
    \centering
    \begin{tikzpicture}
      \begin{feynman}
        \vertex (v);
        \vertex [above left=1.5cm of v] (i1) {\(\phi\)};
        \vertex [below left=1.5cm of v] (i2) {\(\phi\)};
        \vertex [above right=1.5cm of v] (f1) {\(\phi\)};
        \vertex [below right=1.5cm of v] (f2) {\(\phi\)};
        \diagram*{
          (i1) -- [scalar] (v),
          (i2) -- [scalar] (v),
          (v)  -- [scalar] (f1),
          (v)  -- [scalar] (f2),
        };
      \end{feynman}
      \node at (v) [right=6pt] {$\lambda$};
    \end{tikzpicture}
    \caption{Tree level: $\mathcal{O}(\lambda)$}
    \label{fig:sm:tree}
  \end{subfigure}
  \hfill
  \begin{subfigure}[b]{0.44\textwidth}
    \centering
    \begin{tikzpicture}
      \begin{feynman}
        \vertex (v1);
        \vertex (v2) [right=2.5cm of v1];
        \vertex [above left=1.5cm of v1] (i1) {\(\phi\)};
        \vertex [below left=1.5cm of v1] (i2) {\(\phi\)};
        \vertex [above right=1.5cm of v2] (f1) {\(\phi\)};
        \vertex [below right=1.5cm of v2] (f2) {\(\phi\)};
        \diagram*{
          (i1) -- [scalar] (v1),
          (i2) -- [scalar] (v1),
          (v1) -- [scalar, half left, looseness=1.5] (v2),
          (v2) -- [scalar, half left, looseness=1.5] (v1),
          (v2) -- [scalar] (f1),
          (v2) -- [scalar] (f2),
        };
      \end{feynman}
      \node at (v1) [left=6pt] {$\lambda$};
      \node at (v2) [right=6pt] {$\lambda$};
    \end{tikzpicture}
    \caption{One-loop correction: $\mathcal{O}(\lambda^2)$}
    \label{fig:sm:loop}
  \end{subfigure}
  \caption{Feynman diagrams for $\phi\phi \to \phi\phi$ scattering in scalar $\phi^4$ theory.
    (a) The tree-level contact interaction: a single four-point vertex carrying coupling $\lambda$, contributing at order $\lambda$ to the amplitude $\mathcal{M}$.
    (b) The leading one-loop correction: two four-point vertices, each carrying $\lambda$, connected by two internal propagators forming a closed loop, contributing at order $\lambda^2$.
    Each additional loop is suppressed by a further factor of $\lambda/(16\pi^2)$ relative to the preceding order.}
  \label{fig:sm:tree_loop}
\end{figure}
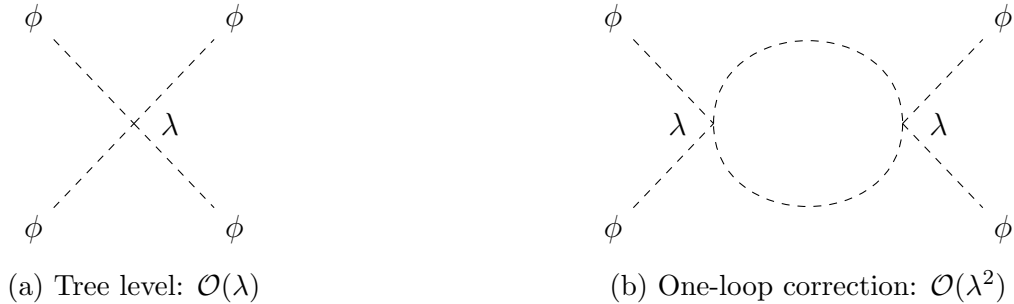

FIG.~\ref{fig:sm:tree_loop} illustrates this concept for $\phi\phi \to \phi\phi$ scattering in the scalar $\phi^4$ theory of Eq.~\eqref{eq:sm:phi4}.
The tree-level diagram is a single four-point contact vertex with coupling $\lambda$.
The one-loop diagram contains two such vertices, each carrying a factor $\lambda$, connected by a pair of internal propagators forming a closed loop.

\subsection*{Ultraviolet Divergences}

The integration over loop momenta exposes a structural difficulty.
In the diagram of FIG.~\ref{fig:sm:tree_loop}(b), the loop momentum $k$ runs over all values from zero to infinity:
\begin{equation}
    \Pi(s) \sim \lambda^2 \int \frac{\mathrm{d}^4k}{(2\pi)^4}
    \frac{1}{\bigl(k^2 - m^2\bigr)\bigl[(k-p)^2 - m^2\bigr]},
    \label{eq:sm:loop_integral}
\end{equation}
where $p^2 = s$ is the centre-of-mass energy squared and $m$ the scalar mass.
At large $k \gg m,\,p$, the integrand scales as $k^{-4}$.
The four-dimensional integration measure contributes $k^3\,\mathrm{d}k$, so the integral behaves as $\int \mathrm{d}k/k \sim \ln(\Lambda/m)$, diverging logarithmically as the ultraviolet cutoff $\Lambda \to \infty$.
These \emph{ultraviolet} (UV) divergences do not invalidate the theory.
They reflect the fact that quantum fluctuations are being naively extrapolated to arbitrarily short distances, where the theory provides no physical description.

\subsection{Renormalisation and Running Coupling}
\label{sec:sm:renorm}

The resolution is \emph{renormalisation}.
UV divergences are absorbed into redefinitions of the physical parameters evaluated at a chosen momentum scale $\mu$, the \emph{renormalisation scale}.
Measuring a coupling at scale $\mu$ implicitly absorbs all virtual corrections from momenta up to $\mu$.
Shifting $\mu$ redistributes these contributions between the explicit value of the coupling and the remaining loop corrections.

Since physical observables cannot depend on the arbitrary choice of $\mu$, any shift in scale must be compensated by a corresponding change in the coupling.
This constraint is the \emph{renormalisation group} (RG), and the rate of change is the \emph{beta function} $\beta$,
\begin{equation}
    \beta(\alpha) \equiv \mu\,\frac{\mathrm{d}\alpha}{\mathrm{d}\mu}.
    \label{eq:sm:beta}
\end{equation}
The trajectory traced by a coupling as $\mu$ varies is the \emph{RG flow}.
Running couplings are therefore not a theoretical artefact.
The effective strength of an interaction genuinely depends on the scale at which it is probed, as higher-energy measurements resolve shorter-distance quantum fluctuations.
The sign of $\beta$ determines whether a coupling grows or shrinks as a function of energy, with concrete consequences in QED and QCD discussed in Section~\ref{sec:sm:running}.

\section{Fundamental Interactions}

\begin{figure}[h]
  \centering
  \includegraphics[width=0.65\textwidth]{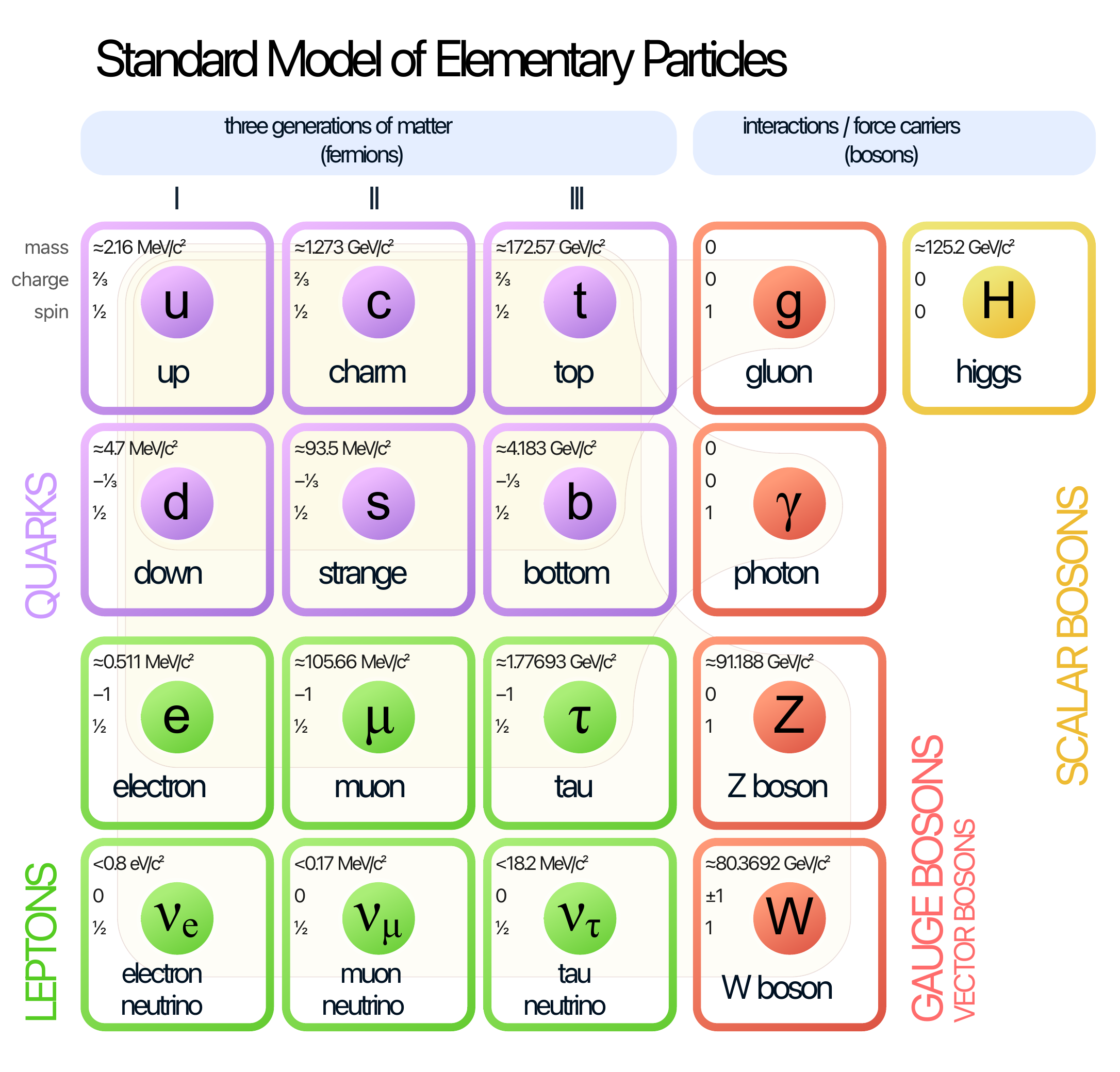}
  \caption{The elementary particles of the Standard Model~\cite{MissMJ:2006sm}.
    Fermions (spin-$1/2$) are divided into quarks (purple) and leptons (green), each arranged in three generations of
    increasing mass. The gauge bosons (spin-$1$, red) mediate the strong ($g$), electromagnetic
    ($\gamma$) and weak ($W^{\pm}$, $Z$) forces. The Higgs boson (spin-$0$, yellow) is responsible for electroweak symmetry breaking
    and mass generation. Each particle cell shows its mass, electric charge and spin.}
  \label{fig:sm:particles}
\end{figure}

The gauge group in Eq.~\eqref{eq:sm_gauge_group} determines how SM fields transform and interact. 
The fields are classified by their transformation properties (representations) under these three factor groups:
\begin{itemize}
    \item $SU(3)_{\mathrm{C}}$ (colour): governs QCD, interactions between quarks and gluons.
    \item $SU(2)_{\mathrm{L}}$ (isospin): governs the weak force, which acts only on left-handed fermions and the Higgs field.
    \item $U(1)_{\mathrm{Y}}$ (hypercharge): governs electromagnetic interactions (after symmetry breaking).
\end{itemize}
An overview of all known elementary particles --- three generations of matter fermions (quarks and leptons), the gauge bosons mediating the three SM forces and the scalar Higgs boson --- is shown in FIG.~\ref{fig:sm:particles}. The gauge and matter fields are described in the subsections that follow; the Higgs field, which plays a qualitatively different role by driving spontaneous symmetry breaking (SSB), is treated in Section~\ref{sec:sm:higgs}.

\subsection{The Gauge Boson Sector}

The gauge fields are spin-1 vector bosons that arise directly from the local gauge invariance requirement and mediate the forces between matter particles. With integer spin, they obey Bose-Einstein statistics: any number of gauge bosons may occupy the same quantum state, as exploited for instance in coherent states of the electromagnetic field. The number of gauge fields equals the number of generators of the associated symmetry group:
\begin{itemize}
    \item $U(1)_{\mathrm{Y}}$: 1 generator $\implies$ 1 gauge field ($B_{\mu}$).
    \item $SU(2)_{\mathrm{L}}$: $2^2 - 1 = 3$ generators $\implies$ 3 gauge fields ($W_{\mu}^{a}$, $a=1, 2, 3$).
    \item $SU(3)_{\mathrm{C}}$: $3^2 - 1 = 8$ generators $\implies$ 8 gauge fields ($G_{\mu}^{A}$, $A=1, \dots, 8$).
\end{itemize}
In total, the SM features 12 gauge bosons.

The generators of each group are the objects that encode the symmetry transformations and whose algebra determines the dynamics.
For $SU(2)_{\mathrm{L}}$, the three generators in the fundamental (doublet) representation are $T^{a} = \tau^{a}/2$, where $\tau^{a}$ are the Pauli matrices,
\begin{equation}
  \tau^{1} = \begin{pmatrix} 0 & 1 \\ 1 & 0 \end{pmatrix}, \qquad
  \tau^{2} = \begin{pmatrix} 0 & -i \\ i & \phantom{-}0 \end{pmatrix}, \qquad
  \tau^{3} = \begin{pmatrix} 1 & \phantom{-}0 \\ 0 & -1 \end{pmatrix}.
  \label{eq:sm:pauli}
\end{equation}
They satisfy the $SU(2)$ Lie algebra,
\begin{equation}
  \left[ \frac{\tau^{a}}{2},\, \frac{\tau^{b}}{2} \right] = i\,\varepsilon^{abc}\,\frac{\tau^{c}}{2},
  \label{eq:sm:su2_algebra}
\end{equation}
where $\varepsilon^{abc}$ is the Levi-Civita symbol ($\varepsilon^{123} = +1$, fully antisymmetric).
The non-commutativity $[\tau^{a}, \tau^{b}] \neq 0$ is precisely what makes $SU(2)_{\mathrm{L}}$ non-Abelian and leads to cubic and quartic self-interactions among the $W$ bosons.

For $SU(3)_{\mathrm{C}}$, the eight generators in the fundamental (triplet) representation are $T^{a} = \lambda^{a}/2$, where $\lambda^{a}$ are the Gell-Mann matrices, satisfying
\begin{equation}
  \left[ \frac{\lambda^{a}}{2},\, \frac{\lambda^{b}}{2} \right] = i\,f^{abc}\,\frac{\lambda^{c}}{2},
  \label{eq:sm:su3_algebra}
\end{equation}
with $f^{abc}$ the totally antisymmetric $SU(3)$ structure constants.
Two of the eight generators are diagonal and commute with each other, forming the Cartan subalgebra,
\begin{equation}
  \lambda^{3} = \begin{pmatrix} 1 & 0 & 0 \\ 0 & -1 & 0 \\ 0 & 0 & 0 \end{pmatrix}, \qquad
  \lambda^{8} = \frac{1}{\sqrt{3}}\begin{pmatrix} 1 & 0 & 0 \\ 0 & 1 & 0 \\ 0 & 0 & -2 \end{pmatrix};
  \label{eq:sm:gellmann_diag}
\end{equation}
their eigenvalues label the colour charges of the three quark states (red, green, blue).
The remaining six Gell-Mann matrices are off-diagonal and generate transitions between colour states, in direct analogy with $\tau^{1}$ and $\tau^{2}$ raising and lowering weak isospin.

The structure constants $f^{abc}$ enter the field strength tensor for a non-Abelian gauge group,
\begin{equation}
  F^{a}_{\mu\nu} = \partial_{\mu}W^{a}_{\nu} - \partial_{\nu}W^{a}_{\mu}
    + g\,f^{abc}\,W^{b}_{\mu}\,W^{c}_{\nu},
  \label{eq:sm:field_strength}
\end{equation}
where the last term, absent in QED, encodes the self-interactions of the gauge bosons.
The kinetic term for all non-Abelian gauge fields is then the Yang-Mills Lagrangian,
\begin{equation}
    \mathcal{L}_{\mathrm{YM}} = -\frac{1}{4} F^{a}_{\mu\nu} F^{\mu\nu, a},
    \label{eq:ym_lagrangian}
\end{equation}
summed over all generators of $SU(2)_{\mathrm{L}}$ and $SU(3)_{\mathrm{C}}$.

\subsection{The Fermion Sector}
\label{sec:sm:fermions}

The matter content consists of 12 fundamental fermions (and their antiparticles), organised into three generations. As spin-$1/2$ particles, fermions obey Fermi-Dirac statistics and the Pauli exclusion principle: no two identical fermions may share the same quantum state, a constraint that underlies the stability of matter and the structure of the periodic table. In QFT, fermions are represented as Weyl spinors, decomposed into left-handed ($\psi_{\mathrm{L}}$) and right-handed ($\psi_{\mathrm{R}}$) components under Lorentz transformations. This distinction is vital, as the $SU(2)_{\mathrm{L}}$ weak interaction only couples to left-handed fields. 
The representation of each field under the SM group in Eq.~\eqref{eq:sm_gauge_group} determines its charge and interaction.

The dynamics of the matter fields follow from a single gauge-invariant kinetic term,
\begin{equation}
  \mathcal{L}_{\mathrm{fermion}} = i\,\bar{\psi}\,\slashed{D}\,\psi,
  \qquad \slashed{D} \equiv \gamma^{\mu} D_{\mu},
  \label{eq:sm:fermion_kinetic}
\end{equation}
where $\gamma^{\mu}$ are the Dirac matrices and the Feynman slash abbreviates their contraction with the covariant derivative. Acting on a general fermion, this covariant derivative collects the couplings to all three gauge sectors,
\begin{equation}
  D_{\mu} = \partial_{\mu}
    + i g_{s}\,\frac{\lambda^{A}}{2}\,G^{A}_{\mu}
    + i g\,\frac{\tau^{a}}{2}\,W^{a}_{\mu}
    + i g'\,\frac{Y}{2}\,B_{\mu},
  \label{eq:sm:fermion_covariant}
\end{equation}
generalising the Abelian QED case of Eq.~\eqref{eq:covariant_derivative} to the full gauge group of Eq.~\eqref{eq:sm_gauge_group}; the generators are the Gell-Mann and Pauli matrices of Eqs.~\eqref{eq:sm:gellmann_diag} and~\eqref{eq:sm:pauli}, and $Y$ is the hypercharge. Each fermion couples only to the factors under which it is charged: a right-handed charged lepton, being a colour and weak-isospin singlet, feels only the $U(1)_{\mathrm{Y}}$ term, whereas a left-handed quark transforms under all three factors. Expanding Eq.~\eqref{eq:sm:fermion_kinetic} thus reproduces every fermion--gauge-boson interaction of the SM---the electromagnetic and strong couplings together with the charged- and neutral-current weak interactions---with their strengths fixed entirely by each field's gauge representation.

Table~\ref{tab:sm_fermions} explicitly displays the quantum numbers of all fundamental fermionic fields under $SU(3)_{\mathrm{C}} \times SU(2)_{\mathrm{L}} \times U(1)_{\mathrm{Y}}$ for a single generation.

\begin{table}[h]
\centering
\caption{Fermionic content of the Standard Model. The entries under $SU(3)_{\mathrm{C}}$ and $SU(2)_{\mathrm{L}}$ denote the representation dimensions, while $U(1)_{\mathrm{Y}}$ indicates hypercharge quantum numbers.}
\label{tab:sm_fermions}
\begin{tabular}{l c c c c}
\hhline{=====}
Field & Symbol & $SU(3)_{\mathrm{C}}$ & $SU(2)_{\mathrm{L}}$ & $U(1)_{\mathrm{Y}}$ \\
\hline
Lepton doublet & $L_{\mathrm{L}}$ & $\mathbf{1}$ & $\mathbf{2}$ & $-1/2$ \\
Electron singlet & $e_{\mathrm{R}}$ & $\mathbf{1}$ & $\mathbf{1}$ & $-1$ \\
Quark doublet & $Q_{\mathrm{L}}$ & $\mathbf{3}$ & $\mathbf{2}$ & $+1/6$ \\
Up-type singlet & $u_{\mathrm{R}}$ & $\mathbf{3}$ & $\mathbf{1}$ & $+2/3$ \\
Down-type singlet & $d_{\mathrm{R}}$ & $\mathbf{3}$ & $\mathbf{1}$ & $-1/3$ \\
\hhline{=====}
\end{tabular}
\end{table}

The physical electric charge $Q$ is derived from the weak isospin third component ($T_3$) and the hypercharge ($Y$) via the Gell-Mann--Nishijima relation,
\begin{equation}
    Q = T_3 + Y.
    \label{eq:gm_nishijima}
\end{equation}
Here $T_3$ is the eigenvalue of the generator $T^{3} = \tau^{3}/2$ introduced in Eq.~\eqref{eq:sm:pauli}: for an $SU(2)_{\mathrm{L}}$ doublet, $\tau^3$ is diagonal with eigenvalues $+1$ and $-1$, so the upper and lower components of any doublet carry $T_3 = +1/2$ and $T_3 = -1/2$ respectively. Right-handed fields are $SU(2)_{\mathrm{L}}$ singlets and do not transform under $SU(2)_{\mathrm{L}}$, so $T_3 = 0$ for them. For a single generation, the left-handed lepton and quark doublets are written explicitly as
\begin{equation*}
    L_{\mathrm{L}} = \begin{pmatrix} \nu_{e} \\ e \end{pmatrix}_{\mathrm{L}}, \qquad
    Q_{\mathrm{L}} = \begin{pmatrix} u \\ d \end{pmatrix}_{\mathrm{L}},
\end{equation*}
where each entry is a two-component $SU(2)_{\mathrm{L}}$ doublet with the neutrino/up-type quark in the upper position ($T_3 = +1/2$) and the charged lepton/down-type quark in the lower position ($T_3 = -1/2$). Applying Eq.~\eqref{eq:gm_nishijima} with $Y = -1/2$ for $L_{\mathrm{L}}$ gives $Q(\nu_e) = 0$ and $Q(e) = -1$; with $Y = +1/6$ for $Q_{\mathrm{L}}$ it gives $Q(u) = +2/3$ and $Q(d) = -1/3$, as indicated in Table~\ref{tab:sm_fermions}.
A notable consequence of Table~\ref{tab:sm_fermions} is that all three lepton generations carry identical gauge quantum numbers; the generations differ only in their Yukawa couplings and hence their masses.
This generation-universality of the gauge couplings is known as \emph{Lepton Flavour Universality} (LFU): the $W^\pm$, $Z$ and $\gamma$ bosons couple with identical strength to electrons, muons and taus, so the SM predicts equal partial widths for processes differing only by the lepton generation involved.

\subsection{Running Couplings and QCD Dynamics}
\label{sec:sm:running}

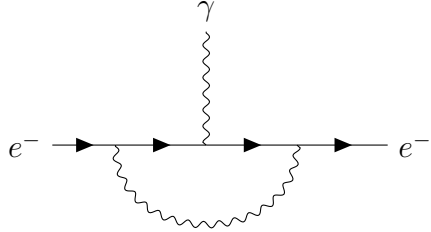
\begin{figure}[h!]
  \centering
  \begin{tikzpicture}
    \begin{feynman}
      \vertex (i) {\(e^-\)};
      \vertex [right=1.2cm of i] (v1);
      \vertex [right=1.2cm of v1] (v);
      \vertex [right=1.2cm of v] (v2);
      \vertex [right=1.2cm of v2] (f) {\(e^-\)};
      \vertex [above=1.5cm of v] (g) {\(\gamma\)};
      \diagram*{
        (i) -- [fermion] (v1) -- [fermion] (v) -- [fermion] (v2) -- [fermion] (f),
        (g) -- [photon] (v),
        (v1) -- [photon, half right] (v2),
      };
    \end{feynman}
  \end{tikzpicture}
  \caption{One-loop correction to the electron--photon vertex.
    A virtual photon arc connects two points on the electron line, shifting the gyromagnetic ratio from the tree-level Dirac prediction $g_e = 2$ and generating the leading contribution to the anomalous magnetic moment $a_e$.}
  \label{fig:sm:vertex_corr}
\end{figure}

The RG framework of Section~\ref{sec:sm:perturbation} has concrete, measurable consequences in each of the SM's interactions.
In QED, the gyromagnetic ratio $g_e$ relates the electron's magnetic moment to its spin angular momentum via $\vec{\mu}_e = g_e\frac{e}{2m_e}\vec{S}$; the Dirac equation predicts $g_e = 2$ at tree level.
The leading one-loop correction to the electron--photon vertex, illustrated in FIG.~\ref{fig:sm:vertex_corr}, shifts this value.
The resulting \emph{anomalous magnetic moment},
\begin{equation}
    a_e \equiv \frac{g_e - 2}{2} = \frac{\alpha_\mathrm{em}}{2\pi} + \mathcal{O}(\alpha_\mathrm{em}^2),
    \label{eq:sm:g2}
\end{equation}
extended to five-loop order in QED and combined with hadronic and electroweak corrections, agrees with experiment to 12 significant figures~\cite{ParticleDataGroup:2024cfk}---the most precisely-tested prediction in all of physics, and a striking demonstration of the efficacy of perturbative QFT.
The electromagnetic coupling itself runs with energy: fermion loops screen the electric charge at long distances, causing $\alpha_\mathrm{em}$ to grow from $1/137$ at low momentum transfer ($Q \to 0$) to approximately $1/128$ at $Q = m_Z$, in line with a positive beta function.

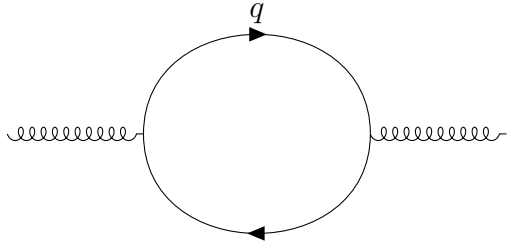
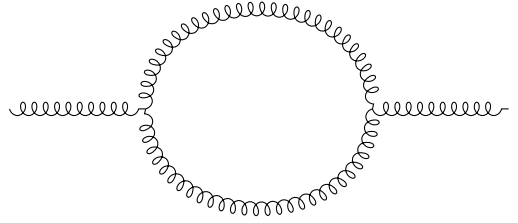
\begin{figure}[htbp]
  \centering
  \begin{subfigure}[b]{0.44\textwidth}
    \centering
    \begin{tikzpicture}
      \begin{feynman}
        \vertex (a);
        \vertex [right=1.8cm of a] (b);
        \vertex [right=3.0cm of b] (c);
        \vertex [right=1.8cm of c] (d);
        \diagram*{
          (a) -- [gluon] (b),
          (b) -- [fermion, half left, looseness=1.5, edge label=$q$] (c),
          (c) -- [fermion, half left, looseness=1.5] (b),
          (c) -- [gluon] (d),
        };
      \end{feynman}
    \end{tikzpicture}
    \caption{Quark loop: colour screening}
    \label{fig:sm:vac_pol_quark}
  \end{subfigure}
  \hfill
  \begin{subfigure}[b]{0.44\textwidth}
    \centering
    \begin{tikzpicture}
      \begin{feynman}
        \vertex (a);
        \vertex [right=1.8cm of a] (b);
        \vertex [right=3.0cm of b] (c);
        \vertex [right=1.8cm of c] (d);
        \diagram*{
          (a) -- [gluon] (b),
          (b) -- [gluon, half left, looseness=1.5] (c),
          (c) -- [gluon, half left, looseness=1.5] (b),
          (c) -- [gluon] (d),
        };
      \end{feynman}
    \end{tikzpicture}
    \caption{Gluon loop: anti-screening}
    \label{fig:sm:vac_pol_gluon}
  \end{subfigure}
  \caption{One-loop corrections to the gluon propagator.
    (a) Virtual quark--antiquark pairs screen the colour charge.
    (b) Gluon self-energy loops, unique to non-Abelian QCD, have the opposite sign and dominate for $n_f \leq 16$, driving the RG flow of $\alpha_s$ towards asymptotic freedom.}
  \label{fig:sm:vac_pol}
\end{figure}

In QCD the situation is different.
Each factor of the gauge group in Eq.~\eqref{eq:sm_gauge_group} comes with its own coupling: $g_s$ for $SU(3)_{\mathrm{C}}$, $g$ for $SU(2)_{\mathrm{L}}$ and $g'$ for $U(1)_{\mathrm{Y}}$.
For the strong coupling $\alpha_s = g_s^2/(4\pi)$, the one-loop beta function of Eq.~\eqref{eq:sm:beta} evaluates to
\begin{equation}
    \beta(\alpha_s) = -\frac{\alpha_s^2}{2\pi}\!\left(\frac{11}{3}N_c - \frac{2}{3}n_f\right),
    \label{eq:sm:qcd_beta}
\end{equation}
where $N_c = 3$ is the number of colours and $n_f$ the number of active quark flavours.
Two contributions compete: the $\frac{2}{3}n_f$ term from fermion loops screens the colour charge, analogously to QED, whilst the $\frac{11}{3}N_c$ term from gluon self-interactions acts in the opposite direction.
This second contribution is a direct consequence of the non-Abelian $SU(3)$ algebra: unlike photons, gluons carry colour charge and interact amongst themselves, generating additional loop diagrams with no QED counterpart.
For $n_f \leq 6$, the gluon term dominates, $\beta < 0$, and $\alpha_s$ \emph{decreases} with increasing energy---\emph{asymptotic freedom}~\cite{Gross:1973id,Politzer:1973fx}.
FIG.~\ref{fig:sm:vac_pol} shows representative one-loop contributions to this running.

Solving Eq.~\eqref{eq:sm:beta} to one loop gives
\begin{equation}
    \alpha_s(Q^2) = \frac{2\pi}{\!\left(\dfrac{11N_c}{3} - \dfrac{2n_f}{3}\right)\ln\!\left(Q/\Lambda_{\mathrm{QCD}}\right)},
    \label{eq:sm:alphas}
\end{equation}
where $\Lambda_{\mathrm{QCD}} \approx 200\,\mathrm{MeV}$ is the \emph{QCD scale}, an intrinsic energy scale arising dynamically from the theory itself.
At $Q \gg \Lambda_{\mathrm{QCD}}$, $\alpha_s$ is small and quarks and gluons---collectively called \emph{partons}---interact weakly; perturbation theory is reliable and hard processes can be computed order by order.
As $Q \to \Lambda_{\mathrm{QCD}}$, $\alpha_s$ grows large and the perturbative expansion breaks down entirely.
This strong-coupling régime is responsible for \emph{confinement}: the energy stored in a colour flux tube stretching between separating quarks grows linearly with distance.
It is therefore always energetically-favourable to create a new quark--antiquark pair from the vacuum rather than pull free colour charges apart.
Partons are therefore never observed in isolation; they are always bound inside colour-neutral \emph{hadrons}, and the process by which energetic partons assemble into these bound states is called \emph{hadronisation}.

These two faces of QCD---asymptotic freedom at high $Q$ and confinement at low $Q$---are directly relevant to collider experiments.
Hard scatterings at the LHC occur at $Q \gg \Lambda_{\mathrm{QCD}}$, where $\alpha_s$ is small and perturbative calculations of partonic production cross-sections are reliable.
After the collision, the energetic final-state partons lose energy and hadronise into colour-neutral bound states---mesons and baryons---which are observed experimentally as collimated hadronic jets.

\section{Electroweak Symmetry Breaking}
\label{sec:sm:higgs}

The gauge structure of the electroweak sector, $SU(2)_{\mathrm{L}} \times U(1)_{\mathrm{Y}}$,
gives rise to four massless gauge fields before symmetry breaking:
three weak isospin fields $W^{a}_{\mu}$ ($a = 1, 2, 3$) associated with $SU(2)_{\mathrm{L}}$,
and one hypercharge field $B_{\mu}$ associated with $U(1)_{\mathrm{Y}}$.
These pre-symmetry-breaking fields carry no mass, in agreement with the gauge invariance
requirement but in apparent conflict with the observed short-range nature of the weak force.
A massive mediator of mass $m$ produces a force with the Yukawa potential,
\begin{equation}
  V(r) \sim \frac{e^{-mr}}{r},
  \label{eq:sm:yukawa_potential}
\end{equation}
which falls off exponentially beyond the range $r \sim 1/m$, in contrast to the infinite-range Coulomb potential of the massless photon.
The observed $W^\pm$ and $Z$ masses of order $80$--$91\,\GeV$ thus imply a force range of ${\sim}10^{-18}\,\mathrm{m}$, consistent with experiment. Yet the gauge invariance of Eq.~\eqref{eq:ym_lagrangian} forbids adding an explicit mass term $m^2 W^{a}_{\mu}W^{\mu,a}$, which would transform inhomogeneously under the local symmetry and break it.
The Higgs mechanism resolves this tension through SSB.

\subsection{The Scalar Potential and Spontaneous Symmetry Breaking}

\begin{figure}[htbp]
  \centering
  \includegraphics[width=0.55\textwidth]{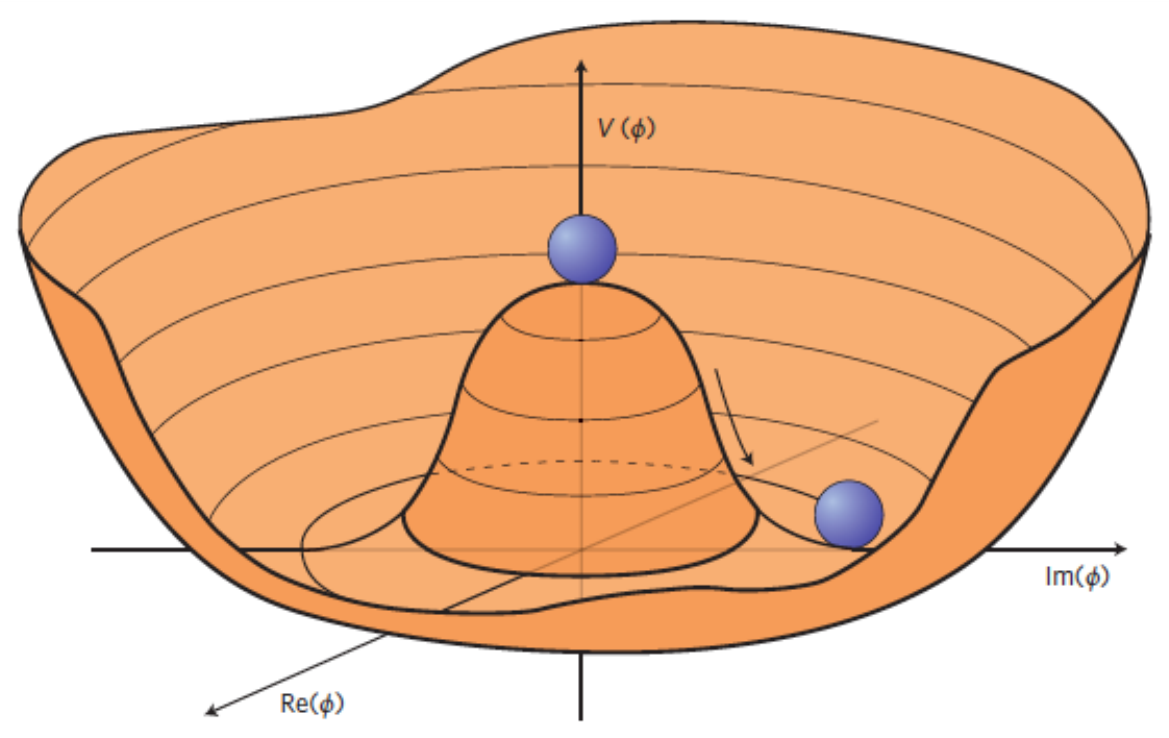}
  \caption{The Mexican hat Higgs potential $V(\phi) = \mu^2|\phi|^2 + \lambda|\phi|^4$
    with $\mu^2 < 0$ and $\lambda > 0$, as a function of the real and imaginary parts of
    a complex scalar field $\phi$~\cite{Lorentey:2012}.
    The hilltop at $\phi = 0$ is an unstable symmetric point; the continuous circle of global
    minima at $|\phi| = v/\sqrt{2}$ constitutes the vacuum.
    Spontaneous symmetry breaking occurs when the field settles into any one point on this
    circle, breaking the $U(1)$ rotational symmetry of the potential while remaining
    degenerate across all choices of vacuum.}
  \label{fig:sm:mexican_hat}
\end{figure}

The Higgs doublet $\Phi$ is a complex $SU(2)_{\mathrm{L}}$ doublet with hypercharge $Y = +1/2$,
\begin{equation}
  \Phi = \begin{pmatrix} \phi^{+} \\ \phi^{0} \end{pmatrix},
  \label{eq:sm:higgs_doublet}
\end{equation}
governed by the potential
\begin{equation}
  V(\Phi) = \mu^{2}\,|\Phi|^{2} + \lambda\,|\Phi|^{4},
  \label{eq:sm:higgs_potential}
\end{equation}
where $|\Phi|^{2} \equiv \Phi^{\dagger}\Phi$; the first term, $\mu^{2}|\Phi|^{2}$, is the \emph{mass term}, whose sign controls the shape of the potential, and the second, $\lambda|\Phi|^{4}$, is the \emph{quartic self-coupling}, with $\lambda > 0$ required for the potential to be bounded from below.
When $\mu^{2} > 0$, the unique minimum is at $\Phi = 0$, and the symmetry is preserved.
The physically-relevant case is $\mu^{2} < 0$ with $\lambda > 0$:
the potential then has the characteristic \emph{Mexican hat} shape illustrated in FIG.~\ref{fig:sm:mexican_hat}, with a circle of degenerate minima at
\begin{equation}
  |\Phi|^{2} = \frac{-\mu^{2}}{2\lambda} \equiv \frac{v^{2}}{2},
  \label{eq:sm:vev}
\end{equation}
where $v \approx 246\,\GeV$ is the electroweak vacuum expectation value (VEV).

The field acquires a VEV corresponding to one point on this degenerate circle.
Choosing the electrically neutral direction,
\begin{equation}
  \langle \Phi \rangle_{0} = \frac{1}{\sqrt{2}} \begin{pmatrix} 0 \\ v \end{pmatrix},
  \label{eq:sm:higgs_vev}
\end{equation}
breaks the electroweak symmetry spontaneously:
$SU(2)_{\mathrm{L}} \times U(1)_{\mathrm{Y}} \to U(1)_{\mathrm{EM}}$.
Three of the four degrees of freedom in $\Phi$ become the longitudinal polarisations
(the Goldstone bosons) of the massive gauge bosons; the fourth becomes the physical Higgs boson $h$, a spin-0 scalar that, like the gauge bosons, obeys Bose-Einstein statistics.

\subsection{Mass Generation for Gauge Bosons}

The kinetic term of the Higgs doublet,
\begin{equation}
  \mathcal{L}_{\mathrm{kin}} = (D^{\mu}\Phi)^{\dagger}(D_{\mu}\Phi),
  \qquad
  D_{\mu} = \partial_{\mu} + ig\,\frac{\tau^{a}}{2}W^{a}_{\mu} + ig'\frac{Y}{2}B_{\mu},
  \label{eq:sm:higgs_kinetic}
\end{equation}
generates mass terms when evaluated at the VEV.
Recall, here $g$ and $g'$ are the $SU(2)_{\mathrm{L}}$ and $U(1)_{\mathrm{Y}}$ gauge couplings,
and $\tau^a$ are the Pauli matrices given in Eq.~\eqref{eq:sm:pauli}.
The physical mass eigenstates are the charged $W$ bosons, obtained from the pre-SSB fields by
\begin{equation}
  W^{\pm}_{\mu} = \frac{1}{\sqrt{2}}\bigl(W^{1}_{\mu} \mp i\,W^{2}_{\mu}\bigr),
  \label{eq:sm:w_fields}
\end{equation}
and the neutral $Z$ boson and photon $A$, obtained by rotating $W^{3}_{\mu}$ and $B_{\mu}$
through the Weinberg mixing angle $\theta_W$:
\begin{equation}
  \begin{pmatrix} Z_{\mu} \\ A_{\mu} \end{pmatrix}
  =
  \begin{pmatrix} \cos\theta_W & -\sin\theta_W \\ \sin\theta_W & \cos\theta_W \end{pmatrix}
  \begin{pmatrix} W^{3}_{\mu} \\ B_{\mu} \end{pmatrix},
  \label{eq:sm:ew_rotation}
\end{equation}
where $\cos\theta_W = g/\sqrt{g^2 + g^{\prime\,2}}$ and $\sin\theta_W = g'/\sqrt{g^2+g^{\prime\,2}}$.
Substituting the expression for the VEV from Eq.~\eqref{eq:sm:higgs_vev} into Eq.~\eqref{eq:sm:higgs_kinetic} yields the gauge boson masses:
\begin{equation}
  m_W = \frac{1}{2}\,g\,v,
  \qquad
  m_Z = \frac{1}{2}\sqrt{g^2 + g^{\prime\,2}}\;v = \frac{m_W}{\cos\theta_W},
  \qquad
  m_A = 0.
  \label{eq:sm:gauge_masses}
\end{equation}
The photon $A_{\mu}$ remains massless, confirming that the electromagnetic $U(1)_{\mathrm{EM}}$
symmetry is unbroken.

\subsection{Fermion Masses via Yukawa Couplings}

Explicit Dirac mass terms of the form $m\bar{\psi}_{\mathrm{L}}\psi_{\mathrm{R}}$ are
forbidden by the $SU(2)_{\mathrm{L}}$ gauge invariance of Eq.~\eqref{eq:ym_lagrangian}, since left-handed fermions transform as doublets
and right-handed fermions as singlets.
Instead, masses arise from gauge-invariant Yukawa interactions,
\begin{equation}
  \mathcal{L}_{\mathrm{Yukawa}} = -y_d \bar{Q}_{\mathrm{L}} \Phi\, d_{\mathrm{R}}
    - y_u \bar{Q}_{\mathrm{L}} \tilde{\Phi}\, u_{\mathrm{R}}
    - y_e \bar{L}_{\mathrm{L}} \Phi\, e_{\mathrm{R}} + \text{h.c.},
  \label{eq:sm:yukawa}
\end{equation}
where $Q_{\mathrm{L}} = (u_{\mathrm{L}},\, d_{\mathrm{L}})^{\top}$ and $L_{\mathrm{L}} = (\nu_{\mathrm{L}},\, e_{\mathrm{L}})^{\top}$ are the left-handed quark and lepton $SU(2)_{\mathrm{L}}$ doublets --- capitalised to indicate a two-component multiplet --- while $d_{\mathrm{R}}$, $u_{\mathrm{R}}$ and $e_{\mathrm{R}}$ are the corresponding right-handed singlets, which carry no $SU(2)_{\mathrm{L}}$ index and are written in lower case. The quantities $y_d$, $y_u$ and $y_e$ are the dimensionless Yukawa coupling constants for down-type quarks, up-type quarks and charged leptons, respectively. The conjugate doublet $\tilde{\Phi} = i\tau_{2}\Phi^{*}$, where $\tau_2$ is the second Pauli matrix of Eq.~\eqref{eq:sm:pauli}, carries hypercharge $Y = -1/2$ (opposite to $\Phi$), enabling a gauge-invariant coupling to the up-type singlet $u_{\mathrm{R}}$; $\Phi$ itself couples to the down-type and lepton terms.
After SSB, each term becomes $-(y_f v/\sqrt{2})\bar{\psi}_f\psi_f + \text{h.c.}$, yielding
fermion masses $m_f = y_f v / \sqrt{2}$ proportional to the Yukawa couplings.
The top quark, with $m_t \approx 173\,\GeV$, has $y_t \approx 1$; the electron, with
$m_e \approx 0.511\,\mathrm{MeV}$, has $y_e \approx 2 \times 10^{-6}$. 

Generalising Eq.~\eqref{eq:sm:yukawa} to all three generations promotes each Yukawa coupling $y_f$ to a complex $3 \times 3$ matrix $Y_f^{ij}$ in generation space ($i, j = 1, 2, 3$),
\begin{equation}
  \mathcal{L}_{\mathrm{Yukawa}} = -Y^{ij}_{d}\,\bar{Q}^{i}_{\mathrm{L}}\,\Phi\,d^{j}_{\mathrm{R}}
    - Y^{ij}_{u}\,\bar{Q}^{i}_{\mathrm{L}}\,\tilde{\Phi}\,u^{j}_{\mathrm{R}}
    - Y^{ij}_{e}\,\bar{L}^{i}_{\mathrm{L}}\,\Phi\,e^{j}_{\mathrm{R}} + \mathrm{h.c.}
  \label{eq:sm:yukawa_gen}
\end{equation}
After SSB, these give rise to quark mass matrices $M_{u,d} = Y_{u,d}\,v/\sqrt{2}$, which are in general non-diagonal and are independently diagonalised by bi-unitary transformations:
\begin{equation}
  V_{fL}^{\dagger}\,M_{f}\,V_{fR} = M_{f}^{\mathrm{diag}} \equiv \mathrm{diag}(m_{f_1},\,m_{f_2},\,m_{f_3}),
  \qquad f \in \{u,\,d\}.
  \label{eq:sm:mass_diag}
\end{equation}
The physical consequence of this mismatch appears in the charged-current weak interaction, where the $W^{\pm}$ boson couples left-handed up-type quarks to left-handed down-type quarks. Rotating to the mass eigenbasis introduces the combination $V_{uL}^{\dagger}V_{dL}$, which cannot be absorbed into field redefinitions. This is the Cabibbo--Kobayashi--Maskawa (CKM) matrix,
\begin{equation}
  V_{\mathrm{CKM}} = V_{uL}^{\dagger}\,V_{dL},
  \label{eq:sm:ckm_def}
\end{equation}
so that the charged-current Lagrangian becomes
\begin{equation}
  \mathcal{L}_{W^{+}} \supset \frac{g}{\sqrt{2}}\,\bar{u}^{i}_{L}\,\gamma^{\mu}\,V^{ij}_{\mathrm{CKM}}\,d^{j}_{L}\,W^{+}_{\mu} + \mathrm{h.c.}
  \label{eq:sm:cc_ckm}
\end{equation}
Being unitary, $V_{\mathrm{CKM}}$ is parameterised by three real mixing angles ($\theta_{12}$, $\theta_{13}$, $\theta_{23}$) and one complex phase $\delta$. 
This phase is the source of CP violation in the quark sector of the SM. 
The experimentally measured magnitudes of the matrix elements are~\cite{ParticleDataGroup:2024cfk}:
\begin{equation}
  |V_{\mathrm{CKM}}| \approx
  \begin{pmatrix}
    |V_{ud}| & |V_{us}| & |V_{ub}| \\
    |V_{cd}| & |V_{cs}| & |V_{cb}| \\
    |V_{td}| & |V_{ts}| & |V_{tb}|
  \end{pmatrix}
  \approx
  \begin{pmatrix}
    0.974 & 0.225 & 0.004 \\
    0.225 & 0.973 & 0.041 \\
    0.009 & 0.040 & 0.999
  \end{pmatrix},
  \label{eq:sm:ckm_values}
\end{equation}
revealing a strong hierarchical structure: transitions within a generation are strongly preferred over cross-generation ones.
This hierarchy is made systematic by the Wolfenstein parameterisation~\cite{Wolfenstein:1983yz}, an expansion in powers of the Cabibbo angle $\lambda \approx 0.225$:
\begin{equation}
  V_{\mathrm{CKM}} =
  \begin{pmatrix}
    1 - \tfrac{\lambda^{2}}{2} & \lambda & A\lambda^{3}(\rho - i\eta) \\[4pt]
    -\lambda & 1 - \tfrac{\lambda^{2}}{2} & A\lambda^{2} \\[4pt]
    A\lambda^{3}(1 - \rho - i\eta) & -A\lambda^{2} & 1
  \end{pmatrix}
  + \mathcal{O}(\lambda^{4}),
  \label{eq:sm:wolfenstein}
\end{equation}
where $\lambda$ and $A$ are real parameters, whilst $\rho$ and $\eta$ jointly encode the CP-violating phase $\delta$ through the combination $\rho - i\eta \propto e^{-i\delta}$; a non-zero $\eta$ is therefore equivalent to $\delta \neq 0, \pi$. 
Measured values are $A \approx 0.82$, $\rho \approx 0.15$ and $\eta \approx 0.35$~\cite{ParticleDataGroup:2024cfk}.
The unitarity of $V_{\mathrm{CKM}}$ imposes relations amongst its elements, visualised as \emph{unitarity triangles} in the complex $(\bar{\rho},\,\bar{\eta})$ plane.

\section{Synthesis: The Standard Model Lagrangian}
\label{sec:sm:lagrangian}

The individual sectors developed throughout this chapter combine into a single, compact Lagrangian density that defines the SM,
\begin{equation}
\begin{aligned}
  \mathcal{L}_{\mathrm{SM}} = {}
    &-\underbrace{\tfrac{1}{4}\,F^{a}_{\mu\nu} F^{\mu\nu,a}}_{\text{gauge}}
     + \underbrace{i\,\bar{\psi}\,\slashed{D}\,\psi}_{\text{fermion}}
     + \underbrace{(D_{\mu}\Phi)^{\dagger}(D^{\mu}\Phi)}_{\text{Higgs kinetic}} \\
    &- \underbrace{V(\Phi)}_{\text{Higgs potential}}
     - \underbrace{\bigl(\bar{\psi}_{\mathrm{L}}\, Y\, \psi_{\mathrm{R}}\, \Phi + \mathrm{h.c.}\bigr)}_{\text{Yukawa}}.
\end{aligned}
  \label{eq:sm:full_lagrangian}
\end{equation}
Reading the terms in turn: the gauge term is the Yang--Mills Lagrangian of Eq.~\eqref{eq:ym_lagrangian}, summed over the generators of $SU(3)_{\mathrm{C}} \times SU(2)_{\mathrm{L}} \times U(1)_{\mathrm{Y}}$; the fermion term, Eq.~\eqref{eq:sm:fermion_kinetic}, carries the matter kinetic energy and, through the covariant derivative, every gauge coupling; the two Higgs terms are the scalar kinetic term of Eq.~\eqref{eq:sm:higgs_kinetic}, which sources the weak-boson masses, and the potential of Eq.~\eqref{eq:sm:higgs_potential}, which drives electroweak symmetry breaking; and the Yukawa term, Eq.~\eqref{eq:sm:yukawa_gen}, generates the fermion masses and CKM mixing. Compact as it is, Eq.~\eqref{eq:sm:full_lagrangian} carries 19 free parameters: three gauge couplings, two Higgs-potential parameters, nine charged-fermion masses, four CKM parameters and the strong-CP phase. This number grows to 28 once neutrino masses and mixing are included. None of these values is predicted by theory and hence, all must be measured. Even once they are fixed, the SM remains incomplete. It accommodates neither gravity nor a dark-matter candidate, and it gives neutrinos no mass without further extension. Such shortcomings necessitate physics beyond the SM, some of which is elaborated in the following chapter.

\clearpage
\clearpage\chapter{Beyond the Standard Model}
\label{chp:bsm}

The SM, introduced in Chapter~\ref{chp:sm}, is one of the most precisely-tested theories in science.
Yet, it cannot be a complete description of Nature: cosmological observations, precision measurements and theoretical considerations all point to physics beyond the Standard Model (BSM).
A large number of BSM theories have been proposed to address these shortcomings, spanning a wide range of energy scales and experimental signatures.
This chapter introduces only a small selection, chosen specifically because they motivate the experimental searches described in the chapters that follow; dark matter and neutrino oscillations are noted in the next two paragraphs as two of the most consequential BSM phenomena, though neither is directly targeted in this thesis.

Among the most observationally-robust pieces of evidence for BSM physics hint at the existence of
\emph{dark matter} (DM): a form of matter that gravitates but neither emits nor absorbs
light.
Galactic rotation curves remain flat far beyond the luminous disc~\cite{Rubin:1980zd},
the Bullet Cluster (1E~0657$-$558) reveals a spatial offset between the X-ray gas and
the gravitational mass inferred from weak lensing~\cite{Clowe:2006eq} and Planck CMB
measurements~\cite{Planck:2018vyg} imply that DM constitutes $\sim 27\%$ of the cosmic energy budget while
ordinary baryons account for only $\sim 5\%$.
The SM contains no stable, neutral, non-baryonic particle that can account for these observations.

Neutrino oscillation experiments have established that neutrinos are massive~\cite{Super-Kamiokande:1998kpq,SNO:2002tuh}, contradicting the SM prediction discussed in Section~\ref{sec:sm:higgs}.
The SM contains no right-handed neutrino singlet, so no gauge-invariant mass term exists.
Any extension must introduce a lepton mixing matrix and a mass-generation mechanism, neither of which the SM provides.
Neither DM nor neutrino masses are directly targeted by the searches in this thesis, but both are unambiguous proof that the SM is incomplete.

The sections that follow cover three BSM scenarios directly connected to the experimental searches in this thesis.
Section~\ref{sec:bsm:hierarchy} discusses the hierarchy problem and extra-dimensional solutions, which motivate the searches for quantum black holes in Chapter~\ref{chp:qbh} and Clockwork/Linear Dilaton graviton spectra in Chapter~\ref{chp:clockwork}.
Section~\ref{sec:bsm:flavour} covers deviations from lepton flavour universality in rare $B$-meson decays, which motivate the search for a $Z'$ boson in dilepton plus $b$-jets final states in Chapter~\ref{chp:zprime}.
The models discussed serve as benchmarks rather than exclusive targets: the primary aim of each search is to probe a specific final-state topology, and the sensitivity may extend to virtually any BSM predicting those signatures.

\section{The Hierarchy Problem}
\label{sec:bsm:hierarchy}

The discovery of the Higgs boson at the LHC in 2012~\cite{ATLAS:2012yve} confirmed the final missing piece of the SM. 
Its mass was measured to be $m_H = 125.09 \pm 0.24\,\GeV$~\cite{ATLAS:2015yey}. 
While representing a triumph of the theory, this measurement immediately confronts a deep theoretical puzzle.
The SM has no natural cutoff and is expected to remain valid up to the Planck scale $M_{\mathrm{Pl}} \equiv \sqrt{\hbar \mathrm{c} / \mathrm{G}} \sim 10^{19}\,\GeV$ --- the unique energy scale built from Newton's constant $G$, at which the gravitational interaction becomes as strong as the gauge interactions.
Put differently, the vast hierarchy $m_H \ll M_{\mathrm{Pl}}$ is equivalent to the observation that gravity is extraordinarily weak compared to the other fundamental forces at accessible energies.
Quantum corrections to $m_H$ from virtual particles are proportional to the highest energy scale in the theory, so one would naturally expect $m_H \sim M_{\mathrm{Pl}}$.
Yet the measured value is $m_H \approx 125\,\GeV$ --- 13 orders of magnitude smaller. Why?

\subsection{Radiative Corrections and Fine-Tuning}

Unlike fermion masses which are protected from large additive renormalisation by chiral symmetry, the mass of a fundamental scalar receives corrections that are
\emph{quadratically} sensitive to the UV cutoff $\Lambda$ of the theory.
At one loop, the dominant contributions to $m_H^2$ arise from the top quark,
the gauge bosons and the Higgs self-coupling~\cite{Koren:2020pio}:
\begin{equation}
  \delta m_H^2 \approx \frac{\Lambda^2}{16\pi^2}
    \Bigl[
      -6\lambda_t^2
      + \tfrac{3}{4}(3g^2 + g^{\prime\,2})
      + 3\lambda
    \Bigr] + \cdots,
  \label{eq:bsm:higgs_correction}
\end{equation}
where $\lambda_t$, $g$, $g'$ and $\lambda$ are the top Yukawa, $SU(2)_L$, $U(1)_Y$ and Higgs quartic couplings, respectively.
If the SM is valid up to the Planck scale, $\Lambda \sim M_\text{Pl} \approx 1.22 \times 10^{19}\,\GeV$, then $\delta m_H^2 \sim \mathcal{O}(M_\text{Pl}^2) \approx 10^{34}\,m_H^2$. 
Obtaining the observed Higgs mass requires the bare parameter $m_{H,0}^2$ and the radiative corrections to cancel to one part in $\sim 10^{34}$ --- a degree of fine-tuning with no natural explanation within the SM. 
This is the \emph{hierarchy problem}. 

\begin{figure}[htbp]
  \centering
  \includegraphics[width=0.95\textwidth]{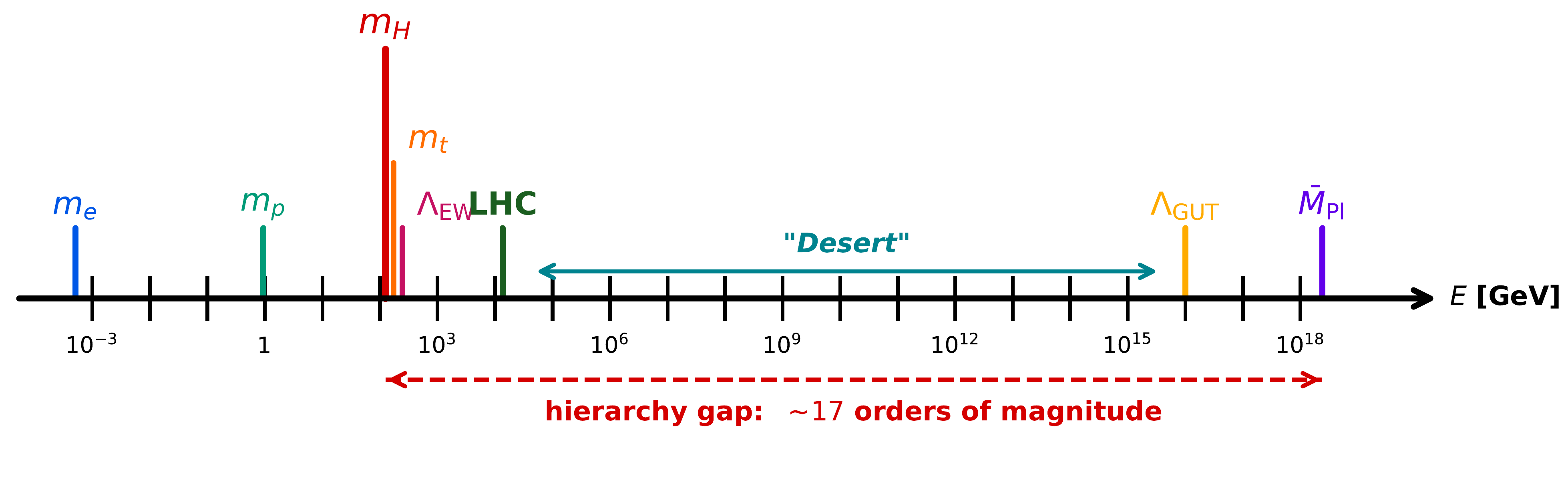}
  \caption{Logarithmic scale of particle physics energies from the electron mass to
    the Planck scale.
    The Higgs boson mass $m_H \approx 125\,\GeV$ and top-quark mass
    $m_t \approx 173\,\GeV$ sit 17 orders of magnitude below the Planck
    scale $M_{\mathrm{Pl}} \sim 10^{19}\,\GeV$.
    Because quantum corrections to $m_H^2$ are proportional to $\Lambda^2$, any UV
    completion of the SM at $\Lambda \sim M_{\mathrm{Pl}}$ requires an
    extraordinary cancellation between the bare mass and radiative corrections.
    The largely unexplored range between the TeV scale and the GUT scale is
    commonly called the ``Desert''.}
  \label{fig:bsm:scales}
\end{figure}

The severity of the problem is made vivid by the energy-scale diagram in FIG.~\ref{fig:bsm:scales}, which illustrates the enormous separation between the
electroweak and Planck scales. 
The Higgs mass sits at $\sim 10^2\,\GeV$, while the Planck scale $M_{\mathrm{Pl}} \sim 10^{19}\,\GeV$ lies 17 orders of magnitude higher.
The top quark, with mass $m_t \approx 173\,\GeV$, is also shown: being the heaviest SM fermion, its Yukawa coupling $\lambda_t \approx 1$ drives the dominant loop correction in Eq.~\eqref{eq:bsm:higgs_correction}, and sits close to the Higgs mass on the logarithmic scale despite being cosmically distant from $M_{\mathrm{Pl}}$. 

\subsection{Extra-Dimensional Solutions}
\label{sec:bsm:extradim}

A geometrical class of solutions to the hierarchy problem introduces additional spatial dimensions, compactified at some characteristic scale, in which gravity is permitted to propagate whilst SM fields remain confined to the familiar four-dimensional spacetime.
The apparent weakness of gravity --- the Planck hierarchy itself --- is then a consequence of the geometry rather than a fundamental trait of couplings.
Many such extra-dimensional models exist in the literature, differing in the number, size and geometry of the extra dimensions (EDs); the two discussed here --- ADD and RS --- have become the standard benchmarks for collider searches.
Both were proposed before the LHC era, both reduce the hierarchy problem to a small number of experimentally-accessible parameters and both predict qualitatively distinct, calculable signatures at TeV-scale energies.
These properties have made them the targets of a sustained, broad experimental programme: unlike more elaborate constructions, their simplicity means that searches in many different final states can be interpreted within a single, well-defined parameter space.

\subsubsection*{The ADD Model: Large Flat Extra Dimensions}

Arkani-Hamed, Dimopoulos and Dvali (ADD)~\cite{Nima,Antoniadis:1998ig} proposed that $n$ additional spatial dimensions, each flat and compactified to a common radius $R$, are accessible to gravity but invisible to SM gauge interactions.
In this picture, SM fields are localised on a four-dimensional \emph{brane}, whilst gravitons propagate freely throughout the $(4+n)$-dimensional \emph{bulk}.
The bulk metric is a direct product of the four-dimensional spacetime metric with a flat $n$-torus,
\begin{equation}
  ds^2 = g_{\mu\nu}\,dx^\mu\,dx^\nu + \delta_{mn}\,dy^m\,dy^n\,,
  \label{eq:bsm:add_metric}
\end{equation}
where $y^m$ ($m = 1,\ldots,n$) are the periodic extra-dimensional coordinates, each with period $2\pi R$, and $\delta_{mn}$ reflects the absence of any warp factor.
Gauss' law in $4+n$ dimensions relates $M_{\mathrm{Pl}}$ to $M_D$, the fundamental Planck mass of the $(4+n)$-dimensional theory, through the compactification volume $R^n$:
\begin{equation}
  M_\text{Pl}^2 \sim M_D^{2+n}\, R^n\,.
  \label{eq:bsm:add}
\end{equation}
Here $M_{\mathrm{Pl}}$ is the effective gravitational mass scale governing Newton's law as measured by four-dimensional observers, while $M_D$ is the scale at which gravity becomes strong in the bulk.
For $M_D \sim \mathcal{O}(1\,\TeV)$, this relation can be satisfied with compactification
radii ranging from $R \sim 10^{-4}$~mm (for $n = 2$) down to $R \sim 10^{-12}$~mm (for $n = 6$), the latter inaccessible to direct gravitational measurements but well within the reach of collider energies.
The hierarchy $M_\text{Pl} \gg M_D$ is thus reinterpreted as a statement about the geometric volume of the extra dimensions, not an unexplained coupling ratio.
Compactifying the extra dimensions forces graviton momentum along the compact directions to be quantised, generating a discrete tower of massive four-dimensional spin-2 states called Kaluza--Klein (KK) gravitons, with masses set by the compactification scale $1/R$.
The experimental appeal of ADD lies in its minimal parameterisation and broad predictive power. With only $M_D$ and $n$ as free inputs, the model yields calculable cross sections for an entire family of signatures: real graviton emission into the bulk (appearing as large missing transverse energy in mono-jet or mono-photon events), virtual KK graviton exchange (distorting high-mass dilepton and diphoton spectra) and quantum black hole production near threshold $M_{\textrm{th}} \sim M_D$.
Each signature probes the same underlying parameters, so independent searches across many final states combine into stringent, complementary constraints on $M_D$.

\subsubsection*{The RS Model: A Single Warped Extra Dimension}

Randall and Sundrum (RS1)~\cite{Randall} offered a complementary solution with a single extra dimension of size $r_c$, but now with a curved (\emph{warped}) geometry.
Two three-dimensional branes --- the ``Planck brane'' and the ``TeV brane'' --- sit at the two endpoints of this dimension.
The key feature is an exponential warp factor in the bulk metric,
\begin{equation}
  ds^2 = e^{-2k|y|}\,g_{\mu\nu}\,dx^\mu\,dx^\nu - dy^2\,,
  \label{eq:bsm:rs_metric}
\end{equation}
where $y$ is the coordinate along the extra dimension ($0 \leq y \leq \pi r_c$) and $k$ sets the curvature scale.
All fundamental mass parameters are of order $M_\text{Pl}$ on the Planck brane, but physical masses observed on the TeV brane are exponentially suppressed by the warp factor:
\begin{equation}
  m_\text{EW} \sim M_\text{Pl}\, e^{-k\pi r_c}\,.
  \label{eq:bsm:rs}
\end{equation}
For $kr_c \approx 12$, the exponential factor $e^{-12\pi} \sim 10^{-16}$ generates the entire electroweak--Planck hierarchy from a modestly-tuned geometric parameter, with only one extra dimension of size $r_c \sim M_\text{Pl}^{-1}$.
The compactification of the extra dimension gives rise to a tower of massive KK graviton excitations; the lightest, $G^{(1)}_\text{KK}$, is a narrow spin-2 resonance whose mass and coupling to SM particles are determined entirely by $m_{G^{(1)}}$ and the dimensionless ratio $k/\bar{M}_\text{Pl}$.
Because a spin-2 particle couples to the energy--momentum tensor of the SM, it decays into virtually every pair of SM particles --- $\ell^+\ell^-$, $\gamma\gamma$, $WW$, $ZZ$, $hh$, $t\bar{t}$ --- producing a narrow, high-mass resonance in each channel.
This combination of theoretical economy (two free parameters), a sharp and unambiguous resonance signature and sensitivity across a wide range of final states has made RS1 one of the most extensively searched-for models at ATLAS and CMS.

\subsection{quantum black holes}
\label{sec:bsm:qbh}

Both ADD and RS models lower $M_D$ to the TeV scale, where a new gravitational
signature becomes accessible: the \emph{quantum black hole}
(QBH)~\cite{Meade:2007sz,Gingrich:2009hj}.
A black hole forms classically when the impact parameter $b$ between two colliding
partons falls within the $(4+n)$-dimensional Schwarzschild radius~\cite{Gingrich:2009hj}:
\begin{equation}
  r_{\textrm{S}}(M) = \frac{1}{\sqrt{\pi}\,M_D}
    \left(\frac{M}{M_D}\right)^{\!\frac{1}{1+n}}
    \left[\frac{8\,\Gamma\!\left(\frac{n+3}{2}\right)}{n+2}\right]^{\!\frac{1}{1+n}},
  \label{eq:bsm:rs_schwarz}
\end{equation}
where $M$ is the invariant mass of the colliding system and $n$ the number of extra
dimensions.
The threshold mass is set to $M_{\textrm{th}} = M_D$: a QBH forms as soon as the
centre-of-mass energy reaches the fundamental Planck scale.
Near this threshold, quantum effects dominate over classical Hawking evaporation,
producing an object whose properties are governed by quantum gravity rather than
semiclassical thermodynamics.

The QBH production cross section is approximated geometrically: any two-parton system
whose impact parameter satisfies $b \leq r_{\textrm{S}}$ is assumed to form a QBH with unit
probability~\cite{Gingrich:2006hm}.
The partonic cross section is therefore
\begin{equation}
  \hat{\sigma}_{ij}(\hat{s}) = \pi\,r_{\textrm{S}}^2\!\left(\sqrt{\hat{s}}\right)\,
    \Theta\!\left(\sqrt{\hat{s}} - M_{\textrm{th}}\right),
  \label{eq:bsm:qbh_partonic}
\end{equation}
where $\Theta$ is the Heaviside step function enforcing the threshold.
Substituting Eq.~\eqref{eq:bsm:rs_schwarz}, the partonic cross section scales as
$\hat{\sigma} \propto M_D^{-2}\,(M_{\textrm{th}}/M_D)^{2/(n+1)}$, making it sensitive to both
$M_D$ and $n$.
The hadronic cross section is obtained by convolving with parton distribution functions
(PDFs),
\begin{equation}
  \sigma(pp \to \text{QBH}+X) =
    \sum_{i,j}\int_{\tau_0}^{1} dx_1 \int_{\tau_0/x_1}^{1} dx_2\;
    f_i(x_1,\mu_F)\,f_j(x_2,\mu_F)\;
    \hat{\sigma}_{ij}(x_1 x_2 s),
  \label{eq:bsm:qbh_hadronic}
\end{equation}
where $\tau_0 = M_{\textrm{th}}^2/s$ and the factorisation scale is set to the inverse
gravitational radius, $\mu_F = 1/r_{\textrm{S}}$~\cite{Gingrich:2009da}.
The cross section falls steeply with increasing $M_{\textrm{th}}$ and is larger for ADD than RS
at equal $M_{\textrm{th}}$, reflecting the different $r_{\textrm{S}}$ dependence on $n$.

A notable consequence of Eq.~\eqref{eq:bsm:qbh_hadronic} is the steep dependence of the QBH
production cross section on the collider centre-of-mass energy $\sqrt{s}$.
The key is that the QBH partonic cross section is not flat: substituting
Eq.~\eqref{eq:bsm:rs_schwarz} into Eq.~\eqref{eq:bsm:qbh_partonic} gives
\begin{equation}
    \hat{\sigma}_{\mathrm{QBH}}(\hat{s}) \;\propto\; \hat{s}^{\,1/(n+1)},
    \label{eq:bsm:qbh_parton_scaling}
\end{equation}
so the partonic cross section \emph{rises} with parton energy.
The origin is geometric: a heavier QBH has a larger Schwarzschild radius and therefore a larger
black-disc cross section.
For perturbative processes --- SM or fixed-mass BSM resonances such as a $Z'$ boson --- the
partonic cross section does not grow this way.

The PDFs in Eq.~\eqref{eq:bsm:qbh_hadronic} suppress the integrand and dominate the outcome:
the $pp$ cross section still falls steeply with increasing QBH mass.
What changes with $\sqrt{s}$ is how strongly the PDFs suppress it.
Raising $\sqrt{s}$ lowers the integration lower bound $\tau_0 = M_{\rm th}^2/s$, shifting
the integral to smaller parton momentum fractions $x$ where PDFs are larger
(see Figure~\ref{fig:QBH_XS_Run2_vs_Run3} in Chapter~\ref{chp:qbh}).
The PDF suppression of the rising partonic cross section is therefore reduced.
For a fixed-mass BSM resonance, this PDF shift at $x = M^2/s$ is the \emph{only} handle on $\sqrt{s}$.
For QBH production, the combination of a rising partonic cross section \emph{and} a growing PDF
at smaller $x$ makes the $\sqrt{s}$ dependence considerably steeper.

\begin{figure}[htbp]
  \centering
  \includegraphics[width=0.45\textwidth]{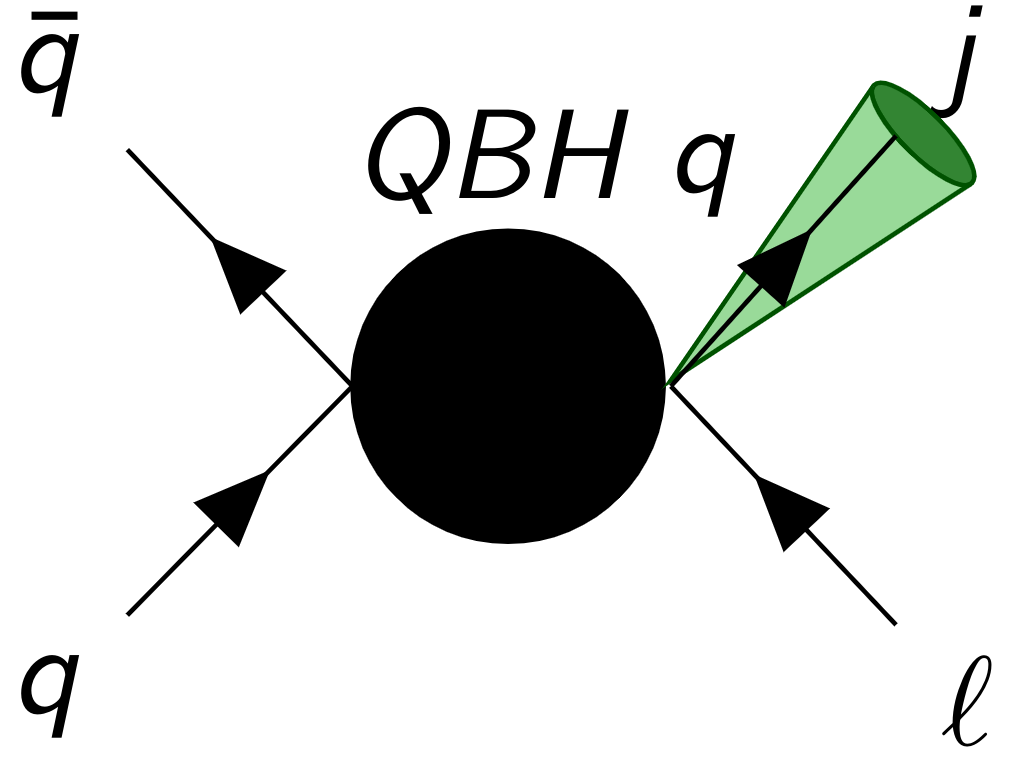}
  \caption{Schematic diagram of quantum black hole production and decay.
    Lacking a coherent perturbative theory for quantum black hole interactions, a conventional Feynman
    diagram cannot be drawn; the black disc represents the non-perturbative quantum black hole state
    formed near threshold $M_{\textrm{th}} \sim M_D$.
    The diagram illustrates one of several possible two-body final states --- here a
    lepton--quark pair --- to which quantum black holes can decay while conserving charge, colour
    and angular momentum~\cite{Gingrich:2009hj}.}
  \label{fig:bsm:qbh_feynman}
\end{figure}

The models postulate conservation of total angular momentum, colour and electric charge
in QBH production and decay, while global SM symmetries such as baryon and lepton number
need not be conserved~\cite{Gingrich:2009hj}.
Unlike their semiclassical counterparts, which thermalise into high-multiplicity final
states via Hawking radiation, QBHs at threshold are predicted to decay predominantly
into \emph{two-particle final states}.
The final-state composition depends on the quantum numbers of the initial partons:
different quark-pair configurations produce QBH states with distinct electric charges,
each yielding a specific two-body final state.
Many such final states are possible; Figure~\ref{fig:bsm:qbh_feynman} illustrates one
example --- a lepton and a jet --- which is also the channel investigated in this thesis,
as described in Chapter~\ref{chp:qbh}.

\subsection{The Clockwork/Linear Dilaton Model}
\label{sec:bsm:cwld}

Another possible consequence of compactifying extra dimensions is the appearance of
KK graviton towers: the higher-dimensional graviton wavefunction
decomposes into a discrete set of four-dimensional spin-2 modes whose mass spectrum
is set by the compactification geometry.
In the ADD model, the KK mode spacing $\Delta m \sim 1/R$ is sub-eV for $n \geq 2$,
so the tower is effectively continuous and its collective exchange modifies gravitational
scattering without producing resolvable resonances.
In RS1, by contrast, the exponential warp factor opens a mass gap of order
$ke^{-k\pi r_c} \sim \mathcal{O}(\TeV)$: the lightest KK graviton $G_\text{KK}^{(1)}$
is a distinct TeV-scale resonance that can be produced and searched for experimentally.
The KK graviton couples to the energy-momentum tensor and decays to many final states,
including quark pairs, $W$ and $Z$ bosons, Higgs bosons, leptons and photons; the
$G_\text{KK} \to \ell^+\ell^-$ and $\gamma\gamma$ channels are particularly well-suited
to narrow-width resonance searches owing to the clean experimental signatures and
precise mass reconstruction they afford.

\begin{figure}[h!]
  \centering
  \begin{subfigure}[b]{0.68\textwidth}
    \centering
    \includegraphics[width=\textwidth]{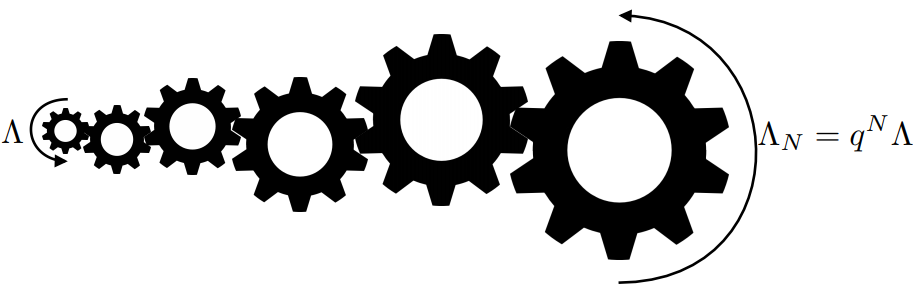}
    \caption{The Clockwork gear mechanism~\cite{Giudice:2016yja}: a chain of gears of
      sequentially-increasing size amplifies a small coupling $\Lambda$ at the input
      end to $\Lambda_N = q^N \Lambda$ at the output end.
      Each gear contributes only an $\mathcal{O}(1)$ ratio $q$, yet the collective
      effect exponentiates the Planck--electroweak hierarchy.}
    \label{fig:bsm:cw_gears}
  \end{subfigure}
  \hfill
  \begin{subfigure}[b]{0.24\textwidth}
    \centering
    \includegraphics[width=0.55\textwidth]{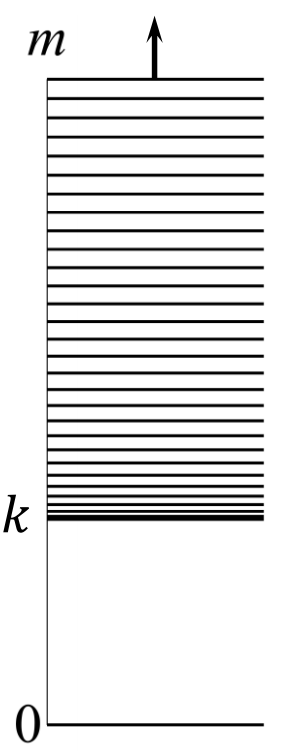}
    \caption{Schematic Clockwork/Linear Dilaton graviton mass spectrum: a mass gap $m_0 \sim k$ separates
      the zero mode from the quasi-equally-spaced tower above.}
    \label{fig:bsm:cw_tower}
  \end{subfigure}
  \caption{The Clockwork/Linear Dilaton model.
    (left) The gear-chain mechanism that generates the large hierarchy from
    $\mathcal{O}(1)$ input parameters.
    (right) The resulting graviton mass spectrum, with a mass gap $k$ and a
    dense tower of equally-spaced resonances whose collective contribution to
    cross sections produces a broad, chirp-like excess.}
  \label{fig:bsm:cw_combined}
\end{figure}

The Clockwork/Linear Dilaton (CW/LD) model~\cite{Giudice:2016yja,Giudice:2017fmj} provides a geometrically-distinct alternative. 
In the five-dimensional Linear Dilaton (5D LD) geometry, the bulk is described by a dilaton field with a linear profile along the extra dimension. 
A key feature of the CW/LD framework is its mechanism for generating large hierarchies
from modest input parameters: a chain of $N$ gears with gear ratio $q > 1$
exponentiates a small coupling $\Lambda$ at one end into a large scale
$\Lambda_N = q^N \Lambda$ at the other, reproducing the Planck--weak hierarchy with
natural $\mathcal{O}(1)$ parameters at every link.
Both the gear-chain mechanism and the resulting KK graviton spectrum are illustrated in
FIG.~\ref{fig:bsm:cw_combined}.

The model is characterised by two parameters: $k$, a mass scale that sets both the
onset of the KK spectrum and the spacing between modes, and $M_5$, the five-dimensional
reduced Planck mass that sets the overall coupling of KK gravitons to SM fields.
The KK graviton mass spectrum takes the form
\begin{equation}
  m_n^2 \approx k^2 + \frac{n^2}{R^2}, \quad n = 1, 2, \ldots,
  \label{eq:bsm:cw_spectrum}
\end{equation}
where $R$ is the size of the extra dimension~\cite{Giudice:2017fmj}.
This gives a mass gap $m_0 \sim k$ separating the massless zero mode from the first KK excitation, followed by a tower of modes with
quasi-equal spacing $\Delta m \sim k/(\pi R)$ that decreases for higher modes.
Unlike RS1, where only the lightest KK mode is phenomenologically relevant, the
entire CW/LD tower contributes collectively to the observable cross section.
Since each KK mode couples to the SM energy--momentum tensor with equal strength
$\sim 1/M_5^{3/2}$, the sum over all modes produces a cross section that scales as
\begin{equation}
  \sigma \propto \frac{1}{M_5^3}\,,
  \label{eq:bsm:cw_xsec}
\end{equation}
making $M_5$ the primary sensitivity parameter of any CW/LD search~\cite{Giudice:2016yja}.
The collective tower contribution produces a distinctive broad excess in the invariant-mass spectrum, whose quasi-periodic structure --- a chirp-like
signal of growing amplitude and decreasing fractional mode spacing --- constitutes
the experimental signature.
A search for this signal in the dielectron and diphoton final states with the ATLAS detector is described in Chapter~\ref{chp:clockwork}.

\section{Persisting Flavour Anomalies}
\label{sec:bsm:flavour}

A class of hints for BSM physics arises in the rare decays of $B$-mesons, directly motivating the search presented in Chapter~\ref{chp:zprime}. 
As discussed in Section~\ref{sec:sm:fermions}, the SM predicts LFU: the $W^\pm$, $Z$ and $\gamma$ bosons couple with identical strength to all three charged-lepton generations. 
The SM further requires that flavour-changing neutral currents (FCNCs) are absent at tree level. 
Persistent tensions in $B$-meson observables with both of these predictions have motivated an extensive experimental and theoretical programme. 

\subsection{Rare \texorpdfstring{$b \to s\ell^+\ell^-$}{b to sll} Transitions}

Processes of the form $b \to s\ell^+\ell^-$ are FCNC transitions that cannot occur at tree level in the SM.
At any neutral-current vertex ($\gamma$, $Z$ or $g$), the CKM rotation matrices cancel as $V_\text{CKM}^\dagger V_\text{CKM} = \mathbf{1}$, leaving the coupling strictly diagonal in flavour space.
Flavour-changing transitions between quark generations are therefore only possible through charged-current ($W^\pm$) loops.
These processes receive three distinct suppressions:
\begin{itemize}
    \item \textbf{CKM factors:} the $b \to s$ transition requires two $W^\pm$ vertices carrying off-diagonal CKM elements, contributing a factor $|V_{tb}V_{ts}^*| \sim \lambda^2 \sim \mathcal{O}(10^{-2})$ (using the Wolfenstein parameterisation of Eq.~\eqref{eq:sm:wolfenstein});
    \item \textbf{Loop factor:} each loop integration contributes a factor $g^2/(16\pi^2) \sim \mathcal{O}(10^{-3})$ to the amplitude relative to a tree-level process;
    \item \textbf{GIM mechanism:} CKM unitarity requires $\sum_i V_{ti}V_{si}^* = 0$, so the $u$, $c$ and $t$ contributions in the loop would cancel exactly for degenerate quark masses --- the Glashow--Iliopoulos--Maiani (GIM) cancellation.
    The large top mass breaks this degeneracy; the residual amplitude is $\propto (m_t^2 - m_{u,c}^2)/m_W^2$.
    For $b \to s\ell^+\ell^-$, the top quark dominates and GIM suppression is mild.
    For lighter-quark transitions such as $s \to d$, where the top contribution is also CKM-suppressed, GIM is the dominant additional suppression.
\end{itemize}

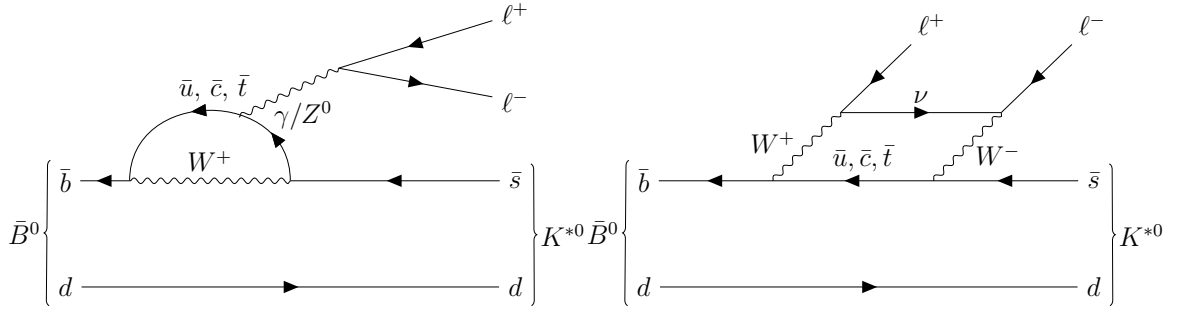
\begin{figure}[h!]
	\centering
	\scalebox{0.85}{\begin{tikzpicture}[baseline=(current bounding box.center)]
	\begin{feynman}
	\vertex (a1) {\(\bar b\)};
	
	\vertex[right=2.5cm of a1] (a2);
	\vertex[right=1.0cm of a1] (l1);
	\vertex[right=2.5cm of l1] (l2);
	\vertex[right=1.7cm of l1](li);
	\vertex[above=0.99cm of li] (l3);
	\vertex[right=7.0cm of a1] (a3) {\(\bar{s}\)};

	\vertex[below=4em of a1] (b1) {\(d\)};
	\vertex[right=3.5cm of b1] (b2);
	\vertex[below=4em of a3] (b3) {\(d\)};
	
	\vertex[above right=6em of a2](c1);
	
	\vertex[above = 3em of a3](c3){\(\ell^{-}\)};
	\vertex[above = 3.3em of c3](c2){\(\ell^{+}\)};

	\vertex[right=2.0cm of a3](z1){\(\bar{b}\)};
	\vertex[right=2.0cm of z1](r1);
	\vertex[right=2.5cm of r1](r2);
	\vertex[above right=1.5cm of r1](r3);
	\vertex[above right=1.5cm of r2](r4);
	\vertex[right=7.0cm of z1](z2){\(\bar{s}\)};
	
	\vertex[above right=1.5cm of r3](o3){\(\ell^{+}\)};
	\vertex[above right=1.5cm of r4](o4){\(\ell^{-}\)};;
	
	\vertex[below=4em of z1](y1){\(d\)};
	\vertex[below=4em of z2](y2){\(d\)};

    \newcommand\tmpda{0.7cm}
	\newcommand\tmpdb{-1.7cm}
	\diagram* {
		{[edges=fermion]
			(l1) -- (a1),
		},

		(l2) -- [with arrow =\tmpda, with arrow =\tmpdb, out=55, in=200, half right] (l1),

		(l1) -- [boson, edge label=\(W^{+}\)] (l2),
		(a3) -- [fermion] (l2),
		(b1) -- [fermion] (b3),
		(l3) -- [boson](c1),
		
		(c2) -- [fermion](c1),
		(c1) -- [fermion](c3),
		
		(r1) -- [fermion](z1),
		(r2) -- [fermion](r1),
		(r1) -- [boson](r3),
		(r2) -- [boson](r4),
		(r3) -- [fermion, edge label=$\nu$](r4),
		(z2) -- [fermion](r2),
		(o3) -- [fermion](r3),
		(o4) -- [fermion](r4),
		
		(y1) -- [fermion] (y2)

	};
	
	\node[vertex, label=above:{$\bar{u}$, $\bar{c}$, $\bar{t}$}] at (2.3, 1.1);
	\node[vertex, label=above:$\gamma/Z^{0}$] at (3.7, 0.65);
	\node[vertex, label=left:$W^{+}$] at (11.5, 0.6);
	\node[vertex, label=left:$W^{-}$] at (15.0, 0.4);
	\node at (12.4, 0.35) {$\bar{u}, \bar{c}, \bar{t}$};
	
	\draw [decoration={brace}, decorate] (b1.south west) -- (a1.north west)
	node [pos=0.5, left] {\(\bar{B}^{0}\)};
	
	\draw [decoration={brace}, decorate] (a3.north east) -- (b3.south east)
	node [pos=0.5, right] {$K^{*0}$};

	\draw [decoration={brace}, decorate] (y1.south west) -- (z1.north west)
	node [pos=0.5, left] {\(\bar{B}^{0}\)};

	\draw [decoration={brace}, decorate] (z2.north east) -- (y2.south east)
	node [pos=0.5, right] {$K^{*0}$};

	\end{feynman}
	\end{tikzpicture}}

	\caption{Standard Model contributions to the $b \to s\ell^+\ell^-$ transition. 
  The left (right) diagrams show the penguin (box) diagram process: a loop involving $W$ bosons and $u/c/t$-quark, with the $b \to s$ flavour change propagating in the outer edges. 
	An off-shell $\gamma/Z$ boson is emitted and decays to $\ell^-\ell^+$. 
	The suppression --- by CKM factors and loop --- makes these processes exquisitely sensitive to new physics contributions that can compete with or exceed the Standard Model amplitude.}
	\label{fig:bsm:sm_penguin}
\end{figure}

These transitions proceed in the SM via electroweak penguin and box diagrams as shown in FIG.~\ref{fig:bsm:sm_penguin}, in which the bottom-quark decays to a strange-quark through a loop involving an up-type quark and a $W^\pm$ boson,
with the lepton pair emitted from a virtual photon or $Z$ boson. 

This rarity makes them very sensitive to BSM physics: any heavy particle coupling to quarks and leptons may contribute to the loops, or, if it couples at tree-level, compete directly with the suppressed SM amplitude.
A natural BSM mediator for such contributions — a new neutral vector boson — and its implications for collider searches are discussed in Section~\ref{sec:bsm:zprime_mediator}.

\subsection{Tensions and Global Fits}
\label{sec:bsm:tensions}

A theoretically-clean test of LFU is provided by the double ratios
\begin{equation}
  R_{K^{(*)}} =
    \frac{\mathcal{B}(B \to K^{(*)}\mu^+\mu^-)}
         {\mathcal{B}(B \to K^{(*)} e^+e^-)},
  \label{eq:bsm:rk}
\end{equation}
integrated over a specified $q^2$ range.
The SM predicts $R_{K^{(*)}} \approx 1$ up to small, well-understood electromagnetic
corrections~\cite{Bordone:2016gaq}.
Hadronic form-factor uncertainties cancel in the double ratio, making it theoretically robust.
LHCb reported $R_{K^*}$ deviating from unity at $2.5\sigma$ in 2017~\cite{RKstar_2017},
and $R_K$ at $3.1\sigma$ in 2021~\cite{LHCb:2021trn}.
Global fits at the time found that a new-physics contribution to the Wilson coefficient
$C_9^\mu$ could simultaneously explain the pattern of
deviations~\cite{Allanach:2022iod,Alguero:2022est}.
In 2022, LHCb released updated measurements of both $R_K$~\cite{LHCb:2022vje} and
$R_{K^*}$~\cite{RKstar_2022} with a larger dataset and an improved analysis technique,
finding results fully consistent with the SM and thereby retracting the LFUV hints.

The $R_{K^{(*)}}$ reversal does not close the book on $B$-anomalies.
Several other observables in $b \to s\mu\mu$ transitions remain in tension with SM
expectations.

\textbf{Angular analyses of $B^0 \to K^{*0}\mu^+\mu^-$} give access to a set of
optimised observables $P_i^{(\prime)}$ constructed to reduce form-factor
uncertainties~\cite{P5prime}.
In the four-body decay $B^0 \to K^{*0}(\to K^+\pi^-)\mu^+\mu^-$, the momenta of the
final-state particles define three independent angles.
The differential decay rate, expressed as a function of these angles and $q^2$,
is a sum of trigonometric terms. The coefficients $S_i$ in this expansion are
measurable quantities that depend on the Wilson coefficients $C_{9,10}^{(\prime)}$
and on hadronic form factors.
The $K^{*0}$ is a spin-1 meson, which carries a definite polarisation: its spin can be
aligned parallel to its flight direction (longitudinal) or perpendicular to it
(transverse).
Each polarisation state produces a distinct angular distribution.
$F_L$ is the fraction of $K^{*0}$ decays produced in the longitudinal state.
$S_5$ is the angular coefficient that receives contributions only from the
interference between the longitudinal and transverse $K^{*0}$ decay amplitudes.
It is directly sensitive to the Wilson coefficient $C_9^\mu$ and therefore to BSM
contributions in $b \to s\ell\ell$.
The optimised observable
\begin{equation}
  P_5' = \frac{S_5}{\sqrt{F_L(1-F_L)}}
  \label{eq:bsm:p5prime}
\end{equation}
is normalised by $\sqrt{F_L(1-F_L)}$ to cancel the leading hadronic form-factor
dependence of $S_5$, yielding a theoretically-clean and experimentally-sensitive probe.
$P_5'$ has shown a persistent $\sim 3\sigma$ discrepancy with SM predictions across
multiple LHCb analyses.
A comprehensive LHCb analysis using the full Run~2 dataset and an independent CMS
measurement have both confirmed these angular tensions~\cite{LHCb:2025Kstmm,CMS:2024angular}.

\textbf{The branching fraction $\mathcal{B}(B^+ \to K^+\mu^+\mu^-)$}~\cite{BK_mumu}
is measured below the SM expectation, with a tension ranging from $\sim 1\sigma$ when
light-cone sum rules are used~\cite{Bharucha:2015bzk} to $4.2\sigma$ when lattice
QCD predictions are employed~\cite{Parrott:2022zte}.

\textbf{The differential branching fraction of $B_s^0 \to \phi\mu^+\mu^-$}~\cite{Bs_phi_mumu}
exhibits a systematic deficit relative to SM predictions across all $q^2$ bins.

Separately from $b \to s\ell\ell$, the ratios
$R_{D^{(*)}} = \mathcal{B}(B \to D^{(*)}\tau\nu)/\mathcal{B}(B \to D^{(*)}\ell\nu)$
have shown persistent deviations from SM predictions in charged-current $b \to c\tau\nu$
transitions, pointing to possible lepton universality violation in a separate
sector~\cite{HFLAV:2022esi}.

Comprehensive reviews and global fits to all available $b \to s\ell\ell$
data~\cite{Capdevila:2023yhq,Alguero:2022est} continue to prefer New Physics over the SM alone, with a shift $\Delta C_9^\mu \approx -0.8$ to $-1.0$ in
the muonic Wilson coefficient preferred at the level of several standard deviations.
Critically, it is demonstrated in~\cite{Allanach:2022iod} that even \emph{after}
incorporating the 2022 LHCb $R_{K^{(*)}}$ reanalysis, a $Z'$ model remains a
competitive explanation of the residual $b \to s\mu^+\mu^-$ anomalies, outperforming
simple leptoquark interpretations on a goodness-of-fit basis.
Their analysis maps the allowed coupling space and shows that the same resonance that
could generate $\Delta C_9^\mu \sim -1$ via loop effects is within reach of direct LHC
searches in dilepton final states.
The complementary study~\cite{Allanach2023} surveys a wider range of $Z'$ model
constructions, identifying a family of ``Plan B'' models consistent with all flavour
constraints, following the latest $R_K$ measurements.
Together, these analyses clarify that the anomalies, although diminished, are not
resolved, and that direct searches for BSM mediators remain warranted.

\subsection{A \texorpdfstring{$Z'$}{Z'} Boson as Mediator}
\label{sec:bsm:zprime_mediator}

At energies well below the mediator mass, any BSM contribution to $b \to s\ell\ell$ can be described by an effective Fermi-like Hamiltonian in which the heavy mediator is integrated out, leaving a set of dimension-6 four-fermion operators with Wilson coefficients $C_{i}$,
\begin{equation}
  \mathcal{H}_\text{eff} \supset -\frac{4G_F}{\sqrt{2}} V_{tb}V_{ts}^*
    \sum_i C_i \mathcal{O}_i\,,
  \label{eq:bsm:heff}
\end{equation}
where $\mathcal{O}_9 \propto (\bar{s}\gamma^\mu P_L b)(\bar\ell\gamma_\mu\ell)$ and
$\mathcal{O}_{10} \propto (\bar{s}\gamma^\mu P_L b)(\bar\ell\gamma_\mu\gamma_5\ell)$
are the dominant semi-leptonic operators.
New Physics shifts these coefficients from their SM values; global fits to the full
$b \to s\ell\ell$ dataset find that a shift $\Delta C_9^\mu \approx -0.8$ to $-1.0$
is preferred at several standard deviations~\cite{Capdevila:2023yhq}.

\begin{figure}[htbp]
  \centering
  \includegraphics[width=0.5\textwidth]{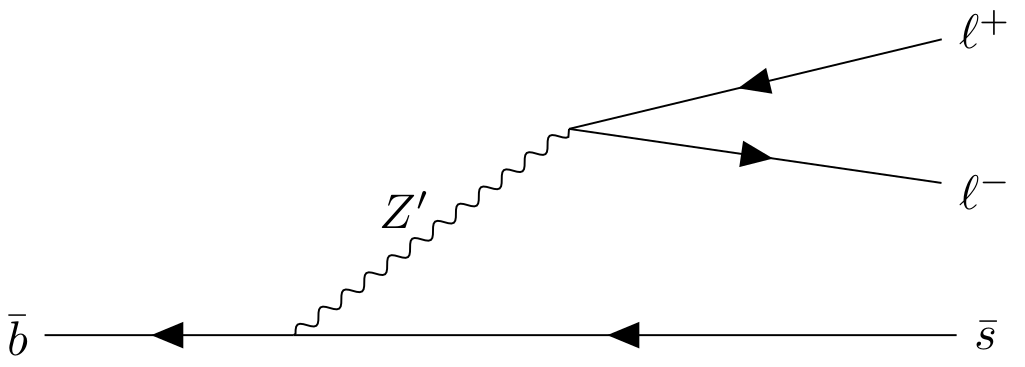}
  \caption{Tree-level $Z'$ contribution to the $b \to s\ell^+\ell^-$ transition.
    A new neutral vector boson $Z'$ couples to the $b$--$s$ quark current and decays
    to a lepton pair, generating a shift in the Wilson coefficient $C_9^\mu$ that can
    partially compensate observed tensions in $b \to s\ell^+\ell^-$ observables.}
  \label{fig:bsm:zprime_bsll}
\end{figure}

A natural mediator for such contributions is a new neutral vector boson $\Zp$ arising
from a $U(1)^\prime$ extension of the SM, coupling to $b$- and $s$-quarks at tree level
and decaying to a lepton pair, as depicted in FIG.~\ref{fig:bsm:zprime_bsll}.
The $\Zp$ contributes to the Wilson coefficients via its tree-level exchange, and for
$m_\Zp \gg \sqrt{q^2}$ this exchange reduces to precisely the Fermi-like contact
interaction of Eq.~\eqref{eq:bsm:heff}.

Several such $U(1)^\prime$ models have been proposed~\cite{Davighi2021,Celis:2015ara,Calibbi:2019lvs,
Alonso:2017uky,Bonilla:2017lsq}, each generating new contributions to the relevant
Wilson coefficients.
Of particular relevance to this thesis is the \emph{less-minimal flavour violation}
benchmark~\cite{Calibbi:2019lvs,Alguero:2022est,Crivellin:2022obd}, in which the
$\Zp$ carries a left-handed $(b\bar{s}, \bar{b}s)$ coupling with arbitrary lepton
flavour.
This model simultaneously accommodates the residual $b \to s\ell^+\ell^-$ anomalies and
can be consistent with the updated $R_{K^{(*)}}$ measurements, since it does not necessarily require
lepton flavour universality violation --- the $\Zp$ couplings to electrons and muons
can be set equal.
The $\Zp$ is therefore produced in association with $b$-quarks at the LHC and searched
for in both dielectron and dimuon final states, providing independent sensitivity in
each lepton channel without assuming LFU conservation.

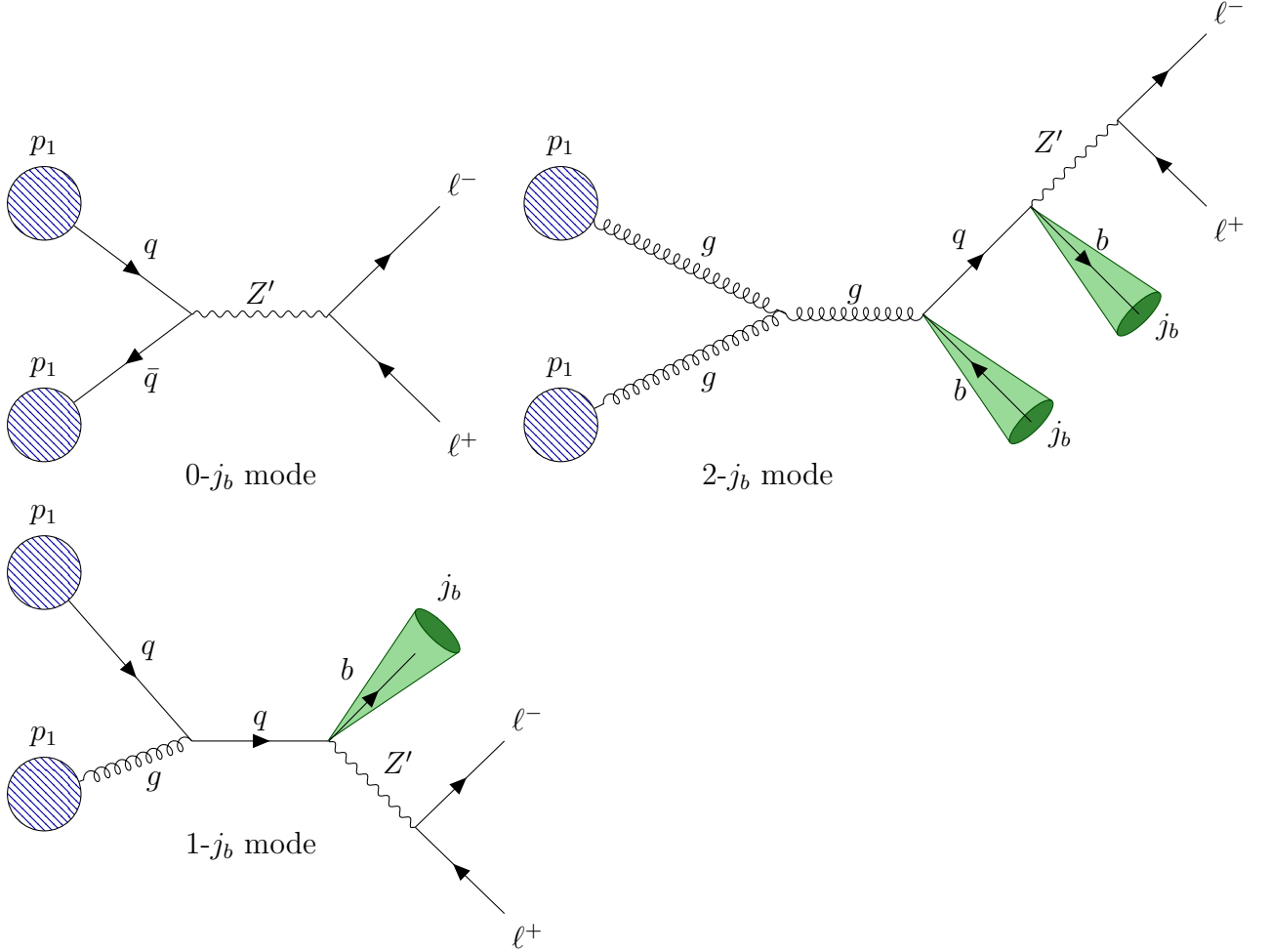
\begin{figure}[htbp]
  \centering
  \begin{tikzpicture}
  \begin{feynman}
  \vertex (c1);
  \vertex [right = 4.5 em of c1] (c2);

  \vertex [below left = 4 em of c1](b1);

  \vertex [above right = 5 em of c2] (a2) {\(\ell^{-}\)};
  \vertex [below right = 5 em of c2] (b2) {\(\ell^{+}\)};

  \vertex[blob,label={above:$p_{1}$}, inner sep=0.35cm, pattern color=blue!65!black] (m) at ( -2, 1.5) {\contour{white}{} };
  \vertex[blob,label={above:$p_{1}$}, inner sep=0.35cm, pattern color=blue!65!black] (n) at ( -2, -1.5) {\contour{white}{} };

  \node[] at (0.8,-2.2) {0-$j_{b}$ mode};

  \vertex [right = 15 em of c2] (k1);
  \vertex [right = 4.5 em of k1] (k2);

  \vertex [below left = 4 em of k1](j1);

  \vertex [above right = 5 em of k2] (i2);
  \vertex [below right = 5 em of k2] (j2);
  \vertex [below right = 5 em of i2] (t1);

  \vertex [above right = 4 em of i2] (t2);

  \vertex [above right = 4 em of t2] (s1) {\(\ell^{-}\)};
  \vertex [below right = 4 em of t2] (s2) {\(\ell^{+}\)};

  \vertex[blob,label={above:$p_{1}$}, inner sep=0.35cm, pattern color=blue!65!black] (o) at ( 5, 1.5) {\contour{white}{} };
  \vertex[blob,label={above:$p_{1}$}, inner sep=0.35cm, pattern color=blue!65!black] (p) at ( 5, -1.5) {\contour{white}{} };

  \jetcone[green!80!black]{i2}{t1}{22}{0.10}
  \jetcone[green!80!black]{k2}{j2}{22}{0.10}

  \node[below right = 0.1 em and 0.7 em of t1, xshift=-0.2cm, yshift=0.2cm] {$j_{b}$};
  \node[below right = 0.1 em and 0.7 em of j2, xshift=-0.2cm, yshift=0.2cm] {$j_{b}$};

  \node[] at (7.8,-2.2) {2-$j_{b}$ mode};

  \vertex[below = 14 em of c1] (x1);
  \vertex [right = 4.5 em of x1] (x2);

  \vertex [above right = 4 em of x2] (y1){\(j_{b}\)};
  \vertex [below right = 4 em of x2] (y2);

  \vertex [above right = 4 em of y2] (z1) {\(\ell^{-}\)};
  \vertex [below right = 4 em of y2] (z2) {\(\ell^{+}\)};

  \vertex[blob,label={above:$p_{1}$}, inner sep=0.35cm, pattern color=blue!65!black] (u) at ( -2, -3.5) {\contour{white}{} };
  \vertex[blob,label={above:$p_{1}$}, inner sep=0.35cm, pattern color=blue!65!black] (w) at ( -2, -6.5) {\contour{white}{} };

  \node[] at (0.8,-7.2) {1-$j_{b}$ mode};

  \jetcone[green!80!black]{x2}{y1}{22}{0.10}

  \node[] at (2.1,-4.8) {$b$};
  \node[] at (3.5,-3.7) {$j_{b}$};

  \diagram* {
      (m) -- [fermion, edge label=\(q\)] (c1),
      (c1) -- [fermion, edge label=\(\bar{q}\)] (n),
      (c1) -- [boson, edge label=\(Z'\)] (c2),
      (b2) -- [fermion] (c2),
      (c2) -- [fermion] (a2),

      (o) -- [gluon, edge label=\(g\)] (k1),
      (k1) -- [gluon, edge label=\(g\)] (p),
      (k1) -- [gluon, edge label=\(g\)] (k2),
      (j2) -- [fermion, edge label=\(b\)] (k2),
      (k2) -- [fermion, edge label=\(q\)] (i2),
      (i2) -- [fermion, edge label=\(b\)] (t1),
      (i2) -- [boson, edge label=\(Z'\)] (t2),
      (t2) -- [fermion] (s1),
      (s2) -- [fermion] (t2),

      (u) -- [fermion, edge label=\(q\)] (x1),
      (x1) -- [gluon, edge label=\(g\)] (w),
      (x1) -- [fermion, edge label=\(q\)] (x2),
      (x2) -- [fermion] (y1);
      (x2) -- [boson, edge label=\(Z'\)] (y2),
      (z2) -- [fermion] (y2);
      (y2) -- [fermion] (z1);
  };
  \end{feynman}
  \end{tikzpicture}
  \caption{Representative Feynman diagrams for the three leading $Z'$ production modes in association
    with $b$-quarks: quark-antiquark annihilation without associated
    $b$-jets (0-$j_b$, top left), quark-gluon scattering with a single $b$-jet (1-$j_b$, bottom),
    and gluon fusion producing a $b\bar{b}$ pair (2-$j_b$, top right). These are the leading
    parton-level subprocesses; production modes defined at the matrix-element level do not map
    trivially onto reconstructed $b$-jet multiplicity categories.}
  \label{fig:bsm:zprime_production}
\end{figure}

A generic parametrisation of the $\Zp$ couplings to SM fermions is given by the Lagrangian
\begin{equation}
  \mathcal{L} = \sum_{f\in\{u,\,d,\,\ell,\,\nu\}}
    \bar{f}_{i}\,\gamma^\mu
    \bigl(\Gamma^{f_L}_{ij} P_L + \Gamma^{f_R}_{ij} P_R\bigr)
    f_j\,Z'_\mu,
\end{equation}
where $P_{L,R} = (1 \mp \gamma_5)/2$ and $\Gamma^{f_{L,R}}_{ij}$ are the left- and
right-handed coupling matrices.
In this benchmark, the $\Zp$ couples dominantly to third-generation quarks via
left-handed coupling matrices of the form
\begin{equation}
  \Gamma^{d_L}_{f_i} \approx
  \begin{pmatrix}
    0 & 0 & 0 \\
    0 & x^2 & x \\
    0 & x   & 1
  \end{pmatrix}
  \mathcal{Q}_3,
  \qquad
  \Gamma^{u_L}_{f_i} \approx
  \begin{pmatrix}
    0 & 0 & 0 \\
    0 & (x - V_{cb})^2 & x - V_{cb} \\
    0 & x - V_{cb}     & 1
  \end{pmatrix}
  \mathcal{Q}_3,
  \qquad
  x \sim \mathcal{O}(V_{cb}),
\end{equation}
from which contributions to the Wilson coefficients $C_9^\mu$ and $C_{10}^\mu$ of
Equation~\eqref{eq:bsm:heff} arise at tree level.
The $\Zp$ is produced at the LHC via three leading subprocesses illustrated in
FIG.~\ref{fig:bsm:zprime_production}: quark-antiquark annihilation with no associated
$b$-jet (0-$j_b$), quark-gluon scattering producing one $b$-jet (1-$j_b$), and gluon
fusion producing a $b\bar{b}$ pair (2-$j_b$).

Although the LFUV hints in $R_{K^{(*)}}$ have been resolved, enduring tensions in
angular observables and exclusive branching fractions continue to motivate direct
collider searches for such resonances.
A search for this $\Zp$ in same-flavour dilepton final states in association with
$b$-quark jets, providing direct sensitivity to the proposed mediators, is presented
in Chapter~\ref{chp:zprime}.
\clearpage
\clearpage\chapter{Experimental Apparatus}
\label{chp:detector}

The experimental programme described in this thesis is performed at the intersection of two cutting-edge machines: the Large Hadron Collider (LHC), which accelerates proton beams to near-light speed and brings them into collision at unprecedented energies and the ATLAS
detector, a general-purpose instrument that records the products of those collisions. 
This chapter describes both in turn. 
Section~\ref{sec:lhc} covers the LHC --- its organisational context, the chain of accelerators that produce and prepare the beam and the key parameters that govern its performance. 
Section~\ref{sec:atlas} then describes the ATLAS detector --- its geometry, coordinate conventions and each of the subsystems that together measure the particles produced at every collision.

\section{The Large Hadron Collider}
\label{sec:lhc}

\subsection*{CERN and the Legacy of the LEP}

The LHC is housed at CERN, the European Organisation for Nuclear Research, which straddles the Franco-Swiss border near Geneva. 
Founded in 1954 with the mandate to promote fundamental research in particle physics and foster international scientific cooperation, CERN has grown into one of the largest scientific
enterprises in the world.
Today, it employs more than 2{,}500 staff and hosts roughly 13{,}000 visiting scientists,
engineers and technicians from over 100 countries, all working together on experiments that
range from precision tests of the SM to searches for BSM physics at the highest accessible energies. 

The LHC occupies the 26.7\,km circular tunnel originally built for its predecessor, the Large Electron--Positron Collider (LEP), which operated from 1989 to 2000. 
LEP achieved centre-of-mass energies up to 209\,\GeV\ and produced extraordinarily-precise measurements of the electroweak sector, measuring the $W$ and $Z$ masses at per-mille-level precision. 
Once LEP had exhausted its physics programme, the tunnel was repurposed for the LHC. 
The electron and positron beams were replaced with two counter-rotating rings of protons. 
This choice opens the door to orders-of-magnitude higher collision energies. 
It comes at the cost of a more complex machine and richer hadronic backgrounds. 

\subsection{The Proton Injection Chain}
\label{sec:lhc:injectors}

\begin{landscape}
\begin{figure}[htbp]
  \centering
  \includegraphics[width=\linewidth,height=0.93\textheight,keepaspectratio]{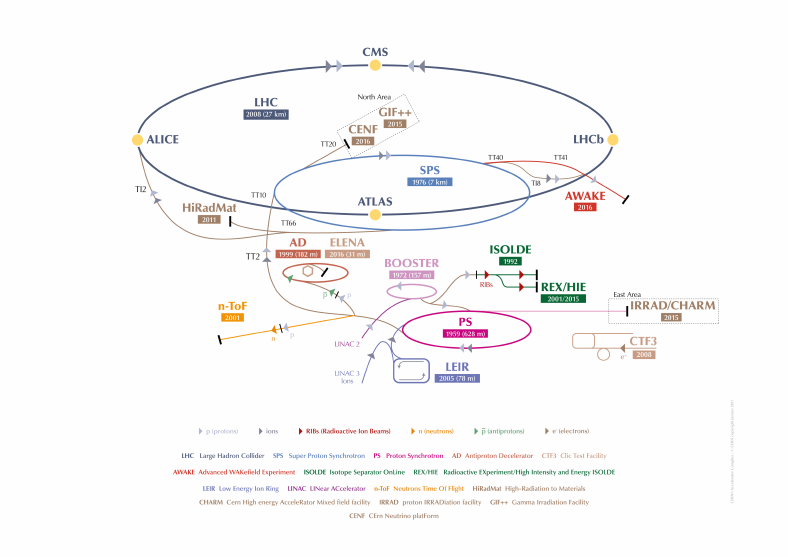}
  \caption{The CERN accelerator complex~\cite{Mobs:2197559}.
    Protons produced from ionised hydrogen gas are successively accelerated by Linac\,4,
    the Proton Synchrotron Booster, the Proton Synchrotron, and the Super
    Proton Synchrotron before injection into the two counter-rotating LHC rings
    at $450\,\GeV$.
    Fixed-target experiments and test beams branch off at various stages of the chain.}
  \label{fig:lhc:complex}
\end{figure}
\end{landscape}

Each LHC proton begins its journey in a small bottle of hydrogen gas.
A duoplasmatron ion source strips the electrons from $\mathrm{H}_2$ molecules, producing a
dense plasma from which a pulsed beam of protons is extracted and pre-accelerated to 100\,keV
by a static electric field.
This beam enters the first stage of CERN's multi-step injector chain, illustrated in
FIG.~\ref{fig:lhc:complex}.

The injector chain comprises four successive
stages~\cite{Bruning:2004ej,Evans:2008zzb}:

\begin{enumerate}

  \item \textbf{Linac\,4} (Linear Accelerator 4): a drift-tube linear accelerator that
    accelerates negatively charged $H^-$ ions to $160\,\mathrm{MeV}$ through a succession
    of RF cavities.
    Linac\,4 replaced its predecessor, Linac\,2, in 2018; it delivers a brighter,
    lower-emittance proton beam by stripping the electrons from the $H^-$ ions at injection
    into the PSB.

  \item \textbf{Proton Synchrotron Booster} (PSB): four superimposed synchrotron rings of
    157\,m circumference that receive the $160\,\mathrm{MeV}$ beam from Linac\,4,
    strip its electrons, and accelerate the resulting proton beam to $2\,\GeV$.
    The four rings can operate independently, providing flexibility in bunch manipulation
    before the merged train is sent onward to the PS.

  \item \textbf{Proton Synchrotron} (PS): a 628\,m-circumference synchrotron that accepts
    the $2\,\GeV$ beam from the PSB and accelerates it to $25\,\GeV$.
    It also sculpts the longitudinal bunch structure --- merging, splitting, and adjusting
    the timing --- to produce the 25\,ns bunch spacing and bunch-train pattern required by
    the LHC.

  \item \textbf{Super Proton Synchrotron} (SPS): with a circumference of 6.9\,km, the SPS
    accelerates the $25\,\GeV$ beam to $450\,\GeV$, the LHC injection energy.
    Protons are transferred from the SPS to the two LHC rings through separate transfer
    tunnels, filling both beams simultaneously in opposite directions.

\end{enumerate}

\subsection{Machine Parameters and Beam Configuration}
\label{sec:lhc:params}

Once inside the LHC, two beams of protons circulate in opposite directions within separate
evacuated beam pipes, kept on their circular orbits by a system of 1{,}232 superconducting
main dipole magnets~\cite{Evans:2008zzb}.
Each dipole is 14.3\,m long and, when cooled to 1.9\,K by superfluid helium, generates a
peak bending field of 8.33\,T --- the highest sustained field of any large-scale
superconducting accelerator.

The bending radius of the proton orbit follows directly from the Lorentz force,
\begin{equation}
  \vec{F} \;=\; q\,\vec{v} \times \vec{B},
  \label{eq:lhc:lorentz}
\end{equation}
where $q$ is the particle charge and $\vec{v}$ its velocity.
For a relativistic proton with momentum $p$ and charge $q = e$ traversing a transverse
magnetic field $B$, the equation of motion gives
\begin{equation}
  p \;=\; q B \rho,
  \label{eq:lhc:rigidity}
\end{equation}
where $\rho$ is the local radius of curvature.
In practical units, this \emph{magnetic rigidity} relation reads
\begin{equation}
  p\;[\mathrm{GeV}/\mathrm{c}] \;=\; 0.2998 \cdot B\;[\mathrm{T}] \cdot \rho\;[\mathrm{m}].
  \label{eq:lhc:rigidity_practical}
\end{equation}
The effective dipole bending radius for the LHC geometry is $\rho \approx 2{,}804$\,m;
inserting the design field of 8.33\,T yields a maximum beam momentum of $p \approx 7\,\TeV$,
corresponding to a proton--proton centre-of-mass energy of $\sqrt{s} = 14\,\TeV$.
During Run~2, the beams were operated at $6.5\,\TeV$ per beam ($\sqrt{s} = 13\,\TeV$),
which required reducing the dipole field to approximately 7.7\,T.
Run~3 operates at $6.8\,\TeV$ per beam ($\sqrt{s} = 13.6\,\TeV$).

Protons are grouped into discrete \emph{bunches}, confined longitudinally by a
400\,MHz superconducting RF system.
The RF frequency $f_\mathrm{RF}$ is related to the revolution frequency $f_\mathrm{rev}$
by an integer harmonic number $h$:
\begin{equation}
  f_\mathrm{RF} \;=\; h \cdot f_\mathrm{rev},
  \label{eq:lhc:rf}
\end{equation}
where $h = 35{,}640$ and
$f_\mathrm{rev} = \beta \mathrm{c} / C \approx 11.245\,\mathrm{kHz}$
for the LHC circumference $C = 26{,}659$\,m at $\beta \approx 1$.
The standard Run~2 bunch spacing of 25\,ns corresponds to bunches occupying every tenth
RF bucket.
Table~\ref{tab:lhc:params} summarises the key LHC parameters for Run~2 operation.

\begin{table}[htbp]
  \centering
  \caption{Selected LHC parameters for Run~2 ($\sqrt{s} = 13\,\TeV$)~\cite{Evans:2008zzb,Bruning:2004ej}.
    Design values refer to the nominal $\sqrt{s} = 14\,\TeV$ configuration.}
  \label{tab:lhc:params}
  \begin{tabular}{lll}
    \hhline{===}
    Parameter                         & Symbol                 & Run~2 value \\
    \hline
    Circumference                     & $C$                    & 26{,}659\,m \\
    Main dipole bending field         & $B$                    & 7.7\,T (8.33\,T design) \\
    Beam energy                       & $E$                    & $6.5\,\TeV$ ($7\,\TeV$ design) \\
    Revolution frequency              & $f_\mathrm{rev}$       & 11.245\,kHz \\
    RF frequency                      & $f_\mathrm{RF}$        & 400.79\,MHz \\
    Harmonic number                   & $h$                    & 35{,}640 \\
    Bunch spacing                     & $\Delta t$             & 25\,ns \\
    Colliding bunch pairs             & $n_b$                  & up to 2{,}556 \\
    Protons per bunch                 & $N$                    & $\sim\!1.15\times10^{11}$ \\
    Peak luminosity at IP1            & $\mathcal{L}_\mathrm{peak}$
                                                               & $2.1\times10^{34}\,\mathrm{cm}^{-2}\mathrm{s}^{-1}$ \\
    \hhline{===}
  \end{tabular}
\end{table}

The transverse beam optics are controlled by quadrupole and higher-order multipole
magnets distributed around the ring.
At each interaction point, a final-focus system squeezes the beam to the smallest possible
transverse size.
At IP1 (the ATLAS interaction point), the transverse r.m.s.\ beam size at the collision
point is $\sigma^* \sim 17\,\mu\mathrm{m}$, achieved with a $\beta$-function value at
the IP of $\beta^* \approx 0.4$\,m during Run~2 peak conditions.
A small beam crossing angle $\theta_c$ is introduced to suppress unwanted long-range
beam--beam interactions at the parasitic encounter points adjacent to the IP.

\subsection{Luminosity}
\label{sec:lhc:lumi}

The rate at which a physics process with cross-section $\sigma$ produces events is
$R = \sigma \cdot \mathcal{L}$, where $\mathcal{L}$ is the \emph{instantaneous luminosity}.
For two Gaussian beams colliding at a single interaction point, the luminosity is~\cite{Evans:2008zzb}
\begin{equation}
  \mathcal{L} \;=\; \frac{n_b\, N_1 N_2\, f_\mathrm{rev}}{4\pi \sigma_x^* \sigma_y^*}
    \cdot F,
  \label{eq:lhc:lumi}
\end{equation}
where $n_b$ is the number of colliding bunch pairs per revolution, $N_{1,2}$ are the
numbers of protons per bunch in the two beams, $\sigma_{x,y}^*$ are the horizontal and
vertical r.m.s.\ beam sizes at the interaction point, and $F$ is a geometric reduction
factor arising from the non-zero crossing angle:
\begin{equation}
  F \;=\; \left[1 + \left(\frac{\theta_c\, \sigma_z}{2\sigma^*}\right)^2\right]^{-1/2},
  \label{eq:lhc:geom}
\end{equation}
with $\sigma_z$ the r.m.s.\ longitudinal bunch length and $\sigma^*$ the transverse beam
size (assuming $\sigma_x^* = \sigma_y^* \equiv \sigma^*$).

The cumulative dataset available for physics analysis is characterised by the
\emph{integrated luminosity}:
\begin{equation}
  L \;=\; \int \mathcal{L}(t)\, dt,
  \label{eq:lhc:intlumi}
\end{equation}
measured in inverse femtobarns ($\mathrm{fb}^{-1}$).
The expected number of events produced by any process during a data-taking period is therefore
$N = \sigma \cdot L$, making the precise determination of $L$ essential for cross-section
measurements and signal yield calculations alike.

Luminosity at ATLAS is measured primarily by the LUCID-2 Cherenkov detector~\cite{LUCID2},
located 17\,m from IP1 along the beam axis on each side.
LUCID-2 counts the mean number of inelastic interactions per bunch crossing and converts
this to an absolute luminosity via van der Meer (vdM) scans~\cite{DAPR-2021-01}, in which
the two beams are swept transversely through each other to determine $\sigma_{x,y}^*$
directly from the rate of observed interactions.
Over the full Run~2 dataset (2015--2018, $\sqrt{s} = 13\,\TeV$), ATLAS recorded an
integrated luminosity of $L = 140\,\mathrm{fb}^{-1}$ with an overall uncertainty of
1.7\%~\cite{DAPR-2021-01}.
Run~3 ($\sqrt{s} = 13.6\,\TeV$), the data-taking period exploited by the analyses in this thesis, began in 2022 and is expected to conclude in 2026.
The integrated luminosity recorded during 2022--2024 amounts to
$L = 164\,\mathrm{fb}^{-1}$~\cite{ATLAS:2024run3lumi}.

A direct consequence of the high instantaneous luminosity is \emph{pileup}: the occurrence
of multiple simultaneous inelastic proton--proton interactions within a single 25\,ns bunch
crossing (\emph{in-time pileup}), as well as residual signals from adjacent bunch crossings
that persist in slow detector elements (\emph{out-of-time pileup}).
The mean number of simultaneous interactions per bunch crossing, $\langle\mu\rangle$,
averaged approximately 34 during Run~2 peak conditions and exceeded 60 in exceptional fills,
posing a significant challenge for object reconstruction and background modelling.
Pileup suppression strategies are discussed in Chapter~\ref{chp:objects}.

\subsection{Proton--Proton Collisions and Parton Distribution Functions}
\label{sec:lhc:pdf}

Although the LHC accelerates and collides \emph{protons}, the fundamental hard scattering
processes that produce heavy particles are mediated by the proton's constituents --- quarks
and gluons, collectively called \emph{partons}.
The proton is a colour-neutral bound state held together by the strong force; far from a
collision, its constituents are permanently confined within a radius of order $1\,\mathrm{fm}$
and cannot be observed as free particles, a consequence of \emph{confinement} and the
related process of \emph{hadronisation} discussed in Chapter~\ref{chp:sm}.
However, the running of the QCD coupling constant $\alpha_s(\mu)$ means the interaction
weakens at short distances --- the property of \emph{asymptotic freedom} (see
Chapter~\ref{chp:sm}).
At momentum transfers $|Q|^2 \gg \Lambda_{\mathrm{QCD}}^2$, the partons taking part in
the hard scatter therefore behave as effectively free on the timescale of the interaction.
This separation of scales is formalised by the QCD \emph{factorisation theorem}.

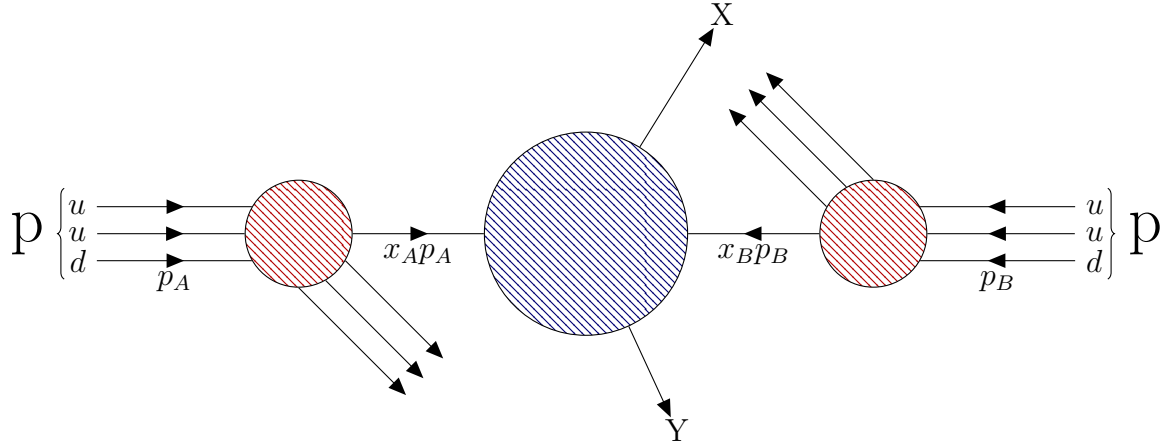
\begin{figure}[htbp]
\centering
\begin{tikzpicture}
  \begin{feynman}
      \vertex[blob,label={}, inner sep=0.5cm, pattern color=red!70!black] (m) at (-1.8, 0) {};
      \path (m. 180) ++ (00:-1.95) node[vertex, label=left:$u$] (a);
      \path (m. 150) ++ (00:-2.05) node[vertex, label=left:$u$] (a1);
      \path (m. 210) ++ (00: -2.05) node[vertex, label=left:$d$] (a2);

      \path (m.270) ++ (-45:1.88) node[vertex] (b);
      \path (m.300) ++ (-45:1.70) node[vertex] (b1);
      \path (m.330) ++ (-45:1.70) node[vertex] (b2);

      \vertex[blob,label={}, inner sep=0.95cm, pattern color=blue!45!black] (n) at (2, 0) {};
      \vertex (e) at (5,0);
      \vertex (f) at (3.8, 2.9){X};
      \vertex (g) at (3.2, -2.6){Y};

      \vertex[blob,label={}, inner sep=0.5cm, pattern color=red!70!black] (l) at (5.8, 0) {};
      \vertex (h) at (8,0);
      \path (l. 0) ++ (00: 1.95) node[vertex, label=right:$u$](h);
      \path (l. 30) ++ (00: 2.05) node[vertex, label=right:$u$](h1);
      \path (l. 330) ++ (00: 2.05) node[vertex, label=right:$d$](h2);

      \path (l.90)  ++ (135:1.88) node[vertex] (c);
      \path (l.120) ++ (135:1.70) node[vertex] (c1);
      \path (l.150) ++ (135:1.70) node[vertex] (c2);

      \diagram* {
        (a) -- [with arrow = 0.52](m)-- [fermion,edge label'=$x_{A}p_{A}$](n) -- [with arrow = 1](f),
        (a1) -- [fermion] (m.150),
        (a2) -- [fermion, edge label'=$p_{A}$](m.210),

        (m.270) -- [with arrow = 1](b),
        (m.300) -- [with arrow = 1](b1),
        (m.330) -- [with arrow = 1](b2),

        (n) -- [with arrow = 1](g),
        (h) -- [with arrow = 0.52](l) -- [fermion ,edge label=$x_{B}p_{B}$](n),
        (h1) -- [fermion](l.30),
        (h2) -- [fermion, edge label = $p_{B}$](l.330),

        (l.90)  -- [with arrow = 1](c),
        (l.120) -- [with arrow = 1](c1),
        (l.150) -- [with arrow = 1](c2),
      };

      \draw [decoration={brace}, decorate] (-4.9, -0.6) -- (-4.9, 0.6)
        node [pos=0.5, left=0.125cm] {\huge p};

      \draw [decoration={brace}, decorate] (8.9, 0.6) -- (8.9, -0.6)
        node [pos=0.5, right=0.125cm] {\huge p};

  \end{feynman}
\end{tikzpicture}
\caption{Schematic proton--proton collision diagram.
  At high $|Q|$ the valence quarks of protons $A$ and $B$ (red blobs) can be treated as
  asymptotically free partons that effectively participate in the scattering.
  Each interacting parton carries momentum fraction $x_i p_i$ ($i \in \{A, B\}$)
  as prescribed by the parton distribution function.
  The hard scatter (blue blob) produces final-state particles $X$ and $Y$; the
  remaining partons form the underlying event.}
\label{fig:lhc:ppfeynman}
\end{figure}

The cross-section for a proton--proton collision producing final-state particles $X$ and
$Y$ can therefore be written as a convolution of \emph{parton distribution functions}
(PDFs) with the partonic hard-scattering cross-section:
\begin{equation}
\begin{split}
  &\sigma\bigl(\mathrm{p}(p_A) + \mathrm{p}(p_B) \!\to\! X + Y\bigr) = \\
  &\quad\int_0^1 \mathrm{d}x_A\,\mathrm{d}x_B
  \sum_{i,j} f_i(x_A, Q^2)\,f_j(x_B, Q^2)\,
  \hat{\sigma}\bigl(\psi_i(x_A p_A) + \psi_j(x_B p_B) \to X + Y\bigr),
\end{split}
\label{eq:lhc:factorisation}
\end{equation}
where $f_i(x, Q^2)$ is the PDF giving the probability of finding parton species $i$
carrying momentum fraction $x$ of the proton at factorisation scale $Q^2$, and
$\hat{\sigma}$ is the perturbatively calculable partonic cross-section.
The sum runs over all contributing parton species $i, j$.
A schematic illustration of the collision at the parton level is given in
FIG.~\ref{fig:lhc:ppfeynman}.

\begin{figure}[htbp]
  \centering
  \includegraphics[width=0.82\textwidth]{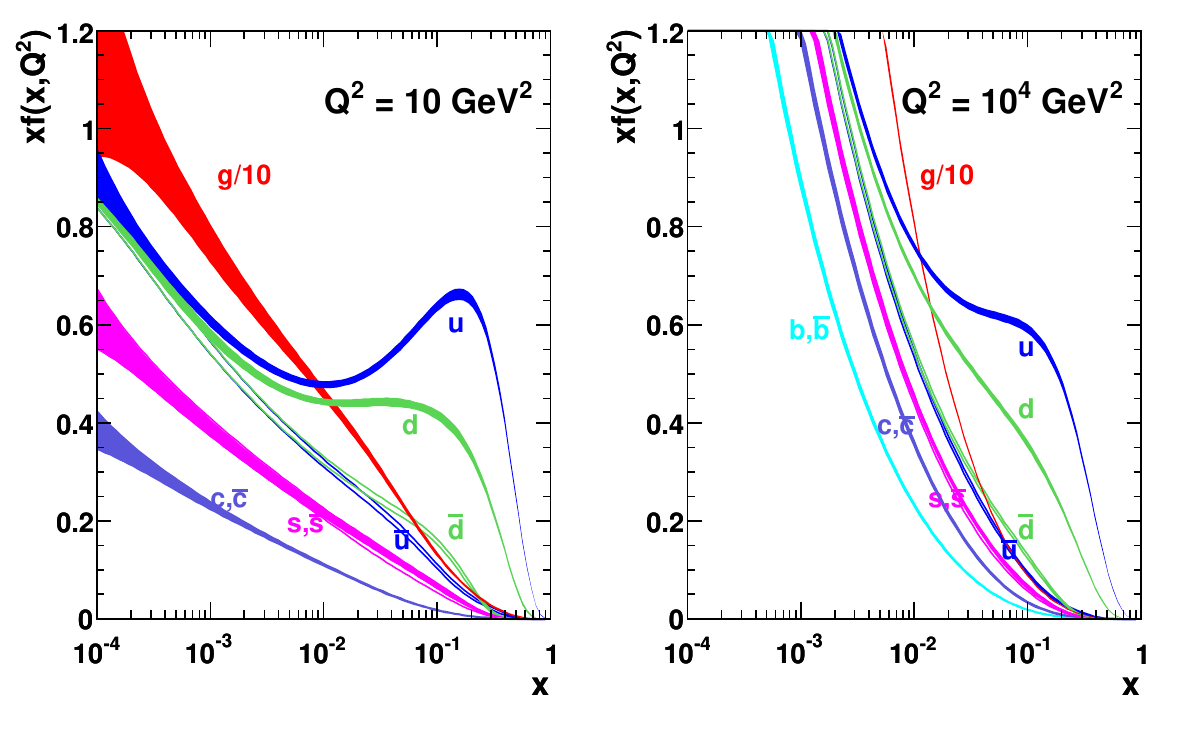}
  \caption{Parton distribution functions of the proton at two factorisation scales,
    $Q^2 = 10\,\mathrm{GeV}^2$ (left) and $Q^2 = 10^4\,\mathrm{GeV}^2$ (right),
    from the MSTW~2008 NLO fit~\cite{Martin:2009iq}.
    The valence $u$- and $d$-quarks peak near $x \approx 0.2$, while sea quarks
    ($\bar{u}$, $\bar{d}$, $s$, $c$, \ldots) and especially gluons (shown as $g/10$
    for readability) rise steeply toward small $x$.
    The marked increase of all densities at small $x$ and higher $Q^2$ reflects
    DGLAP evolution.}
  \label{fig:lhc:pdf}
\end{figure}

The PDFs are non-perturbative objects that cannot be computed from first principles;
they are instead extracted from global fits to a wide range of experimental data ---
deep-inelastic scattering, Drell--Yan, and jet production measurements --- and evolved
in $Q^2$ using the Dokshitzer--Gribov--Lipatov--Altarelli--Parisi (DGLAP) evolution equations.
FIG.~\ref{fig:lhc:pdf} shows the MSTW~2008 NLO PDF set~\cite{Martin:2009iq} at two
representative scales.

The proton contains three \emph{valence quarks} ($uud$) that carry its quantum numbers, as
well as a sea of virtual quark--antiquark pairs and gluons generated by QCD radiation.
At low momentum fractions $x \lesssim 0.01$, the PDF is dominated by \emph{sea quarks}
($\bar{u}$, $\bar{d}$, $s$, $\bar{s}$, $c$, $\bar{c}$, \ldots) and gluons, all of
which rise steeply as $x \to 0$ --- a reflection of the proliferation of soft parton
splittings.
At large $x \gtrsim 0.1$, the valence $u$- and $d$-quarks dominate, as seen in the
characteristic peak in $xf(x, Q^2)$ near $x \approx 0.2$.
At the LHC centre-of-mass energy of $\sqrt{s}=13.6\,\TeV$, the momentum fractions accessed
by hard processes producing particles at the TeV scale are of order
$x \sim 2M/\sqrt{s} \sim 10^{-1}$, placing the interaction squarely in the valence-quark
regime.
For lower-mass or forward production, $x$ can fall well below $10^{-2}$, where sea
quarks and gluons are the primary initiators.
The strong $Q^2$ evolution of the gluon density, particularly visible between the two panels
of FIG.~\ref{fig:lhc:pdf}, reflects DGLAP splitting and underlies the large
hard-scattering cross-sections accessible at the LHC.

\section{The ATLAS Detector}
\label{sec:atlas}

\subsection*{The ATLAS Collaboration}

A Toroidal LHC ApparatuS (ATLAS) is a general-purpose particle physics detector operated by
an international collaboration of approximately 5{,}900 physicists, engineers, and technicians
from more than 180 institutions across 42 countries.
Established in 1992, the collaboration grew continuously as the detector was designed, built,
and installed.

\begin{figure}[htbp]
  \centering
  \includegraphics[width=0.92\textwidth]{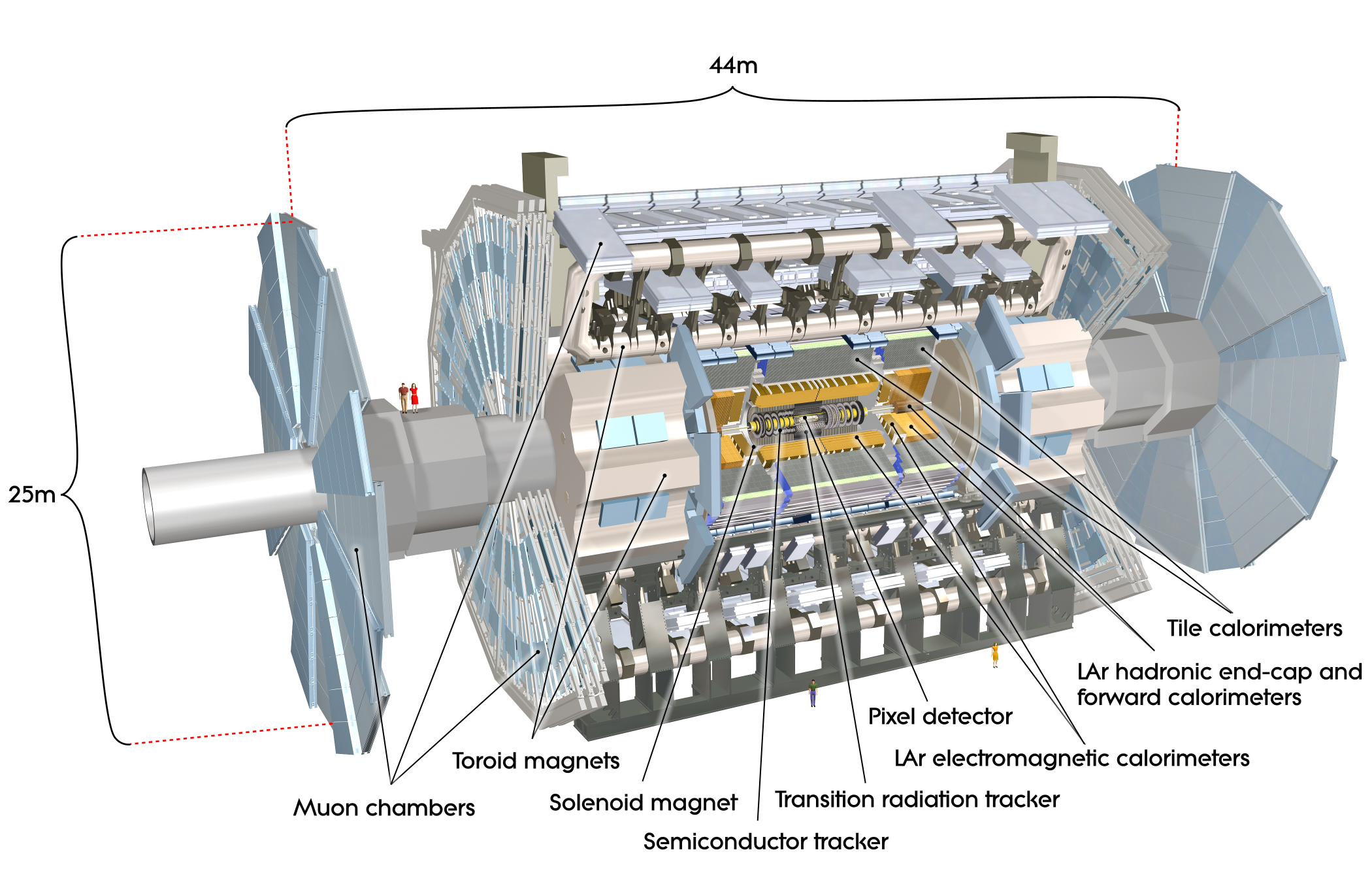}
  \caption{Cut-away view of the ATLAS detector~\cite{ATLAS:2008xda}.
    The detector is approximately 46\,m long and 25\,m in diameter.
    Observing outward from the beam axis: the Inner Detector enclosed by the central solenoid;
    the electromagnetic and hadronic calorimeters; and the Muon Spectrometer embedded in the
    large toroidal magnet system.}
  \label{fig:atlas:overview}
\end{figure}

The detector itself is described in detail in Refs.~\cite{ATLAS:2008xda,ATLAS:2023dns}.
It is installed in a cavern 100\,m underground at IP1 on the LHC ring, is approximately 46\,m
in length and 25\,m in diameter, and weighs some 7{,}000\,tonnes --- one of the largest and
most complex scientific instruments ever constructed.
A cutaway overview is shown in FIG.~\ref{fig:atlas:overview}.

\subsection{Geometry and Coordinate System}
\label{sec:atlas:coords}

ATLAS uses a right-handed coordinate system with its origin at the nominal interaction point
at the centre of the detector.
The $z$-axis lies along the beam direction; the $x$-axis points from the IP towards the
centre of the LHC ring; and the $y$-axis points upward.
The azimuthal angle $\phi$ is measured in the transverse ($x$--$y$) plane around the beam
axis, while the polar angle $\theta$ is measured from the positive $z$-axis.
The side of the detector with $z > 0$ is labelled side-A and the side with $z < 0$ is
labelled side-C.

Because particle production in high-energy hadron collisions is approximately invariant under
boosts along the beam axis, it is convenient to replace $\theta$ with the \emph{pseudorapidity}:
\begin{equation}
  \eta \;=\; -\ln\tan\frac{\theta}{2},
  \label{eq:atlas:eta}
\end{equation}
which equals zero at $\theta = 90^\circ$ (perpendicular to the beam) and tends to
$\pm\infty$ as $\theta \to 0^\circ$ or $180^\circ$ (along the beam axis).
For massive particles, the Lorentz-invariant generalisation is the \emph{rapidity}:
\begin{equation}
  y \;=\; \frac{1}{2}\ln\frac{E + p_z}{E - p_z},
  \label{eq:atlas:rapidity}
\end{equation}
where $E$ is the particle energy and $p_z$ is its momentum component along the beam.
In the ultrarelativistic (massless) limit $E \approx |\vec{p}\,|$, the rapidity reduces
to the pseudorapidity: $y \to \eta$.
Differences in rapidity are Lorentz-invariant under longitudinal boosts, making $\eta$
and $\phi$ the natural coordinates for characterising angular distributions at a hadron
collider.

The angular separation between two objects is quantified by
\begin{equation}
  \Delta R \;=\; \sqrt{(\Delta\eta)^2 + (\Delta\phi)^2},
  \label{eq:atlas:deltar}
\end{equation}
which is used throughout this thesis for jet reconstruction, isolation requirements, and
object--track matching.
The transverse momentum $\pT$ and transverse energy $\ET$ are the projections of
momentum and energy onto the plane perpendicular to the beam axis.

\subsection{The Inner Detector}
\label{sec:atlas:id}

The Inner Detector (ID) provides precision tracking of charged particles at
$|\eta| < 2.5$, enclosed within the 2\,T axial field of the central solenoid.
A charged particle traversing the solenoid follows a helical trajectory whose transverse
curvature is inversely proportional to its transverse momentum:
\begin{equation}
  \pT\;[\mathrm{GeV}/\mathrm{c}] \;=\; 0.3 \cdot B\;[\mathrm{T}] \cdot R\;[\mathrm{m}],
  \label{eq:id:pt}
\end{equation}
where $R$ is the radius of curvature of the helix projected onto the transverse plane.
In practice, curvature is extracted from the \emph{sagitta} of the reconstructed track:
the perpendicular deviation of the track's midpoint from a straight line connecting its
endpoints.
For a track of chord length $L$ in a field $B$,
\begin{equation}
  s \;=\; \frac{L^2}{8R}
  \quad\implies\quad
  \pT \;=\; \frac{0.3\, B\, L^2}{8\, s},
  \label{eq:id:sagitta}
\end{equation}
so the relative momentum resolution $\sigma(\pT)/\pT \propto \sigma(s)/s \propto \pT$
degrades at high $\pT$ as the track becomes increasingly straight and the sagitta
shrinks towards the spatial resolution of individual sensors.

\begin{figure}[htbp]
  \centering
  \includegraphics[width=0.80\textwidth]{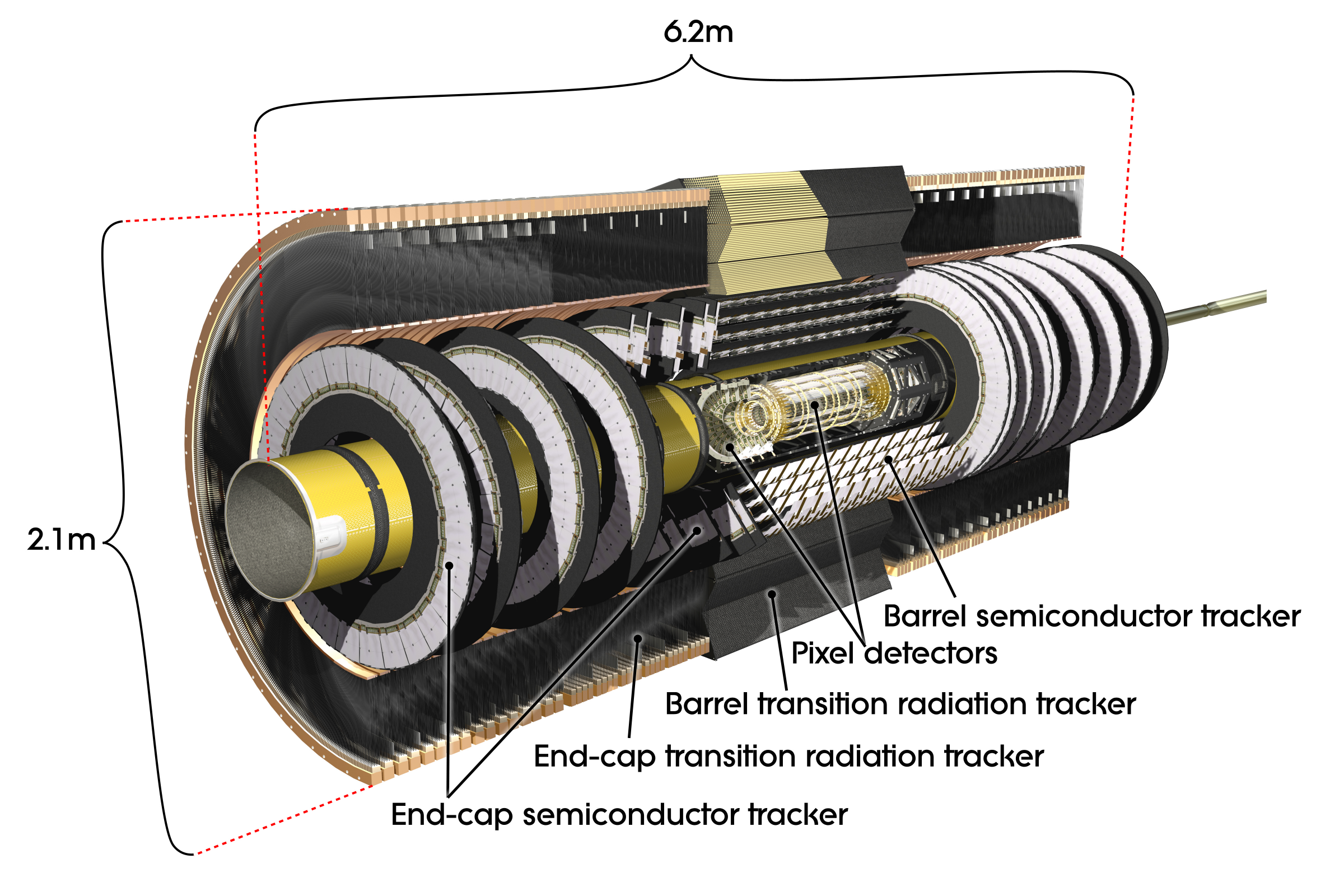}
  \caption{Cutaway schematic of the ATLAS Inner Detector~\cite{ATLAS:2008xda}.
    From innermost to outermost: the three pixel barrel layers,
    four Semiconductor Tracker barrel layers, and the Transition Radiation Tracker straw-tube
    volume, all enclosed by the 2\,T central solenoid.
    An additional innermost pixel layer --- the Insertable B-Layer at
    $r = 33.25$\,mm --- was installed during Long Shutdown~1~\cite{ATL-TDR-019}.}
  \label{fig:atlas:id}
\end{figure}

The ID comprises three complementary detector technologies, illustrated in
FIG.~\ref{fig:atlas:id}.

\paragraph{Pixel Detector and Insertable B-Layer.}
The innermost tracking system consists of silicon pixel sensors segmented into
$400 \times 50\,\mu\mathrm{m}^2$ ($z \times \phi$) cells, arranged in three barrel layers
at radii of 50.5, 88.5, and 122.5\,mm, with three disc layers per end-cap.
For Run~2, a fourth innermost layer --- the Insertable B-Layer
(IBL)~\cite{ATL-TDR-019} --- was installed at $r = 33.25$\,mm inside a newly fabricated
narrow beam pipe, made possible by a complete replacement of the beam pipe during the first
long shutdown.
IBL planar sensors use finer $250 \times 50\,\mu\mathrm{m}^2$ pixels, providing the
highest-precision space-points at the smallest radius.
Together, the four pixel layers deliver a transverse impact parameter resolution of
$\sigma(d_0) \lesssim 15\,\mu\mathrm{m}$ for high-$\pT$ tracks, essential for
resolving displaced secondary vertices from $b$-hadron decays.

\paragraph{Semiconductor Tracker \textmd{(SCT)}.}
Surrounding the pixel system are four barrel layers and nine end-cap disc layers per side,
each comprising pairs of silicon microstrip modules bonded back-to-back with a stereo angle
of $\pm 40$\,mrad between the two strip directions.
The 80\,$\mu$m strip pitch gives a spatial resolution of approximately 17\,$\mu$m in the
bending ($R$--$\phi$) plane and 580\,$\mu$m along $z$.
The SCT contributes four additional high-precision space-points per track in the barrel,
extending the tracking lever arm beyond the pixel volume.

\paragraph{Transition Radiation Tracker \textmd{(TRT)}.}
The outermost ID component consists of approximately 300{,}000 thin-walled polyimide
\emph{straw drift tubes} of 4\,mm diameter, filled with a Xe-based gas mixture.
Up to 36 drift-time measurements per track are provided with an intrinsic resolution of
$\sim\!130\,\mu\mathrm{m}$ in $R$--$\phi$, and the large radial extent
($51 \lesssim r \lesssim 105$\,cm in the barrel) contributes substantially to the overall
momentum determination through the extended lever arm.
In addition, the TRT exploits \emph{transition radiation}: highly relativistic particles
--- predominantly electrons with Lorentz factor $\gamma \gtrsim 1000$ --- crossing
interfaces between materials of different dielectric constant emit X-ray photons.
These produce characteristically high signal amplitudes in neighbouring straws,
providing powerful electron--pion separation supplementing the
shower-based identification in the electromagnetic calorimeter.

The full ID achieves a transverse momentum resolution of
$\sigma(\pT)/\pT \approx (0.05\%\cdot\pT\,[\GeV]) \oplus 1\%$
for isolated tracks in the barrel~\cite{ATLAS:2008xda}.

\subsubsection*{Primary and Secondary Vertices}

A central output of ID tracking is the three-dimensional reconstruction of interaction
vertices.
The \emph{primary vertex} (PV) of the hard scattering is the convergence point of the
tracks produced in the hardest collision of the bunch crossing; in pileup-rich conditions,
$\mathcal{O}(30\text{--}70)$ vertices may be reconstructed per event.
The PV is selected from this set as the vertex with the largest $\sum \pT^2$ of associated
tracks, as described in Chapter~\ref{chp:objects}.
The signed transverse distance of a track from the PV --- the \emph{transverse impact
parameter} $\dzero$ --- and its longitudinal projection $z_0 \sin\theta$ quantify the
degree of track displacement from the primary interaction.

\emph{Secondary vertices} arise from the delayed decays of long-lived particles:
$b$-hadrons have typical displacements $c\tau \sim 400$--500\,$\mu$m, $c$-hadrons
$c\tau \sim 100$--300\,$\mu$m, while strange hadrons and photon conversions produce
vertices at even larger radii.
The IBL's proximity to the interaction point is critical for resolving SVs displaced by
only a few hundred microns from the PV.

\subsection{The Calorimeter System}
\label{sec:atlas:calo}

\begin{figure}[htbp]
  \centering
  \includegraphics[width=0.80\textwidth]{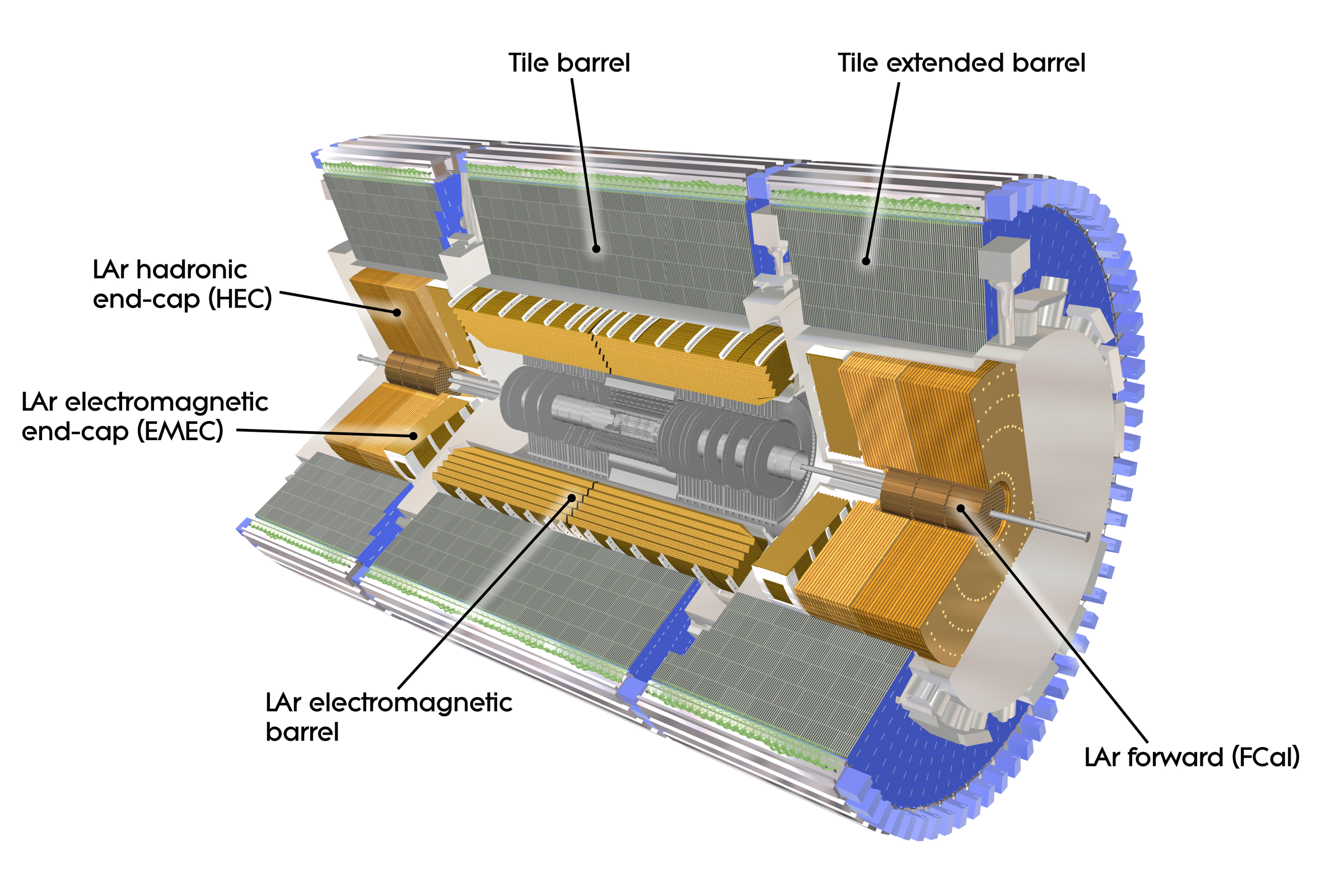}
  \caption{Cutaway schematic of the ATLAS calorimeter system~\cite{ATLAS:2008xda}.
    The lead--LAr electromagnetic calorimeter (LAr barrel and end-cap, EMEC) covers
    $|\eta| < 3.2$.
    The steel-scintillator Tile calorimeter covers $|\eta| < 1.7$.
    The hadronic end-cap (HEC) and forward calorimeters (FCal), both liquid-argon based,
    extend the hadronic coverage to $|\eta| < 4.9$.}
  \label{fig:atlas:calo}
\end{figure}

Surrounding the solenoid is the ATLAS calorimeter system, which measures the energies of
electrons, photons, and hadrons through the total absorption of their associated particle
showers.
The system spans $|\eta| < 4.9$ and is divided into an electromagnetic (EM) calorimeter,
which stops electrons and photons, and a hadronic calorimeter, which absorbs
strongly-interacting particles.
Both employ a \emph{sampling} design, in which dense passive absorber layers interleave
with thinner active detector layers.
A schematic of the full system is shown in FIG.~\ref{fig:atlas:calo}.

\subsubsection*{Electromagnetic Calorimeter}

Electrons and photons traversing dense material initiate \emph{electromagnetic showers}:
cascades of bremsstrahlung photons and electron--positron pairs that multiply rapidly until
the average particle energy falls below the material-dependent critical energy $E_{\mathrm{c}}$, at
which point ionisation and excitation losses overtake radiation losses and the shower terminates.
The longitudinal extent of the shower grows logarithmically as $\sim\ln(E/E_{\mathrm{c}})$ in units
of the radiation length $X_0$.

The mean energy deposited per unit length by a charged particle traversing matter is
described by the Bethe--Bloch formula~\cite{ParticleDataGroup:2024cfk}:
\begin{equation}
  -\!\left\langle\frac{dE}{dx}\right\rangle \;=\;
    K z^2 \frac{Z}{A} \frac{1}{\beta^2}
    \left[
      \frac{1}{2}\ln\frac{2 m_e c^2 \beta^2 \gamma^2 T_{\mathrm{max}}}{I^2}
      - \beta^2 - \frac{\delta(\beta\gamma)}{2}
    \right],
  \label{eq:atlas:bethe}
\end{equation}
where $K = 0.307\,\mathrm{MeV\,mol}^{-1}\,\mathrm{cm}^2$ is a universal constant;
$z$ is the projectile charge in units of $e$; $Z/A$ is the absorber's atomic number-to-mass
ratio; $\beta$ and $\gamma$ are the standard relativistic factors; $m_e$ is the electron
mass; $T_{\mathrm{max}}$ is the maximum kinetic energy transferable to a free electron in
a single collision; $I$ is the mean excitation energy of the absorber; and
$\delta(\beta\gamma)$ is a density-effect correction important at high energies.
By surrounding the active liquid argon with a sufficient number of radiation lengths of a dense absorber, the calorimeter contains the shower entirely and the integrated ionisation
signal becomes a precise measurement of the incident particle's energy.

The ATLAS EM calorimeter uses \emph{lead} as absorber
($\rho = 11.35\,\mathrm{g\,cm}^{-3}$, $X_0 = 5.6$\,mm) and \emph{liquid argon} (LAr)
as the active sampling medium.
The electrodes are shaped into an \emph{accordion} geometry, providing complete $\phi$
coverage without azimuthal cracks and allowing fast signal read-out from both ends.
The system is organised as follows:
\begin{itemize}
  \item \textbf{EM Barrel} ($|\eta| < 1.475$): depth exceeds $22\,X_0$, segmented
    longitudinally into three layers.
    The first (strip) layer has the finest $\eta$-granularity
    ($\Delta\eta \approx 0.003$) for resolving pairs of closely spaced photons from
    $\pi^0$ decays; the middle layer collects the bulk of the shower energy; the back
    layer samples the shower tail for containment corrections.
  \item \textbf{EM End-Cap} (EMEC, $1.375 < |\eta| < 3.2$): two coaxial wheels per side,
    similarly segmented into three longitudinal layers within the precision region
    $|\eta| < 2.5$.
  \item \textbf{Presampler} ($|\eta| < 1.8$): a thin active LAr layer upstream of the
    accordion absorber, correcting for energy lost in the solenoid and cryostat material
    before the active volume.
\end{itemize}
The transition region $1.37 < |\eta| < 1.52$, where the barrel and end-cap cryostats
overlap, suffers from increased inactive material and degraded resolution; objects
in this region are typically excluded from precision analyses.
The EM energy resolution is parametrised as
\begin{equation}
  \frac{\sigma(E)}{E} \;=\; \frac{a}{\sqrt{E\,[\GeV]}} \oplus b,
  \label{eq:atlas:em_res}
\end{equation}
with a sampling term $a \approx 10\%/\sqrt{E}$ and a constant term $b \approx 0.7\%$,
the latter arising from non-uniformities in detector response and calibration~\cite{ATLAS:2008xda}.

\subsubsection*{Hadronic Calorimeter}

Quarks and gluons produced in hard interactions manifest as collimated sprays of
hadrons --- \emph{jets} --- that initiate \emph{hadronic showers} upon striking nuclei
in the absorber.
Hadronic showers involve nuclear breakup, spallation, and secondary hadronic interactions,
making them broader and longer than EM showers; they are characterised by the nuclear
interaction length $\lambda_I \gg X_0$.
Containing them requires substantially-greater material: ATLAS provides more than
$7\lambda_I$ of total calorimeter depth in the barrel region.

The \textbf{Tile calorimeter} covers $|\eta| < 1.7$ and uses \emph{steel} as absorber
and \emph{plastic scintillating tiles} as the active medium.
Wavelength-shifting fibres collect the scintillation light and guide it to
photomultiplier tubes at the calorimeter's outer edge.
The detector is divided into a central barrel ($|\eta| < 1.0$) and two extended barrels
($0.8 < |\eta| < 1.7$), with a small geometric overlap between the two extended barrels
and the barrel to ensure coverage continuity.
The hadronic energy resolution for pions is approximately
\begin{equation}
  \frac{\sigma(E)}{E} \;\approx\; \frac{50\%}{\sqrt{E\,[\GeV]}} \oplus 3\%.
  \label{eq:atlas:had_res}
\end{equation}

Two LAr-based detectors, also shown in FIG.~\ref{fig:atlas:calo}, extend the hadronic
coverage to larger pseudorapidities.
The \textbf{Hadronic End-Cap} (HEC, $1.5 < |\eta| < 3.2$) uses copper absorber plates
with LAr-filled active gaps, arranged in two independent wheels per side.
The \textbf{Forward Calorimeter} (FCal, $3.1 < |\eta| < 4.9$) is a high-density,
radiation-hard system with copper absorbers in the first (EM) module and tungsten
absorbers in the second and third (hadronic) modules, surrounding narrow coaxial
LAr-filled tubes chosen for their short radiation length and resistance to radiation
damage in the high-flux forward region.

Together, the EM and hadronic calorimeters provide the hermetic energy measurement
required to reconstruct jets and missing transverse momentum $\Etmiss$ --- the
momentum imbalance in the transverse plane that serves as the experimental signature of
weakly interacting particles escaping the detector, whether neutrinos in SM processes or
new BSM candidates.

\subsection{The Muon Spectrometer}
\label{sec:atlas:ms}

\begin{figure}[htbp]
  \centering
  \includegraphics[width=0.80\textwidth]{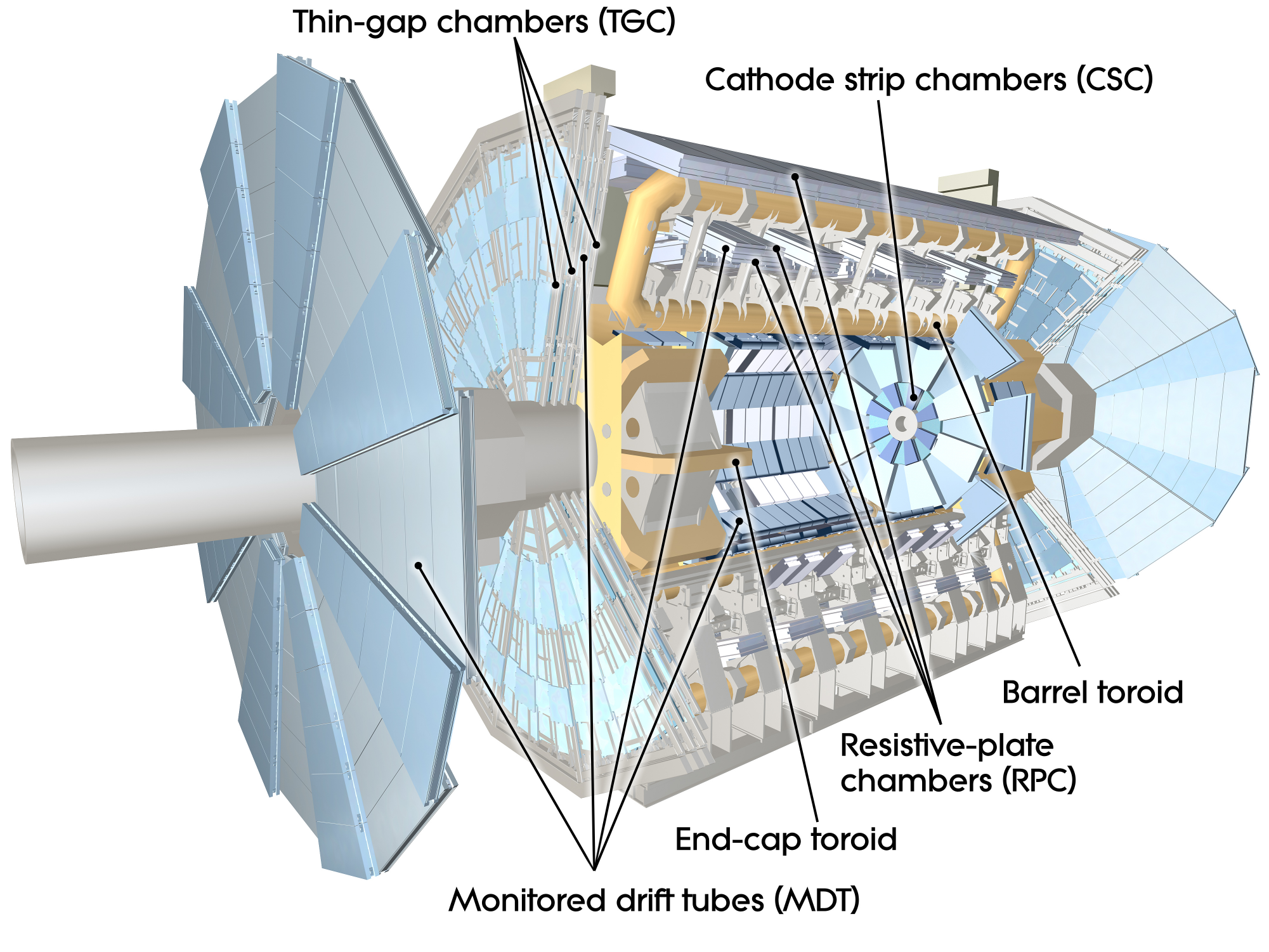}
  \caption{Three-dimensional cutaway render of the ATLAS Muon Spectrometer~\cite{ATLAS:2008xda}.
    Three stations of Monitored Drift Tube precision-tracking chambers are embedded in the
    toroidal field.
    Cathode Strip Chambers occupy the innermost end-cap layer at $|\eta| > 2.0$.
    Resistive Plate Chamber and Thin Gap Chamber trigger detectors provide the Level-1
    muon trigger in the barrel and end-cap regions, respectively.
    The original Small Wheel (innermost end-cap station) was replaced by the New Small
    Wheel for Run~3.}
  \label{fig:atlas:muon}
\end{figure}

\begin{figure}[htbp]
  \centering
  \includegraphics[width=0.92\textwidth]{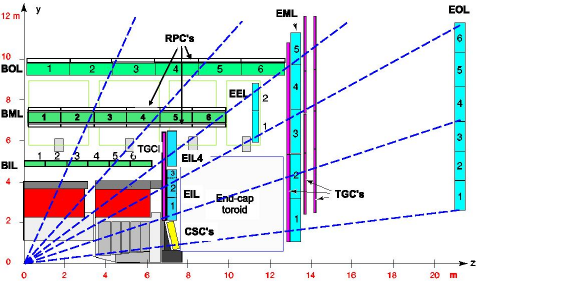}
  \caption{Cross-section of the ATLAS Muon Spectrometer in the $R$--$z$ plane (bending
    plane)~\cite{ATLAS:2008xda}.
    Dashed lines indicate trajectories of high-momentum muon tracks, defining the
    $\eta$ coverage of each station.
    In the barrel, three concentric cylindrical shells of Monitored Drift Tube chambers
    (labelled BIL/BIS, BML/BMS, BOL/BOS) span $|\eta| \lesssim 1.05$, with Resistive
    Plate Chamber trigger detectors interleaved.
    In the end-cap, three wheels of Monitored Drift Tube chambers (EIL, EML, EOL) cover
    $1.05 \lesssim |\eta| \lesssim 2.0$; beyond $|\eta| > 2.0$ the innermost
    end-cap layer, the Small Wheel, uses Cathode Strip Chambers for their higher rate
    capability.
    Thin Gap Chamber wheels provide the end-cap Level-1 muon trigger.
    The Small Wheel was replaced by the New Small Wheel for Run~3.}
  \label{fig:atlas:muon:rzxsec}
\end{figure}

Muons are minimum-ionising particles that traverse the entire calorimeter system
depositing only a few GeV through ionisation and excitation before emerging into the outermost layer
of the detector.
The Muon Spectrometer (MS)~\cite{ATLAS:2008xda} provides standalone momentum measurement
for muons with $|\eta| < 2.7$ and trigger coverage up to $|\eta| < 2.4$.
A three-dimensional cutaway of the full system is shown in FIG.~\ref{fig:atlas:muon};
the chamber layout in the bending plane ($R$--$z$) is illustrated in
FIG.~\ref{fig:atlas:muon:rzxsec}, which makes the $\eta$ coverage of each
station type explicit.

\subsubsection*{Momentum Measurement via Sagitta}

The MS measures muon momentum through the deflection of muon tracks in the toroidal
magnetic field described in Section~\ref{sec:atlas:magnets}, applying the same sagitta principle
introduced for the ID in Eq.~\eqref{eq:id:sagitta}.
A muon traversing the spectrometer experiences an integrated bending field
$\int B\,\mathrm{d}l$ (in T$\cdot$m); the resulting sagitta measured across three precision
tracking stations separated by a lever arm $L$ is
\begin{equation}
  s \;=\; \frac{0.3 \cdot \bigl(\int B\,\mathrm{d}l\bigr) \cdot L}{8\;\pT},
  \label{eq:ms:sagitta}
\end{equation}
with $\pT$ in $\mathrm{GeV}/\mathrm{c}$, $\int B\,\mathrm{d}l$ in T$\cdot$m, and $L$, $s$ in metres.
For a 1\,\TeV\ muon, the sagitta is of order 0.5\,mm over a typical station separation
of $\sim\!5$\,m, placing stringent demands on the precision and alignment of the tracking
chambers.
The standalone MS momentum resolution is $\sigma(\pT)/\pT \sim 10\%$ at 1\,\TeV; combined
with the ID measurement, the resolution improves to $\sigma(\pT)/\pT \approx 2$--$3\%$
over the bulk of the kinematic range relevant to this thesis~\cite{PERF-2015-10}.

\subsubsection*{Detector Technologies}

Four technologies collectively provide precision tracking and fast triggering across the
full pseudorapidity acceptance:

\paragraph{Monitored Drift Tubes \textmd{(MDT)}.}
The primary precision tracking element in both the barrel ($|\eta| < 2.0$) and most of the
end-cap ($|\eta| < 2.7$).
MDTs are aluminium drift tubes of 34\,mm diameter and up to 6\,m length, filled with an
Ar--CO$_2$ gas mixture (93:7) at 3\,bar absolute pressure.
A central tungsten-rhenium sense wire held at 3{,}080\,V collects the ionisation avalanche.
Measured drift times give the particle's minimum distance to the wire with a single-tube
resolution of $\sim\!80\,\mu\mathrm{m}$; combining many tubes across three stations yields
a track sagitta precision of $\lesssim 50\,\mu\mathrm{m}$.

\paragraph{Cathode Strip Chambers \textmd{(CSC)}.}
In the innermost end-cap layer ($2.0 < |\eta| < 2.7$), where the muon flux is highest,
MDTs are replaced by multi-wire proportional chambers with segmented cathode strip readout.
Charge interpolation across adjacent strips gives an intrinsic bending-plane resolution of
$\sim\!60\,\mu\mathrm{m}$, together with the sub-25\,ns response time needed to manage
the high occupancy of this forward region.

\paragraph{Resistive Plate Chambers \textmd{(RPC)}.}
Covering the barrel region $|\eta| < 1.05$, RPCs are parallel-plate gaseous detectors
with resistive Bakelite electrodes operated in avalanche mode.
They provide the Level-1 barrel muon trigger with a spatial resolution of $\sim\!1$\,cm
and a timing resolution better than 1\,ns, sufficient to unambiguously associate trigger
candidates with the correct 25\,ns bunch crossing.

\paragraph{Thin Gap Chambers \textmd{(TGC)}.}
Multi-wire proportional chambers in which the wire-to-cathode gap is smaller than the
wire pitch, providing fast signal collection.
TGCs cover $1.05 < |\eta| < 2.4$ and deliver the Level-1 end-cap muon trigger; a
separate innermost TGC layer also provides precise $\phi$ and $\eta$ measurements used
for the precision end-cap tracking alignment.

\subsection{The New Small Wheel}
\label{sec:atlas:nsw}

During Run~2, the inner end-cap stations of the Muon Spectrometer --- known as the
\emph{Small Wheel} --- consisted of one CSC layer for precision tracking and three
TGC wheels for triggering, covering $1.3 < |\eta| < 2.4$.
At the high luminosities delivered in Run~2 and, prospectively, at the even higher
luminosities of Run~3 and the High-Luminosity LHC (HL-LHC), this arrangement
suffered from a critical limitation.
The TGC trigger in the Small Wheel could not independently verify that a Level-1
muon candidate pointed back towards the interaction point: any charged particle
traversing the TGC wheels at the correct momentum threshold would generate a trigger
primitive regardless of its true origin.
This admitted large rates of \emph{fake} Level-1 muon candidates arising from
punch-through hadrons, cavern neutron background converted to low-energy electrons,
and secondary particles produced in material upstream of the Small Wheel ---
collectively consuming a disproportionate share of the Level-1 trigger bandwidth
and limiting the usable $p_\mathrm{T}$ thresholds.

\begin{figure}[htbp]
  \centering
  \includegraphics[width=0.70\textwidth]{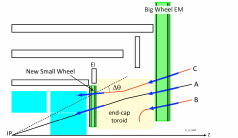}
  \caption{Schematic of the ATLAS muon end-cap showing the position of the
    New Small Wheel (NSW) relative to the outer Big Wheel and the end-cap
    toroid~\cite{Tsigaridas:2025nsw}.
    Three track types illustrate the trigger-pointing requirement: track~A originates
    from the interaction point (IP) and is accepted by both the NSW and the Big Wheel;
    track~B is a fake candidate that satisfies the Big Wheel threshold but fails the
    NSW pointing cut and is rejected; track~C points away from the IP and is
    rejected by the NSW.}
  \label{fig:atlas:nsw:schematic}
\end{figure}

To address these limitations, the \textbf{New Small Wheel} (NSW)~\cite{ATL-TDR-020}
was designed, constructed, and installed during Long Shutdown~2 (LS2, 2019--2022),
replacing both Small Wheels before the start of Run~3 in 2022.
The NSW occupies the same geometric position as its predecessor, at $|z| \approx 7$\,m
from the interaction point, covering $1.3 < |\eta| < 2.7$ (with trigger coverage
to $|\eta| < 2.4$).
A schematic illustrating the NSW in the context of the muon end-cap is shown in
FIG.~\ref{fig:atlas:nsw:schematic}.

Each NSW incorporates two distinct detector technologies, chosen for their complementary
strengths in precision tracking and fast triggering:

\paragraph{Micromegas \textmd{(MM)}.}
A gaseous detector comprising a thin metallic mesh suspended $\sim\!0.128$\,mm above
a segmented copper strip cathode.
Ionisation electrons produced in the drift region above the mesh are amplified in this
narrow gap; charge induction on the strips gives a spatial resolution of
$\lesssim\!100\,\mu\mathrm{m}$ per layer in the precision ($\eta$) direction.
Four MM layers per NSW provide precision track segments for both reconstruction
and triggering.

\paragraph{Small-strip Thin Gap Chambers \textmd{(sTGC)}.}
Multi-wire proportional chambers with a wire-to-cathode gap (1.4\,mm) smaller than the
wire pitch (1.8\,mm), providing fast charge collection.
Segmented cathode pads deliver fast trigger primitives within $\sim\!25$\,ns, while
finer cathode strips yield precision position measurements.
Four sTGC layers per NSW complement the MM tracking layers and supply the
bunch-crossing-identified coincidence signals required for the Level-1 trigger.

\begin{figure}[htbp]
  \centering
  \includegraphics[width=0.92\textwidth]{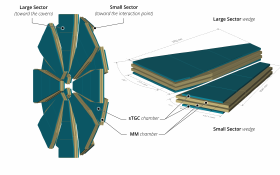}
  \caption{Illustration of one New Small Wheel showing the sector geometry~\cite{Tsigaridas:2025nsw}.
    (Left) Face-on view of the New Small Wheel, with eight large sectors (toward the cavern)
    and eight small sectors (toward the interaction point) arranged with 16-fold
    $\phi$-symmetry.
    (Right) Exploded view of a large-sector wedge (3702\,mm) and a small-sector
    wedge (3547\,mm), showing the small-strip Thin Gap Chamber and Micromegas chamber
    layers that together provide eight active detection planes per sector.}
  \label{fig:atlas:nsw:wheel}
\end{figure}

Each NSW consists of two mechanically independent wheels, each comprising eight large
and eight small sectors arranged with 16-fold symmetry in $\phi$, as illustrated
in FIG.~\ref{fig:atlas:nsw:wheel}.
Each sector contains a sandwich of two MM quadruplets and two sTGC triplets,
providing eight active MM layers and eight active sTGC layers per NSW.
Both wheels were successfully integrated into the ATLAS data-acquisition, trigger,
and reconstruction systems, and have been operational from the start of Run~3
data-taking in 2022~\cite{Tsigaridas:2025nsw}.

The key physics benefit of the NSW is the \emph{trigger-pointing} requirement.
By demanding that a Level-1 muon candidate produce a consistent track segment in the NSW
pointing towards the interaction point --- in addition to satisfying the Big Wheel
coincidence --- the rate of fake single-muon Level-1 triggers in the end-cap is
substantially suppressed, allowing the $p_\mathrm{T}$ thresholds to be maintained at
values appropriate for the physics programme of Run~3.
The integration of the NSW strip readout into the Level-1 muon trigger architecture,
and a measurement of the resulting trigger efficiency and background rejection in
Run~3 data, are described in Appendix~\ref{app:nsw_trigger}.

\subsection{The Magnet System}
\label{sec:atlas:magnets}

Two complementary superconducting magnet systems provide the bending fields for the
ID and the MS~\cite{ATLAS:2008xda}.

\paragraph{Central Solenoid.}
A single-layer solenoid wound from an Al-stabilised NbTi conductor surrounds the Inner
Detector at a mean radius of 1.22\,m.
Operating at 7.7\,kA, it generates an axial field of 2.0\,T within its 5.8\,m-long,
2.46\,m-bore volume.
To minimise the material budget presented to particles before they enter the calorimeter,
the solenoid is engineered to contribute only $\sim\!0.66\,X_0$ at normal incidence.
The coil is integrated into the same cryostat as the barrel EM calorimeter, eliminating
a separate cryostat wall between the ID and the calorimeter.

\paragraph{Toroid System.}
The toroidal magnetic field pervading the MS is generated by three
separate superconducting toroids --- one barrel and two end-caps --- each operating as
an air-core system without an iron flux return.

The \emph{barrel toroid} consists of eight flat superconducting coils, each 25.3\,m long,
arranged with eight-fold azimuthal symmetry around the calorimeters.
At 20.5\,kA, the stored energy reaches 1.08\,GJ and the peak field in the coil is 3.85\,T.
The integrated bending field experienced by a muon traversing the barrel ranges from
approximately 2.0 to 6.0\,T$\cdot$m depending on $\eta$ and $\phi$, with a mean of
$\sim\!2.5$\,T$\cdot$m.

The two \emph{end-cap toroids}, each housing eight coils in a common cryostat, are
inserted into the ends of the barrel toroid and rotated 22.5$^\circ$ to optimise field
uniformity in the barrel--end-cap overlap region.
Each stores 250\,MJ and provides an integrated field of 4.0--8.0\,T$\cdot$m in the
end-cap muon chambers.

Because the toroid is an open, air-core structure, the field is non-uniform in both
$\eta$ and $\phi$.
A network of several thousand Hall probes distributed throughout the muon volume monitors
the field, and the resulting field map is cross-checked against reconstructed
$Z \to \mu^+\mu^-$ events to validate the momentum calibration at the TeV scale.

\subsection{Trigger and Data Acquisition}
\label{sec:atlas:tdaq}

At a bunch-crossing rate of 40\,MHz, the LHC delivers proton--proton interactions far
faster than any detector system can fully read out and store to disk.
The overwhelming majority of these interactions are soft inelastic scatters with low
momentum transfer, of no direct relevance to searches for rare, high-mass processes.
The ATLAS trigger and data-acquisition (TDAQ) system is responsible for identifying the
small fraction of events containing potentially interesting hard-scattering signatures and
rejecting the rest in real time~\cite{ATLAS:2008xda}.

ATLAS employs a two-level trigger hierarchy:

\paragraph{Level-1 (L1) Trigger.}
The L1 trigger is implemented in custom hardware electronics and operates with a fixed
latency of 2.5\,$\mu$s.
It analyses coarse-granularity calorimeter information --- trigger towers of
$\Delta\eta \times \Delta\phi \approx 0.1 \times 0.1$ --- and hit patterns from the
muon trigger detectors (RPCs and TGCs).
Candidate electrons, photons, muons, taus, jets, and $\Etmiss$ above configurable
energy thresholds are identified; events satisfying any active trigger item are flagged
for full detector read-out.
The L1 accept rate is limited to approximately 100\,kHz by the detector front-end
electronics.

\paragraph{High Level Trigger \textmd{(HLT)}.}
L1-accepted events are passed to the HLT, a software-based system running on a processor
farm of approximately 40{,}000 CPU cores.
The HLT has access to the full detector read-out and applies reconstruction algorithms
approaching offline quality: tracking within regions of interest, calorimeter clustering
at full granularity, and topological selections closely mirroring the offline analysis
requirements.
The HLT reduces the event rate to approximately 1--2\,kHz, the bandwidth available for
permanent storage.

The trigger menu --- the set of physics signatures and associated thresholds active at
any given time --- evolved throughout Run~2 in response to the increasing LHC luminosity.
The precise trigger requirements used in each search presented in this thesis are described
in the corresponding analysis chapters.
\clearpage
\clearpage\chapter{Common Object Definition}
\label{chp:objects}

The particles recorded by the ATLAS detector are reconstructed from raw detector signals into
the physics objects --- electrons, muons, photons, jets and missing transverse momentum ---
used in the analyses presented in this thesis.
Reconstruction algorithms combine information from multiple detector subsystems, and dedicated
calibrations anchor the reconstructed quantities to known physics processes.
A central concept throughout is the \textit{working point} (WP): a specific set of
identification and isolation requirements that balances signal efficiency against background
rejection.
Tighter WPs accept fewer background candidates at the cost of lower signal acceptance,
while looser ones maximise efficiency but admit more fakes or misidentified objects.
The WPs chosen for each analysis in this thesis are detailed in the respective analysis
chapters.

\section{Tracks and Primary Vertices}
\label{sec:reco:tracks}

Charged-particle trajectories in the ID (Section~\ref{sec:atlas:id}) are the primary input
to lepton reconstruction and to pileup suppression throughout the event.
Track reconstruction begins by forming seeds from triplets of space-points in the pixel
detector and SCT.
Seeds are extrapolated outward through the SCT and TRT and refined by a global $\chi^2$ fit.
A complementary outside-in pass recovers secondary tracks from particles produced beyond the
innermost layers, such as photon conversions and $K_S^0$ decays.

Each reconstructed track is described by five parameters evaluated at the point of closest
approach to the beam axis: the transverse impact parameter $d_0$, the longitudinal impact
parameter $z_0$, the azimuthal angle $\phi$, the polar angle $\theta$, and the signed
charge-to-momentum ratio $q/p$.
Two derived quantities are used extensively in object quality selections:
\begin{itemize}
    \item the transverse impact parameter significance
          $|d_0^{\mathrm{BL}}(\sigma)| \equiv |d_0^{\mathrm{BL}}|/\sigma(d_0^{\mathrm{BL}})$,
          where the superscript BL denotes that $d_0$ is measured relative to the beam-line;
    \item the longitudinal projection $|z_0^{\mathrm{BL}}\sin\theta|$, which quantifies the
          displacement along the beam axis from the primary vertex.
\end{itemize}
Requiring small values of both quantities suppresses tracks originating from pileup vertices
or secondary decays.

The PV of each event is selected from the set of reconstructed vertices as
the one with the largest scalar sum of squared transverse momenta of associated tracks,
$\sum p_\text{T}^2$, where only tracks with $p_\text{T} > 0.4$~GeV are
considered~\cite{ATL-PHYS-PUB-2015-026}.
At least two associated tracks are required.

\section{Electrons}
\label{sec:reco:electrons}

\subsection{Reconstruction}

Electron candidates are formed by matching tracks reconstructed in the ID to energy clusters
in the EM calorimeter (Section~\ref{sec:atlas:calo})~\cite{EGAM-2018-01,EGAM-2021-01,EGAM-2021-02}.
The cluster finding uses a sliding-window algorithm that scans the EM calorimeter in windows
of $3\times5$ cells (in $\eta\times\phi$) and retains local maxima above a threshold.
For Run~2 and later data, a supercluster algorithm is applied: satellite clusters from
bremsstrahlung photons emitted along the electron's trajectory are merged with the primary
cluster, recovering energy that would otherwise be lost.
The merged cluster is then matched to the closest ID track compatible with originating from
the PV and with a bending in $\phi$ consistent with energy loss.

Electron candidates are reconstructed in the region $|\eta| < 2.47$, excluding the
calorimeter transition region $1.37 < |\eta| < 1.52$ where the energy resolution is
degraded.
The reconstruction efficiency exceeds 97\% for electrons with $\ET > 15$~GeV.

\subsection{Identification}

Electron identification is performed using a likelihood (LH) discriminant constructed from
a set of input variables characterising the shower shape in the EM calorimeter, the quality
and multiplicity of ID hits, and the match between the electron track and cluster.
The shower-shape variables exploit the fact that genuine electrons deposit most of their
energy in a narrow core, whilst hadronic fakes produce broader and less regular showers.
Track-quality variables (number of hits in the pixel detector and SCT, TRT hit fraction)
and the ratio $E/p$ between cluster energy and track momentum further distinguish prompt
electrons from photon conversions and hadronic backgrounds.

\begin{figure}[ht]
  \centering
  \begin{subfigure}[b]{0.48\textwidth}
    \includegraphics[width=\textwidth]{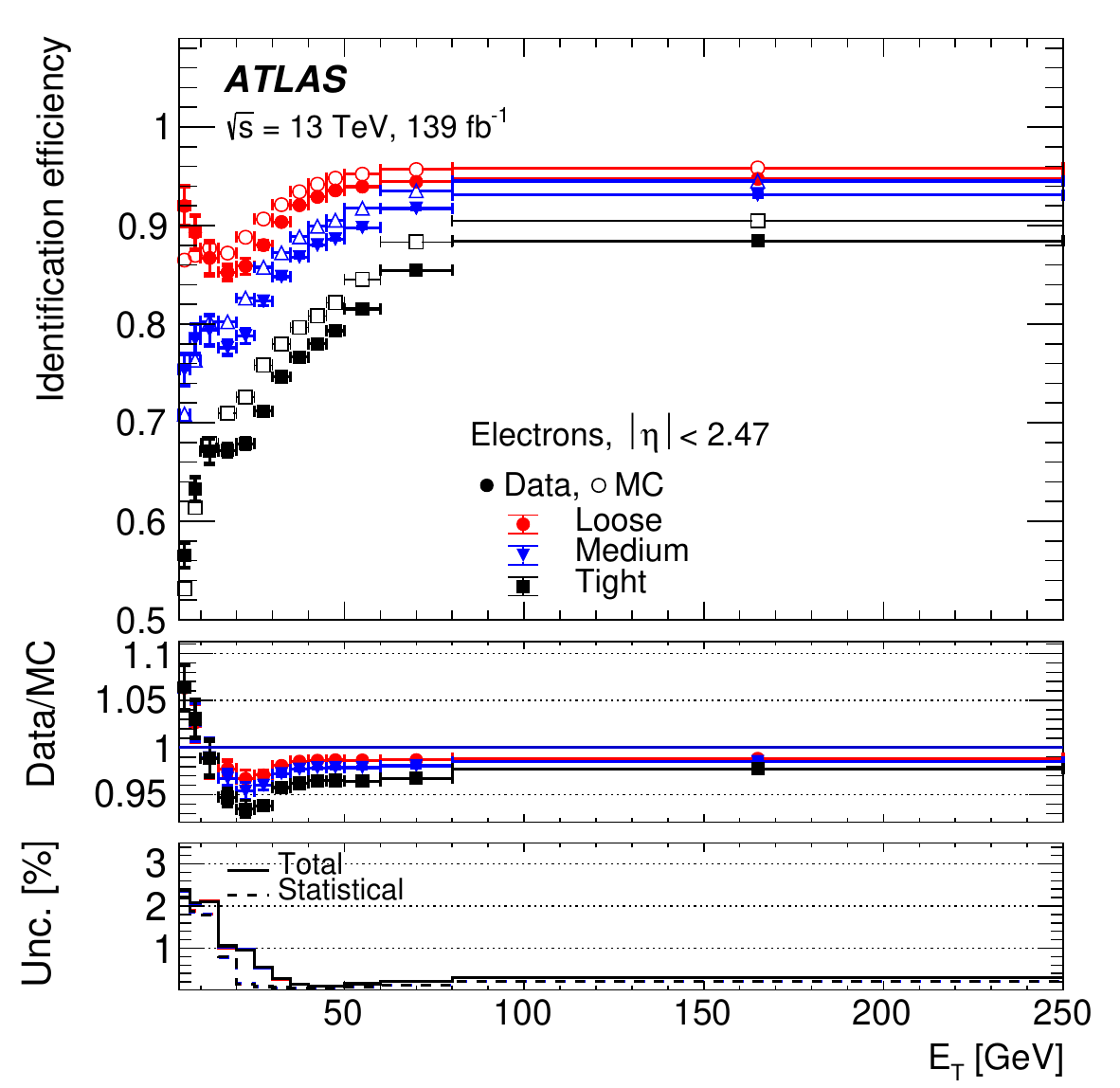}
    \caption{As a function of $\ET$.}
  \end{subfigure}
  \hfill
  \begin{subfigure}[b]{0.48\textwidth}
    \includegraphics[width=\textwidth]{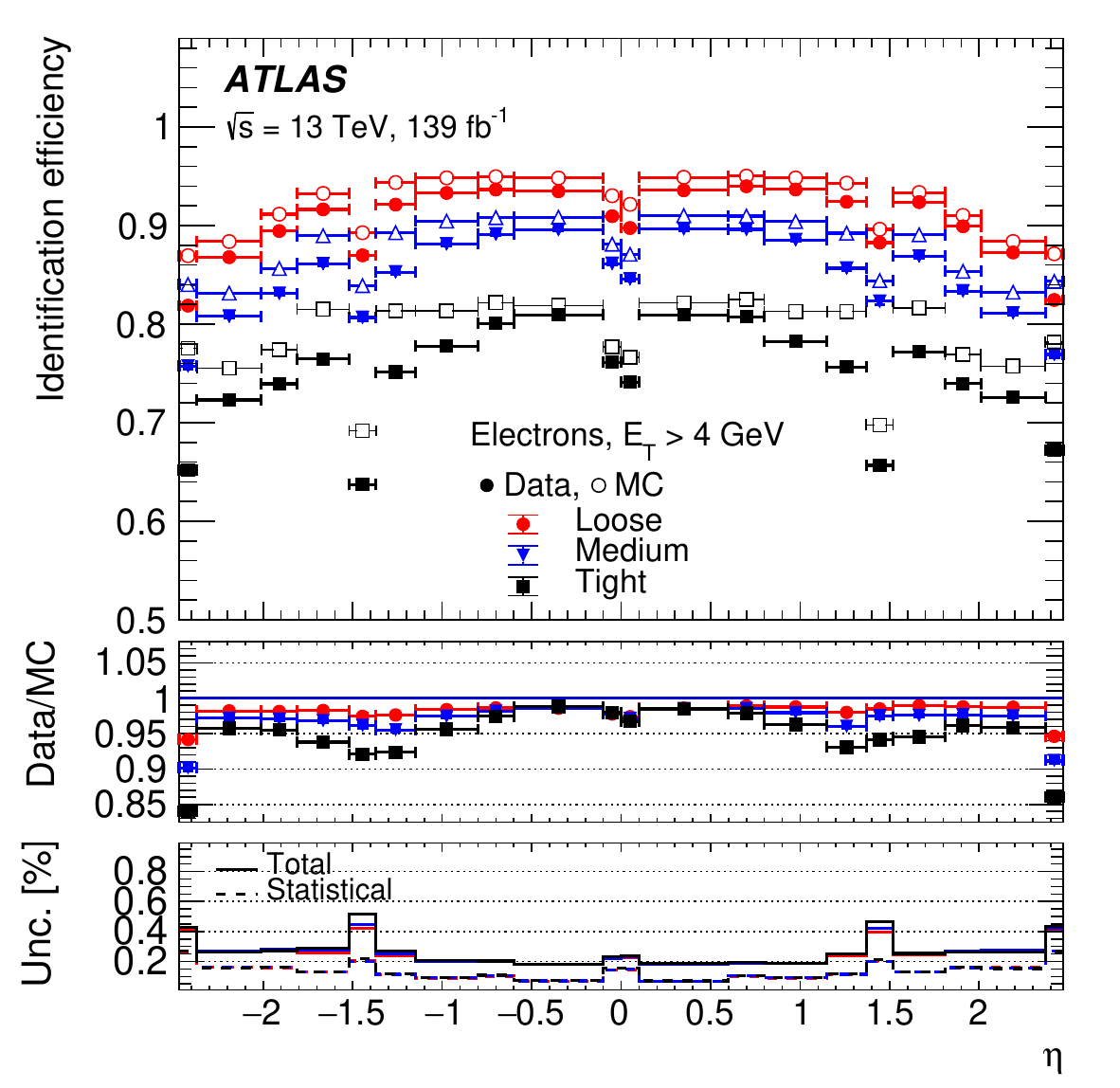}
    \caption{As a function of $\eta$.}
  \end{subfigure}
  \caption{Electron identification efficiency for the Medium and Tight likelihood working points,
    measured in $Z \to ee$ tag-and-probe data and compared to simulation~\cite{EGAM-2021-01}.
    Scale factors correcting residual data--MC differences are derived from the ratio.}
  \label{fig:objects:electron_eff}
\end{figure}

Four identification WPs are defined, in order of increasing purity:
\begin{itemize}
    \item \textbf{Loose} (\texttt{LH\_Loose}): loosest selection, primarily used for
          background studies.
          The \texttt{LooseBL} variant additionally requires at least one hit in the
          innermost pixel layer (B-layer), which suppresses photon conversions producing
          fake electron tracks.
    \item \textbf{Medium} (\texttt{LH\_Medium}): tighter shower-shape and track-matching
          requirements.
          Achieves an identification efficiency of approximately 87--90\% for
          $\ET \gtrsim 15$~GeV, measured in $Z \to ee$ decays~\cite{EGAM-2018-01,EGAM-2021-01}.
    \item \textbf{Tight} (\texttt{LH\_Tight}): the most stringent WP; imposes strict
          requirements on all discriminating variables.
          Achieves an identification efficiency of approximately 78--82\% for
          $\ET \gtrsim 15$~GeV, with the highest rejection of hadronic
          fakes~\cite{EGAM-2018-01,EGAM-2021-01}.
          The default choice for searches requiring high electron purity at high $\pT$.
\end{itemize}
Efficiencies are measured in data using tag-and-probe methods in $Z \to ee$ and
$J/\psi \to ee$ samples and compared to simulation; scale factors are derived to correct any
residual discrepancy~\cite{EGAM-2018-01,EGAM-2021-01}.
The efficiency for the Medium and Tight WPs as a function of $\ET$ and $\eta$ is shown in
Figure~\ref{fig:objects:electron_eff}.

Table~\ref{tab:electron_id_wps} summarises the main properties of each WP.

\begin{table}[htbp]
    \centering
    \caption{Summary of electron identification WPs. Efficiencies are approximate,
      measured for electrons from $Z \to ee$ decays with $\ET \gtrsim 15$~GeV~\cite{EGAM-2018-01,EGAM-2021-01}.}
    \label{tab:electron_id_wps}
    \begin{tabular}{lll}
        \hhline{===}
        WP & Approx.\ Efficiency & Typical Use Case \\
        \hline
        Loose (LooseBL) & $>95\%$       & Background studies, loose preselection \\
        Medium          & $87$--$90\%$  & General purpose \\
        Tight           & $78$--$82\%$  & High-purity searches \\
        \hhline{===}
    \end{tabular}
\end{table}

\subsection{Isolation}

Prompt electrons from $W$ or $Z$ boson decays are separated from non-prompt electrons by
isolation requirements.
Non-prompt electrons arise from three main sources: hadronic jets misidentified as electrons
(primarily due to $\pi^0 \to \gamma\gamma$ decays producing overlapping EM showers), photon
conversions in the detector material ($\gamma \to e^+e^-$), and semi-leptonic decays of
$b$- or $c$-hadrons inside jets, which produce real but displaced electrons.
Two complementary quantities are used:
\begin{itemize}
    \item \textbf{Calorimeter isolation}: the sum of transverse energy deposited in a cone
          of $\Delta R = 0.2$ centred on the electron cluster, excluding the electron's own
          contribution.
    \item \textbf{Track isolation}: the scalar sum of $\pT$ of all tracks within a cone of
          $\Delta R = 0.2$ (or a variable-radius cone), excluding the electron track.
\end{itemize}
Several standard isolation WPs are available~\cite{EGAM-2018-01}:
\begin{itemize}
    \item \textbf{FCLoose / FCTight} (Fixed Cut): both track and calorimeter isolation
          variables are required to be below fixed $\pT$-independent thresholds.
          For \texttt{FCTight}, the calorimeter isolation must satisfy
          $E_\mathrm{T}^\mathrm{cone20}/\ET < 0.06$ and the track isolation
          $p_\mathrm{T}^\mathrm{cone20}/\ET < 0.06$; \texttt{FCLoose} applies looser
          thresholds of $0.20$ and $0.15$ respectively.
          \texttt{FCTight} provides higher purity at the cost of some efficiency loss.
    \item \textbf{HighPtCaloOnly}: applies only a calorimeter-based isolation requirement
          and omits the track-based criterion.
          Designed for very high-$\pT$ electrons, where the narrow opening angle between
          bremsstrahlung photons and nearby activity can cause the variable-cone track sum to
          fail even for genuine electrons.
    \item \textbf{PLImproved} variants: profile-likelihood-based, deriving isolation
          thresholds as a function of $\pT$ and $\eta$ to maintain an approximately flat
          efficiency of $\sim$98\% across the full kinematic range; typically used in
          analyses requiring a precise and uniform efficiency.
\end{itemize}

\subsection{Energy Calibration}

The electron energy scale and resolution are calibrated using a multivariate regression
trained on simulated samples and validated with $Z \to ee$, $J/\psi \to ee$, and
$W \to e\nu$ decays in data~\cite{EGAM-2018-01,EGAM-2021-02}.
Residual corrections to the energy scale and resolution are derived in bins of $\eta$ and
$\ET$ and applied as scale factors.
These corrections are encapsulated in named calibration schemes, whose recommended versions
depend on the data-taking period and software release.

\subsection{Impact Parameter Requirements}

To suppress electrons from pileup interactions or from secondary particle decays
(non-prompt electrons), signal-quality electrons are required to satisfy requirements on
the impact parameters $d_0$ and $z_0$ (defined in Section~\ref{sec:reco:tracks}):
\begin{equation*}
    |z_0^{\mathrm{BL}}\sin\theta| < 0.5~\mathrm{mm}, \qquad |d_0^{\mathrm{BL}}(\sigma)| < 5.
\end{equation*}
The longitudinal cut suppresses electrons from vertices displaced along $z$; the significance
cut on $d_0$ retains prompt electrons (which have $d_0 \approx 0$ with some detector
resolution) whilst rejecting electrons from $b$-hadron decays or photon conversions, which
have genuinely non-zero $d_0$.

\section{Photons}
\label{sec:reco:photons}

\subsection{Reconstruction}

Photon candidates are reconstructed from EM calorimeter clusters that lack a matched
reconstructed track consistent with a prompt charged particle~\cite{EGAM-2018-01}.
They arise in two classes:
\begin{itemize}
    \item \textbf{Unconverted photons}: clusters with no associated conversion vertex.
          These are the most straightforward case --- the photon deposits its energy directly
          in the EM calorimeter.
    \item \textbf{Converted photons}: clusters associated with a conversion vertex
          ($\gamma \to e^+ e^-$) reconstructed in the ID.
          The conversion can be single-track (only one conversion electron is reconstructed)
          or double-track.
          The reconstruction algorithm identifies electron-pair vertices consistent with a
          photon originating from the PV, and classifies the cluster accordingly.
\end{itemize}
Photon reconstruction is performed in the range $|\eta| < 2.37$, again excluding the
barrel-endcap transition region $1.37 < |\eta| < 1.52$.

\subsection{Identification}

Photon identification uses shower-shape variables from the first and second layers of the
EM calorimeter.
Since photons are electrically neutral, no track-based variables enter the discriminant.
The lateral width and fine-grained strip-layer profile of the shower are the primary
discriminants against hadrons (mostly $\pi^0 \to \gamma\gamma$, which produce broader
showers from two overlapping EM deposits).

Two standard WPs are defined:
\begin{itemize}
    \item \textbf{Loose}: relies primarily on second-layer shower shapes, with efficiency
          above 90\% for prompt photons.
    \item \textbf{Tight}: adds first-layer (strip) variables, providing much stronger
          $\pi^0$ rejection at a reduced efficiency of approximately 85\% (unconverted)
          and 90\% (converted) for $\ET > 25$~GeV.
\end{itemize}
Identification efficiencies are measured in data using radiative $Z \to \ell\ell\gamma$
decays and an inclusive photon sample enriched via the electron extrapolation technique.

\subsection{Isolation}

Prompt photons from hard-scatter processes are typically well-separated from surrounding
hadronic activity, whereas photons inside jets or from $\pi^0$ decays are embedded in a
dense environment.
Photon isolation is measured in a cone of $\Delta R = 0.4$ around the photon candidate in
the calorimeter, corrected for the photon's own energy leakage and for the ambient pileup
energy density.
A complementary track isolation variable sums $\pT$ of tracks in a $\Delta R = 0.2$ cone
originating from the PV.

Standard fixed-cut WPs (\texttt{FixedCutLoose}, \texttt{FixedCutTight}) set thresholds on
the calorimeter and track isolation variables as a function of $\ET$.

\section{Muons}
\label{sec:reco:muons}

\subsection{Reconstruction}

Muons are reconstructed by combining information from the ID and the MS
(Section~\ref{sec:atlas:ms}), which comprises the MDT chambers and CSC for precision
tracking, and the RPC and TGC for triggering~\cite{PERF-2015-10}.
Several reconstruction strategies are used, giving rise to distinct muon types:

\begin{itemize}
    \item \textbf{Combined} (CB): an independent track fit is performed in the ID and
          in the MS, and the two tracks are then combined in a global fit that uses the full
          hit information from both detectors.
          CB muons provide the best momentum resolution and highest purity, and constitute
          the majority of reconstructed muons in the pseudorapidity range $|\eta| < 2.5$.

    \item \textbf{Segment-Tagged} (ST): an ID track is extrapolated to the MS and
          required to be associated with at least one MS track segment (a local straight-line
          fit through chamber hits), even if a full MS track cannot be formed.
          This strategy recovers muons at low $\pT$ (below the threshold for MS track
          reconstruction) or in regions of reduced MS acceptance.

    \item \textbf{Calorimeter-Tagged} (CT): an ID track is matched to a calorimeter
          energy deposit consistent with a minimum-ionising particle traversing the detector.
          This type is used in the pseudorapidity range $|\eta| < 0.1$, where the MS has a
          gap to accommodate services for the solenoid and calorimeters.
          CT muons have lower purity but recover acceptance in this uninstrumented region.

    \item \textbf{MS Extrapolated} (ME): reconstructed using only the MS track, extrapolated
          back to the interaction point.
          Used in the range $2.5 < |\eta| < 2.7$, beyond the ID acceptance.
          ME muons have no ID-based track to constrain the vertex association, but their
          coverage extends the muon acceptance in forward pseudorapidity.
\end{itemize}

\begin{figure}[ht]
  \centering
  \begin{subfigure}[b]{0.49\textwidth}
    \includegraphics[width=\textwidth]{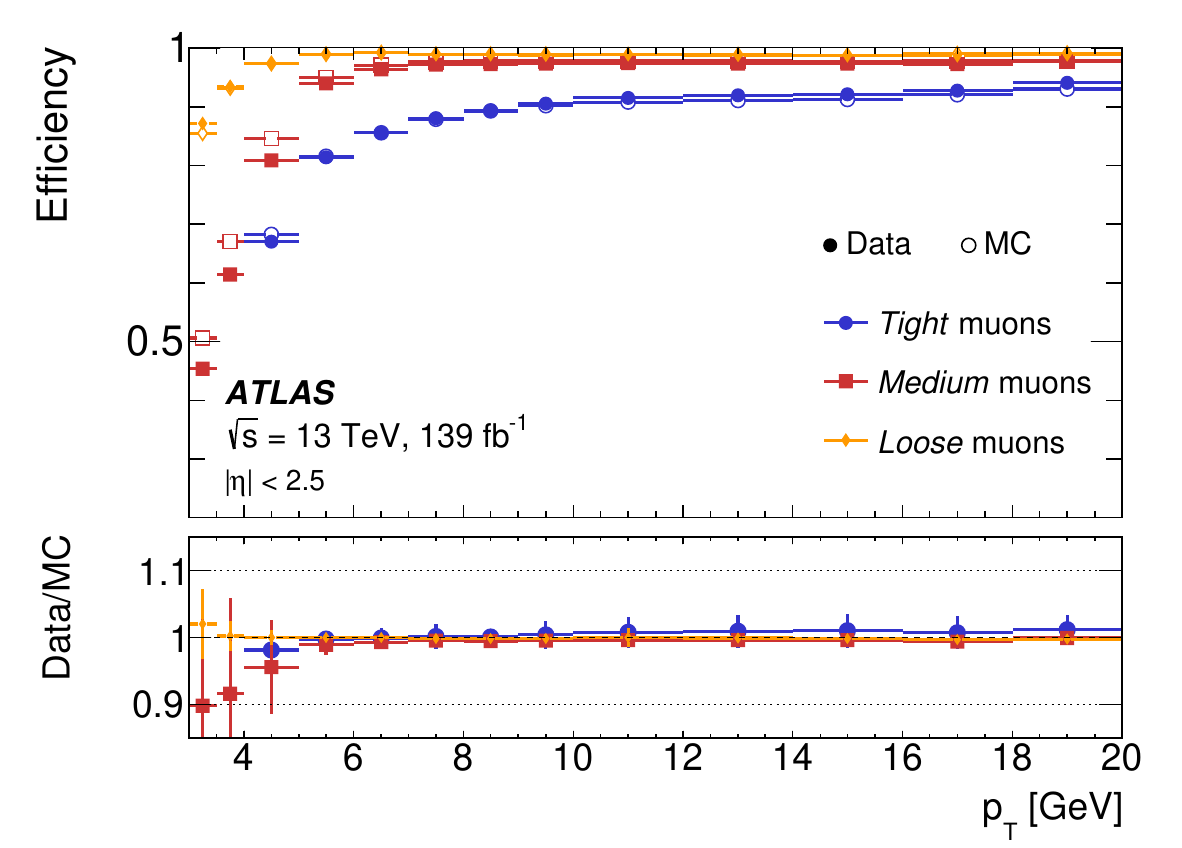}
    \caption{As a function of $\pT$.}
  \end{subfigure}
  \hfill
  \begin{subfigure}[b]{0.49\textwidth}
    \includegraphics[width=\textwidth]{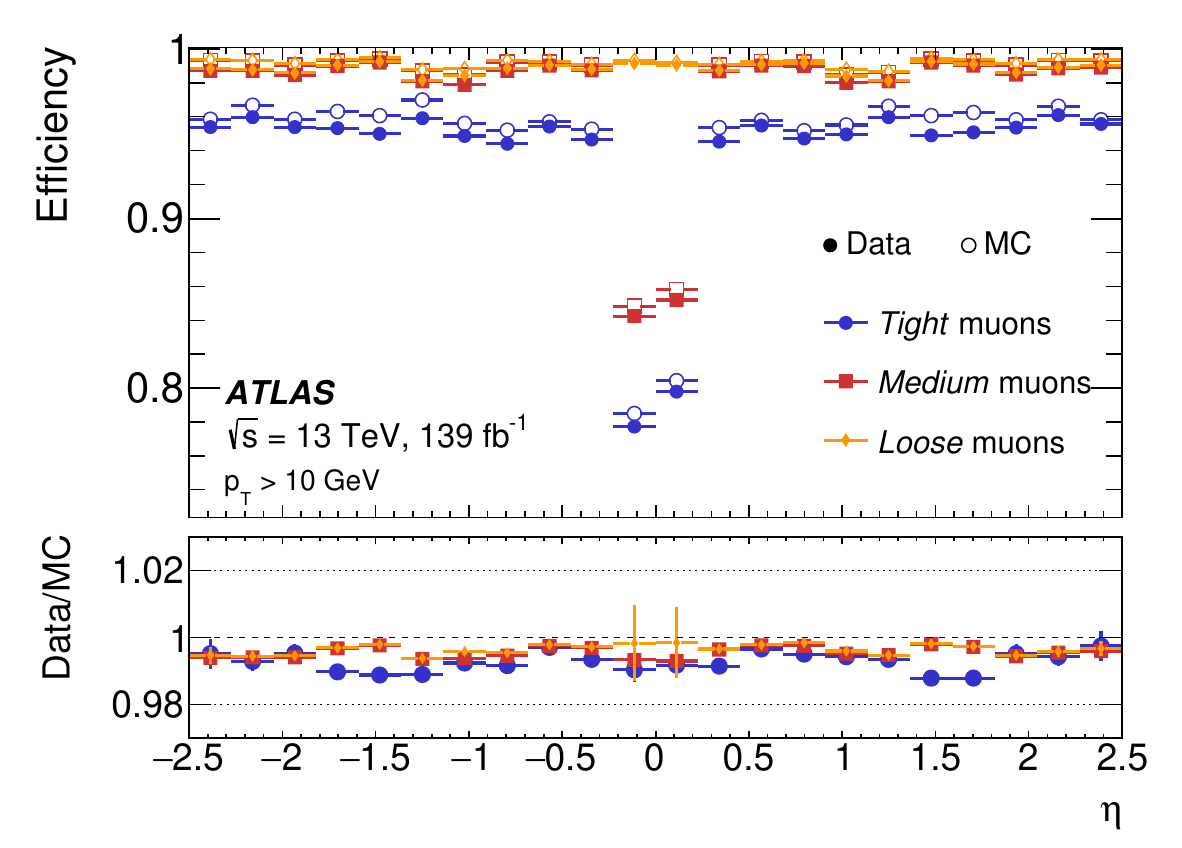}
    \caption{As a function of $\eta$.}
  \end{subfigure}
  \caption{Muon reconstruction and identification efficiency for the Loose, Medium, and Tight
    working points, measured in $Z \to \mu\mu$ tag-and-probe data using the full Run~2
    dataset~\cite{MUON-2018-03}.}
  \label{fig:objects:muon_eff}
\end{figure}

Figure~\ref{fig:objects:muon_eff} illustrates the muon reconstruction efficiency as a
function of $\pT$ and $\eta$ for the Loose, Medium, and Tight working points, measured with
the full Run~2 dataset.

\subsection{Identification}

Four identification WPs are defined, each selecting a subset of muon types with progressively
stricter quality requirements~\cite{PERF-2015-10}:

\begin{itemize}
    \item \textbf{Loose}: accepts CB, ST, CT (in $|\eta| < 0.1$), and ME
          (in $2.5 < |\eta| < 2.7$) muons.
          Provides the highest efficiency, approximately 97.5\% for muons from $Z$ decays.
          Recommended for analyses needing maximum acceptance, such as those with
          low-multiplicity final states at moderate $\pT$.

    \item \textbf{Medium}: the default WP for most ATLAS analyses.
          Accepts CB and ME muons with a minimum number of precision hits in the MS.
          Achieves approximately 96\% efficiency.
          The looser MS hit requirements compared to Tight reduce sensitivity to inactive MS
          regions without significant purity loss.

    \item \textbf{Tight}: a strict subset of Medium, requiring CB muons with matched
          segments in at least two MS stations and tighter track-quality criteria.
          Efficiency is approximately 92\%.
          Recommended when low fake rates are essential.

    \item \textbf{High-$\boldsymbol{p}_\mathrm{T}$} (\texttt{HighPt}): specifically designed for muons in the
          multi-hundred-GeV momentum range, as relevant to high-mass resonance searches.
          Requires hits in all three precision MS stations, providing excellent momentum
          resolution at high $\pT$ where bending angles become small.
          The WP also applies stricter requirements on the ID--MS momentum match, quantified
          by the $q/p$ significance:
          \begin{equation}
              \sigma_{q/p} \;\equiv\;
              \frac{\left|(q/p)^{\mathrm{ID}} - (q/p)^{\mathrm{MS}}\right|}
                   {\sqrt{\,\sigma^2_{q/p,\mathrm{ID}} + \sigma^2_{q/p,\mathrm{MS}}\,}},
              \label{eq:qp_sig}
          \end{equation}
          which measures the consistency between the ID and MS charge-to-momentum
          measurements; the HighPt WP requires $\sigma_{q/p} < 7$~\cite{PERF-2015-10}.
          The efficiency is approximately 79--85\% at $\pT = 100$~GeV, but the momentum
          resolution is substantially better than the Medium WP, which is critical when
          reconstructing narrow resonances at high invariant mass.
\end{itemize}

Table~\ref{tab:muon_id_wps} summarises the main properties of each WP.

\begin{table}[htbp]
    \centering
    \caption{Summary of muon identification WPs.
      Efficiencies are approximate, measured for muons from $Z \to \mu\mu$ decays.
      CT muons are included only in the region $|\eta| < 0.1$;
      ME muons only in $2.5 < |\eta| < 2.7$.
      Combined (CB), Segment-Tagged (ST), Calorimeter-Tagged (CT), MS Extrapolated (ME).}
    \label{tab:muon_id_wps}
    \begin{tabular}{llcl}
        \hhline{====}
        WP         & Muon Types      & Approx.\ Efficiency & Typical Use Case \\
        \hline
        Loose      & CB, ST, CT, ME  & 97.5\%              & Maximum acceptance \\
        Medium     & CB, ME          & 96\%                & General purpose \\
        Tight      & CB (2 stations) & 92\%                & High purity required \\
        High-$\pT$ & CB (3 stations) & 79--85\%            & High-mass searches \\
        \hhline{====}
    \end{tabular}
\end{table}

\subsection{Isolation}

Isolated muons from $W/Z$ boson or resonance decays are separated from non-prompt muons
(primarily from heavy-flavour decays inside jets) using isolation requirements.
Muon isolation is measured using track-based quantities, and optionally calorimeter-based
quantities:
\begin{itemize}
    \item \textbf{Track isolation}: the scalar sum of $\pT$ of ID tracks within a cone
          around the muon, excluding the muon's own track.
          A variable-radius cone, $\Delta R = \min(10~\mathrm{GeV}/\pT^{\mu},\; 0.3)$,
          shrinks at higher $\pT$ to avoid contamination from nearby jets in boosted
          configurations.
    \item \textbf{Particle-flow isolation}: uses particle-flow track objects for improved
          pileup suppression.
    \item \textbf{Calorimeter isolation}: less commonly used for muons but available as a
          complement.
\end{itemize}
Standard isolation WPs include track-only fixed-cut and variable-cone variants, and
particle-flow-based variants; the latter are the recommended default for Run~3 analyses,
where particle-flow reconstruction benefits from improved pileup suppression under the higher
instantaneous luminosity conditions.

\subsection{Momentum Calibration}

Muon momentum is calibrated \textit{in situ} using $Z \to \mu\mu$ decays, which provide a
sharp invariant mass peak whose width is sensitive to the momentum resolution.
Separate corrections are derived for the ID and MS measurements, and the combined muon
momentum scale is anchored to the known $Z$ boson mass~\cite{PERF-2015-10}.
The calibration is provided by the ATLAS Muon Combined Performance (MCP) group in named
calibration schemes reflecting the data-taking conditions.
For analyses using the \texttt{HighPt} WP, a sagitta bias correction may or may not be
applied depending on the analysis strategy, since at very high $\pT$ the curvature
measurement is dominated by ID--MS alignment effects.

Impact parameter requirements for signal muons mirror those for electrons:
\begin{equation*}
    |z_0^{\mathrm{BL}}\sin\theta| < 0.5~\mathrm{mm}, \qquad |d_0^{\mathrm{BL}}(\sigma)| < 3.
\end{equation*}
The slightly tighter $d_0$ significance cut compared to electrons reflects the lower rate of
fake muons from secondary decays and the better ID momentum resolution available at high
$\pT$.

\subsection{Run~2 \textit{vs.}\ Run~3 Definition}
\label{sec:reco:muons:run2run3}

The three physics analyses presented in this thesis span two distinct data-taking periods,
and the muon reconstruction accordingly uses different calibration releases.
The $Z'$ search (Chapter~\ref{chp:zprime}) and the Clockwork search
(Chapter~\ref{chp:clockwork}) are both performed on Run~2 data
($\sqrt{s}=13$~TeV, 2015--2018), and apply the MCP recommendations and calibration scheme
validated for that period.
The QBH search (Chapter~\ref{chp:qbh}) uses Run~3 data ($\sqrt{s}=13.6$~TeV,
2022--2024), and applies the updated Run~3 MCP calibrations, which incorporate
re-optimised momentum corrections and an improved sagitta-bias correction benefiting
high-$\pT$ muons.

In terms of WPs, all three analyses employ the \texttt{HighPt} WP for signal muons, as it
provides the best momentum resolution for multi-hundred-GeV tracks.
For isolation, the Run~2 analyses ($Z'$ and Clockwork) use fixed-cut track-based WPs, while
the Run~3 QBH analysis adopts a particle-flow-based WP, the recommended default for Run~3,
which offers improved pileup suppression under the higher instantaneous luminosity conditions
of that period.
Efficiency scale factors are derived separately for each run period from $Z\to\mu\mu$
tag-and-probe samples and applied to simulated events to correct residual data--MC
discrepancies in identification and isolation efficiencies.

\section{Jets}
\label{sec:reco:jets}

\subsection{Particle Flow and Jet Inputs}

Jets in ATLAS are built from \texttt{particle-flow objects}, which combine ID tracks and
calorimeter energy deposits (Section~\ref{sec:atlas:calo}) to achieve better energy
resolution and pileup suppression than using calorimeter deposits
alone~\cite{PERF-2014-07}.
The particle-flow algorithm proceeds as follows: for each ID track, the expected calorimeter
energy deposit from that charged particle is subtracted from the matched cluster, and the
track momentum is used instead.
The remaining calorimeter energy (from neutral particles or particles not matched to any
track) is retained in the form of topological clusters.
The resulting set of charged-particle tracks and neutral clusters provides the inputs to jet
clustering and is denoted \texttt{EMPFlow}.

Topological clusters are formed by the \textit{topological clustering} algorithm, which
seeds cells with signal-to-noise ratio above 4, then aggregates neighbouring cells above a
threshold of 2, and finally includes all cells immediately surrounding the cluster boundary.
This noise-suppressing procedure produces irregular three-dimensional energy blobs
(\textit{topo-clusters}) in the calorimeter.

\subsection[Jet Clustering: the Anti-$k_t$ Algorithm]{Jet Clustering: the Anti-$\bm{k_t}$ Algorithm}

\begin{figure}[ht]
  \centering
  \includegraphics[width=0.85\textwidth]{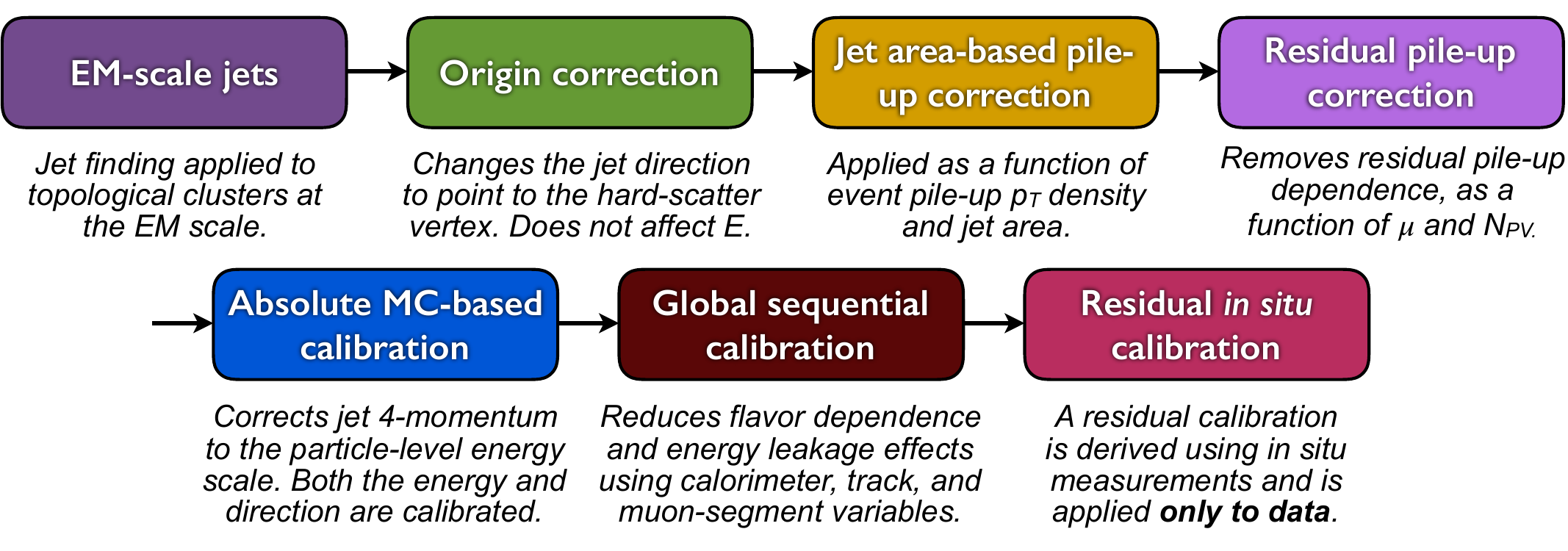}
  \caption{Illustration of jet clustering with the anti-$k_t$ algorithm~\cite{Cacciari:2008gp}
    ($R = 0.4$) applied to a simulated event, showing the approximately conical jet boundaries
    that result from the inverse-$k_T$ distance metric.}
  \label{fig:objects:antikt}
\end{figure}

Jet clustering uses the anti-$k_t$ algorithm~\cite{Cacciari:2008gp}, implemented in the
FastJet library~\cite{Fastjet}.
The algorithm is a sequential recombination procedure defined by the distance metrics
\begin{align}
    d_{ij} &= \min\!\left(k_{\text{T}i}^{-2},\, k_{\text{T}j}^{-2}\right)
               \frac{\Delta R_{ij}^2}{R^2}, \\
    d_{iB} &= k_{\text{T}i}^{-2},
\end{align}
where $\Delta R_{ij} = \sqrt{(\Delta\eta)^2 + (\Delta\phi)^2}$, $R$ is the jet radius
parameter, and $k_{\text{T}i}$ is the transverse momentum of object $i$.
At each step, the minimum among all $d_{ij}$ and $d_{iB}$ determines the next action: if
$d_{ij}$ is smallest, objects $i$ and $j$ are merged; if $d_{iB}$ is smallest, object $i$
is declared a jet and removed from the list.
The inverse-$k_T$ weighting causes soft particles to cluster around hard ones first,
producing approximately conical jets whose boundaries are robust against soft radiation.
The algorithm is infrared and collinear safe, and produces the approximately conical jet
shapes illustrated in Figure~\ref{fig:objects:antikt}.

The canonical small-$R$ jet collection used in ATLAS analyses is built with $R = 0.4$,
providing good separation between individual parton showers in most configurations.
Large-$R$ jets ($R = 1.0$) are used for boosted object reconstruction but are not employed
in the analyses presented in this thesis.

\subsection{Jet Energy Scale Calibration}

The raw jet energy measured by the calorimeter differs from the true particle-level jet
energy due to detector effects, out-of-cone radiation, and pileup contributions.
A multi-step calibration procedure restores the jet energy scale (JES)~\cite{PERF-2014-02,
PERF-2016-04}:

\begin{enumerate}
    \item \textbf{Origin correction}: the jet four-vector is recalculated pointing from the
          hard-scatter PV rather than the nominal detector centre, improving the $\eta$
          measurement.
    \item \textbf{Pileup subtraction}: an area-based correction subtracts the estimated
          pileup contribution $\rho \times A$ from the jet $\pT$, where $\rho$ is the median
          $\pT$ density of the event (estimated from $k_t$ jets) and $A$ is the jet area.
          A residual correction depending on the number of PVs $N_\text{PV}$ and the average
          number of $pp$ interactions per bunch crossing $\langle\mu\rangle$ removes
          remaining pileup sensitivity.
    \item \textbf{Absolute MC-based JES}: a multiplicative correction derived from
          simulation matches the reconstructed jet $\pT$ to the truth particle-level jet
          $\pT$, as a function of jet $\pT$ and $\eta$.
    \item \textbf{$\eta$ intercalibration}: corrections that ensure a uniform response
          across $\eta$ are derived from dijet balance, where a central jet recoiling
          against a forward jet provides the forward response relative to the well-measured
          central region.
    \item \textbf{Global Sequential Calibration} (GSC): a sequence of multiplicative
          corrections that reduce sensitivity to the jet flavour composition, punch-through
          into the MS, and other topology-dependent effects.
          Each step uses a different track-based or calorimeter-based variable, \textit{e.g.}\
          number of tracks or fraction of energy in the EM layers.
    \item \textbf{\textit{In situ} calibration} (data only): residual corrections derived by
          balancing jets against well-calibrated reference objects --- $Z$ or $\gamma$ bosons
          in $Z/\gamma$+jet events, and calibrated jets in multi-jet events.
          These corrections, applied only to data, absorb residual data--MC differences in
          the absolute scale.
    \item \textbf{Smearing} (MC only): a Gaussian smearing correction is applied to
          simulated jet energies, derived from dijet asymmetry measurements in data, to
          ensure the simulated resolution matches that observed in data.
\end{enumerate}

\subsection{Pileup Jet Rejection}

At high instantaneous luminosity, a significant fraction of reconstructed jets originate from
pileup interactions rather than the hard scatter.
The Jet Vertex Tagger (\texttt{JVT}) discriminates pileup jets from hard-scatter jets using
track-based variables~\cite{PERF-2014-03}:
\begin{itemize}
    \item \textbf{Corrected jet vertex fraction} (corrJVF): the fraction of the scalar
          $\sum \pT$ of tracks associated to the jet that belong to tracks pointing to the
          PV, corrected for the number of pileup tracks.
    \item \textbf{$R_{p_T}$}: ratio of the scalar sum of $\pT$ of tracks from the PV
          associated with the jet to the fully calibrated jet $\pT$.
\end{itemize}
The multivariate \texttt{JVT} score is applied to central jets ($|\eta| < 2.5$) below a
$\pT$ threshold (typically 60~GeV), above which pileup contamination becomes negligible.
A neural-network variant, \texttt{NNJvt}, with improved performance is the default for
Run~3 analyses.

For jets in the forward region ($2.5 < |\eta| < 4.5$), where no ID tracks are available, a
forward \texttt{fJVT} algorithm exploits the correlation between forward jet activity and
central soft pileup activity to identify pileup-induced forward jets.

\subsection{Jet Quality and Cleaning}

Non-collision backgrounds --- cosmic-ray muons, calorimeter noise bursts, beam-induced
backgrounds --- can produce spurious jets that mimic genuine hard-scatter activity.
The \texttt{LooseBad} cleaning criterion~\cite{ATLAS-CONF-2015-029} identifies these by
examining the pulse shape (out-of-time signals), the fraction of energy in the EM or
hadronic layers, and the jet timing (arrival relative to the bunch crossing).
Events containing at least one jet failing the \texttt{LooseBad} criterion are vetoed.

\section[\texorpdfstring{$b$}{b}-Jet Identification]{\texorpdfstring{$\bm{b}$}{b}-Jet Identification}
\label{sec:reco:btag}

Jets originating from the hadronisation of $b$-quarks can be identified via a process coined
$b$-tagging, by exploiting the relatively long lifetime of $b$-hadrons
($c\tau \approx 450~\mu$m), which causes their decay products to be displaced from the PV
by a measurable amount resolvable with the pixel detector (Section~\ref{sec:atlas:id}).
The primary distinguishing signatures are:
\begin{itemize}
    \item the presence of a secondary vertex displaced from the PV by a few millimetres;
    \item the $d_0$ and $z_0$ distributions of tracks within the jet
          (defined in Section~\ref{sec:reco:tracks}), which are significantly more displaced
          from the PV than those of light-quark jets;
    \item the semi-leptonic decay signature (a soft lepton within the jet) in a subset of
          cases.
\end{itemize}

The \texttt{DL1r} tagger is a deep-learning algorithm that combines the outputs of three
baseline taggers --- one based on track impact parameters, one on secondary vertex
reconstruction, and one on the full $b$-hadron decay-chain topology --- into a single
discriminant~\cite{b_tagging_ATLAS}.
The discriminant is the log-likelihood ratio
\begin{equation}
    D_{\mathrm{DL1r}} = \ln\!\left(\frac{p_b}{f_c \cdot p_c + (1-f_c) \cdot p_u}\right),
    \label{eq:dl1r}
\end{equation}
where $p_b$, $p_c$, $p_u$ are the $b$-, $c$- and light-jet probabilities output by the
network, and $f_c$ is a tunable charm-fraction parameter.
Four fixed WPs are defined by their inclusive $b$-jet efficiency in simulated $t\bar{t}$
events, listed in Table~\ref{tab:btag_wps}.

\begin{table}[htbp]
    \centering
    \caption{Standard $b$-tagging Working Points (WPs) for the \texttt{DL1r} algorithm
      in $\sqrt{s}=13$~TeV samples, showing approximate efficiencies in $t\bar{t}$ events
      and rejection factors for $c$- and light-quark jets.}
    \label{tab:btag_wps}
    \begin{tabular}{cccc}
        \hhline{====}
        WP   & $b$-jet efficiency & $c$-jet rejection & Light-jet rejection \\
        \hline
        60\% & 60\%              & $\sim$34           & $\sim$1500 \\
        70\% & 70\%              & $\sim$12           & $\sim$380  \\
        77\% & 77\%              & $\sim$6            & $\sim$110  \\
        85\% & 85\%              & $\sim$3            & $\sim$30   \\
        \hhline{====}
    \end{tabular}
\end{table}

The 60\% WP provides the highest background rejection and is used when purity is paramount,
at the cost of reduced $b$-jet efficiency.
The 85\% WP provides the highest efficiency and is used when $b$-jet samples need to be
collected with high completeness, accepting moderate contamination from charm and
light-quark jets.
The 77\% and 70\% WPs offer increasingly better purity.
Efficiency scale factors, correcting for data--MC differences in tagger performance, are
derived from $t\bar{t}$ control samples using a combinatorial likelihood method and applied
as weights to simulated events.

\section{Missing Transverse Momentum}
\label{sec:reco:met}

\subsection{Definition and Reconstruction}

The total $\pT$ of the initial $pp$ state is approximately zero, since the colliding partons
carry negligible $\pT$ relative to the beam direction.
By momentum conservation, the vector sum of the transverse momenta of all final-state
particles must also be zero.
Any significant imbalance --- $\boldsymbol{p}_\mathrm{T}^\mathrm{miss}$ --- signals the
presence of particles that escape without interacting in the detector: in the SM,
neutrinos; in many BSM scenarios, DM candidates or other weakly interacting particles.

The $\MET$ is reconstructed as the negative vector sum of the transverse momenta of all
identified and calibrated physics objects, supplemented by a soft term from low-$\pT$
activity not associated with any hard object:
\begin{equation}
    \boldsymbol{p}_\mathrm{T}^\mathrm{miss} = -\left(
      \sum_{\text{electrons}} \boldsymbol{p}_\mathrm{T}^{e}
    + \sum_{\text{muons}} \boldsymbol{p}_\mathrm{T}^{\mu}
    + \sum_{\text{photons}} \boldsymbol{p}_\mathrm{T}^{\gamma}
    + \sum_{\text{jets}} \boldsymbol{p}_\mathrm{T}^{j}
    + \sum_{\tau\text{-lep.}} \boldsymbol{p}_\mathrm{T}^{\tau}
    + \boldsymbol{p}_\mathrm{T}^{\,\text{soft}} \right),
\end{equation}
whose magnitude $\MET = |\boldsymbol{p}_\mathrm{T}^\mathrm{miss}|$ is the scalar missing
transverse energy.
The hard terms use the fully calibrated and identified objects described in the preceding
sections.
The \textbf{Track Soft Term} (TST) collects all ID tracks originating from the PV that are
not matched to any reconstructed hard object, providing a pileup-robust estimate of soft
hadronic activity~\cite{PERF-2016-07, ATLAS:2018ghb}.

The \texttt{EMPFlow} algorithm for $\MET$ reconstruction uses particle-flow jets and leptons
as the hard-term inputs.
The \texttt{Tight} operating point applies additional track-quality requirements to the TST
to further suppress pileup contributions.

\subsection{Missing Transverse Momentum Significance}

A useful discriminant between genuine and fake $\MET$ is the MET significance,
$\metsig$, which accounts for the expected $\MET$ resolution given the event's
hard-object configuration:
\begin{equation}
    \metsig \equiv \mathcal{S}(\pT^\text{miss}) =
    \frac{\MET}{\sqrt{\sigma_L^2\left(1 - \rho_{LT}^2\right)}},
    \label{eq:metsig}
\end{equation}
where $\sigma_L$ is the expected $\MET$ resolution along the
$\boldsymbol{p}_\mathrm{T}^\mathrm{miss}$ direction, and $\rho_{LT}$ is the correlation
between the longitudinal and transverse resolution components.
Events with a large $\metsig$ are unlikely to have their $\MET$ generated by jet
mismeasurement or detector noise, and are more consistent with genuine invisible particles.

\section{Overlap Removal}
\label{sec:reco:overlap}

The reconstruction algorithms described above operate independently, so the same energy
deposit or track can be claimed by more than one object type.
For example, an electron deposits its energy in the EM calorimeter and will also be clustered
into a nearby jet; a muon traversing the calorimeter can mimic a low-$\pT$ soft jet.
The overlap removal (OR) procedure resolves these ambiguities in a defined sequence, using
the angular distance
\begin{equation}
    \Delta R = \sqrt{(\Delta\phi)^2 + (\Delta y)^2},
\end{equation}
computed with the rapidity $y$ (Eq.~\eqref{eq:atlas:rapidity} in
Section~\ref{sec:atlas:coords}) rather than the pseudorapidity $\eta$
(Eq.~\eqref{eq:atlas:eta}), as used for the general ATLAS coordinate $\Delta R$
definition (Eq.~\eqref{eq:atlas:deltar}), for greater Lorentz-invariance at high $\pT$.

The baseline objects --- those passing the looser selection criteria --- are passed through
the OR algorithm in the following order of priority (lepton-favoured scheme):

\begin{enumerate}
    \item \textbf{Muon \textit{vs.}\ electron}: if a muon is of type CT and shares an ID
          track with an electron, the muon is rejected.
          This handles cases where a genuine electron radiates a soft photon that the
          calorimeter-tagging algorithm associates with a fake muon signature.
    \item \textbf{Electron \textit{vs.}\ muon}: if an electron shares an ID track with a
          muon, the electron is rejected.
          This occurs when a genuine muon passes through the EM calorimeter and a
          bremsstrahlung photon is misidentified as an electron.
    \item \textbf{Jet \textit{vs.}\ electron}: jets within $\Delta R < 0.2$ of an electron
          candidate are removed, as the electron is almost certainly the source of the jet
          energy.
    \item \textbf{Electron \textit{vs.}\ jet}: electrons within $\Delta R < 0.4$ of a
          remaining jet are removed, since they are likely to be non-prompt electrons from
          heavy-flavour decays within the jet.
    \item \textbf{Jet \textit{vs.}\ muon}: jets with fewer than three associated tracks that
          lie within $\Delta R < 0.2$ of a muon are removed; such jets are likely induced by
          the muon's calorimeter energy deposits.
    \item \textbf{Muon \textit{vs.}\ jet}: muons within $\Delta R < 0.4$ of a remaining jet
          are removed, as they are likely non-prompt muons from $b$- or $c$-hadron decays
          within the jet.
\end{enumerate}

Table~\ref{tab:or_criteria} summarises the OR criteria applied in the QBH analysis, which
are representative of the general ATLAS procedure.

\begin{table}[htbp]
  \centering
  \caption{Overlap removal criteria. $\Delta R$ is computed using rapidity throughout.}
  \label{tab:or_criteria}
  \begin{tabular}{lll}
    \hhline{===}
    Reject   & Against  & Criterion \\
    \midrule
    Muon     & Electron & CT muon sharing an ID track \\
    Electron & Muon     & Shared ID track \\
    Jet      & Electron & $\Delta R < 0.2$ \\
    Electron & Jet      & $\Delta R < 0.2$ \\
    Jet      & Muon     & $\texttt{NumTrack} < 3$ \& (ghost-associated or $\Delta R < 0.2$) \\
    Muon     & Jet      & $\Delta R < 0.4$ \\
    \hhline{===}
  \end{tabular}
\end{table}

The specific OR configuration --- including the exact $\Delta R$ thresholds, whether the
electron-\textit{vs.}-jet step is applied, and whether rapidity or pseudorapidity is used
--- may differ slightly between analyses.
The configurations adopted for each analysis in this thesis are specified in the respective
analysis chapters alongside the analysis-specific object selections.
\clearpage
\clearpage\chapter{Statistical Inference}
\label{chp:stats}


The dominant statistical framework used in high energy physics (HEP) searches is built on
frequentist hypothesis testing and likelihood-based methods.
This section establishes the key concepts and tools that underpin the results
reported in Chapters~\ref{chp:qbh}--\ref{chp:clockwork}.
The treatment largely follows Refs.~\cite{Cowan:2010js,ParticleDataGroup:2024cfk}
and the original $\mathrm{CL}_s$ papers~\cite{Read:2002hq,Junk:1999kv}.

\section{Probability Model and Hypotheses}
\label{ssec:stats:hypotheses}

An experiment observes a dataset $\mathbf{x}$, typically a histogram of event
counts $\mathbf{n} = (n_1, \ldots, n_N)$ in $N$ bins of some discriminating
variable, \textit{e.g.}\ invariant mass.
The probability model predicts the expected bin contents
$\boldsymbol{\nu}(\mu, \boldsymbol{\theta})$, which depend on:
\begin{itemize}
  \item the \emph{parameter of interest} (POI) $\mu$ — conventionally a signal
    strength that multiplies the expected signal yield, with $\mu = 0$
    corresponding to the background-only (SM) hypothesis and $\mu = 1$ to a
    specific signal hypothesis; and
  \item \emph{nuisance parameters} $\boldsymbol{\theta} = (\theta_1, \ldots,
    \theta_k)$ — quantities that are not of direct interest but must be accounted
    for, such as the systematic uncertainties on background normalisation,
    energy scale, or luminosity.
\end{itemize}
Each bin count is modelled as a Poisson random variable, so the full likelihood
function is
\begin{equation}
  \mathcal{L}(\mu, \boldsymbol{\theta})
    = \prod_{i=1}^{N} \frac{\nu_i^{n_i}\, e^{-\nu_i}}{n_i!}
      \cdot \prod_{j=1}^{k} p(\tilde{\theta}_j \mid \theta_j),
  \label{eq:stats:likelihood}
\end{equation}
where the second product encodes the constraint from auxiliary measurements
$\tilde{\boldsymbol{\theta}}$ on the nuisance parameters, typically modelled as
Gaussian penalty terms $p(\tilde{\theta}_j \mid \theta_j) = G(\tilde{\theta}_j;
\theta_j, \sigma_j)$~\cite{Cowan:2010js}.

Two hypotheses are contrasted in a search:
\begin{description}
  \item[$H_0$ (null hypothesis)] The background-only hypothesis, $\mu = 0$.
    This represents the SM prediction with no new physics.
  \item[$H_1$ (alternative hypothesis)] The signal-plus-background hypothesis,
    $\mu > 0$.
    In a discovery context $H_1$ is the claim of a new signal; in an exclusion
    context $H_0$ plays the role of the signal model being tested.
\end{description}
According to the Neyman--Pearson lemma~\cite{ParticleDataGroup:2024cfk}, the
most powerful test statistic for distinguishing $H_0$ from a simple alternative
$H_1$ is the likelihood ratio
\begin{equation}
  \lambda =
    \frac{\mathcal{L}(H_1)}{\mathcal{L}(H_0)}.
  \label{eq:stats:lr}
\end{equation}
In practice, signal hypotheses are composite (parameterised by $\mu$) and
nuisance parameters must be accounted for; the likelihood ratio is therefore
generalised to the profile likelihood ratio described in
Section~\ref{ssec:stats:profile}.

\section[$p$-value, Significance, and Error Rates]{$\bm{p}$-value, Significance, and Error Rates}
\label{ssec:stats:pvalue}

Given a test statistic $t(\mathbf{x})$ and the observed value $t_\text{obs}$,
the \emph{$p$-value} of hypothesis $H$ is defined as the probability of
obtaining a result at least as extreme as the observation, assuming $H$ is true:
\begin{equation}
  p = \int_{t_\text{obs}}^{\infty} f(t \mid H)\, \mathrm{d}t,
  \label{eq:stats:pvalue}
\end{equation}
where $f(t \mid H)$ is the probability density of $t$ under $H$.
A small $p$-value indicates that the data are unlikely under $H$, providing
evidence against it.

The \emph{size} $\alpha$ of a test is the pre-specified threshold below which
the $p$-value must fall to reject $H_0$.
Choosing $\alpha = 0.05$ yields a 5\% probability of falsely rejecting $H_0$
when it is true (a \emph{Type~I error}).
Conversely, the probability of failing to reject a false $H_0$ is the
\emph{Type~II error} rate $\beta$; the test's \emph{power} is $1 - \beta$.

\begin{figure}[ht]
  \centering
  \includegraphics[width=0.6\textwidth]{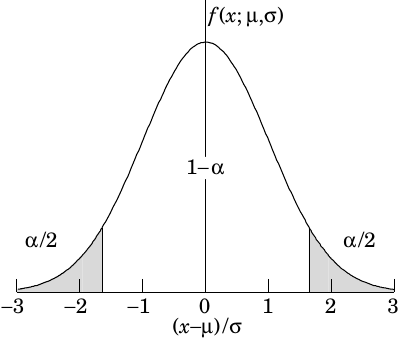}
  \caption{Standard normal probability density $f(Z)$ showing the relationship
    between significance $Z$ and $p$-value~\cite{ParticleDataGroup:2024cfk}.
    The shaded tail area corresponds to the $p$-value: the probability of
    observing a fluctuation at least as large as $Z$ under $H_0$.
    A value of $Z = 5$ corresponds to $p_0 \approx 2.87 \times 10^{-7}$.}
  \label{fig:stats:pvalue_gaussian}
\end{figure}

In HEP, $p$-values are conventionally converted to a \emph{significance} $Z$
(or $S$) in units of standard deviations of a standard Gaussian:
\begin{equation}
  Z = \Phi^{-1}(1 - p_0),
  \label{eq:stats:significance}
\end{equation}
where $\Phi$ is the cumulative distribution of the standard normal and $p_0$
denotes the $p$-value under $H_0$ (the background-only hypothesis).
The significance $Z = 5$ corresponds to $p_0 \approx 2.87 \times 10^{-7}$
and is the conventional threshold for claiming a discovery.
A significance of $Z = 3$ ($p_0 \approx 1.35 \times 10^{-3}$) is taken as
evidence of a new signal.
Figure~\ref{fig:stats:pvalue_gaussian} illustrates the relationship between the
significance $Z$ and the $p$-value as the tail probability of the standard Gaussian.

\subsection{The Profile Likelihood Ratio}
\label{ssec:stats:profile}

To properly account for nuisance parameters, a \emph{profile likelihood} is
constructed by maximising the likelihood over $\boldsymbol{\theta}$ for each
fixed value of $\mu$:
\begin{equation}
  \hat{\mathcal{L}}(\mu)
    = \mathcal{L}\!\left(\mu,\, \hat{\hat{\boldsymbol{\theta}}}(\mu)\right),
  \label{eq:stats:profL}
\end{equation}
where $\hat{\hat{\boldsymbol{\theta}}}(\mu)$ denotes the conditional maximum-likelihood
estimator of $\boldsymbol{\theta}$ for a given $\mu$.
The global maximum of the likelihood is achieved at $(\hat{\mu},
\hat{\boldsymbol{\theta}})$.
The \emph{profile likelihood ratio} is then
\begin{equation}
  \lambda(\mu) = \frac{\hat{\mathcal{L}}(\mu)}{\mathcal{L}(\hat{\mu},\hat{\boldsymbol{\theta}})}.
  \label{eq:stats:plr}
\end{equation}
By construction $0 \leq \lambda(\mu) \leq 1$, with $\lambda(\mu) \to 1$
indicating good agreement between the data and the hypothesis $\mu$.

For a \emph{discovery} test (testing $\mu = 0$), the test statistic is
\begin{equation}
  q_0 = \begin{cases}
    -2\ln\lambda(0) & \hat{\mu} \geq 0, \\
    0               & \hat{\mu} < 0,
  \end{cases}
  \label{eq:stats:q0}
\end{equation}
where the condition $\hat{\mu} < 0$ (a downward fluctuation) is not interpreted
as evidence against the background-only hypothesis.
For \emph{upper limits} on $\mu$, the test statistic is
\begin{equation}
  q_\mu = \begin{cases}
    -2\ln\lambda(\mu) & \hat{\mu} \leq \mu, \\
    0                 & \hat{\mu} > \mu.
  \end{cases}
  \label{eq:stats:qmu}
\end{equation}
The $p$-value for the hypothesis $\mu$ is then
$p_\mu = \int_{q_{\mu,\text{obs}}}^{\infty} f(q_\mu \mid \mu)\, \mathrm{d}q_\mu$.

\paragraph{Nuisance parameters and systematic uncertainties:}
Nuisance parameters encode the experimental and theoretical uncertainties that
affect the expected signal and background yields.
Each nuisance parameter $\theta_j$ is constrained by an auxiliary measurement
$\tilde{\theta}_j$ with an associated uncertainty (often taken as $\pm 1$);
the penalty term in Eq.~(\ref{eq:stats:likelihood}) ensures that $\theta_j$
is not pulled far from its nominal value without cost to the likelihood.
The profiling procedure of Eq.~(\ref{eq:stats:plr}) marginalises over
$\boldsymbol{\theta}$, incorporating the effect of systematic uncertainties
into the final result.
Nuisance parameters that are pulled more than $\sim\!1\sigma$ from their nominal
values or that are significantly constrained (reduced uncertainty compared to
the prior) indicate a tension between the data and the model in the region
sensitive to that systematic.

\paragraph{Asimov dataset:}
An \emph{Asimov dataset}~\cite{Cowan:2010js} is a representative dataset in
which all observed quantities are set to their expected values under a given
hypothesis ($\mathbf{n} = \boldsymbol{\nu}$, no statistical fluctuations, and
nuisance parameters at their nominal values).
It provides an analytic handle for computing the \emph{expected} (median)
sensitivity of an analysis without the need for pseudo-experiments.

\section{Asymptotic Approximations}
\label{ssec:stats:asymptotic}

For large samples, Wilks' theorem states that $-2\ln\lambda(\mu)$ follows a
chi-squared ($\chi^2$) distribution with one degree of freedom.
Asymptotic approximations for the distributions of $q_0$ and $q_\mu$ that hold
in the general case of nuisance parameters were derived in Ref.~\cite{Cowan:2010js}.
Under these approximations,
\begin{equation}
  Z \approx \sqrt{q_0},
  \label{eq:stats:asympZ}
\end{equation}
and the expected median significance for a signal with strength $\mu'$ is
\begin{equation}
  Z_\text{exp} = \sqrt{q_{\mu,A}},
  \label{eq:stats:Zexp}
\end{equation}
where $q_{\mu,A}$ is the test statistic evaluated on the Asimov dataset.
The asymptotic formulae are validated against pseudo-experiments in the analyses
described in this thesis and found to be in good agreement.

The same chi-squared framework underlies the Pearson goodness-of-fit test, which
provides a direct measure of the compatibility between an observed distribution
$\{n_i\}$ and a prediction $\{\nu_i\}$:
\begin{equation}
  \chi^2 = \sum_{i=1}^{N} \frac{(n_i - \nu_i)^2}{\nu_i}.
  \label{eq:stats:chi2}
\end{equation}
For a histogram with $N$ bins and $k$ freely fitted parameters, this statistic
approximately follows a $\chi^2$ distribution with $N - k$ degrees of freedom,
provided bin contents are sufficiently large ($\nu_i \gtrsim 5$), and the
goodness-of-fit $p$-value is $p = P(\chi^2_{N-k} \geq \chi^2_\text{obs})$.
A ratio $\chi^2/\mathrm{dof} \approx 1$ indicates that the model provides a
good description of the data.
This test is used throughout the analyses in this thesis to validate background
models and assess data--MC agreement in control regions.

\section{Confidence Intervals and Feldman--Cousins Construction}
\label{ssec:stats:ci}

A $1 - \alpha$ \emph{confidence interval} (CI) for a parameter $\mu$ has the
frequentist property that it contains the true value of $\mu$ in at least a
fraction $1 - \alpha$ of repeated experiments.
The Neyman construction achieves this by inverting hypothesis tests: the
confidence interval is the set of $\mu$ values for which $p_\mu \geq \alpha$.

A well-known problem arises when a measurement fluctuates below the expected
background: the Neyman construction may yield an unphysical negative lower
bound, or one may be tempted to quote a one-sided upper limit rather than a
two-sided interval, leading to a violation of coverage.
Feldman and Cousins~\cite{Feldman:1997qc} proposed a \emph{unified approach}
that eliminates this ambiguity by constructing a single confidence belt from
the ordering principle
\begin{equation}
  R(\mathbf{x}) = \frac{f(\mathbf{x} \mid \mu)}
    {f(\mathbf{x} \mid \hat{\mu}_\text{best}(\mathbf{x}))},
  \label{eq:stats:FC}
\end{equation}
where $\hat{\mu}_\text{best}$ is the physically allowed value of $\mu$ that best
fits the data.
This ratio orders data outcomes for each hypothesised $\mu$, automatically
transitioning from two-sided intervals when the data are consistent with a
signal to one-sided upper limits when they are not.
Exact coverage is maintained by construction.
The Feldman--Cousins method is widely used in HEP for setting limits on physical
parameters, particularly in cases of small signals or where the Gaussian
approximation fails.

\section{Exclusion Limits}
\label{ssec:stats:limits}

When no significant excess above the background-only hypothesis is observed in
data, upper limits on the signal strength $\mu$ (or equivalently on a model
parameter such as cross-section, coupling or mass) are set at the 95\% confidence level.
This means that hypotheses with $p_\mu < 0.05$ are excluded at 95\% CL.

Both \emph{expected} and \emph{observed} limits are reported:
\begin{itemize}
  \item The \emph{observed} limit is derived from data.
  \item The \emph{expected} (median) limit is derived from the Asimov dataset
    and represents the sensitivity of the analysis in the absence of a signal.
    Uncertainty bands at $\pm 1\sigma$ and $\pm 2\sigma$ represent the range
    of expected limits across pseudo-experiments.
\end{itemize}
Expected limits that are stronger (weaker) than the observed ones indicate that
the data contain a downward (upward) fluctuation relative to the background
prediction.

\section[The $\mathrm{CL}_{s}$ Method]{The $\bm{\mathrm{CL}_{s}}$ Method}
\label{ssec:stats:cls}

A well-known deficiency of the simple $p$-value procedure for exclusion is that
a downward statistical fluctuation in the observed event count can lead to the
exclusion of signal models to which the experiment has essentially no
sensitivity — since the data appear more background-like than expected, almost
any signal model appears excluded by the standard test.
To protect against such spurious exclusions, the modified frequentist
$\mathrm{CL}_s$ method is used~\cite{Read:2002hq,Junk:1999kv}.

Let $p_{s+b}$ denote the $p$-value of the signal-plus-background hypothesis
($\mu = 1$) derived from the test statistic $q_\mu$, and let $p_b = 1 -
p_{b,\text{obs}}$ denote the analogous quantity under the background-only
hypothesis.
The $\mathrm{CL}_s$ statistic is defined as
\begin{equation}
  \mathrm{CL}_s = \frac{p_{s+b}}{1 - p_b}
  = \frac{p_{s+b}}{p_b'},
  \label{eq:stats:cls}
\end{equation}
where $p_b' = 1 - p_b$ is the $p$-value of the background-only hypothesis
computed with the same test statistic.
A signal model is excluded at 95\% CL when $\mathrm{CL}_s < 0.05$.

The denominator $1 - p_b$ acts as a normalisation that prevents the exclusion
of signal models when the observed data fluctuate far below the background
prediction, \textit{i.e.}\ when $1 - p_b$ is small.
The $\mathrm{CL}_s$ method yields more conservative limits than the standard
$p$-value approach~\cite{Read:2002hq}, at the cost of a slight loss of
exclusion power in the absence of any downward fluctuation.
It is the standard approach in ATLAS and CMS for setting upper limits and
exclusion contours.

In combination with the profile likelihood test statistic of
Section~\ref{ssec:stats:profile} and the asymptotic approximations of
Section~\ref{ssec:stats:asymptotic}, the $\mathrm{CL}_s$ method allows exclusion
limits to be derived efficiently across one-dimensional parameter spaces,
as in the quantum black hole search (Chapter~\ref{chp:qbh}) and the $Z'$ search
(Chapter~\ref{chp:zprime}), as well as two-dimensional parameter spaces,
as in the Clockwork/Linear Dilaton model (Chapter~\ref{chp:clockwork}).
The expected sensitivity can be computed analytically using the Asimov dataset,
while pseudo-experiments are used to validate the asymptotic approximations.

\medskip
\noindent In the analysis chapters of this thesis, the statistical methods
described above are applied to specific search contexts.
In each case, only the analysis-specific choices (test statistic definition,
systematic uncertainty model, and results) are presented, with the underlying
formalism taken from this chapter.
The reader is referred to the review in Ref.~\cite{Cowan:2010js} for a thorough
derivation of the asymptotic results and to Ref.~\cite{ParticleDataGroup:2024cfk}
for a broader overview of statistical methods in particle physics.
\clearpage
\clearpage\chapter{Search for New Physics in \texorpdfstring{$\bm{\ell j}$}{\textit{l}j} Final States at \texorpdfstring{$\bm{\sqrt{s}=13.6}$}{sqrt(s)=13.6}~TeV}
\label{chp:qbh}

\section{Introduction}
\label{sec:qbh:intro}

The hierarchy problem --- the immense gulf between the electroweak scale and Planck scale --- may be resolved by theories with EDs that lower the fundamental $D$-dimensional scale of gravity, $M_{\textrm{D}}$, to the TeV range, as described in Section~\ref{sec:bsm:extradim}.
In the ADD model~\cite{Nima,Antoniadis:1998ig}, gravity propagates through $n$ flat EDs (where $D=4+n$), while SM fields are restricted to four-dimensional space-time.
Alternatively, the RS model~\cite{Randall} employs one warped extra dimension to generate this scale hierarchy.
These theories predict a plethora of new phenomena, some of which are accessible at the LHC.
One such prediction is the production of QBHs~\cite{Gingrich:2009hj,Meade:2007sz,Calmet:2008dg,Gingrich:2009da,Gingrich:2015yda} at threshold masses $M_{\textrm{th}} \sim M_{\textrm{D}}$; see Section~\ref{sec:bsm:qbh} for a detailed discussion.
In this benchmark framework, it is postulated that the QBH parton--parton cross-section follows a power-law increase with partonic centre-of-mass energy, with the power determined by the number of extra dimensions.
For higher proton-beam energy, the suppression of the partonic cross-section arising from the PDF at high Bjorken-$x$ (the fraction of the proton's momentum carried by the interacting parton) is reduced, resulting in a significant increase in the QBH production cross-section, $\sigma_{\textrm{QBH}}$.
Unlike semi-classical black holes, which decay thermally via Hawking radiation into high-multiplicity final states~\cite{Anchordoqui:2001cg,ATLAS:2015yln,ATLAS:2018rvc,CMS:2017boz,CMS:2018ozv,CMS-EXO-24-028}, these QBHs decay into two-particle final states, where lepton- and baryon-number conservation may be violated.
This search focuses on diquark production of QBHs that can decay to lepton+antiquark $(\ell+\bar{q})$ final states:
\begin{equation}
  uu \to \mathrm{QBH}^{4/3} \to \bar{d}\ell^{+}, \quad
  ud \to \mathrm{QBH}^{1/3} \to \bar{u}\ell^{+}, \quad
  dd \to \mathrm{QBH}^{-2/3} \to \bar{d}\ell^{-},
  \label{eq:QBH_processes}
\end{equation}
where the superscript indicates the QBH electric charge and the processes are sorted by descending $\sigma_{\textrm{QBH}}$, left-to-right.
The notation $u$ and $d$ marks the up-type and down-type quarks.
The branching ratios for the processes ordered as in Eq.~\eqref{eq:QBH_processes} are 11\%, 5.6\% and 6.7\%, to each lepton flavour~\cite{Gingrich:2009hj}.

\begin{figure}[htbp]
  \centering
  \includegraphics[width=0.95\textwidth]{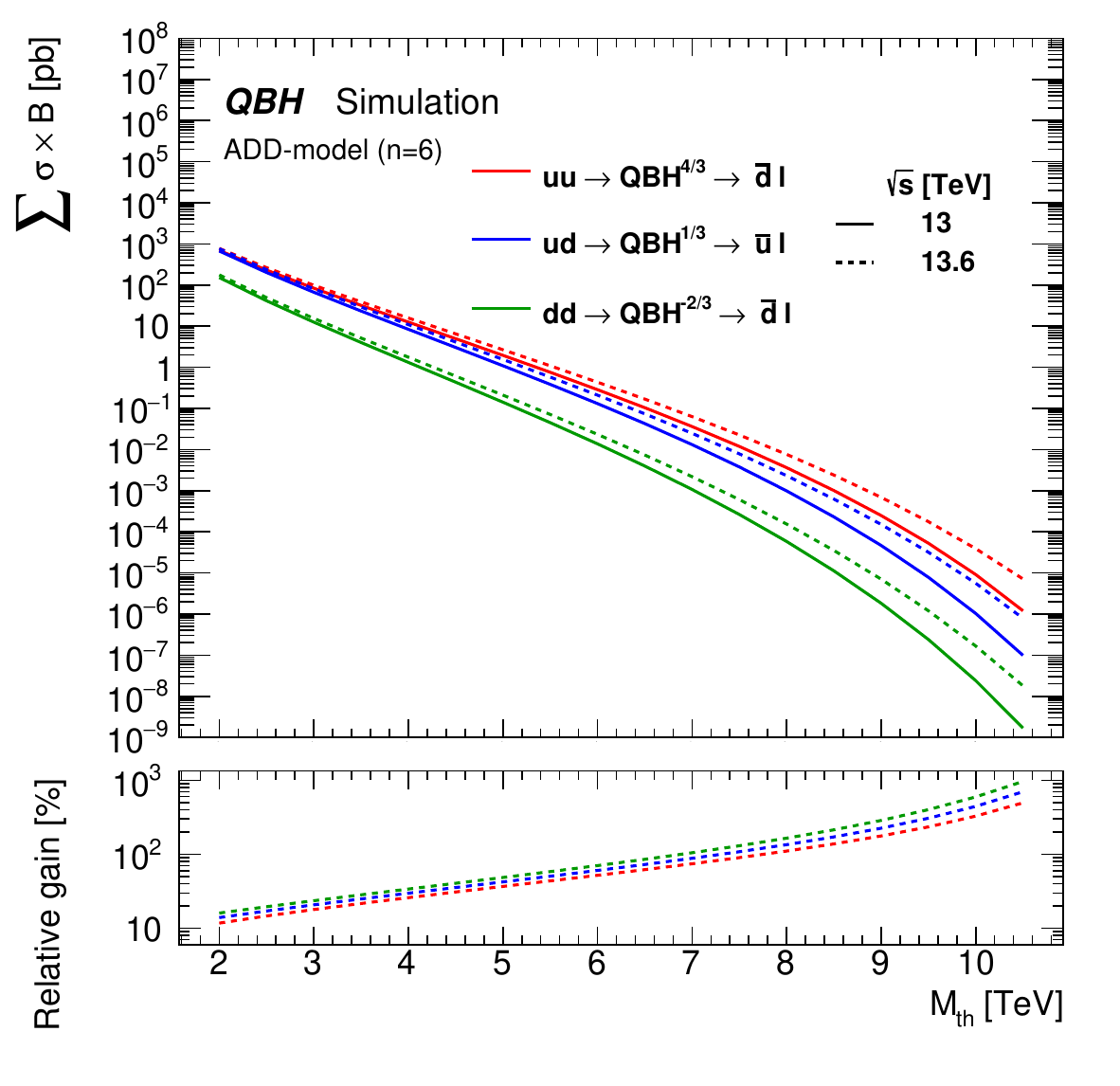}
  \caption{Summed production cross sections times branching fractions of QBHs as a
    function of threshold mass $M_\text{th}$ for the ADD model with $n=6$ extra
    dimensions, for the six QBH states that can decay to a lepton-plus-jet final state.
    Solid and dashed lines correspond to $\sqrt{s} = 13~\TeV$ (Run~2) and
    $13.6~\TeV$ (Run~3), respectively.
    The lower panel shows the relative increase in cross section times branching
    fraction from Run~2 to Run~3.
    The RS model behaviour is consistent with the ADD results shown.
    Cross sections are computed with the \texttt{QBH v3.02} event
    generator~\cite{Gingrich:2009da}.}
  \label{fig:QBH_XS_Run2_vs_Run3}
\end{figure}

A key feature that sets this model apart from virtually all other new-physics scenarios, and motivates this search already with a partial run dataset, is the production cross-section's strong dependence on the proton--proton centre-of-mass energy, $\sqrt{s}$.
This dependence causes the increase in $\sqrt{s}$ from Run~2 of the LHC ($13~\TeV$) to Run~3 ($13.6~\TeV$) to enhance $\sigma_{\textrm{QBH}}$ by approximately 100\% at $M_{\textrm{th}} = 6~\TeV$ up to about 1000\% at $M_{\textrm{th}} = 10.5~\TeV$~\cite{Gingrich:2009hj}, as discussed in Section~\ref{sec:bsm:qbh} and illustrated in FIG.~\ref{fig:QBH_XS_Run2_vs_Run3}.

This analysis supersedes the previous ATLAS search performed at $\sqrt{s} = 13~\TeV$ using the full $140~\textrm{fb}^{-1}$ Run~2 dataset~\cite{ATLAS:2023vat}.
The same search was done by ATLAS at $\sqrt{s} = 8~\TeV$ with $20.3~\textrm{fb}^{-1}$ of data~\cite{ATLAS:2013wgh}.
Complementary searches were also performed in dijet, dilepton and photon+jet final states by ATLAS~\cite{ATLAS:2015nsi,ATLAS:2016loq,ATLAS:2019fgd,ATLAS:2013ylb,ATLAS:2015esi,ATLAS:2017dpx} and CMS~\cite{CMS:2017caz,CMS:2018hnz,CMS-EXO-19-014,CMS:2023twl}.
In comparison with the other final states, the lepton+jet ones benefit from significantly lower backgrounds, which compensates for their smaller cross-section times branching ratio.
Lepton+jet final states are of course not unique to QBH production, and other searches probing similar regions of phase space do not necessarily provide an explicit interpretation in terms of QBHs.
For example, ATLAS has published a lepton+jet search for leptoquarks with a combination of the full Run~2 dataset and a partial ($56~\textrm{fb}^{-1}$) Run~3 dataset~\cite{ATLAS:2025upm}.

The present search uses $164~\textrm{fb}^{-1}$ of $\sqrt{s} = 13.6~\TeV$ proton--proton collision data, recorded by the ATLAS detector during the 2022--2024 period of Run~3.
Even with a Run~3 dataset similar in size to the full Run~2 dataset, this analysis yields considerably higher QBH exclusion limits thanks to the significant increase in $\sigma_{\textrm{QBH}}$.
Electron+jet and muon+jet final states with an invariant mass above $3~\TeV$ are targeted separately in this search.
Explicit contributions from QBH decays with $\tau$-leptons are not considered, since they are expected to be negligible compared to those from the other two lepton+jet final states, owing to the $\tau \to \ell\bar{\nu}_{\ell}\nu_{\tau}$ branching ratio and the softer electron and muon transverse momentum ($\pT$) spectra due to the presence of neutrinos in $\tau$ decays.
The main background processes include $V$+jets ($V=W,Z$), $VV$ and events with top quarks.
These are modelled using MC simulations.
Events containing non-prompt electrons, or electrons from photon conversions or jets misidentified as electrons, collectively referred to as ``fakes'', are estimated using data-driven techniques.
This is the dominant background in the electron channel, while $W$+jets production contributes the most background in the muon channel.

\section{Analysis Strategy}
\label{sec:qbh:strategy}
QBHs decaying to lepton-quark pairs are searched for. The two-body decay is reconstructed from exactly one energetic electron or muon and one energetic jet. High-$\pT$ requirements are set on the leading lepton and jet, in addition to geometrical cuts which select back-to-back two-body decays. The discriminant used is the invariant mass of the leading signal lepton and signal jet, $m_{\ell j}$. The analysis strategy is similar for the two lepton flavours, except that in the $e+j$ channel the multijet background is dominant whereas in the $\mu+j$ channel it is nearly negligible. This is because jets are more often misidentified as electrons than as muons. In the electron channel it is estimated from data, and from dijet MC in the muon channel. The analysis regions are described in Section~\ref{sec:EvSel}.

The analysis regions differ between the $e+j$ and $\mu+j$ channels in that the former includes the fake-electron background as a controlled and validated background source. The MET significance $\mathcal{S}(\Etmiss)$ (Section~\ref{sec:reco:met}, Eq.~\eqref{eq:metsig}) is used to separate the fake CR and VR from the $W$+jets CR and VR. For the $\mu+j$ channel, a simpler strategy is devised, controlling and validating only the $W$+jets and $Z$+jets backgrounds. The $Z$+jets CR and VR are common to both channels and are enriched by requiring the presence of a second baseline lepton. Further details of all analysis regions are given in Section~\ref{sec:EvSel}.

To enhance sensitivity in the SRs, angular cuts on the lepton-jet system are applied, selecting events where the lepton and jet are approximately back-to-back in the transverse plane while maintaining high signal efficiency. Further details of these cuts are given in Section~\ref{sec:EvSel}.

Using a profile-likelihood fit, normalisation factors are derived for the $W$+jets and $Z$+jets background estimates, used to produce an Asimov dataset in the post-fit SR. Systematic uncertainties from detector simulation on the analysis objects, electrons, muons, jets and $\Etmiss$, as well as theoretical uncertainties on the PDF choice, QCD scales and others are considered as nuisance parameters in the fit. A limit on the signal strength, $\mu$, calculated at each signal point is used to place exclusion limits on the QBH production cross-section times decay branching fraction.

\section{Data and Simulation}
\label{sec:data}
\section{Code and Analysis Version}
\label{CodeAna}

The data, signal and background samples are described in Sections~\ref{sec:data}, \ref{sec:signal} and \ref{sec:background}. The analysis is performed with Athena, the ATLAS software framework used for event reconstruction and physics analysis. In practical terms, Athena reads detector data and simulated events, reconstructs physics objects (electrons, muons, jets and missing transverse momentum), and provides a reproducible environment to apply calibrations, selections and algorithms across large datasets. For this thesis we use the standard Run~3 software releases and produce compact “ntuples” (lightweight summary trees) derived from centrally prepared ATLAS datasets to enable fast iteration in the final analysis stage.
For Monte Carlo samples, the effect of pileup, i.e., multiple $pp$ interactions in the same or neighboring bunch crossings, is included in all simulated event samples. Signal samples use \texttt{Pythia 8.8309} and multijet, $ttV$ and top samples use \texttt{Pythia 8.230} to simulate the pileup in collisions, using the ATLAS A3 set of tuned parameters and the NNPDF3.0 PDF set \cite{Ball:2014uwa}. These are weighted to reproduce the average number of pileup interactions per bunch crossing observed in data. The generated background events are passed through a full detector simulation \cite{SOFT-2010-01} based on Geant4 \cite{Agostinelli:2002hh}. For the simulated QBH event samples, for the calorimeter, a fast parametrization response \cite{ATL-PHYS-PUB-2010-013} is used. The other detector systems used Geant4.

\section{Collision Data}

This analysis uses $\sqrt{s} = 13.6~\textrm{TeV}$ proton–proton collision data recorded between 2022--2024 in Run~3 of the LHC, corresponding to a total integrated luminosity of $164~\textrm{fb}^{-1}$.

The data were collected by the ATLAS detector during stable beam conditions with all detector systems operating normally. Events are selected using the lowest unprescaled single-lepton triggers, which depend on the data-taking period. An overview of the triggers used for each period is presented in Table~\ref{tab:datasets::triggers}.

For the electron channel, events are required to pass at least one of the three single-electron triggers, with the same Level 1 requirement of an EM object with transverse momentum greater than 22~GeV and High Level Trigger requirements of: $p_{\textrm{T}}$ threshold of 26~GeV with a tight likelihood requirement accompanied by a loose isolation requirement; or a higher $p_{\textrm{T}}$ threshold of 60~GeV and a medium likelihood condition; or 140~GeV transverse momentum with a looser identification criteria.  The L1 requirements for these triggers depend on the data-taking period, as shown in Table~\ref{tab:datasets::triggers}.

The estimation of fake-electron background, described in Section~\ref{sec:fakes}, uses the loosest available unprescaled single-electron trigger in each data-taking period: a 140~GeV threshold with only a loose identification criterion and no isolation requirement. This is the most permissive of the single-electron triggers and is chosen deliberately to maximise the acceptance of fake-electron candidates in the matrix method.

In the muon channel, events are recorded if they pass one of two single-muon triggers: either one with a $p_{\textrm{T}}$ requirement of at least 24~GeV and a medium identification criterion, or a trigger with a 50~GeV $p_{\textrm{T}}$ requirement.

The full trigger list for each data-taking period is given in Appendix~\ref{app:qbh:data} (Table~\ref{tab:datasets::triggers}).
Because the quantum black hole signals targeted here decay to objects with very large invariant masses, the resulting lepton \pT is well above all trigger thresholds; in practice, essentially all selected events are already captured by the loosest single-lepton chain, which both ensures near-unity trigger efficiency for signal and defines the well-controlled trigger sample used for the fake-electron estimation.

To ensure data quality, events with noise bursts or coherent noise in the calorimeters are removed. All events are required to satisfy the official ATLAS Good Runs Lists (GRL) for each data-taking year. The GRL files correspond to a total integrated luminosity of $164~\textrm{fb}^{-1}$ and are listed in Appendix~\ref{app:qbh:data} (Table~\ref{tab:GRL}).

\section{Signal Simulation}
\label{sec:signal}
\begin{figure}
    \subfloat[]{
      \includegraphics[width=0.5\textwidth]{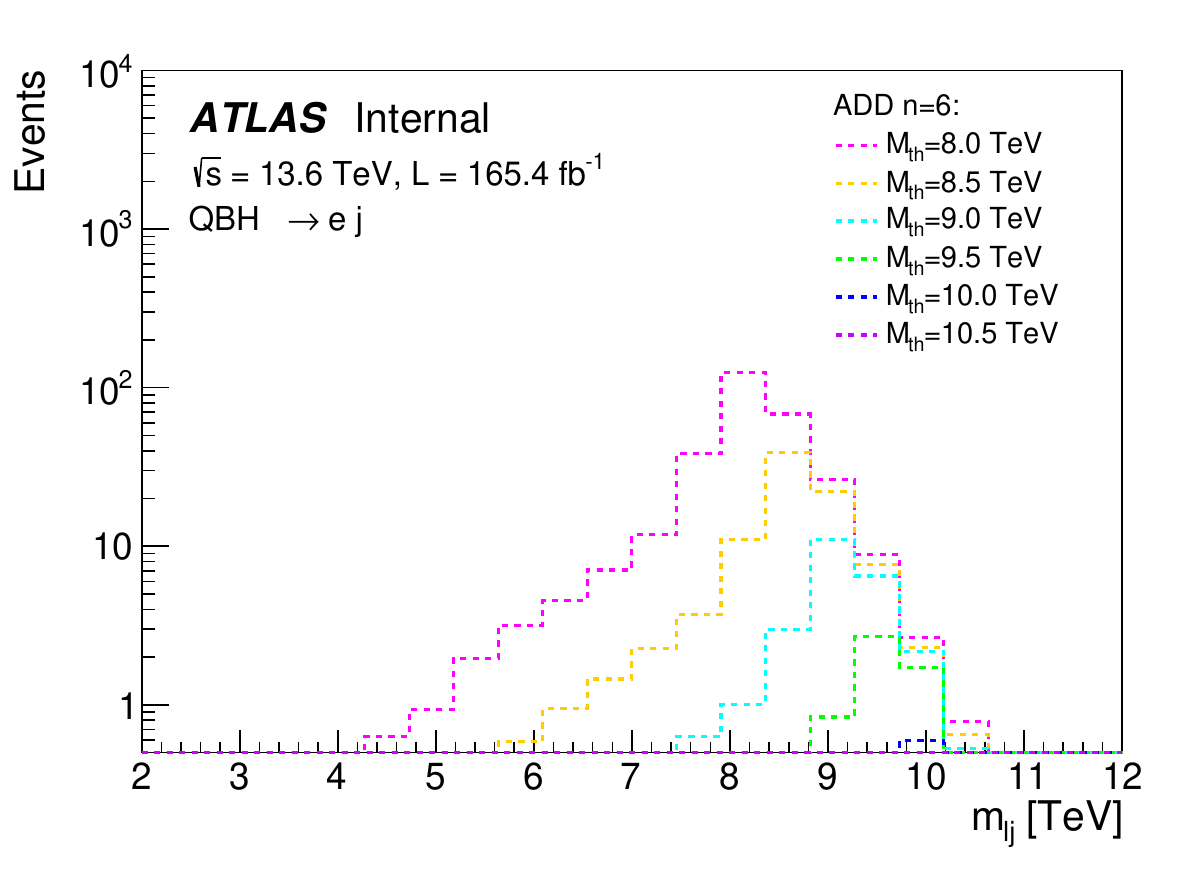}
    }
    \subfloat[]{
      \includegraphics[width=0.5\textwidth]{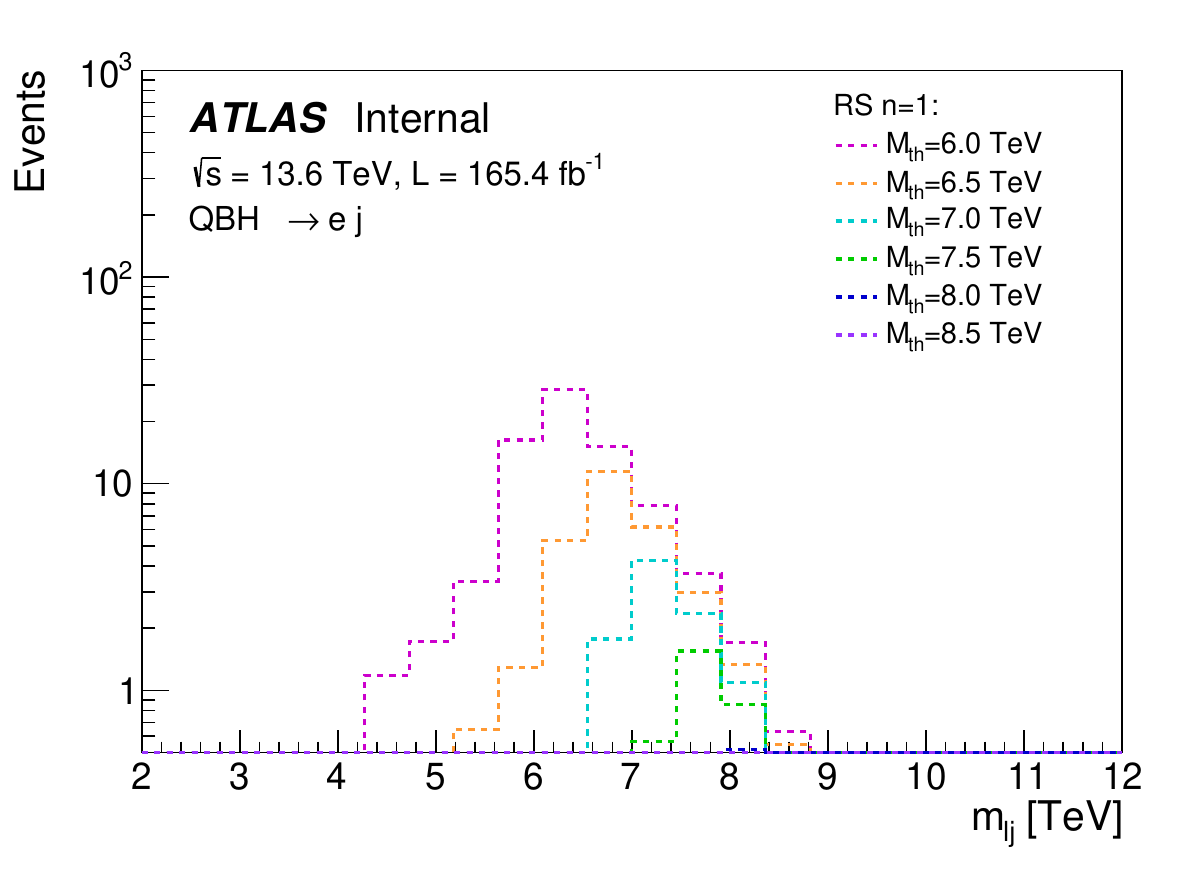}
    }
\caption{Distribution of signal samples over invariant mass of the lepton + jet shown for ADD (left) and RS (right) models with $n=6$ and $n=1$ extra dimensions, respectively. Variation of threshold masses, $M_{\textrm{th}}$, indicated by different colours. Both examples shown for QBH decaying to electron + jet final state at centre-of-mass energy $\sqrt{s}=13.6\;\textrm{TeV}$.}    
\label{fig:QBH_Sig_mlj}
\end{figure}

QBH production and decay is simulated using a MC generator of black holes in $pp$ collisions \cite{Gingrich:2009da}, \texttt{QBHv3.02}. The signals produced couple universally to all quarks, leptons and gauge bosons. Lorentz invariance is assumed so only angular momentum-conserving decays are permitted. Baryon and lepton numbers are violated in the final states. In the LHC, QBHs can be produced via quarks and gluons such that nine electric charge states are possible: $\pm 4/3, \pm 1, \pm 2/3, \pm 1/3$ and $0$. The $+4/3$-charge state can be formed by quark pairs, $+2/3$ by either an antiquark pair or a quark-gluon combination, $+1/3$ by a quark pair or antiquark-gluon combination, $+1$ by a quark-antiquark pair and $0$ by either a quark-antiquark or a gluon-gluon pairs. By analogy with above compositions of the black hole the negative charge states can be enumerated in a similar way. Therefore, the six initial states producing QBHs which are allowed to decay to a lepton-quark pair, as well as their corresponding charge, are \cite{Gingrich:2009hj}:
\begin{enumerate}
    \item $uu \rightarrow \textrm{QBH}^{+4/3}_{uu} \rightarrow e^{+}\bar{d}, \mu^{+}\bar{d}$
    \item $\bar{d}\bar{d} \rightarrow \textrm{QBH}^{+2/3}_{\bar{d}\bar{d}} \rightarrow e^{+}d, \mu^{+}d$
    \item $ud \rightarrow \textrm{QBH}^{+1/3}_{ud} \rightarrow e^{+}\bar{u}, \mu^{+}\bar{u}$
    \item $\bar{u}\bar{d} \rightarrow \textrm{QBH}^{-1/3}_{\bar{u}\bar{d}} \rightarrow e^{-}u, \mu^{-}u$
    \item $dd \rightarrow \textrm{QBH}^{-2/3}_{dd} \rightarrow e^{-}\bar{d}, \mu^{-}\bar{d}$
    \item $\bar{u}\bar{u} \rightarrow \textrm{QBH}^{-4/3}_{\bar{u}\bar{u}} \rightarrow e^{-}d, \mu^{-}d$
\end{enumerate}
where $u$, $d$ denotes all up-, down-type quarks, respectively. The branching fractions for $\textrm{QBH}^{+4/3}_{uu} \rightarrow e^{+}\bar{d}, \mu^{+}\bar{d}$, $\textrm{QBH}^{+2/3}_{\bar{d}\bar{d}} \rightarrow e^{+}d, \mu^{+}d$ and $\textrm{QBH}^{+1/3}_{ud} \rightarrow e^{+}\bar{u}, \mu^{+}\bar{u}$ are 11\%, 6.7\% and 5.6\% for each lepton flavour. The \texttt{QBHv3.02} generator is used to simulate the hard-scatter and compute the $QBH$ production cross-sections. Events are then interfaced to \texttt{Pythia8} for parton showering and hadronisation modelling. The QBH is assumed to have zero angular momentum and the total angular momentum is conserved through the production and decay process. The PDF set \texttt{CTEQ6L1} is used with QCD scale equal to the inverse gravitational radius; while newer PDF sets are available, \texttt{CTEQ6L1} has been adopted by all previous QBH searches and is retained here to preserve comparability with the existing literature. The conditions $M_{\textrm{th}} = M_{D}$ and $M_{\textrm{th}} < 3\times M_{D}$ are used to keep non-thermal decay. Only two-body decays are considered. The production cross-sections of quantum black holes in ADD ($D=10$) and RS ($D=5$) at $\sqrt{s}=13.6 \; \textrm{TeV}$ are shown in Appendix~\ref{app:qbh:sig} (FIG.~\ref{fig:QBH_XS_Run3_ADD_RS}).
The production cross-section of the $\textrm{QBH}_{qq}$ state is several orders of magnitude higher than that of its charge-conjugated state (FIG.~\ref{fig:QBH_XS_Run3_ADD_RS}), and hence the latter is not considered in the analysis. MC samples are produced simulating QBHs in ADD model with $n=2,4$ and $6$ EDs and in RS model with $n=1$ ED. Samples are produced in $0.5 \; \textrm{TeV}$ steps through a range of masses, $M_{\textrm{th}} \in [8, 10.5], [6, 8.5]\; \textrm{TeV}$ for RS, ADD models, respectively. The mass range lower bounds choices follow exclusion limits set by ATLAS search for QBHs in the $\ell j$ final state with full Run2 dataset \cite{ATLAS:2023vat} at 6.8 TeV, 9.2 TeV for RS, ADD models, respectively. Signal samples are simulated in three ATLAS MC campaigns, each matching the detector conditions of a different data-taking year (2022, 2023 and 2024), as described fully in Table~\ref{tab:QBH_Sig_Production_Table} for both the electron and muon channels. The cross-sections and branching fractions are the same for both lepton channels, and the same number of events was generated for each, the only difference being the DSIDs. Two examples of signal sample distributions in invariant lepton+jet mass are shown in FIG. \ref{fig:QBH_Sig_mlj} to demonstrate the signal width. 

The full list of signal samples with dataset identifiers and production campaigns is given in Appendix~\ref{app:qbh:sig} (Table~\ref{tab:QBH_Sig_Production_Table}).

In this analysis iteration, the signal generation is streamlined by considering only the direct decays of QBHs into electron+jet and muon+jet final states, consistent with previous studies from Run1~\cite{ATLAS:2013wgh} and Run2~\cite{ATLAS:2023vat}. This approach neglects contributions from leptonic tau-lepton decays ($\tau \to e/\mu$). The leptonic tau decays are expected to contribute approximately 17\% of the inclusive signal yield to the electron and muon channels, albeit with a characteristically softer $p_{\textrm{T}}$ spectrum. The neglect of these contributions yields a conservative cross-section limit, as including them would most likely increase the effective signal yield and thus strengthen the derived limits.

\section{Background Simulation}
\label{sec:background}
Background events, in this analysis, are those which contain a high-$p_{T}$ lepton and one or more jets, but do not originate from a QBH. Such signatures arise from electroweak processes including vector boson production with additional jets ($W$/$Z$+jets), dibosons ($WW$, $WZ$ and $ZZ$), top-quark pair ($t\bar{t}$) and single-top-quark production, and multijet processes including nonprompt leptons from hadron decays, photon conversions and jets misidentified as leptons.

The expected contributions of various SM processes as well as possible QBH signals are modelled using MC simulation, either directly or adjusted by a fit to data in dedicated control regions. The multijet background is measured directly in data, collected using a set of unprescaled single-lepton triggers with different $\pT$-thresholds.

For MC samples, the effect of pileup, \textit{i.e.}, multiple $pp$ interactions in the same or neighboring bunch crossings, is included in all simulated event samples. Multijet, $ttV$ and top samples use \texttt{Pythia 8.230} \cite{Sjostrand:2014zea} to simulate the pileup in collisions, using the ATLAS A3 set of tuned parameters and the NNPDF3.0 PDF set \cite{Ball:2014uwa}. These are weighted to reproduce the average number of pileup interactions per bunch crossing observed in data. The generated background events are passed through a full detector simulation \cite{SOFT-2010-01} based on Geant4 \cite{Agostinelli:2002hh}.

\subsection*{$\bm{V}$+jets}


The $W$+jets and $Z$+jets samples are simulated with the \texttt{Sherpa 2.2.14} generator~\cite{Bothmann:2019yzt}. The strong production of $Z$ bosons in association with multiple jets is generated with up to 2 additional partons at NLO and up to 5 additional partons at LO using the NNPDF3.0nnlo PDF set~\cite{Ball:2014uwa}. The samples are enhanced in $\max(H_{\textrm{T}}, p_{\textrm{TV}})$ and sliced in quark flavour content. The $W$+jets and $Z$+jets sample details are listed in Appendix~\ref{app:qbh:bkg} (Tables~\ref{tab:wjets_MC} and~\ref{tab:zjets_MC}).

\subsection*{Diboson}

Samples of diboson final states ($VV$) are simulated with the \texttt{Sherpa 2.2.14} generator~\cite{Bothmann:2019yzt}. Fully leptonic final states and semileptonic final states, where one boson decays leptonically and the other hadronically, are generated using matrix elements at NLO accuracy in QCD for up to one additional parton and at LO accuracy for up to three additional parton emissions. Electroweak production (``vector-boson fusion'') of diboson is generated with up to one additional parton in the fully leptonic final state. The NNPDF3.0nnlo PDF set~\cite{Ball:2014uwa} is used, along with the dedicated set of tuned parton-shower parameters developed by the Sherpa authors. The diboson sample details are listed in Appendix~\ref{app:qbh:bkg} (Table~\ref{tab:vv_MC}).

\subsection*{Top}

Table~\ref{tab:top_MC} lists the top samples used in this analysis. The $t\bar{t}$ and single-$t$ background events are generated as follows.

The $t\bar{t}$ events are modelled using the \texttt{Powheg Box v2}~\cite{Alioli:2010xd} generator at NLO with the NNPDF3.0nlo PDF set~\cite{Ball:2014uwa} and the $h_{\textrm{damp}}$ parameter set to $1.5\,m_{\textrm{top}}$. These events are interfaced to \texttt{Pythia 8.230}~\cite{Sjostrand:2014zea} to model the parton shower, hadronisation, and underlying event, with parameters set according to the A14 tune using the NNPDF3.0 PDF set~\cite{Ball:2014uwa}. EvtGen 1.6.0 is used to perform the decays of bottom and charm hadrons. The $t\bar{t}$ samples are generated in three mutually exclusive sub-samples — single-lepton, all-hadronic and dilepton filtered — which are combined in the analysis.

The associated production of top quarks with $W$ bosons ($tW$) is modelled by the \texttt{Powheg Box v2} generator at NLO in QCD using the five-flavour scheme and the NNPDF3.0nlo PDF set. To remove interference and overlap with $t\bar{t}$ production, the diagram removal scheme is employed. Events are interfaced to \texttt{Pythia 8.230} using the A14 tune and the NNPDF3.0 PDF set.

The single-top $t$-channel ($s$-channel) production is modelled using the \texttt{Powheg Box v2} generator at NLO in QCD using the four-flavour (five-flavour) scheme and the corresponding NNPDF3.0nlo PDF set. These events are also interfaced with \texttt{Pythia 8.230} using the A14 tune and the NNPDF3.0 PDF set.

The associated production of top-quark pairs with a vector boson ($t\bar{t}V$) is simulated using \texttt{Sherpa 2.2.14}~\cite{Bothmann:2019yzt}. Other $t\bar{t}V$ samples are produced using the \texttt{MadGraph5\_aMC@NLO} 2.3.3~\cite{Alwall:2014hca} generator with the NNPDF3.0nlo PDF set, interfaced with \texttt{Pythia 8.210}~\cite{Sjostrand:2014zea} using the A14 tune and the NNPDF3.0 PDF set. The top-quark and $t\bar{t}V$ sample details are listed in Appendix~\ref{app:qbh:bkg} (Tables~\ref{tab:top_MC} and~\ref{tab:ttv_MC}).

\subsection*{Dijet} 

Monte Carlo dijet samples are used for the muon+jet background channel and, in the electron+jet channel, as a cross-check of the data-driven fake-electron background estimation and to derive one of its associated systematic uncertainties. These samples are listed in Table \ref{tab:jj_MC}. Contribution of this background in muon channel is $<1\%$ in SR ($m_{\ell j}>3$~TeV, see FIG.~\ref{fig:Pie_Charts_SR}). For the electron+jet channel this background is much more significant in SR ($=40\%$, see FIG.~\ref{fig:Pie_Charts_SR}) and is estimated using data-driven fake estimation method, described in Chapter~\ref{sec:fakes}. These samples were produced with \texttt{Pythia 8.230} using the A14 tune and the NNPDF3.0 set of PDFs. The dijet sample details are listed in Appendix~\ref{app:qbh:bkg} (Table~\ref{tab:jj_MC}).

\section{Event Selection}
\label{sec:EvSel}

Each event must contain exactly one signal lepton --- either a signal electron or a signal muon --- together with at least one signal jet. Signal electrons are required to have $\pT > 20~\GeV$, satisfy the \texttt{Tight} likelihood identification working point~\cite{EGAM-2018-01}, and pass the \texttt{HighPtCaloOnly} isolation criterion, within $|\eta| < 2.47$ (excluding the calorimeter transition region $1.37 < |\eta| < 1.52$); see Section~\ref{sec:reco:electrons}. Signal muons are required to have $\pT > 20~\GeV$, satisfy the \texttt{HighPt} identification working point~\cite{PERF-2015-10}, and satisfy a track- and calorimeter-based isolation requirement that tightens with increasing jet activity (see Section~\ref{sec:reco:muons}), within $|\eta| < 2.5$. Signal jets are reconstructed with the \Antikt\ algorithm ($R = 0.4$)~\cite{Cacciari:2008gp} from particle-flow constituents and must satisfy $\pT > 20~\GeV$ and $|\eta| < 2.5$ (Section~\ref{sec:reco:jets}). The \MET\ significance $\metsig$ (Section~\ref{sec:reco:met}, Equation~\ref{eq:metsig}) is used to distinguish $W$+jets and dijet backgrounds from the signal, as described below.

Simulated events are weighted to correct the MC simulation to enhance compatibility with data. Corrections are applied event-by-event to efficiencies of the electron and muon trigger, reconstruction, identification and isolation, the jet vertex tagging and pileup rejection by applying the respective weights.

\begin{table}[htbp]
    \centering
    \caption{Analysis regions definitions. The control (validation) regions used to estimate the leading backgrounds, fake electrons and $V$+jets, are indicated. The label $ej$ ($\mu j$) refers to the electron (muon) channel. The ``signal'' subscript denotes the different lepton and jet signal definitions in the text. The dash (---) indicates a selection criterion not applied, where N/A indicates a criterion cannot be applied, in regions which require exactly one signal lepton. Selections on $m_{\ell j}$ and $\metsig$ orthogonalise analysis regions from one another. The $f$CR and $f$VR are only relevant in the context of the electron channel. The $m_{\ell\ell}$ criterion in the $f$CR is only applied in the presence of a second electron.}
    \label{tab:selection}
    \begin{tabular}{c c c c | c}
    \hhline{=====}
    \multirow{2}{*}{Variable}                        & \multirow{2}{*}{$W$CR ($W$VR)} & $f$CR ($f$VR)                    & \multirow{2}{*}{$Z$CR ($Z$VR)} & \multirow{2}{*}{SR}  \\
                                                      &                                & \small{(electron channel only)}  &                                &                      \\ \hline
    $N^{j}_{\textrm{signal}}$                        & $\geq 1$      & $\geq 1$                & $\geq 1$           & $\geq 1$       \\ \Tstrut\Bstrut
    $N^{\ell}_{\textrm{signal}}$                     & 1             & 1                       & 2                  & 1              \\ \Tstrut\Bstrut
    $m_{\ell j}$ [TeV]                               & 1.0--2.0 (2.0--3.0) & 1.0--3.0          & 1.0--2.0 (2.0--3.0) & $>3.0$        \\ \Tstrut\Bstrut
    $p^{\ell}_{\textrm{T}}$ [GeV]                   & $>150$        & $>150$                  & $>150$             & $>150$         \\ \Tstrut\Bstrut
    $p^{j}_{\textrm{T}}$ [GeV]                      & $>130$        & $>130$                  & $>130$             & $>130$         \\ \Tstrut\Bstrut
    $\metsig$~$ej$, $\mu j$                          & $>5.0$, $3.0$ & $<3.0$ ($3.0$--$5.0$)  & ---                & ---            \\ \Tstrut\Bstrut
    $m_{\ell\ell}$ [GeV]                             & N/A           & $>120$                  & 60--120            & N/A            \\ \Tstrut\Bstrut
    $\Delta\eta_{\ell j}$                            & $<3.25$       & ---                     & ---                & $<3.25$        \\ \Tstrut\Bstrut
    $\Delta\phi_{\ell j}$                            & $>2.8$        & ---                     & ---                & $>2.8$         \\ \hhline{=====}
    \end{tabular}
\end{table}

In the $e+j$ channel, the fake-electron background is included as a controlled and validated background source.
Its control region requires $\mathcal{S}(\Etmiss) < 3.0$ to suppress $W$+jets contributions, and its validation region sets $3.0 < \mathcal{S}(\Etmiss) < 5.0$ as a middle ground between fake enrichment and $W$+jets contamination; the resulting fake fraction in the fake CR ranges from 60--90\% for the baseline population and 30--45\% for the signal population.
To maintain orthogonality, the $W$+jets CR (VR) begins at $\mathcal{S}(\Etmiss) = 5.0$ in the $e+j$ channel.
The purity of $W$+jets background in the $W$+jets CR and VR is 69\% in the electron channel.
Further details of the fake-electron background estimation are given in Section~\ref{sec:fakes}.

For the $\mu+j$ channel, the analysis strategy is analogous but simpler, controlling only $W$+jets and $Z$+jets backgrounds.
The $W$+jets CR begins at $\mathcal{S}(\Etmiss) = 3.0$; the fake-muon contribution is less than 1\% and is not separately estimated.
The purity of $W$+jets background in the $W$+jets CR and VR is 72\% in the muon channel.

The $Z$+jets CR (VR), common to both channels, is enriched by requiring a second baseline lepton alongside the signal lepton, with dilepton invariant mass within $[60,\,120]~\GeV$.
Its purity reaches 95\% in both channels.
Subdominant backgrounds from top-quark and diboson processes each contribute less than 10\% in the SR and are not assigned dedicated control regions.
As detailed in Table~\ref{tab:selection}, the $Z$+jets CR (VR) requires exactly two baseline leptons while the $W$+jets CR (VR) requires one, ensuring orthogonality between the two regions. 

\begin{figure}[h!]
  \captionsetup[subfigure]{labelformat=empty}
  \subfloat[(a)]{
    \includegraphics[width=0.5\textwidth]{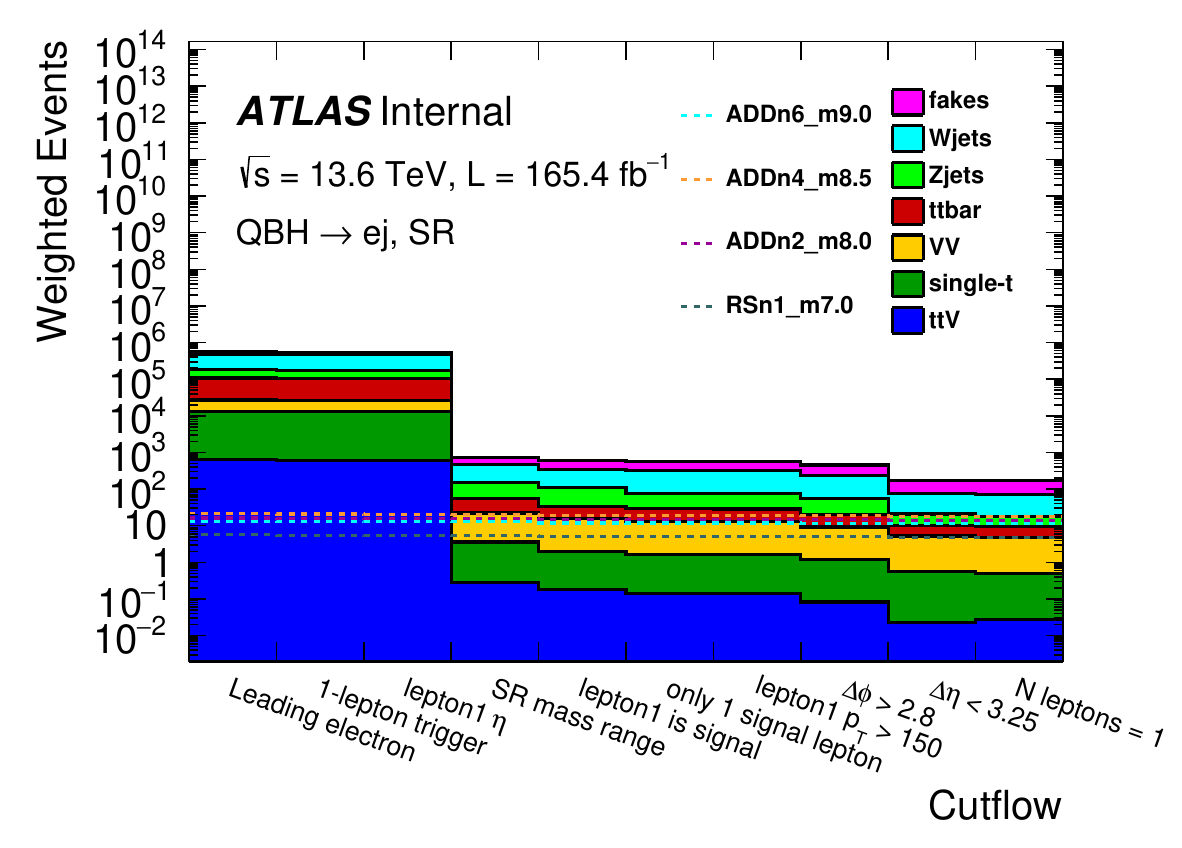}
  }
  \hfill
  \subfloat[(b)]{
    \includegraphics[width=0.5\textwidth]{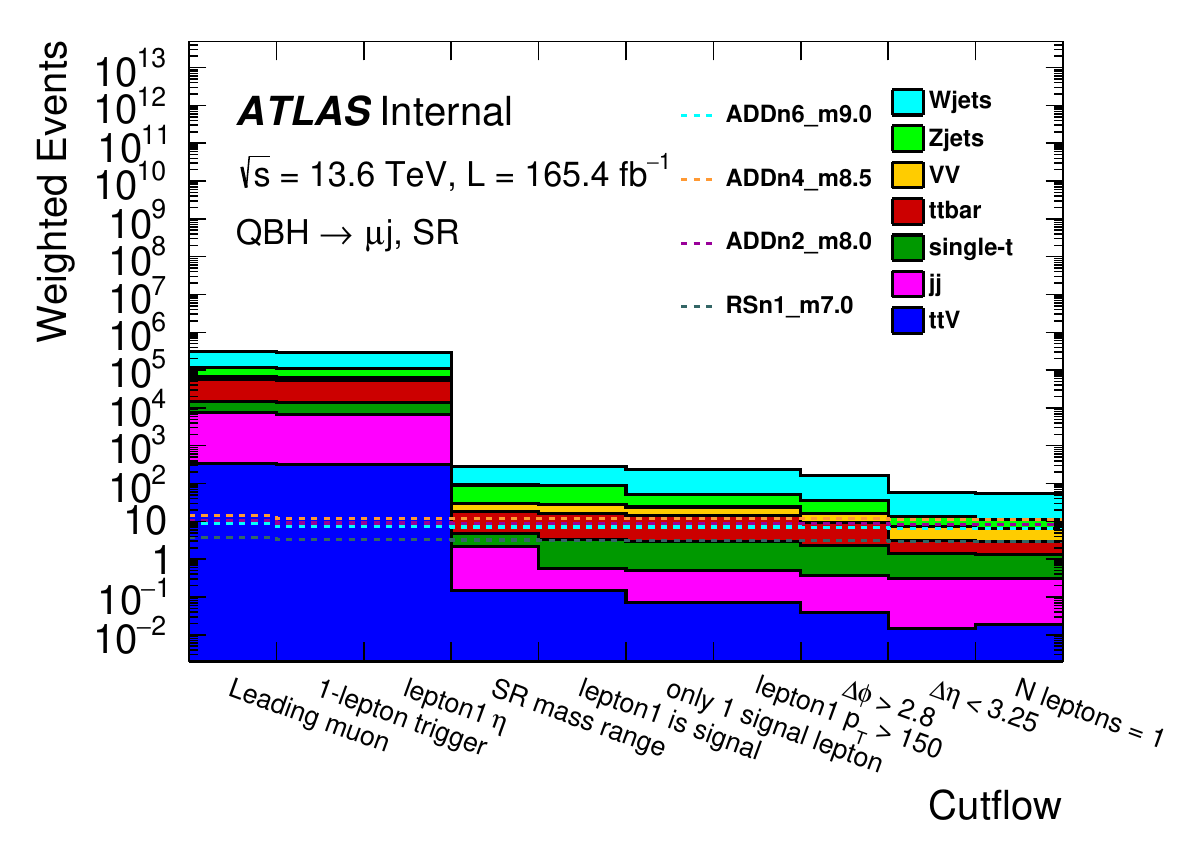}
  }
  \hfill
  \subfloat[(c)]{
    \includegraphics[width=0.5\textwidth]{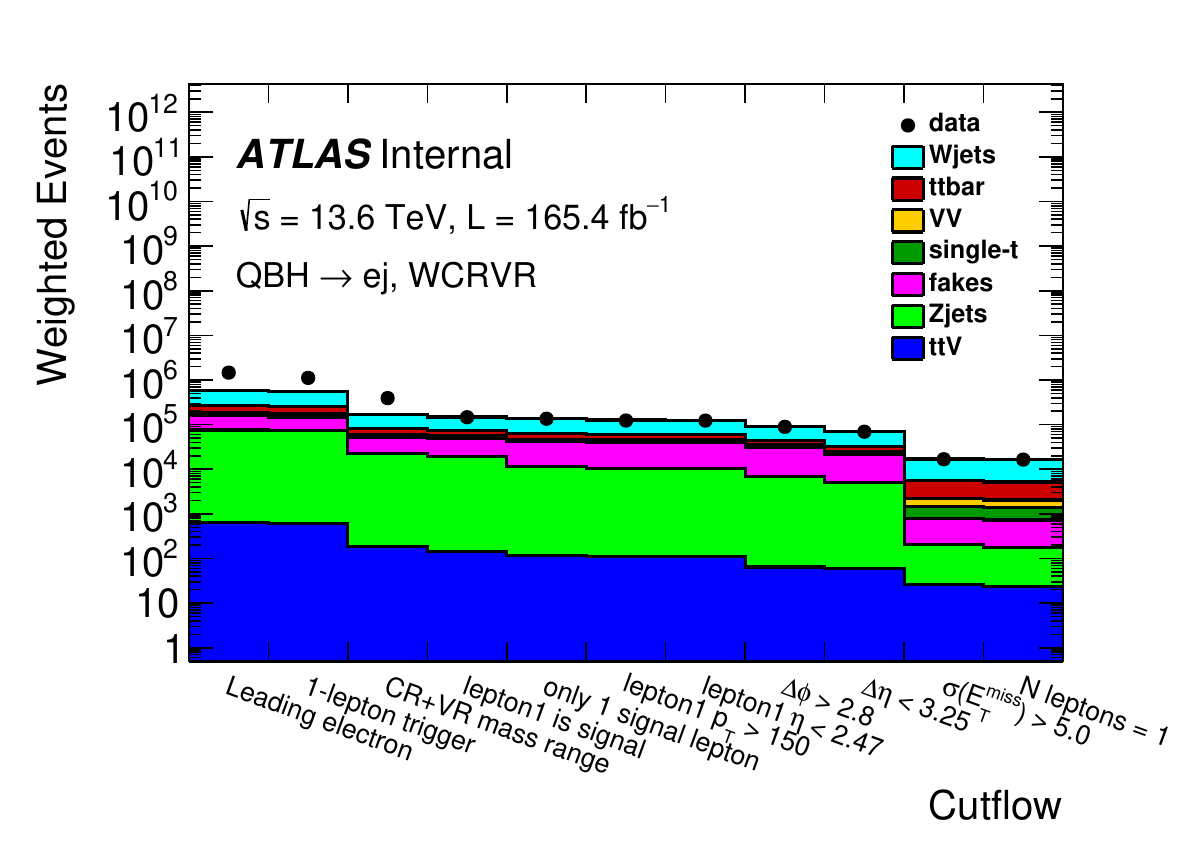}
  }
  \hfill
  \subfloat[(d)]{
    \includegraphics[width=0.5\textwidth]{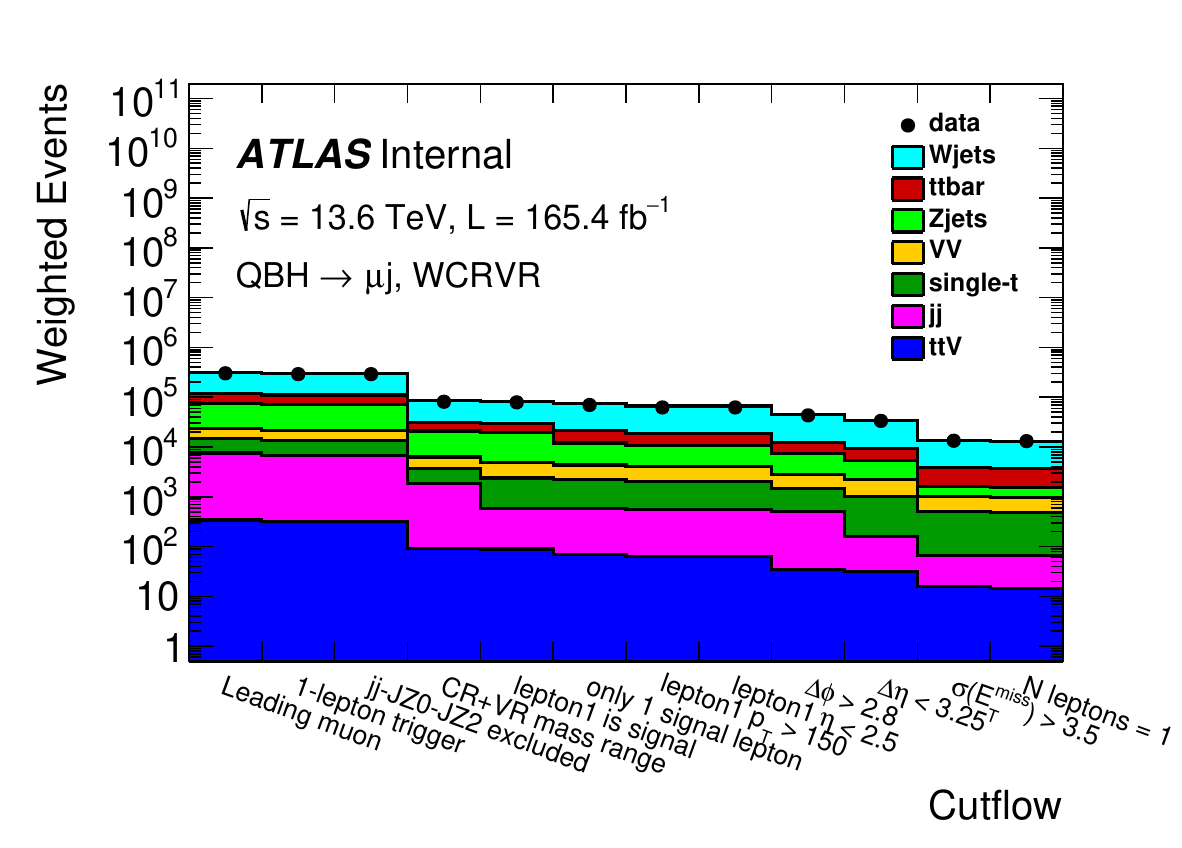}
  }
  \caption{Cutflow of weighted events for the $e+j$ channel (left) and $\mu+j$ channel (right). Top row (a--b): signal region. Bottom row (c--d): $W$+jets CR and VR.}
  \label{fig:cutflows}
\end{figure}

FIG.~\ref{fig:cutflows} shows cutflow plots of weighted events after each selection cut is applied, for the signal region and $W$+jets CR and VR.

\subsection*{Angular cuts}
\begin{figure}[h!]
    \captionsetup[subfigure]{labelformat=empty}
    \subfloat[(a)]{
      \includegraphics[width=0.5\textwidth]{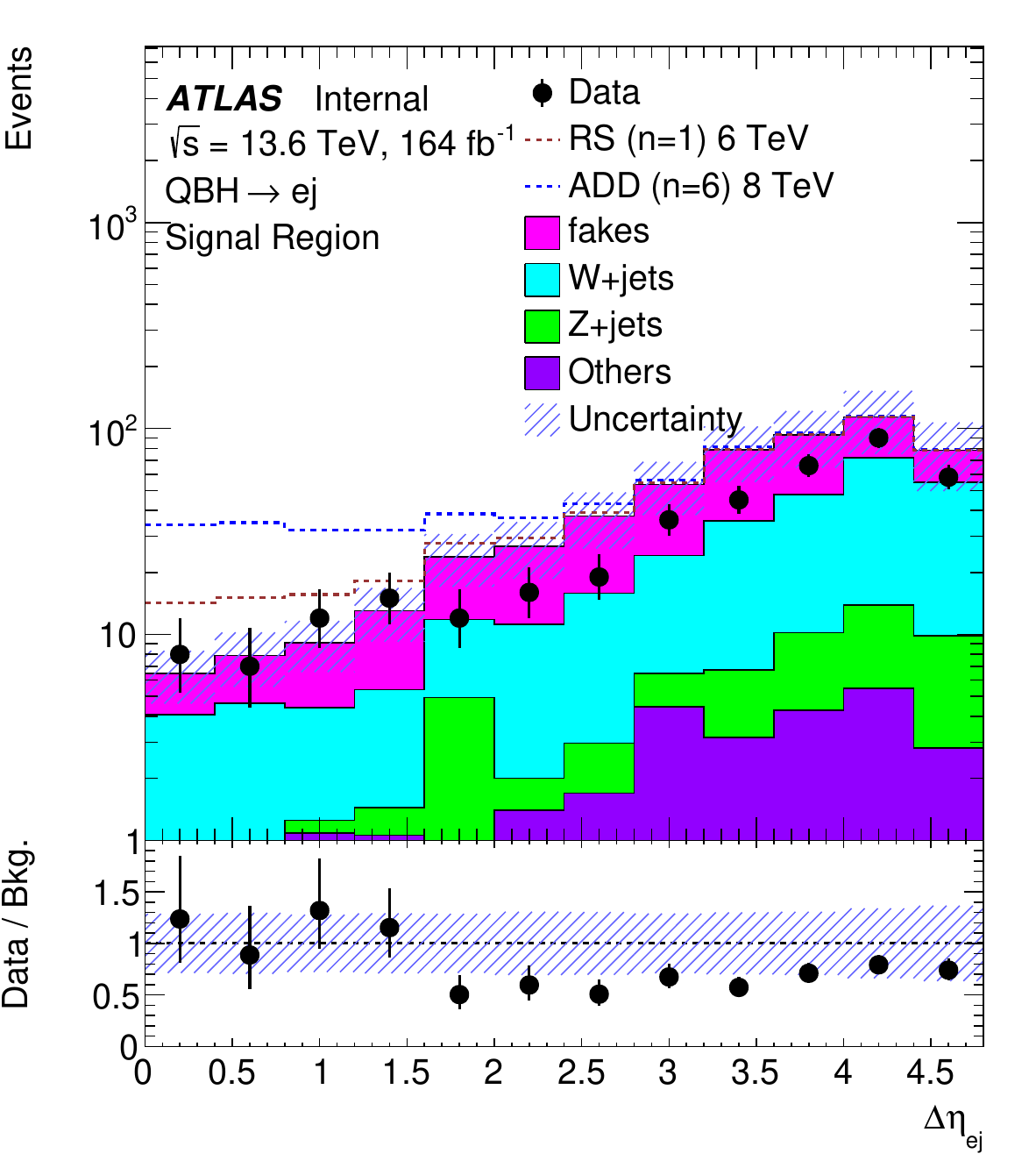}
    }
    \hfill
    \subfloat[(b)]{
      \includegraphics[width=0.5\textwidth]{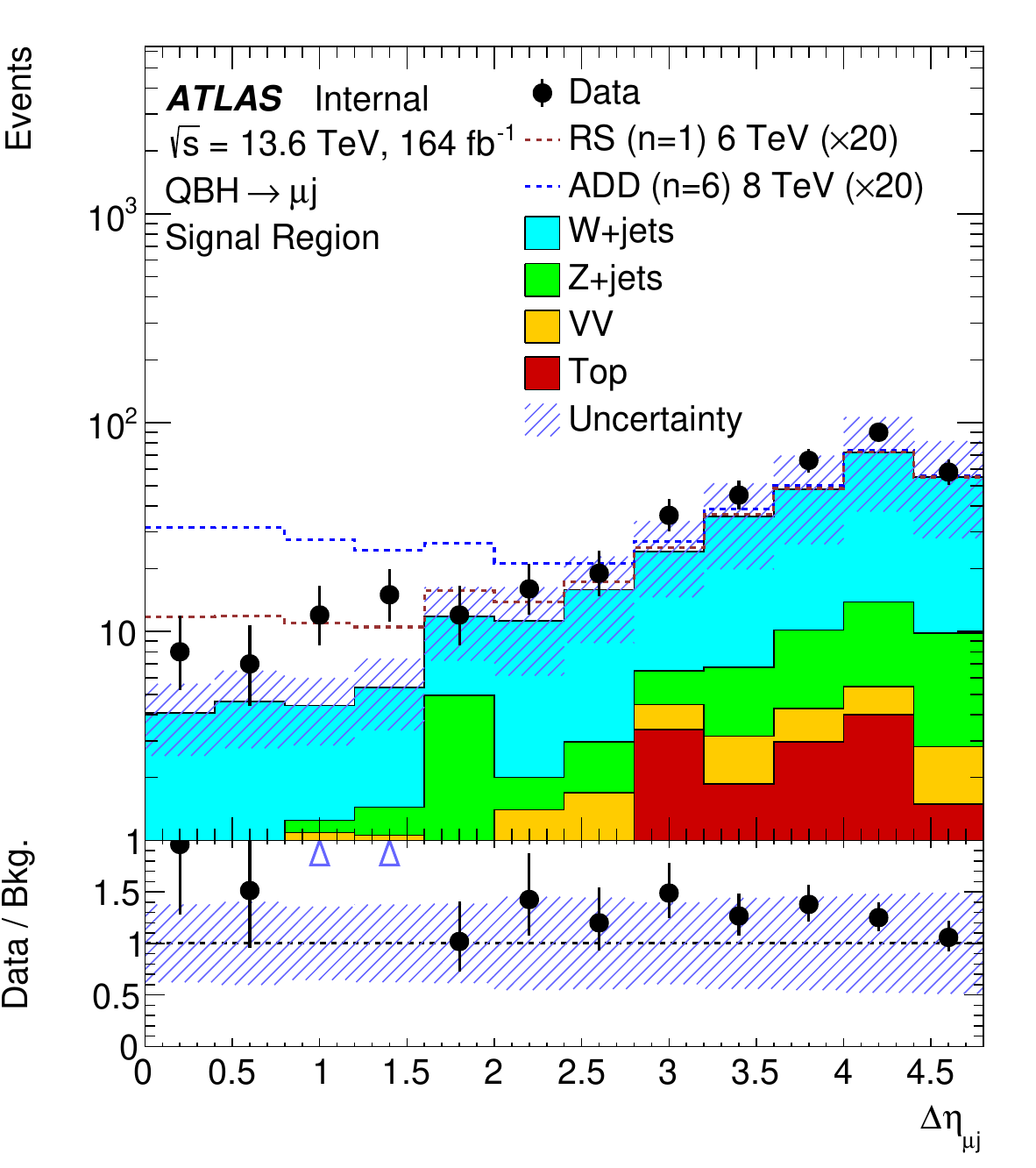}
    }
    \hfill
    \subfloat[(c)]{
      \includegraphics[width=0.5\textwidth]{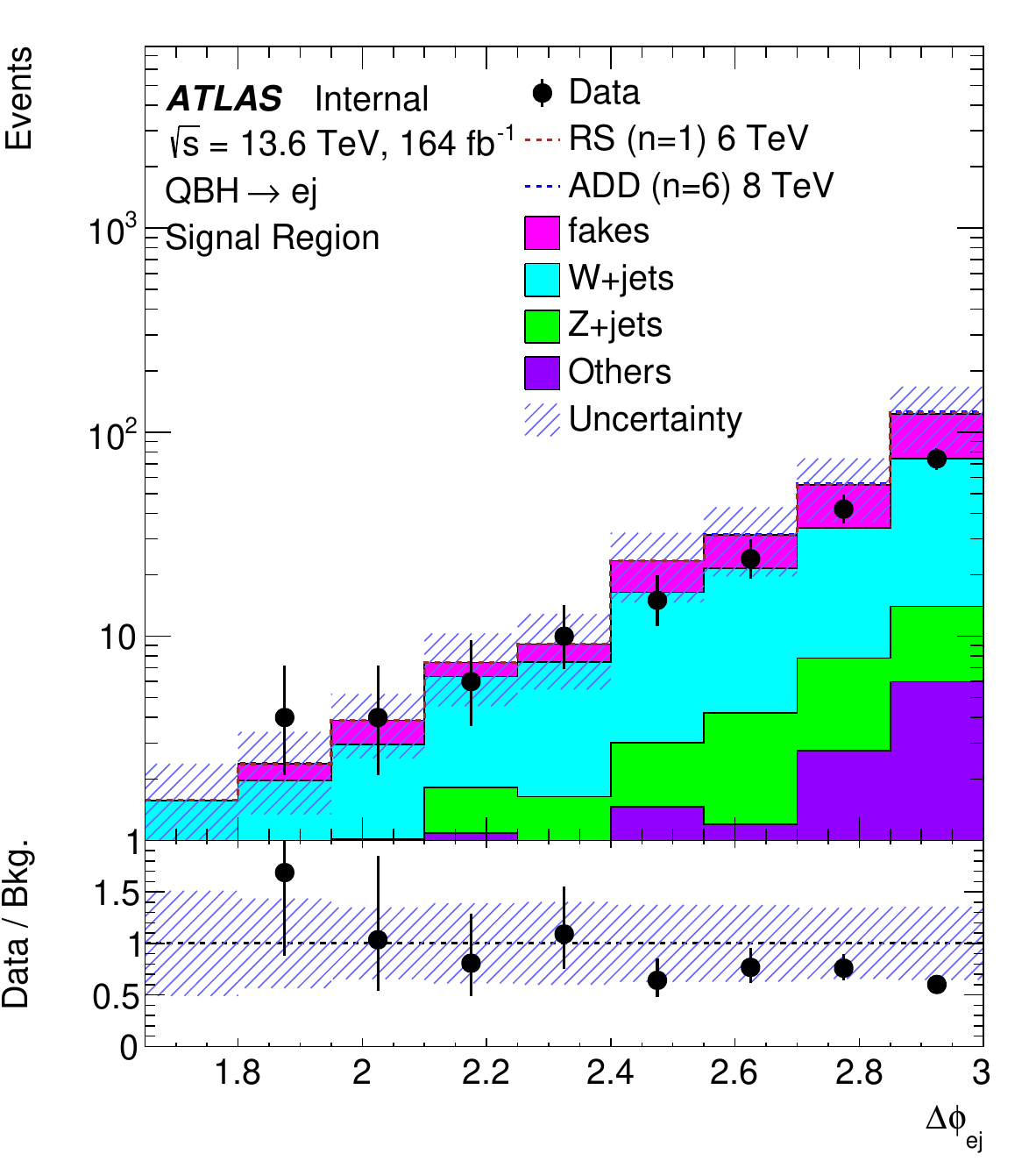}
    }
    \hfill
    \subfloat[(d)]{
      \includegraphics[width=0.5\textwidth]{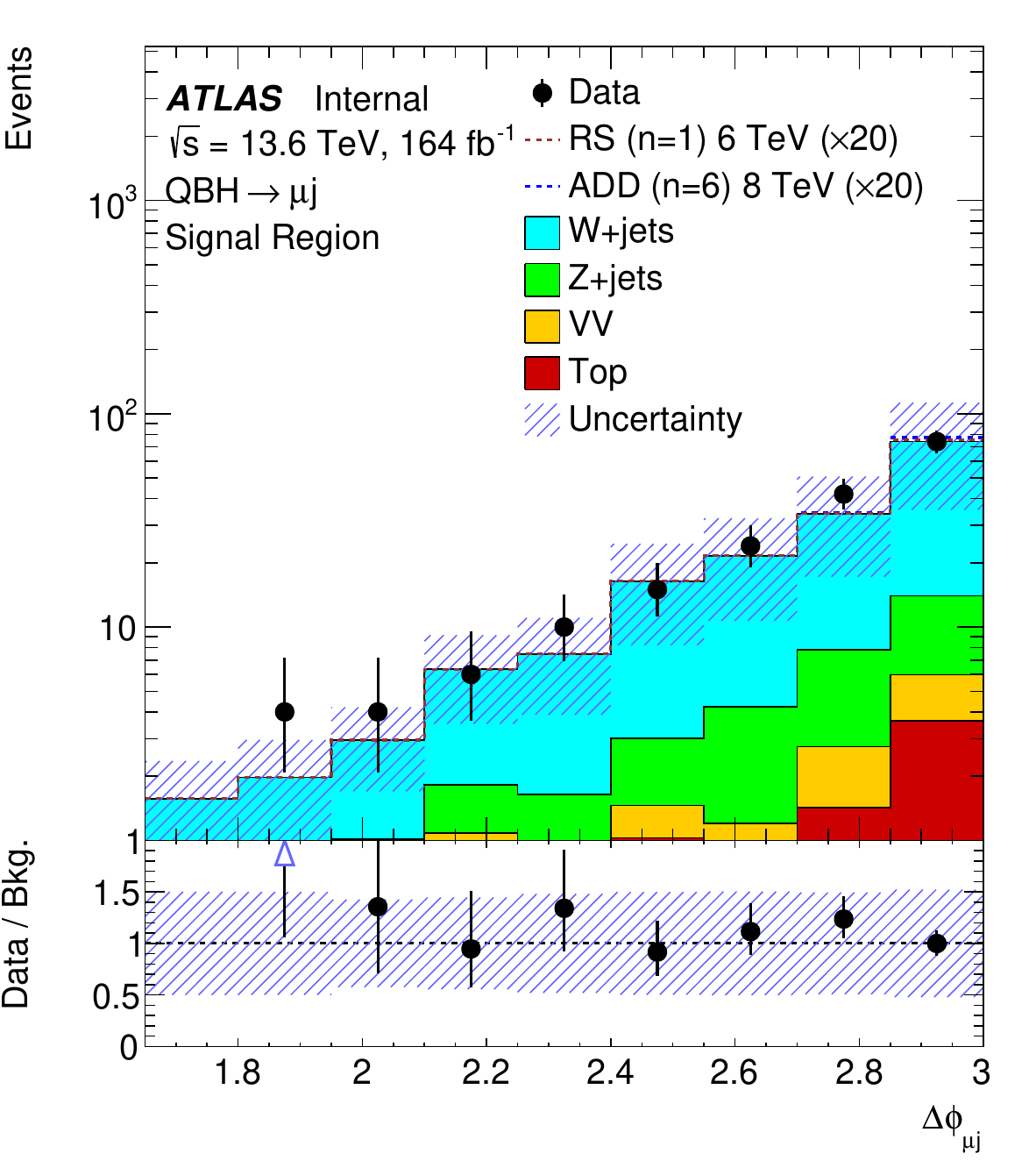}
    }
  \caption{Topological separation of the lepton-jet final-state components. Distributions of $\Delta\eta_{\ell j}$ (top, a--b) and $\Delta\phi_{\ell j}$ (bottom, c--d) in the signal region. The $e+j$ channel is shown on the left and the $\mu+j$ channel on the right. The shaded band in the data-to-background ratio represents the total systematic and statistical uncertainty.}
  \label{fig:QBH:angular_cuts}
\end{figure}

To enhance sensitivity in the SRs, angular cuts on the lepton-jet system are applied, leveraging the expected event topology of $\textrm{QBH}\rightarrow\ell+j$ decays. Specifically, requiring $\Delta\phi_{\ell j} > 2.8$ and $\Delta\eta_{\ell j} < 3.25$ selects events where the lepton and jet are approximately back-to-back in the transverse plane while maintaining high signal efficiency. FIG.~\ref{fig:QBH:angular_cuts} shows distributions of these variables with all SR cuts applied except the angular ones; the shaded band in each ratio panel reflects the total systematic and statistical uncertainty, with the systematic component described in Section~\ref{sec:systematic_uncertainties}. These cuts efficiently suppress background contamination while retaining most of the signal. The angular cuts are optimised to maximise signal sensitivity while maintaining high signal efficiency.

\newpage

\section{Fake Lepton Estimation}
\label{sec:fakes}
This section describes the manner in which the fake electron background is obtained using the data-driven Matrix Method (MM) \cite{ATLAS:2022swp}. Fake electron background consists of objects misidentified as signal electrons, namely jets and non-prompt electrons. The latter include electrons stemming primarily from photon conversions and heavy-flavour decays. 
Fake electrons are largely found within a jet due to decays in the showering process, misidentification and interactions with the detector. These backgrounds are poorly modelled with available MC simulations of QCD processes. They are therefore estimated using a data-driven approach, which relies on the full dataset and MC simulations to account for the contribution of real electrons originating from electroweak processes.

\subsection{Matrix Method}
To assess the number of fake electrons contaminating the signal sample as defined in Section~\ref{sec:reco:electrons}, the MM is used.
This tool relates between the probabilities that a real, fake electron meets the signal criteria, $r, f$, respectively, and the total number of fake electrons in the \textit{baseline} sample, $F$. The baseline sample definition is an analysis choice, set here following the definition in Section~\ref{sec:reco:electrons} albeit passing the electron \texttt{LH\_ID} WP \textit{Medium}. This baseline is defined to avoid a trigger bias which may have been introduced if a looser ID likelihood requirement were used. The fake-electron background is estimated by using the loosest available unprescaled single-electron trigger in each data-taking period (see Table~\ref{tab:datasets::triggers}): a trigger requiring $p_\mathrm{T} > 140$~GeV with only a loose likelihood identification criterion (Section~\ref{sec:reco:electrons}), with the Level-1 seed changing between 2022 and 2023--2024 following the ATLAS hardware calorimeter trigger upgrade.
The single-electron MM defines the following relation

\begin{equation}
    \overbrace{\begin{pmatrix}
        s \\
        b  
        \end{pmatrix}}^{\textrm{detected}}
    = \begin{pmatrix}
        r & f \\
        1-r & 1-f 
        \end{pmatrix} 
        \overbrace{\begin{pmatrix}
            R \\
            F  
        \end{pmatrix}}^{\textrm{inferred}},
    \label{Eq_Matrix_Method}
\end{equation}
where $s, b$ is the number of electrons in the \textit{signal}, \textit{baseline}-not-\textit{signal} samples, respectively, and $R, F$ the true number of real, fake electrons in the \textit{baseline} sample, respectively. Although $s, b$ can be directly measured, $R, F$ can only be evaluated by solving Eq. (\ref{Eq_Matrix_Method}) for them. The entries in the matrix are referred to as real efficiency and fake efficiency, given by
\begin{equation}
r = \frac{N^{\textrm{real}}_{\textrm{signal}}}{N^{\textrm{real}}_{\textrm{baseline}}}, \quad \quad f = \frac{N^{\textrm{fake}}_{\textrm{signal}}}{N^{\textrm{fake}}_{\textrm{baseline}}},
\label{Eq_Real_Fake_Effs}
\end{equation}
where $N^{\textrm{real}}_{\textrm{signal, baseline}}$ is the number of real electrons in the signal, baseline sample and $N^{\textrm{fake}}_{\textrm{signal, baseline}}$ the number of fake electrons in the signal, baseline sample. In other words, $r$ ($f$) reflects the probability for a real (fake) electron in the baseline sample to be found also in the signal sample. By construction, the total number of signal electrons equals
\begin{equation}
s = rR + fF,
\end{equation}
where the first term contains the real electron contribution and the second term that of fake electrons. Inverting the matrix in Eq. (\ref{Eq_Matrix_Method}), an equation for the truth variables is found as

\begin{equation}
    \begin{pmatrix}
        R \\
        F  
    \end{pmatrix} = 
    \frac{1}{r(1-f)-f(1-r)}
    \begin{pmatrix}
        1-f & -f \\
        r-1 & r 
        \end{pmatrix}
        \begin{pmatrix}
            s \\
            b  
            \end{pmatrix},
\end{equation}
from which the total number of fake electrons in the signal sample can be determined as
\begin{equation}
fF = \frac{f}{r-f}[rb - (1-r)s],
\label{Eq_Fake_Contribution}
\end{equation}
containing only measurable quantities. The real, fake efficiencies are measured as a two-dimensional function of the $p_{\textrm{T}}$ and $|\eta|$ in dedicated regions where the real-, fake-electron purity is high. In this respect, the quantities $N^{\textrm{real (fake)}}_{\textrm{signal,baseline}}$ in Eq. (\ref{Eq_Real_Fake_Effs}) indicate the number of real (fake) events measured in those individual regions. Once the rates have been reliably measured and encoded in two-dimensional histograms of $p_{\textrm{T}}$ and $|\eta|$, the expression in Eq. (\ref{Eq_Fake_Contribution}) is used to construct a \texttt{fakeWeight} variable,

\begin{align}
    \begin{split}
    \texttt{fakeWeight}^i &= \\ w^i_f(p^i_{\text{T}}, \eta^i) 
    &= 
    \begin{cases}
        \frac{f^i}{r^i - f^i}(r^i - 1) & | \ e^i \text{ is signal}\phantom{\text{ baseline-not}} \quad (s = 1, b = 0) \\[8pt]
        \frac{f^i}{r^i - f^i}r^i, & | \ e^i \text{ is baseline-not-signal} \quad (s = 0, b = 1)
    \end{cases}
    \end{split}
    \label{eq:fakeWeight}
\end{align}

where $i$ indicates the event index. This weight is then applied event-by-event to the baseline data sample, obtaining a fake background estimate,

\begin{equation}
    N^{\textrm{bkg.}}_{\textrm{fake}} = \sum^{N^{\textrm{data}}_{\textrm{baseline}}}_{i} w^{i}_{f}(p^{i}_{\textrm{T}}, \eta^{i}) \times 1,
    \label{eq:fakeWeight_application}
\end{equation}
where $\sum^{N^{\textrm{data}}_{\textrm{baseline}}}_{i} 1$ is the total number of baseline data events. Description of how the real-enriched and fake-enriched regions are constructed and thus, the real and fake efficiencies measured, is given in the next sections.

Both the real and fake rates are binned in $|\eta|$ following detector geometry: two bins in the barrel region ($0<|\eta|<1.37$) and two bins in the endcap region, one with TRT coverage ($1.52<|\eta|<2.01$) and one without TRT coverage ($2.01<|\eta|<2.47$). Another bin, set in the barrel/endcap transition region ($1.37<|\eta|<1.52$), remains empty following rejection of electrons with these $|\eta|$ values. $p_{\textrm{T}}$ binning widens logarithmically, consistent with lowering statistics in high-$p_{\textrm{T}}$ values.

\subsection{Real rate calculation}
To compute the real electron efficiency, $r$, a real-enriched region is constructed where real electron candidates passing signal and baseline requirements are counted, following Eq. (\ref{Eq_Real_Fake_Effs}). Aiming to maximise the purity of real electrons, MC simulations of Drell-Yan processes, $Z\rightarrow \ell\ell$, are used. Description of the samples is given in Table~\ref{tab:zjets_MC} in Section~\ref{sec:background}. In addition, MC samples whose \texttt{bornMass} variable, the invariant mass of the two leading truth leptons, is filtered above 105 GeV, are used to obtain a consistently-high $r$ throughout the electron $p_{\textrm{T}}$ spectrum. Avoiding double-counting of events, the inclusive samples are filtered at $\texttt{bornMass}<105 \; \textrm{GeV}$ when used together with the high-mass samples. As high-mass samples are not yet fully available for Run3 and hence the following plots shows real-rate calculation using Run2 simulations. A rate of above 90\% is expected regardless of which run is considered.

\begin{figure}[h!]
  \captionsetup[subfigure]{labelformat=empty}
    \subfloat[(a)]{
      \includegraphics[width=0.5\textwidth]{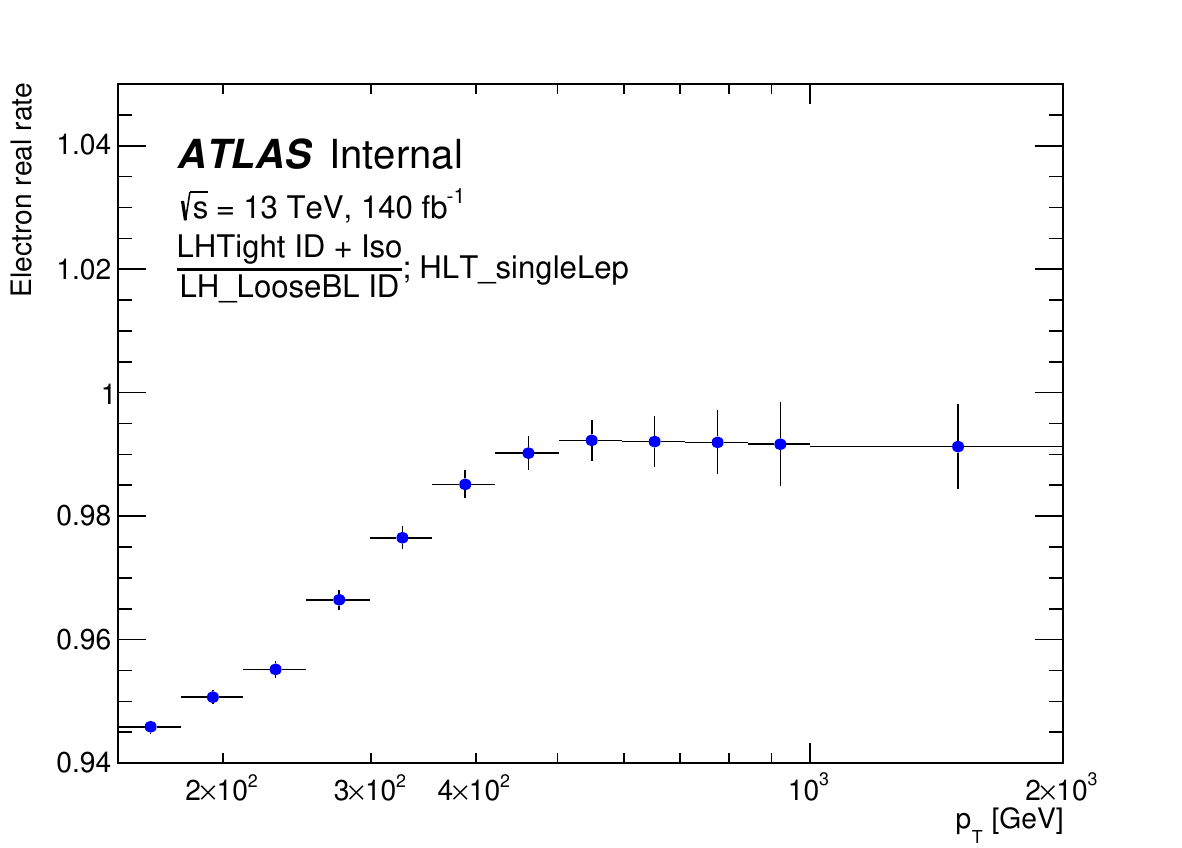}
    }
    \hfill
    \subfloat[(b)]{
      \includegraphics[width=0.5\textwidth]{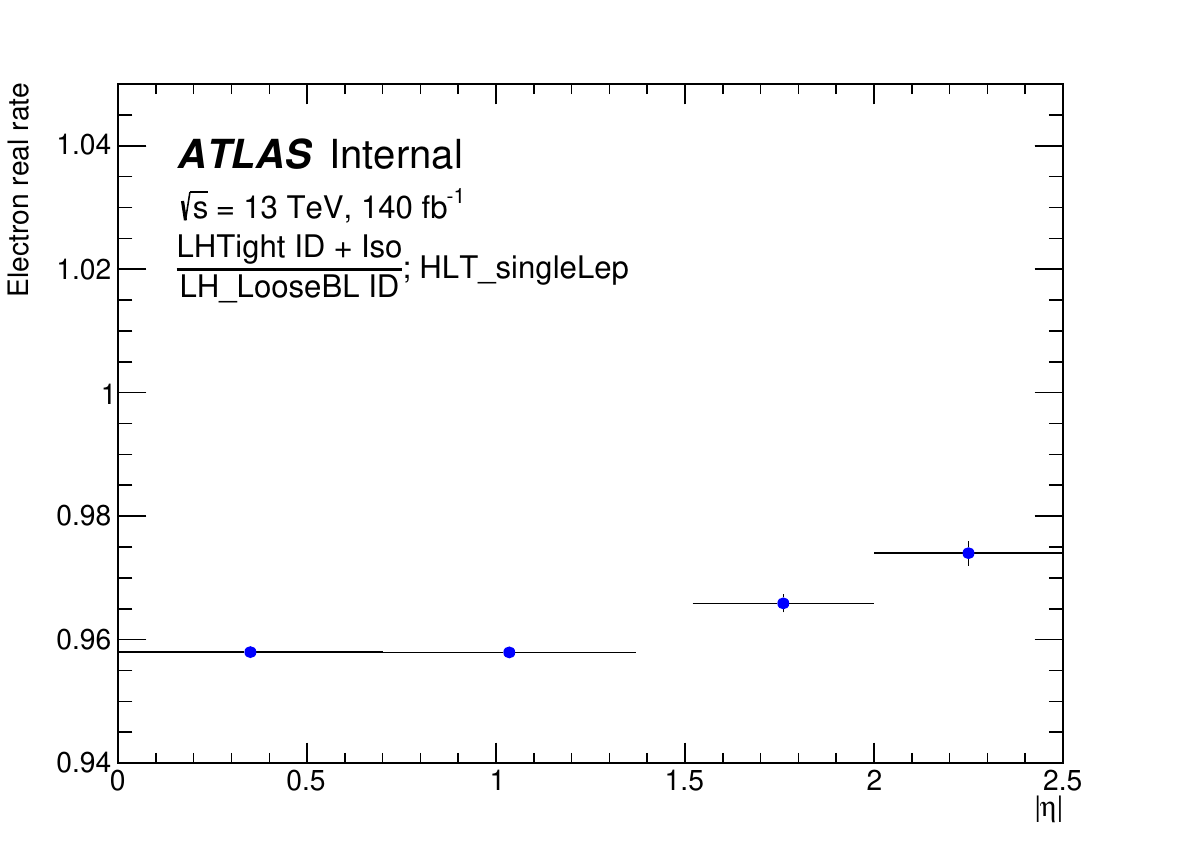}
    }
\caption{Real electron efficiency as a function of (a) $p_{\textrm{T}}$ and (b) $|\eta|$.}
\label{fig:Real_Rates}
\end{figure}

Events are selected using the recommended single-electron trigger chain as specified in Table~\ref{tab:datasets::triggers}. Every event must contain exactly two truth-matched, opposite sign electrons which lie within the detector's fiducial region, $0<|\eta|<2.47$, and outside the barrel/endcap transition region. Furthermore, every electron must satisfy $p_{\textrm{T}}>150 \; \textrm{GeV}$, to remain within the trigger plateau. FIG.~\ref{fig:Real_Rates} shows the real efficiencies binned in $p_{\textrm{T}}$ and $|\eta|$.

\begin{figure}    
    \centering
    \includegraphics[width=0.5\textwidth]{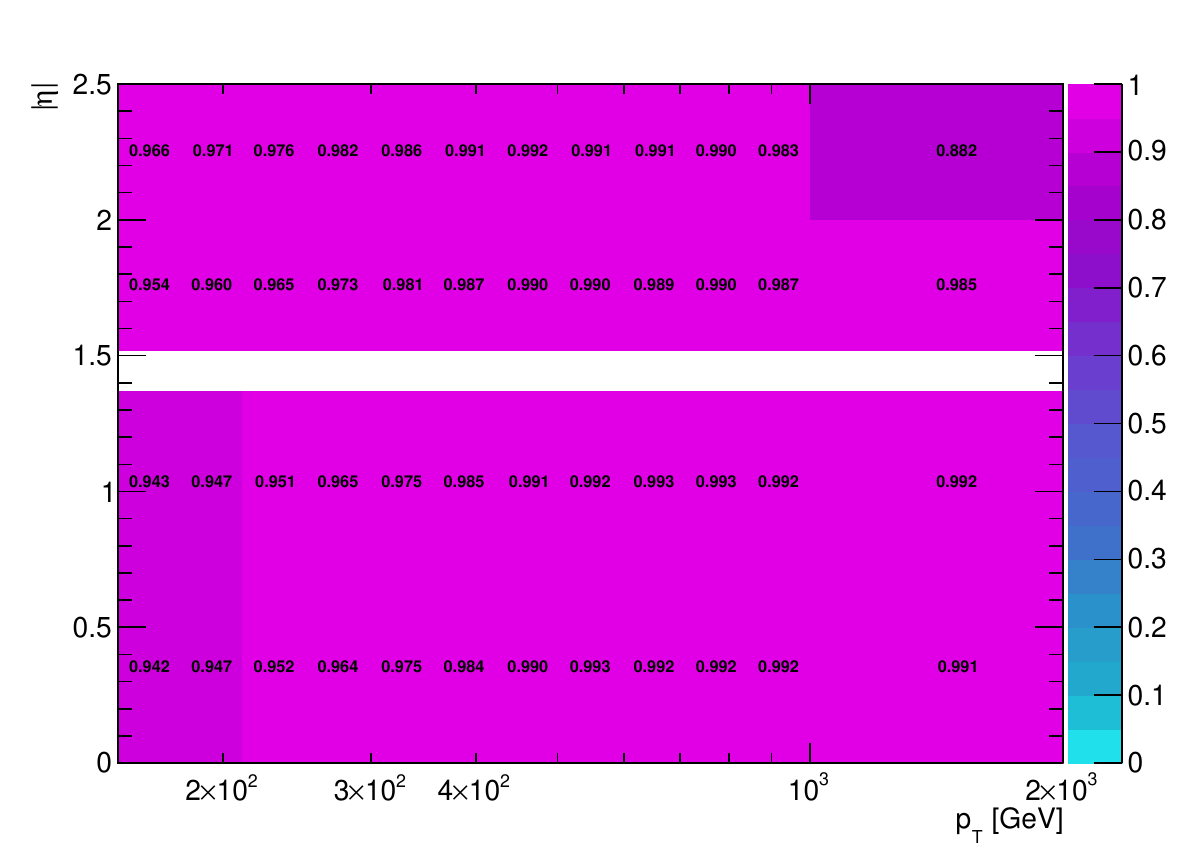}
    \caption{Real electron efficiency as a function of $p_{\textrm{T}},|\eta|$}    
    \label{fig:Real_Rate_2D}
\end{figure}

As depicted, the real rate remains high throughout the phase-space, between 90\%--99\%. FIG.~\ref{fig:Real_Rate_2D} gives the two-dimensional encoding of the real rate, using the same binning.

\subsection{Fake rate calculation}

The fake efficiency, $f$, represents the probability that a baseline fake electron satisfies the signal criteria. Since $f$ cannot be reliably determined solely from simulation, a data-driven approach is used for its calculation. This calculation is performed in a region where the QCD content is high, referred to as the fake-enriched region. The fake-background estimate is then validated in a similar (but orthogonal) fake validation region. Both the fake-enriched (control) and fake-validation regions are orthogonal to the analysis regions. This strategy is also noted and illustrated schematically under Section~\ref{sec:qbh:strategy}. The fake control region is defined by cuts designed to suppress electroweak processes, leaving a QCD-enriched sample: 

\begin{itemize}
    \item Reduction of $W$ decays
    \begin{itemize}
    \item [\textopenbullet] $\mathcal{S}(E^{\textrm{miss}}_{\textrm{T}})<3.0$
    \end{itemize}
    \item Reduction of $Z$ decays
    \begin{itemize}
    \item [\textopenbullet] $m_{ee} > 120~\textrm{GeV} \ | \ N_{\textrm{baseline}}=2$
    \item [\textopenbullet] $N_{\textrm{signal e}} < 2$
    \end{itemize}
\end{itemize}

\begin{figure}[h!]
  \captionsetup[subfigure]{labelformat=empty}
    \subfloat[(a)]{
      \includegraphics[width=0.5\textwidth]{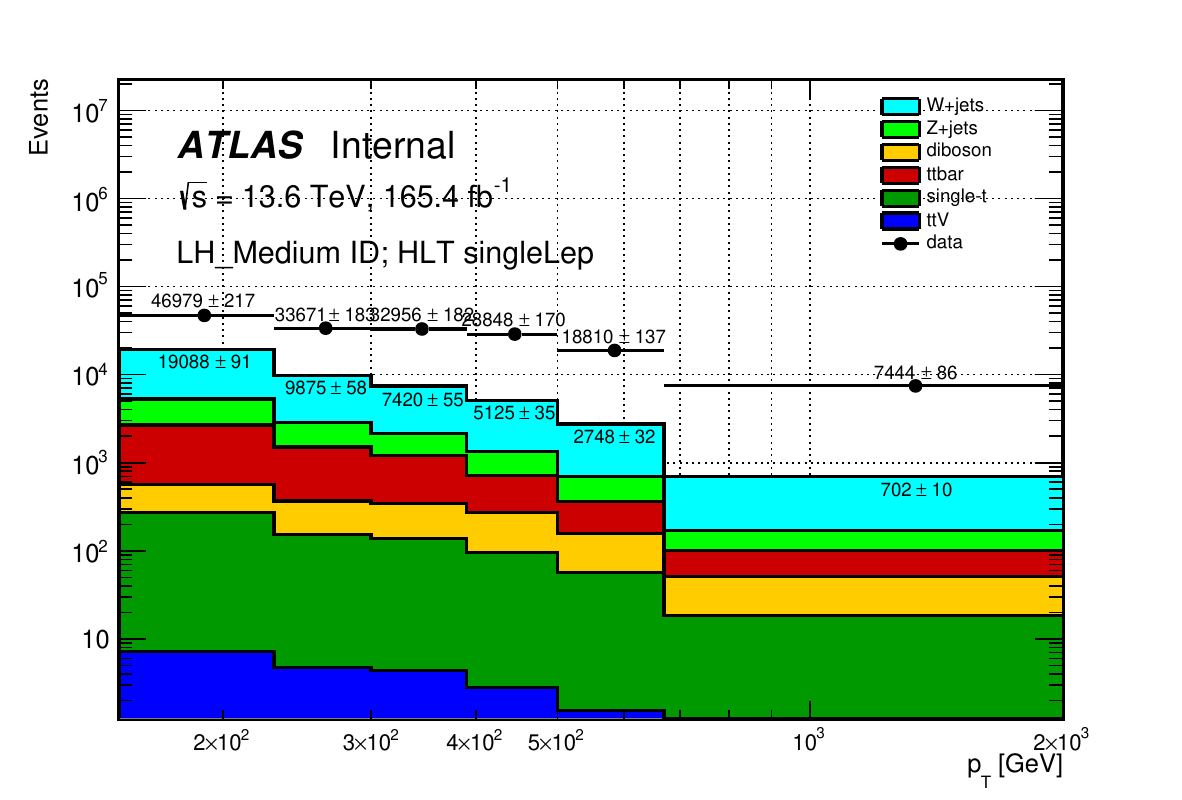}
    }
    \hfill
    \subfloat[(b)]{
      \includegraphics[width=0.5\textwidth]{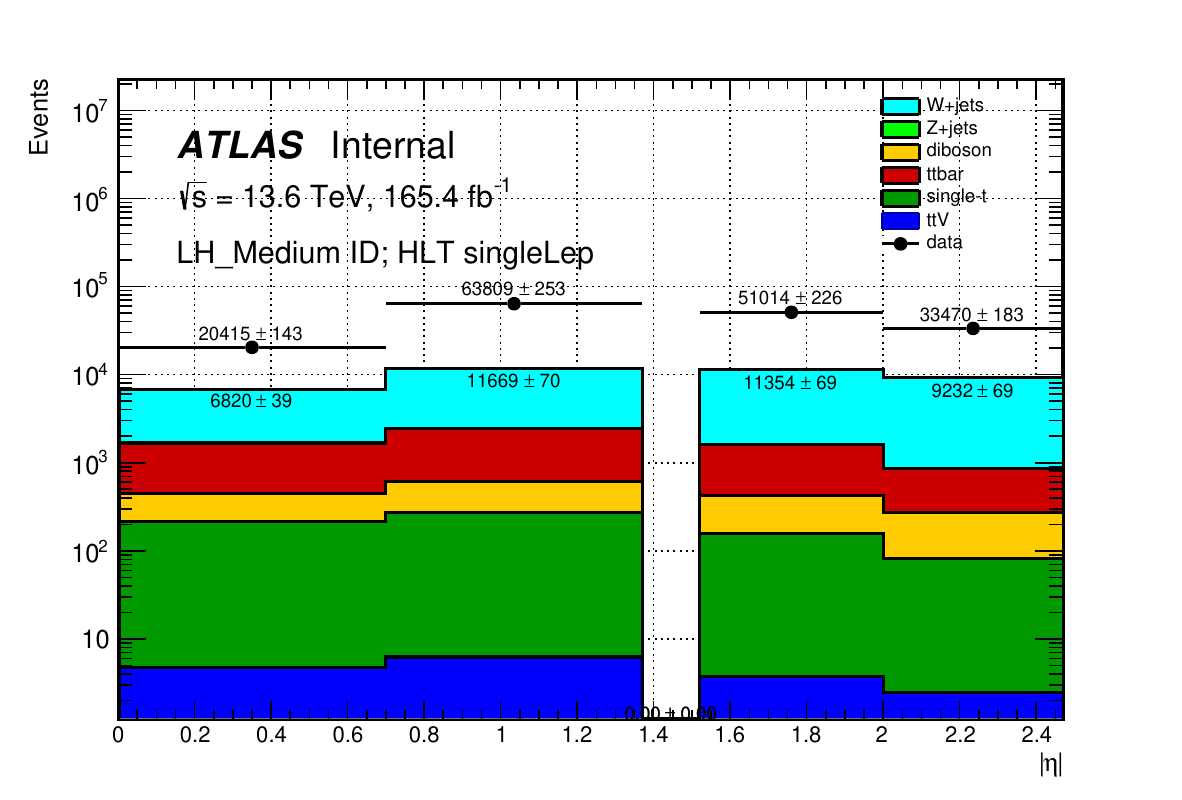}
    }
    \hfill
    \subfloat[(c)]{
      \includegraphics[width=0.5\textwidth]{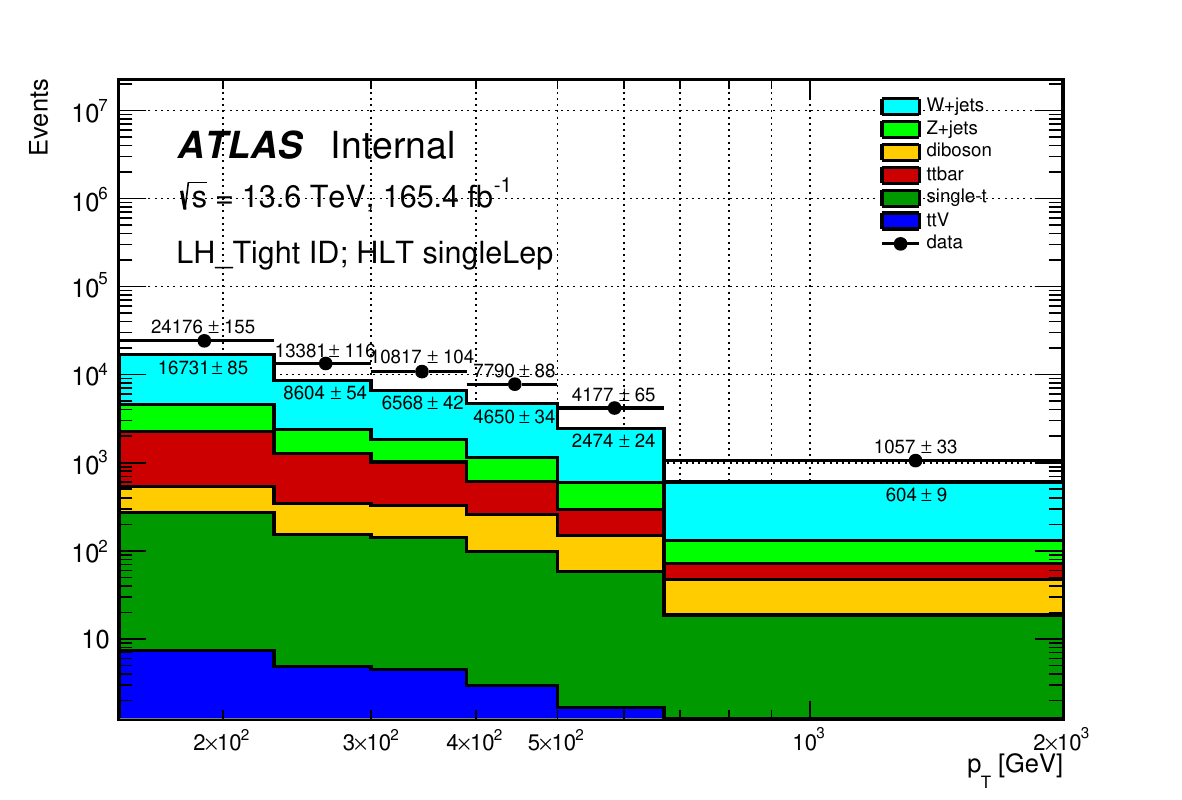}
    }
    \hfill
    \subfloat[(d)]{
      \includegraphics[width=0.5\textwidth]{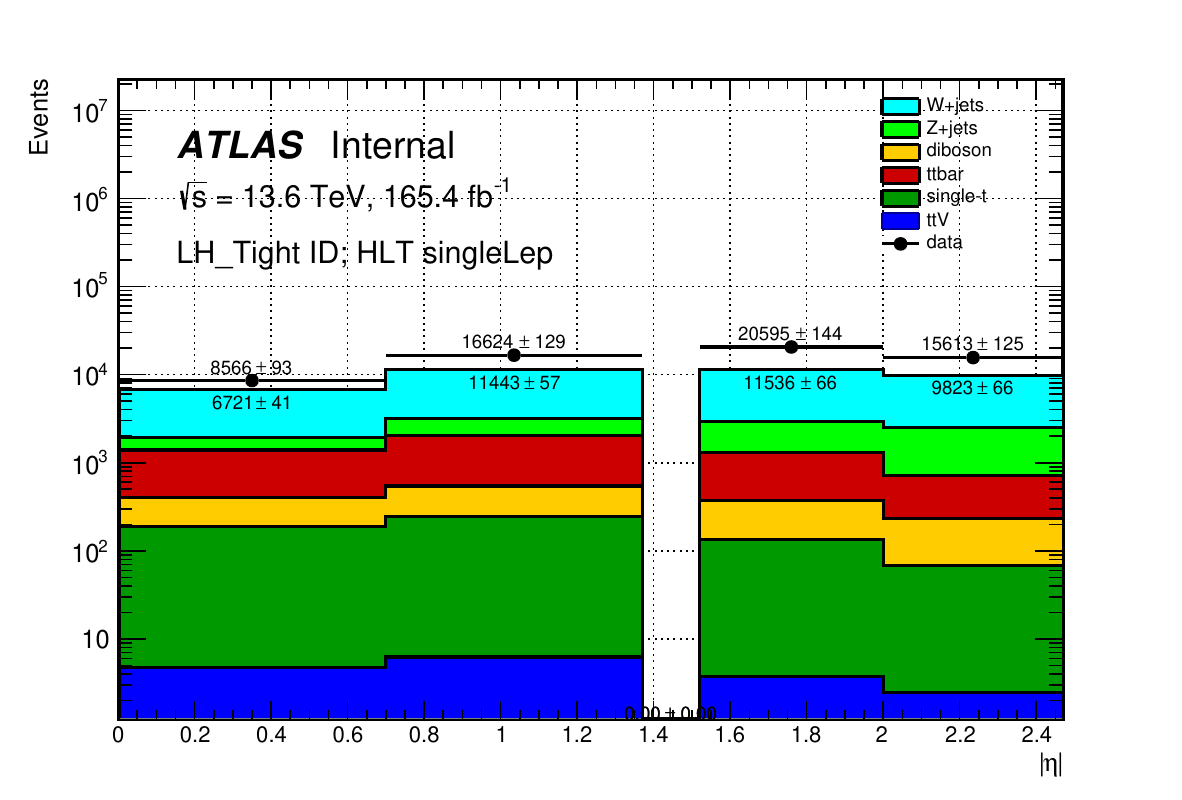}
    }
\caption{Distributions of the leading electron $p_{\textrm{T}}$ (left) and $|\eta|$ (right) following the signal (top, a--b) and baseline (bottom, c--d) requirements in the fake-enriched region. The full dataset is plotted on top of the electroweak backgrounds modelled via MC simulations. The hollow areas between the data and electroweak backgrounds is where the fake background estimate is needed.}
\label{fig:Fake_Enriched_Region_Dilution}
\end{figure}

These cuts are placed in addition to the single-electron triggers defined in the previous section. The real-electron dilution in this region following those selections is modelled using the same MC simulations as described in Section~\ref{sec:background}. This contribution is subtracted from the data for the calculation of $f$. The behaviour of the data and MC-modelled background in the fake-enriched region is shown as a function of the leading electron $p_{\textrm{T}}$ and $|\eta|$ for both the signal and baseline criteria is shown in FIG.~\ref{fig:Fake_Enriched_Region_Dilution}.

\begin{figure}[h!]
  \captionsetup[subfigure]{labelformat=empty}
    \subfloat[(a)]{
      \includegraphics[width=0.5\textwidth]{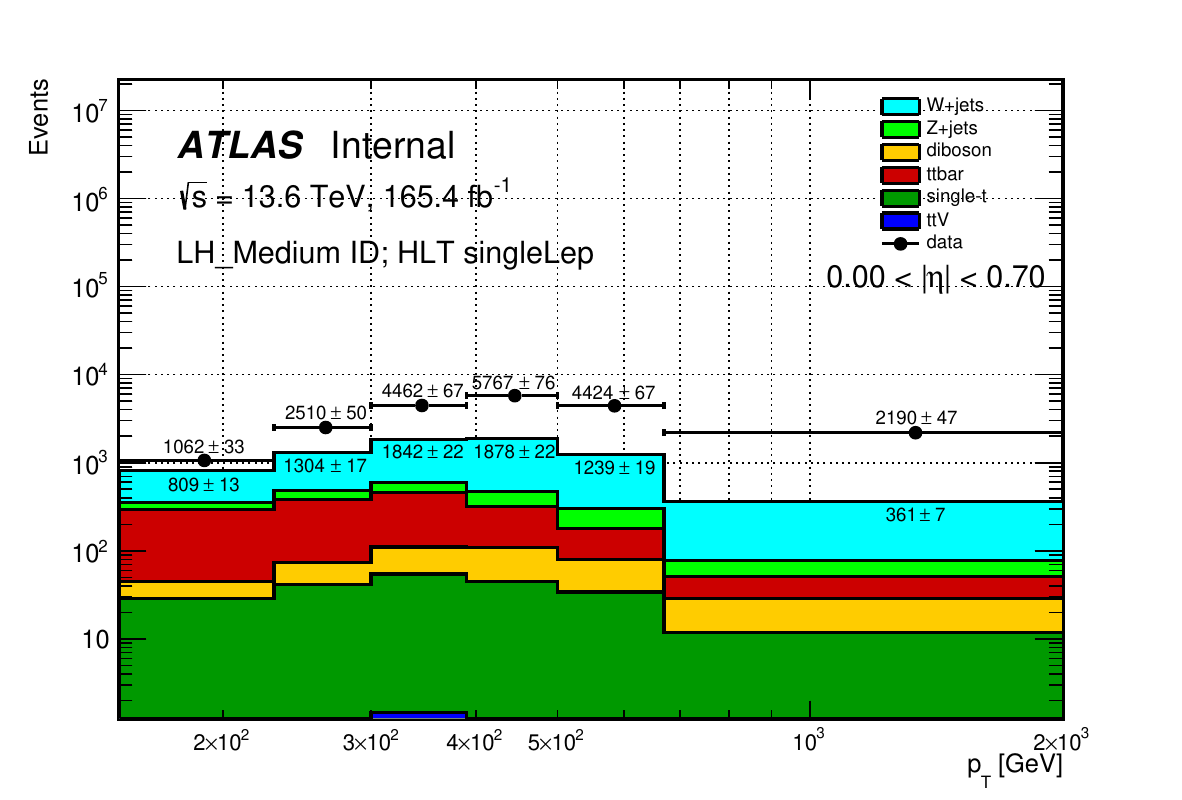}
    }
    \hfill
    \subfloat[(b)]{
      \includegraphics[width=0.5\textwidth]{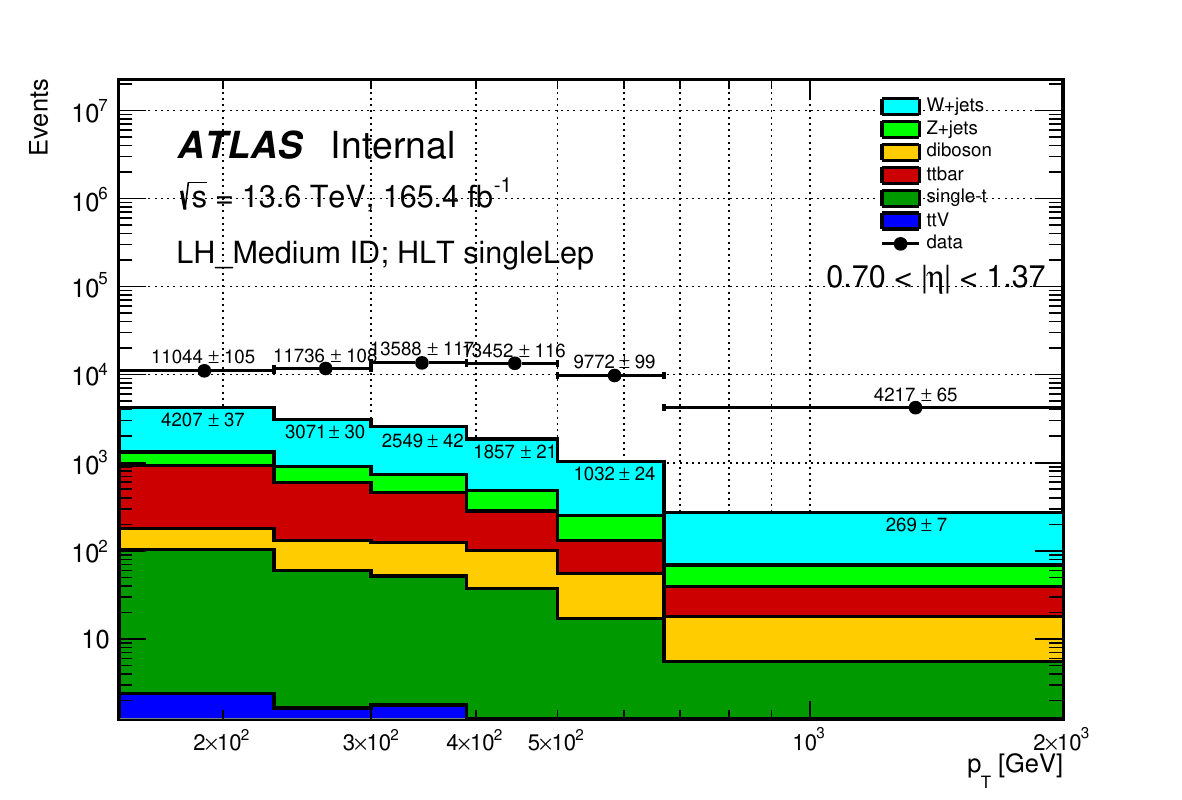}
    }
    \hfill
    \subfloat[(c)]{
      \includegraphics[width=0.5\textwidth]{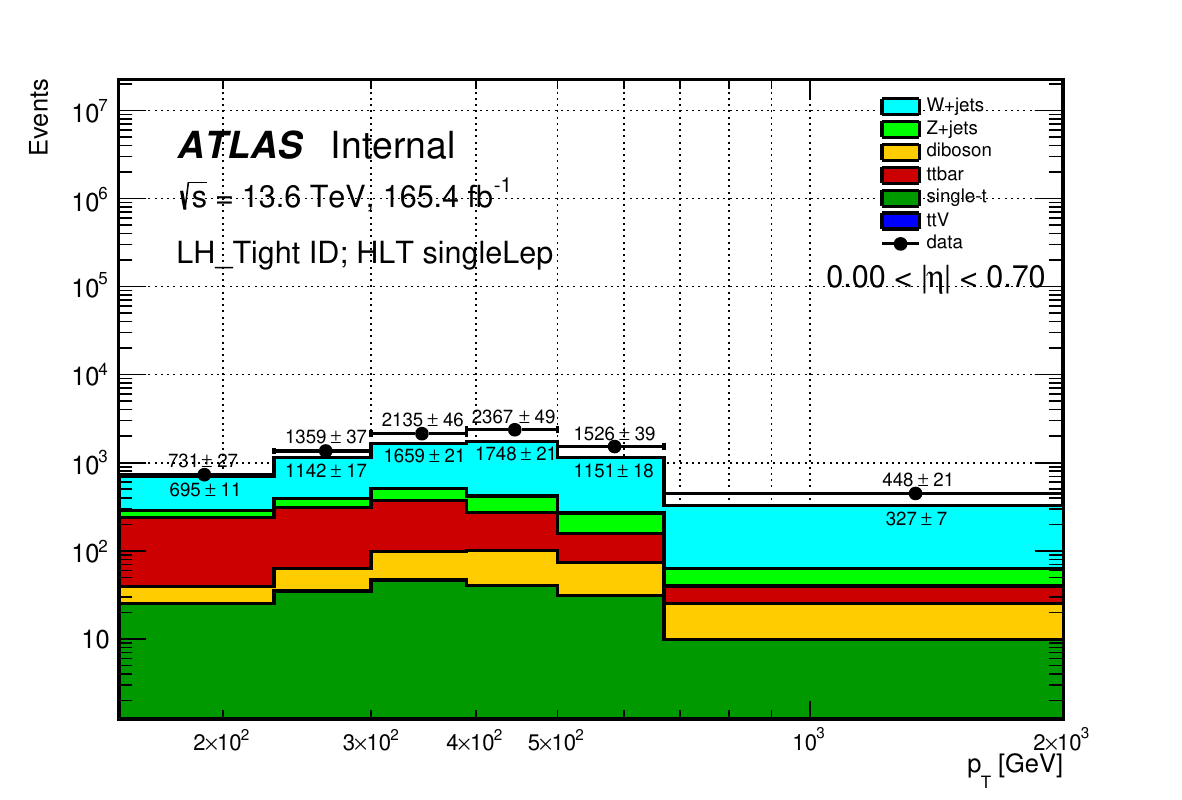}
    }
    \hfill
    \subfloat[(d)]{
      \includegraphics[width=0.5\textwidth]{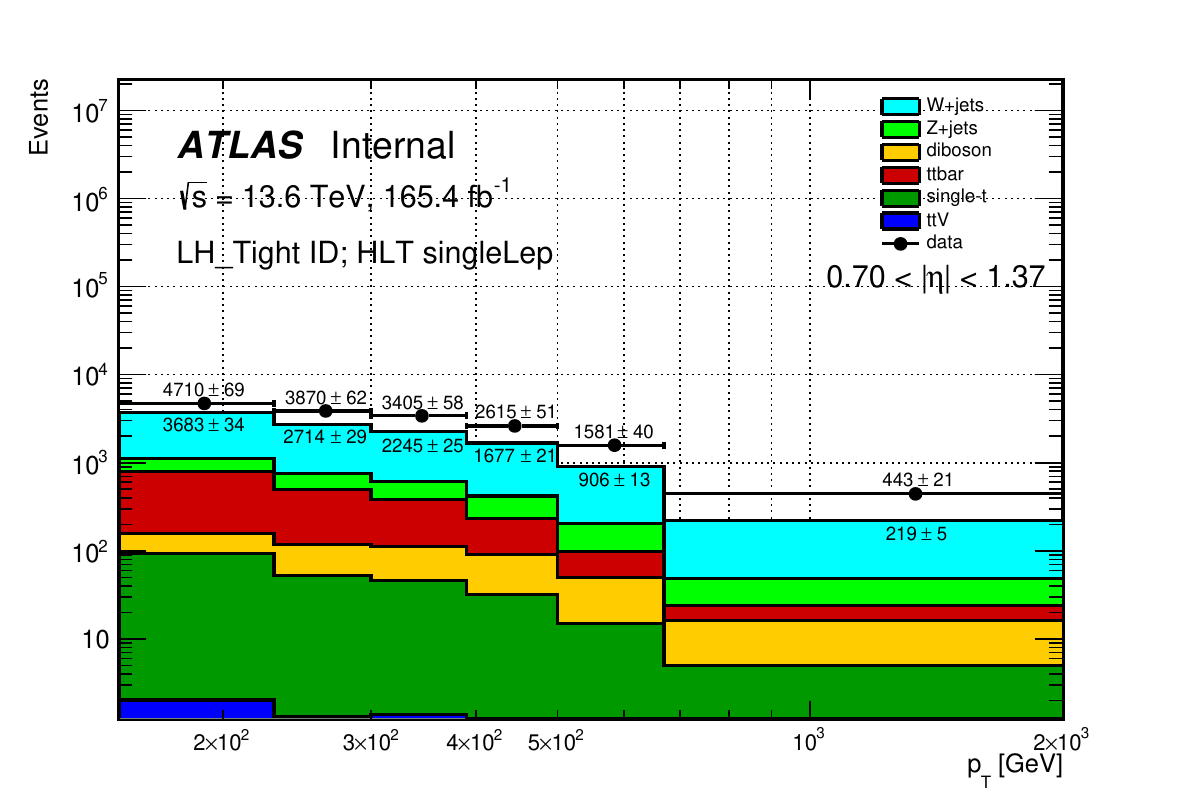}
    }
\caption{Distributions of the leading electron $p_{\textrm{T}}$ following the signal (top, a--b) and baseline (bottom, c--d) requirements in the fake-enriched region. (a,c) $0.00<|\eta|<0.70$; (b,d) $0.70<|\eta|<1.37$. The full dataset is plotted on top of the electroweak backgrounds modelled via MC simulations.}
\label{fig:Fake_Enriched_Region_Dilution_0.00_1.37}
\end{figure}

\begin{figure}[h!]
  \captionsetup[subfigure]{labelformat=empty}
    \subfloat[(a)]{
      \includegraphics[width=0.5\textwidth]{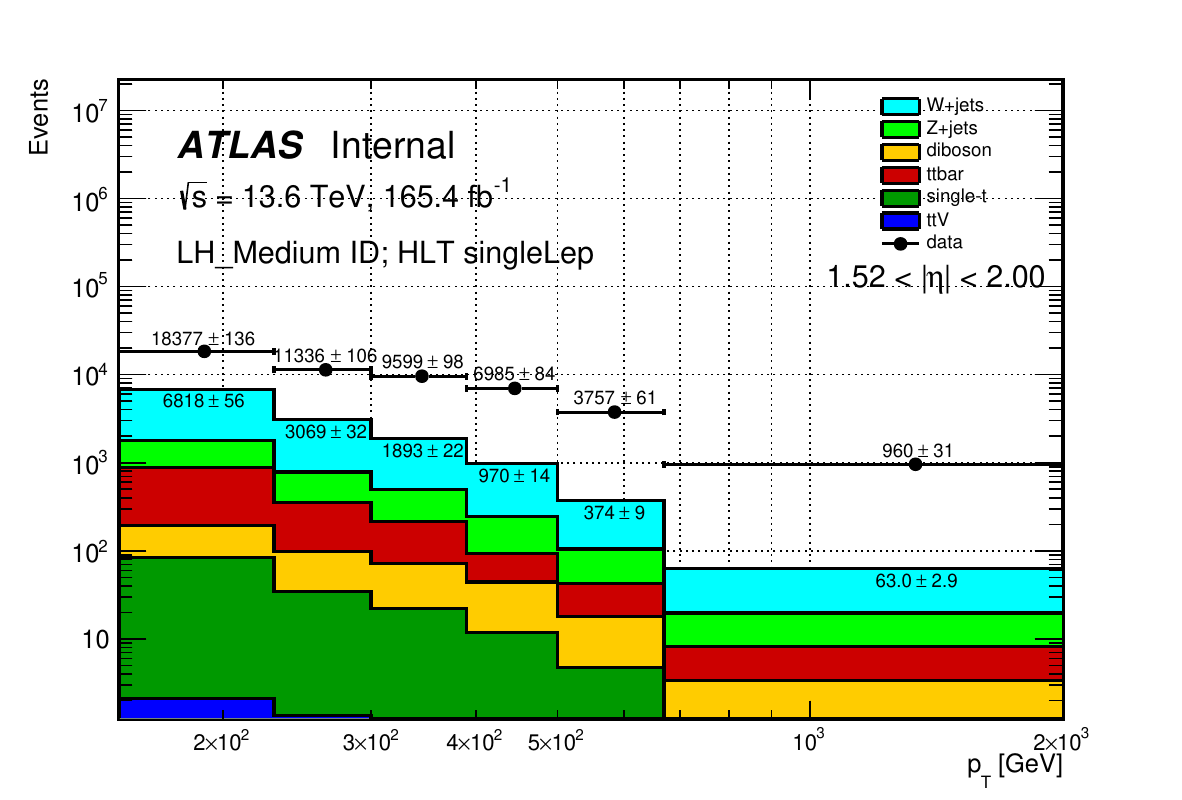}
    }
    \hfill
    \subfloat[(b)]{
      \includegraphics[width=0.5\textwidth]{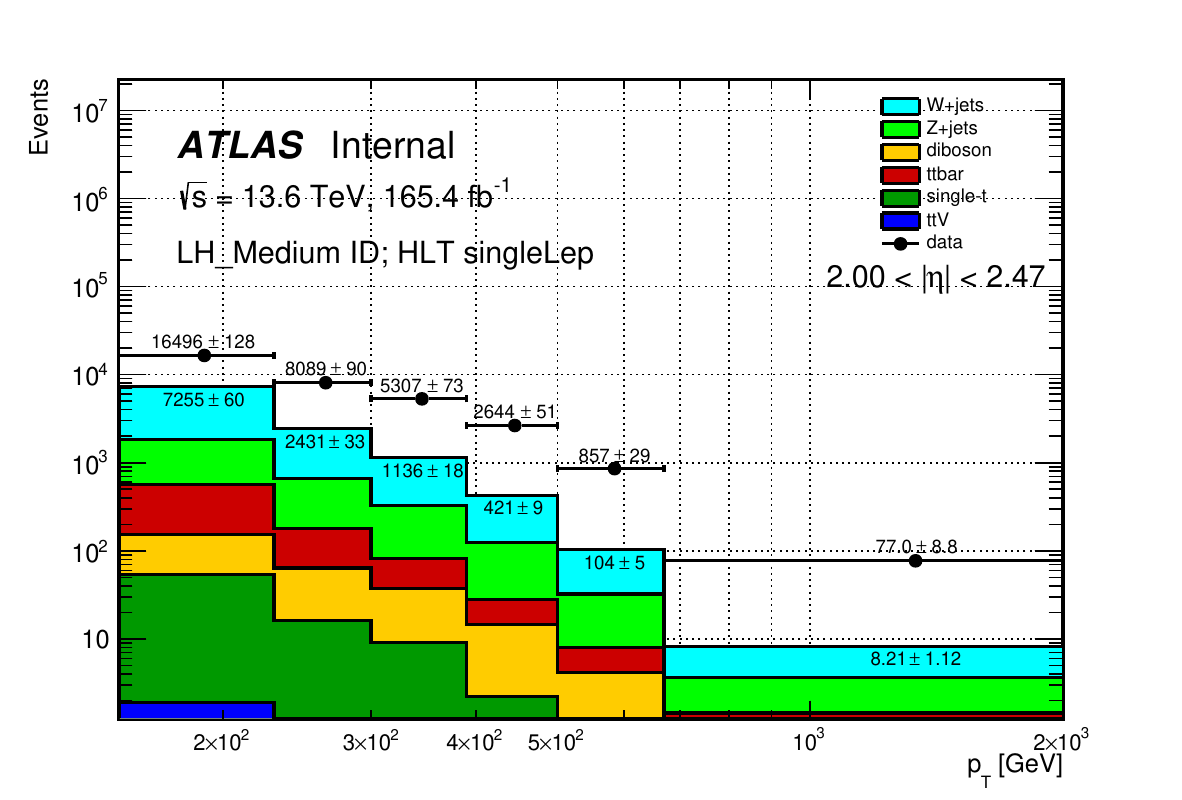}
    }
    \hfill
    \subfloat[(c)]{
      \includegraphics[width=0.5\textwidth]{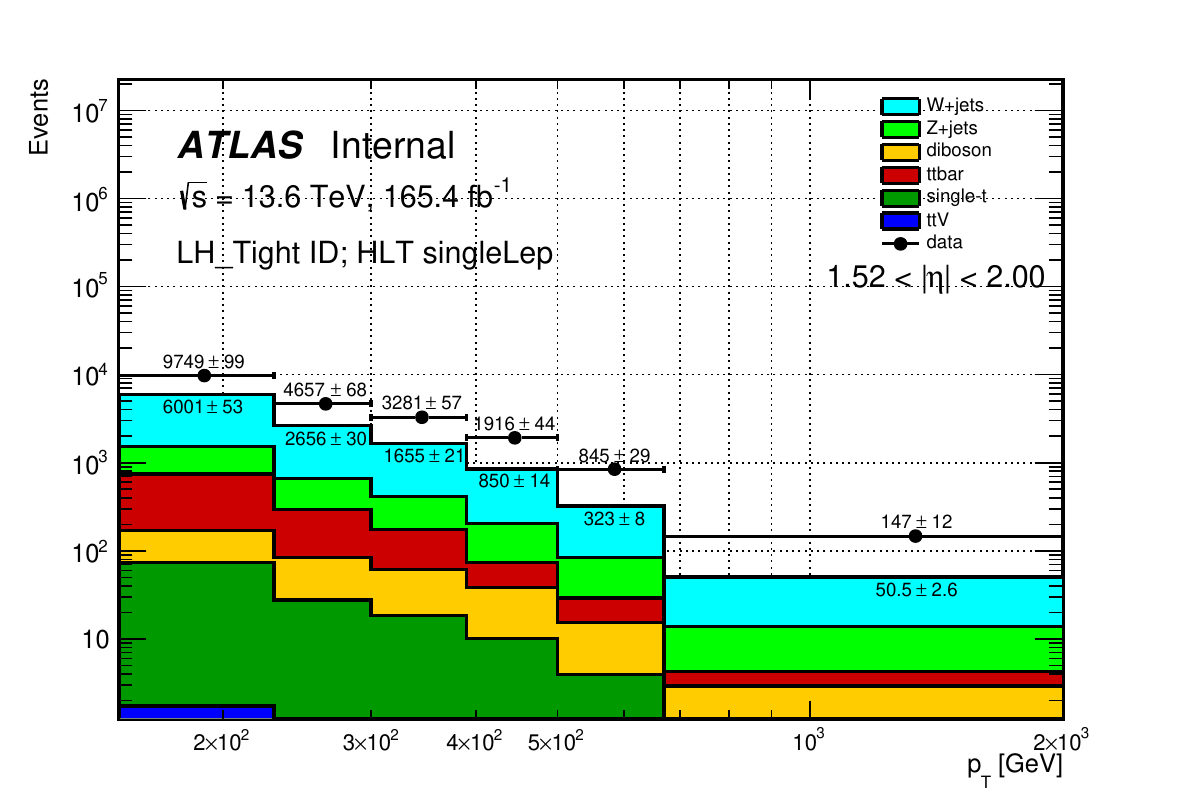}
    }
    \hfill
    \subfloat[(d)]{
      \includegraphics[width=0.5\textwidth]{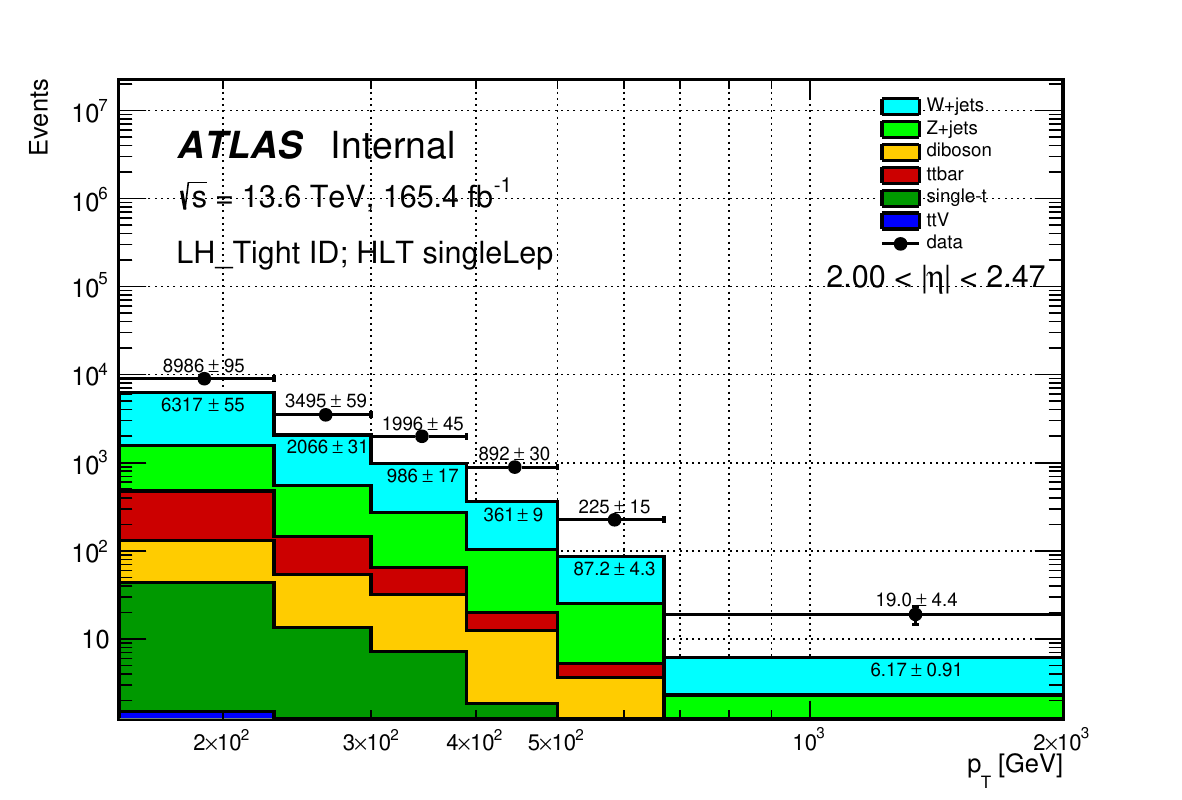}
    }
\caption{Distributions of the leading electron $p_{\textrm{T}}$ following the signal (top, a--b) and baseline (bottom, c--d) requirements in the fake-enriched region. (a,c) $1.52<|\eta|<2.00$; (b,d) $2.00<|\eta|<2.47$. The full dataset is plotted on top of the electroweak backgrounds modelled via MC simulations.}
\label{fig:Fake_Enriched_Region_Dilution_1.52_2.47}
\end{figure}

\begin{figure}[h!]
  \captionsetup[subfigure]{labelformat=empty}
    \subfloat[(a)]{
      \includegraphics[width=0.5\textwidth]{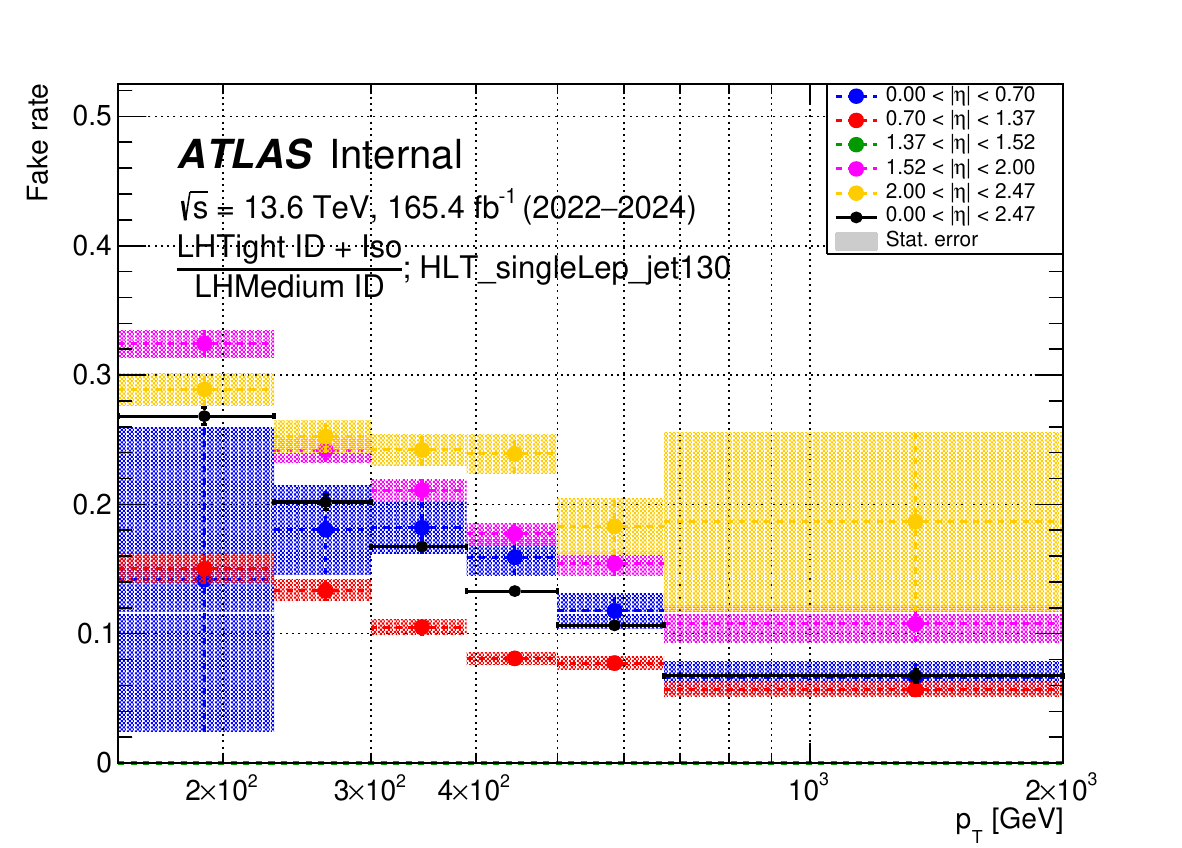}
    }
    \hfill
    \subfloat[(b)]{
      \includegraphics[width=0.5\textwidth]{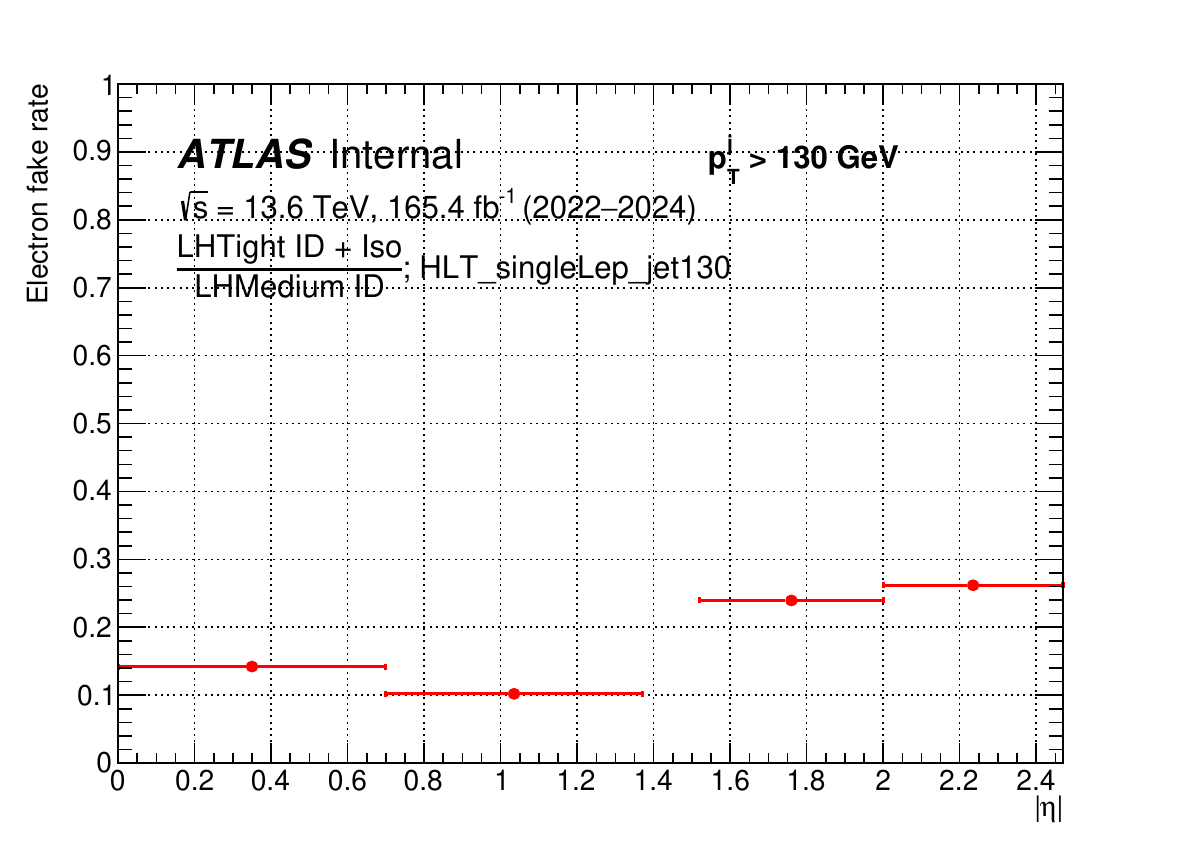}
    }
\caption{Fake electron efficiency as a function of (a) $p_{\textrm{T}}$ (per $|\eta|$ bin) and (b) $|\eta|$. The statistical uncertainty is shown in the hatched boxes of (a).}
\label{fig:Fake_Rates}
\end{figure}

Distributions like those in FIG.~\ref{fig:Fake_Enriched_Region_Dilution} are eventually used for the calculation of $f$. This is done in bins of $p_{\textrm{T}}$ and $|\eta|$. Accordingly, FIG.~\ref{fig:Fake_Enriched_Region_Dilution_0.00_1.37}--\ref{fig:Fake_Enriched_Region_Dilution_1.52_2.47} show the $p_{\textrm{T}}$ distributions of the signal and baseline electron in different bins of $|\eta|$.
To obtain a fake-enriched sample, the total backgrounds stacked in FIG.~\ref{fig:Fake_Enriched_Region_Dilution_0.00_1.37}--\ref{fig:Fake_Enriched_Region_Dilution_1.52_2.47} are subtracted from the data. By extension of Eq. (\ref{Eq_Real_Fake_Effs}), the fake-enriched ratio is given by
\begin{equation}
    f = \frac{N^{\textrm{fake}}_{\textrm{signal}}}{N^{\textrm{fake}}_{\textrm{baseline}}} 
    = \frac{N^{\textrm{Data}}_{\textrm{signal}} - N^{\textrm{EWK MC}}_{\textrm{signal}}}{N^{\textrm{Data}}_{\textrm{baseline}} - N^{\textrm{EWK MC}}_{\textrm{baseline}}},
    \label{Eq_Fake_Eff_Calc}
\end{equation}
which is the fake rate. The distributions of $f$ in $p_{\textrm{T}}$ and $|\eta|$ are given in FIG.~\ref{fig:Fake_Rates}.

\begin{figure}[h!]
  \captionsetup[subfigure]{labelformat=empty}
    \centering
    \includegraphics[width=0.5\textwidth]{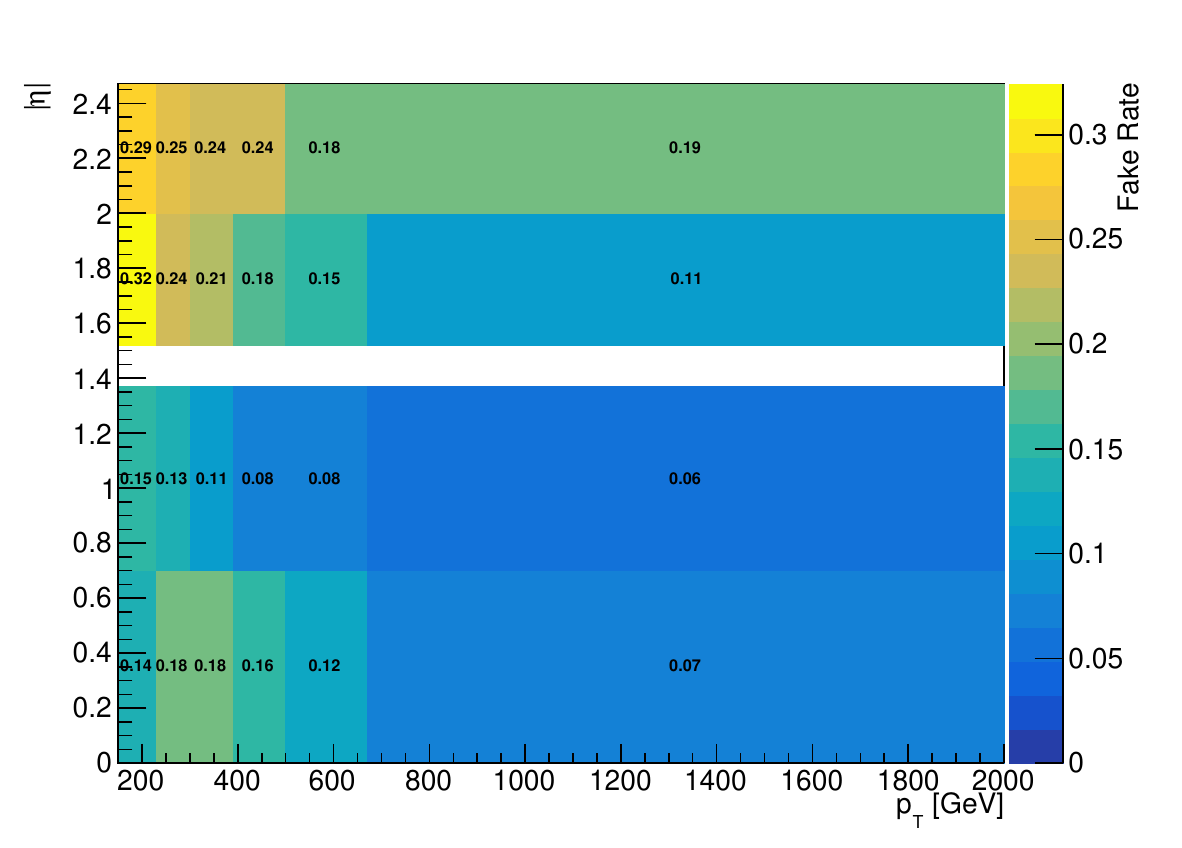}
    \caption{Fake electron efficiency as a function of $p_{\textrm{T}},|\eta|$}    
    \label{fig:Fake_Rate_2D}
\end{figure}

Combining the $p_{\textrm{T}}$ and $|\eta|$ with the binning shown in FIG.~\ref{fig:Fake_Rates}, the two-dimensional distribution follows, as shown in FIG.~\ref{fig:Fake_Rate_2D}.

Using the two-dimensional distributions of the real and fake rates shown in FIG.~\ref{fig:Real_Rate_2D} and~\ref{fig:Fake_Rate_2D} respectively, following Eqs. (\ref{eq:fakeWeight})--(\ref{eq:fakeWeight_application}), the \texttt{fakeWeight} is calculated and applied on the baseline data sample, yielding the fake background estimate. The next section shows how this estimate is validated.

\newpage

\subsection{Fake background validation}
\begin{figure}[h!]
  \captionsetup[subfigure]{labelformat=empty}
    \subfloat[(a)]{
      \includegraphics[width=0.5\textwidth]{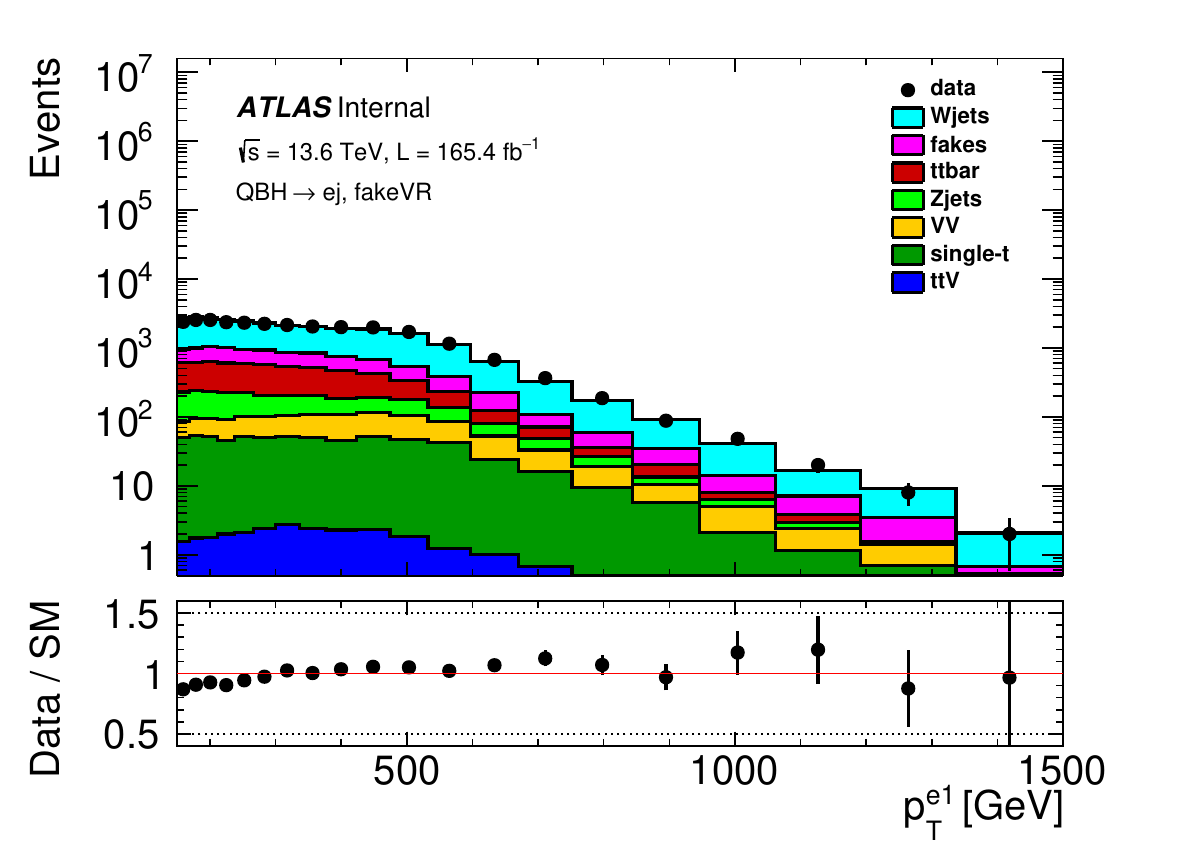}
    }
    \hfill
    \subfloat[(b)]{
      \includegraphics[width=0.5\textwidth]{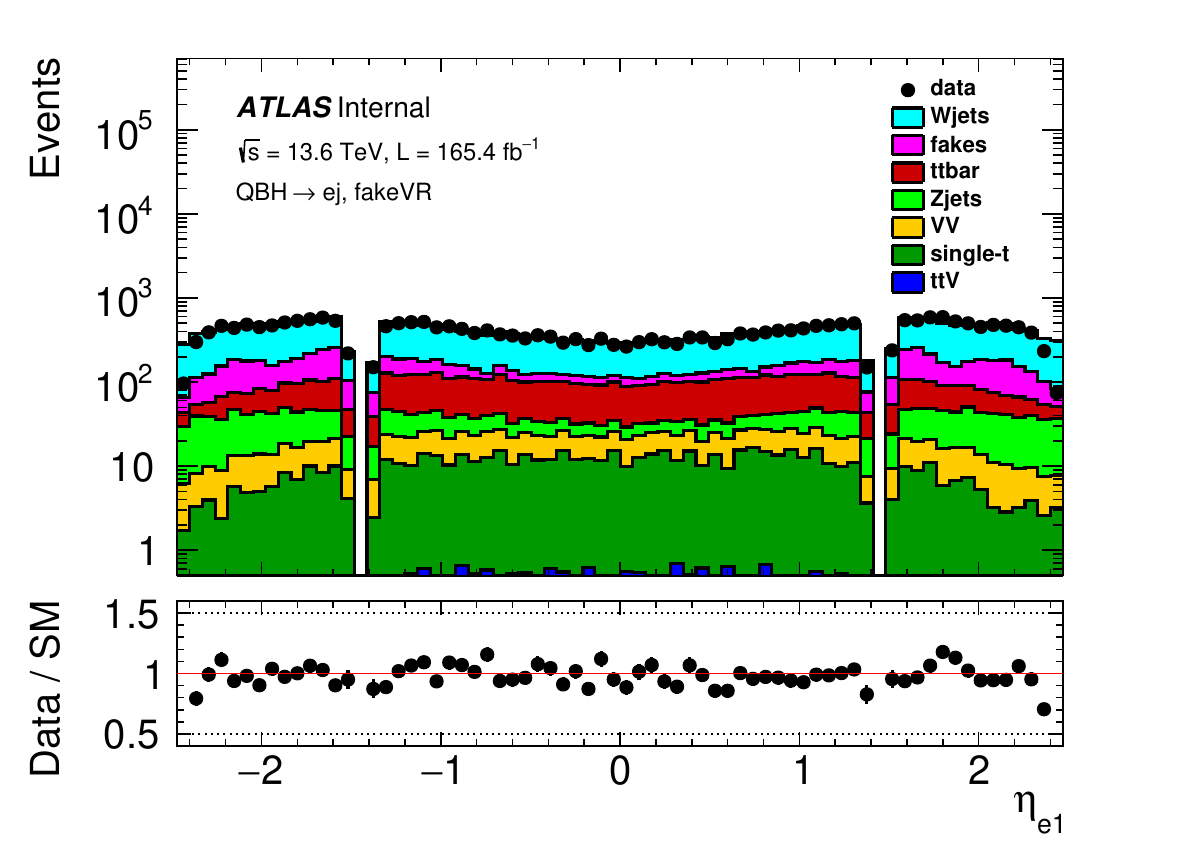}
    }
    \hfill
    \subfloat[(c)]{
      \includegraphics[width=0.5\textwidth]{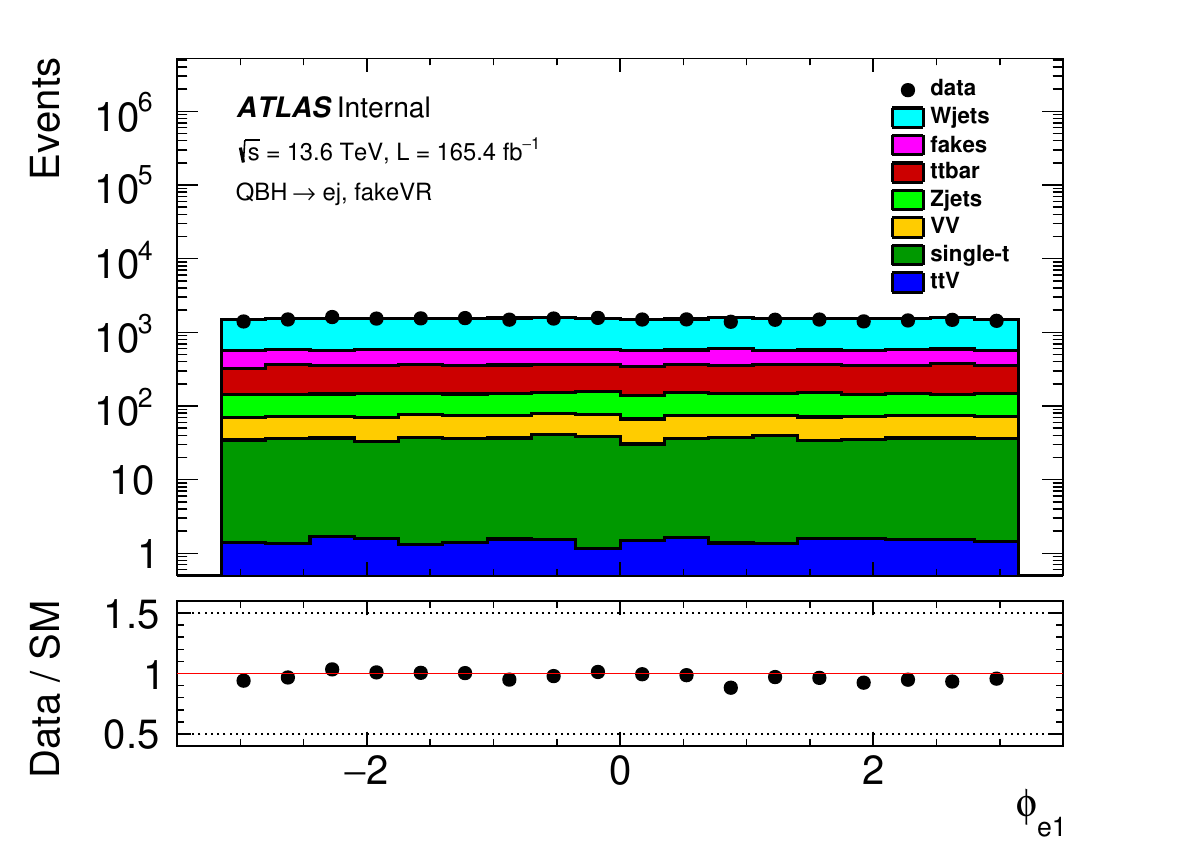}
    }
    \hfill
    \subfloat[(d)]{
      \includegraphics[width=0.5\textwidth]{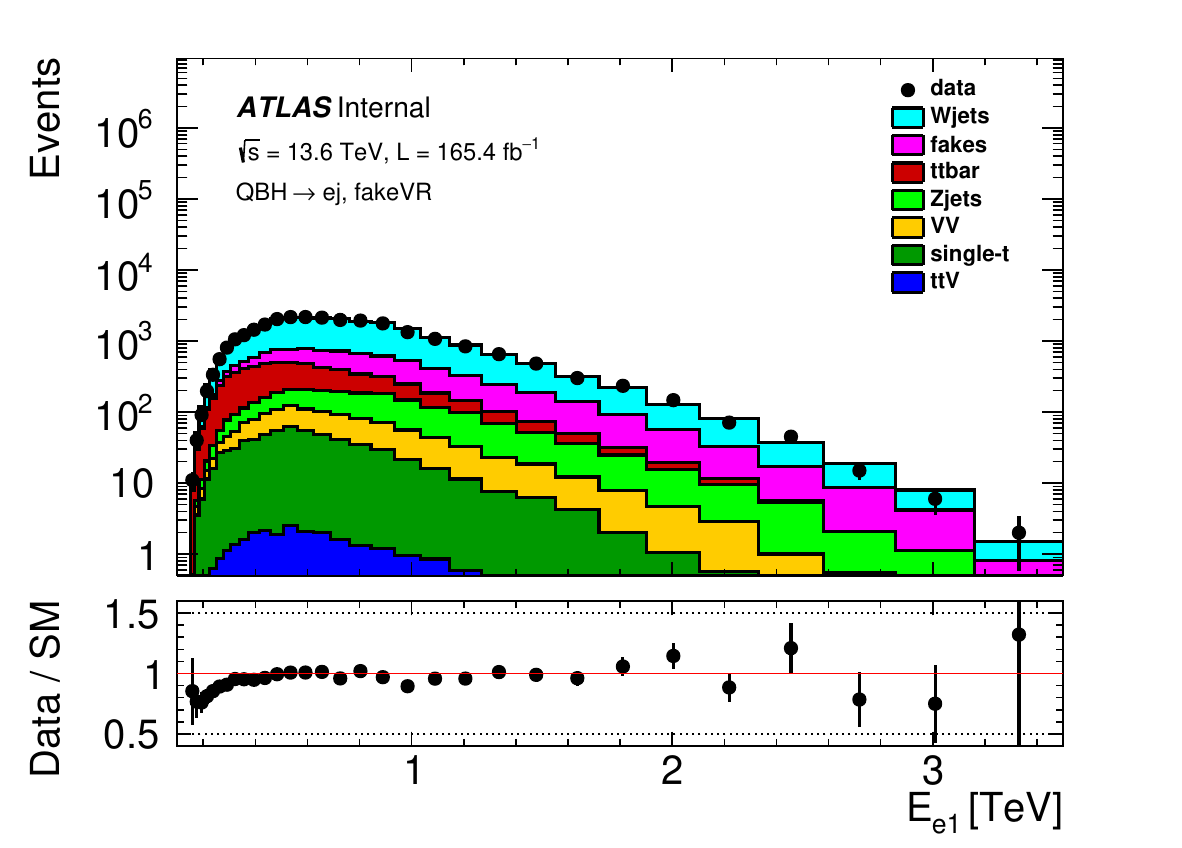}
    }
\caption{Kinematic distributions of the leading electron in the fake validation region. (a) $p_{\textrm{T}}$, (b) $\eta$, (c) $\phi$, (d) energy.}
\label{fig:Fake_Validation_Region_Leading_Electron}
\end{figure}

The fake VR is defined to suppress contributions from electroweak backgrounds whilst being orthogonal to other analysis regions. FIG.~\ref{fig:Fake_Validation_Region_Leading_Electron} shows kinematic distributions of the leading electron in the fake VR.

\begin{figure}[h!]
  \captionsetup[subfigure]{labelformat=empty}
  \subfloat[(a)]{
    \includegraphics[width=0.5\textwidth]{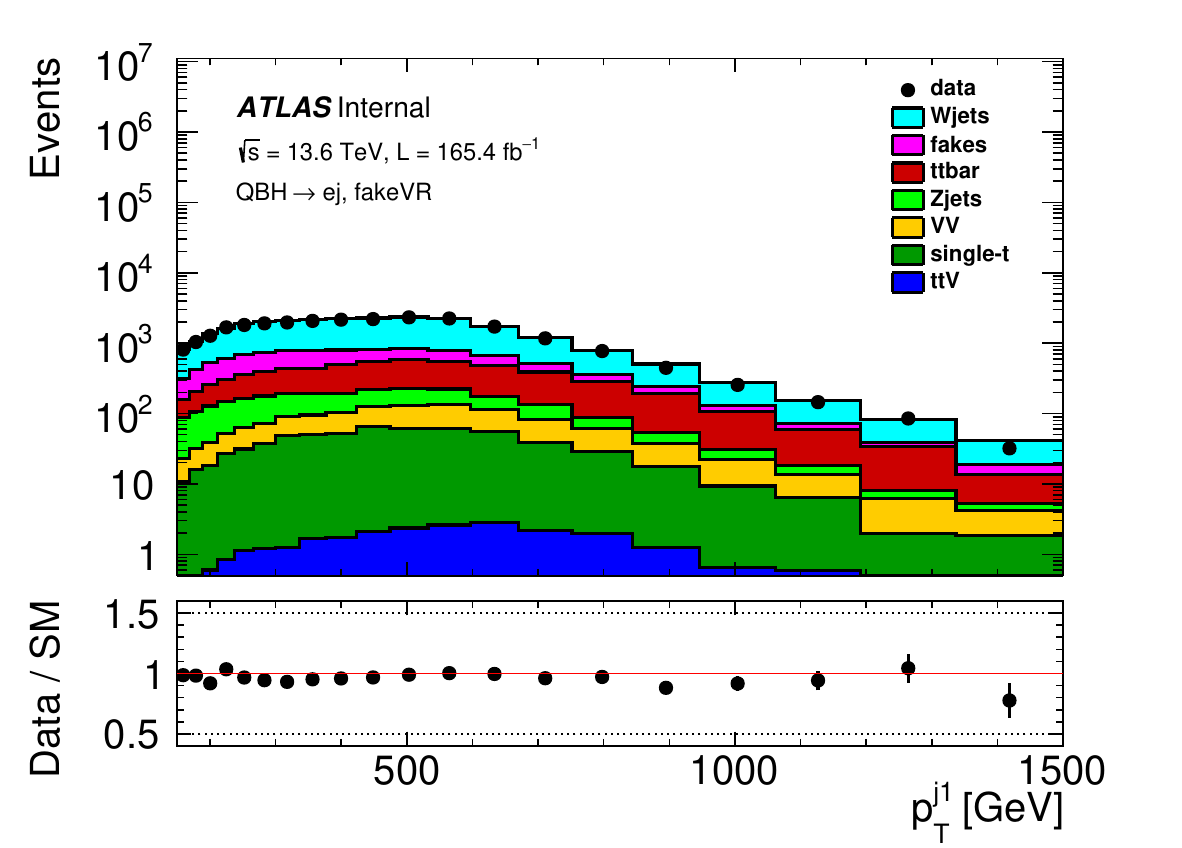}
  }
  \hfill
  \subfloat[(b)]{
    \includegraphics[width=0.5\textwidth]{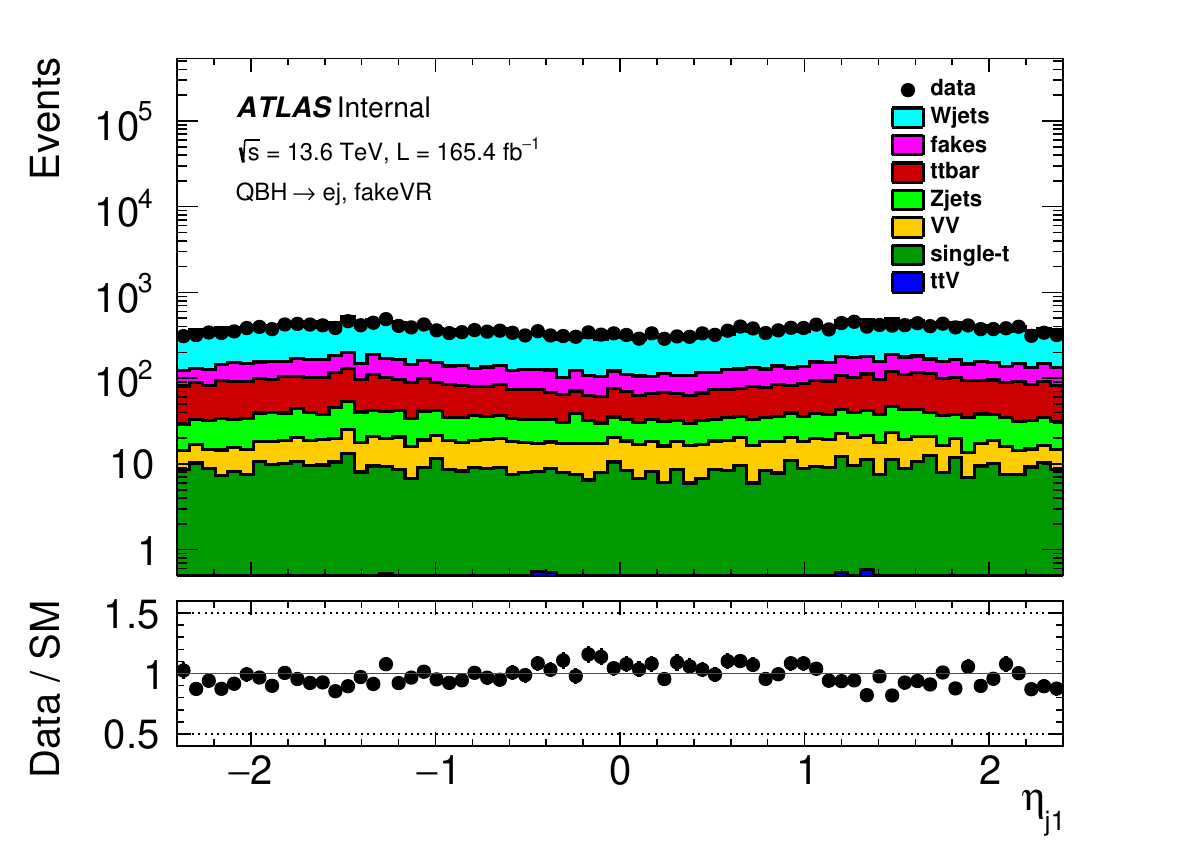}
  }
  \hfill
  \subfloat[(c)]{
    \includegraphics[width=0.5\textwidth]{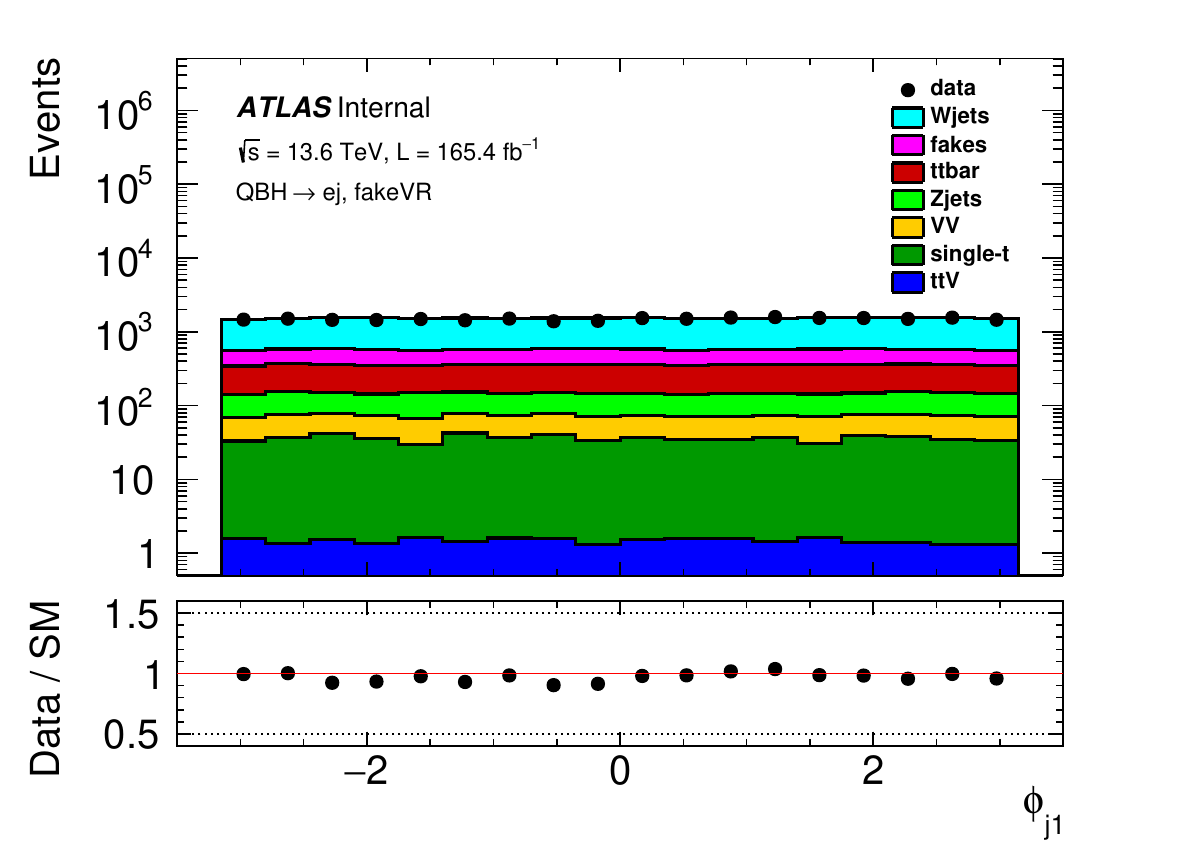}
  }
  \hfill
  \subfloat[(d)]{
    \includegraphics[width=0.5\textwidth]{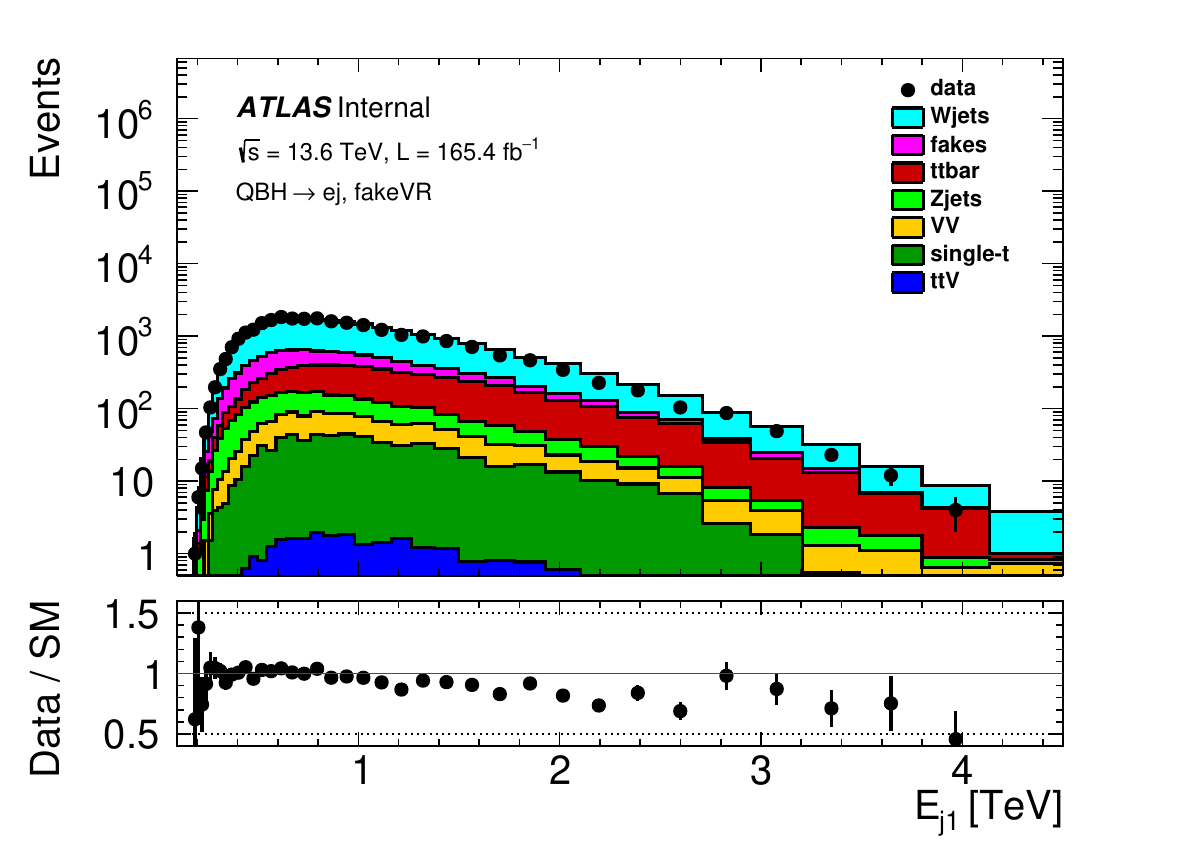}
  }
\caption{Kinematic distributions of the leading jet in the fake validation region. (a) $p_{\textrm{T}}$, (b) $\eta$, (c) $\phi$, (d) energy.}
\label{fig:Fake_Validation_Region_Leading_Jet}
\end{figure}

FIG.~\ref{fig:Fake_Validation_Region_Leading_Jet} shows kinematic distributions of the leading jet in the fake VR.

\begin{figure}[h!]
  \centering
  \includegraphics[width=0.6\textwidth]{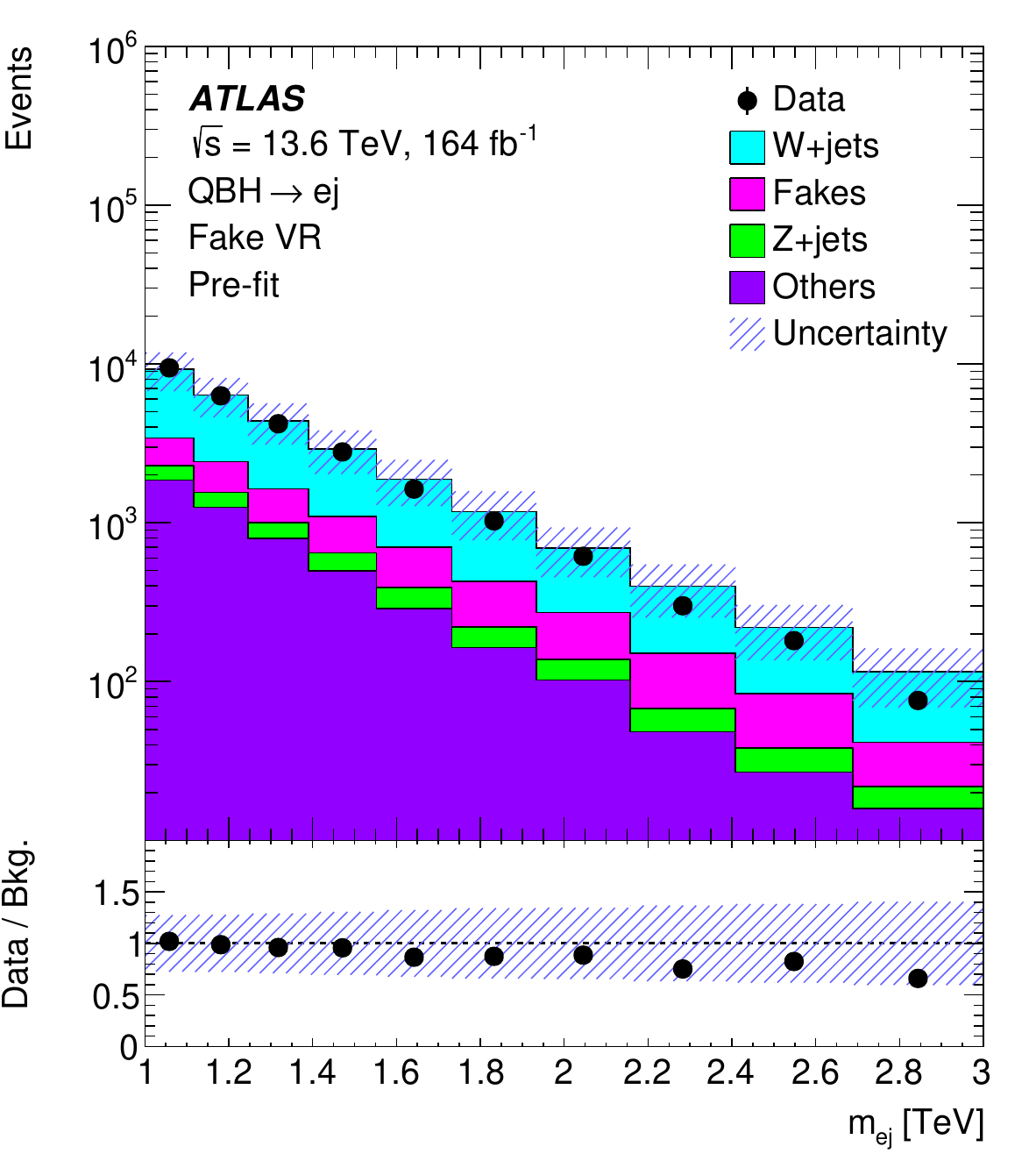}
\caption{Invariant mass of the lepton--jet pair in the fakes validation region. The data-to-background ratio is shown in the bottom panel, and hatched bands indicate the total statistical and systematic uncertainty. ``Others'' denotes subdominant backgrounds including top and diboson processes. The MC contributions are normalised based on their expected cross-section, prior to the likelihood fit.}
\label{fig:Fake_Validation_Region_High_Level_Kinematics}
\end{figure}

The kinematic distributions of the leading electron and jet in the fake-enriched region are shown in FIG.~\ref{fig:Fake_Validation_Region_Leading_Electron}--\ref{fig:Fake_Validation_Region_Leading_Jet}. The $m_{\ell j}$ distribution in the $f$VR is shown in FIG.~\ref{fig:Fake_Validation_Region_High_Level_Kinematics}, where the data and background predictions are seen to be in agreement within the assigned uncertainties.

FIG.~\ref{fig:Fake_Validation_Region_High_Level_Kinematics} shows good agreement between the total background, including the fake background estimate, and the data. This estimate is used in subsequent steps, through the statistical interpretation outlined in Section~\ref{ssec:stats:profile}, and the results are presented in Section~\ref{sec:results}.

\newpage
\clearpage

\subsection{Systematic uncertainties}

The source of systematic uncertainty on $f$ relies upon the error in the MC background used to estimate the electroweak (real) contributions in the fake CR \textit{i.e.} the background subtracted from data to estimate the rate. As a diagnostic, the effect of each systematic source on the total electroweak MC background is examined directly as a function of $m_{ej}$ in the fake CR, to probe the general size of each contribution; these distributions are shown in Appendix~\ref{app:qbh:fakes_systs} (FIG.~\ref{fig:Fake_systematics_mlj_CR}). The approach taken to propagate these uncertainties to the fake rate itself is more direct, working bin-by-bin in $(p_{\textrm{T}}, |\eta|)$, as described below.

\begin{figure}[h!]
  \captionsetup[subfigure]{labelformat=empty}
  \subfloat[(a)]{
    \includegraphics[width=0.5\textwidth]{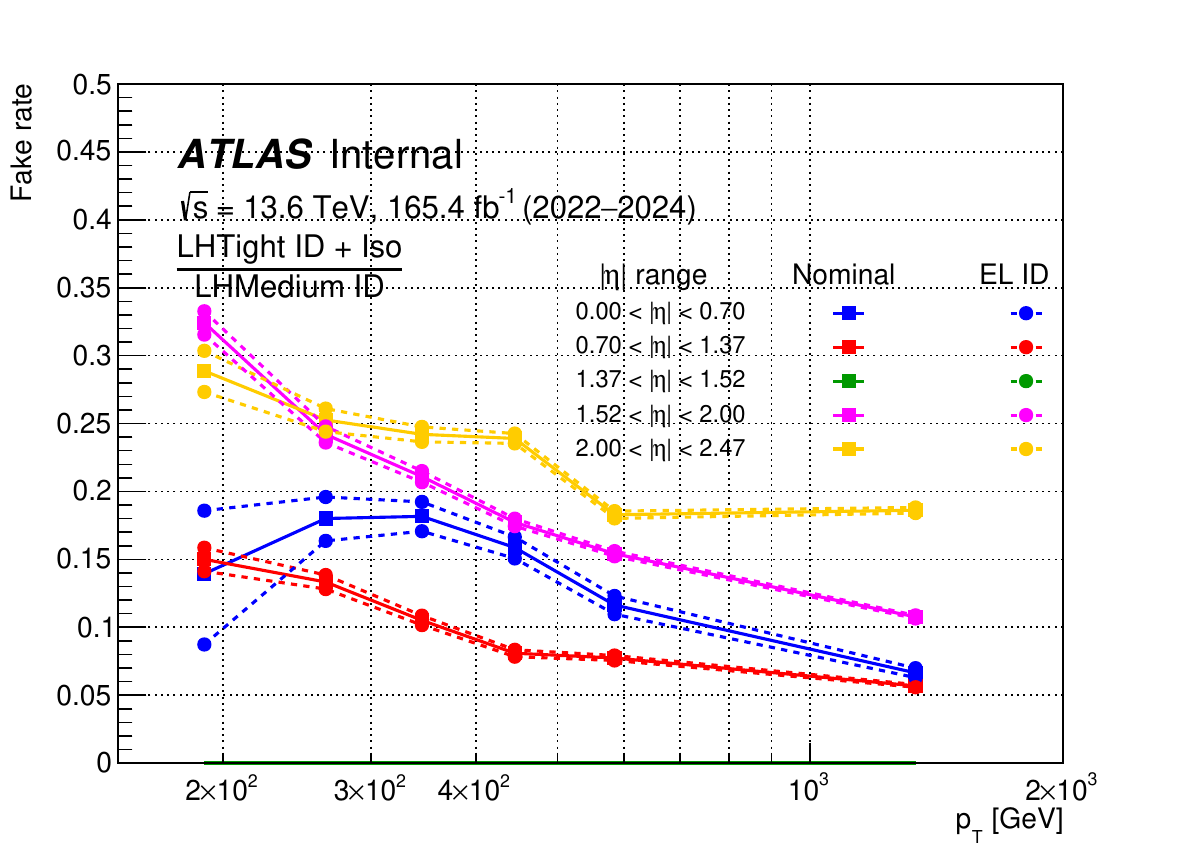}
  }
  \hfill
  \subfloat[(b)]{
    \includegraphics[width=0.5\textwidth]{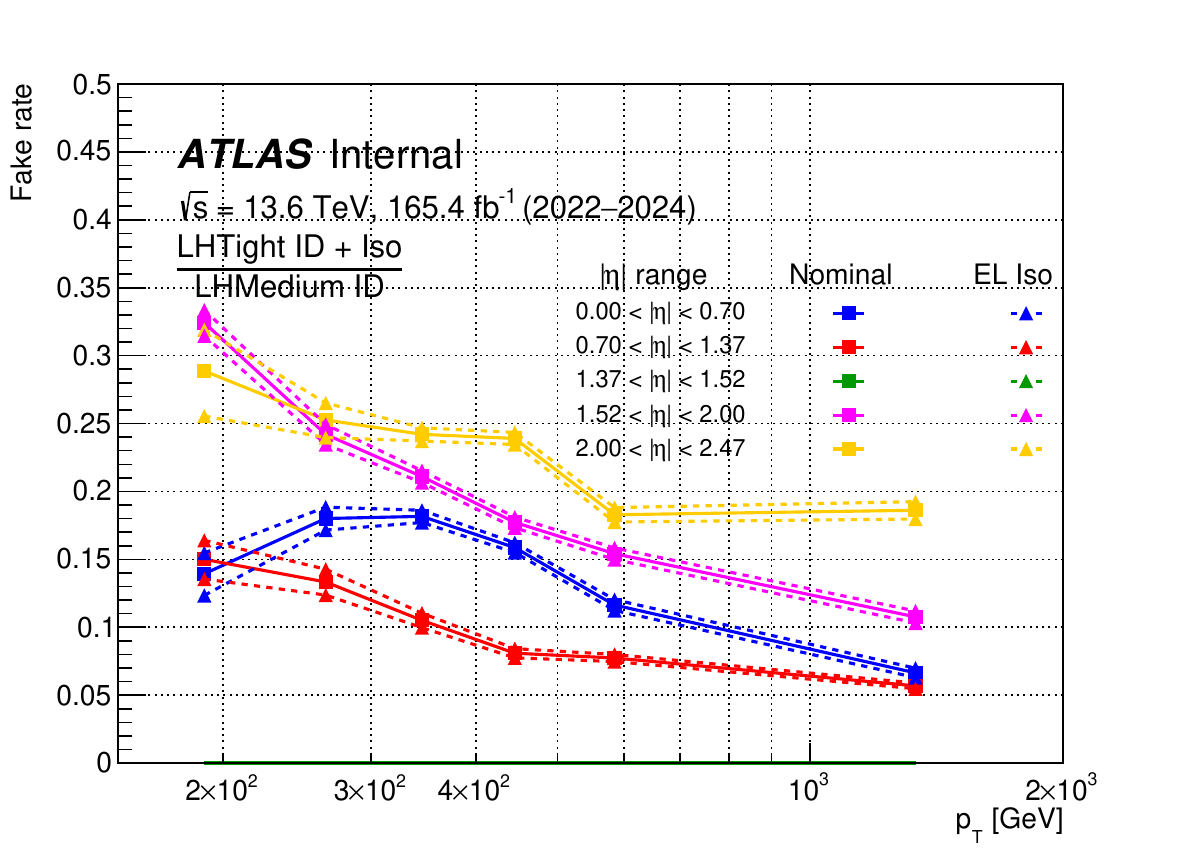}
  }
  \hfill
  \subfloat[(c)]{
    \includegraphics[width=0.5\textwidth]{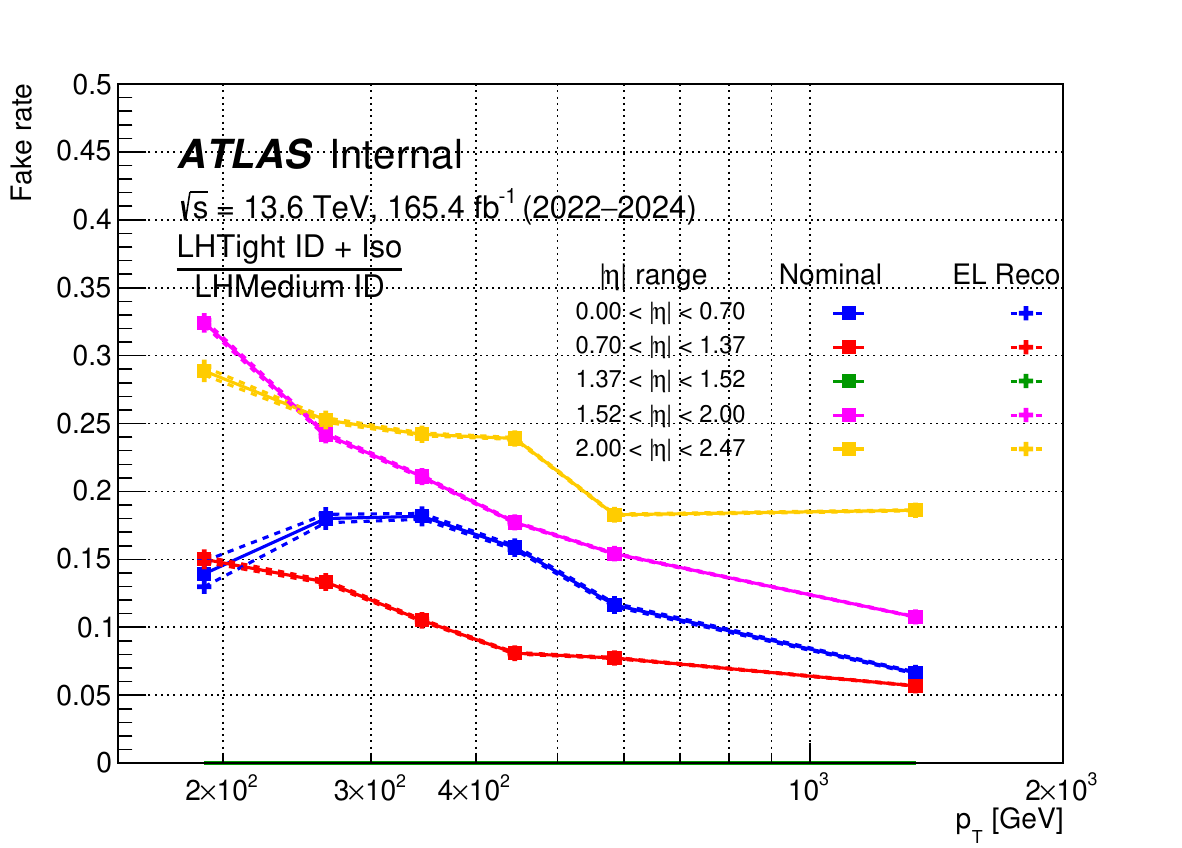}
  }
  \hfill
  \subfloat[(d)]{
    \includegraphics[width=0.5\textwidth]{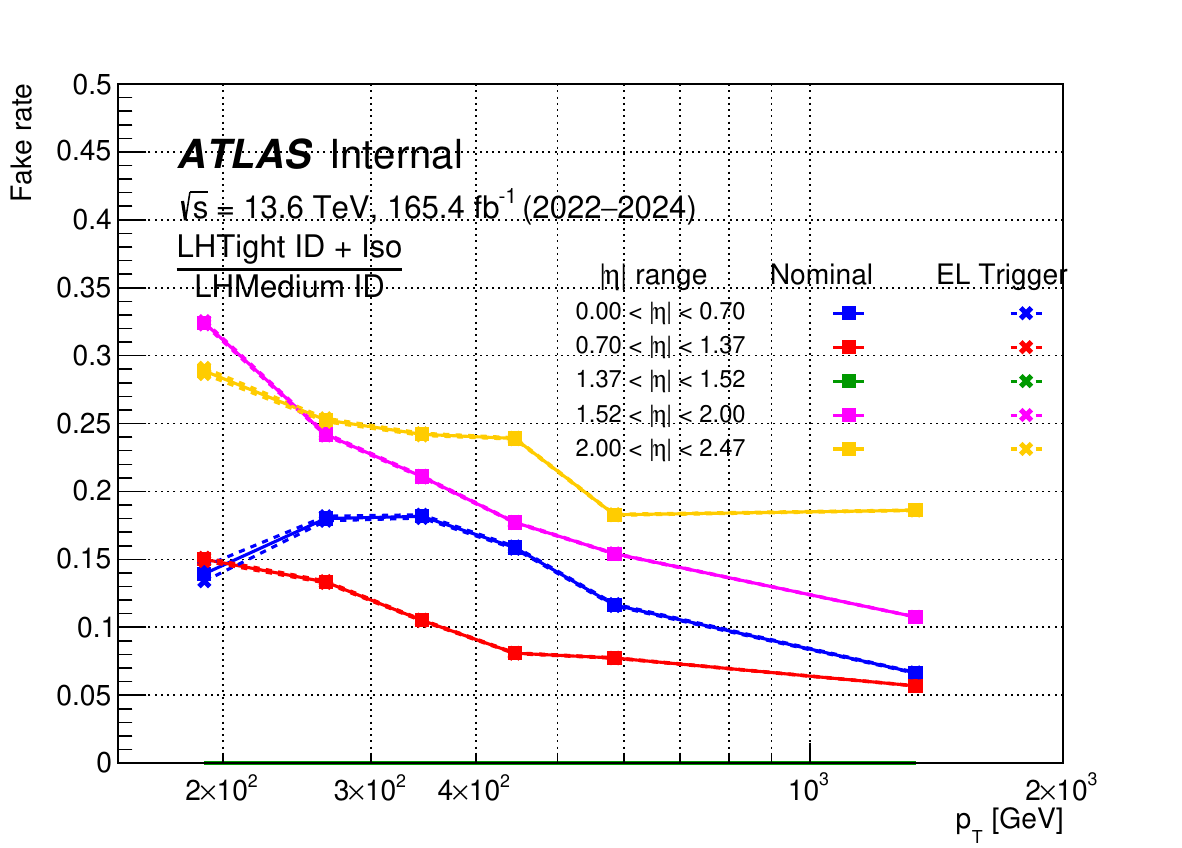}
  }
\caption{Error on the fake rate due to electron efficiency sources. (a) identification; (b) isolation; (c) reconstruction; (d) trigger.}
\label{fig:Fake_systs_FR_electron}
\end{figure}

To estimate the error on $f$, we consider the electron efficiency systematics and the Sherpa theoretical uncertainties. All other sources of systematic uncertainty are found to be negligible. Following Eq. (\ref{Eq_Fake_Eff_Calc}), this is done in bins of $p_{\textrm{T}}$ and $|\eta|$ to yield
\begin{equation*}
  f_{\textrm{syst}}(p_{\textrm{T}},|\eta|) = \frac{N_{\textrm{data}}^{T} - N_{\textrm{MC, syst}}^{T}}{N_{\textrm{data}}^{\textrm{baseline}} - N_{\textrm{MC, syst}}^{\textrm{baseline}}},
\end{equation*}
where each $f_{\textrm{syst}}(p_{\textrm{T}},|\eta|)$ will be used in turn as a systematic template added in quadrature, obtaining a total error on $f$. Correspondingly, the error on the fake rates due to electron uncertainties is shown in FIG.~\ref{fig:Fake_systs_FR_electron}.

\begin{figure}[h!]
  \captionsetup[subfigure]{labelformat=empty}
  \subfloat[(a)]{
    \includegraphics[width=0.5\textwidth]{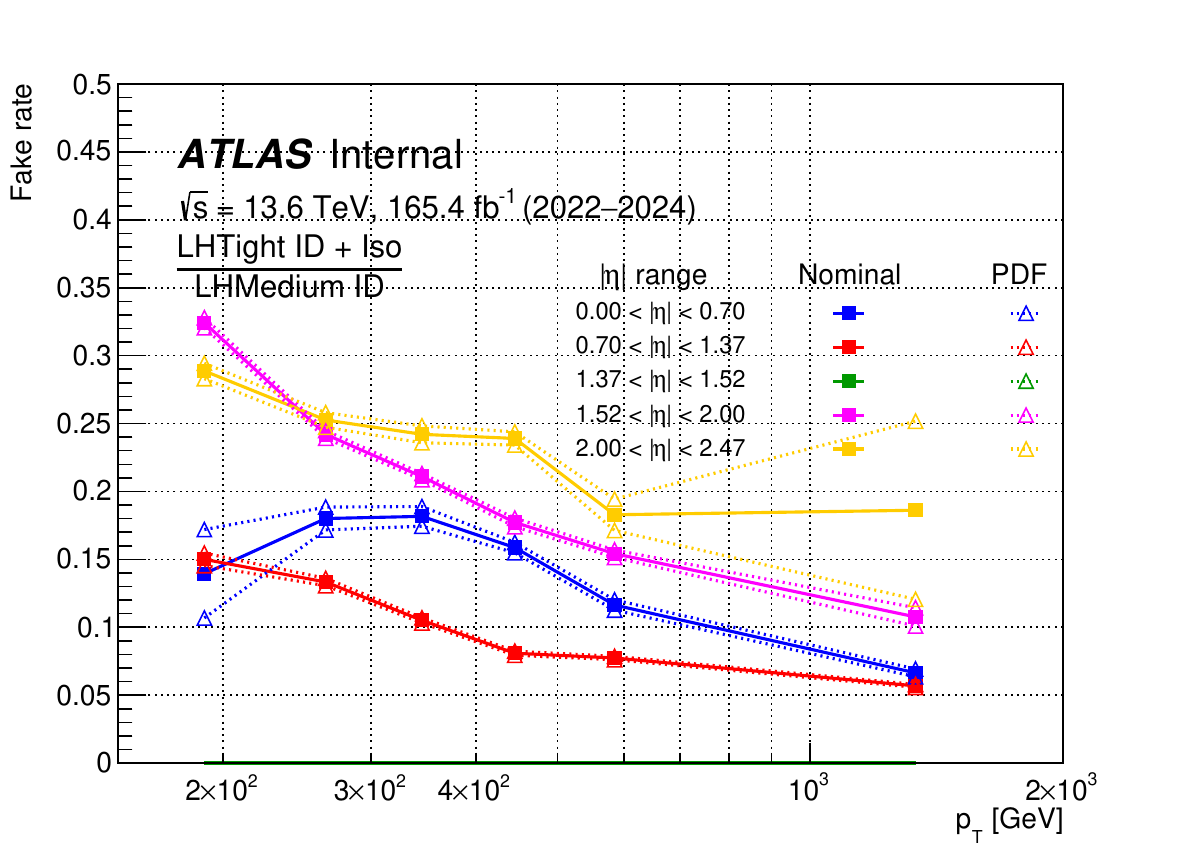}
  }
  \hfill
  \subfloat[(b)]{
    \includegraphics[width=0.5\textwidth]{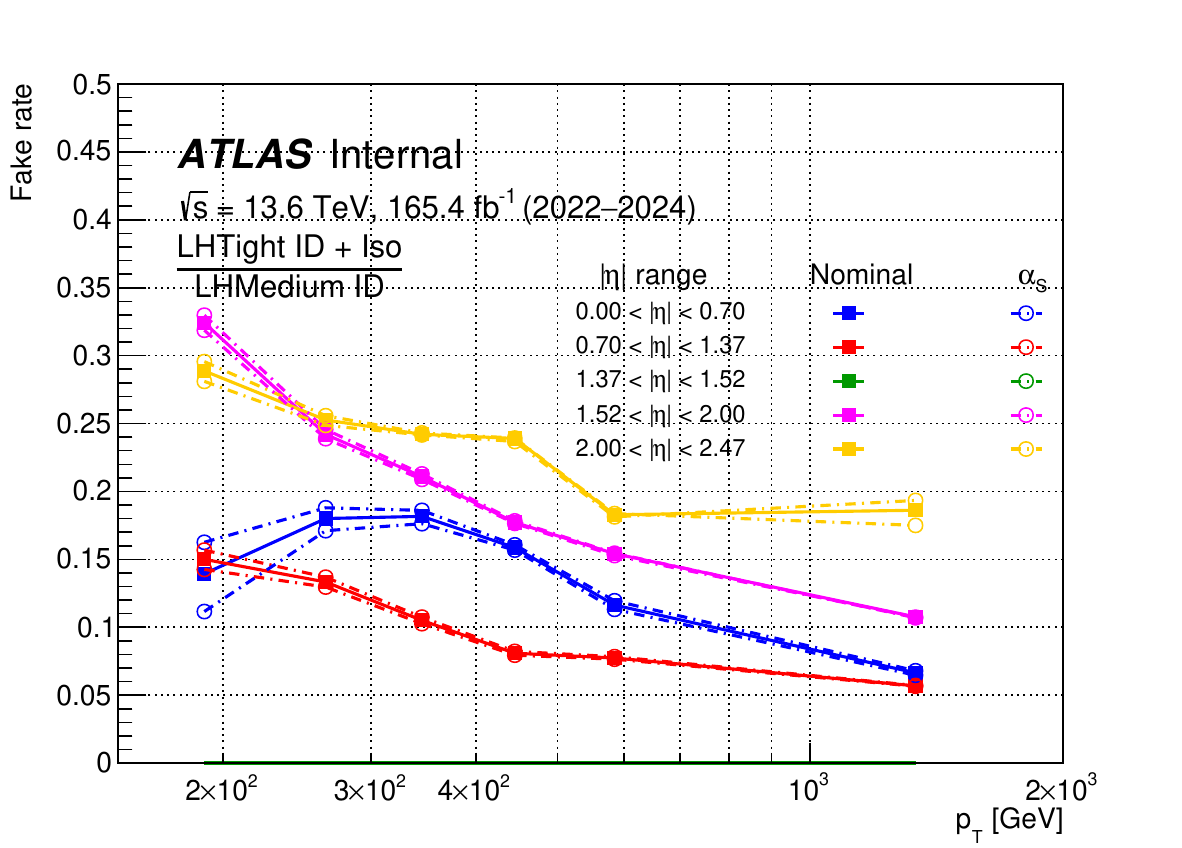}
  }
  \hfill
  \subfloat[(c)]{
    \includegraphics[width=0.5\textwidth]{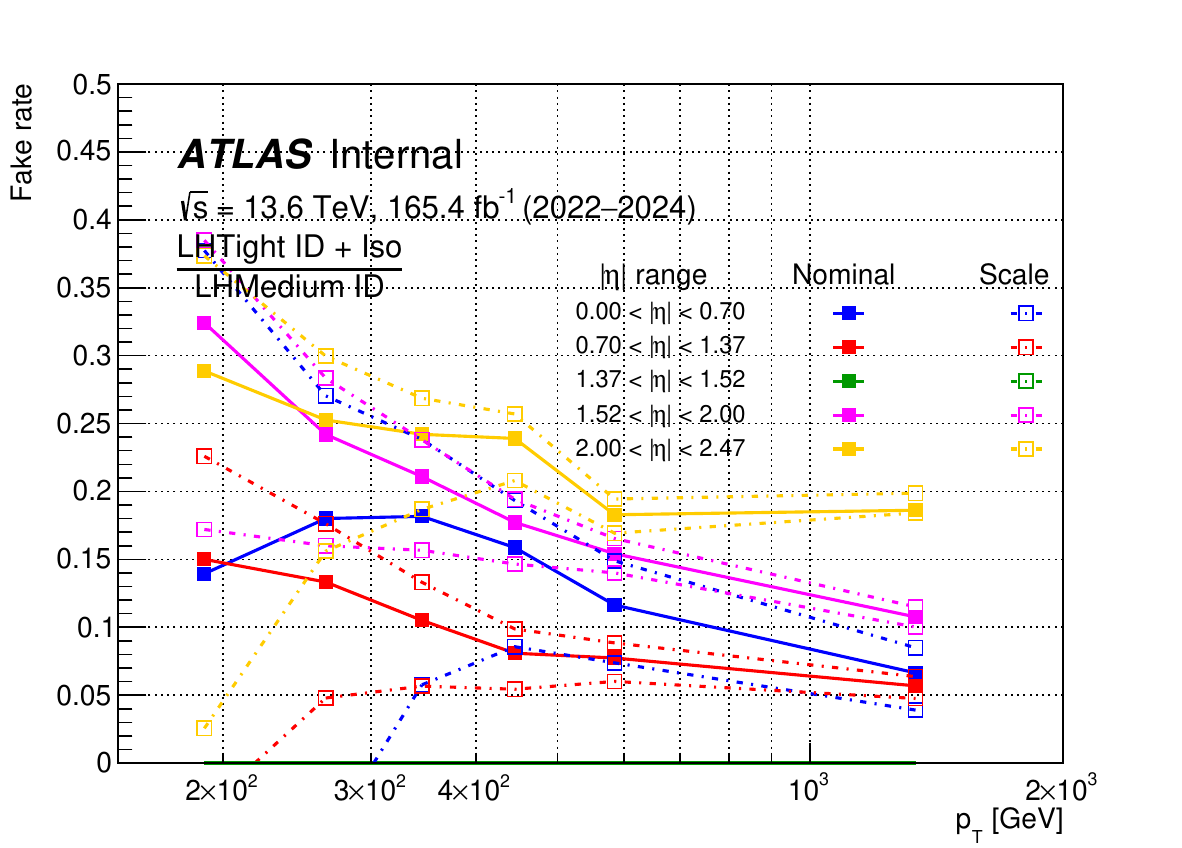}
  }
\caption{Error on the fake rate due to Sherpa generator theoretical uncertainties. (a) combined PDF set; (b) strong coupling; (c) QCD renormalisation and factorisation scales.}
\label{fig:Fake_systs_TH}
\end{figure}

\begin{figure}[h!]
  \captionsetup[subfigure]{labelformat=empty}
  \subfloat[(a)]{
    \includegraphics[width=0.5\textwidth]{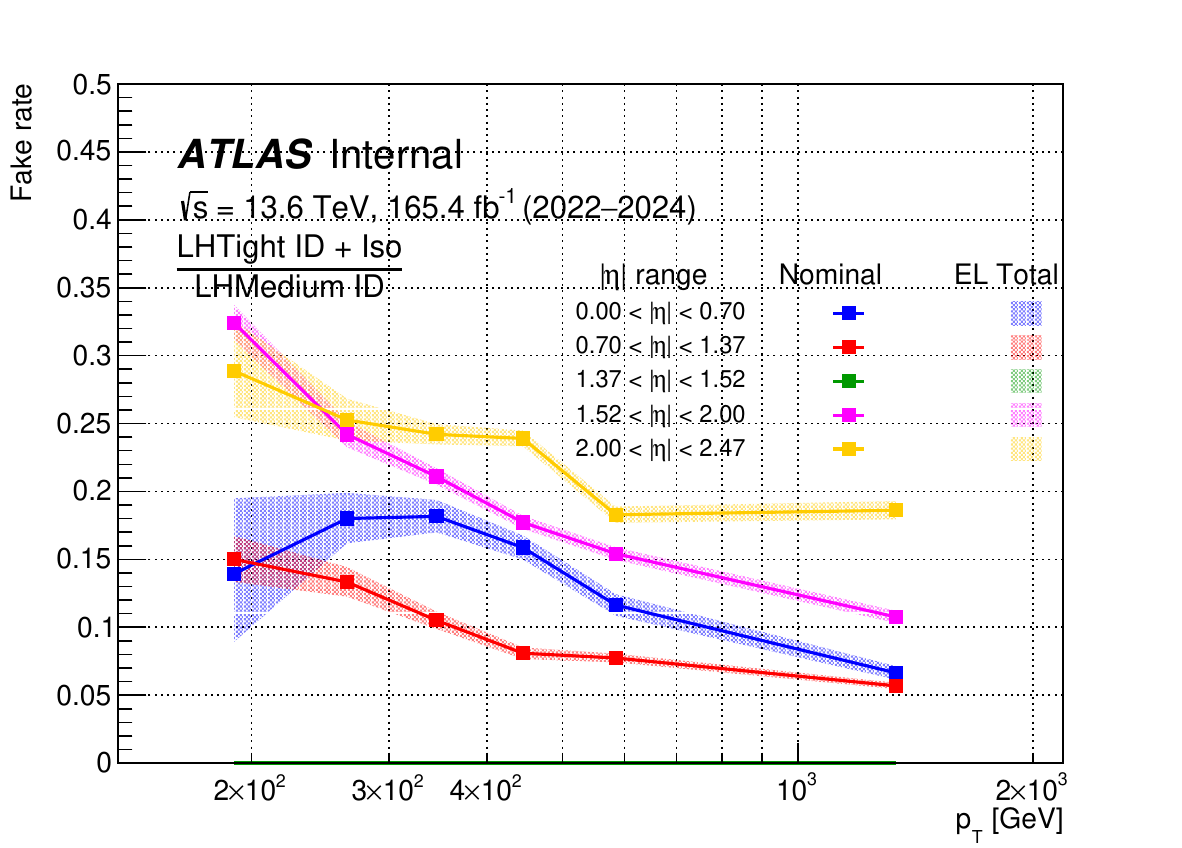}
  }
  \hfill
  \subfloat[(b)]{
    \includegraphics[width=0.5\textwidth]{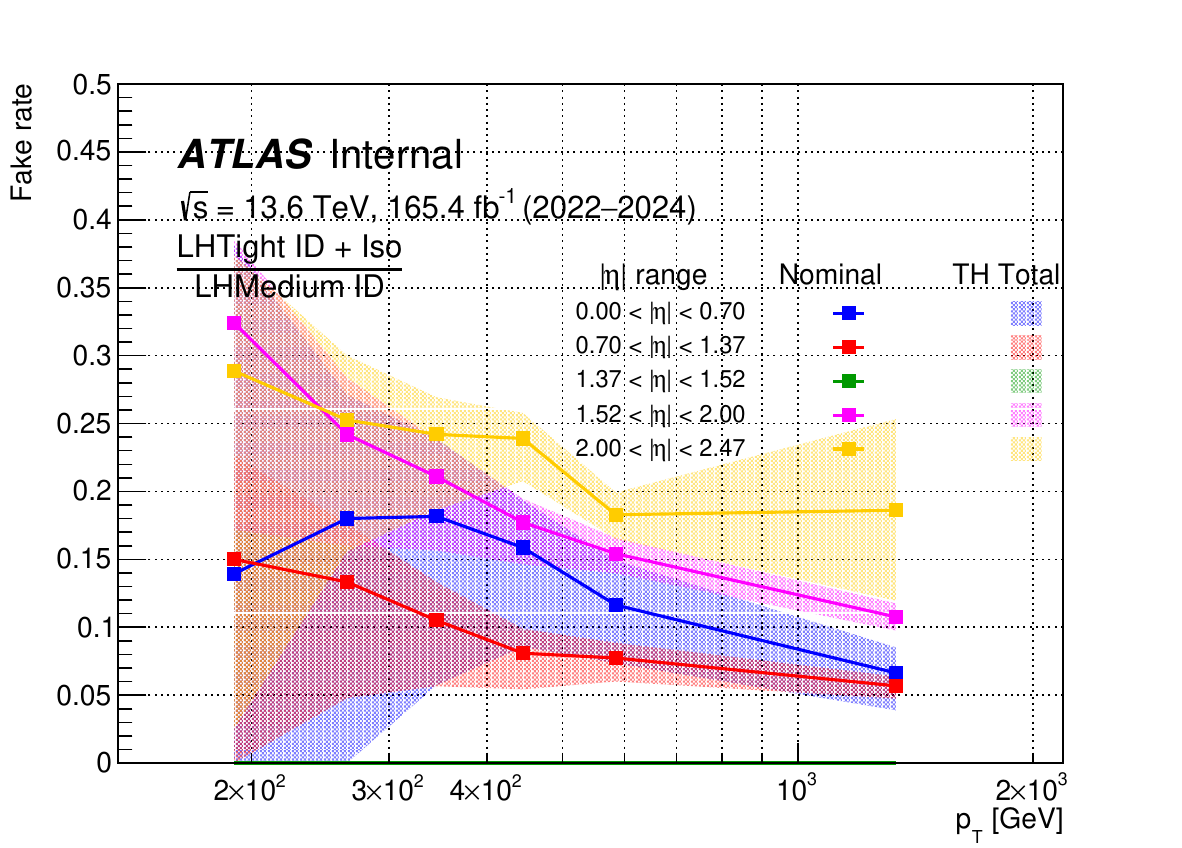}
  }
  \hfill
  \subfloat[(c)]{
    \includegraphics[width=0.5\textwidth]{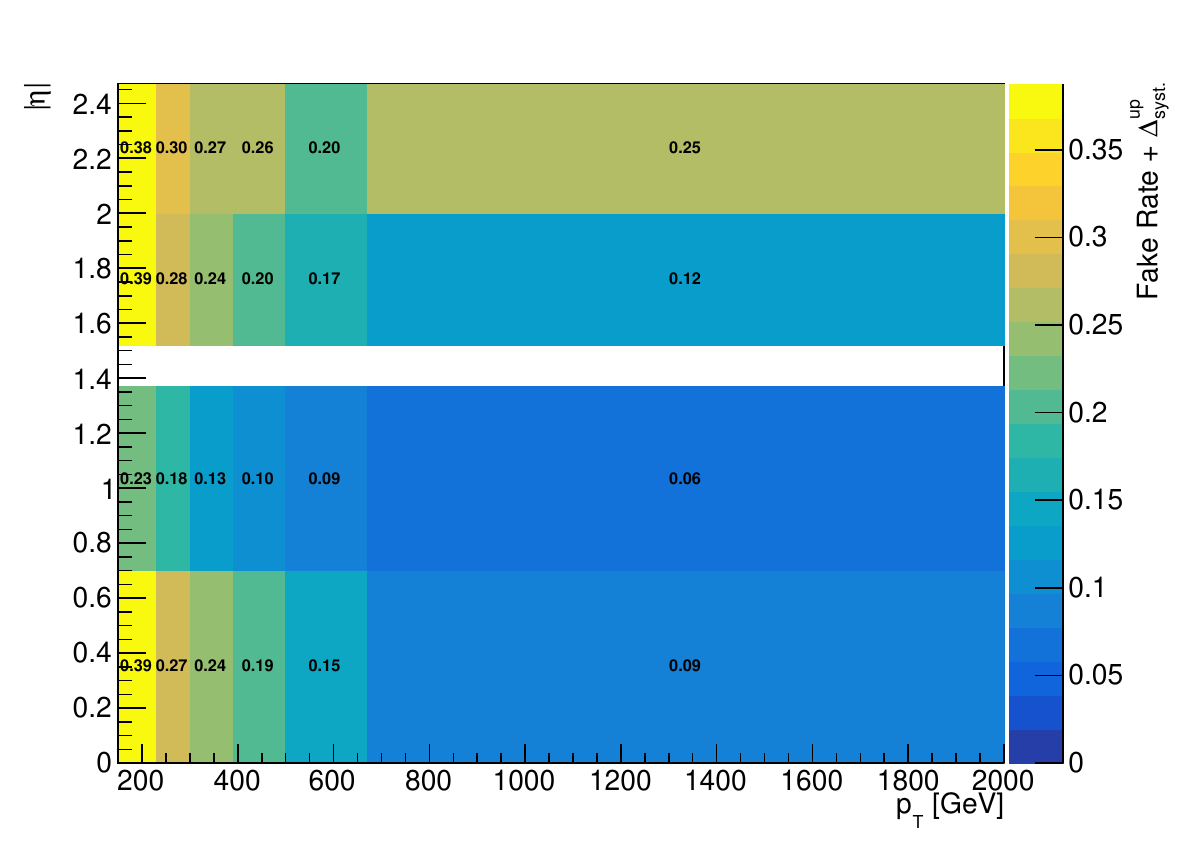}
  }
  \hfill
  \subfloat[(d)]{
    \includegraphics[width=0.5\textwidth]{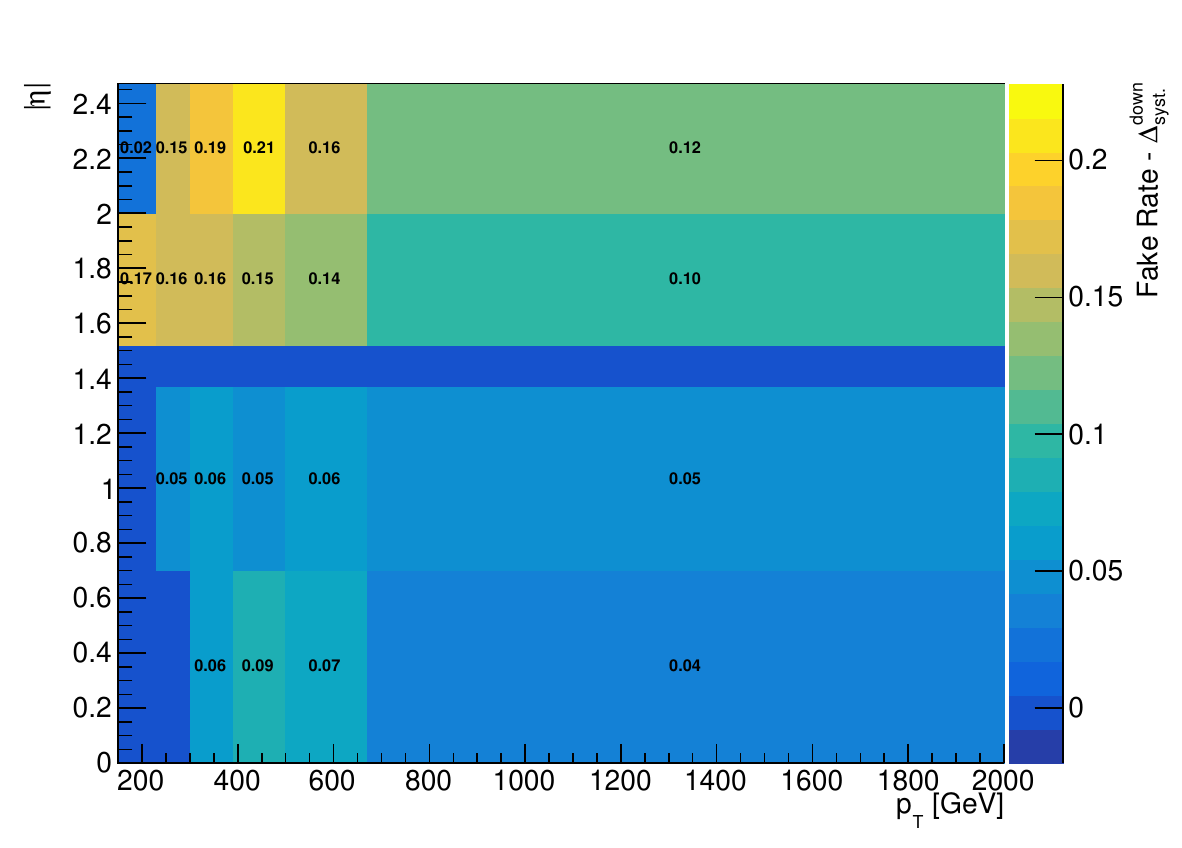}
  }
\caption{Variation on the fake rate due to total systematic uncertainty. (a) total experimental; (b) total Sherpa generator theoretical; (c) up variation on $f(p_{\textrm{T}},|\eta|)$; (d) down variation on $f(p_{\textrm{T}},|\eta|)$.}
\label{fig:Fake_systs_total}
\end{figure}

\begin{figure}[ht]
\centering
\includegraphics[width=0.7\textwidth]{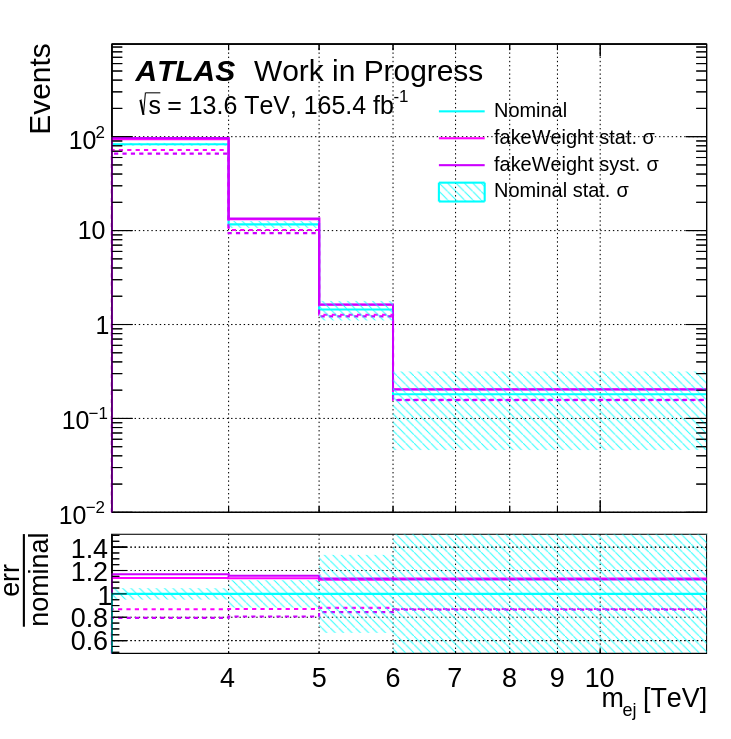}
\caption{Fake electron background distribution in the signal region showing the nominal estimate along with statistical and systematic uncertainty variations. The uncertainties shown are the statistical and systematic uncertainties derived on the fake rate and propagated to the fake background estimate, \texttt{fakeWeight} stat $\sigma$ and \texttt{fakeWeight} syst $\sigma$, respectively. Additionally, the hatched area shows the statistical uncertainty on the baseline data sample on which the fake weight is applied.}
\label{fig:fake_background_sr}
\end{figure}

In the same fashion, $f_{\textrm{syst}}(p_{\textrm{T}},|\eta|)$ is computed for the Sherpa generator theoretical uncertainties, as shown in FIG.~\ref{fig:Fake_systs_TH}.
%
The uncertainty due to PDF choice and QCD scales shown in FIG.~\ref{fig:Fake_systs_TH} are combined as Hessian and envelope, respectively. Eventually, all uncorrelated errors are added in quadrature, bin-by-bin, to obtain a total systematic uncertainty on $f$. FIG.~\ref{fig:Fake_systs_total} shows the total experimental and theoretical error on $f$, as well as the two-dimensional distributions of $f(p_{\textrm{T}},|\eta|)$ corresponding to the total up and down systematic uncertainty.
%
The above two-dimensional variations on $f$, are propgated to up and down variations on \texttt{fakeWeight} which in turn are applied on the baseline dataset to yield the systematic variation on the fake-electron background. FIG.~\ref{fig:fake_background_sr} shows the fake background distribution in the SR, plotted together with the statistical and systematic variations. As shown, the conservative approach outlined here estimates predict the total uncertainty on the fake-electrons background in the SR to lie within 20\%.

These sources of uncertainties are eventually considered as two nuisance parameters, fakes statistical uncertainty and fakes systematic uncertainty, in the likelihood function described in Section~\ref{ssec:stats:profile}.

\newpage

\section{Background Modelling}
\label{sec:background_modelling}
\begin{figure}[h]
  \captionsetup[subfigure]{labelformat=empty}
  \subfloat[(a)]{
    \includegraphics[width=0.5\textwidth]{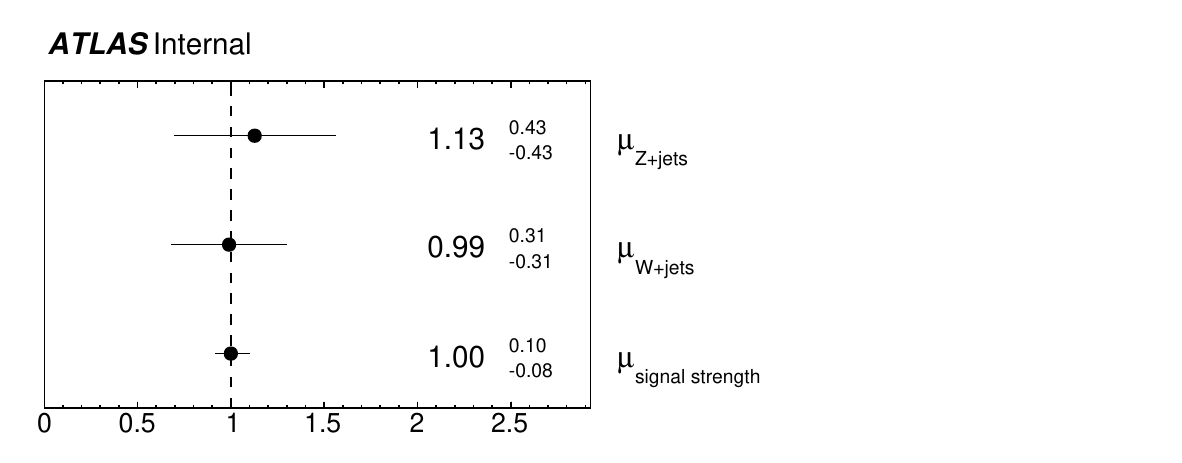}
  }
  \hfill
  \subfloat[(b)]{
    \includegraphics[width=0.5\textwidth]{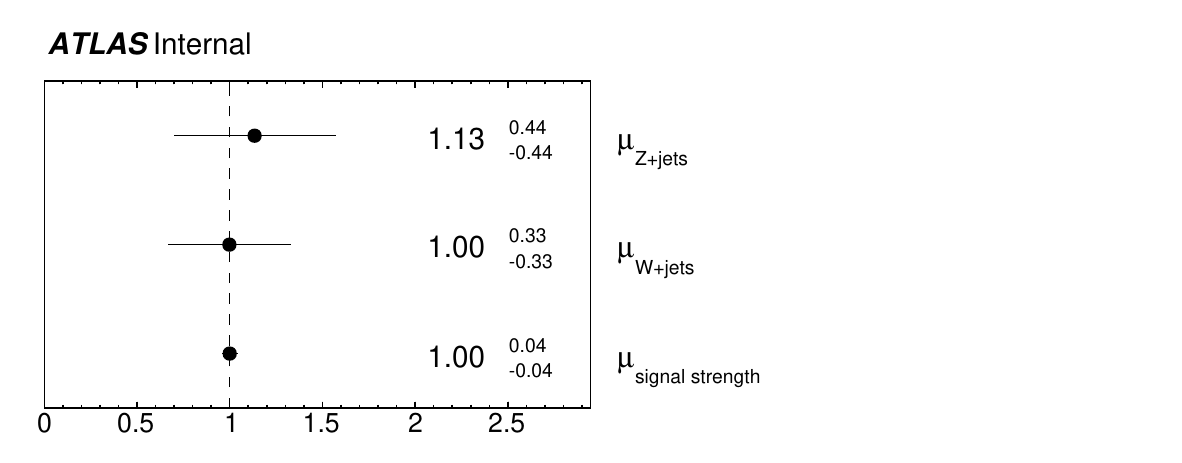}
  }
  \caption{Normalisation factors on the $W$+jets and $Z$+jets backgrounds for (a) the $e+j$ channel and (b) the $\mu+j$ channel.}
  \label{fig:norm_factor}
\end{figure}

The background simulations described in Section~\ref{sec:background} and data-driven fake-electron background estimation described in Section~\ref{sec:fakes} are compared to data in dedicated control and validation regions as outlined in Section~\ref{sec:EvSel}. All systematic uncertainties elaborated in the next chapter are included to estimate the total error on the estimation of the different background contributions. Compatibility of the total background estimation with the data is estimated in these regions within a profile-likelihood fit-to-data, deriving normalisation factors for the leading backgrounds, $W$+jets and $Z$+jets. These are found to be $1.13\pm 0.43$ and $0.99\pm 0.31$, respectively, in the electron channel, and $1.1\pm 0.5$ and $1.0\pm 0.3$, respectively, in the muon channel, as illustrated in FIG.~\ref{fig:norm_factor}. In the electron channel prior to the fit, the fakes component constitutes approximately 56\% of the total SR background, followed by $W$+jets at 32\%, with $Z$+jets, top, and diboson processes contributing the remaining 12\%. In the muon channel, $W$+jets dominates at approximately 80\%, with $Z$+jets, top, and diboson processes accounting for the remaining 20\%.

\subsection[$W$+jets Control and Validation Regions]{$\bm{W}$+jets Control and Validation Regions}
\begin{figure}
  \captionsetup[subfigure]{labelformat=empty}
    \subfloat[(a)]{
      \includegraphics[width=0.5\textwidth]{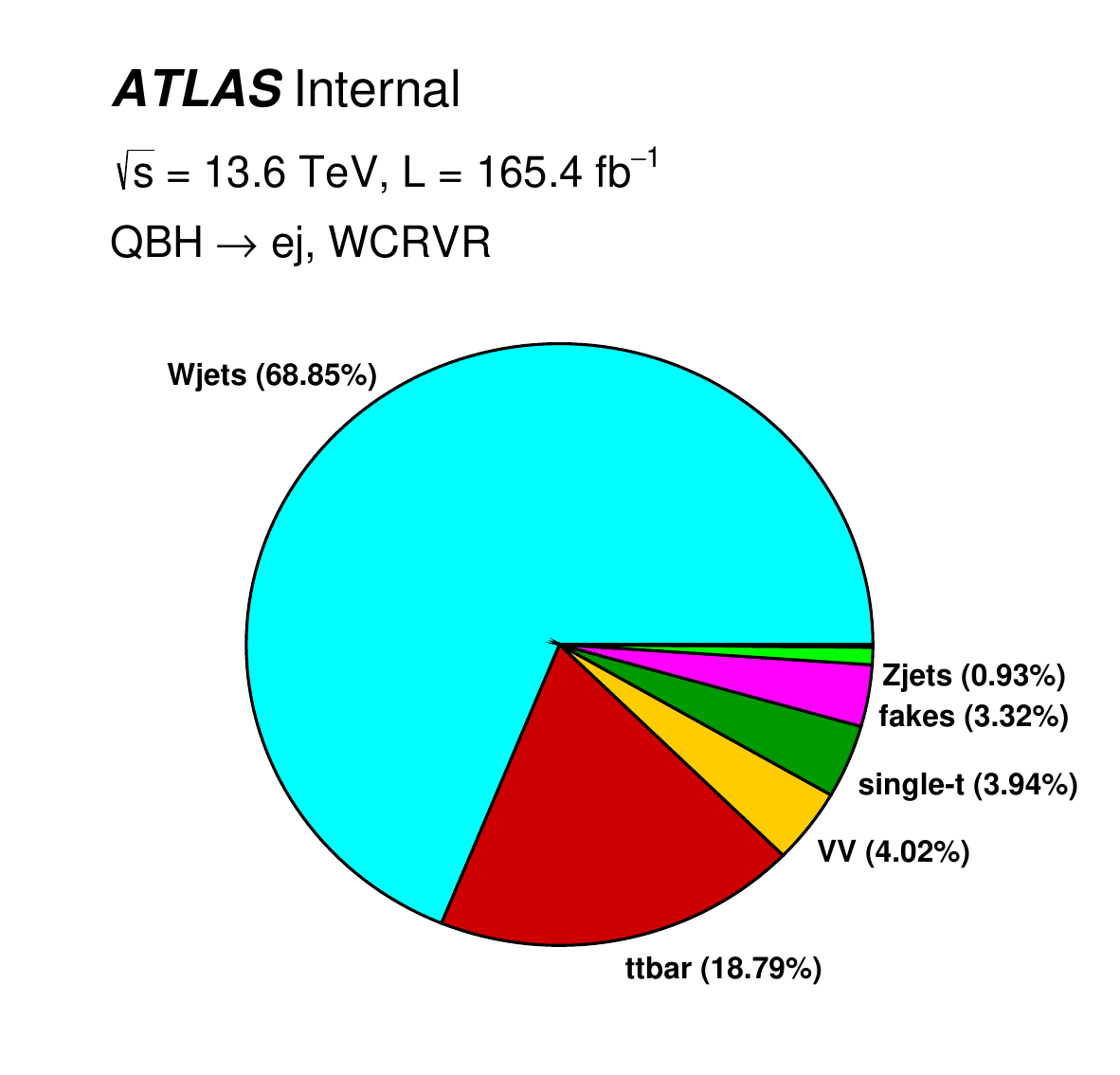}
    }
    \hfill
    \subfloat[(b)]{
      \includegraphics[width=0.5\textwidth]{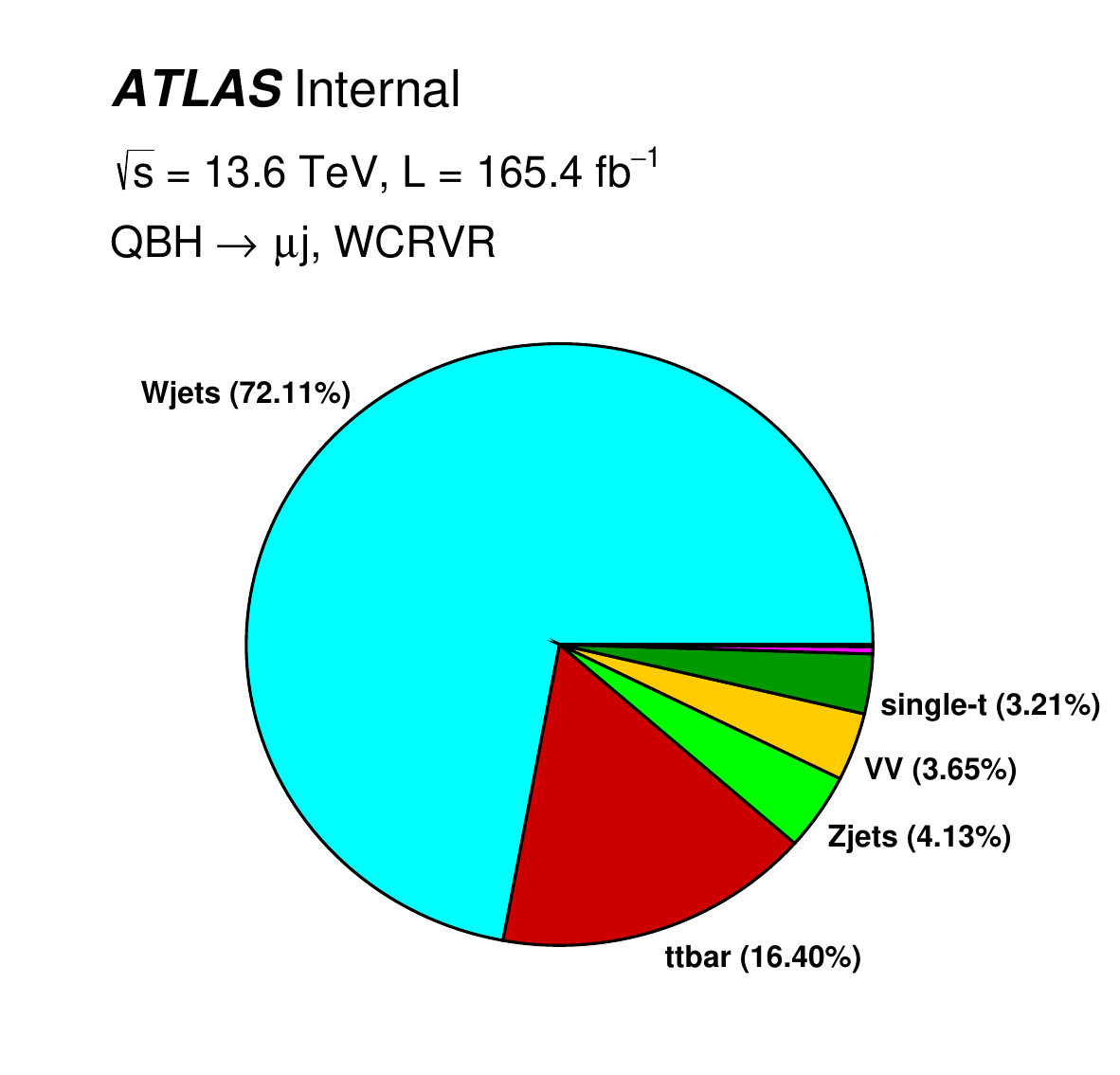}
    }
\caption{Background contributions in the $W$+jets control and validation regions for (a) the $e+j$ channel and (b) the $\mu+j$ channel.}
\label{fig:Pie_Charts_WCRVR}
\end{figure}

This region includes selections to increase the contribution from $W$+jet processes: $\mathcal{S}(E^{\textrm{miss}}_{\textrm{T}}) > 5.0, 3.5$ for the $e+j$, $\mu+j$ channels, respectively. The pie-charts given in FIG. \ref{fig:Pie_Charts_WCRVR} illustrate the contribution of different background sources to the $W$ CR (VR).

\begin{figure}[h!]
  \captionsetup[subfigure]{labelformat=empty}
  \subfloat[(a)]{
    \includegraphics[width=0.5\textwidth]{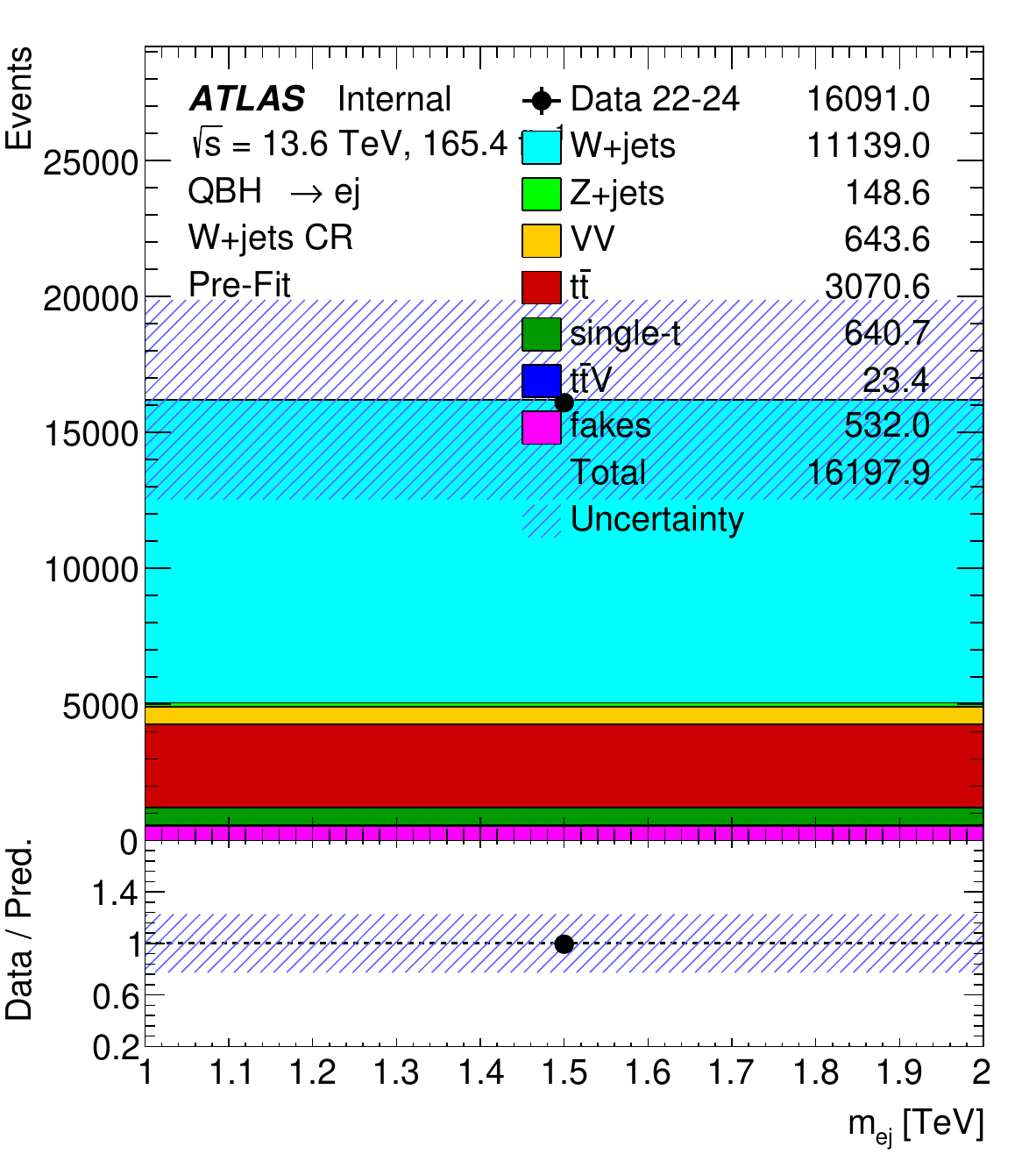}
  }
  \hfill
  \subfloat[(b)]{
    \includegraphics[width=0.5\textwidth]{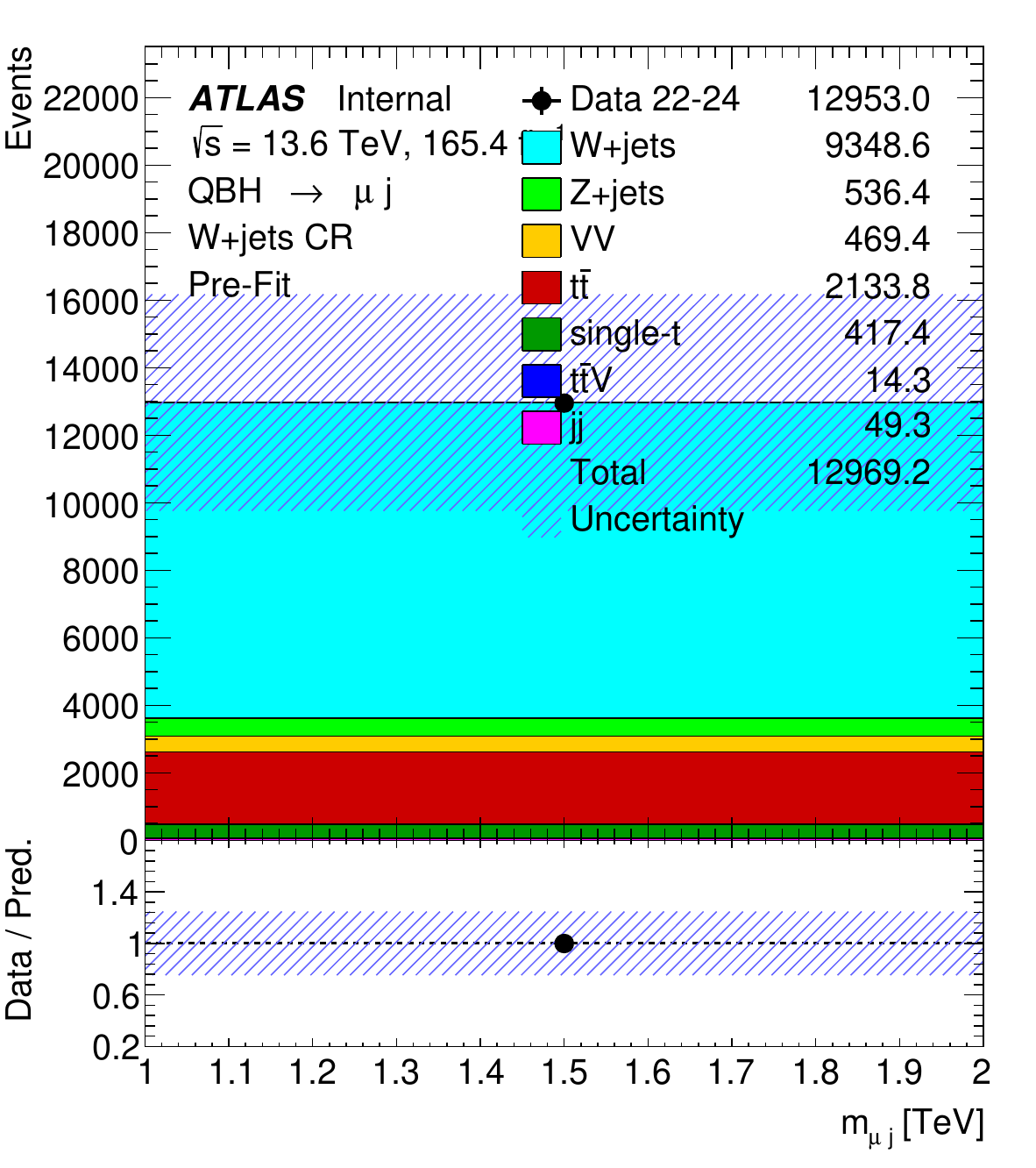}
  }
  \hfill
  \subfloat[(c)]{
    \includegraphics[width=0.5\textwidth]{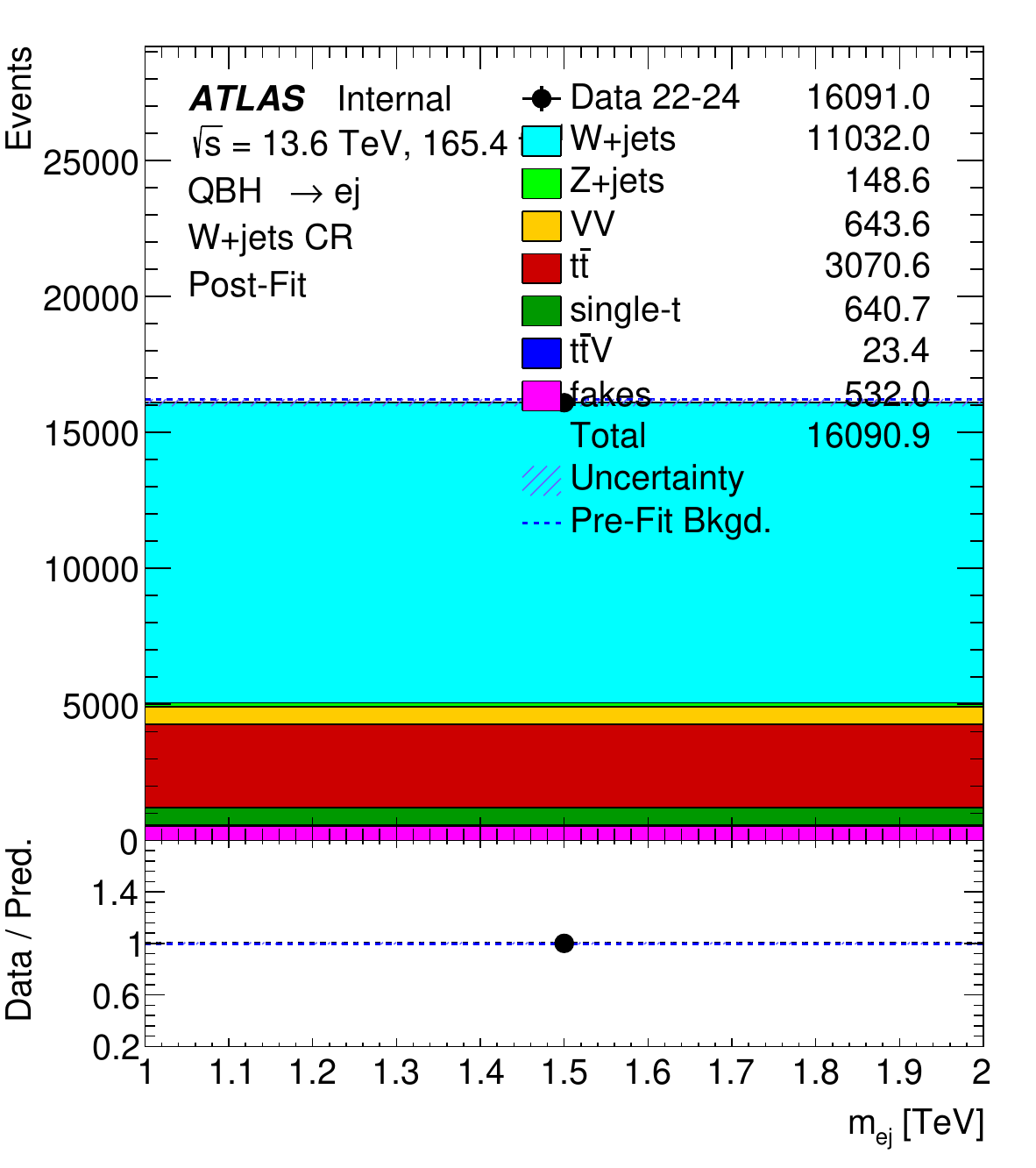}
  }
  \hfill
  \subfloat[(d)]{
    \includegraphics[width=0.5\textwidth]{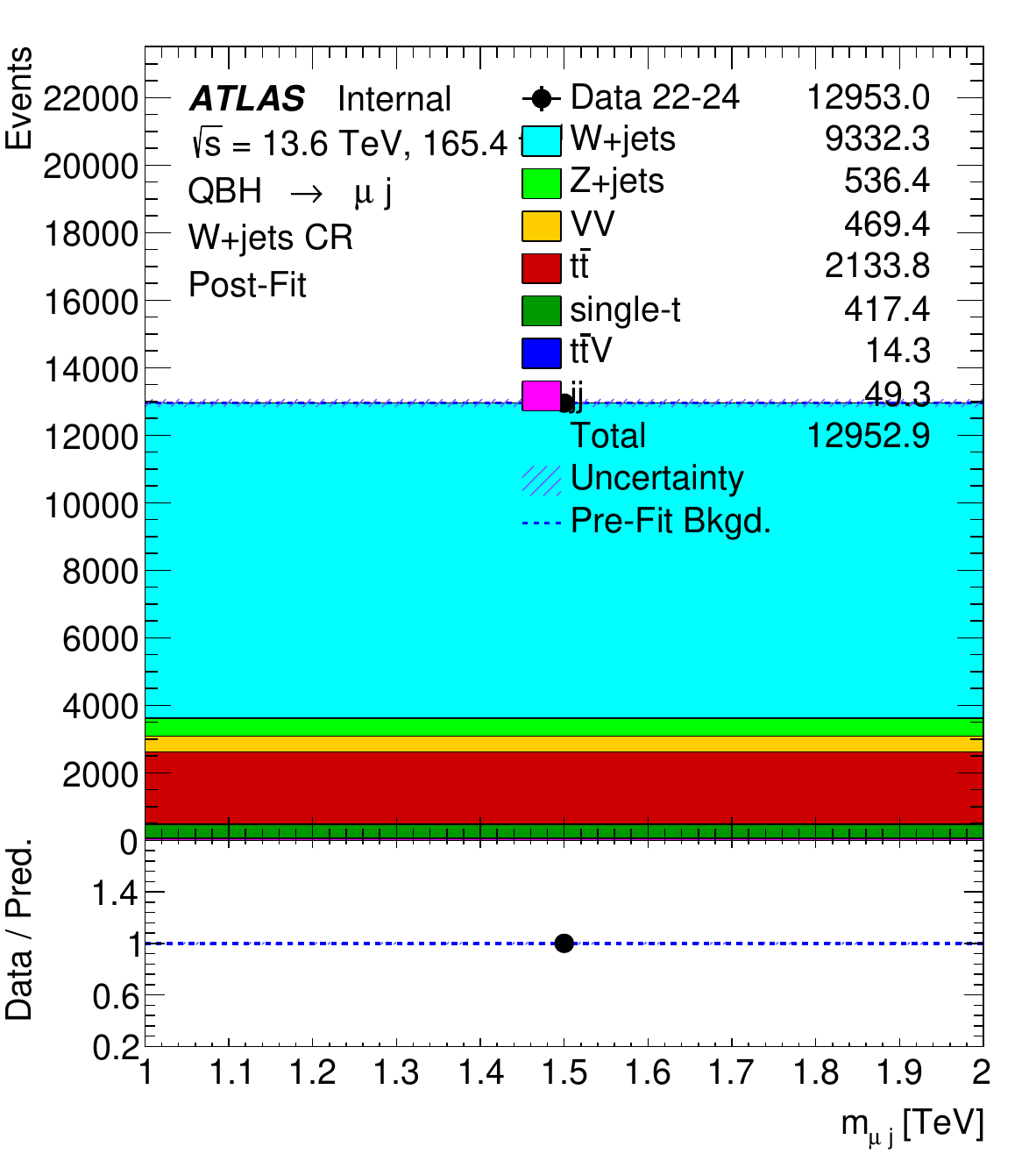}
  }
\caption{Invariant mass of the lepton-jet pair in the $W$+jets Control Region in the $e+j$ (left) and $\mu+j$ (right), pre-fit (top, a--b) and post-fit (bottom, c--d). The data-to-background ratio is shown in the bottom panes. The total uncertainty which includes statistical and systematic errors is shown the hatched bands. The multijet background is estimated from data, MC in the $e+j$, $\mu+j$ channels, respectively.}    
\label{fig:Wjets_CR}
\end{figure}

FIG. \ref{fig:Wjets_CR} shows the single-bin invariant mass of the lepton-jet pair in the $W$+jets CR

\begin{figure}[h!]
  \captionsetup[subfigure]{labelformat=empty}
    \subfloat[(a)]{
      \includegraphics[width=0.5\textwidth]{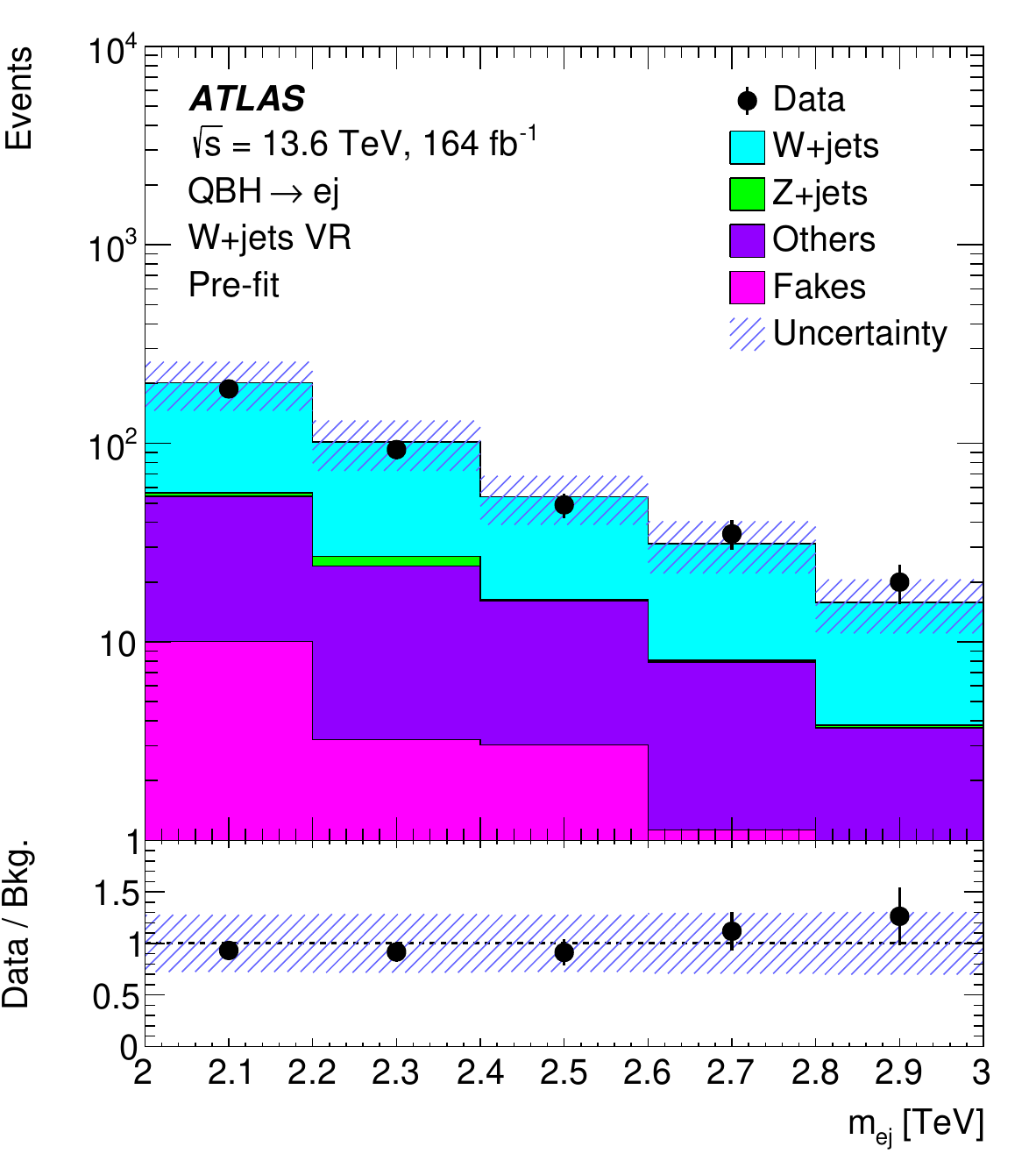}
    }
    \subfloat[(b)]{
      \includegraphics[width=0.5\textwidth]{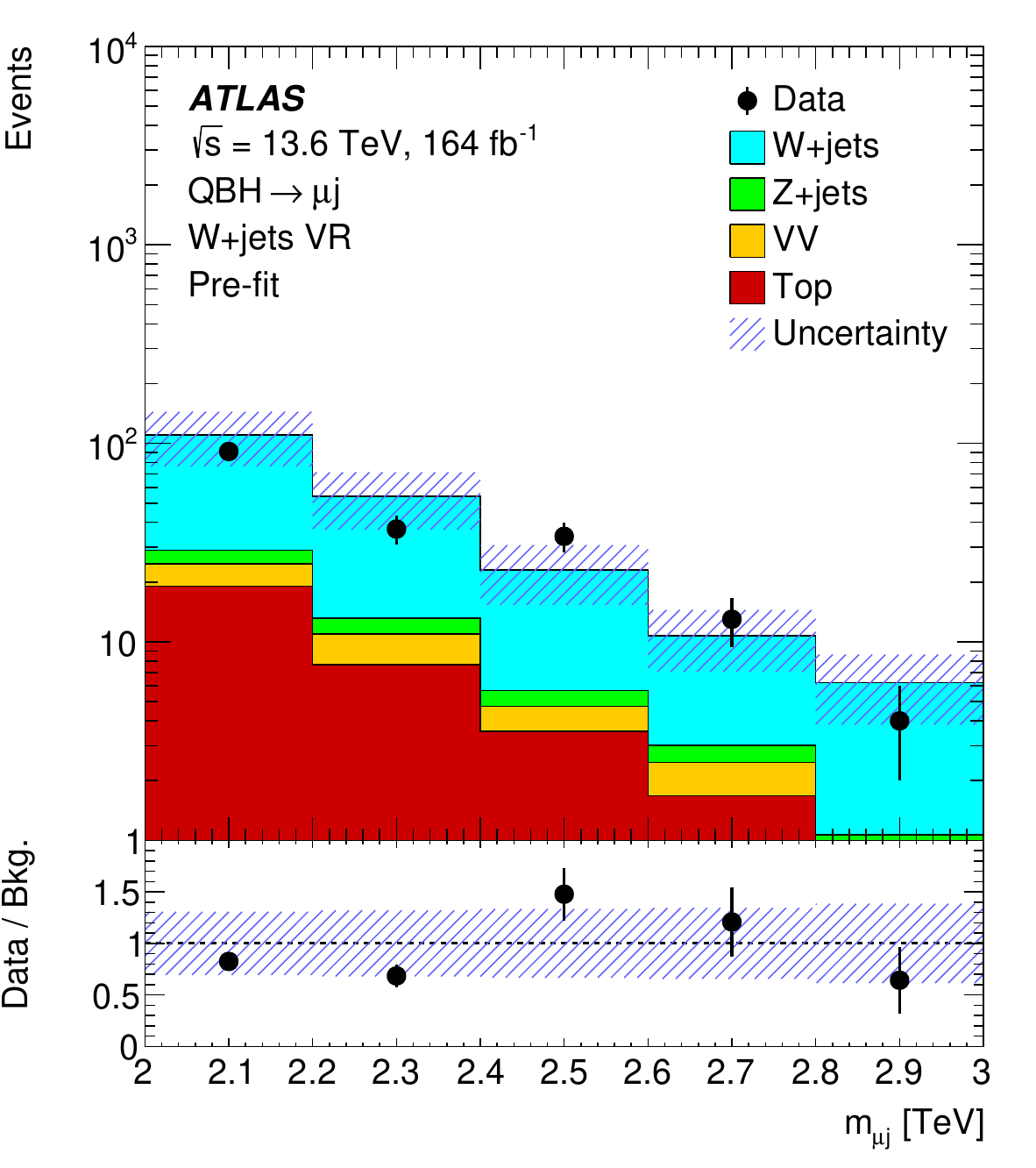}
    }
\caption{Invariant mass of the lepton--jet pair in the $W$+jets validation region for (a) the electron channel and (b) the muon channel. The data-to-background ratio is shown in the bottom panels, and hatched bands indicate the total statistical and systematic uncertainty. In (a), ``Others'' denotes subdominant backgrounds including top and diboson processes. The MC contributions are normalised based on their expected cross-section, prior to the likelihood fit.}
\label{fig:Wjets_VR}
\end{figure}

The $W$+jet background is validated in the VR, which has the same requirements but a higher $m_{\ell j}$ range of 2--3 TeV. FIG.~\ref{fig:Wjets_VR} shows the invariant mass of the lepton-jet pair in the $W$+jets VR for both channels, where the data-to-background ratio is shown in the bottom panels with hatched bands indicating the total statistical and systematic uncertainty.

These regions are used in the profile-likelihood fit described in Section~\ref{sec:results} to derive a normalisation factor on the $W$+jets background, $\mu_{W\textrm{+jets}}$, shown in FIG.~\ref{fig:norm_factor}.

\begin{figure}[h!]
  \captionsetup[subfigure]{labelformat=empty}
  \subfloat[(a)]{
    \includegraphics[width=0.5\textwidth]{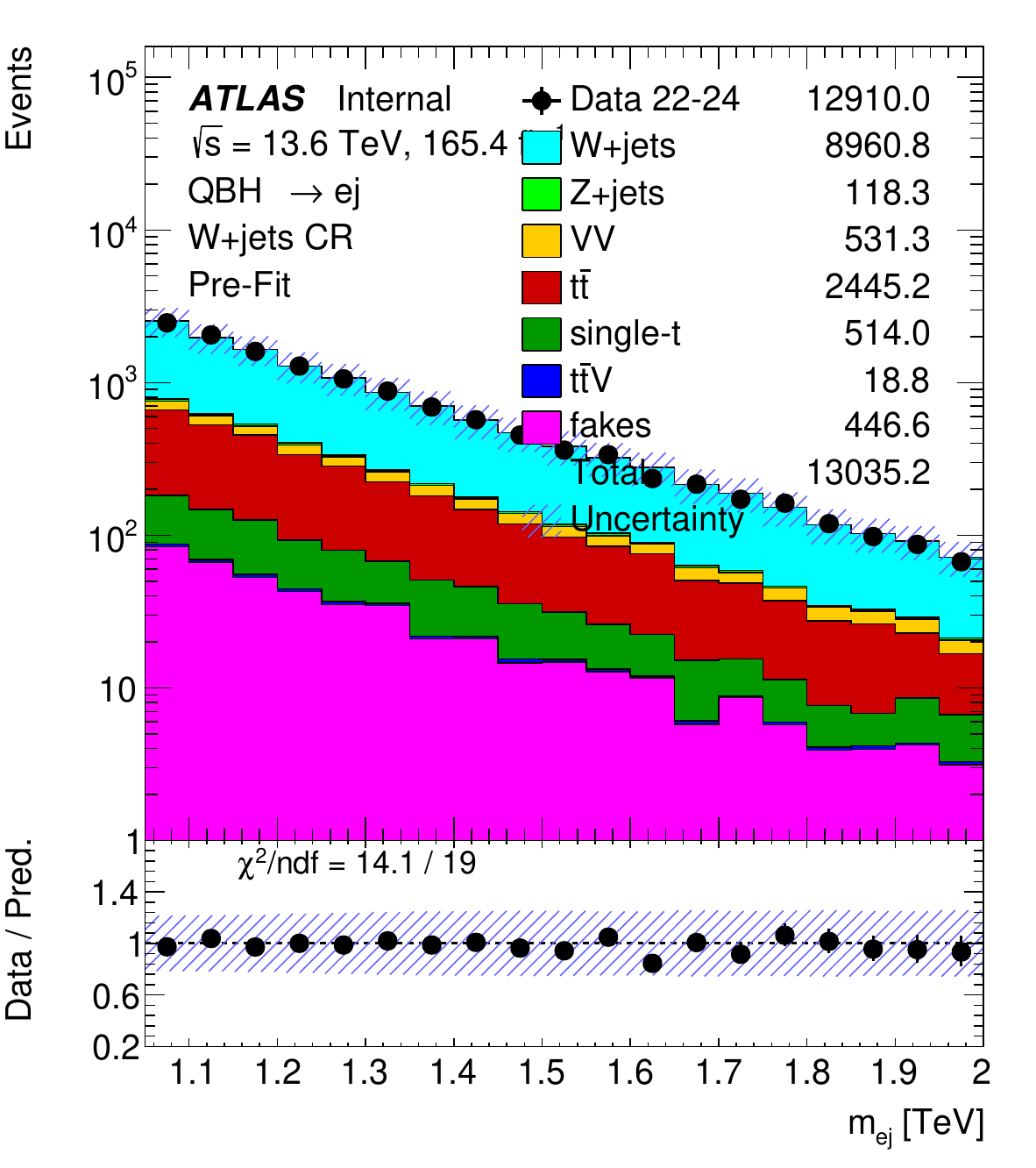}
  }
  \hfill
  \subfloat[(b)]{
    \includegraphics[width=0.5\textwidth]{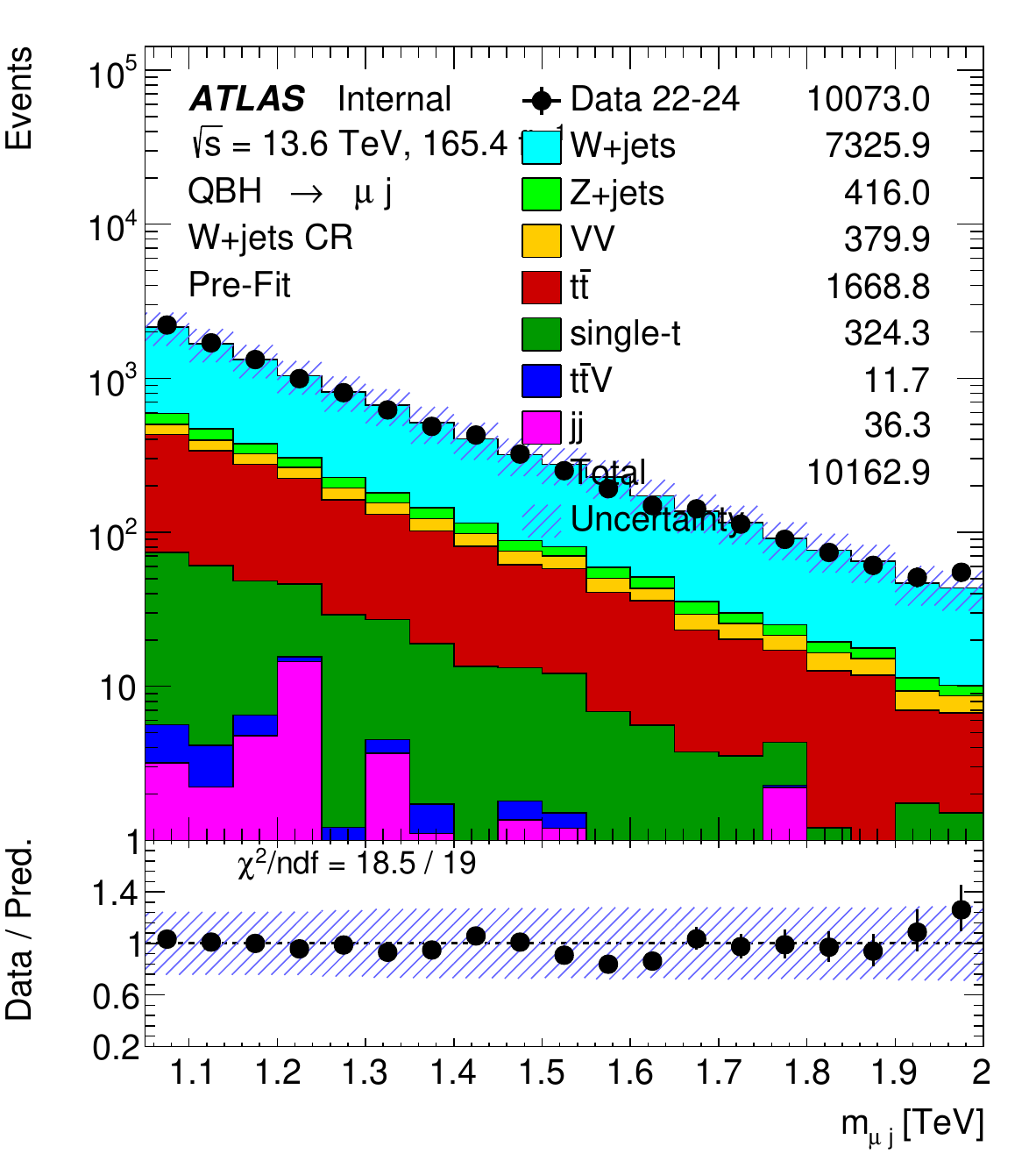}
  }
  \hfill
  \subfloat[(c)]{
    \includegraphics[width=0.5\textwidth]{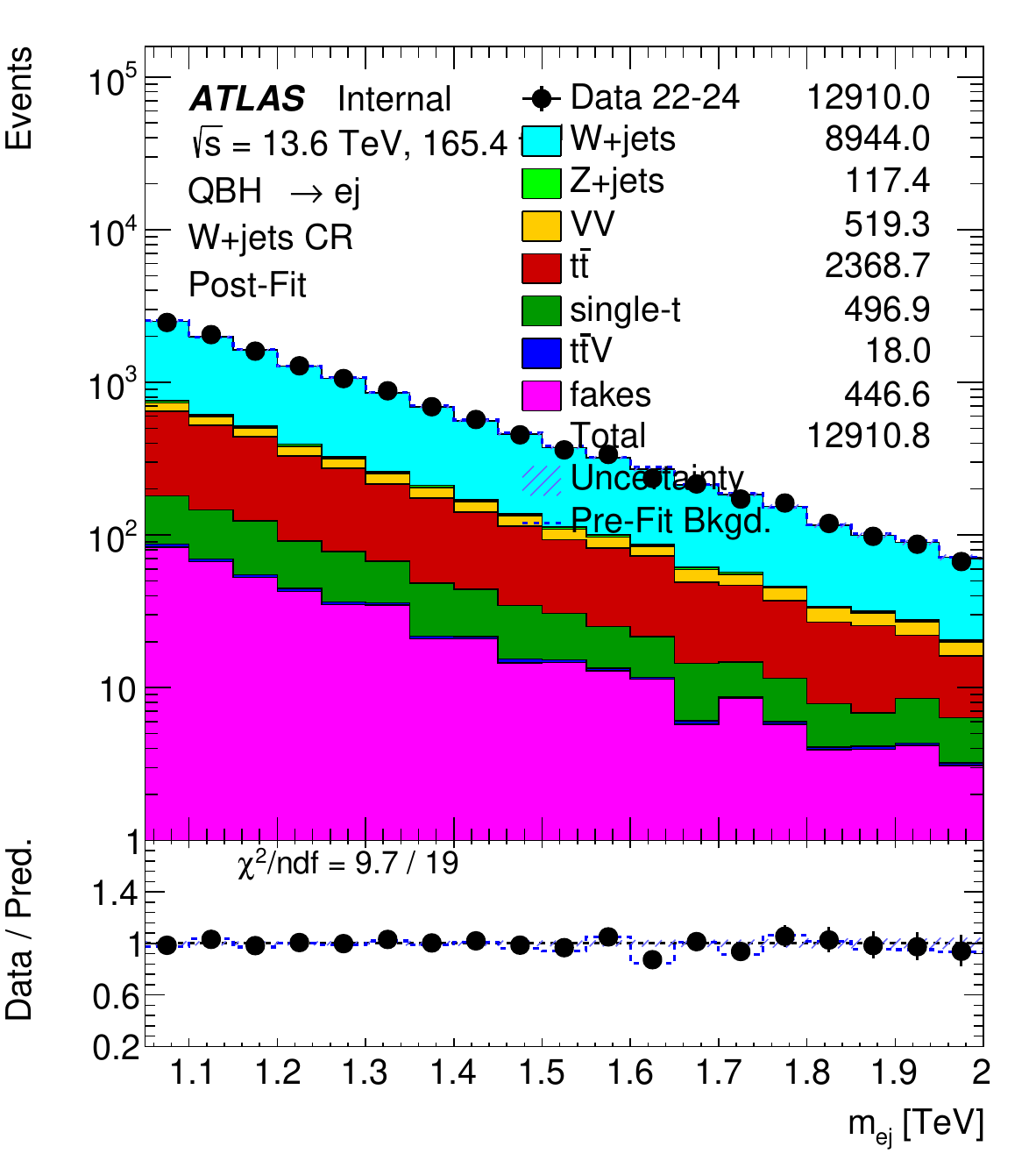}
  }
  \hfill
  \subfloat[(d)]{
    \includegraphics[width=0.5\textwidth]{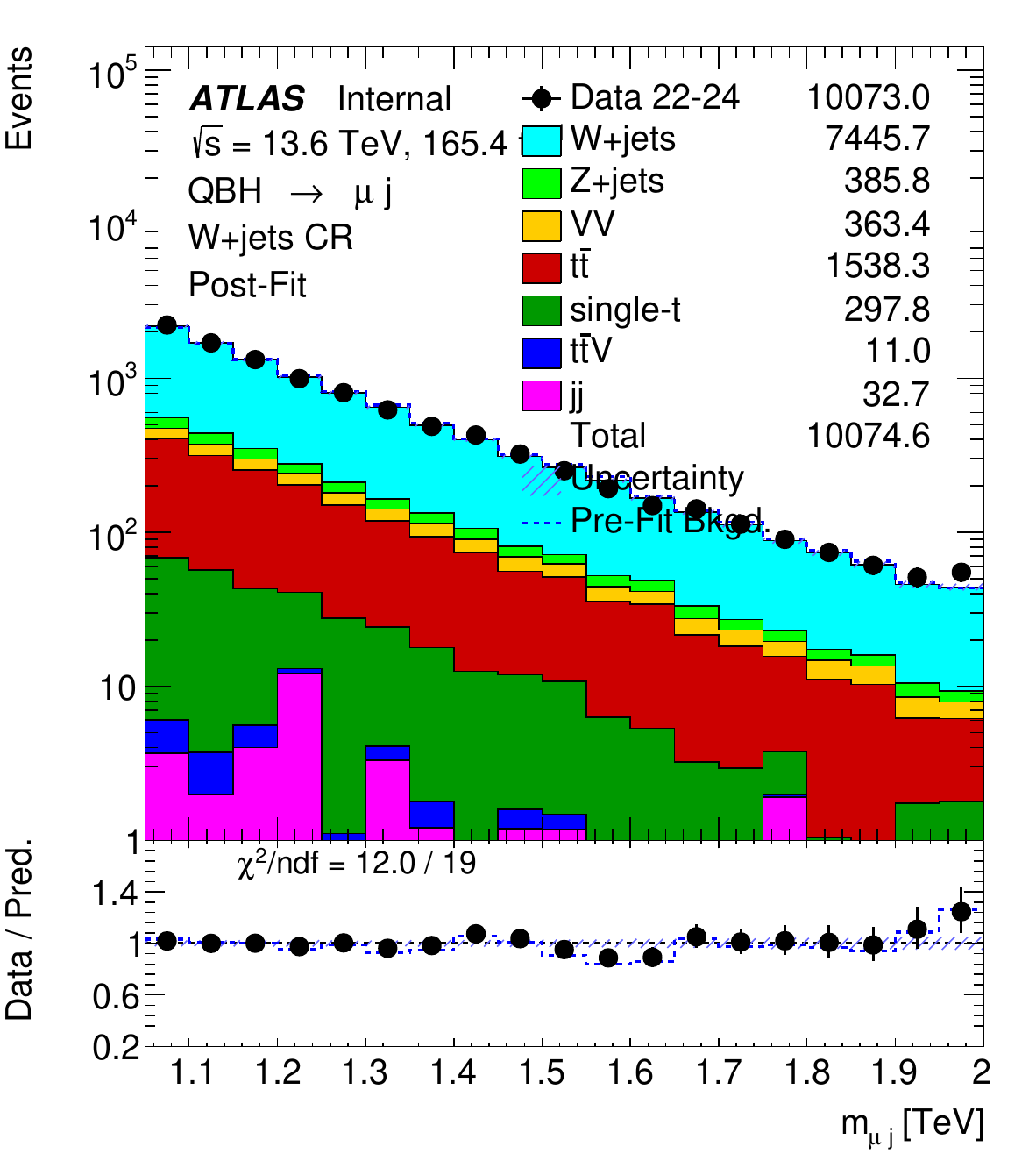}
  }
\caption{Invariant mass of the lepton-jet pair in the unused multi-binned $W$+jets Control Region in the $e+j$ (left) and $\mu+j$ (right), pre-fit (top) and post-fit (bottom). The data-to-background ratio is shown in the bottom panes. The total uncertainty which includes statistical and systematic errors is shown the hatched bands. The multijet background is estimated from data, MC in the $e+j$, $\mu+j$ channels, respectively.}    
\label{fig:Wjets_CR_multibinWCRVR}
\end{figure}

The single-binned $W$ CR and VR were chosen after confirmation of lack of shape effects that would necessitate multi-binned regions. To take one example using an RS1 signal at 6.0~\textrm{TeV}, a signal+background fit is conducted using a multi-binned $W$ CR and VR. FIG.~\ref{fig:Wjets_CR_multibinWCRVR} shows the $W$ CR in this scheme.

\begin{figure}[h!]
  \captionsetup[subfigure]{labelformat=empty}
  \subfloat[(a)]{
    \includegraphics[width=0.5\textwidth]{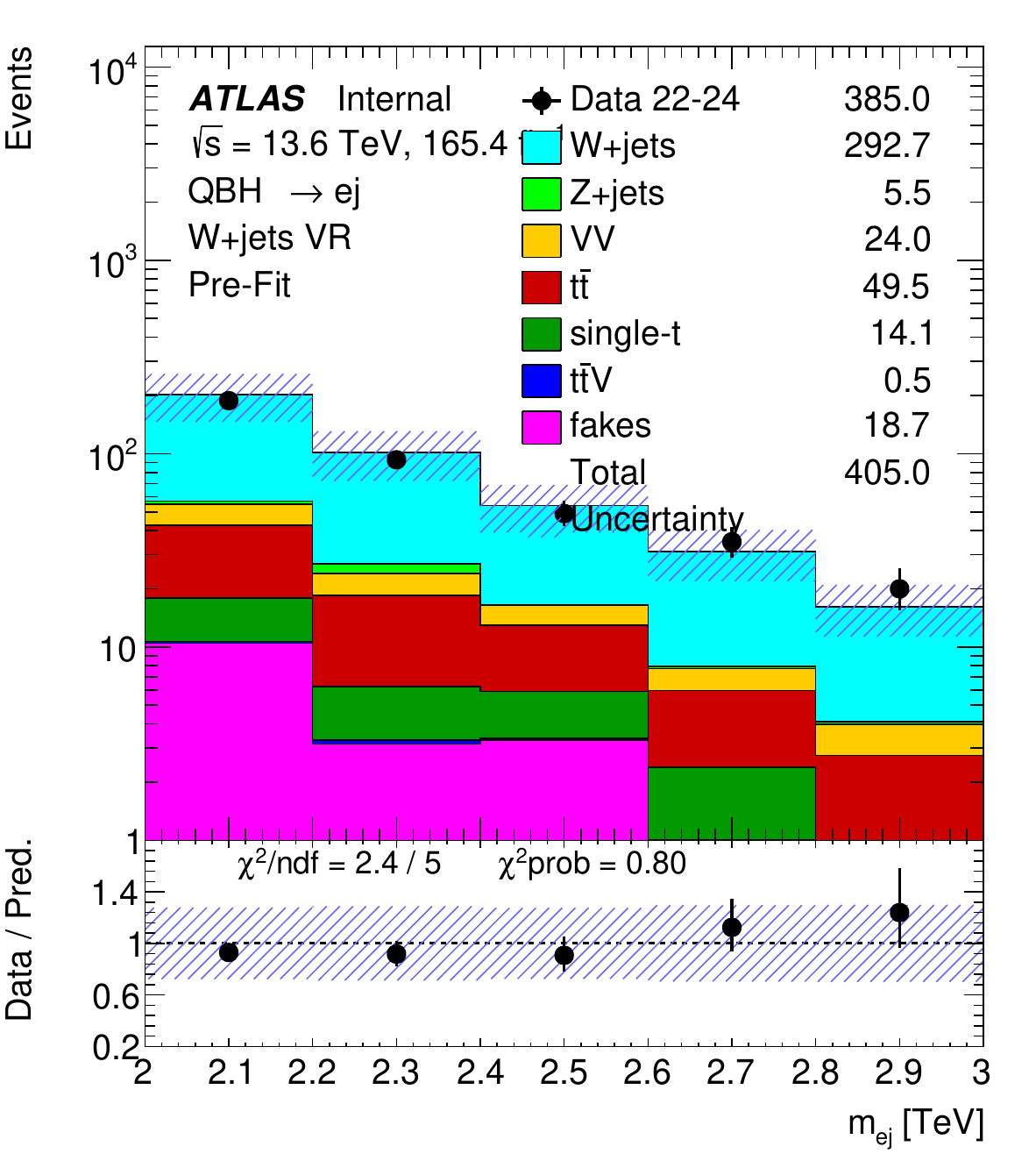}
  }
  \hfill
  \subfloat[(b)]{
    \includegraphics[width=0.5\textwidth]{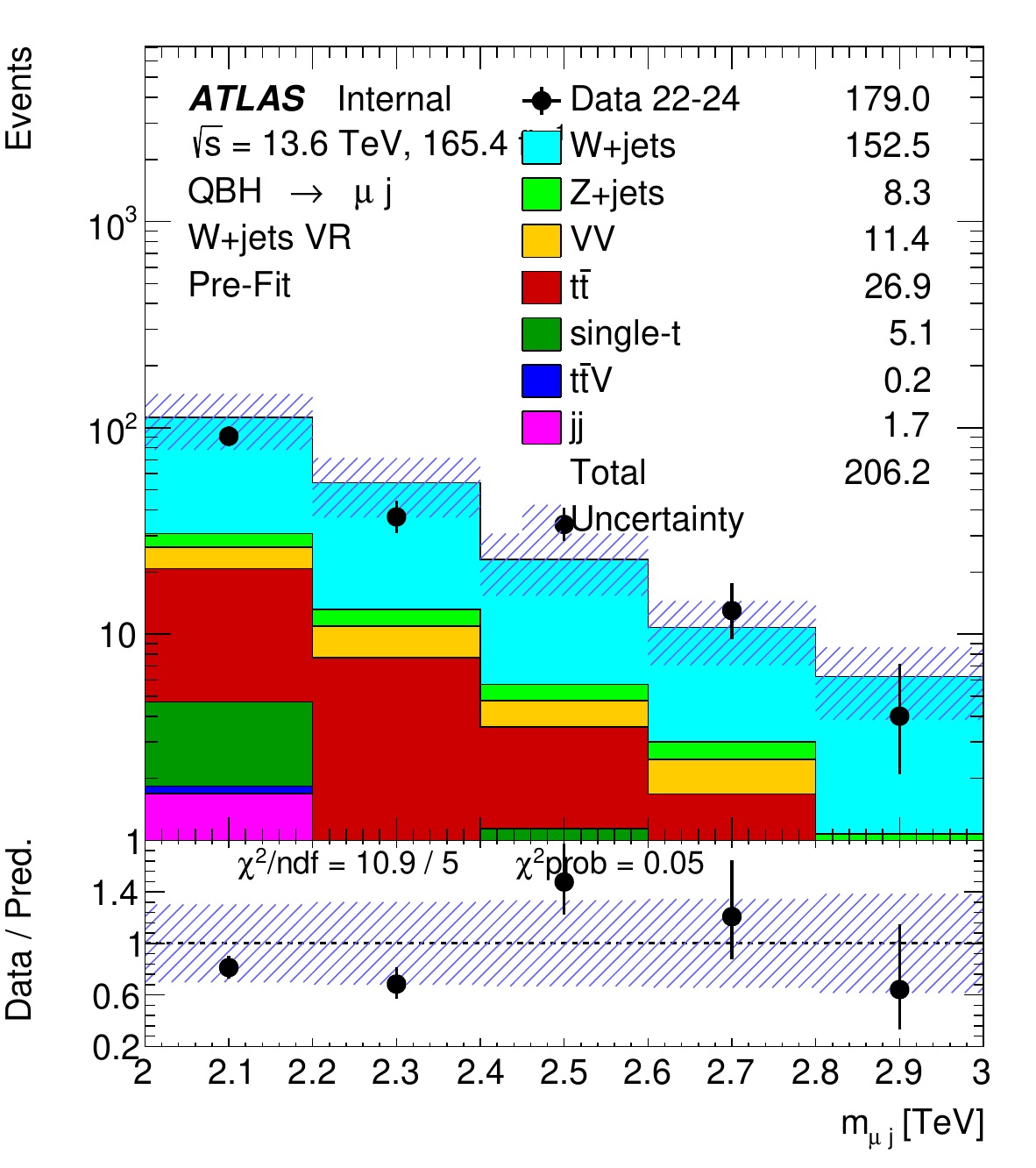}
  }
  \hfill
  \subfloat[(c)]{
    \includegraphics[width=0.5\textwidth]{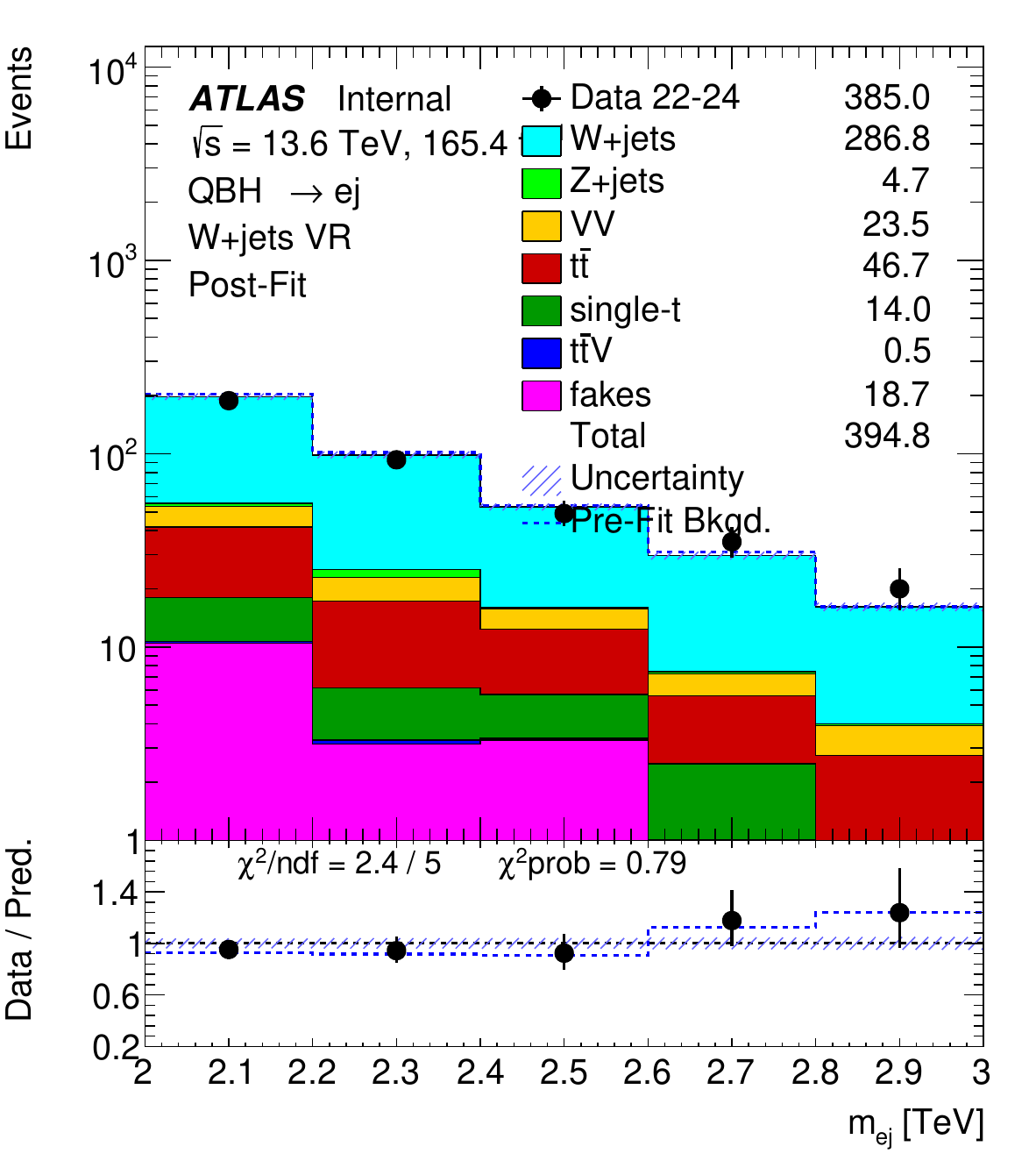}
  }
  \hfill
  \subfloat[(d)]{
    \includegraphics[width=0.5\textwidth]{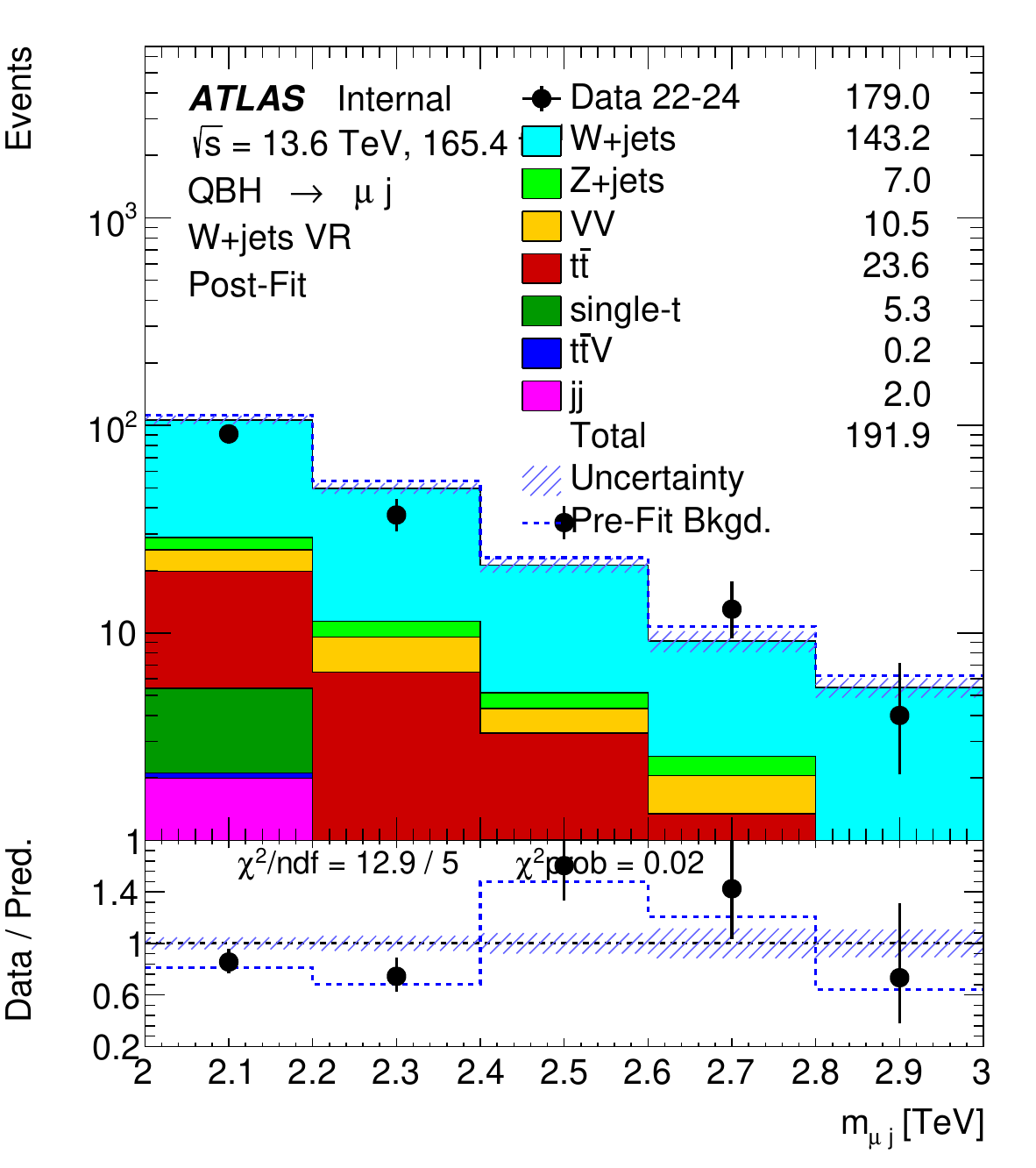}
  }
\caption{Invariant mass of the lepton-jet pair in the unused multi-binned $W$+jets Validation Region in the $e+j$ (left) and $\mu+j$ (right), pre-fit (top) and post-fit (bottom). The data-to-background ratio is shown in the bottom panes. The total uncertainty which includes statistical and systematic errors is shown the hatched bands. The multijet background is estimated from data, MC in the $e+j$, $\mu+j$ channels, respectively.}    
\label{fig:Wjets_VR_multibinWCRVR}
\end{figure}

The plots in FIG.~\ref{fig:Wjets_CR_multibinWCRVR} show no shape effect in the region where the $W$+jets normalisation factor is derived and hence allow this region to be single-bin for simplicity. Similarly, FIG.~\ref{fig:Wjets_VR_multibinWCRVR} shows the same distributions in the $W$ VR.

To reiterate, the VRs are used only to validate the analysis choices made to estimate the leading MC background, $W+$jets, and it does not enter the calculation of $\mu_{W+\mathrm{jets}}$. For clarification, these multi-bin plots are not used in the limit calculation---or in any other part of the analysis---but are shown solely to address potential concerns about unaccounted-for shape effects in a single-bin region. Although the \(\mu+j\) plots in Fig.~\ref{fig:Wjets_VR_multibinWCRVR} show some mismodelling in individual bins, the relatively flat data-to-background ratios in Fig.~\ref{fig:Wjets_CR_multibinWCRVR} demonstrate that shape effects are not significant.

\subsection[$Z$+jets Control and Validation Regions]{$\bm{Z}$+jets Control and Validation Regions}

This region includes selections to increase the contribution from $Z$+jet processes: $m_{\ell\ell} \in [60,120]\; \textrm{GeV}$. In addition to the signal lepton, the presence of another baseline lepton is required. The $Z$+jets background constitutes above 95\% of the total background in this region (see FIG.~\ref{fig:Pie_Charts_ZCRVR} in Appendix~\ref{app:qbh:zcrvr}). These regions are used in the profile-likelihood fit described in Section~\ref{sec:results} to derive a normalisation factor on the $Z$+jets background, $\mu_{Z\textrm{+jets}}$, shown in FIG.~\ref{fig:norm_factor}. The single-binned CR and VR pre- and post-fit distributions are shown in Appendix~\ref{app:qbh:zcrvr} (FIGS.~\ref{fig:Zjets_CR} and~\ref{fig:Zjets_VR}).

\begin{figure}[h]
  \captionsetup[subfigure]{labelformat=empty}
  \subfloat[(a)]{
    \includegraphics[width=0.5\textwidth]{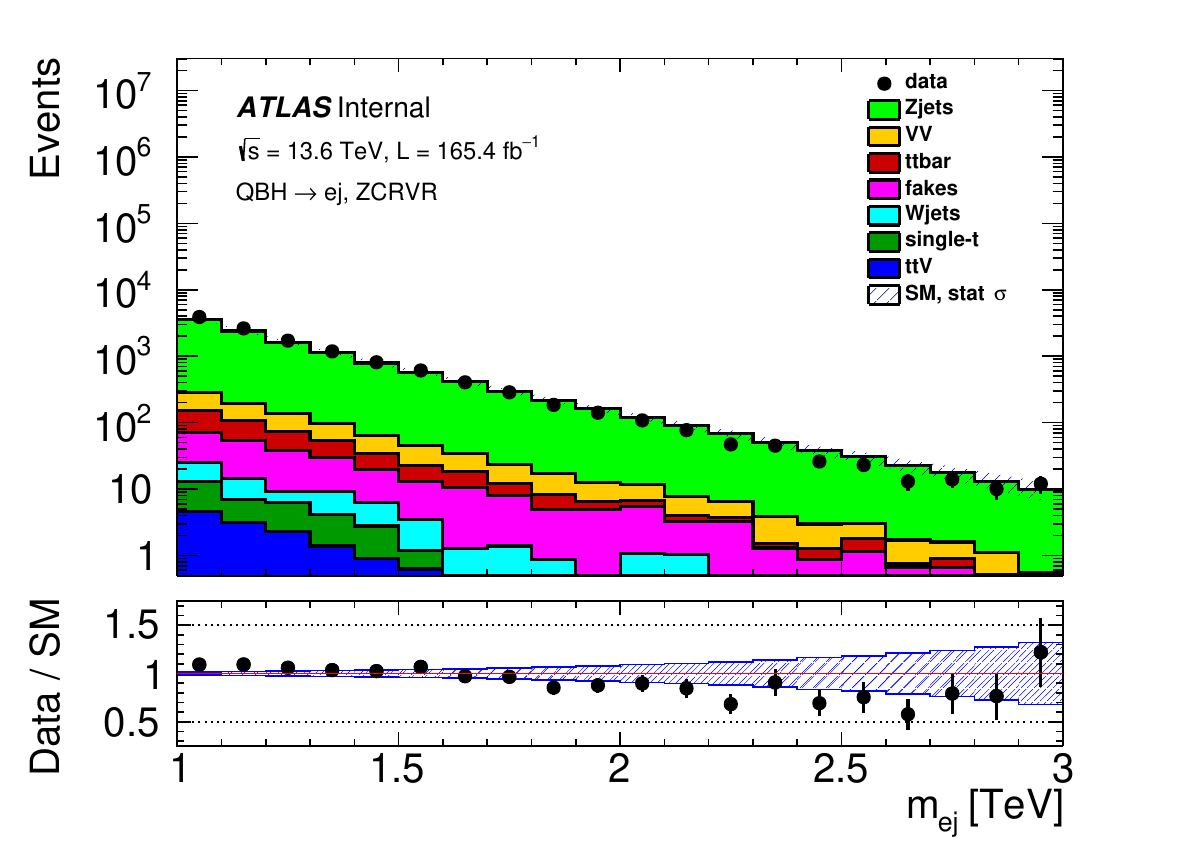}
  }
  \hfill
  \subfloat[(b)]{
    \includegraphics[width=0.5\textwidth]{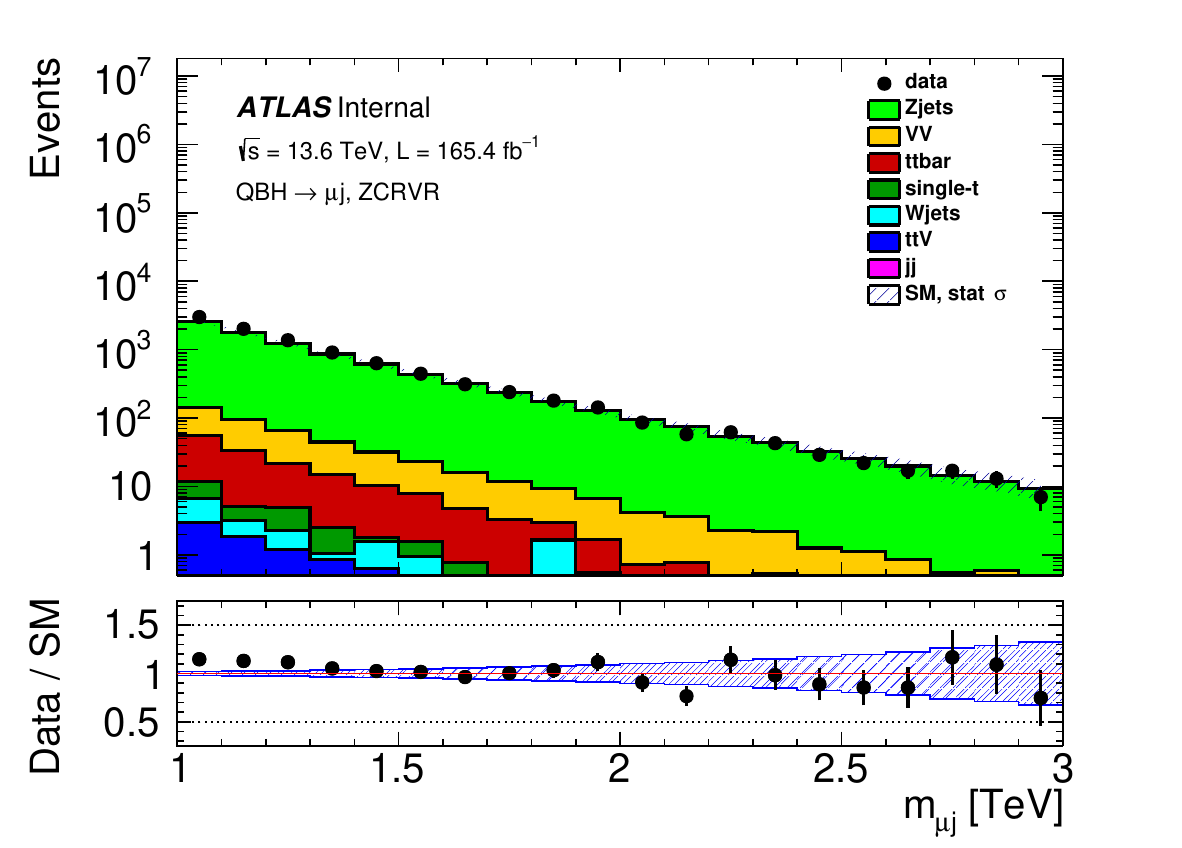}
  }
\caption{Invariant mass of the lepton-jet pair pre-fit distributions across the merged $Z$+jets Control (1--2~TeV) and Validation (2--3~TeV) Regions for (a) the $e+j$ and (b) the $\mu+j$ channel. These bins are shown for illustrative cross-check purposes only; the actual fit uses a single bin per region. The $Z$+jets background dominates throughout (above 95\%, consistent with FIG.~\ref{fig:Pie_Charts_ZCRVR} in the appendix), and the data-to-background ratio is relatively flat within statistical uncertainties, indicating no significant shape effect. The multijet background is estimated from data and from dijet MC in the $e+j$ and $\mu+j$ channels, respectively.}
\label{fig:Zjets_multibinZCRVR}
\end{figure}

As a cross-check, FIG.~\ref{fig:Zjets_multibinZCRVR} shows the pre-fit $m_{\ell j}$ distribution across the merged $Z$+jets CR and VR (1--3~TeV) in multiple bins. Note that these bins are only used here for illustrative purposes and do not represent the actual single-binned regions used in the fit. The figure serves two purposes: it confirms that $Z$+jets dominates above 95\% of the background throughout (consistent with the pie charts in the appendix), and it demonstrates that the data-to-background ratio is relatively flat within statistical uncertainties, indicating no significant shape effect that would require a multi-binned treatment.

\section{Systematic uncertainties}
\label{sec:systematic_uncertainties}
There are two types of systematic uncertainties affecting this analysis: experimental, arising from the reconstruction of physics objects, and theoretical, from modelling the background and signal. Both are discussed and evaluated in each of the analysis regions defined in Section~\ref{sec:EvSel}. For both channels the signal region is dominated by experimental jet uncertainties and theoretical Sherpa QCD scales. The theoretical uncertainties amount to approximately 30\% on the $V$+jets background yield in the SR. The experimental and theoretical uncertainties are also propagated to the fakes background through the prompt electron MC subtracted from data in the $f$CR, as detailed in Section~\ref{sec:fakes}, and are found to amount to approximately 18\% on the fakes background throughout the SR. An additional uncertainty arising from the difference in fake origin (light-flavour decays versus photon conversions) between the $f$CR and SR is estimated at 22\% in the SR, while the same quantity computed between the $f$CR and $f$VR is 17\%. The limited size of the data and MC samples in the $f$CR contributes approximately 16\% to the SR. All uncertainties concerning the fake-electron background are collectively referred to as ``fake uncertainties''.

A Systematic Analysis Tool is used to visualise the relevant systematic uncertainties for this analysis. This tool calculates a ratio for each systematic as $\frac{\textrm{Systematic}} {\textrm{Nominal}}$, producing an up and down value per  bin. These are calculated using symmetrisation on the variations provided for each systematic. This is done in one of three ways, depending on the type of variation provided for the systematic:
\begin{itemize}
    \item Up and down variations provided: no symmetrisation needed
    \item Only up variations provided: down calculated as  $y_{\textrm{ratio,down}} = \frac{y_{\textrm{nom.}}- |y_{\textrm{var. up}}-y_{\textrm{nom.}}|}{y_{\textrm{nom.}}}$
    \item Neither provided: If single variation (not up/down), $y_{\textrm{var.}}$, provided it is used to calculate $y_{\textrm{ratio,down}}$ as before ($y_{\textrm{var.}}$ in place of $y_{\textrm{var. up}}$). And up variation calculated as $y_{\textrm{ratio,up}} = \frac{y_{\textrm{nom.}}+ |y_{\textrm{var.}}-y_{\textrm{nom.}}|}{y_{\textrm{nom.}}}$
\end{itemize}
Where $y_{\textrm{nom.}}$ is the bin contents for the nominal and $y_{\textrm{var.}}$ the bin contents for the variation. The plots produced using this tool show the ratio per systematic or object overlaid.

\subsection{Experimental}
\label{sec:Syst_exp}

Tables~\ref{tab:lep_exp_syst} and~\ref{tab:jet_exp_syst} in Appendix~\ref{app:qbh:exp_systs} summarise the experimental systematic uncertainties considered in this analysis. They are separated into two types, scale factor (SF) and calibration (calib). SFs are used when correcting MC to data such as the simulation of the efficiency of object triggers, reconstruction, identification, and isolation. The uncertainty on these SFs contributes to the experimental systematic uncertainty. Calibration systematics are kinematic-based and arise from the uncertainties of the calibration of physics objects: electrons, muons and jets. These systematic uncertainties are provided by the respective ATLAS Combined Performance (CP) groups.

The electron systematics (see Table~\ref{tab:lep_exp_syst} in the appendix) comprise scale factors on identification, trigger, reconstruction, and isolation efficiency, plus energy scale and resolution calibrations. For the high-\pT\ isolation working point used here, the electron isolation efficiency systematic is treated as a flat 20\% uncertainty, following a solution agreed with the ATLAS EGamma CP group due to a known issue with the calorimeter isolation at high \pT; details are given in Appendix~\ref{sec:app_systematics}.

The muon systematics include isolation, reconstruction, trigger, and track-to-vertex-association efficiency scale factors, alongside energy scale and resolution calibrations (Table~\ref{tab:lep_exp_syst} in the appendix). The reconstruction efficiency systematic is replaced by the Run~2 value plus a cross-check safety factor, following discussions with the ATLAS Muon Combined Performance group regarding an overly conservative extrapolation to high $m_{\ell j}$; details are given in Appendix~\ref{sec:app_systematics}.

Jet systematics (Table~\ref{tab:jet_exp_syst} in the appendix) include the jet vertex tagging efficiency, jet energy scale (JES) and resolution (JER), as well as $\eta$ intercalibration and in-situ calibration corrections. The leading jet systematic is a residual non-closure in the in-situ calibration at high \pT\ prior to the current data-taking period, which the Jet CP group expects at high transverse momenta and covers with a dedicated uncertainty. The $E^{\textrm{miss}}_{\textrm{T}}$ soft-term uncertainties (scale and resolution) are also included and are summarised in Table~\ref{tab:jet_exp_syst} of the appendix.

In addition to these object-based systematics, two further experimental systematic uncertainties are considered: a luminosity uncertainty of 1.9\%, applied as a flat normalisation to all MC processes; and the pile-up reweighting (PRW) uncertainty.

\subsubsection*{Signal Region}

The signal region used for both channels is defined in Section~\ref{sec:EvSel} and uses cuts only on the properties of the jet and respective lepton. Therefore only systematics on these objects are presented, for the total backgrounds.

\begin{figure}[h]
  \captionsetup[subfigure]{labelformat=empty}
  \subfloat[(a)]{
    \includegraphics[width=0.5\textwidth]{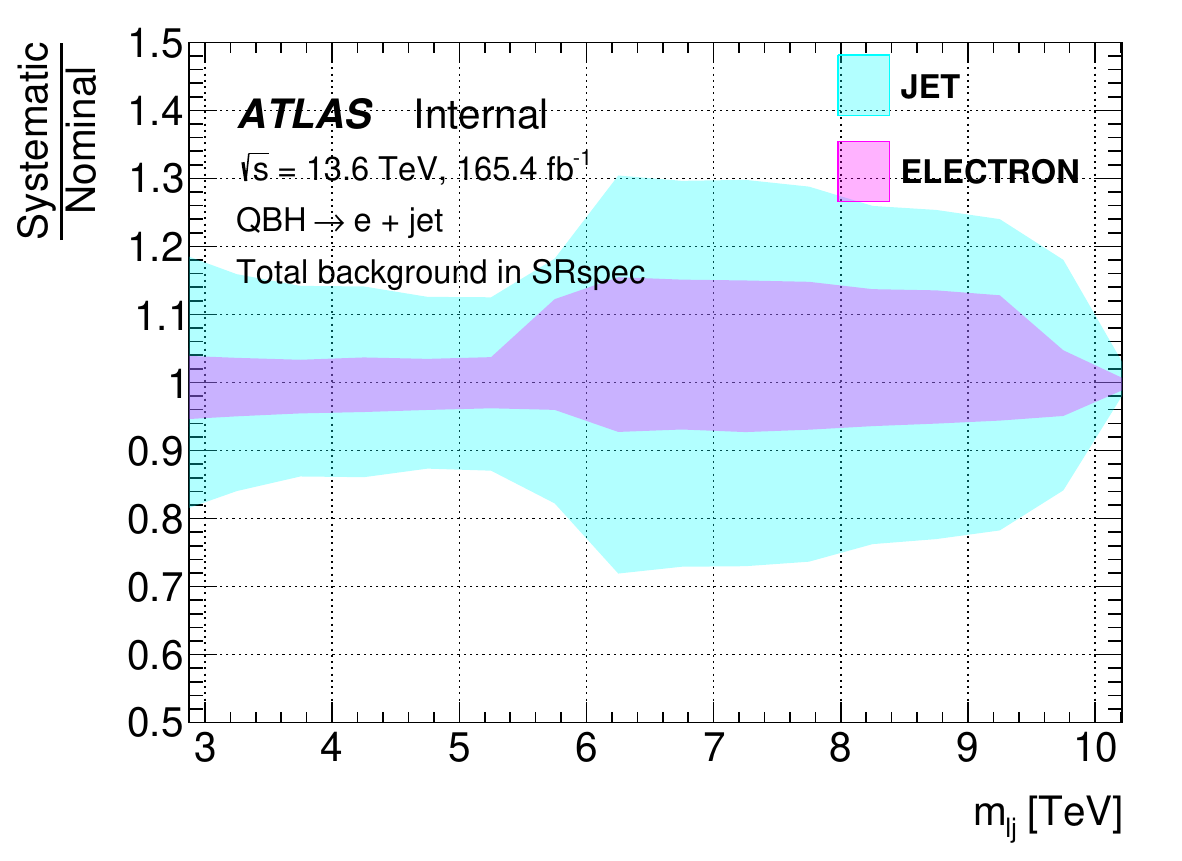}
  }
  \hfill
  \subfloat[(b)]{
    \includegraphics[width=0.5\textwidth]{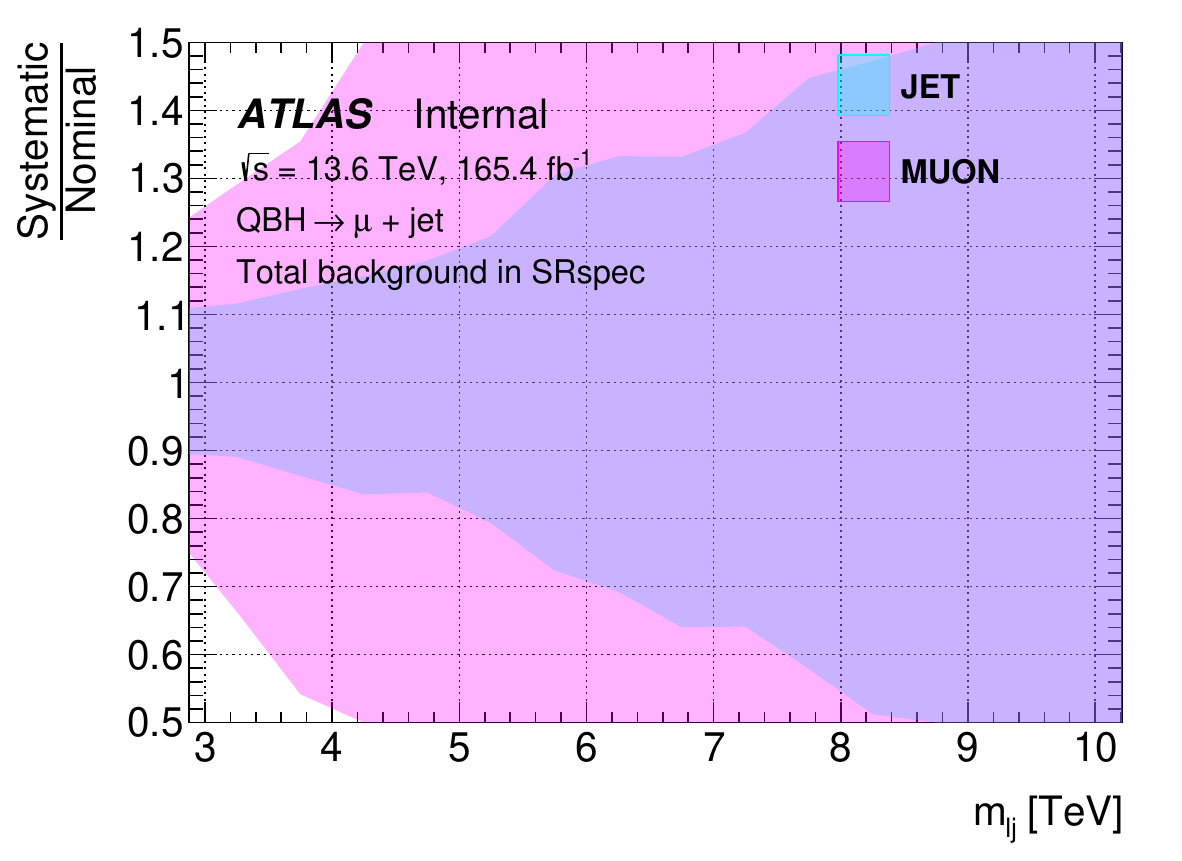}
  }
  \caption{Grouped experimental systematics for total background in the signal region for (a) the $e+j$ channel and (b) the $\mu+j$ channel.}
  \label{fig:SR_Tot_exp_syst}
\end{figure}

FIG.~\ref{fig:SR_Tot_exp_syst} shows the grouped experimental systematic uncertainties per physics object for the total background in both decay channels. The dominant contribution in both channels is from jet calibration uncertainties, growing with $m_{\ell j}$. Detailed per-object breakdowns are provided in Appendix~\ref{app:qbh:exp_syst_plots}.

\subsection[$W$+Jets Control and Validation Regions]{$\bm{W}$+Jets Control and Validation Regions}

\begin{figure}[h]
  \captionsetup[subfigure]{labelformat=empty}
  \subfloat[(a)]{
    \includegraphics[width=0.5\textwidth]{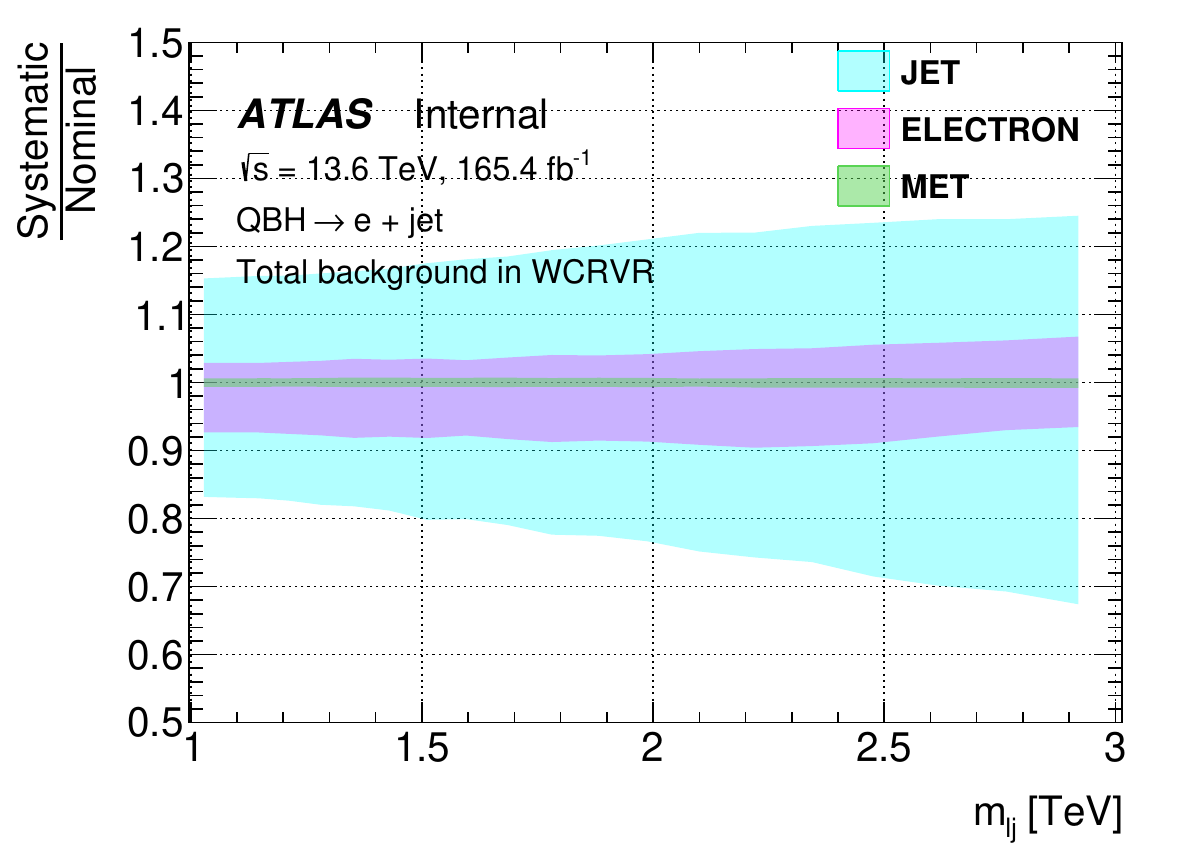}
  }
  \hfill
  \subfloat[(b)]{
    \includegraphics[width=0.5\textwidth]{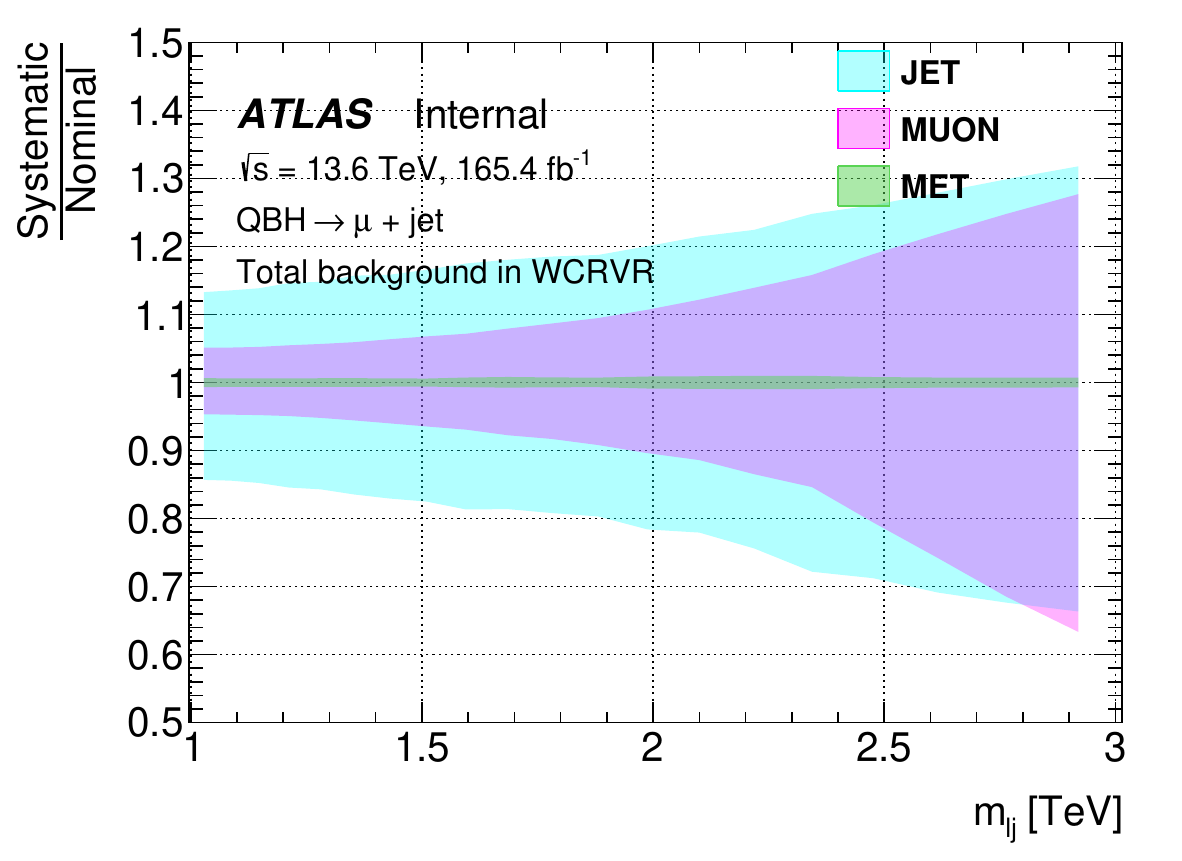}
  }
  \caption{Grouped experimental systematics for total background in the $W$+jets control and validation regions for (a) the $e+j$ channel and (b) the $\mu+j$ channel.}
  \label{fig:WCRVR_Tot_exp_syst}
\end{figure}

The cuts used for the $W$+jets CR and VR, defined in Section~\ref{sec:EvSel}, are on the properties of the jet, respective lepton and the \MET\ significance. FIG.~\ref{fig:WCRVR_Tot_exp_syst} shows the grouped experimental systematic uncertainties for both channels; the dominant contribution is again from jet calibration objects. Detailed per-object breakdowns are shown in Appendix~\ref{app:qbh:exp_syst_plots}.

\subsubsection*{$Z$+Jets Control and Validation Regions}

\begin{figure}[h]
  \captionsetup[subfigure]{labelformat=empty}
  \subfloat[(a)]{
    \includegraphics[width=0.5\textwidth]{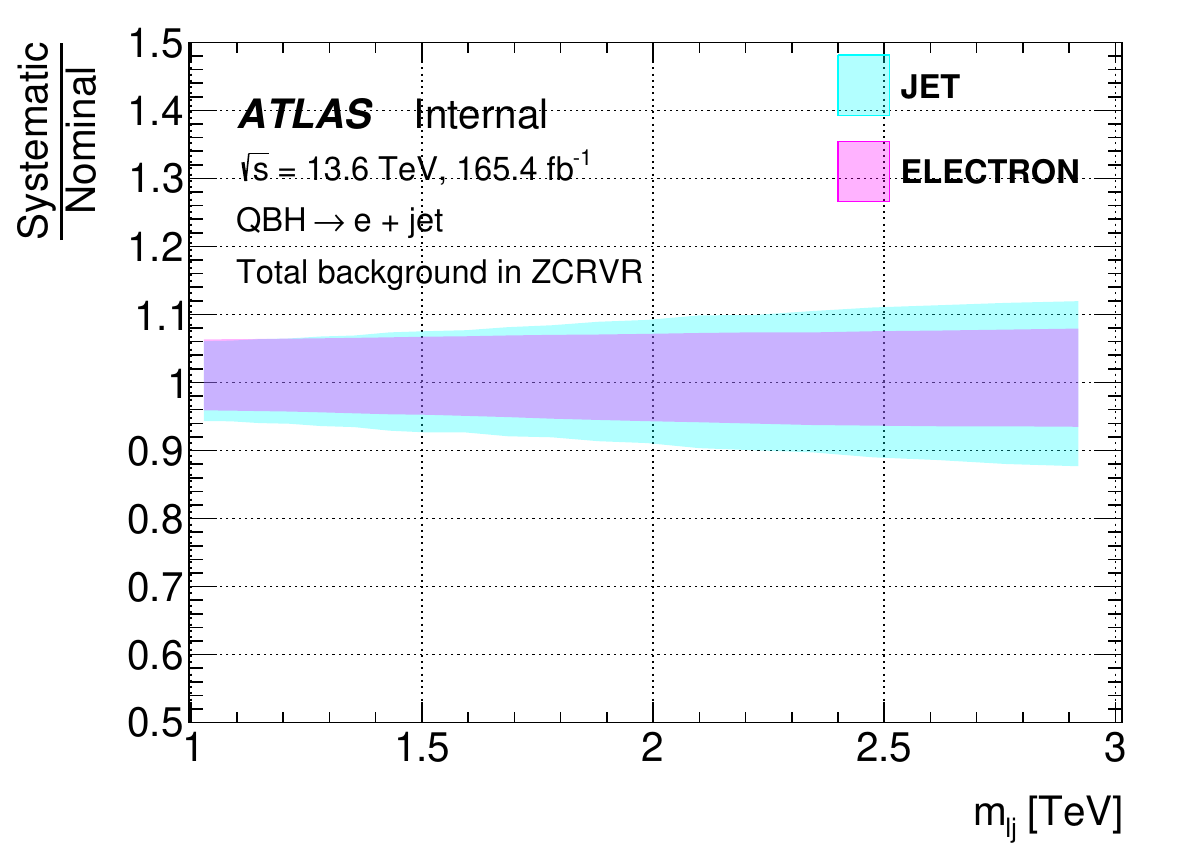}
  }
  \hfill
  \subfloat[(b)]{
    \includegraphics[width=0.5\textwidth]{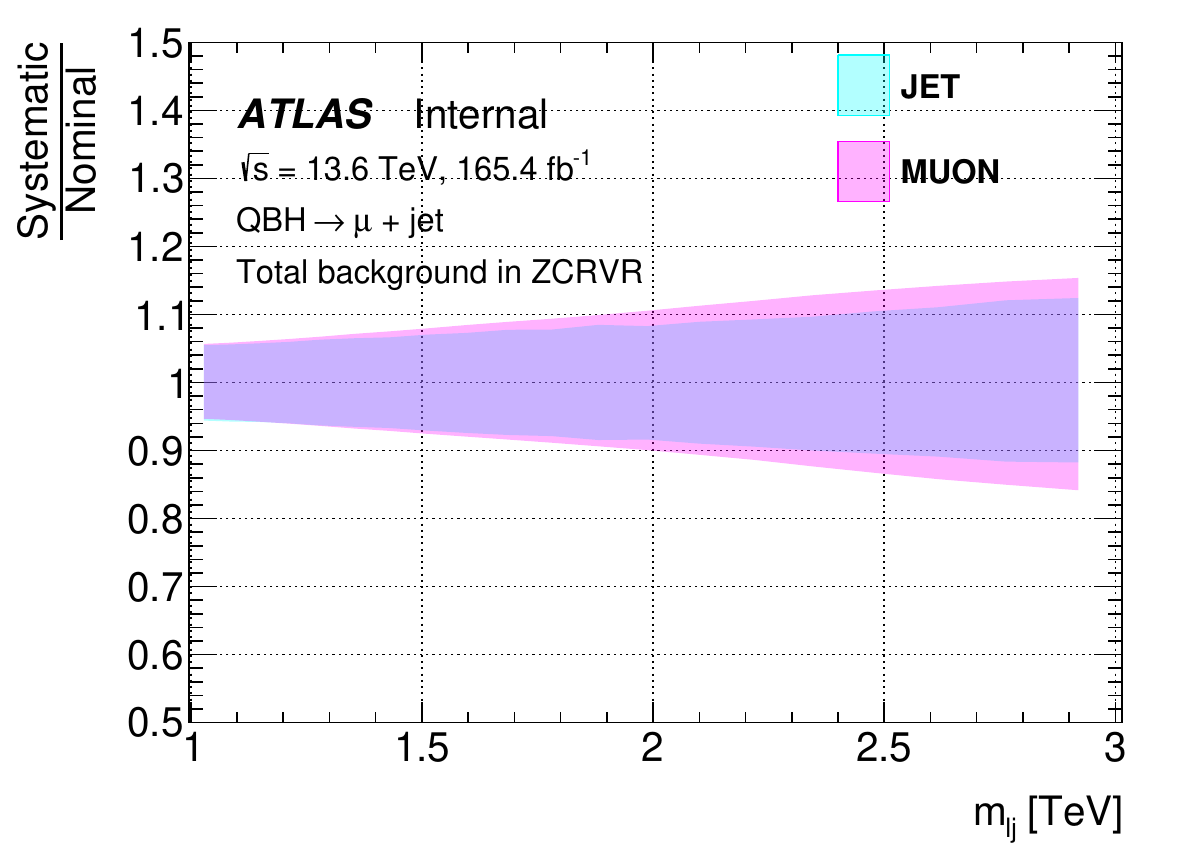}
  }
  \caption{Grouped experimental systematics for total background in the $Z$+jets control and validation regions for (a) the $e+j$ channel and (b) the $\mu+j$ channel.}
  \label{fig:ZCRVR_Tot_exp_syst}
\end{figure}

Similarly to the signal region, the $Z$+Jets CR and VR only cut on properties of the jet and respective lepton. FIG.~\ref{fig:ZCRVR_Tot_exp_syst} shows the grouped systematics; all contributions remain below 20\% in $m_{\ell j}$ for both objects. Detailed per-object breakdowns are shown in Appendix~\ref{app:qbh:exp_syst_plots}.

\clearpage

\subsection{Theoretical}
\label{sec:Syst_th}

Theoretical systematic uncertainties arise from the modelling of background samples. These are estimated for the leading $V$+jets and sub-leading top ($t\bar{t}$ and single top) backgrounds, following standard ATLAS PMG recipes. There are three common variations --- scale, PDF and $\alpha_{\textrm{S}}$ --- which are evaluated from on-the-fly weights available from the MC generator of the leading backgrounds, Sherpa. The theory variations considered are summarised in Table~\ref{tab:syst:theory_overview} in Appendix~\ref{app:qbh:exp_systs}. In the fit, nuisance parameters representing theoretical uncertainties are assigned individually per process, $W$+jets and $Z$+jets.

FIG.~\ref{fig:SR_TH_syst}, \ref{fig:WCRVR_TH_syst} \& \ref{fig:ZCRVR_TH_syst} show for all regions the contribution to the uncertainty from the top theoretical uncertainties is negligible. Therefore, the top uncertainties are not considered in the fit.

\textbf{Scale Uncertainties}:

The two scales that are varied are $\mu_R$, the renormalisation scale, and $\mu_F$, the factorisation scale.  They are varied in the following combinations:
\begin{itemize}
    \setlength\itemsep{-0.5em}
    \item $\mu_{\textrm{R}},\mu_{\textrm{F}}=$ 0.5, 0.5
    \item  $\mu_{\textrm{R}},\mu_{\textrm{F}}=$ 0.5, 1.0
    \item $\mu_{\textrm{R}},\mu_{\textrm{F}}=$ 1.0, 0.5
    \item $\mu_{\textrm{R}},\mu_{\textrm{F}}=$ 1.0, 1.0 (nominal)
    \item $\mu_{\textrm{R}},\mu_{\textrm{F}}=$1.0, 2.0
    \item $\mu_{\textrm{R}},\mu_{\textrm{F}}=$ 2.0, 1.0
    \item $\mu_{\textrm{R}},\mu_{\textrm{F}}=$ 2.0, 2.0
\end{itemize}

These values are varied for the nominal PDF sets PDF303200 for $V$+jets and PDF260000 for top. Individual events are re-weighted with these new scales and the new value per bin is calculated. The ratio to nominal is calculated for each as described in Section~\ref{sec:Syst_exp}. The total of this systematic is calculated by taking the maximum and minimum of each bin.

\textbf{PDF Uncertainties:}

The uncertainties depend on the choice of PDF set, therefore 100 PDF sets (with  $\mu_{\textrm{R}},\mu_{\textrm{F}}=$ 1.0, 1.0) are used, PDF303201--303300 for $V$+jets. These are then combined by taking the standard deviation of the sets:

\begin{equation}
	\Delta X = \sqrt{\frac{1}{N} \sum_i (X_i - X_{\textrm{nom}})^2},
\end{equation}
where $X$ represents the content per bin, $N$ the number of PDF sets used, $i$ referring to the $i$-th PDF set and $\textrm{nom}$ the nominal.

These uncertainities are considered in the fit to affect the shape only, such that the normalisation is fixed to that of the nominal, across all analysis regions.

\textbf{Strong Coupling Constant:}

$\alpha_S$ is the strong coupling constant. The uncertainty of $\alpha_S$ for $V$+jets is calculated by evaluating the PDF set at two values, 0.117 and 0.119, and taking the average. Top $\alpha_S$ variations (ISR/FSR) are found to be negligible and are not included in the fit.

\subsubsection*{Signal Region}

\begin{figure}[h]
  \captionsetup[subfigure]{labelformat=empty}
  \subfloat[(a)]{
    \includegraphics[width=0.5\textwidth]{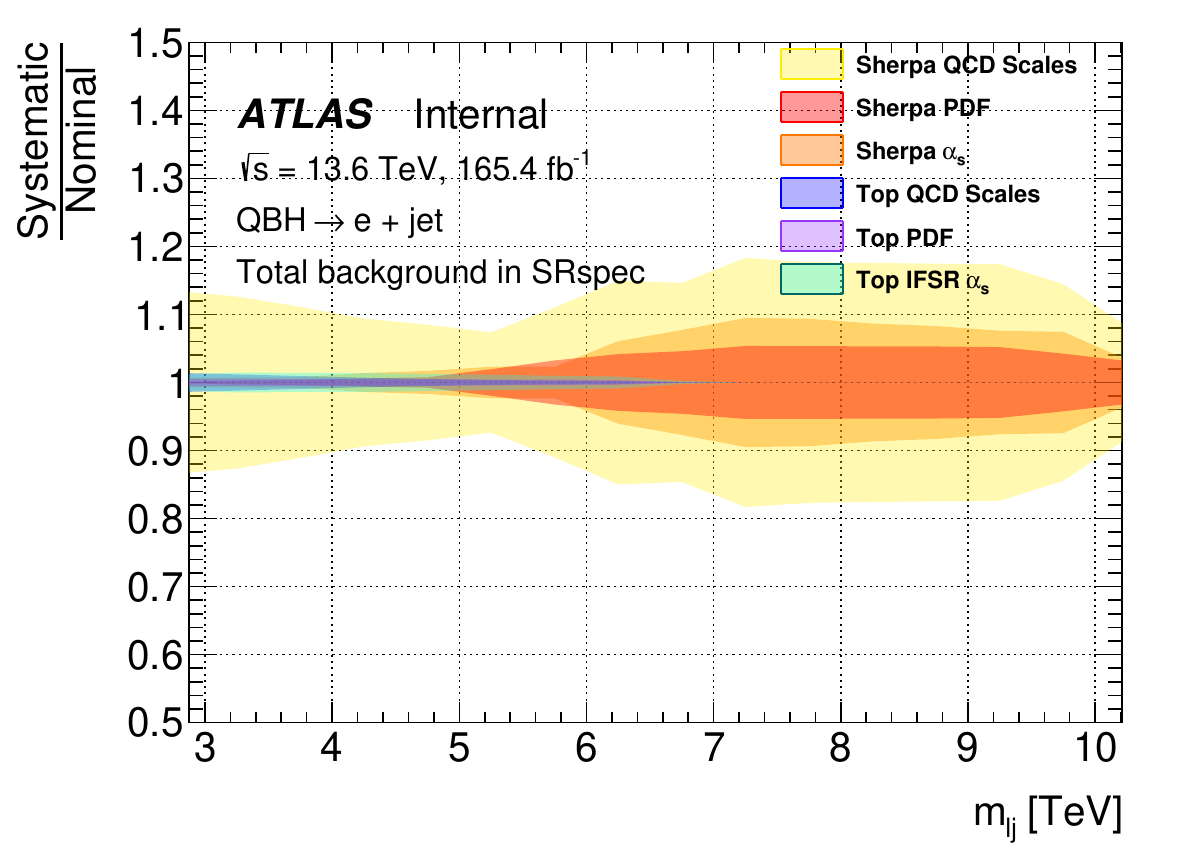}
  }
  \hfill
  \subfloat[(b)]{
    \includegraphics[width=0.5\textwidth]{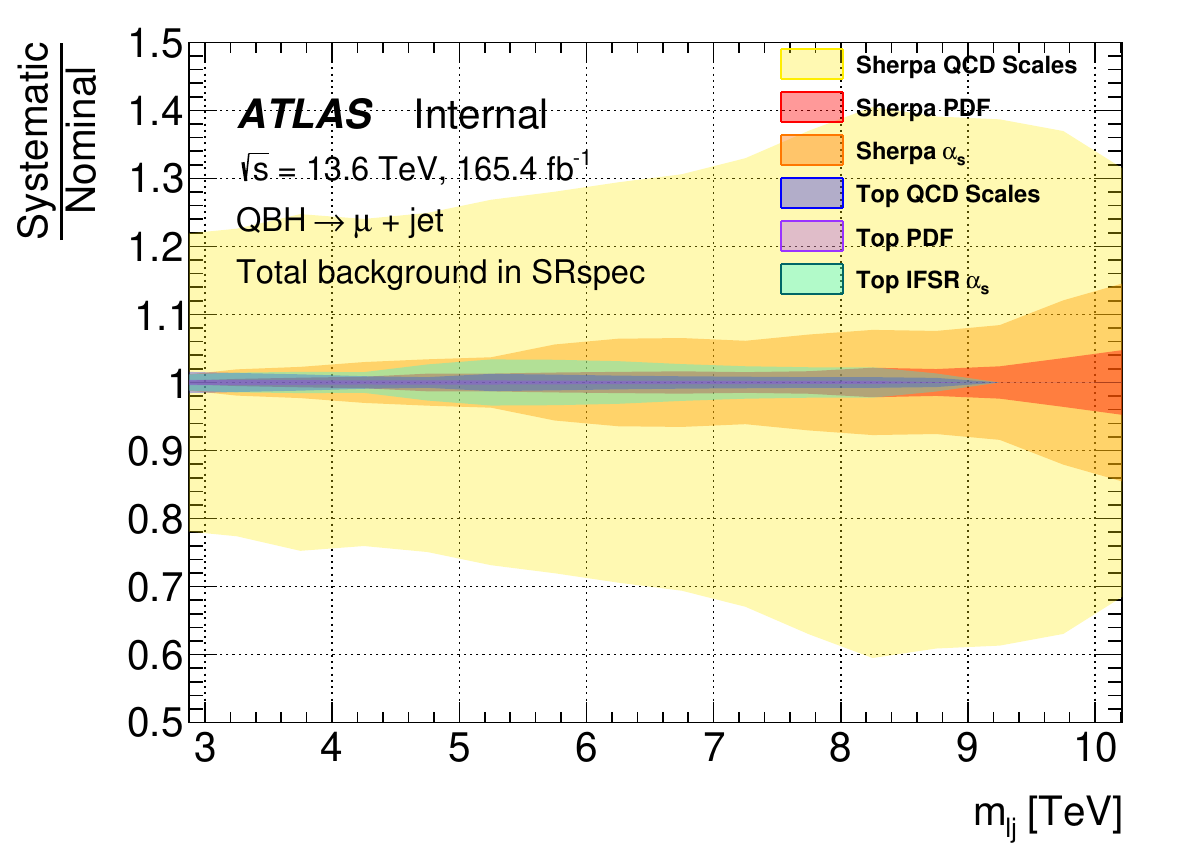}
  }
  \caption{Grouped theoretical uncertainties for total background in the signal region for (a) the $e+j$ channel and (b) the $\mu+j$ channel.}
  \label{fig:SR_TH_syst}
\end{figure}

The signal region used for both channels is defined in Section~\ref{sec:EvSel}. FIG.~\ref{fig:SR_TH_syst} shows the contributions to the theoretical systematic uncertainties of the total background  in the signal region. For both channels the leading contribution is from Sherpa QCD scales, and sub leading from other Sherpa uncertainties, $\alpha_s$ and PDF.

\subsection[\texorpdfstring{$W$+Jets Control and Validation Regions}{W+Jets Control and Validation Regions}]{$W$+Jets Control and Validation Regions}

\begin{figure}[h]
  \captionsetup[subfigure]{labelformat=empty}
  \subfloat[(a)]{
    \includegraphics[width=0.5\textwidth]{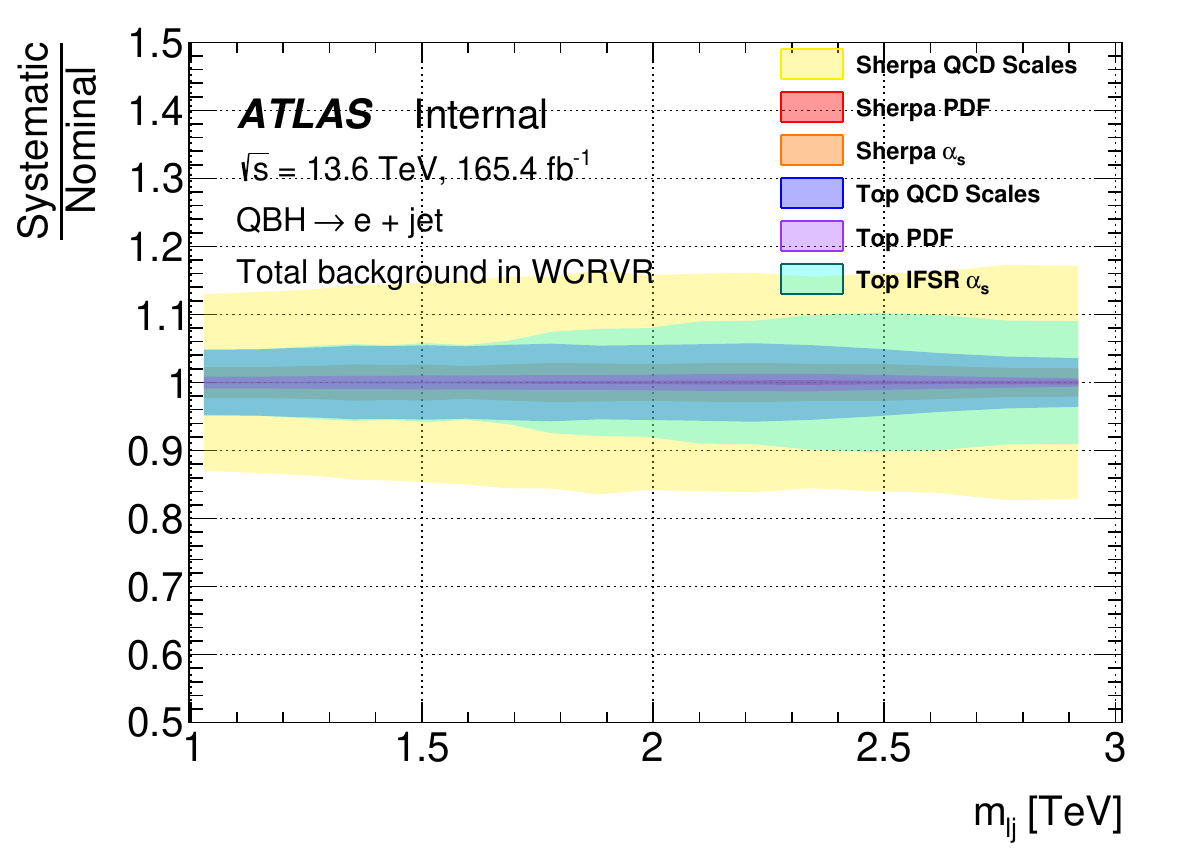}
  }
  \hfill
  \subfloat[(b)]{
    \includegraphics[width=0.5\textwidth]{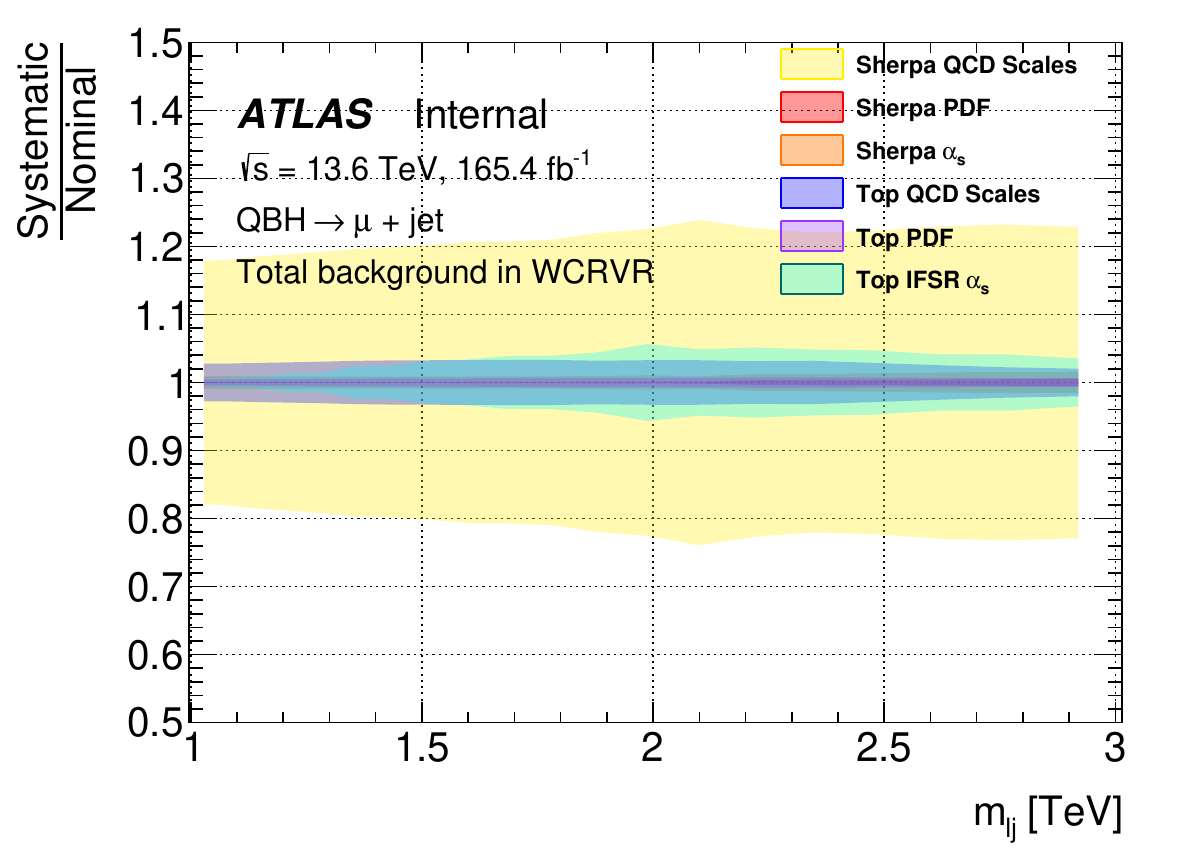}
  }
  \caption{Grouped theoretical uncertainties for total background in the $W$+jets control and validation regions for (a) the $e+j$ channel and (b) the $\mu+j$ channel.}
  \label{fig:WCRVR_TH_syst}
\end{figure}

FIG.~\ref{fig:WCRVR_TH_syst} shows the contributions to the theoretical systematic uncertainties of the total background  in the $W$+jets control and validation regions. For both channels the leading contribution is again from Sherpa QCD scales, and sub leading from from top QCD scales (including ISR/FSR).

\subsection[\texorpdfstring{$Z$+Jets Control and Validation Regions}{Z+Jets Control and Validation Regions}]{$Z$+Jets Control and Validation Regions}

\begin{figure}[h]
  \captionsetup[subfigure]{labelformat=empty}
  \subfloat[(a)]{
    \includegraphics[width=0.5\textwidth]{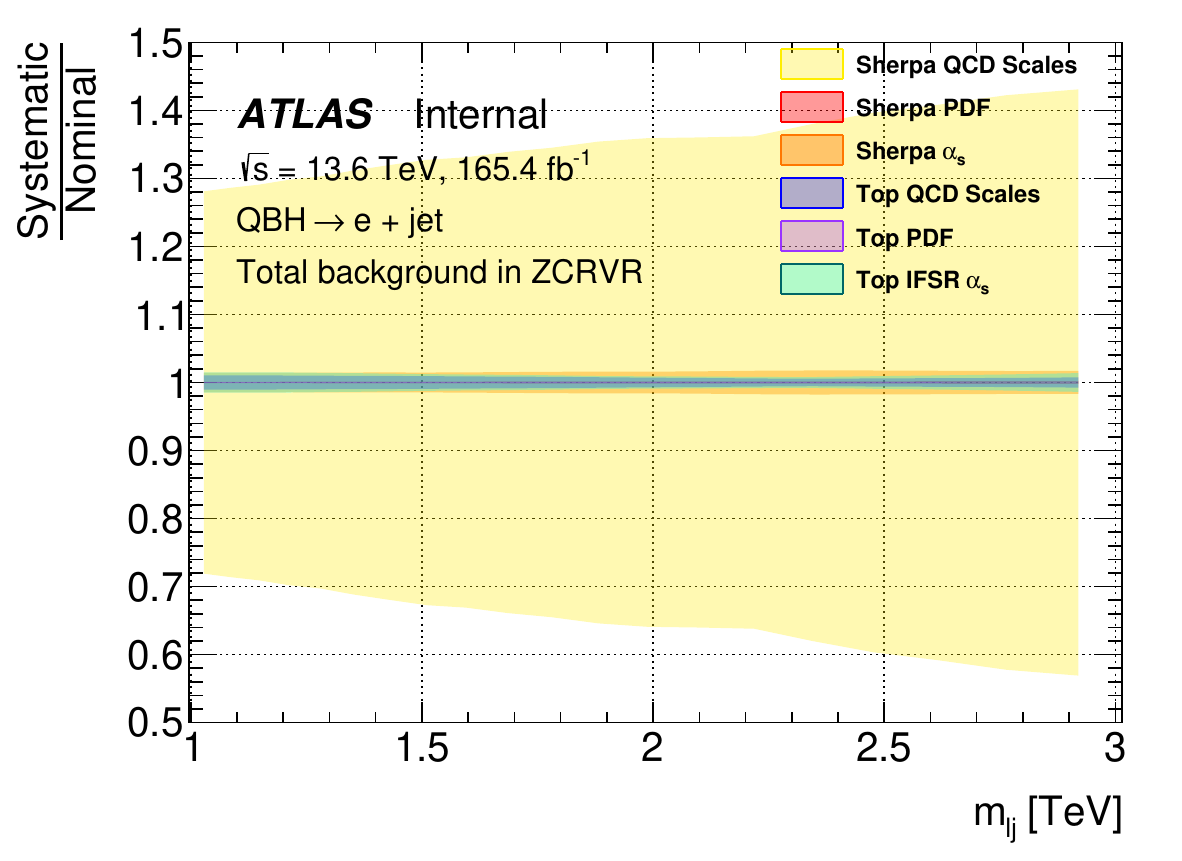}
  }
  \hfill
  \subfloat[(b)]{
    \includegraphics[width=0.5\textwidth]{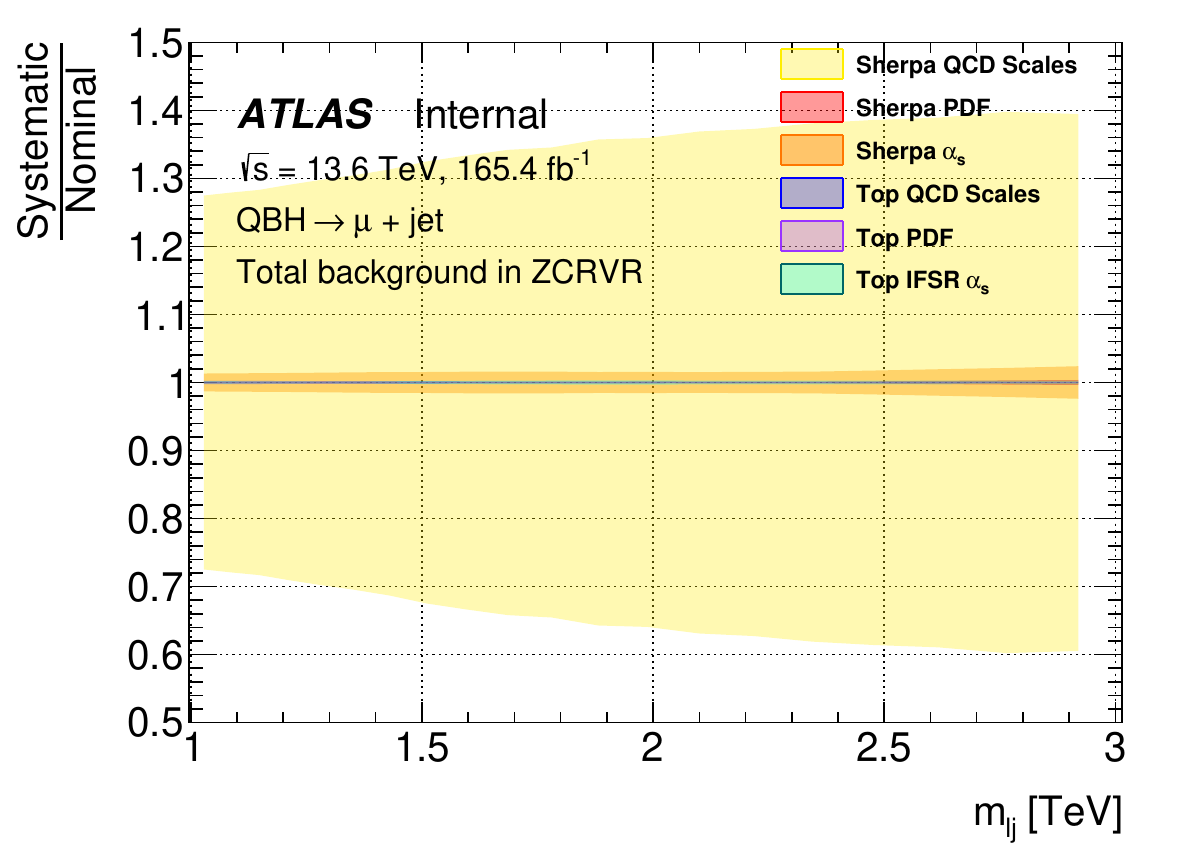}
  }
  \caption{Grouped theoretical uncertainties for total background in the $Z$+jets CR and VR for (a) the $e+j$ channel and (b) the $\mu+j$ channel.}
  \label{fig:ZCRVR_TH_syst}
\end{figure}

FIG.~\ref{fig:ZCRVR_TH_syst} shows the contributions to the theoretical systematic uncertainties of the total background  in the $Z$+jets control and validation regions. Again the leading contribution is from Sherpa QCD scales, which is higher than for other regions, at over 30\% for all values of $m_{lj}$.

\section{Pulls and constraints}

The nuisance parameter (NP) pull-and-constraint plot shows, for each systematic uncertainty included in the fit, its post-fit central value (the ``pull'', expressed in units of its pre-fit uncertainty) and its post-fit uncertainty relative to the pre-fit uncertainty (the ``constraint''). A pull near zero means the data is consistent with the prior on that uncertainty; a constraint significantly below unity means the data has restricted the uncertainty beyond its prior. These plots for both channels are shown in Appendix~\ref{app:qbh:fit_diagnostics} (FIG.~\ref{fig:Pulls_Constraints}). Overall the fit behaves well: most NPs are unconstrained and show no significant pull. A few entries are mildly constrained, and the muon momentum resolution (\texttt{MUON\_CB}) is tightly constrained, consistent with the precision of the combined inner-detector and muon-spectrometer measurement at high \pT.

\subsection{Correlations}
The nuisance parameter correlation matrix, shown in Appendix~\ref{app:qbh:fit_diagnostics} (FIG.~\ref{fig:Correlations}), illustrates the degree to which pairs of systematic uncertainties are correlated with each other and with the parameters of interest in the fit. Large off-diagonal entries would signal degeneracies or shared sensitivities among systematic sources, which would indicate that one systematic is absorbing the effect of another. The matrices shown here are dominated by diagonal entries, confirming that the systematic uncertainties are largely independent.

\subsection{Nuisance parameter impacts}

The impact of nuisance parameters associated with the systematic uncertainties described hitherto is evaluated and ranked. A post-fit error of $\pm 1$ standard deviation is placed on each nuisance parameter, $\theta_{i}\in \boldsymbol{\theta}$. The likelihood function described in Chapter~\ref{chp:stats} (Section~\ref{ssec:stats:profile}), $\mathcal{L}(\mu, \boldsymbol{\theta})$ ($\mu$ is the signal strength in this very specific context, not to be confused with $\mu$ for nuisance parameter impact) is profiled with this fixed nuisance parameter to assess its impact on the Parameter of Interest (PoI), which is the signal strength, $\mu$, in this analysis. Taking all nuisance parameters to have a Gaussian PDF, the values of $\Delta\mu/\mu$ thus obtained constitute the correlation coefficients between the PoI and the nuisance parameter.

In other words, the impact $\Delta \mu_{i}$ of a nuisance parameter $\theta_{i}$ on the PoI is given by the shift in PoI between the nominal fit and a subsequent fit where $\theta_{i}$ is fixed to $\hat{\theta}_{i}\pm x$, where $\hat{\theta}_{i}$ is the maximum-likelihood estimator of $\theta_{i}$ \textrm{i.e.} its post-fit value. Now, $x=\Delta \theta_{i}=1$ induces the pre-fit impact, $\Delta \mu$, encoding an uncertainty on $\theta_{i}$ which shifts the likelihood by $\pm 1$ standard deviation. $\theta_{0}$ is the pre-fit value of $\theta_{i}$. Similarly, $x=\Delta \hat{\theta}_{i}\leq 1$ gives the post-fit impact, where $\Delta \hat{\theta}_{i}$ is the uncertainty on $\hat{\theta}_{i}$. Since $\theta_{i}$ can be constrained, the $\Delta \hat{\theta}_{i}$ may be smaller than $\Delta \theta_{i}$.

\begin{figure}[h]
  \captionsetup[subfigure]{labelformat=empty}
    \subfloat[(a)]{
      \includegraphics[width=0.5\textwidth]{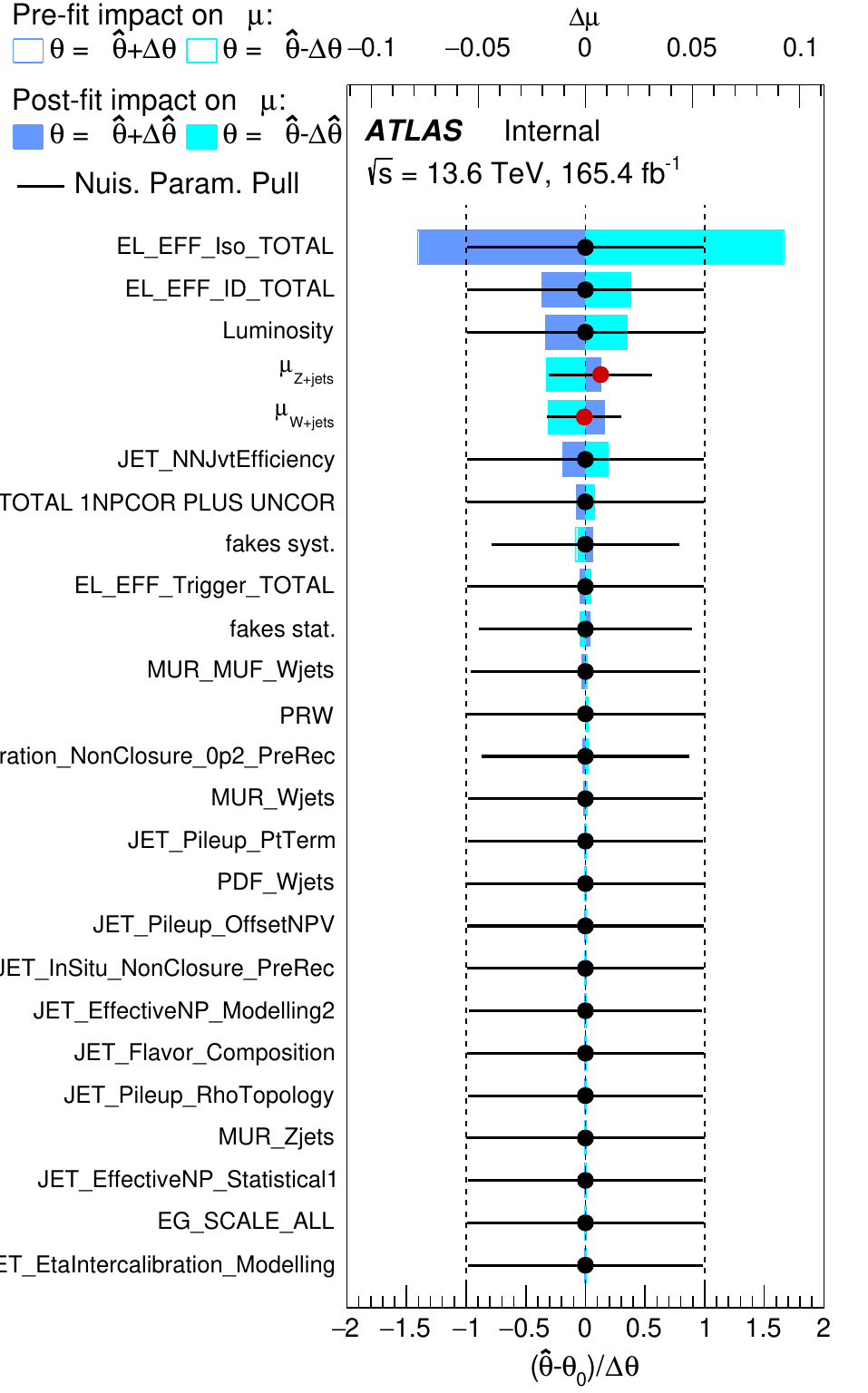}
    }
    \hfill
    \subfloat[(b)]{
      \includegraphics[width=0.5\textwidth]{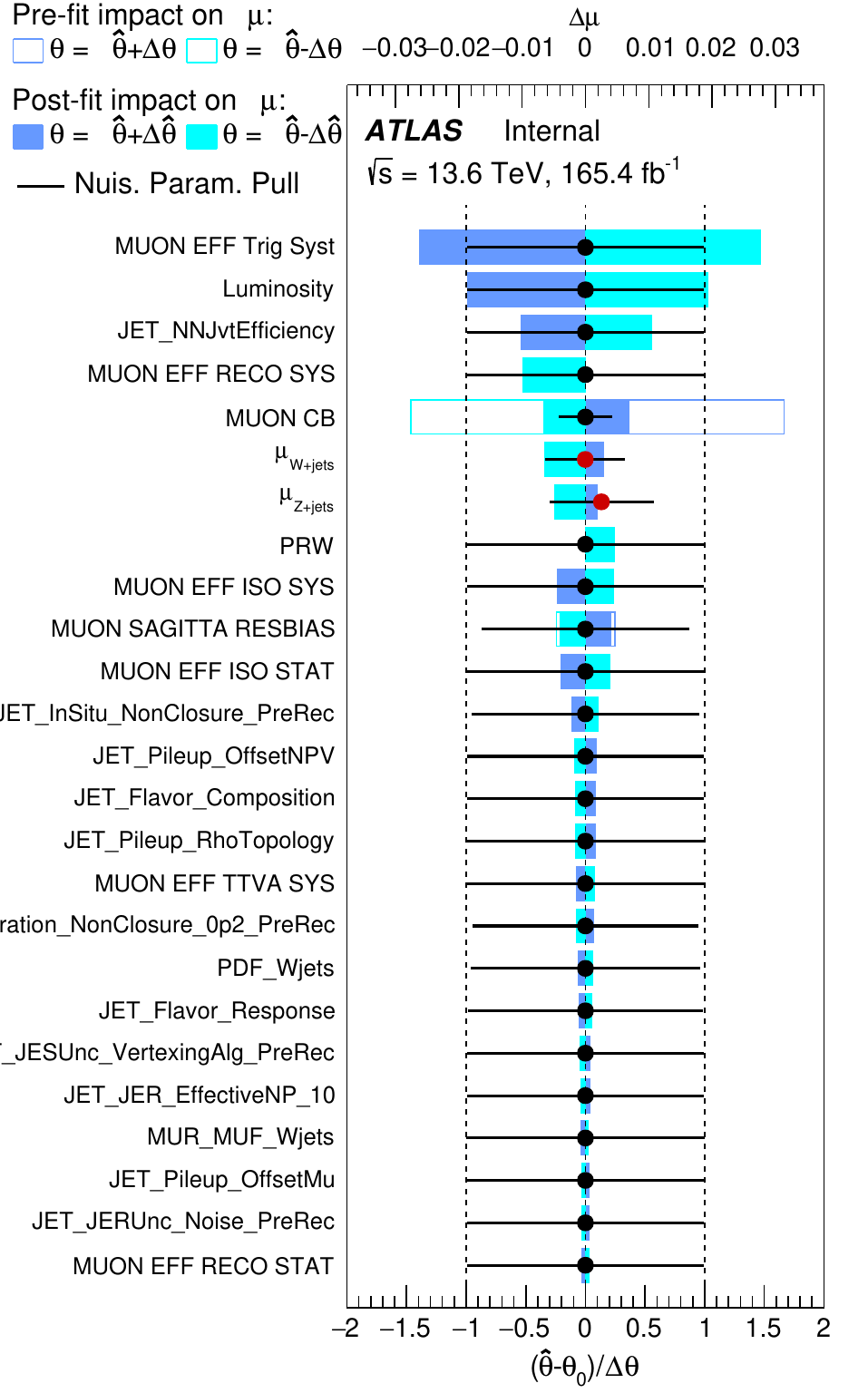}
    }
\caption{Nuisance parameter impacts on the signal strength for (a) the $e+j$ channel and (b) the $\mu+j$ channel. The names of the nuisance parameters shown correspond to the systematic sources listed in Tables~\ref{tab:lep_exp_syst} and~\ref{tab:jet_exp_syst} of Appendix~\ref{app:qbh:exp_systs}.}
\label{fig:Rankings}
\end{figure}

A small positive post-fit pull of $\mu_W$ and $\mu_Z$ is not unexpected.
In this fit configuration, the background predictions in the CRs and SR tend to slightly undershoot the observed data in regions dominated by $W{+}$jets and $Z{+}$jets.
Since $\mu_W$ and $\mu_Z$ act as global normalisation factors for these processes, the fit compensates by increasing them mildly to better reproduce the event yields.
Because the two processes share systematic uncertainties and populate partially overlapping kinematic regions, a coherent upward pull in both parameters is a natural outcome of the fit as it absorbs small data--MC differences. This is observed in FIG.~\ref{fig:Rankings}, especially for $\mu_Z$.

\section{Results}
\label{sec:results}
\subsection{Signal region}
\begin{figure}
    \subfloat{
      \includegraphics[width=0.5\textwidth]{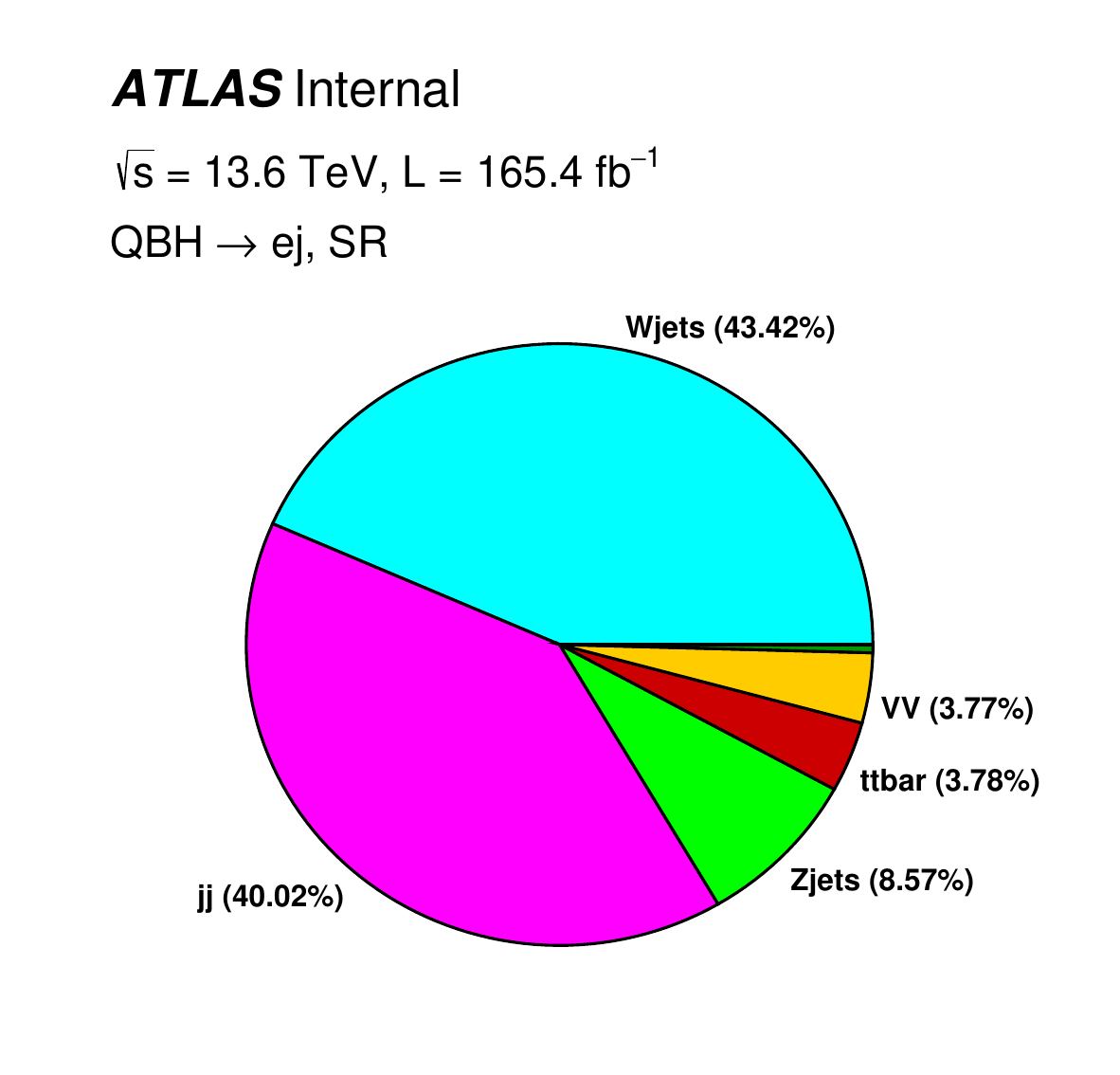}
    }
    \subfloat{
      \includegraphics[width=0.5\textwidth]{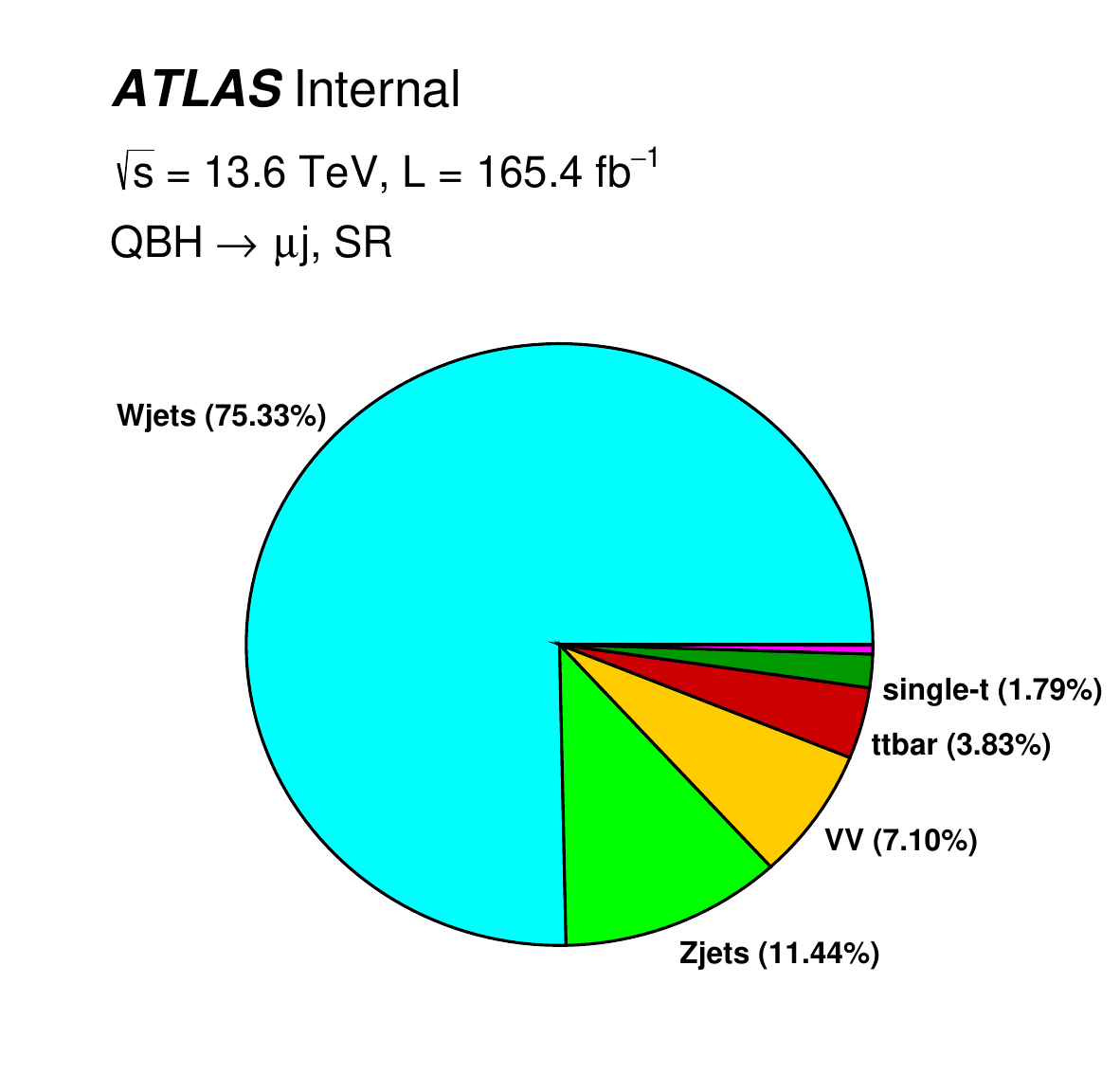}
    }
\caption{Background contributions pie plots in the signal region in $e+j$ channel (left) and $\mu+j$ channel (right).}
\label{fig:Pie_Charts_SR}
\end{figure}

The SR is defined to maximise sensitivity by reducing background contributions whilst maintaining high signal efficiency. It follows the analysis strategy and event selection criteria as defined in Chapter~\ref{sec:EvSel}. The dominant backgrounds in this region are fake-electrons and $W$+jets; however, due to the extremely high $m_{\ell j}$ phase-space considered, most bins have negligible background contamination.
The pie-charts given in FIG.~\ref{fig:Pie_Charts_SR} illustrate the contribution of different background sources to the SR.

The signal-plus-background model is used to fit the observed $m_{\ell j}$ spectra simultaneously in the SR, $W$CR and $Z$CR using the profile-likelihood method~\cite{Cowan:2010js} (see Section~\ref{ssec:stats:profile}).
The expected and observed event count in each $m_{\ell j}$ bin is described as a Poisson-distributed variable, where the systematic uncertainties are accounted for via nuisance parameters.
The sensitivity of the analysis is driven by the highest $m_{\ell j}$ bin of the SR, 5--13~TeV; contributions above 13~TeV are accounted for in this last bin.
In that bin, the number of background events in the electron channel is predicted to be $1.46\pm 0.65_{\textrm{stat.}}\pm 0.38_{\textrm{syst.}}$, where the leading systematic uncertainty is on the fakes estimation.
In the muon channel, the number of background events in the highest bin is predicted to be $3.10\pm 0.95_{\textrm{stat.}}\pm 0.47_{\textrm{syst.}}$, where the leading systematic uncertainty is due to the modelling of the $W$+jets background.
Normalisation factors for the leading MC backgrounds, $W$+jets and $Z$+jets, are determined from the simultaneous fit and discussed in Section~\ref{sec:background_modelling}.

\begin{figure}[htbp]
  \centering
  \includegraphics[width=\textwidth]{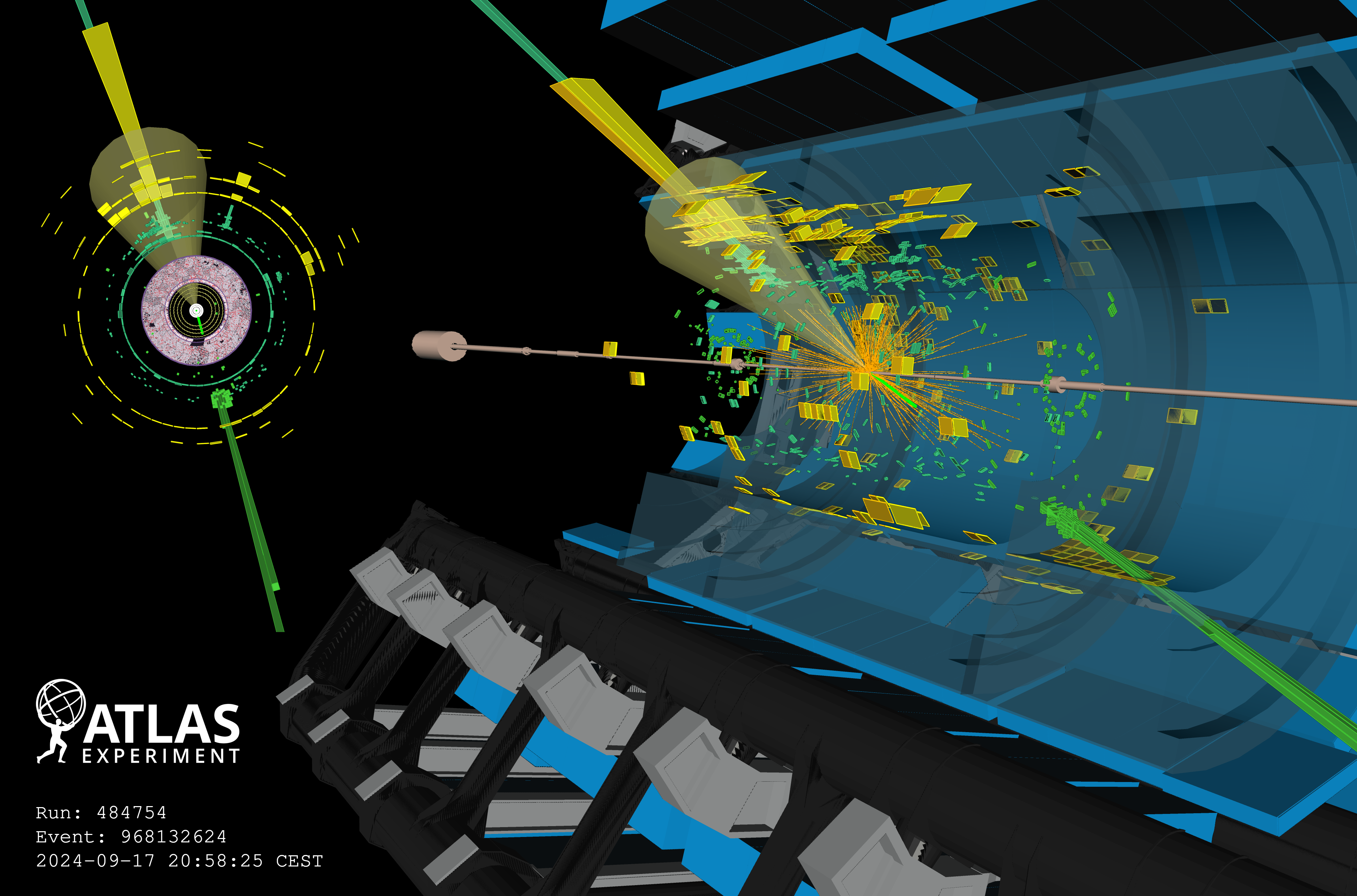}
  \caption{Event data visualisation of a $pp$ collision at $\sqrt{s}=13.6$~TeV recorded by the ATLAS detector on 17 September 2024 (Run~484754, Event~968132624). A high-$\pT$ electron is produced back-to-back with a hard small-$R$ jet. The main panel shows a cut-out 3D view of the detector; Inner Detector tracks with $\pT > 1.2$~GeV are shown as orange lines, LAr and Tile calorimeter deposits above 200~MeV as green/cyan and yellow/orange boxes, the reconstructed jet as a semi-transparent yellow cone, and the electron track and associated EM calorimeter cluster in green/teal. The inlay (left) shows the transverse projection. This event is the highest-mass candidate in the electron+jet channel, with $m_{ej} = 5.3$~TeV; the electron has $(\pT,\,\eta,\,\phi) = (1.2~\textrm{TeV},\,1.55,\,-1.31)$ and the jet has $(\pT,\,\eta,\,\phi) = (1.2~\textrm{TeV},\,-1.29,\,1.92)$, with $\Delta\phi(e,j)=3.05$ and $\Delta\eta(e,j)=2.84$.}
  \label{fig:vp1_ej}
\end{figure}

\begin{figure}[htbp]
  \centering
  \includegraphics[width=\textwidth]{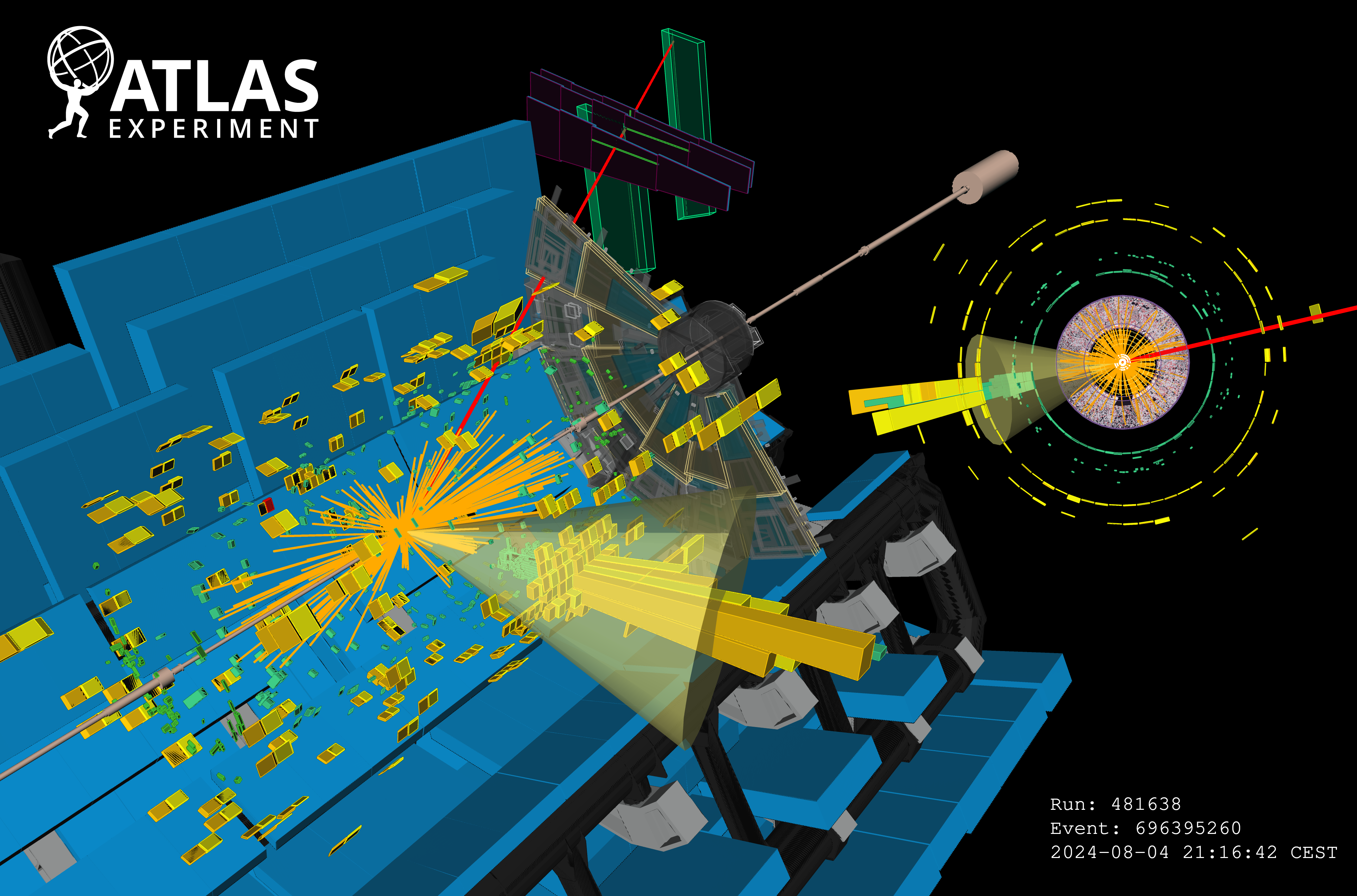}
  \caption{Event data visualisation of a $pp$ collision at $\sqrt{s}=13.6$~TeV recorded by the ATLAS detector on 4 August 2024 (Run~481638, Event~696395260). A high-$\pT$ muon is produced back-to-back with a hard small-$R$ jet. The main panel shows a cut-out 3D view of the detector; Inner Detector tracks with $\pT > 1.2$~GeV are shown as orange lines, calorimeter deposits above 200~MeV as green/cyan and yellow/orange boxes, the jet as a yellow cone, and the muon track as a red line with associated NSW (grey), TGC, and MDT chamber hits (purple and green). The inlay (right) shows the transverse projection. This event is the highest-mass candidate in the muon+jet channel, with $m_{\mu j} = 5.5$~TeV; the muon has $(\pT,\,\eta,\,\phi) = (5.2~\textrm{TeV},\,1.33,\,0.23)$ and the jet has $(\pT,\,\eta,\,\phi) = (1.2~\textrm{TeV},\,0.39,\,-2.94)$, with $\Delta\phi(\mu,j)=3.11$ and $\Delta\eta(\mu,j)=0.94$.}
  \label{fig:vp1_muj}
\end{figure}

\begin{figure}[htbp]
\subfloat[]{
\includegraphics[width=0.5\textwidth]{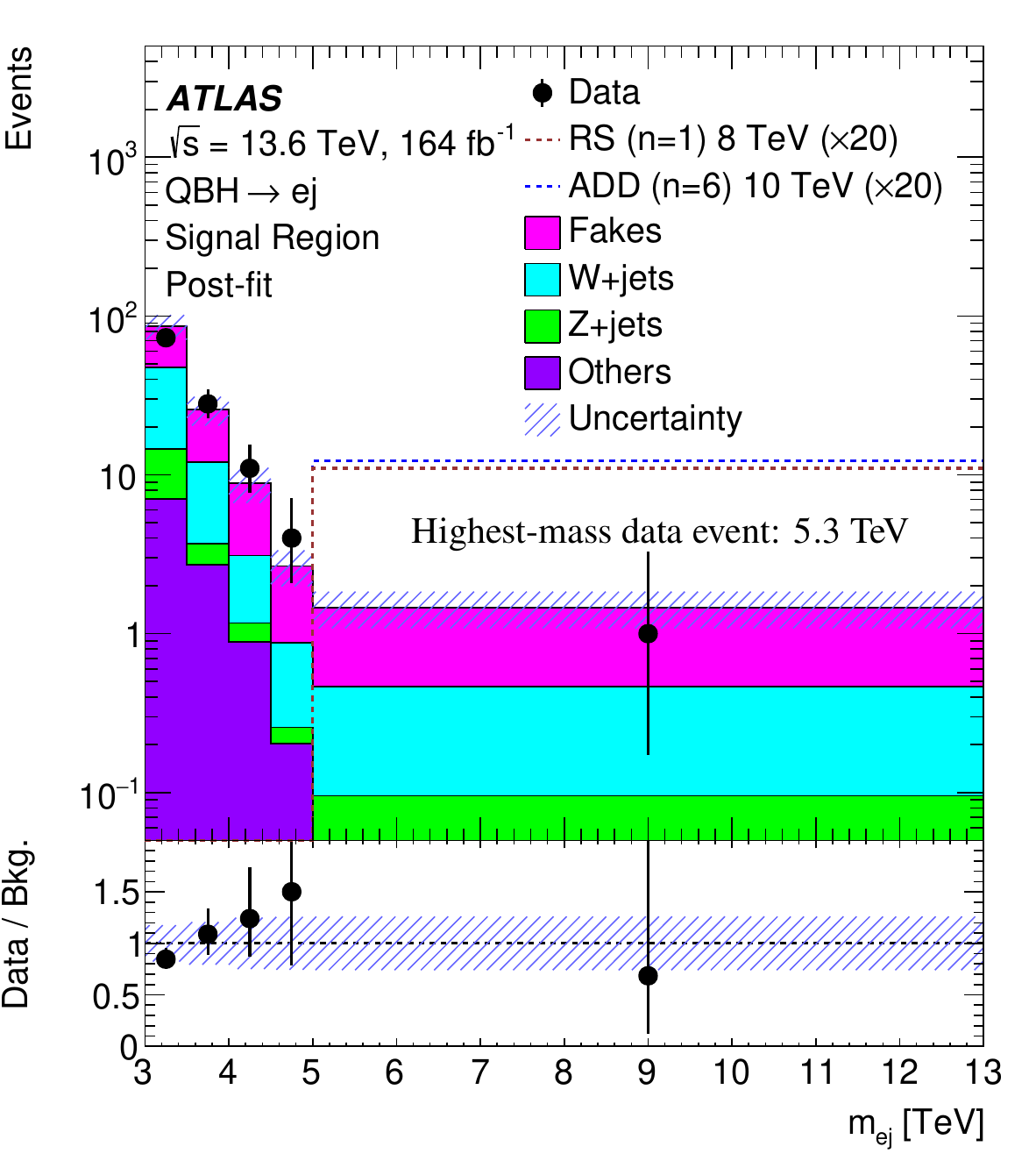}
}
\subfloat[]{
\includegraphics[width=0.5\textwidth]{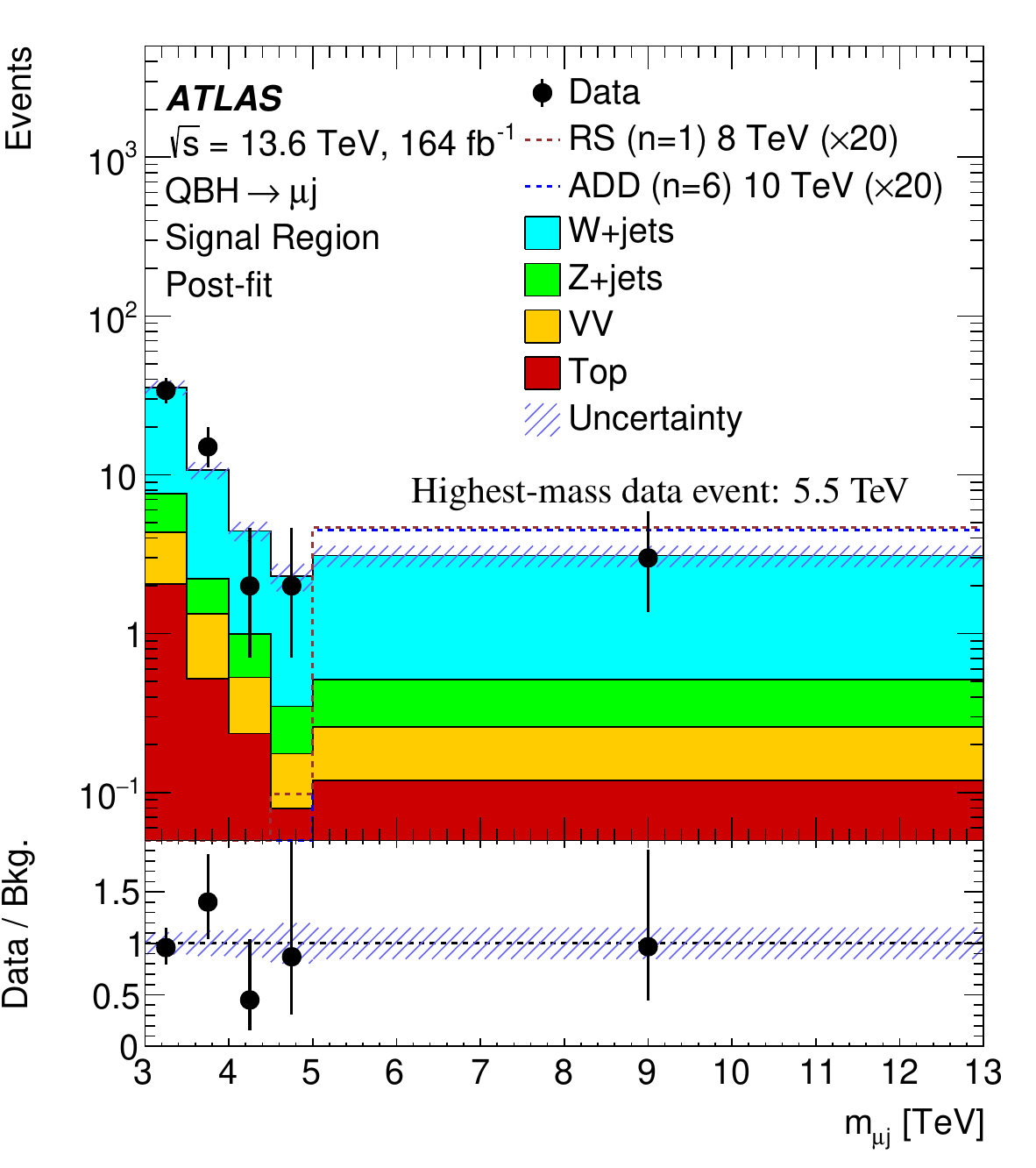}
}
\caption{Distribution of the lepton+jet invariant mass for events passing the full signal selection for (a) the electron channel and (b) the muon channel. Benchmark signals for the RS and ADD models with $n=1,6$ extra dimensions at $M_{\textrm{th}}=8,10~\textrm{TeV}$, respectively, are scaled by a factor of 20 for presentation purposes and overlaid on the total background estimate. The uncertainty band includes the statistical and systematic uncertainties. The highest-mass data events observed in the last bin have $m_{\ell j}$ values of 5.3~(5.0, 5.1, 5.5)~TeV in the electron (muon) channel, and are plotted at the bin's centre. In (a), ``Others'' denotes subdominant backgrounds including top and diboson processes. The MC contributions are normalised based on the likelihood fit.}
\label{fig:SR_mLepJet}
\end{figure}

No significant excess is observed above the SM background.
The $m_{\ell j}$ distributions in the SR of the electron and muon channels following the fit are shown in FIG.~\ref{fig:SR_mLepJet}.
Benchmark signal MC corresponding to the RS and ADD models are overlaid for reference.
The signals considered in this search are generally expected to populate the region above 6~TeV, while the background shape falls steeply between 3 and 6~TeV in both channels.
In the electron channel, the highest-mass event observed in data corresponds to $m_{\ell j} = 5.3~\textrm{TeV}$.
This event features an electron with $(p_{\textrm{T}}, \eta) = (1.2~\textrm{TeV}, 1.55)$ and a jet with $(p_{\textrm{T}}, \eta) = (1.2~\textrm{TeV}, -1.29)$.
The objects are nearly back-to-back in the transverse plane with $\Delta\phi(\ell, j) = 3.05$, with a large rapidity separation $\Delta\eta(\ell, j) = 2.84$ and low $\mathcal{S}(\Etmiss)$ at 2.80.
In the muon channel, the event with the highest invariant mass corresponds to $m_{\ell j} = 5.5~\textrm{TeV}$.
This event features a muon with $(p_{\textrm{T}}, \eta) = (5.2~\textrm{TeV}, 1.33)$ and a jet with $(p_{\textrm{T}}, \eta) = (1.2~\textrm{TeV}, 0.39)$.
The objects are nearly back-to-back in the transverse plane with $\Delta\phi(\ell, j) = 3.11$, while the relatively small rapidity separation $\Delta\eta(\ell, j) = 0.94$ indicates a centrally-produced system, with a low $\mathcal{S}(\Etmiss)$ at 1.43.
ATLAS event displays for these two highest-mass candidates are shown in Figures~\ref{fig:vp1_ej} and~\ref{fig:vp1_muj}.

\begin{figure}[htbp]
\subfloat[]{
\includegraphics[width=0.5\textwidth]{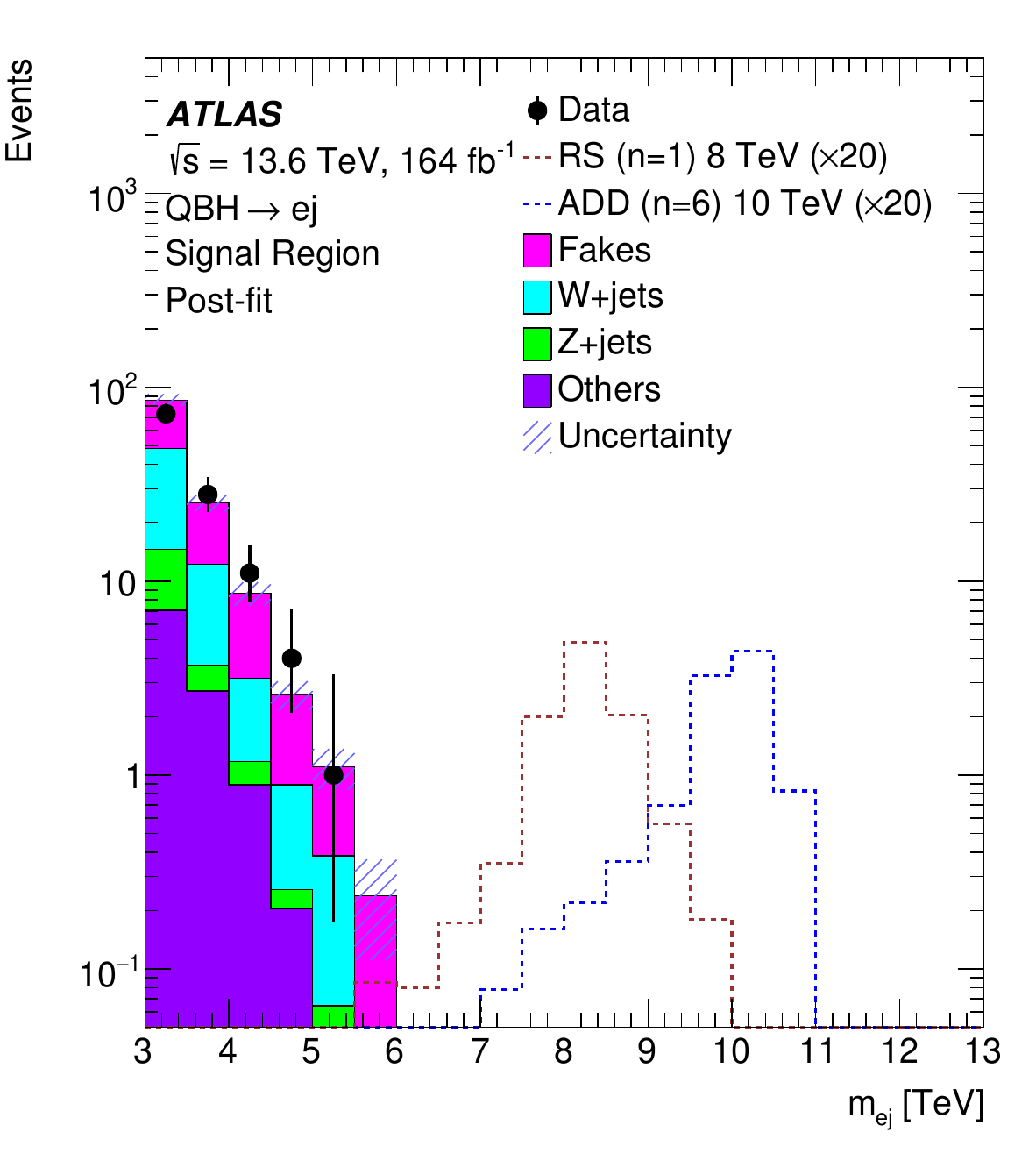}
}
\subfloat[]{
\includegraphics[width=0.5\textwidth]{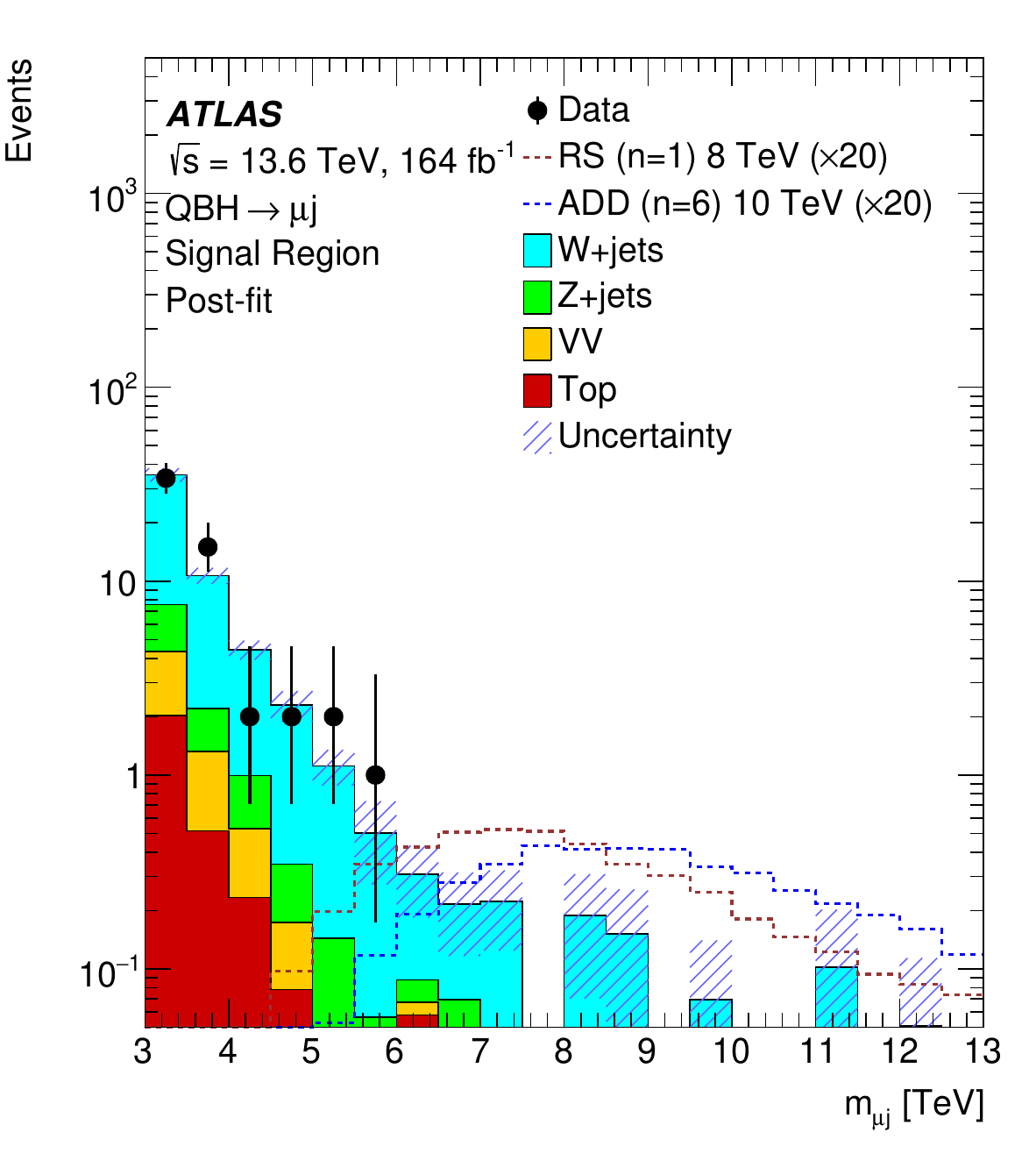}
}
\caption{Distribution of the lepton+jet invariant mass for events passing the full signal selection for (a) the electron channel and (b) the muon channel, shown in 0.5~TeV-wide bins. Benchmark signals for the RS and ADD models with $n=1,6$ extra dimensions at $M_{\textrm{th}}=8,10~\textrm{TeV}$, respectively, are scaled by a factor of 20 for presentation purposes and overlaid on the total background estimate. The highest-mass data event is observed at $5.3~(5.5)$~TeV in the electron (muon) channel. The bin width is chosen solely to reflect the different spectral shapes of the two channels; it differs from the bin choice used in the fit and limit-setting. No uncertainty band is shown, as the background shape at this granularity carries limited systematic understanding and these distributions are not used in the statistical analysis. In (a), ``Others'' denotes subdominant backgrounds including top and diboson processes. The MC contributions are normalised based on the likelihood fit.}
\label{fig:SR_mLepJet_ManyBins}
\end{figure}

FIG.~\ref{fig:SR_mLepJet_ManyBins} shows the same SR distributions with finer, 0.5~TeV-wide bins.
These plots are presented without an uncertainty band, as the background shape at this bin granularity is not used in the fit and is subject to limited understanding of shape uncertainties; they are intended only to illustrate the different spectral shapes of the two channels.

\subsection{Exclusion limits}
\label{sec:exc_limits}
Upper limits at 95\% confidence level (CL) on the QBH production cross-section times branching ratio are set using the $\mathrm{CL}_s$ method~\cite{Read:2002hq,Junk:1999kv} in the asymptotic approximation~\cite{Cowan:2010js} (see Sections~\ref{ssec:stats:cls}--\ref{ssec:stats:limits}). The limits set using the asymptotic approximation are validated against those obtained with pseudo-experiments, showing good agreement.

\begin{figure}[h]
    \subfloat[]{
      \includegraphics[width=0.5\textwidth]{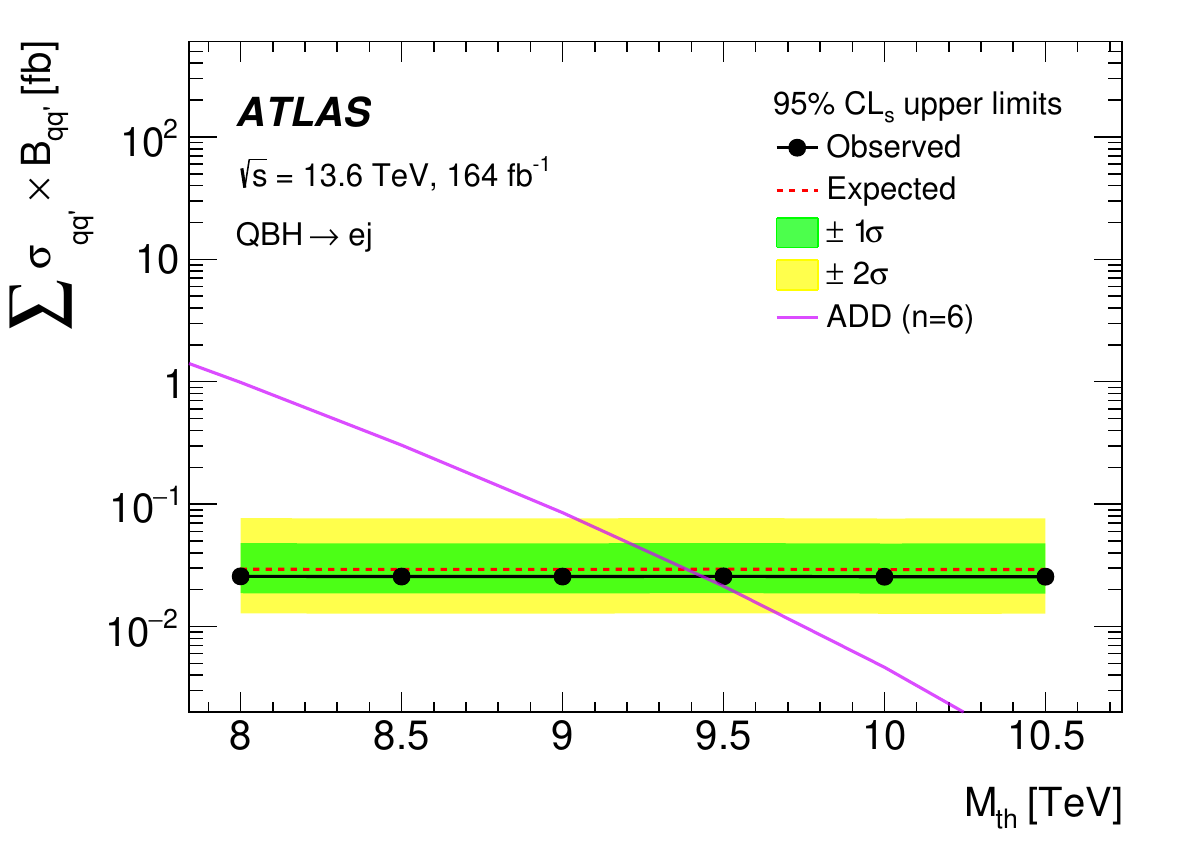}
    }
    \subfloat[]{
      \includegraphics[width=0.5\textwidth]{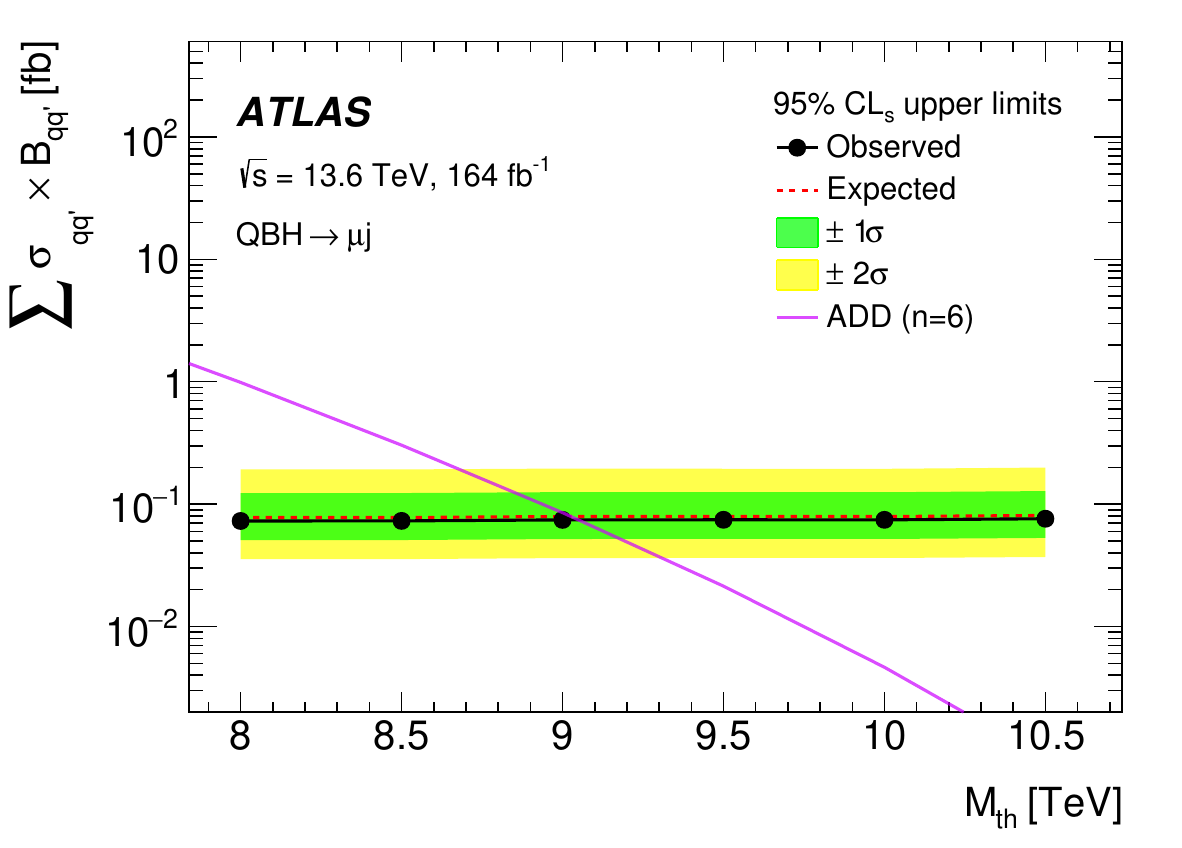}
    }\\
    \subfloat[]{
      \includegraphics[width=0.5\textwidth]{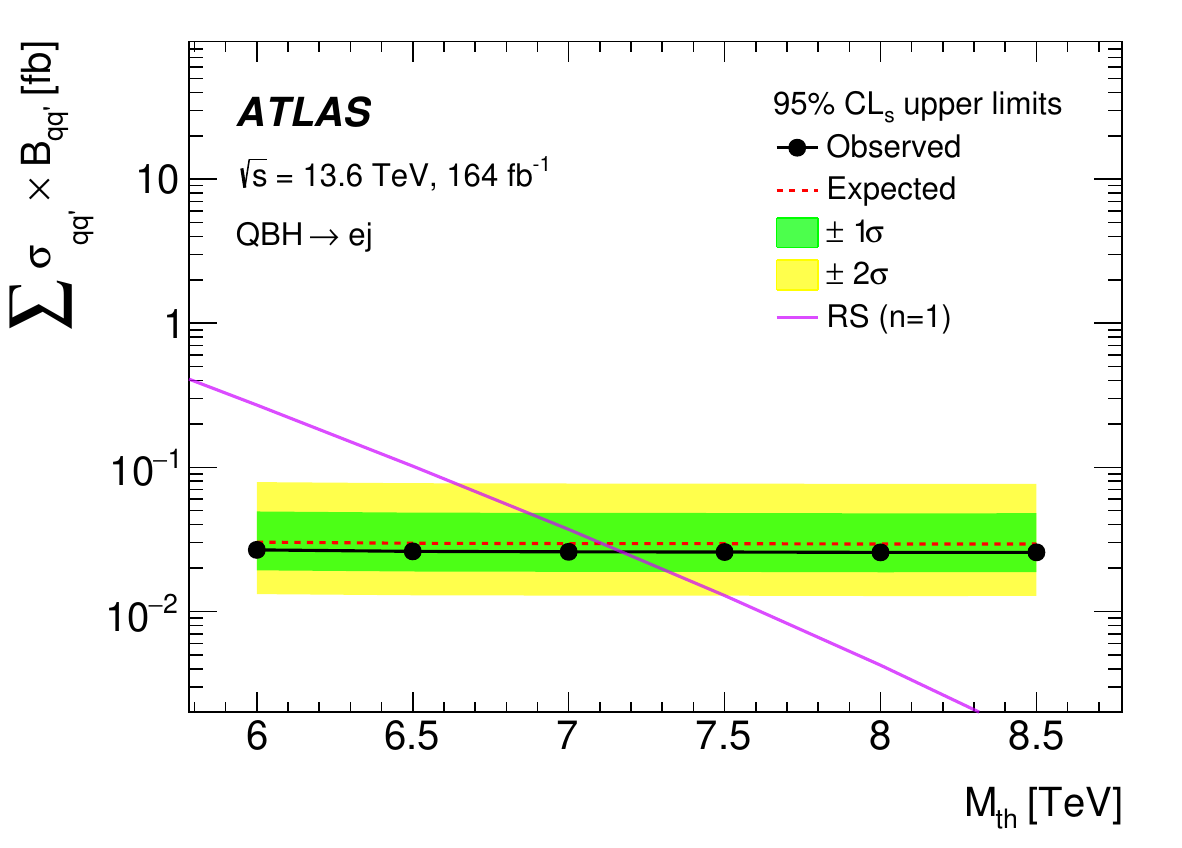}
    }
    \subfloat[]{
      \includegraphics[width=0.5\textwidth]{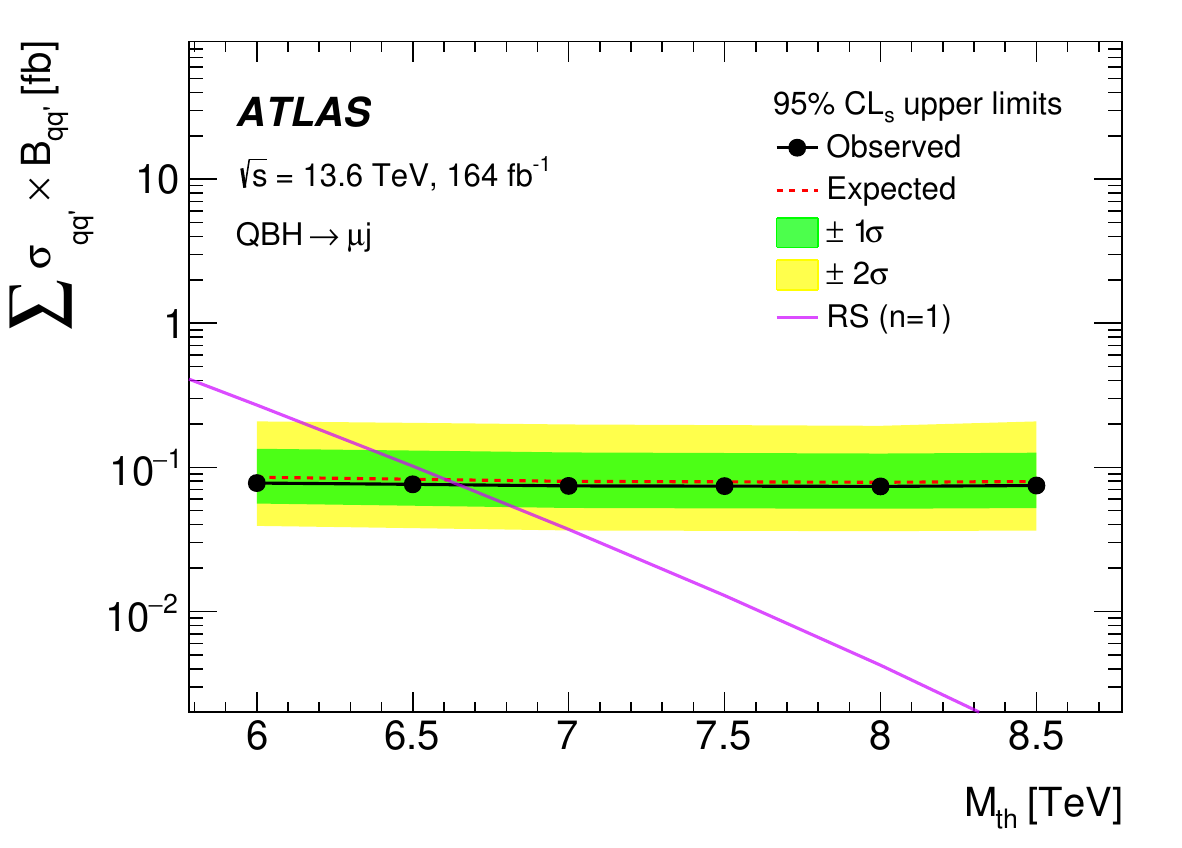}
    }
  \caption{Exclusion limits at 95\% confidence level (CL) on the sum of the production cross-section times branching ratio of quantum black holes in the ADD model with $n=6$ extra dimensions for (a) the electron and (b) the muon channels, and in the RS model with $n=1$ extra dimension for (c) the electron and (d) the muon channels. The $\pm 1\sigma$ and $\pm 2\sigma$ uncertainty bands around the expected limit are shown. The central theoretical values are overlaid. The summation subscript $qq'$ corresponds to the three relevant production channels of the quantum black hole. The theory curve corresponds to one lepton flavour decay.}
\label{fig:limits_with_comparisons}
\end{figure}

Limits on the production cross-section of ADD and RS QBHs with $n=6$ and $n=1$ extra dimensions, respectively, for decays into $ej$ and $\mu j$ final states are shown in FIG.~\ref{fig:limits_with_comparisons}.

The RS model is excluded at 95\% CL for signals with $M_{\textrm{th}} < 7.2~\textrm{TeV}$. Depending on the number of EDs, ADD models with $M_{\textrm{th}}$ below 8.6--9.4~TeV are excluded at 95\% CL. The electron channel surpasses the muon channel in sensitivity by a factor of ${\sim}3$, due to better momentum resolution for electrons and its impact on the acceptance times efficiency, as well as the lower background in the SR. Hence, the two channels are not combined.

The observed and expected exclusion limits at 95\% CL on the threshold mass for all signal models considered are detailed in Table~\ref{tab:intersection_limits}.

\begin{table}[h!]
\centering
\caption{Expected and observed exclusion limits at 95\% confidence level on the QBH mass threshold $M_{\mathrm{th}}$ for RS and ADD models with different number of extra dimensions $n$, in the electron and muon channels, using Run~3 data at a centre-of-mass energy of 13.6 TeV.}
\begin{tabular}{ll|cc|cc|cc|cc}
\hhline{==========}
Model   & & \multicolumn{2}{c|}{RS} & \multicolumn{6}{c}{ADD} \\
\hhline{----------}
$n$     & & \multicolumn{2}{c|}{1}  & \multicolumn{2}{c|}{2} & \multicolumn{2}{c|}{4} & \multicolumn{2}{c}{6} \\
\hhline{==========}
Channel & & $e+j$ & $\mu+j$ & $e+j$ & $\mu+j$ & $e+j$ & $\mu+j$ & $e+j$ & $\mu+j$ \\
\hhline{==========}
Exclusion limit & & \multicolumn{8}{c}{Expected} \\
\hhline{----------}
$M_{\mathrm{th}}$ [TeV] & & 7.1 & 6.6 & 8.5 & 8.1 & 9.1 & 8.7 & 9.4 & 9.0 \\
\hhline{==========}
Exclusion limit   & & \multicolumn{8}{c}{Observed} \\
\hhline{----------}
$M_{\mathrm{th}}$ [TeV] & & 7.2 & 6.6 & 8.6 & 8.2 & 9.1 & 8.7 & 9.4 & 9.0 \\
\hhline{==========}
\end{tabular}
\label{tab:intersection_limits}
\end{table}

\begin{figure}[h]
    \subfloat[]{
      \includegraphics[width=0.5\textwidth]{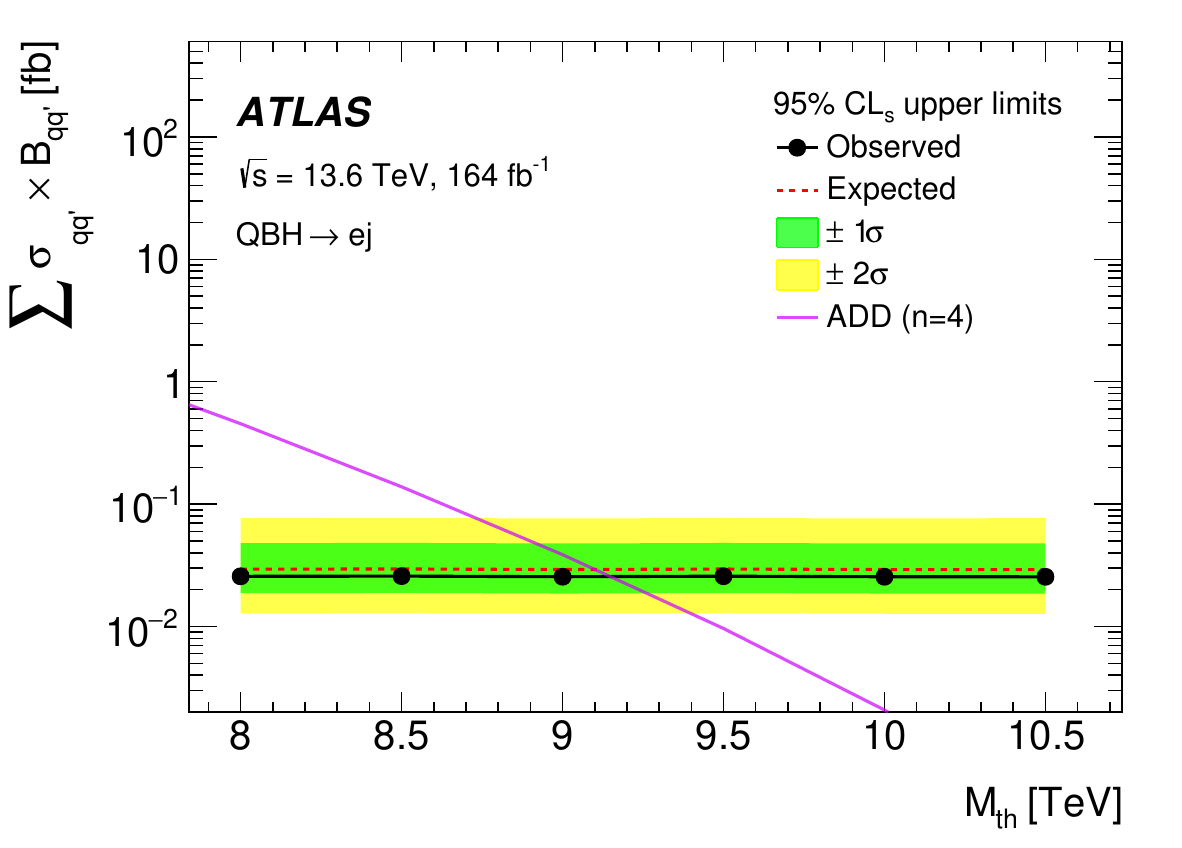}
    }
    \subfloat[]{
      \includegraphics[width=0.5\textwidth]{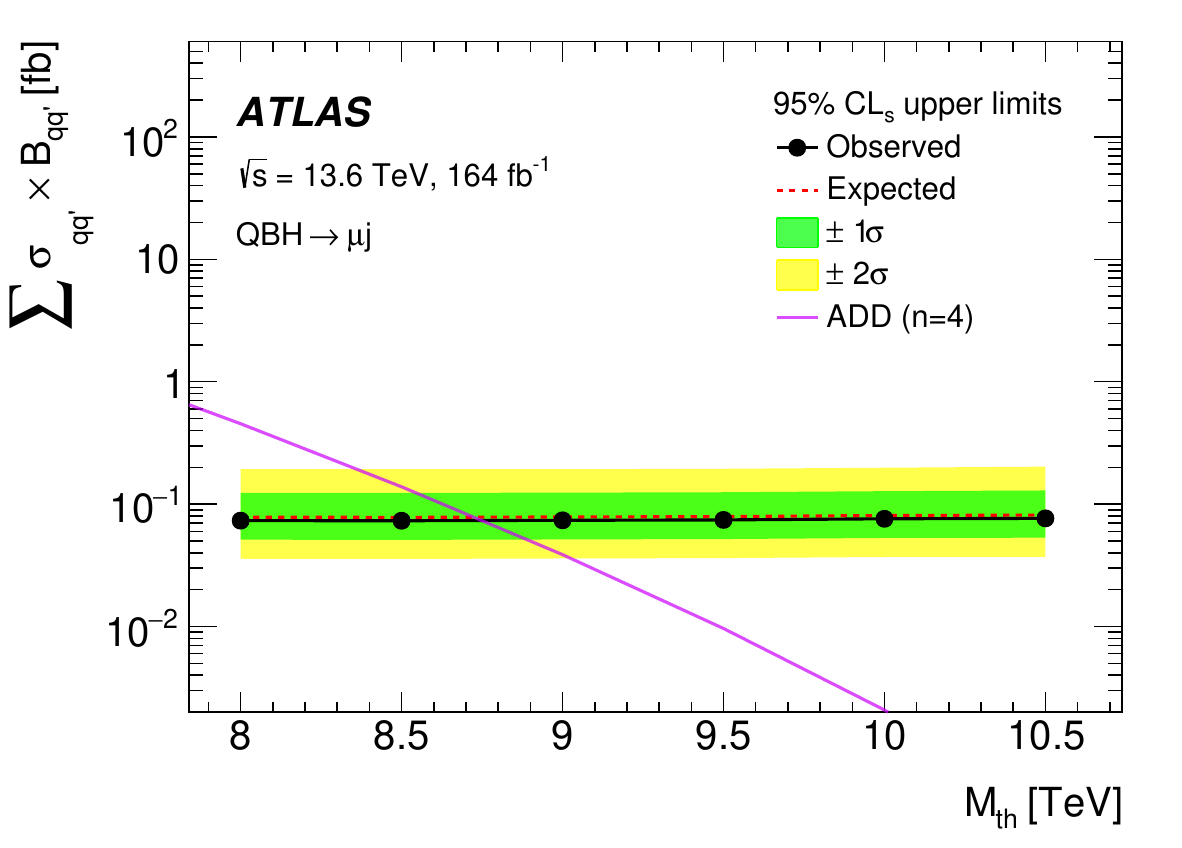}
    }\\
    \subfloat[]{
      \includegraphics[width=0.5\textwidth]{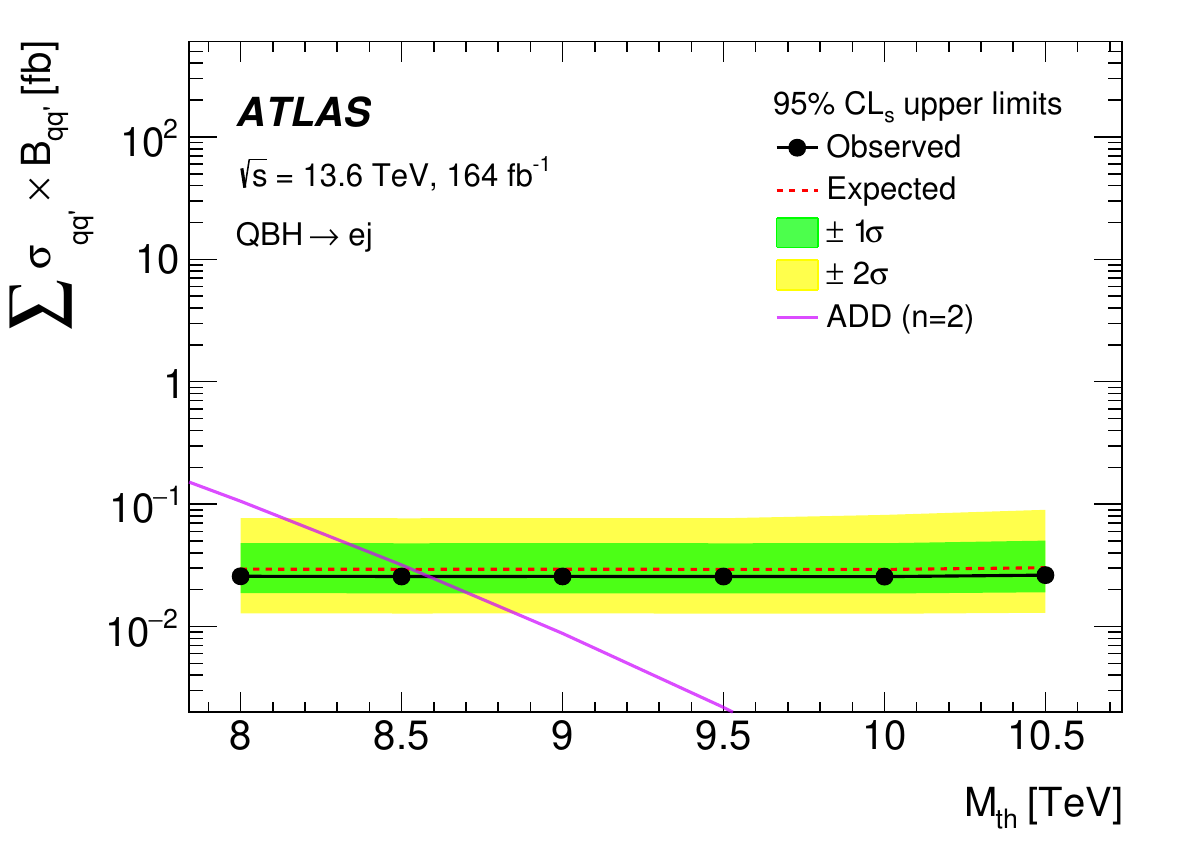}
    }
    \subfloat[]{
      \includegraphics[width=0.5\textwidth]{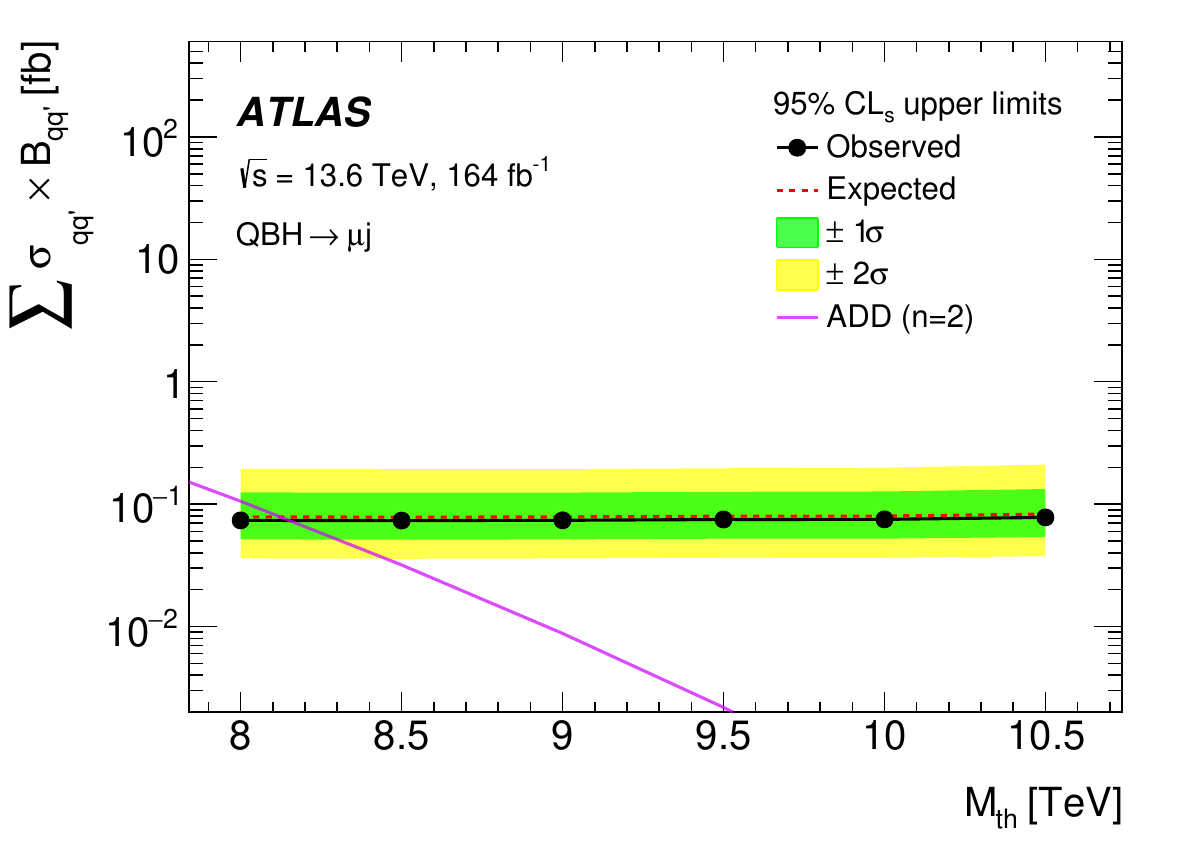}
    }
  \caption{Exclusion limits at 95\% confidence level (CL) on the sum of the production cross-section times branching ratio of quantum black holes in the ADD model with $n=4$ (top) and $n=2$ (bottom) extra dimensions, for the electron (left) and muon (right) channels. The $\pm 1\sigma$ and $\pm 2\sigma$ uncertainty bands around the expected limit are shown. The theory curve corresponds to one lepton flavour decay at a centre-of-mass energy of 13.6~TeV.}
\label{fig:limits_no_comparisons}
\end{figure}

Additional ADD modelling scenarios with $n=4$ and $n=2$ extra dimensions are also considered; these cases have not been included in previous QBH searches in this final state and are presented here for the first time. The limits for these models are shown in FIG.~\ref{fig:limits_no_comparisons}.

FIG.~\ref{fig:limits_no_comparisons} shows a similar trend to that observed in FIG.~\ref{fig:limits_with_comparisons}, with consistently improved sensitivity in the electron channel. In the higher-mass region where the background approaches zero, a flat sensitivity is observed for ADD results at masses above ${\sim}8$~TeV.

\section{Summary and Status}
\label{sec:summary}
A search for quantum black holes decaying to lepton+jet final states has been performed using $164~\textrm{fb}^{-1}$ of proton--proton collision data at $\sqrt{s}=13.6~\textrm{TeV}$ recorded by the ATLAS detector during 2022--2024. The observed invariant lepton+jet mass spectrum is consistent with Standard Model expectations. Upper limits at 95\% confidence level are set on the production cross-section times branching fraction for QBH signals in both the ADD and RS extra-dimensional frameworks. The RS model is excluded for threshold masses below $7.2~\textrm{TeV}$, while ADD models with $n=6$, $4$ and $2$ extra dimensions are excluded below $9.4$, $9.1$ and $8.6~\textrm{TeV}$, respectively. These results significantly surpass those of the preceding ATLAS search in this channel at $\sqrt{s}=13~\textrm{TeV}$~\cite{ATLAS:2023vat}, exploiting both the increased centre-of-mass energy and the full Run~3 dataset.

This analysis is complete and public; the corresponding ATLAS paper has been submitted to Physics Letters B~\cite{ATLAS:2026hzb}.
\clearpage
\clearpage\chapter{Search for New Resonances in \texorpdfstring{$\bm{\ell^{+}\ell^{-}+b}$}{\textit{ll}+b} Final States at \texorpdfstring{$\bm{\sqrt{s}=13}$}{sqrt(s)=13}~TeV}
\label{chp:zprime}

\section{Introduction}
\label{sec:zprime_intro}
Persistent anomalies in $b \to s\ell^+\ell^-$ transitions motivate direct collider
searches for a new neutral vector boson $\Zp$ arising from a $U(1)'$ extension of
the SM, as elaborated in Section~\ref{sec:bsm:flavour}.
The signal benchmark used in this search is the less-minimal flavour violation
model~\cite{Calibbi:2019lvs,Alguero:2022est,Crivellin:2022obd} (Section~\ref{sec:bsm:zprime_mediator}), in which the $\Zp$
carries a left-handed $(b\bar{s},\,\bar{b}s)$ coupling with arbitrary lepton flavour
and is produced in association with $b$-quarks at the LHC.
This model simultaneously accommodates the residual $B$-meson tensions in angular
observables and exclusive branching fractions (Section~\ref{sec:bsm:tensions}) while remaining consistent with the
$R_{K^{(*)}}$ measurements, and motivates searches in both dielectron and dimuon
final states without assuming lepton flavour universality.
The dimuon final state provides the primary sensitivity given the original muon-sector
anomaly hints; the dielectron final state serves as an independent cross-check and
enables combination of the two channels.
Global fit analyses comparing LFU and LFUV new-physics models to data find both
more compatible with measurements than the SM alone~\cite{Allanach:2022iod,Allanach2023},
further motivating these searches.

Direct searches for new resonances in inclusive dilepton final states have been
performed by both ATLAS~\cite{ATLAS_dilepton_2019} and CMS~\cite{CMS_dilepton_2018},
but these searches are dominated by the large Drell-Yan (DY) background and are
insensitive to a $\Zp$ that couples preferentially to $b$-quarks.
ATLAS has also searched for scalar resonances in dimuon
final states with $b$-jets~\cite{ATLAS_scalar_dilepton_2019}.
CMS performed the first dedicated LHC search for a high-mass dimuon resonance
produced in association with multiple $b$-jets~\cite{CMS_Zprimebb}, using the
$B_3 - L_2$ model~\cite{Alonso:2017uky,Bonilla:2017lsq} as the signal benchmark.
That search required at least one $b$-jet and applied a minimum muon--$b$-jet
invariant mass requirement $\min(m_{\mu b}) > 175~\GeV$ to reduce the $\ttbar$
background, excluding $350 < m_{\Zp} < 500~\GeV$ over most of the allowed coupling
parameter space.

This search is the primary effort within an umbrella of exclusive dilepton searches at ATLAS targeting the same broad new-physics landscape. A complementary ATLAS search for a new neutral vector boson decaying leptonically in association with missing transverse energy ($Z' \to \ell\ell + \Etmiss$) has been performed using Run~2 data~\cite{ATLAS-CONF-2023-045}, probing models with invisible $Z'$ decay modes. A further complementary effort targets contact interactions in the $b\bar{b}\ell^+\ell^-$ final state~\cite{Afik:2019htr}. The search was unblinded in May 2025 and the combined publication, incorporating all these exclusive dilepton efforts with the $Z' \to \ell\ell + b$-jets search presented here as the primary focus, is currently in preparation.

\section{Analysis Strategy}
\label{sec:zprime_strategy}
A summary of key strategy points of the analysis is given below.


\begin{itemize}
	\item \textbf{Physics target}: search for resonant signatures in final states including two same-flavour leptons and $b$-jets, namely ($ee$ or $\mu\mu$) with 0, 1 or at least 2 $b$-jets.
	\item \textbf{Observable}: the dilepton invariant mass ($m_{\ell\ell}$) spectrum. The statistical analysis is conducted in bins of $m_{\ell\ell}$.
	\item \textbf{Strategy}: the signal and background are estimated from MC simulations using a set of event generators. The search is performed in a $\Zp$ pole mass range of 0.5 TeV to 4 TeV for resonances of narrow width (about 2\%) and wide width (about 8\%). The width is controlled by the coupling constant of $\Zp$ to fermions, $g_{\Zp}=0.5,1.0$.
\end{itemize}

The fake background consists of events where at least one of the selected final-state leptons is due to a misidentified jet and its contribution is estimated by the Matrix Method using data. Dominant processes included in the fake background are $W$+jets and multijet, yielding one fake lepton or two fake leptons, respectively.

Signals are modelled via MC simulation which interfaces the hard-scatter process with shower, hadronisation and detector response. For both dimuon and dielectron signals, two $\Zp$ coupling constants are used which induce resonance distributions of different widths. The hard process includes three subprocesses, corresponding to the number of QCD matrix elements in the amplitude: zero, one and two.

Control regions with high purity in the leading backgrounds are defined. These primarily test the $t\bar{t}$ and DY estimations against data and are used to fit the backgrounds in the invariant mass spectrum and determine their normalisation for the signal regions (SR). Validation regions test the normalisations obtained from the control regions in the $m_{\ell\ell}$ range near the lower edge of the signal region, up to $500\;\textrm{GeV}$. Three signal regions are defined by the $b$-jet multiplicity: zero, one, and at least two. This determines the signal and background efficiencies and the background composition. A fit to the dilepton invariant mass distribution is then used to set limits on the $\Zp$ production cross-section as a function of $m_{\Zp}$. The analysis regions are detailed in Table~\ref{tab:analysis_regions} .

\begin{table}[h]
\centering
\caption{Analysis regions. Key are the dilepton and $b$-jet multiplicity requirements, in conjunction with the dilepton invariant mass range.}
\label{tab:analysis_regions}
\scalebox{0.85}{%
    \begin{tabular}{c|c|c|c|c|c|c|c}
    \hhline{========}
    Type                        & Description                            & Dilepton                            & $N_{b\textrm{-jets}}$     & $\sigma(E^{\textrm{miss}}_{\textrm{T}})$ & $E^{\textrm{miss}}_{\textrm{T}} \; \textrm{[GeV]}$ & $\text{min}(m_{\ell b})\; \textrm{[GeV]}$ & $m_{\ell\ell}\; \textrm{[GeV]}$ \\ \hline
    \multirow{3}{*}{Signal}     & 0 $b$-jets                          & \multirow{5}{*}{$e^{+}e^{-}/\mu^{+}\mu^{-}$} & 0                         & \multirow{3}{*}{$< 5.0$}    &    \multirow{4}{*}{N/A}  &     N/A                            & \multirow{3}{*}{300--5000}    \\
	    \cline{7-7}
                                & 1 $b$-jets                             &                                     & 1                         &                                          &                                                  &  \multirow{2}{*}{$> 155$}   &                          \\
                                & $\geq 2$ $b$-jets                    &                                     & $\geq 2$                  &                                          &                                                  &    &                           \\ \cline{1-2} \cline{4-5} \cline{7-8} 
    \multirow{3}{*}{Control}    & Z+LF                     &                                     & 0                         & \multirow{5}{*}{N/A}   & \multirow{6}{*}{N/A}                   &                                                  & \multirow{2}{*}{130--300}     \\ \cline{4-4} \cline{6-6}
                                & Z+HF                      &                                     & \multirow{2}{*}{$\geq 1$} &                                          & $< 20$                                           &                    &           \\ \cline{3-3} \cline{6-6}  \cline{8-8}
                                & Top                                    & $e\mu$                              &                           &                                          & & \multirow{2}{*}{N/A}                             & 130--2000                    \\ \cline{1-4} \cline{8-8} 
    Validation   & Z+LF                     & \multirow{2}{*}{$e^{+}e^{-}/\mu^{+}\mu^{-}$} & 0                         &                                          &   &                                               & \multirow{2}{*}{300--500}     \\ \cline{6-6}
                                (pre-unblinding)& {Z+HF} &                                     & $\geq 1$                  &             
                                                           &                $< 20$  &                                &                               \\ 
\hhline{========}
    \end{tabular}%
}
\end{table}

The effect of experimental and theoretical systematic uncertainties on the cross section in the $m_{\ell\ell}$ spectrum is evaluated. For presentation, theoretical uncertainties are grouped by category following Physics Modelling Group recommendations. For the fit and limit-setting, each systematic uncertainty is treated as an individual nuisance parameter with pruning applied. Finally, the sensitivity to the $\Zp$ is estimated by setting limits on its inclusive production cross-section as a function of $m_{\ell\ell}$. Limits are also set on the fiducial cross section and on the signal strength $\mu$ for each SR.

\section{Signal Simulation}
\label{sec:zprime_signal_model}
The theoretical motivation, generic $\Zp$ Lagrangian, coupling matrices, and production modes for the less-minimal flavour violation benchmark are described in Section~\ref{sec:bsm:zprime_mediator}.
The specific MC signal generation, coupling parametrisation, decay width calculation, and fiducial selections used in this analysis are detailed in the following section.

Signal samples are generated from MC simulation following the phenomenology of a benchmark $\Zp$ decaying to a lepton pair in association with $b$-quarks~\cite{Calibbi:2019lvs}. Using \texttt{MadGraph v.2.9.3}, hard-scatter events are produced. The electroweak parameters used in the MG5\_aMC@NLO signal production are taken directly from the parameter configuration file. The $Z$ boson mass is set to $m_Z = 91.1876$~GeV and the Cabibbo angle via $\lambda = 0.227736$. No other EW parameters are explicitly modified. Decay of the $\Zp$ to two muons or two electrons is modelled separately, such that each mode has $BR=1$. Coupling of the $\Zp$ to quarks is a model parameter for which two values are considered, corresponding to different resonance widths. The specific Lagrangian considered for the model is

\begin{align}
\begin{split}
\mathcal{L}_{\textrm{LH}} &= g_{Z'\mu} \mu \gamma {}^{\nu } P_L \mu \Zp_{\nu} + g_{Z'e} e\gamma {}^{\nu } P_L e \Zp_{\nu} + g_{Z'q} \bar{b} \gamma {}^{\nu } P_L b \Zp_{\nu } \\ &+ g_{\Zp q} \bar{t} \gamma^{\nu } P_L t \Zp_{\nu} + |V_{cb}|^{2} g_{\Zp q} \bar{s} \gamma {}^{\nu } P_L s \Zp_{\nu } + |V_{cb}| g_{Z'q} \bar{b} \gamma {}^{\nu } P_L s \Zp_{\nu},
\label{specific_model_langrangian}
\end{split}
\end{align}
where the first two terms define the $\Zp$ couplings to electrons and muons; the coupling is set such that only one lepton flavour is active at a time.  Correspondingly, the hard-scatter modelling is initiated based on the generalised process,

\begin{equation}
p^{i}p^{i} \rightarrow Z' + (0/b/bb) \rightarrow \ell^{+}\ell^{-},
\end{equation}
where $p^{i} \ni \{g, u, d, c, s, b, \bar{u}, \bar{d}, \bar{c}, \bar{s}, \bar{b} \}$ (five-flavour scheme) and $(0/b/bb)$ corresponds to the number of QCD vertices included in the matrix element, generated inclusively; the corresponding production-mode Feynman diagrams are shown in FIG.~\ref{fig:bsm:zprime_production}.

Hard-process generation at matrix-element level is illustrated in FIG.~\ref{fig:bsm:zprime_production}. However, production modes defined at parton level do not map straightforwardly onto the signal regions, which are categorised by the number of reconstructed $b$-jets. For example, an event generated in the 0-$b$ mode may involve an initial-state quark that radiates a gluon, which then splits into a $b\bar{b}$ pair. If these $b$-quarks are reconstructed as $b$-jets, the event may populate the 1-$b$ or even $\geq$2-$b$ signal regions. Conversely, in the 1-$b$ mode, the generated $b$-quark might fail to be reconstructed as a $b$-jet, causing the event to appear in the 0-$b$ category. This illustrates the ambiguity in connecting parton-level production modes with reconstructed final-state observables.

Free parameters include $m_{\Zp}$ and $\Zp$ couplings which are set accordingly to the theoretical model,

\begin{equation}
g_{Z', b\bar{b}} = g; \; g_{Z', b\bar{s}/s\bar{b}} = g \times |V_{cb}|; \; g_{Z', s\bar{s}} = g \times |V_{cb}|^{2},
\end{equation}
where $g \in \{0.5, 1.0\}$ corresponds to narrow and wide resonance decay modes. The $\Zp$ mediates the transition of a $b$-quark to an $s$-quark at leading order (LO) and is taken to be left-handed. Its dominant coupling is to $b$-quarks, while $s$-quark couplings are CKM-suppressed. Coupling of $\Zp$ to leptons is taken to be equal to its $b\bar{b}$ coupling, namely $g_{Z', e^{-}e^{+}/\mu^{-}\mu^{+}} = g$. The free mass parameter $m_{\Zp}$ is produced at 22 values in the range $[500, 4000] \; \textrm{GeV}$ for each value of g, for each dilepton decay mode, resulting in $22 \times 2 \times 2 = 88$ signal samples altogether. The frequency of $m_{\Zp}$ points decreases at higher masses, where the cross-section is low.

The hard-process modelling is followed by simulation of hadronisation and parton shower using Pythia v.8.245. Multileg parton-level events are matched to Pythia using CKKW-L merging. The ATLAS detector response is simulated using the ATLAS fast simulation framework (AtlFast-II). The Parton Distribution Function (PDF) taken is \texttt{NNPDF30\_lo\_as\_0130}.

The decay width of $\Zp$ is computed from the model Lagrangian in Eq.~\ref{specific_model_langrangian} and input to \texttt{MadGraph} as a signal parameter:

\begin{align}
\begin{split}
\Gamma_{Z'} &= \frac{1}{8\pi} \Biggr[\frac{1}{3}\frac{g_{\Zp \ell}^{2}}{m^{2}_{\Zp}} (m^{2}_{\Zp} - m^{2}_{\ell})\sqrt{m^{2}_{\Zp} - 4m^{2}_{\ell}} + \frac{g_{\Zp q}^{2}}{m^{2}_{\Zp}} (m^{2}_{\Zp} - m^{2}_{b})\sqrt{m^{2}_{\Zp} - 4m^{2}_{b}}\Biggr] \\ &+ \frac{1}{8\pi} \Biggr[\frac{V_{cb}^{4} g_{Z'q}^{2}}{m_{Z'}^{2}} (m_{Z'}^{2} - m_{s}^{2}) \sqrt{m_{Z'}^2-4m_{s}^2} + \frac{g_{\Zp q}^{2}}{m^{2}_{\Zp}} (m^{2}_{\Zp} - m^{2}_{t})\sqrt{m^{2}_{\Zp} - 4m^{2}_{t}}\Biggr] \\ &+
\frac{2}{16\pi} \frac{(g_{Z'q}V_{cb})^2}{m_{Z'}^3} \Big(\frac{m_{Z'}^2}{2}-\frac{m_{b}^2}{4}-\frac{m_{s}^2}{4}-\frac{m_{b}^4}{4m_{Z'}^2}+\frac{m_{b}^2m_{s}^2}{2m_{Z'}^2}-\frac{m_{s}^4}{4m_{Z'}^2}\Big) \\
&\quad \times \sqrt{m_{b}^4-2m_{b}^2m_{s}^2+m_{s}^4-2m_{b}^2m_{Z'}^2-2m_{s}^2m_{Z'}^2+m_{Z'}^4},
\label{eq:signal_Gamma}
\end{split}
\end{align}
where the first term accounts for coupling to leptons, where $g_{\Zp e}$ is off (set to zero) when $g_{\Zp\mu}$ is on (set arbitrarily to 0.5 or 1.0). The second, third and fourth terms account for $\Zp$'s coupling to $b\bar{b}, s\bar{s}$ and $t\bar{t}$, respectively. The last term accounts for the coupling to $s\bar{b}$ and its conjugate (hence the factor of 2). Signal events are generated with $\Zp$ decaying exclusively to either two electrons or two muons, with quarks produced only in association. The fractional width, $\Gamma_{\Zp}/m_{\Zp}$, ranges between 1.9\%--2.3\% and 7.8\%--9.2\% for g = 0.5, 1.0, respectively.

\subsection{Fiducial signal definition}
\label{sec:fiducial_signal_cuts}

Signals generated with TeV-scale pole mass tend to contain an enhanced low-mass tail due to parton luminosity effects. Despite having a dilepton mass distribution, $m_{\ell\ell}$ strongly peaking around the pole mass, $m_{\Zp}$, the probability to find two partons with sufficient momenta needed for the generation of a high-mass $\Zp$ is strongly suppressed according to the PDF. Multi-TeV resonances consequently develop \textit{parton luminosity tails} from the relative increase in off-shell production. As a result, such signals appear broader, with a significant fraction of the cross-section falling outside the core resonance, reducing sensitivity especially at high masses. This can be addressed by applying cuts that focus the search on the core resonance, filtering out contributions affected by the parton luminosity tail.

\begin{figure}[h!]
	\captionsetup[subfigure]{labelformat=empty}
	\subfloat[(a)]{
		\includegraphics[width=0.5\textwidth]{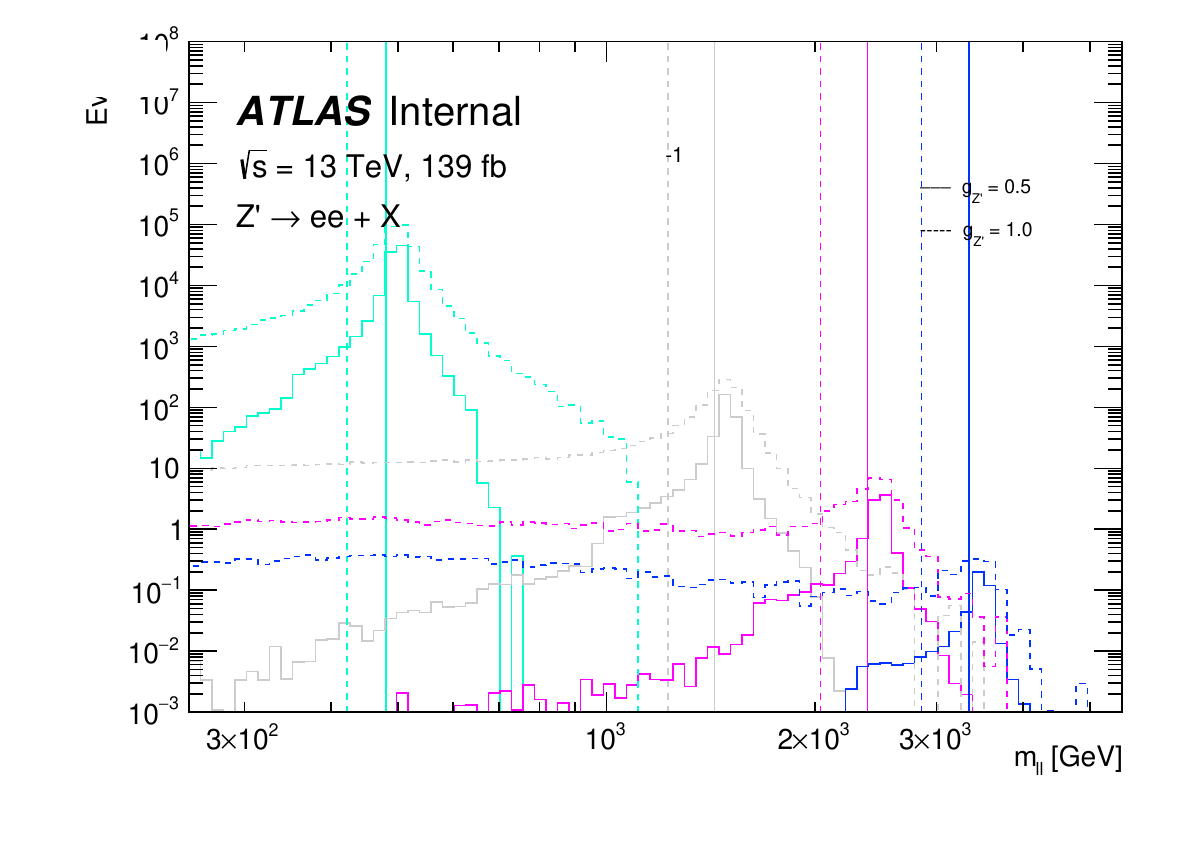}
	}
	\hfill
	\subfloat[(b)]{
		\includegraphics[width=0.5\textwidth]{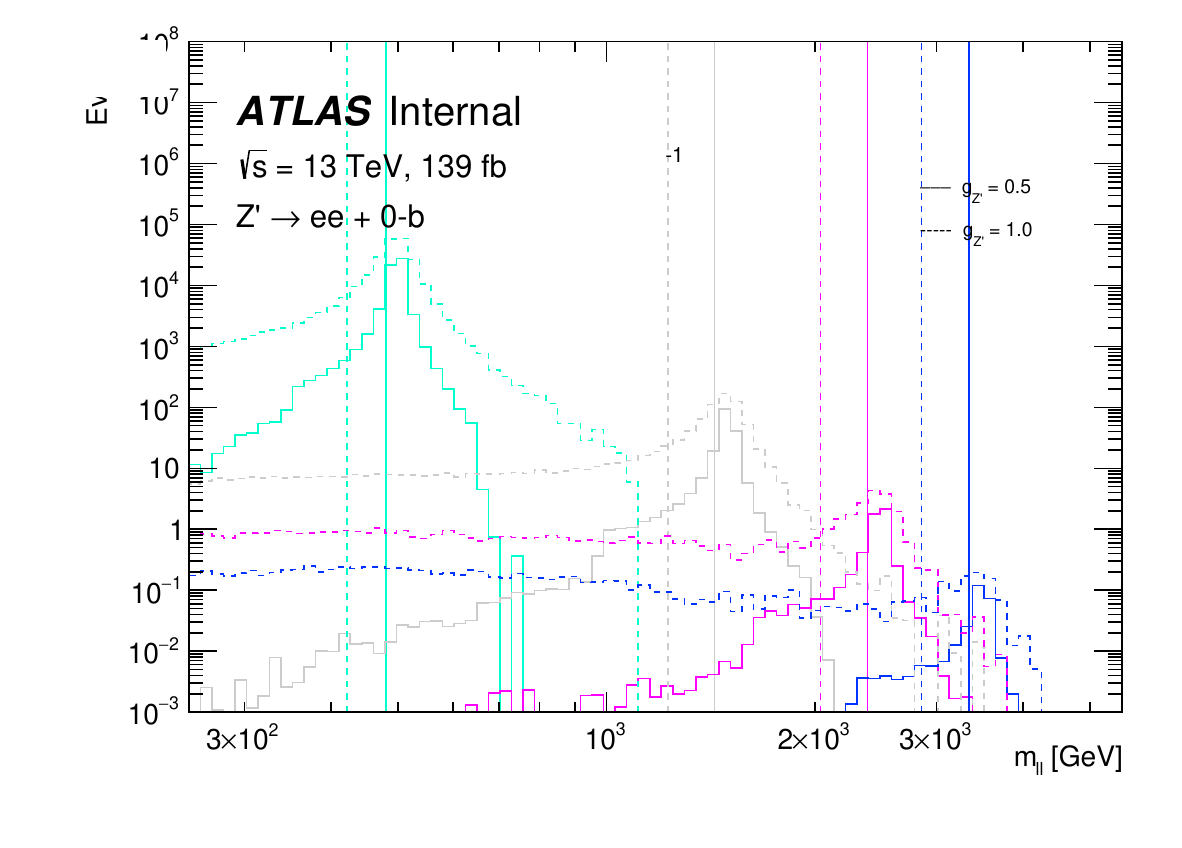}
	}
	\hfill
	\subfloat[(c)]{
		\includegraphics[width=0.5\textwidth]{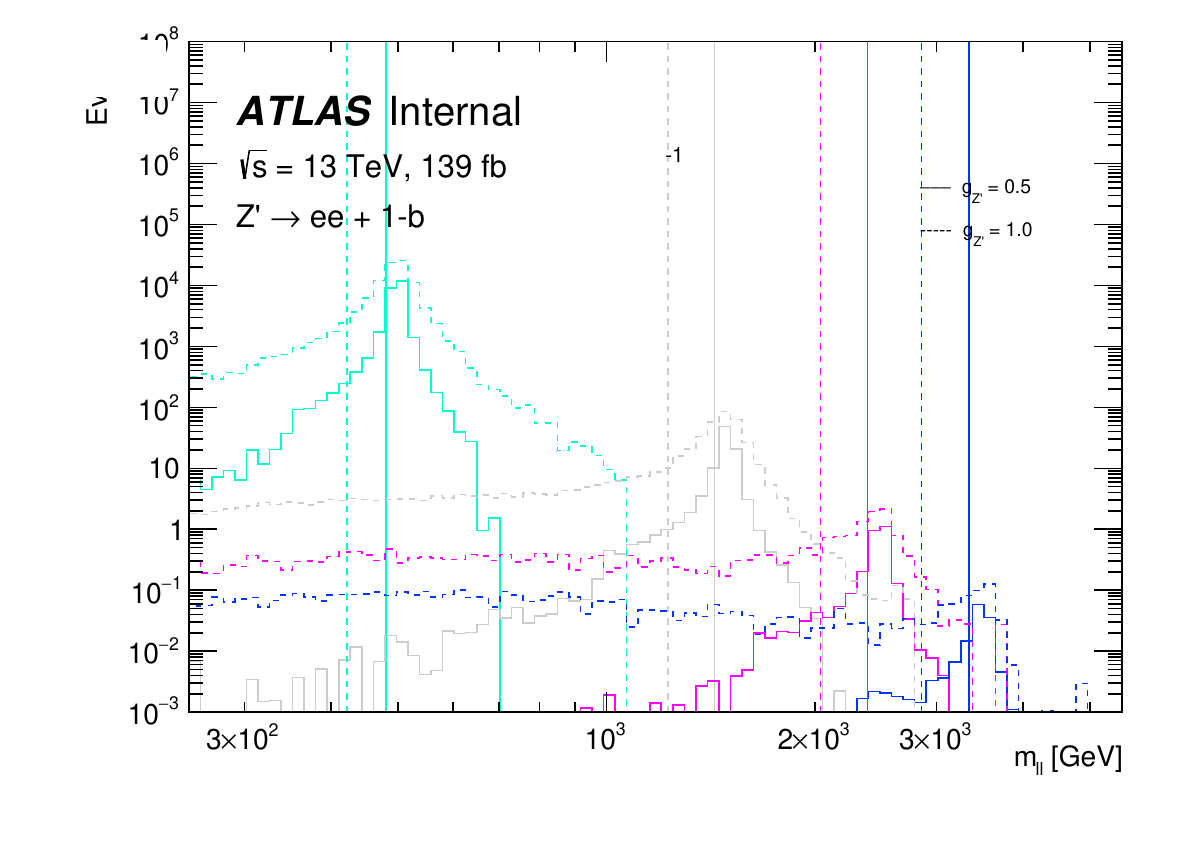}
	}
	\hfill
	\subfloat[(d)]{
		\includegraphics[width=0.5\textwidth]{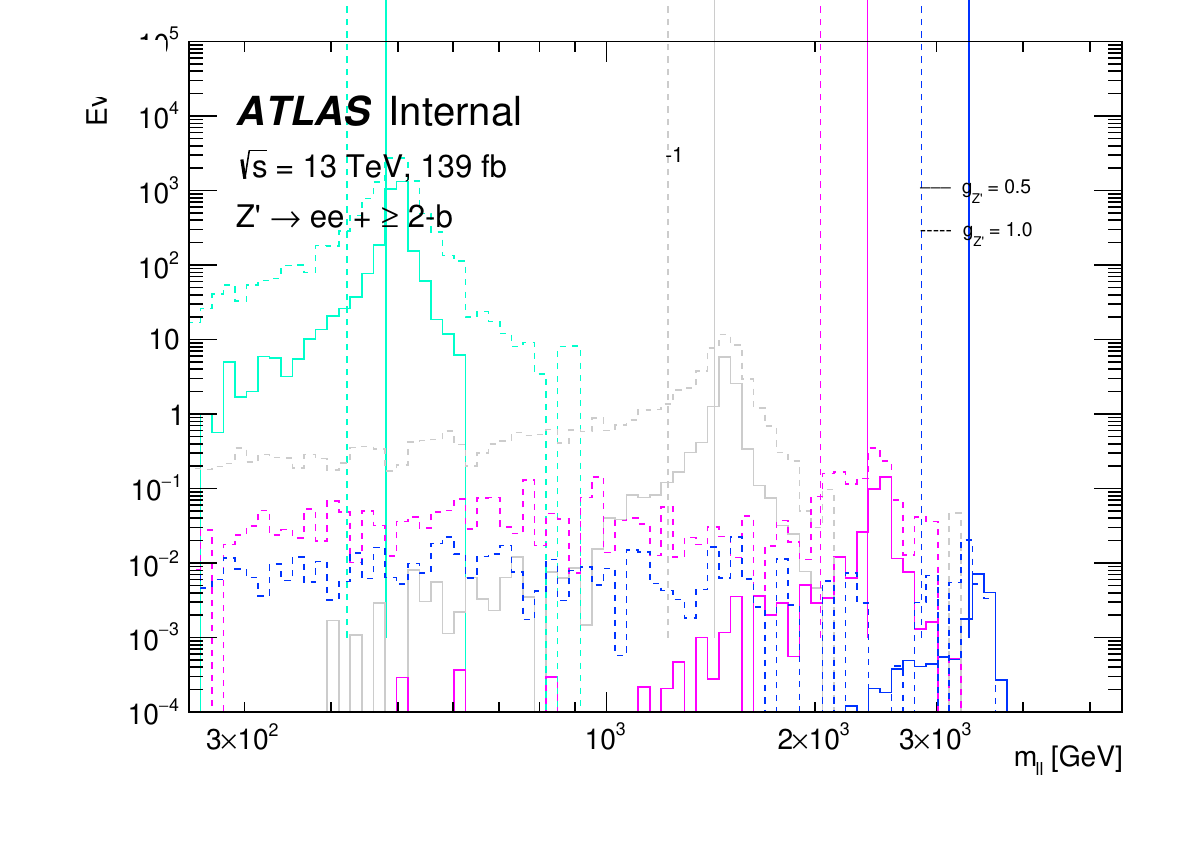}
	}
	\caption{Dielectron reconstructed invariant mass of $\Zp$ signals at various points in the 500--4000 GeV production range (cyan: $m_{Z'}=500\,\text{GeV}$, grey: $m_{Z'}=1500\,\text{GeV}$, pink: $m_{Z'}=2500\,\text{GeV}$, blue: $m_{Z'}=3500\,\text{GeV}$). The vertical lines indicate corresponding fiducial mass cuts applied at truth level to avoid the signals being dependent on parton luminosity tail effects, whereas the individual distributions are shown before fiducial cuts are applied. (a) no requirements on $b$-jet multiplicity. (b) final state required to contain no $b$-jets. (c) final state required to contain exactly one $b$-jet. (d) final state required to contain at least two $b$-jets.}
	\label{fig:Dielectron_Signal_Mass_no_fid_cuts}
\end{figure}

\begin{figure}[h!]
	\captionsetup[subfigure]{labelformat=empty}
	\subfloat[(a)]{
		\includegraphics[width=0.5\textwidth]{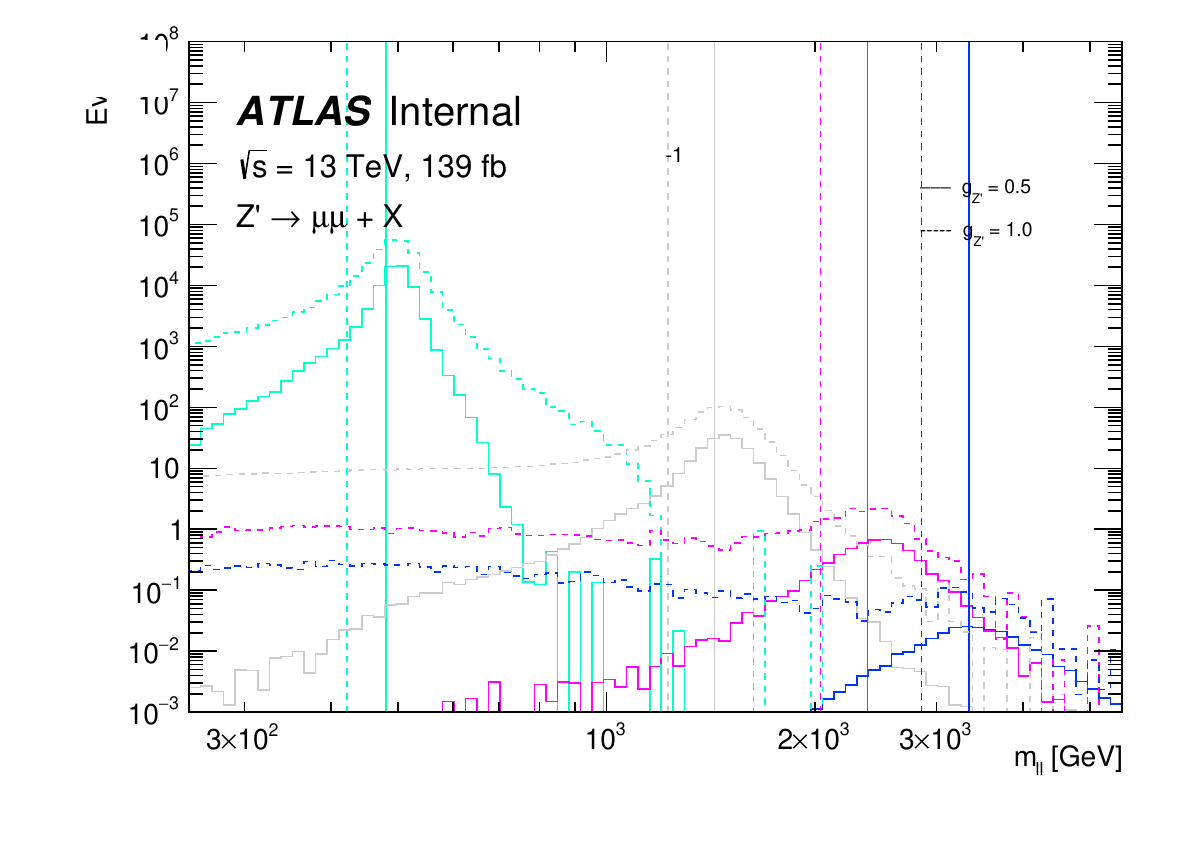}
	}
	\hfill
	\subfloat[(b)]{
		\includegraphics[width=0.5\textwidth]{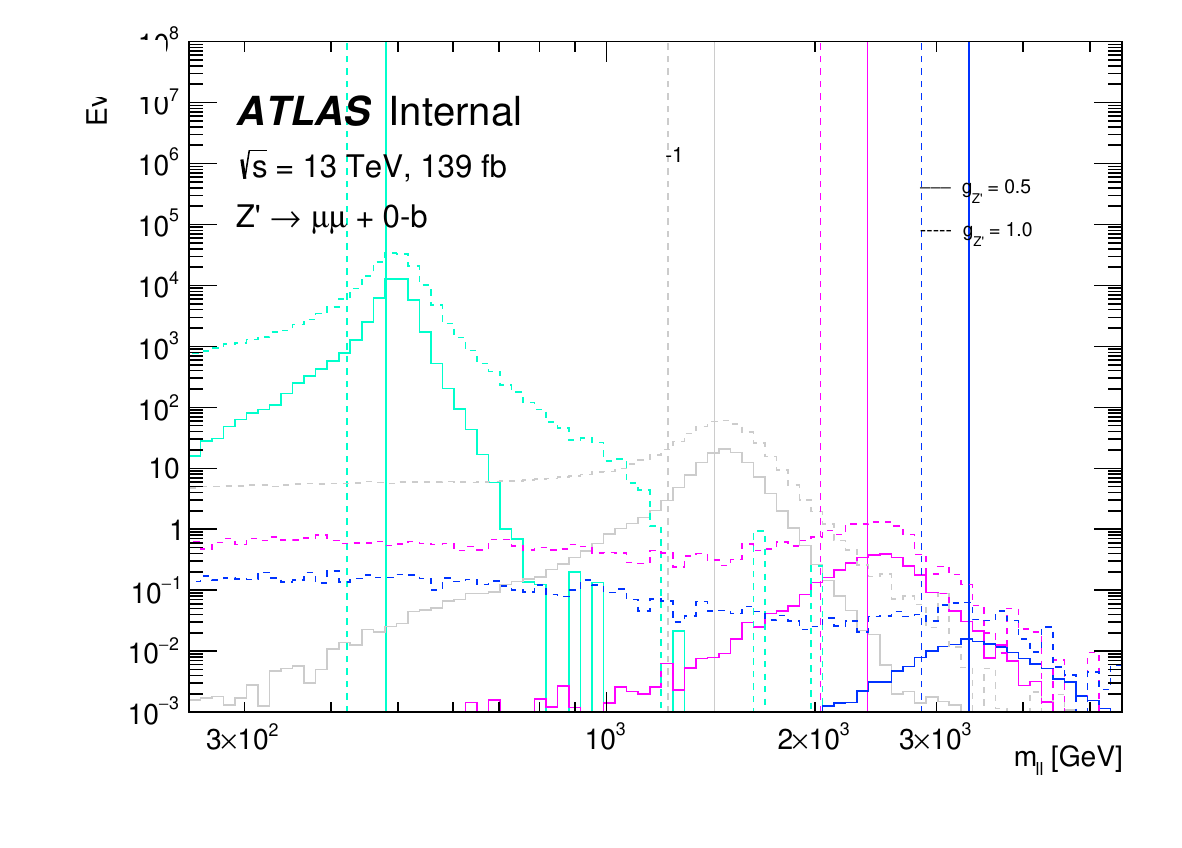}
	}
	\hfill
	\subfloat[(c)]{
		\includegraphics[width=0.5\textwidth]{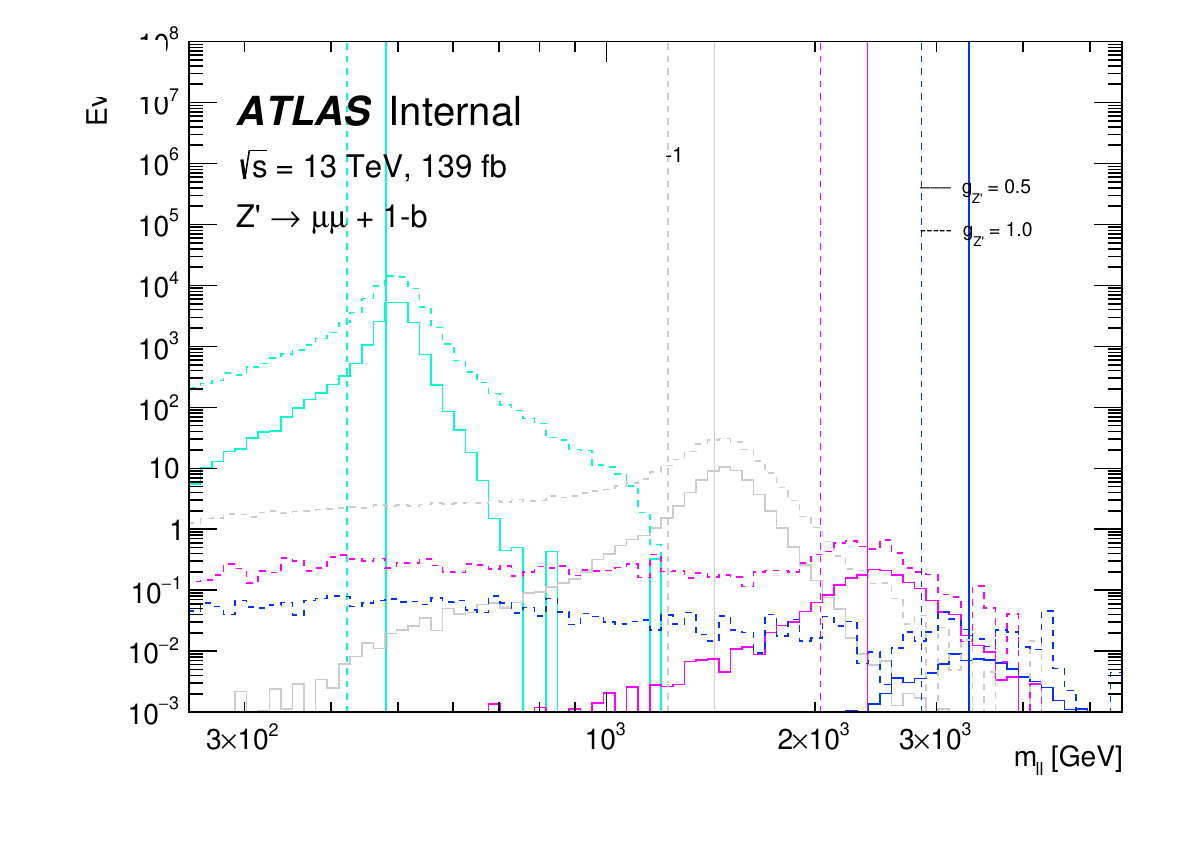}
	}
	\hfill
	\subfloat[(d)]{
		\includegraphics[width=0.5\textwidth]{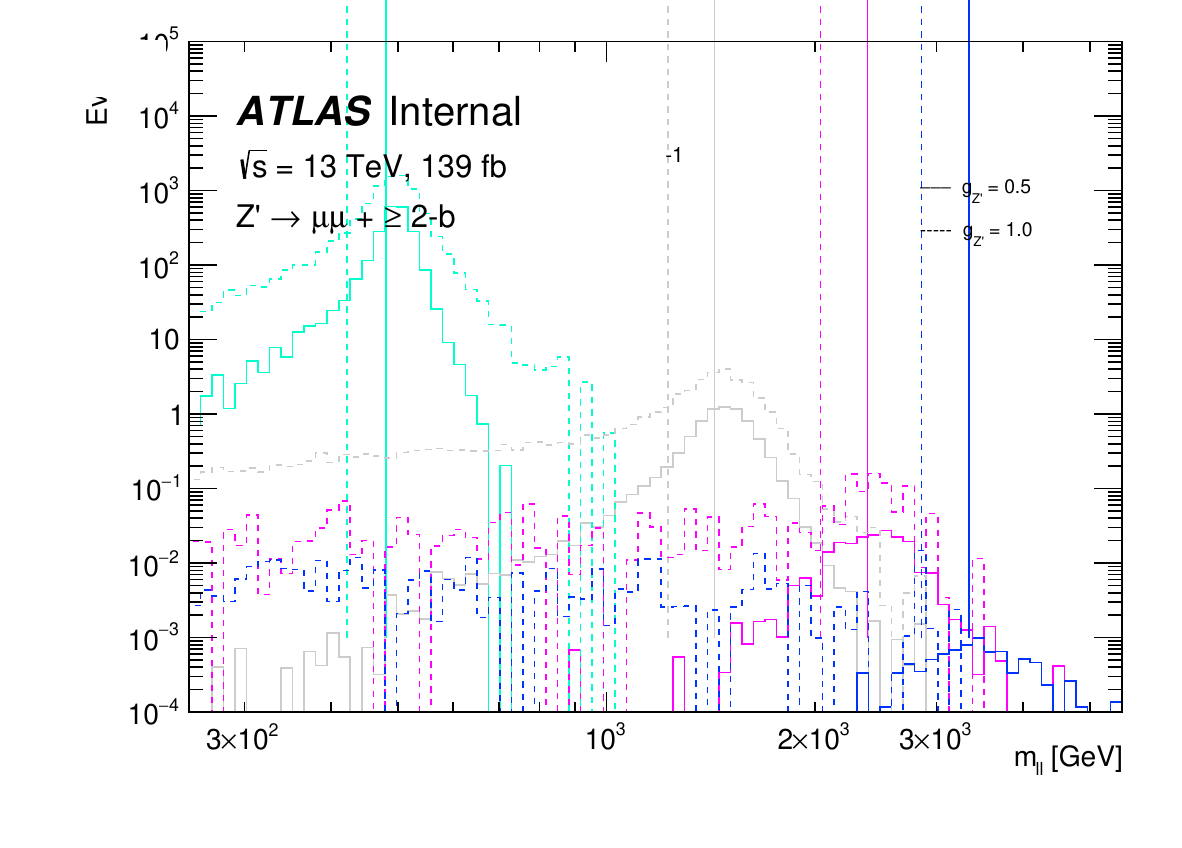}
	}
	\caption{Dimuon reconstructed invariant mass of $\Zp$ signals at various points in the 500--4000 GeV production range (cyan: $m_{Z'}=500\,\text{GeV}$, grey: $m_{Z'}=1500\,\text{GeV}$, pink: $m_{Z'}=2500\,\text{GeV}$, blue: $m_{Z'}=3500\,\text{GeV}$). The vertical lines indicate corresponding fiducial mass cuts applied at truth level to avoid the signals being dependent on parton luminosity tail effects, whereas the individual distributions are shown before fiducial cuts are applied. (a) no requirements on $b$-jet multiplicity. (b) final state required to contain no $b$-jets. (c) final state required to contain exactly one $b$-jet. (d) final state required to contain at least two $b$-jets.}
	\label{fig:Dimuon_Signal_Mass_no_fid_cuts}
\end{figure}

Setting fiducial limits is also consistent with previous dilepton searches~\cite{ATLAS_dilepton_2019,ATLAS:2016cyf}, enabling generic, model-independent limits on the $\Zp$ cross-section. The tail extent is model-dependent; here the $\Zp$ couples primarily to $b\bar{b}$, whereas previous searches used $u\bar{u}$ pairs. By default no cut is applied to this tail. To keep results broadly applicable and consistent with the inclusive search~\cite{ATLAS_dilepton_2019}, truth-level fiducial cuts are applied in one mode of the limit-setting procedure:

\begin{align}
	\text{individual truth lepton:} & \
	\begin{cases}
		\ \ \ \pT^\mathrm{truth} > 30\ \mathrm{GeV} \\
		\ \left|\eta^\mathrm{truth}\right| < 2.5  \\
	\end{cases} \label{eq:fidCuts1and2} \\
	\text{truth lepton pair:}& \ \ \ \ \ \ m_{\ell\ell}^\mathrm{truth} > m_{\Zp} - 2\Gamma_{\Zp}, \label{eq:fidCuts3}
\end{align}
where the first two cuts are applied on individual leptons at truth level, whereas the latter on the invariant mass of the two leading truth leptons in the event which had passed the individual cuts. In FIG. \ref{fig:Dielectron_Signal_Mass_no_fid_cuts} and \ref{fig:Dimuon_Signal_Mass_no_fid_cuts}, the reconstructed dilepton invariant mass distributions for several $m_{\Zp}$ values and both coupling values are shown before the fiducial truth-level cuts are applied. The reconstruction-level $m_{\ell\ell}$ values corresponding to the cut of Eq.~(\ref{eq:fidCuts3}) are indicated.

\FloatBarrier

Signal kinematic distributions after the full analysis selection are presented in Appendix~\ref{app:zprime:kinematics}.

\section{Data and Simulation}
\label{sec:zprime_data}
Data from proton--proton collisions at a centre-of-mass energy of $\sqrt{s}=13\,\TeV$ is used in this analysis. The data was collected by the ATLAS detector in Run~2 of the LHC between 2015 and 2018 corresponding to a total integrated luminosity of $140\,\mathrm{fb^{-1}}$.
The data were collected using single-lepton and dilepton triggers, which vary by run period and lepton channel. The full trigger lists for the electron, muon, and electron--muon channels are given in Appendix~\ref{app:zprime:data} (Tables~\ref{tab:triggersEl}--\ref{tab:triggersElMu}).
Only datasets from the Good Run List (GRL) are used, meaning that these datasets satisfy data-quality requirements for stable beam and detector conditions.

A summary of the generator chains for each background and signal process is given in Appendix~\ref{app:zprime:mc} (Table~\ref{tab:MCsamples}).

The dominant background in final states with one or more $b$-jets is $\ttbar$ production, simulated with Powheg-Box v2 interfaced with Pythia8 for showering, using the NNPDF3.0nlo PDF set. Non-all-hadronic $\ttbar$ samples with $H_T$-slices are used instead of purely dileptonic samples to avoid low MC statistics at high dilepton invariant masses; further details are given in Appendix~\ref{app:zprime:mc}.
The neutral Drell-Yan (DY) process $Z/\gamma^* \to \ell\ell$, dominant in final states without $b$-tagged jets, is simulated with Sherpa 2.2.11 using the NNPDF3.0NNLO PDF set. Inclusive dilepton mass samples are combined with mass-enhanced samples, the latter generated with a lower invariant mass cut at $120\;\GeV$ to improve statistics at high masses. Both sample types are split into three categories according to the heavy- or light-flavour content of the final state. The DY event yields are corrected with a rescaling that depends on the dilepton invariant mass from NLO to next-to-next-to-leading
order (NNLO) in $\alpha_s$, computed with VRAP 0.9~\cite{PhysRevD.69.094008} and the CT14nnlo PDF set~\cite{Dulat_2016}. Mass-dependent electroweak corrections were computed at NLO with mcsanc 1.20~\cite{BONDARENKO20132343}.
The photon-induced background is generated with Pythia~8 at LO, using the NNPDF3.1NLO LUX QED PDF~\cite{Ball:2017nwa} for the matrix element and NNPDF2.3LO for showering and hadronisation.
Single-top production is simulated with the same generator, parton shower, and PDF set as the $\ttbar$ background.
The diboson background is generated with Sherpa 2.2.11 for semi-leptonic and Sherpa 2.2.12 for fully leptonic final states, using the NNPDF3.0NNLO PDF set.
The fake-lepton background comprises events in which at least one lepton arises from a misidentified hadronic jet; it is estimated with the data-driven Matrix Method. Contributions come from $W$+jets, with one prompt lepton and one fake, and from multijet events, where both leptons are fake.

\section{Event Selection}
\label{sec:zprime_evsel}

Only datasets satisfying the Good Run List (GRL) requirements are used, as specified in Section~\ref{sec:zprime_data}.
Event-level vetoes are applied to reject bad or corrupt events, including LAr noise bursts, Tile calorimeter corrupted events, SCT recovery events, and incomplete events.
Events are also rejected if bad jets or muons are present.


Events are required to have a primary vertex with at least two associated tracks (Section~\ref{sec:reco:jets}).
The primary vertex is selected as the one with the largest \(\Sigma \pT^2\),
where the sum is over all tracks with transverse momentum \(\pT > \SI{0.4}{\GeV}\) that are associated with the vertex.

Signal events are characterised by final states with exactly two opposite-sign, same-flavour leptons (dimuon or dielectron) and at least one $b$-tagged jet. Signal muons satisfy the \texttt{High-pT} identification working point~\cite{PERF-2015-10} with $\pT > 25~\GeV$ and $|\eta|<2.5$ (Section~\ref{sec:reco:muons}); signal electrons satisfy the \texttt{Medium} likelihood identification working point~\cite{EGAM-2018-01} with $\ET > 25~\GeV$ and $|\eta|<2.47$ (excluding $1.37<|\eta|<1.52$) (Section~\ref{sec:reco:electrons}). Jets are reconstructed with the \Antikt\ algorithm ($R=0.4$)~\cite{Cacciari:2008gp} with $\pT>20~\GeV$ (Section~\ref{sec:reco:jets}), and $b$-jets are identified using the \texttt{DL1r} algorithm at the 85\% efficiency working point (\texttt{DL1r} score $>0.665$)~\cite{b_tagging_ATLAS} (Section~\ref{sec:reco:btag}). Full object selection criteria are described in Chapter~\ref{chp:objects}.

An event pre-selection is applied, summarised in Table~\ref{tab:preselection}.
Electrons and muons must pass the di-electron or single-muon trigger and satisfy the object definition requirements. Following ATLAS Combined Performance Group recommendations, a combination of low-$p_{\textrm{T}}$ triggers is used. This is necessary because the trigger scale factors used in the experimental systematic uncertainties are calculated for these trigger combinations, even though a $p_{\textrm{T}}^{\ell}>65\;\textrm{GeV}$ requirement is applied to both signal leptons. In the dielectron channel, loose or very loose ID requirements are needed at trigger level to derive the fake background estimate, and such triggers are only available with low $p_{\textrm{T}}$ thresholds. The single-muon trigger has higher acceptance than the dimuon trigger, so it is used for the dimuon channel. The trigger chains for both channels are consistent with~\cite{ATLAS_dilepton_2019} albeit one update in the 2018 electron trigger. The transverse momentum of the leptons needs to be larger than $65\,\GeV$ and the transverse momentum of the jets should be larger than $20\,\GeV$. In order to exclude the $Z$-peak, the invariant mass of the two leptons is required to be above $130\,\GeV$.
Jet cleaning is applied. Events are vetoed if a muon is flagged as ``bad'' (indicating degraded momentum resolution) or if an electron has a cluster in the calorimeter transition region $1.37 < |\eta_{\mathrm{cluster}}| < 1.52$. Requirements on $\mathcal{S}(E_{\textrm{T}}^{\textrm{miss}})$ and $\min(m_{\ell b})$ apply only to specific analysis regions, as described in Section~\ref{sec:zprime_strategy}. These cuts are further motivated in Section~\ref{sec:zprime_SR}. The \MET\ significance $\metsig$ is defined in Section~\ref{sec:reco:met} (Equation~\ref{eq:metsig}).
The variable
$\min(\min(m_{\ell b}))$ is the minimum invariant mass of any lepton-$b$-jet pair in the event.

\begin{table}[h]
	\centering
	\caption{Event selection for the $\ell^{+}\ell^{-} + j_{b}/j_{b}j_{b}$ analysis. The inclusive and exclusive event selections are clearly marked, where the latter is applied in addition to the former. The $E^{\textrm{miss}}_{\textrm{T}}$ cuts specificied in the last two rows pertain to specific regions only, as described in Section~\ref{sec:zprime_strategy}.} 
	\label{tab:preselection}
	\begin{tabular}{ccc}
		\hhline{===}
		
		Final state & $\mu^{+}\mu^{-}$ & $e^{+}e^{-}$
		
		\\ \hline
		
		Cut title                           & \multicolumn{2}{c}{Requirement}                               \\ \hline
		
		\hspace{0.1cm} & \multicolumn{2}{c}{Inclusive}                                                                                                                                                                   \\ \hline
			
		Jet clean                      & Yes    & Yes                                                                                                                                                            \\
		$N_{\textrm{Signal leptons}}$               & Exactly two  & Exactly two                                                                                                                                                    \\
		Bad lepton veto                & Yes      & No                                                                                                                                                      \\
		$1.37< |\eta|<1.52$ & No & Yes\\
		$\textrm{min}(p_{\textrm{T}}^{\ell})$ & $65 \; \textrm{GeV}$   & $65 \; \textrm{GeV}$                   \\
		$\textrm{min}(m_{\ell\ell})$   & $130 \; \textrm{GeV}$     & $130 \; \textrm{GeV}$                   \\ 
		\hhline{===}
		\hspace{0.1cm} & \multicolumn{2}{c}{Exclusive ($b$-jet channel)}                                                                                                                                                 \\ \hline		
		$\textrm{min}(p^{j}_{\textrm{T}})$ & \multicolumn{2}{c}{$20 \; \textrm{GeV}$}                                                                                                                                           \\
		$\textrm{max}(|\eta|)$ (only for $b$-tagged jets)         & \multicolumn{2}{c}{2.5} \\
		$b$-jet tagging               & \multicolumn{2}{c}{$N(j_{\texttt{DL1r} > 0.665}) = 0, 1, \geq 2$}                                                                                                                                       \\
		\hhline{===}
		\hspace{0.1cm} & \multicolumn{2}{c}{Additional cuts for $b$-jet channel}                                                                                                                                                 \\ \hline		
		$\textrm{max}(E^{\textrm{miss}}_{\textrm{T}})$ & \multicolumn{2}{c}{$20 \; \textrm{GeV}$}                                                                                                                                           \\
		$\textrm{max}(\mathcal{S}(E^{\textrm{miss}}_{\textrm{T}}))$ & \multicolumn{2}{c}{$5.0 $}                                                                                                                                           \\
		$\min(\min(m_{\ell b}))$ & \multicolumn{2}{c}{$155\,\mathrm{GeV} $}                                                                                                                                           \\
		\hhline{===}
	\end{tabular}
	
\end{table}

\section{Background Modelling}
\label{sec:zprime_bkg_modelling}

\subsection{Multijet background}

The multijet background arises from fake leptons --- reconstructed leptons that do not originate from a prompt hard-scatter process, typically from hadronic jets or secondary heavy-flavour decays.
It is estimated using the MM~\cite{ATLAS:2022swp}, following the same approach as described for the QBH analysis in Section~\ref{sec:fakes}.
In that single-lepton case, the tight/loose lepton categories give a $2\times 2$ system of equations~\cite{ATLAS:2022swp}.
Here, the dilepton final state produces four orthogonal categories ($TT$, $TL$, $LT$, $LL$), requiring a $4\times 4$ matrix inversion to extract the fake contribution to the SR yield.
In the electron channel, a mixed fake efficiency combining light- and heavy-flavour estimates is used, with the mixing weight determined from MC as a function of $b$-jet multiplicity.

\FloatBarrier

\subsection{Background composition}

\begin{figure}[h]
	\centering
	\captionsetup[subfigure]{labelformat=empty}
	\subfloat[(a)]{
		\includegraphics[width=0.4\textwidth]{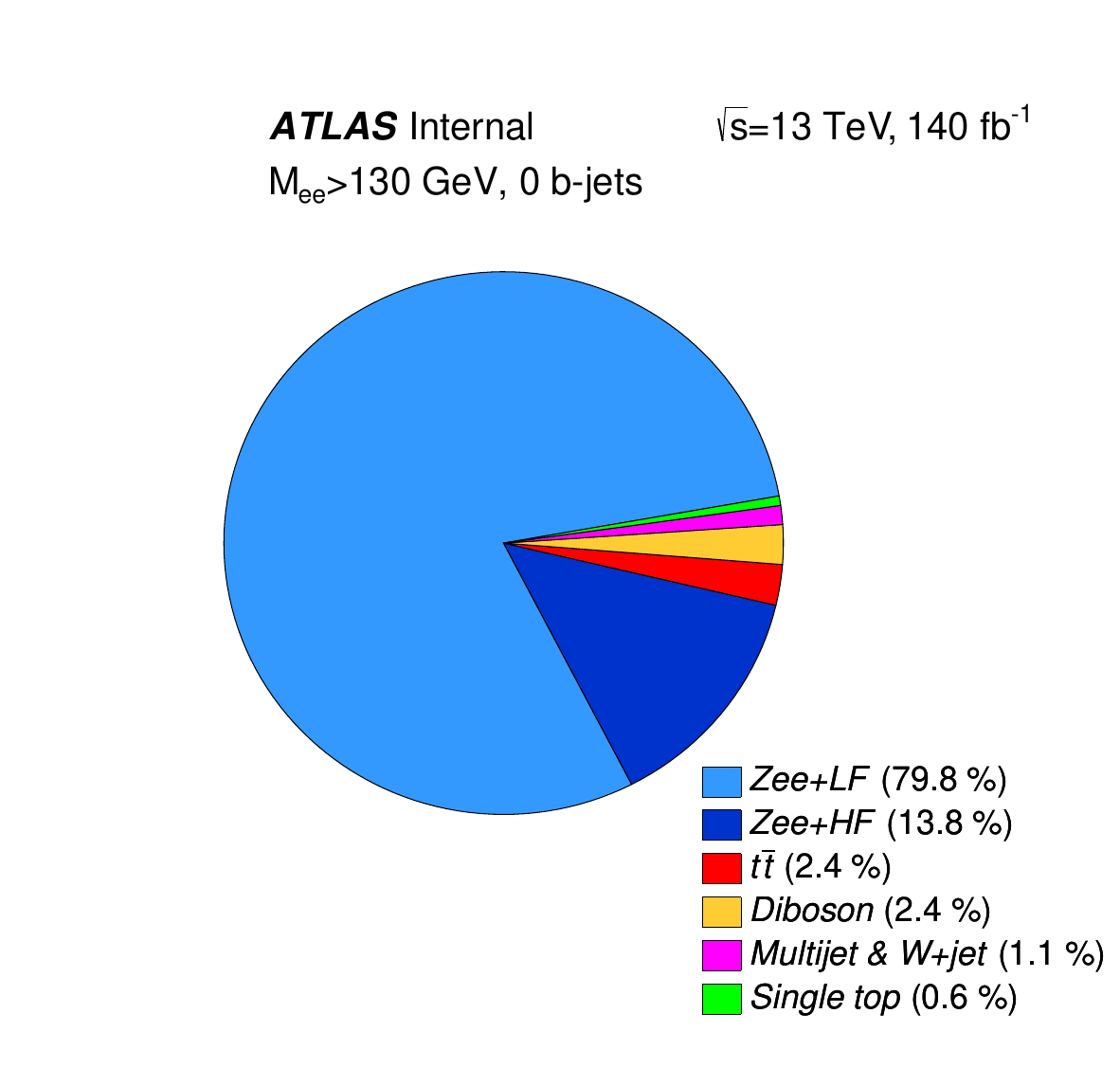}
		\label{fig:bkg_preselection_0b_ele}
	}
	\subfloat[(b)]{
		\includegraphics[width=0.4\textwidth]{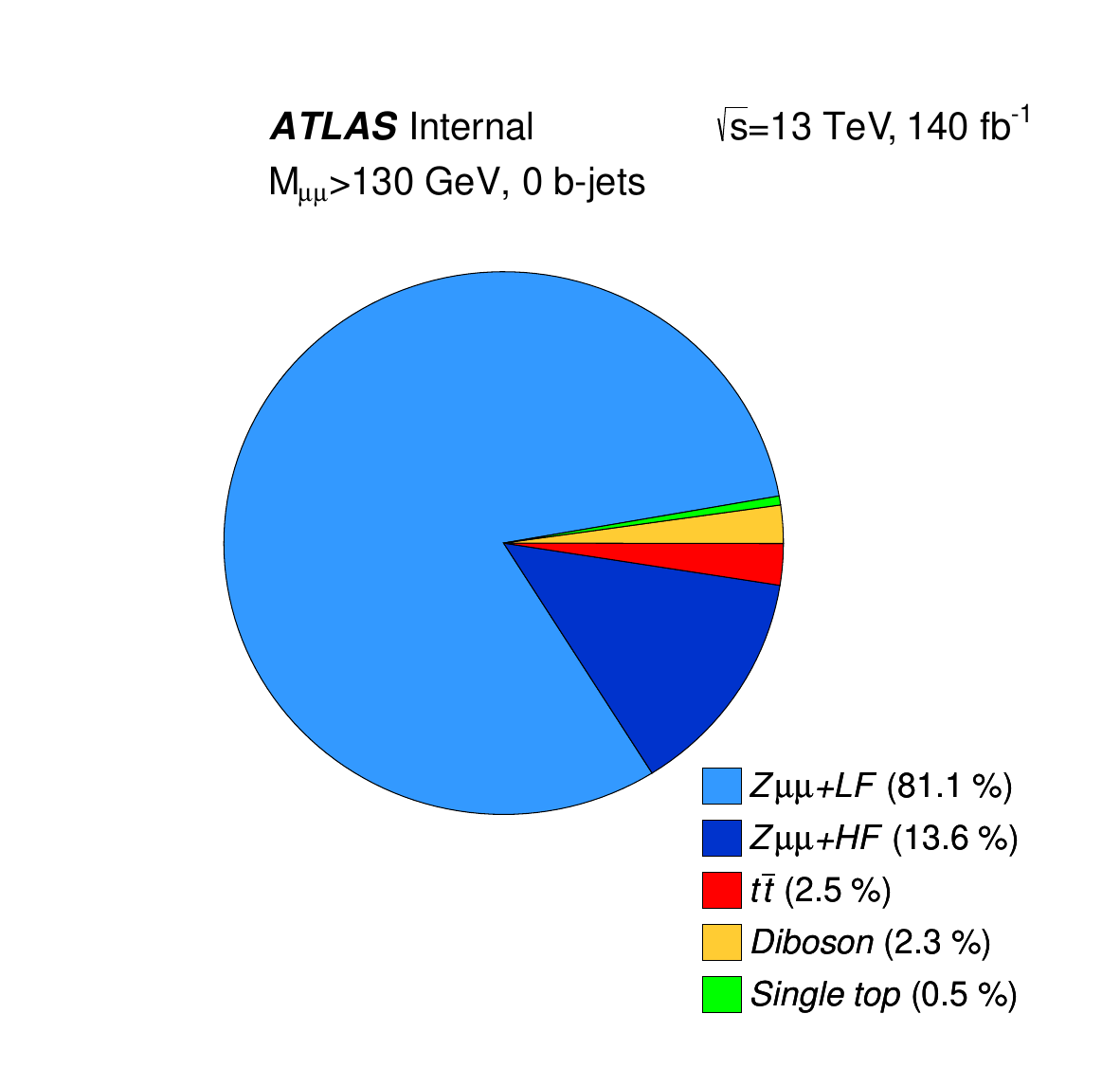}
		\label{fig:bkg_preselection_0b_mu}
	}

	\subfloat[(c)]{
		\includegraphics[width=0.4\textwidth]{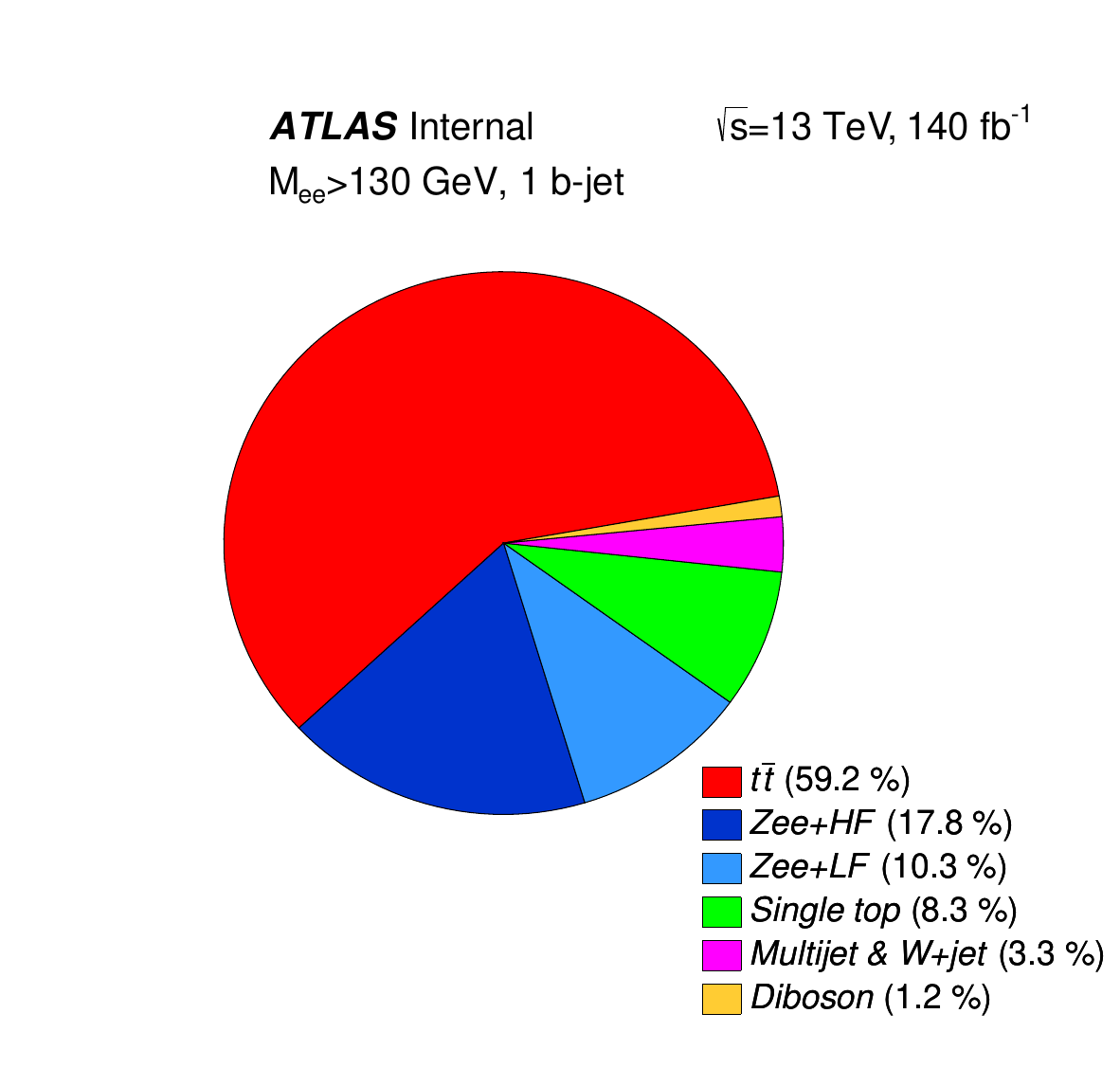}
		\label{fig:bkg_preselection_1b_ele}
	}
	\subfloat[(d)]{
		\includegraphics[width=0.4\textwidth]{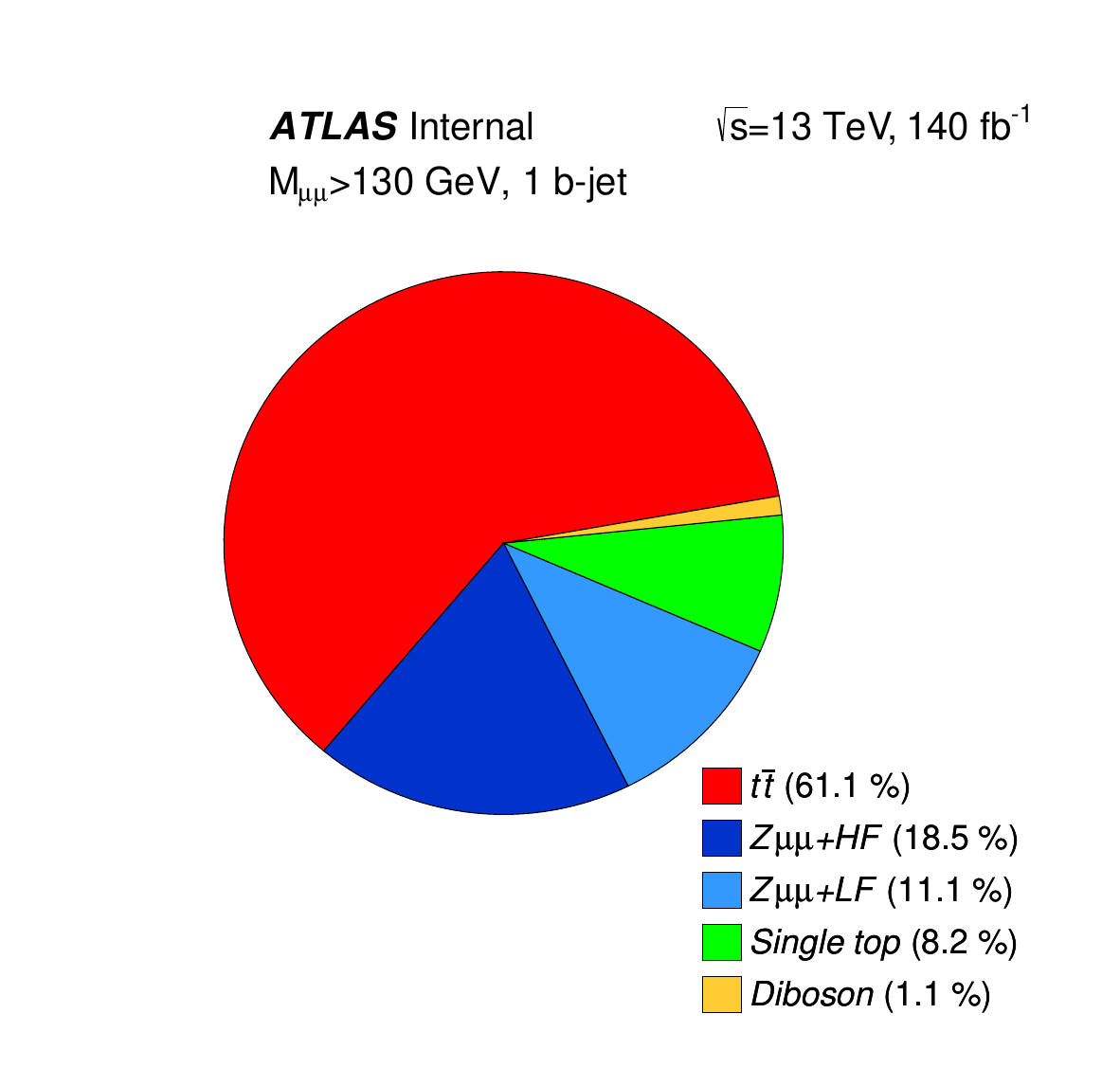}
		\label{fig:bkg_preselection_1b_mu}
	}

	\subfloat[(e)]{
		\includegraphics[width=0.4\textwidth]{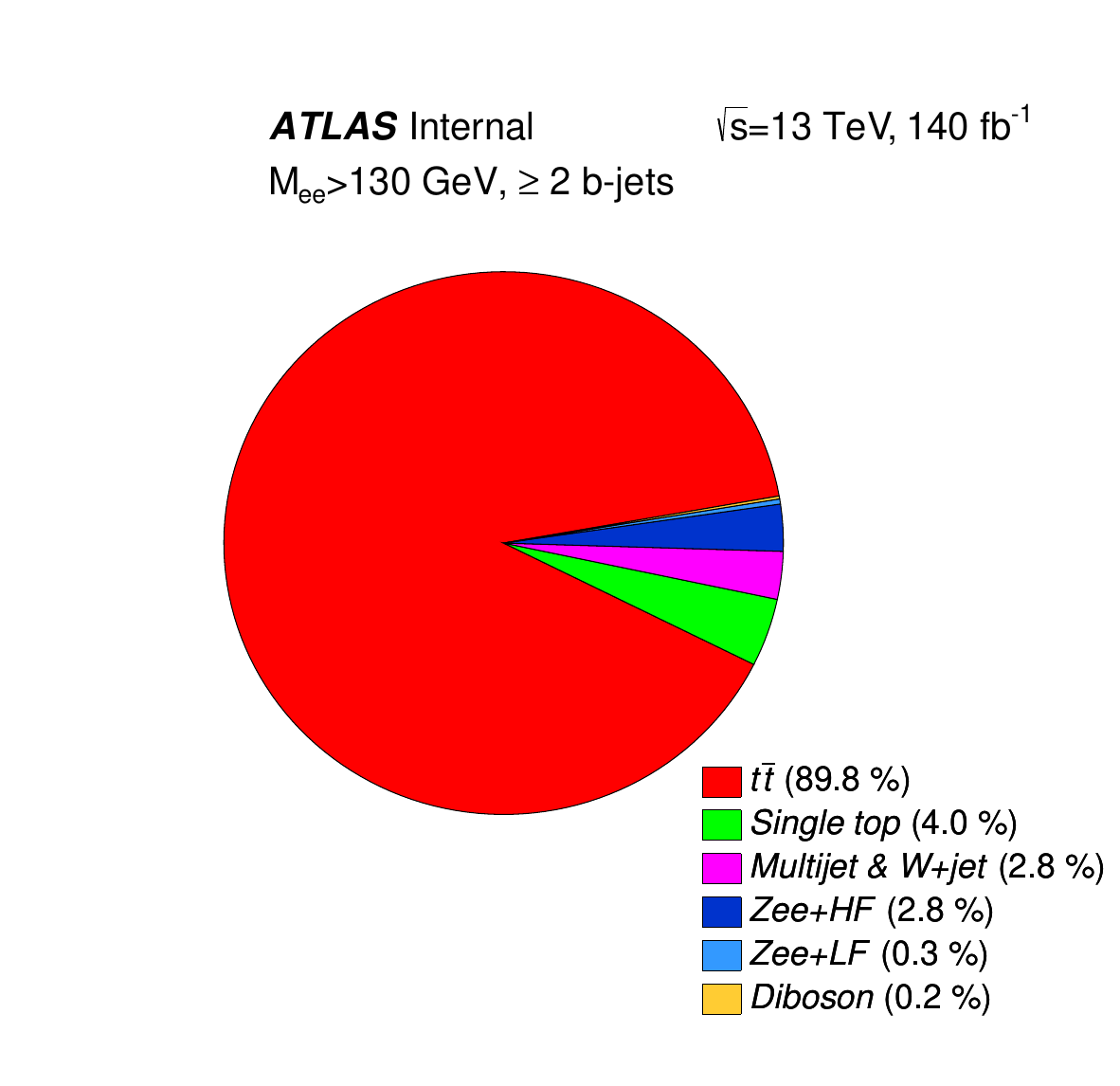}
		\label{fig:bkg_preselection_atleast2b_ele}
	}
	\subfloat[(f)]{
		\includegraphics[width=0.4\textwidth]{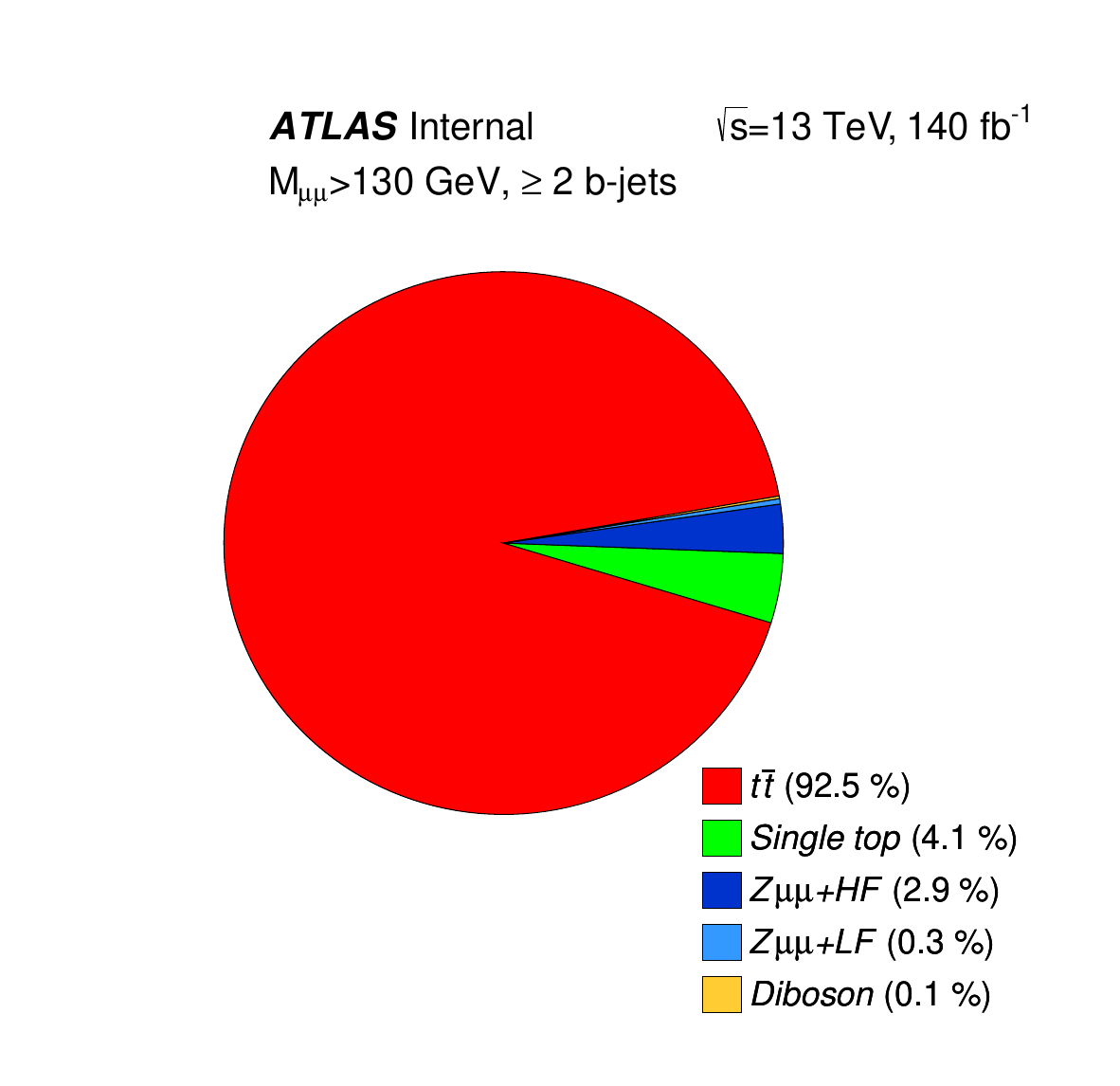}
		\label{fig:bkg_preselection_atlaest2b_mu}
	}

	\caption{Background composition for the event preselection with different $b$-jet multiplicities in the final states: (a) 0 $b$-jets, electron channel; (b) 0 $b$-jets, muon channel; (c) 1 $b$-jet, electron channel; (d) 1 $b$-jet, muon channel; (e) $\geq2$ $b$-jets, electron channel; (f) $\geq2$ $b$-jets, muon channel. The requirements on $\metsig$ and $\min(m_{\ell b})$ are not applied.}
	\label{fig:bkg_preselection}
\end{figure}

FIG.~\ref{fig:bkg_preselection} shows the background composition at pre-selection level for zero, one, and at least two $b$-jets. The DY background dominates in final states without $b$-tagged jets and decreases with increasing $b$-jet multiplicity, while the $\ttbar$ background shows the opposite trend. All other processes contribute below 10\%.

\FloatBarrier

\subsection{Background Control Regions}

Since the analysis searches for an excess above the SM prediction, the background processes must be well modelled in both shape and normalisation. Dedicated control regions, orthogonal to the signal region, are defined for this purpose. For the Z control region, this is reached by limiting the event preselection to $m_{\ell\ell}<300\,\GeV$. For the top control region, a different selection orthogonal to the signal region is chosen, as further explained in the following.

\subsubsection{Top Control Region}

\begin{figure}[H]
    \centering
    \includegraphics[width=0.45\textwidth]{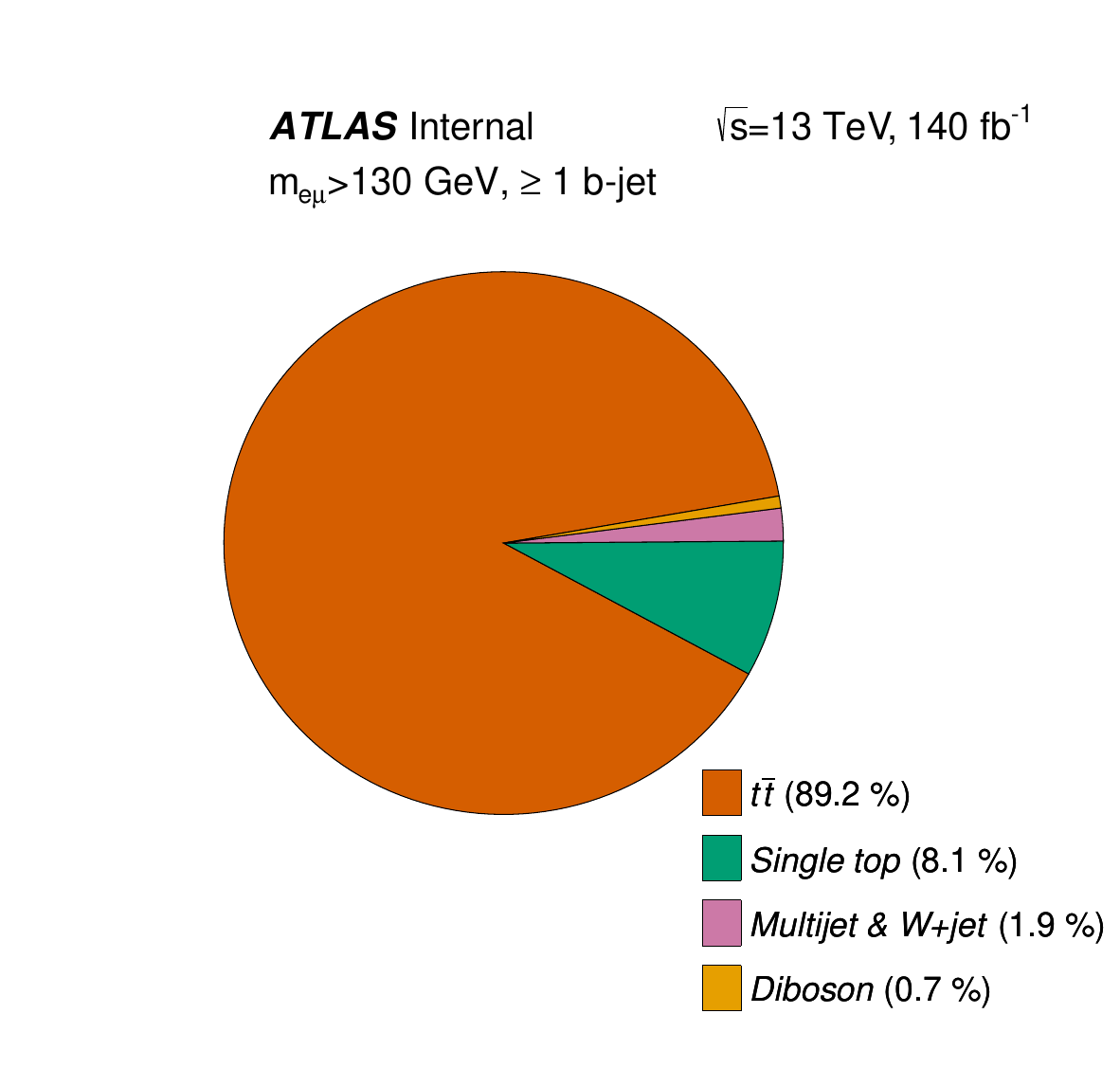}
    \caption{Background composition of the Top Control Region.}
    \label{fig:TopControl_Pie}
\end{figure}

Since $\ttbar$ production dominates the background in regions with at least one $b$-jet, a top control region is defined to validate this background.
For this region the following event selection is applied:

\begin{itemize}
	\item one muon satisfying the High-pT identification working point~\cite{PERF-2015-10} (Section~\ref{sec:reco:muons}),
	\item one electron satisfying the Medium likelihood identification working point~\cite{EGAM-2018-01} (Section~\ref{sec:reco:electrons}),
	\item opposite lepton charge.

\end{itemize}
Requirements regarding the invariant mass of the two leptons, the transverse momenta, passed triggers, jet cleaning and lepton vetos are the same as in the pre-selection (see Table 
\ref{tab:preselection}).
FIG.~\ref{fig:TopControl_Pie} shows the composition of the top control region. The $\ttbar$ background has a contribution of nearly 90\%, so the purity of the targeted
background is high. The single top background contributes with roughly 8\% to the top control region.
Contributions of approximately 1\% come from the multijet and the diboson background each.

\begin{figure}[h]
	\centering
	\captionsetup[subfigure]{labelformat=empty}
	\subfloat[(a)]{
		\includegraphics[width=0.33\textwidth]{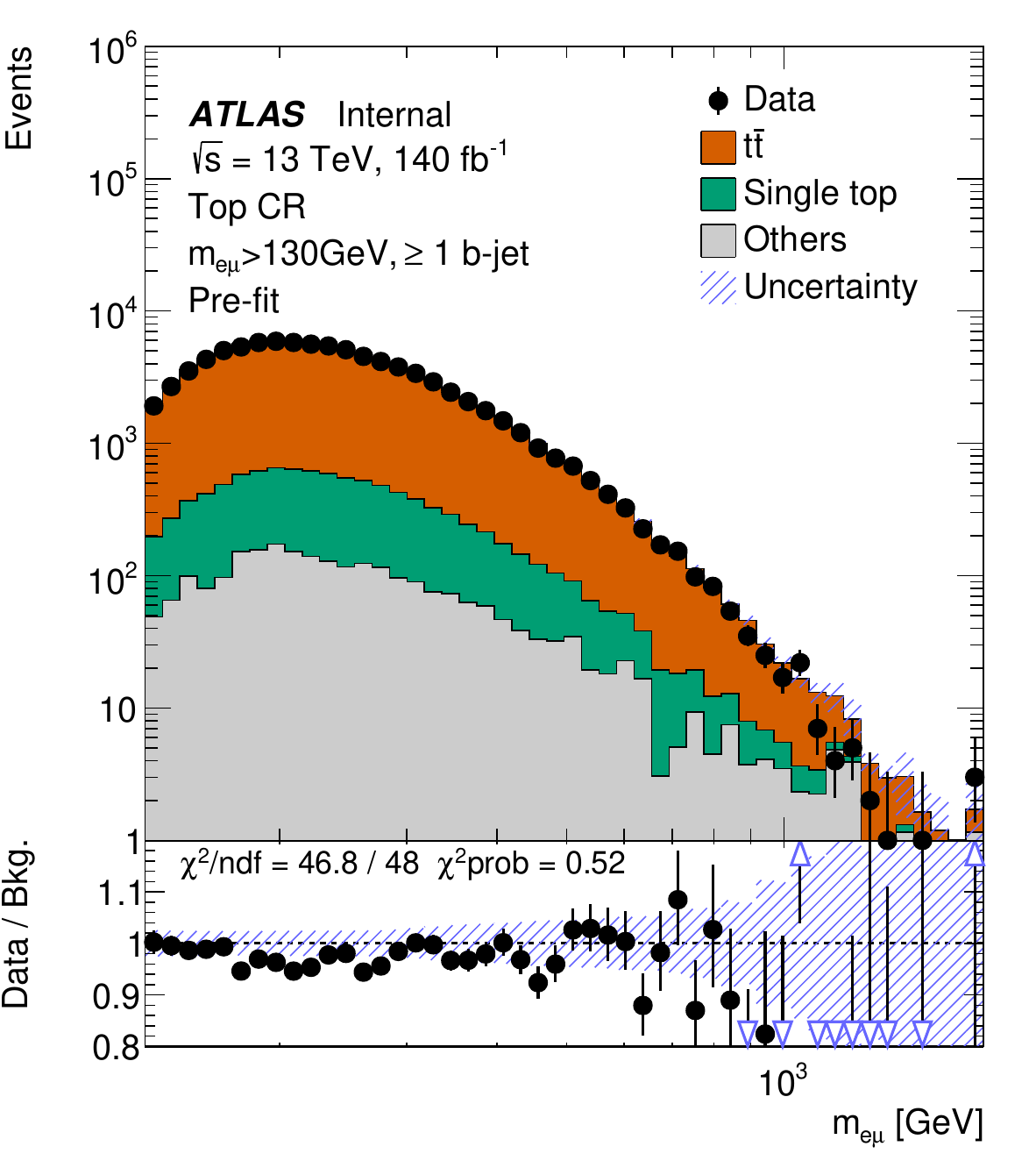}
		\label{fig:TC_inv_mass}
	}
	\subfloat[(b)]{
		\includegraphics[width=0.33\textwidth]{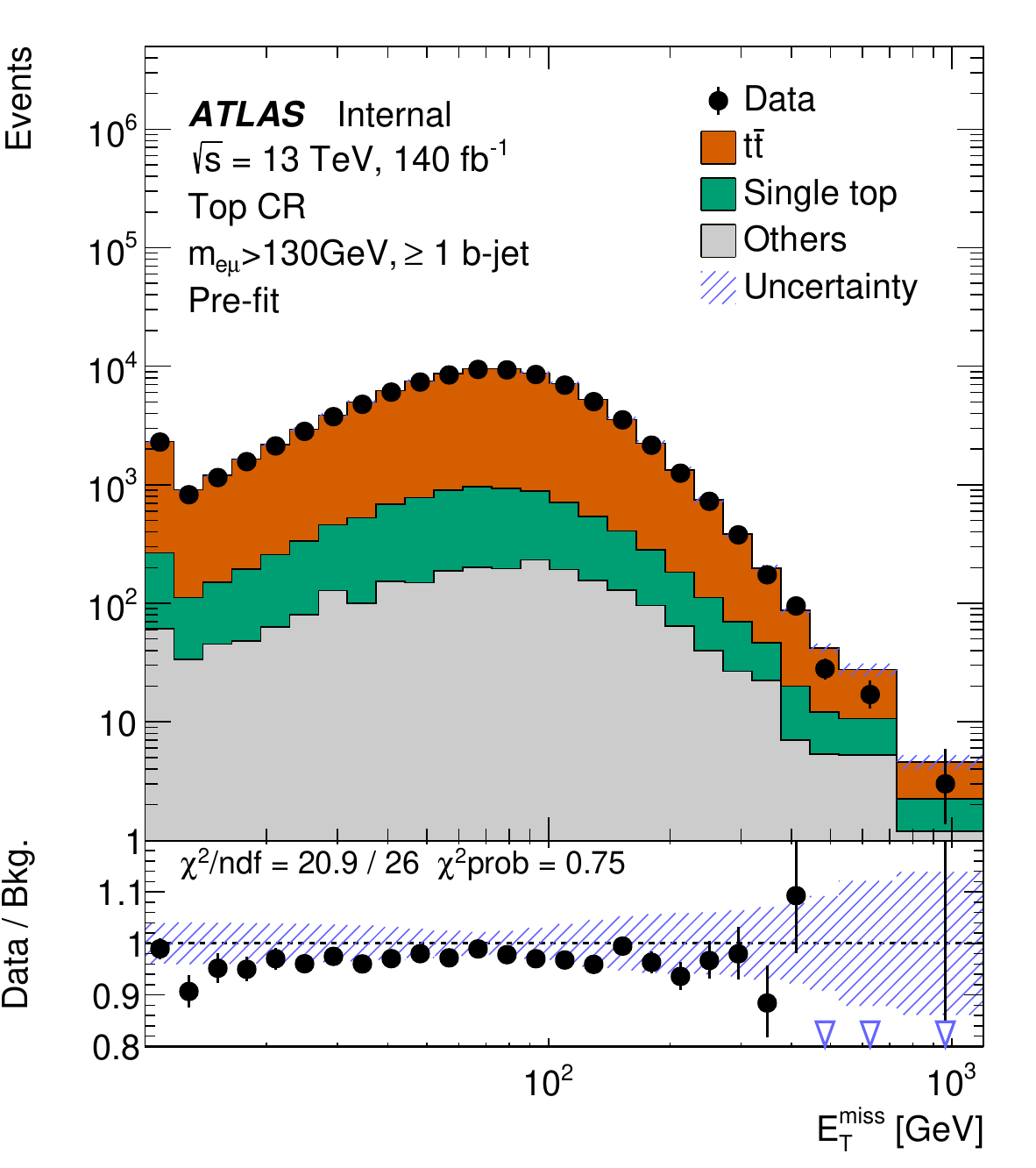}
		\label{fig:TC_met}
	}
	\subfloat[(c)]{
		\includegraphics[width=0.33\textwidth]{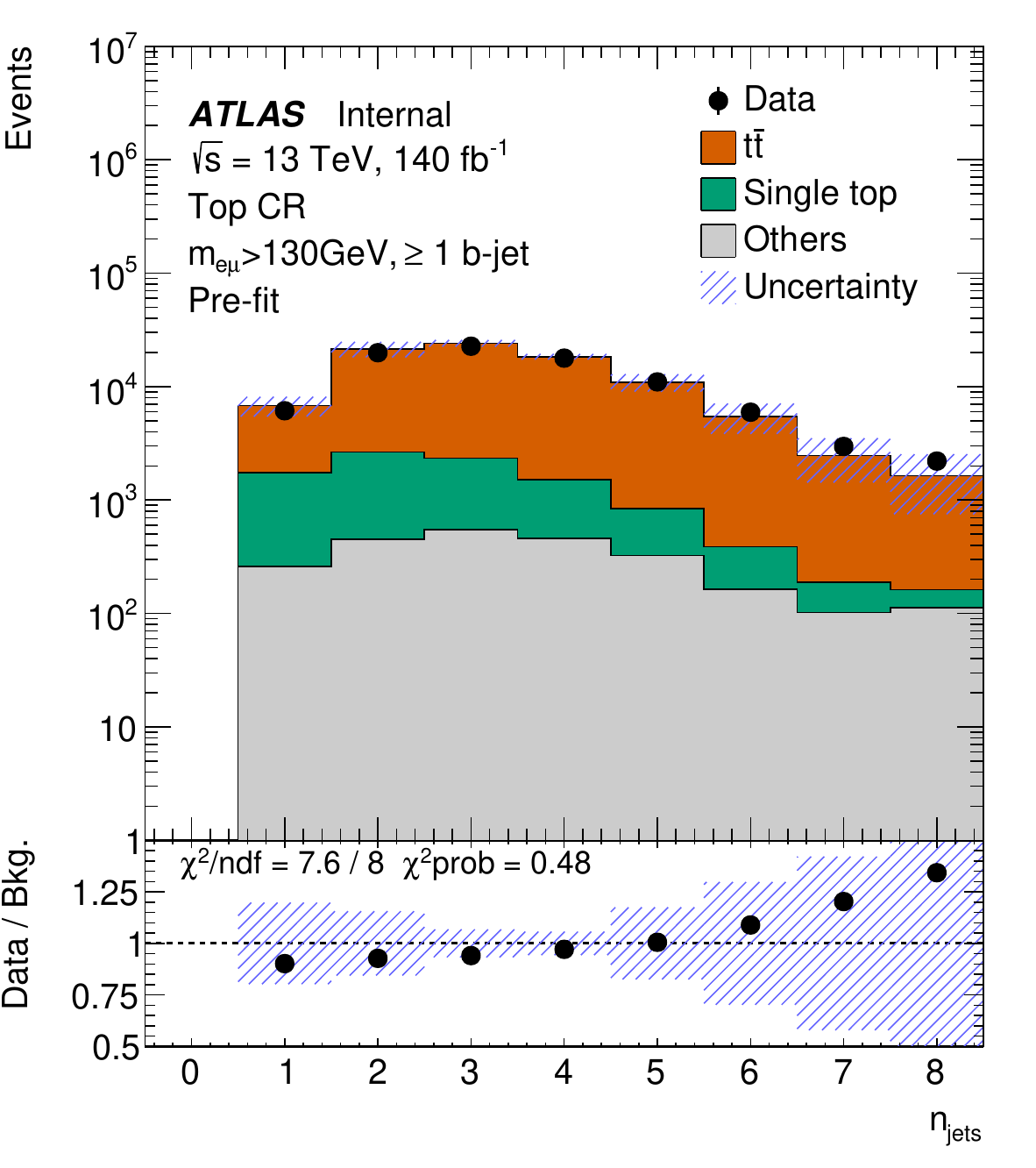}
		\label{fig:TC_njets}
	}
	\hfill
	\subfloat[(d)]{
		\includegraphics[width=0.33\textwidth]{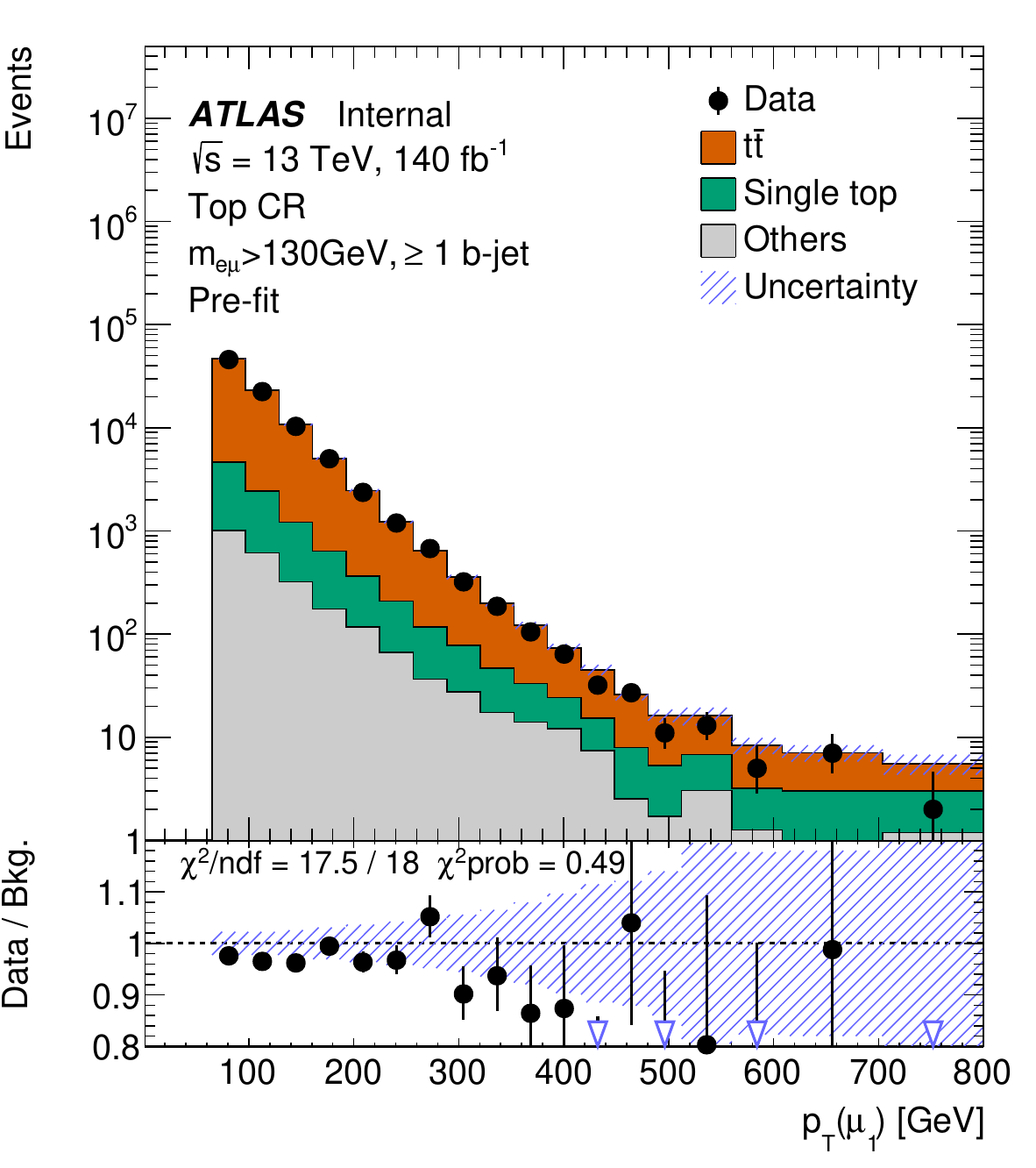}
		\label{fig:TC_mu_pt}
	}
	\subfloat[(e)]{
		\includegraphics[width=0.33\textwidth]{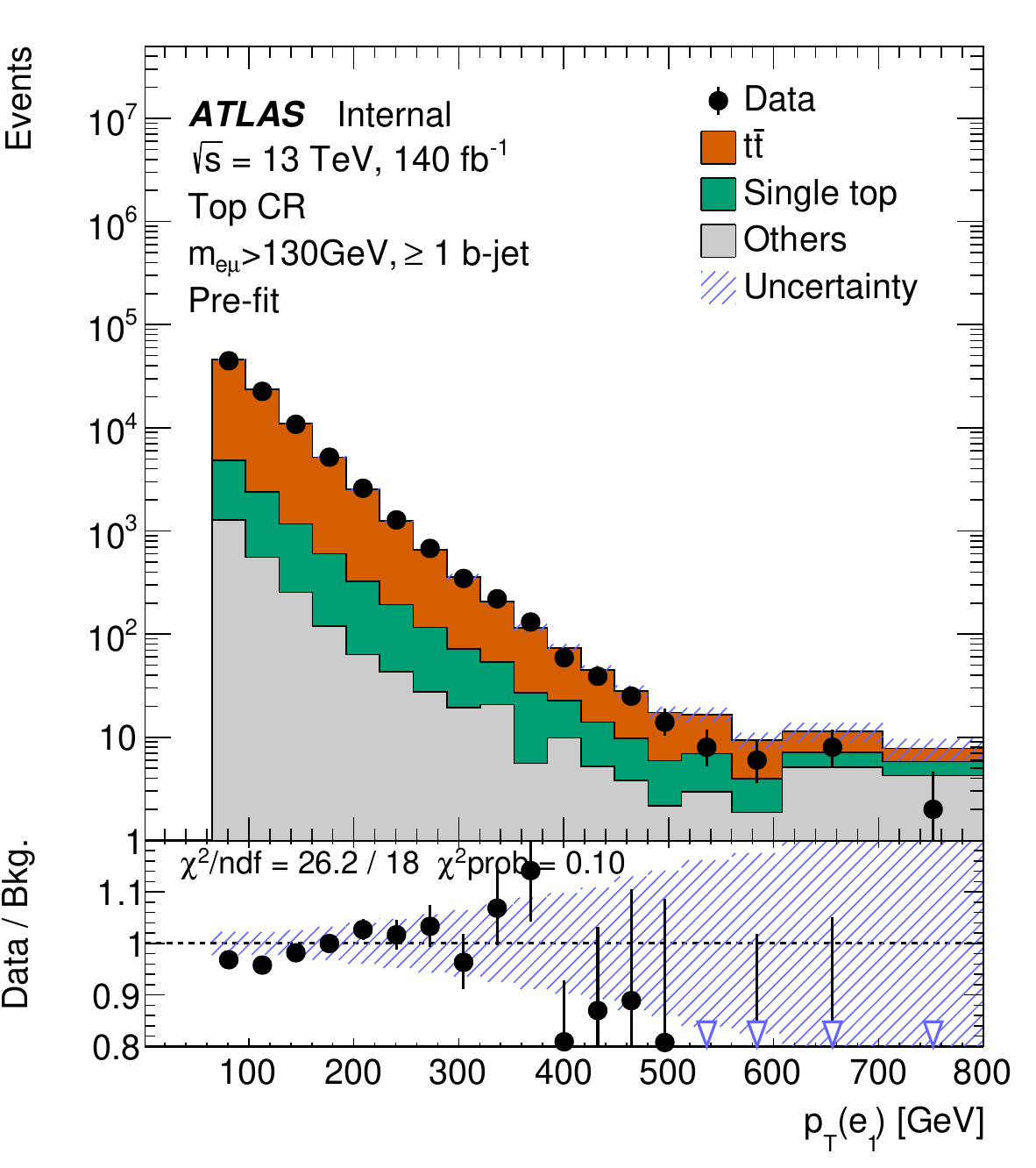}
		\label{fig:TC_e_pt}
	}
	\subfloat[(f)]{
		\includegraphics[width=0.33\textwidth]{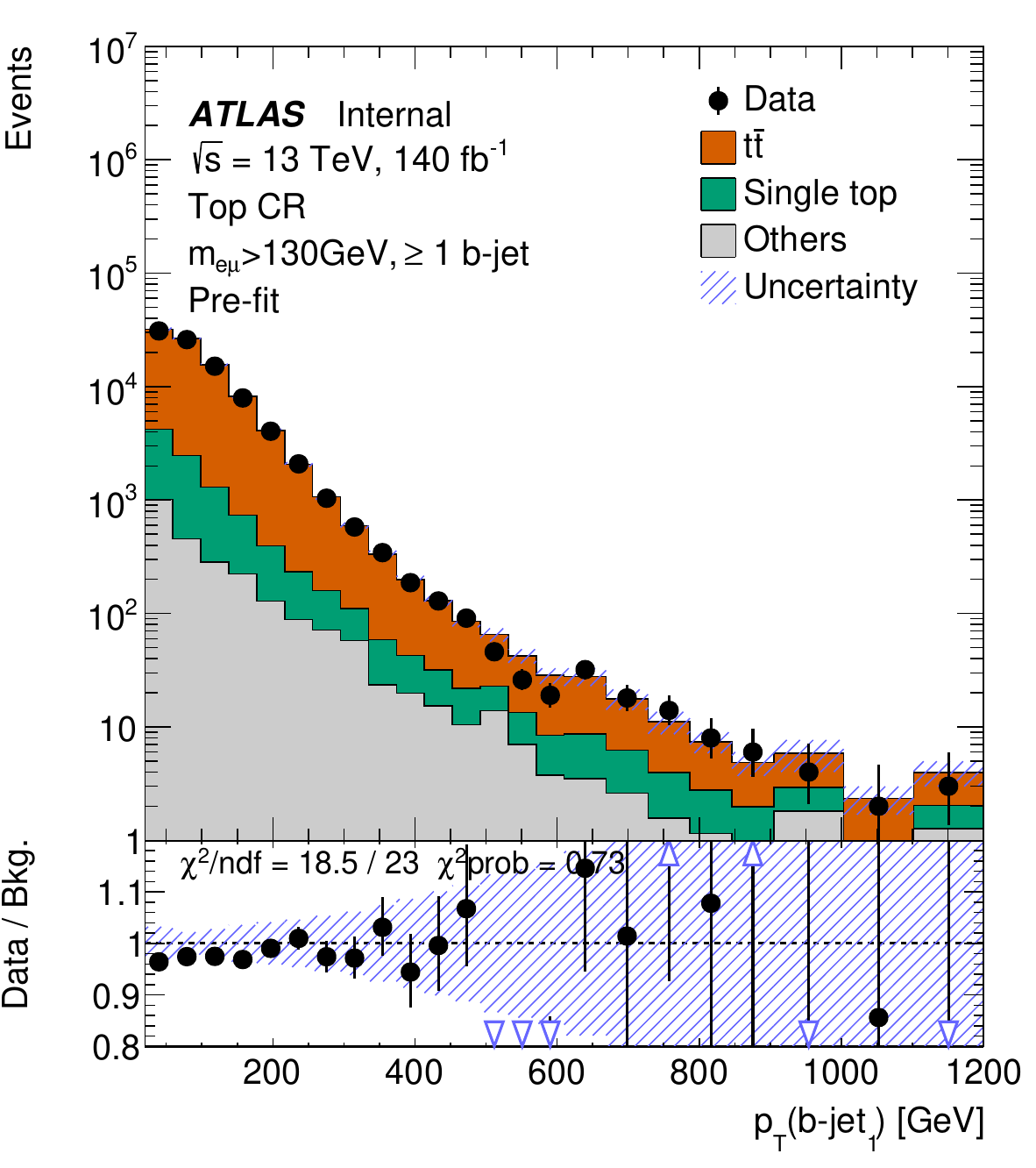}
		\label{fig:TC_bjet_pt}
	}

	\caption{Data/MC comparison in the top Control Region for (a) the invariant dilepton mass, (b) the missing transverse momentum, (c) the jet multiplicity, (d) the muon $\pT$, (e) the electron $\pT$, and (f) the $\pT$-leading $b$-jet $\pT$. The uncertainty band includes statistical and background systematic uncertainties, where only the shape of theoretical modelling uncertainties is taken into account.}
	\label{fig:TC_datamc}
\end{figure}

FIG.~\ref{fig:TC_datamc} shows the data/MC comparison in the top control region for the dilepton invariant mass, $\met$, jet multiplicity, and lepton and $b$-jet $p_T$. Good agreement is observed for $p_T<400\;\text{GeV}$, with some deviations at higher $p_T$ where statistics are limited. Overall, the $\ttbar$ background is well modelled.

\FloatBarrier

\subsubsection{$\bm{Z}$ Control Region}

The $Z\rightarrow\ell\ell$ (DY) background dominates in final states without $b$-jets and decreases with increasing $b$-jet multiplicity. Since events with and without $b$-jets have different kinematics, dedicated control regions are defined separately for the two categories, in both the muon and electron channels.
Events passing the the event preselection criteria in Table \ref{tab:preselection} with an invariant dilepton mass in the range between $130\,\GeV<M_{\ell\ell}<300\,\GeV$ are selected.

\subsubsection*{$Z$+LF CR}

\begin{figure}[h!]
	\centering
	\captionsetup[subfigure]{labelformat=empty}
	\subfloat[(a)]{
		\includegraphics[width=0.45\textwidth]{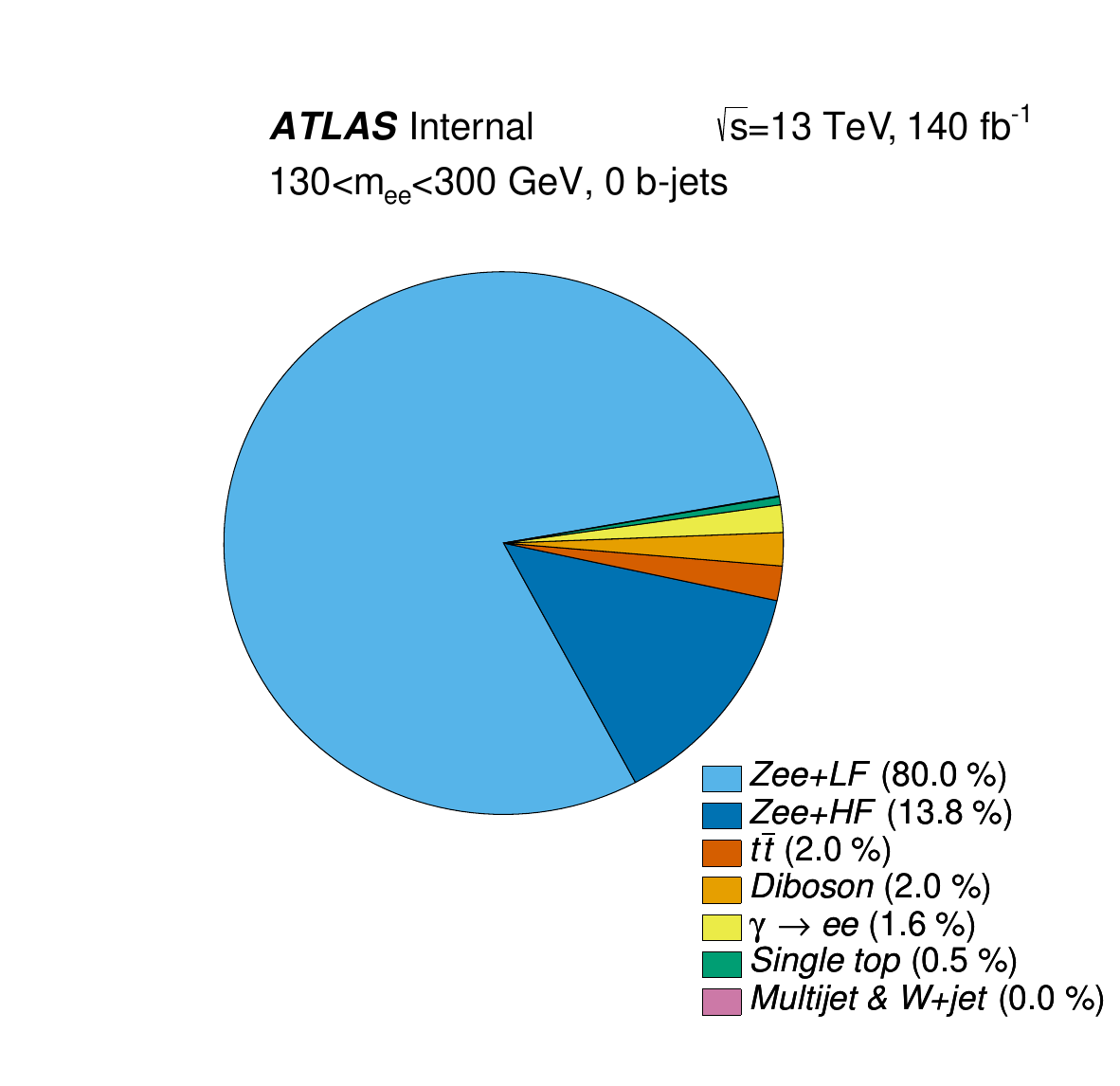}
		\label{fig:ZCR_0b_composition_ele}
	}
	\hfill
	\subfloat[(b)]{
		\includegraphics[width=0.45\textwidth]{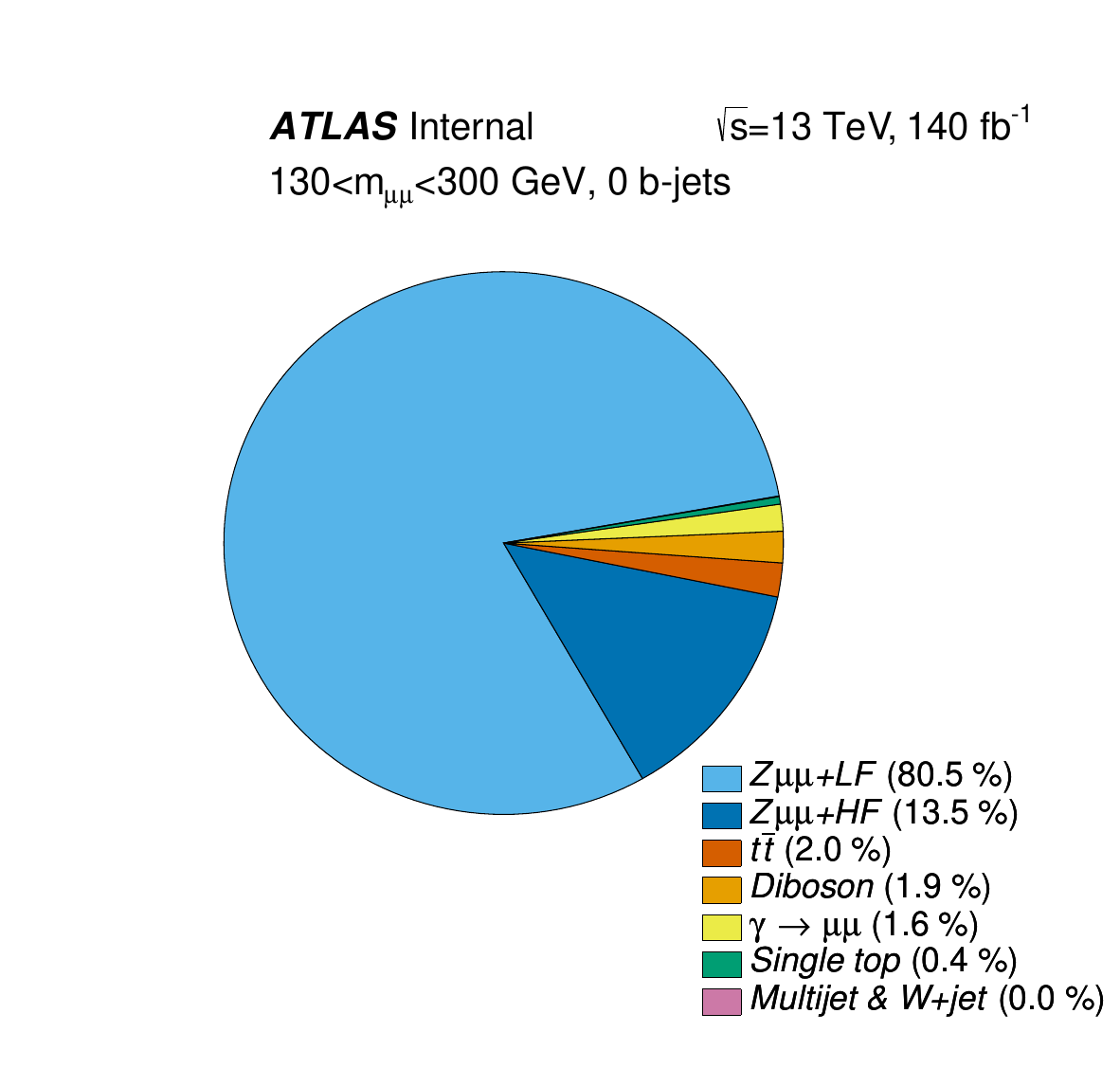}
		\label{fig:ZCR_0b_composition_mu}
	}
	\caption{Background composition of the $Z$+LF Control Region in the (a) electron and (b) muon channel.}
	\label{fig:ZCR_0b_composition}
\end{figure}

First, the $Z$ CR with zero $b$-jets in the final state is investigated. FIG.~\ref{fig:ZCR_0b_composition} shows the background composition of this CR in the muon and electron channel.

It can be seen that the $Z$+jets process is split into a light flavour ($Z$+LF) and a heavy flavour ($Z$+HF) component in order to check the normalisation of both components separately. The splitting is done based on the different filters on generator level for $b$-quarks and $c$-quarks as explained in Chapter \ref{sec:zprime_data}. 

In the zero-$b$-jet CR, $Z$+LF is the largest component at $\sim$80\%, followed by $Z$+HF. All other backgrounds contribute 3\% or less.

\begin{figure}[h]
	\centering
	\captionsetup[subfigure]{labelformat=empty}
	\subfloat[(a)]{
		\includegraphics[width=0.33\textwidth]{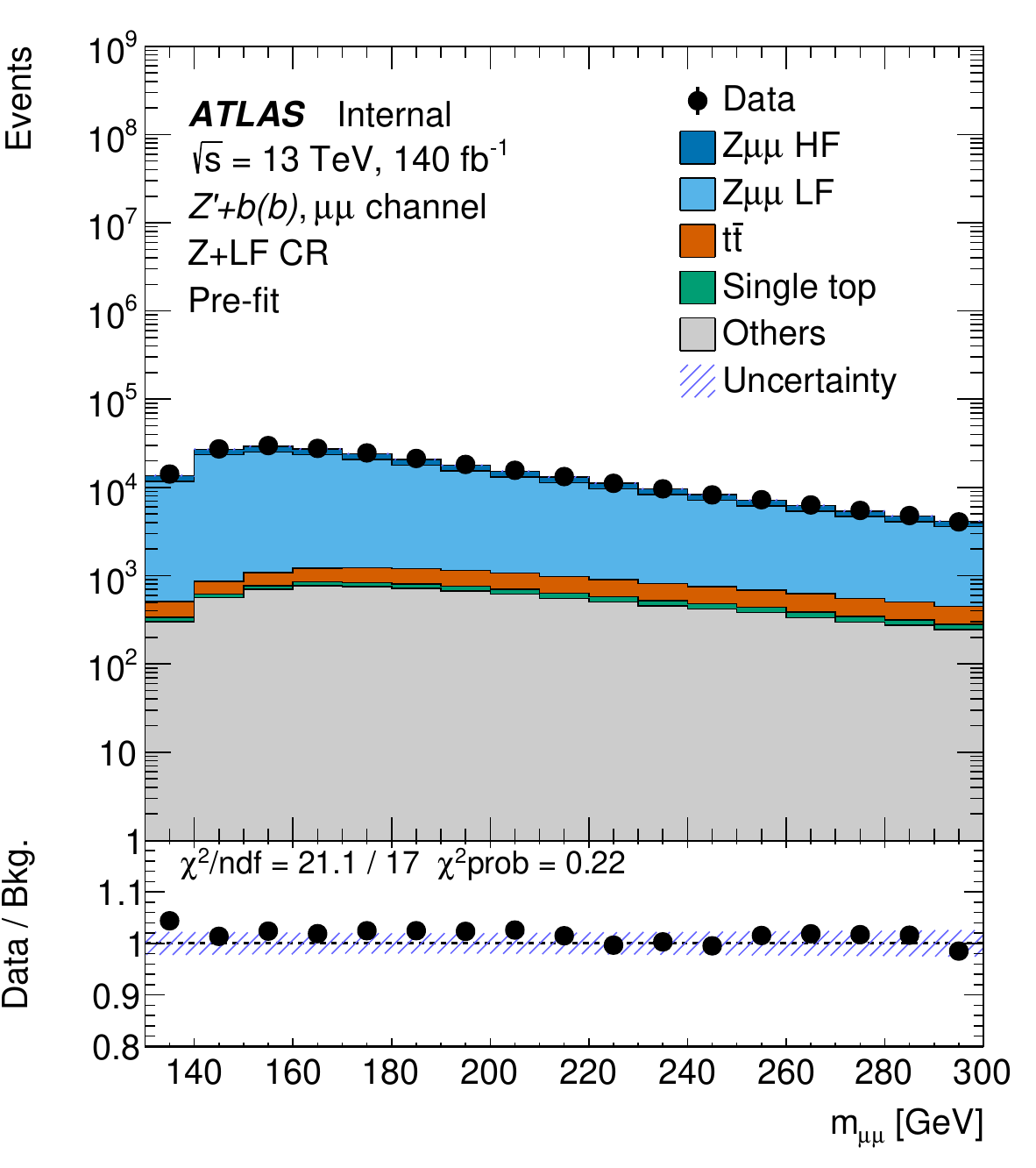}
		\label{fig:ZCR_0b_datamc_inv_mass_mu}
	}
	\subfloat[(b)]{
		\includegraphics[width=0.33\textwidth]{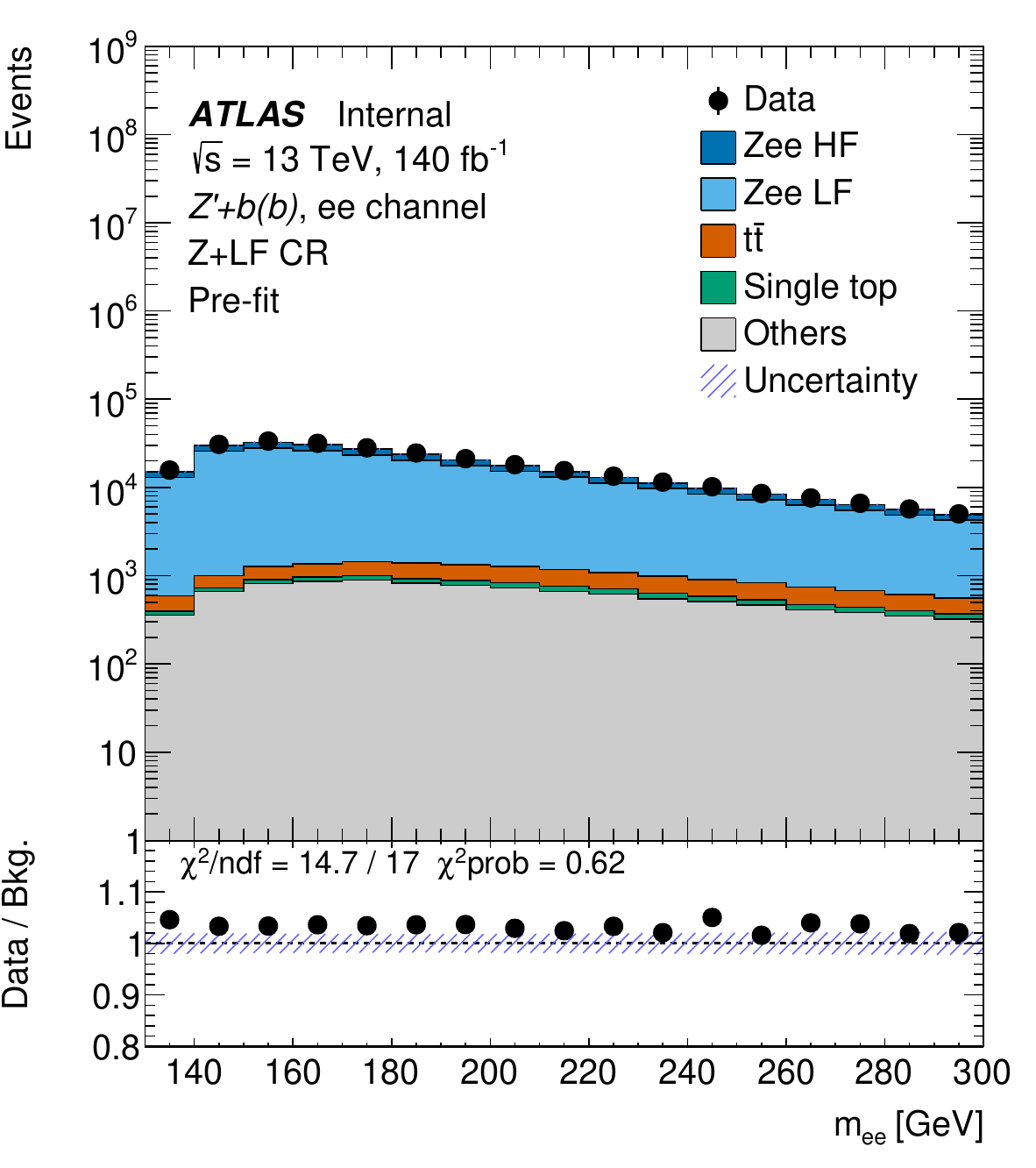}
		\label{fig:ZCR_0b_datamc_inv_mass_ele}
	}

	\subfloat[(c)]{
		\includegraphics[width=0.33\textwidth]{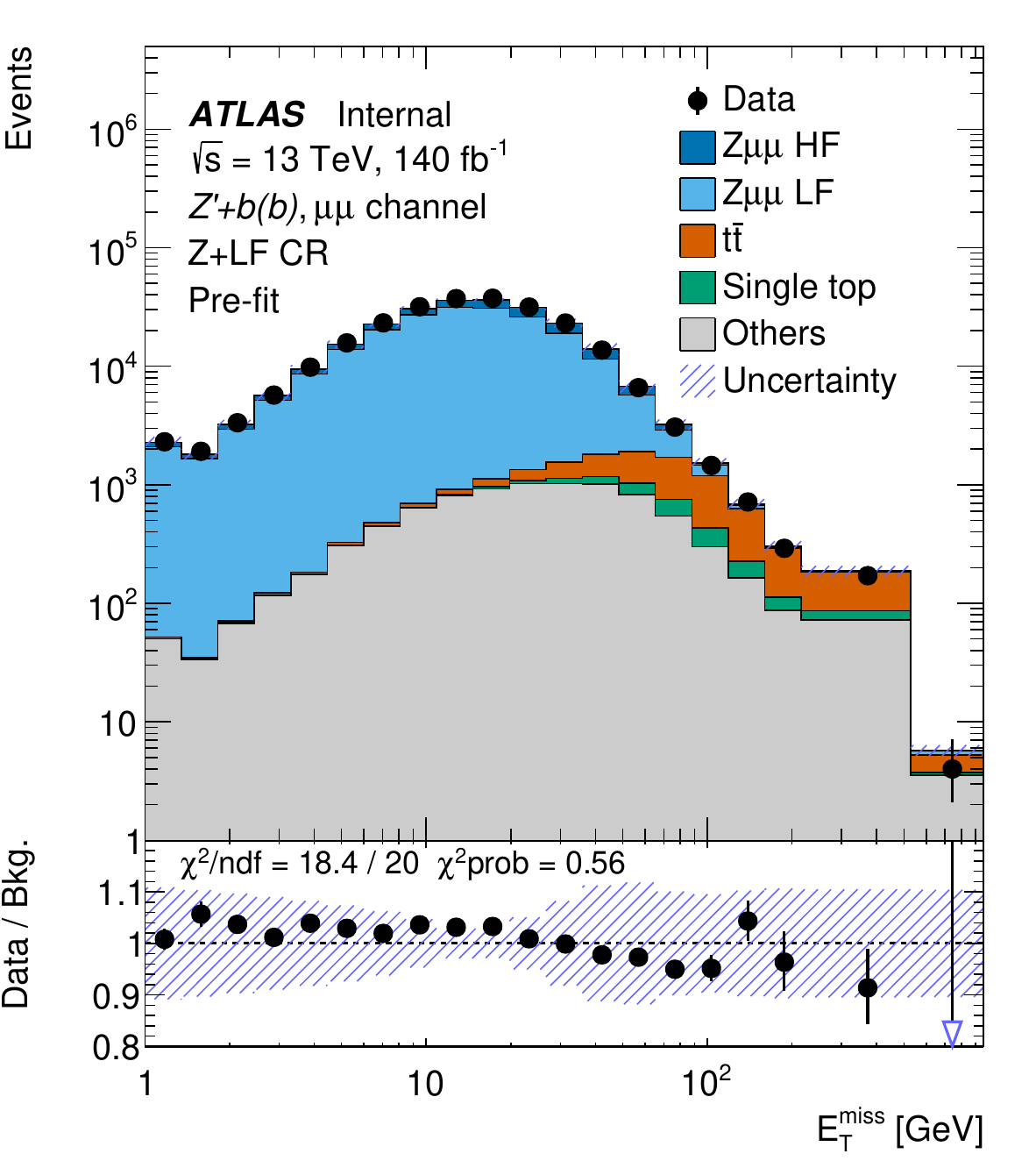}
		\label{fig:ZCR_0b_datamc_met_ele}
	}
	\subfloat[(d)]{
		\includegraphics[width=0.33\textwidth]{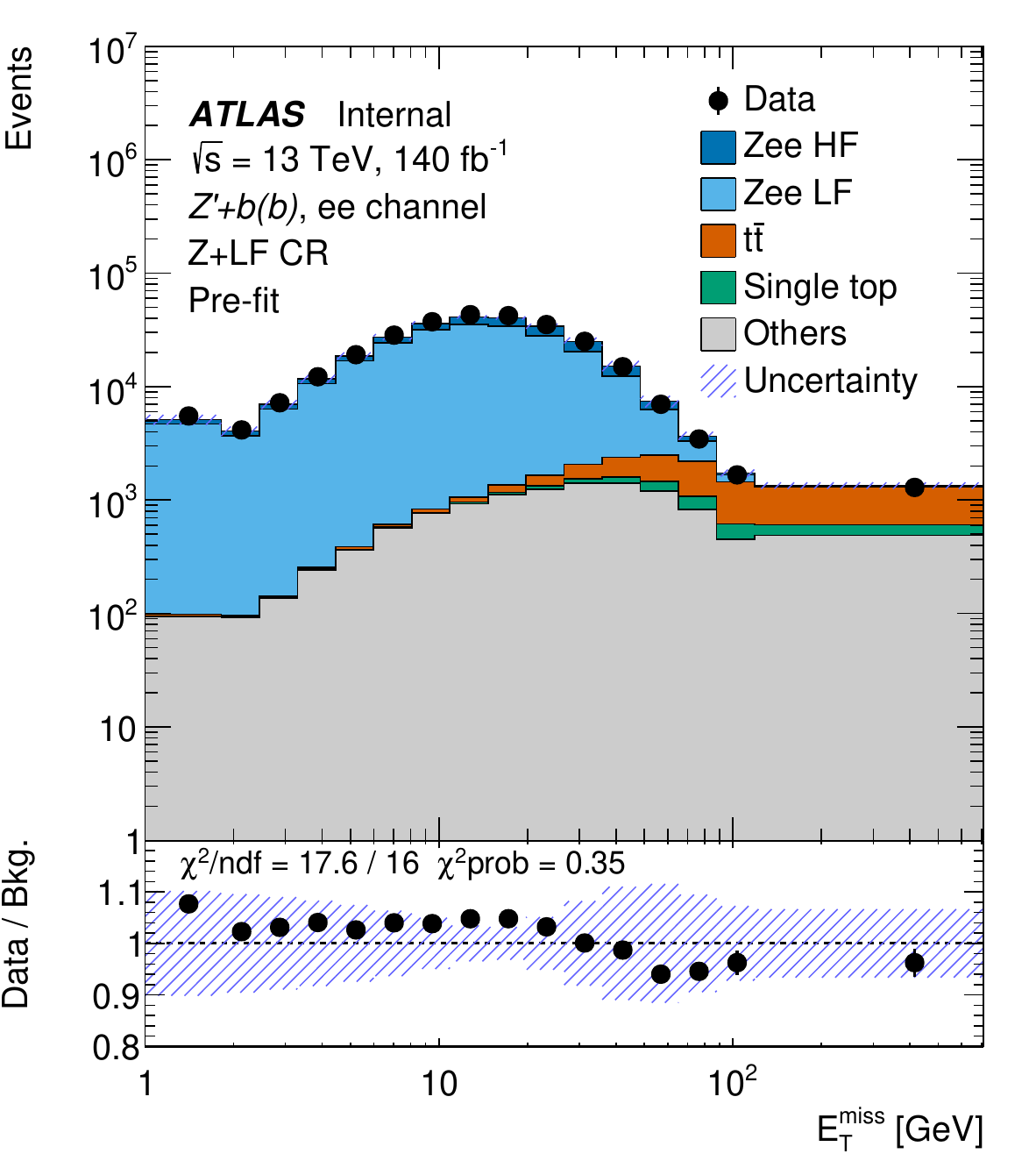}
		\label{fig:ZCR_0b_datamc_met_mu}
	}

	\subfloat[(e)]{
		\includegraphics[width=0.33\textwidth]{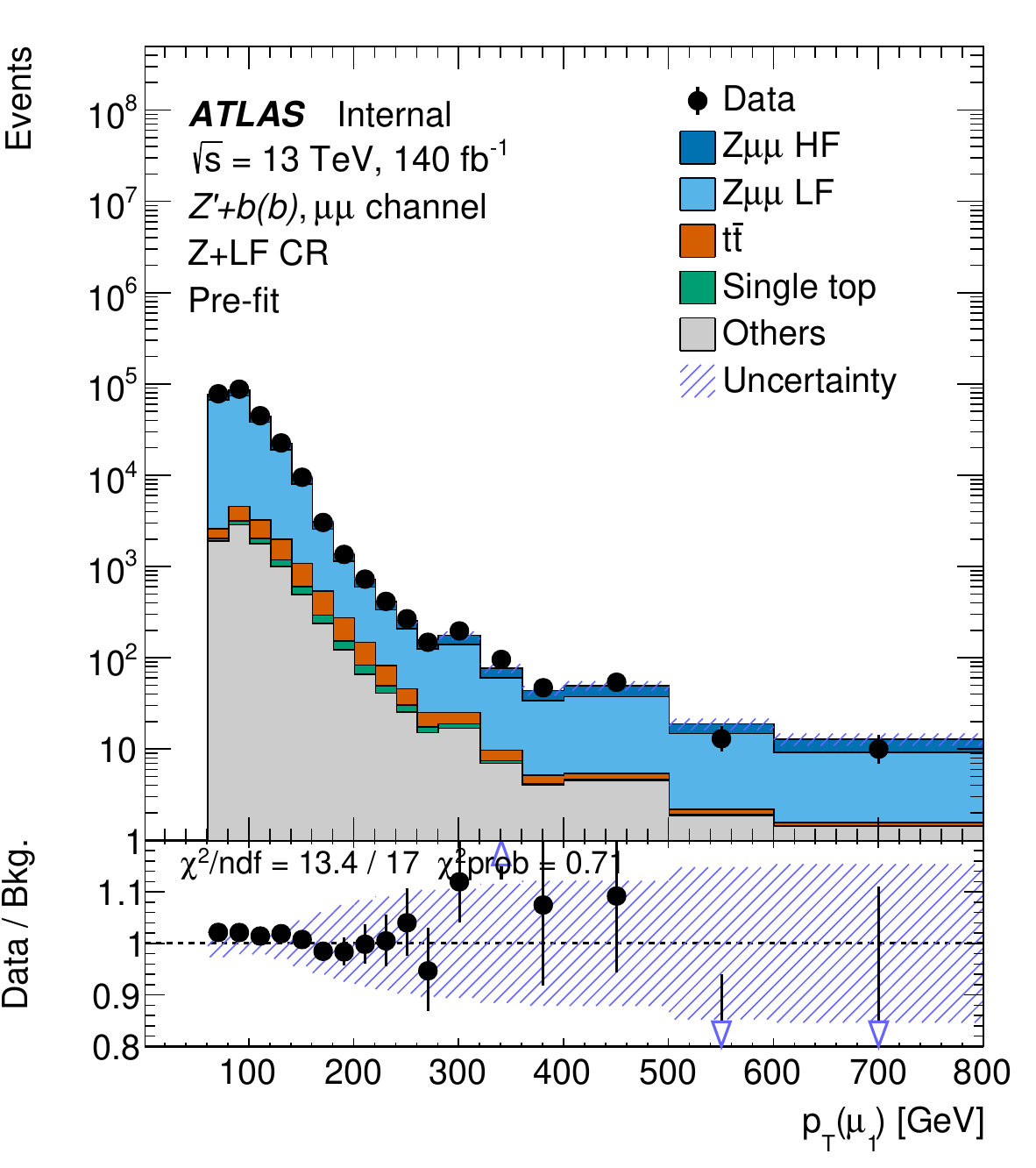}
		\label{fig:ZCR_0b_datamc_pt_mu}
	}
	\subfloat[(f)]{
		\includegraphics[width=0.33\textwidth]{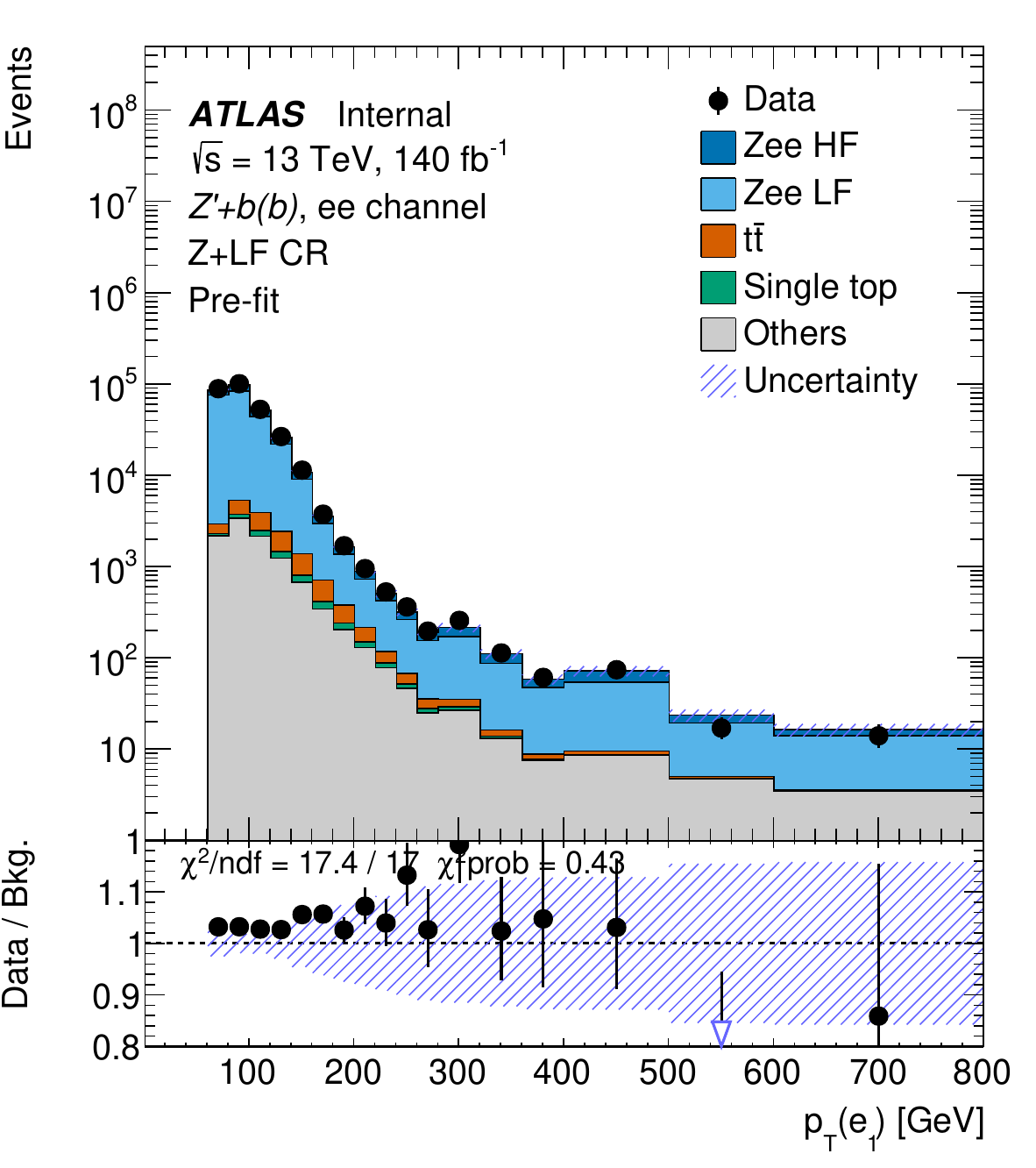}
		\label{fig:ZCR_0b_datamc_pt_ele}
	}

	\caption{Data/MC comparison in the $Z$+LF Control Region for the invariant dilepton mass in the (a) muon and (b) electron channel, the missing transverse momentum in the (c) muon and (d) electron channel, and the $\pT$-leading lepton $\pT$ in the (e) muon and (f) electron channel. The uncertainty band includes statistical and background systematic uncertainties, where only the shape of theoretical modelling uncertainties is taken into account.}
	\label{fig:ZCR_0b_datamc}
\end{figure}

Data/MC comparison plots showing the invariant dilepton mass distribution, the missing transverse momentum and the transverse momentum of the $\pT$-leading lepton are shown in FIG.~\ref{fig:ZCR_0b_datamc} for the muon and electron channel. 


The $\met$ distribution is well described. The $Z\ell\ell$ background dominates at low $\met$, while $\ttbar$ dominates at high $\met$, as expected: $Z$ decays produce no neutrinos whereas dileptonic $\ttbar$ events do.
It can also be seen that the multijet background contributes only for higher values of the missing transverse momentum. For low $\met$, the most dominant contribution to the multijet background is induced by events with two fake leptons in the final state, which are expected to have no real $\met$. Most of these events get a negative weight by the Matrix method. Since bins with a negative event yield are not physical, these bins are set to zero. However, this behaviour, especially the negative weights from the Matrix method, needs to be further investigated and studied.

The distribution of the transverse momentum of the $p_T$-leading lepton looks generally a bit worse compared to the other two distributions discussed before. However, for low values of the transverse momentum, the ratio stays relatively flat and is in agreement with a value of one. Most deviations are covered by the uncertainty band. 

\FloatBarrier

\subsubsection*{$Z$+HF CR}

\begin{figure}[h!]
	\centering
	\captionsetup[subfigure]{labelformat=empty}
	\subfloat[(a)]{
		\includegraphics[width=0.45\textwidth]{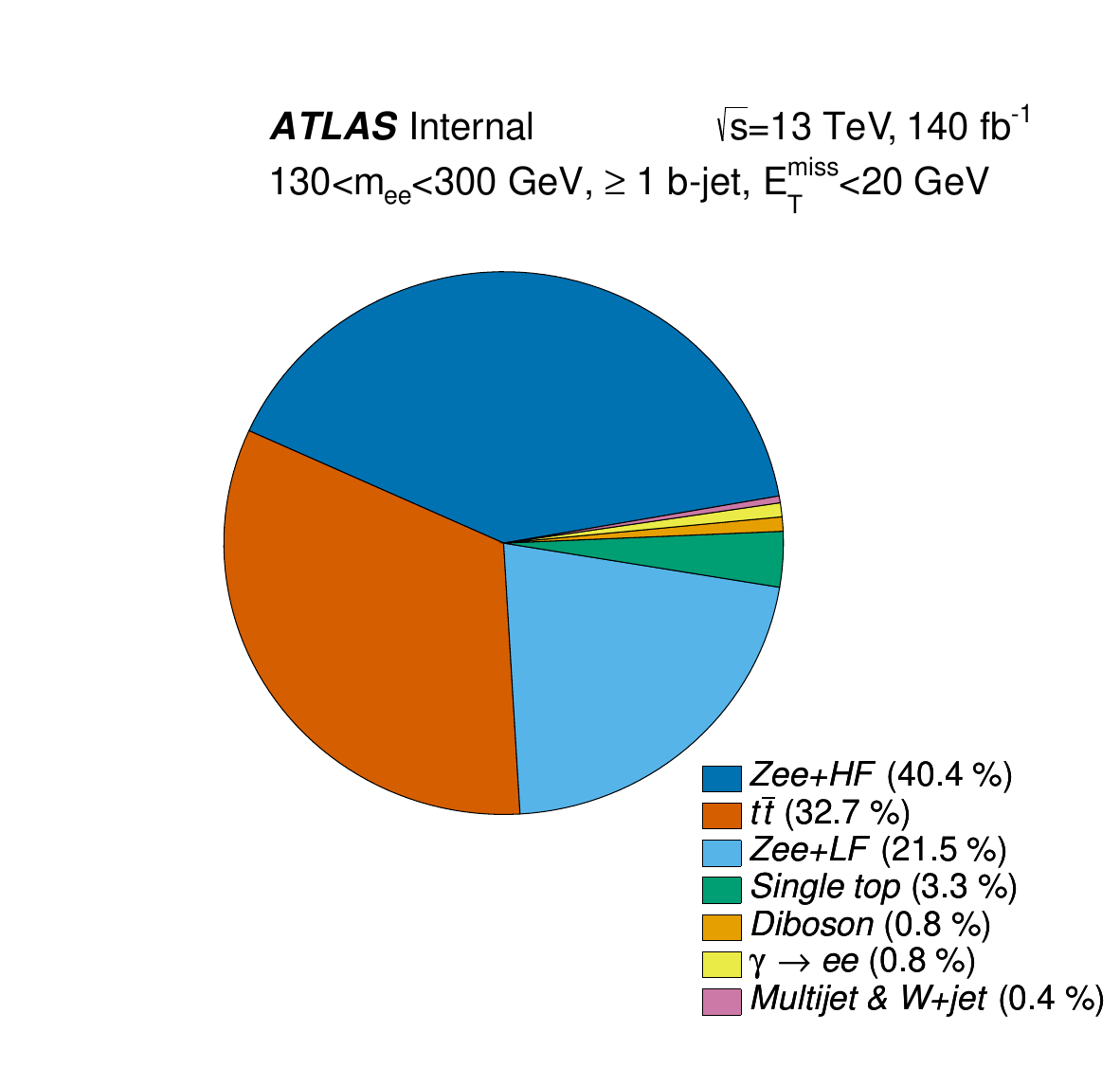}
		\label{fig:ZCR_atleast1b_composition_ele}
	}
	\hfill
	\subfloat[(b)]{
		\includegraphics[width=0.45\textwidth]{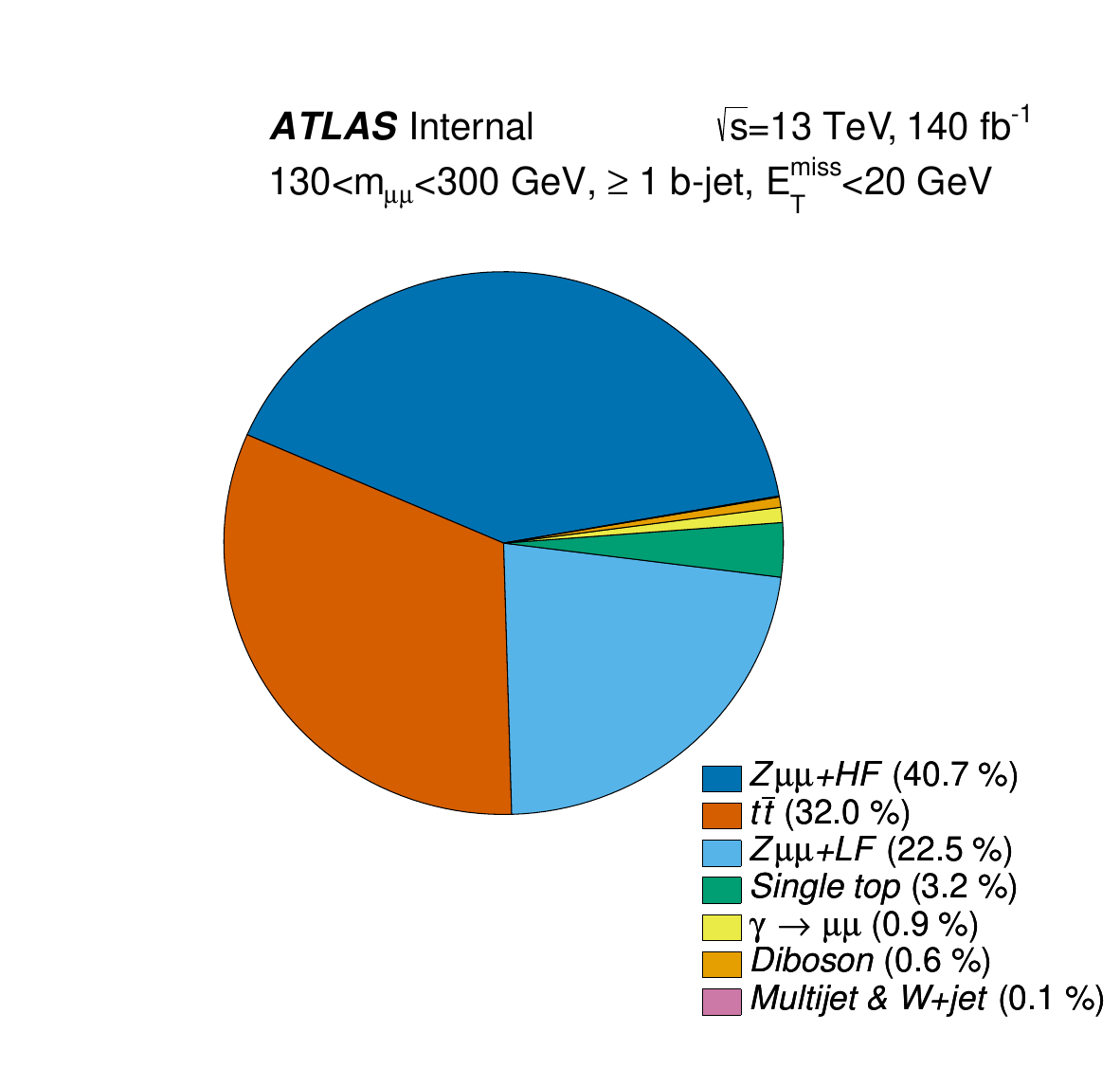}
		\label{fig:ZCR_atleast1b_composition_mu}
	}
	\caption{Background composition of the $Z$+HF Control Region in the (a) electron and (b) muon channel.}
	\label{fig:ZCR_atleast1b_composition}
\end{figure}

The second $Z$ CR contains events with at least one $b$-jet in the final state. This CR aims to validate the shape and normalisation of the $Z$+HF background.
The $\ttbar$ contribution is large in this region due to the $b$-jet requirement. An additional requirement of $\met < 20\;\GeV$ is therefore applied to suppress the $\ttbar$ background and increase Z+HF purity. The resulting background composition is shown in FIG.~\ref{fig:ZCR_atleast1b_composition}.
Although the $\ttbar$ background is strongly reduced by the additional cut on the missing transverse momentum, $\ttbar$ events still contribute with roughly 32\% to this control region. The Z+HF background has the largest contribution with roughly 40\%. The Z+LF background has a contribution of approximately 22\%, while the other backgrounds have contributions below 5\%.

\begin{figure}[h]
	\centering
	\captionsetup[subfigure]{labelformat=empty}
	\subfloat[(a)]{
		\includegraphics[width=0.33\textwidth]{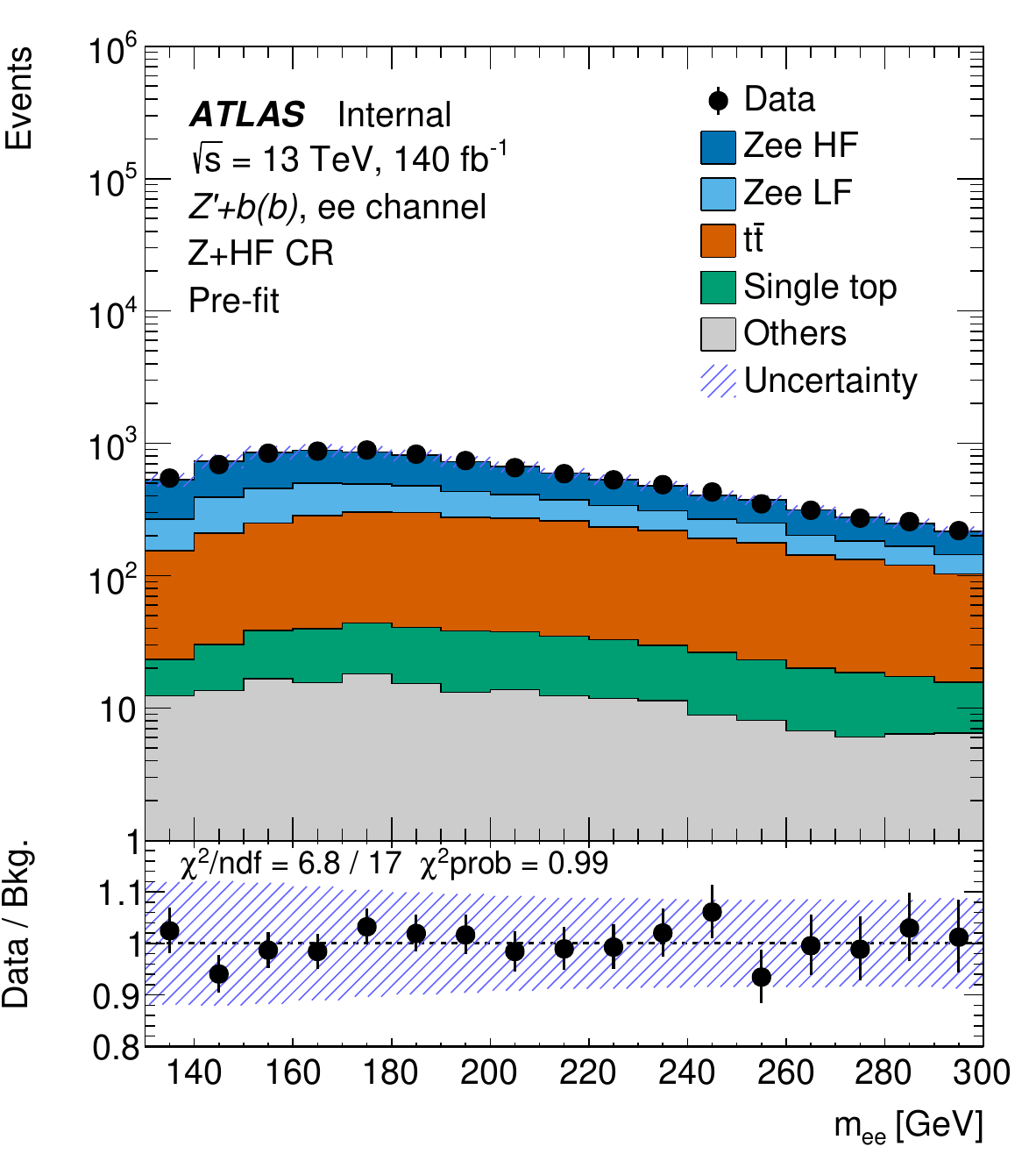}
		\label{fig:ZCR_atleast1b_datamc_inv_mass_ele}
	}
	\subfloat[(b)]{
		\includegraphics[width=0.33\textwidth]{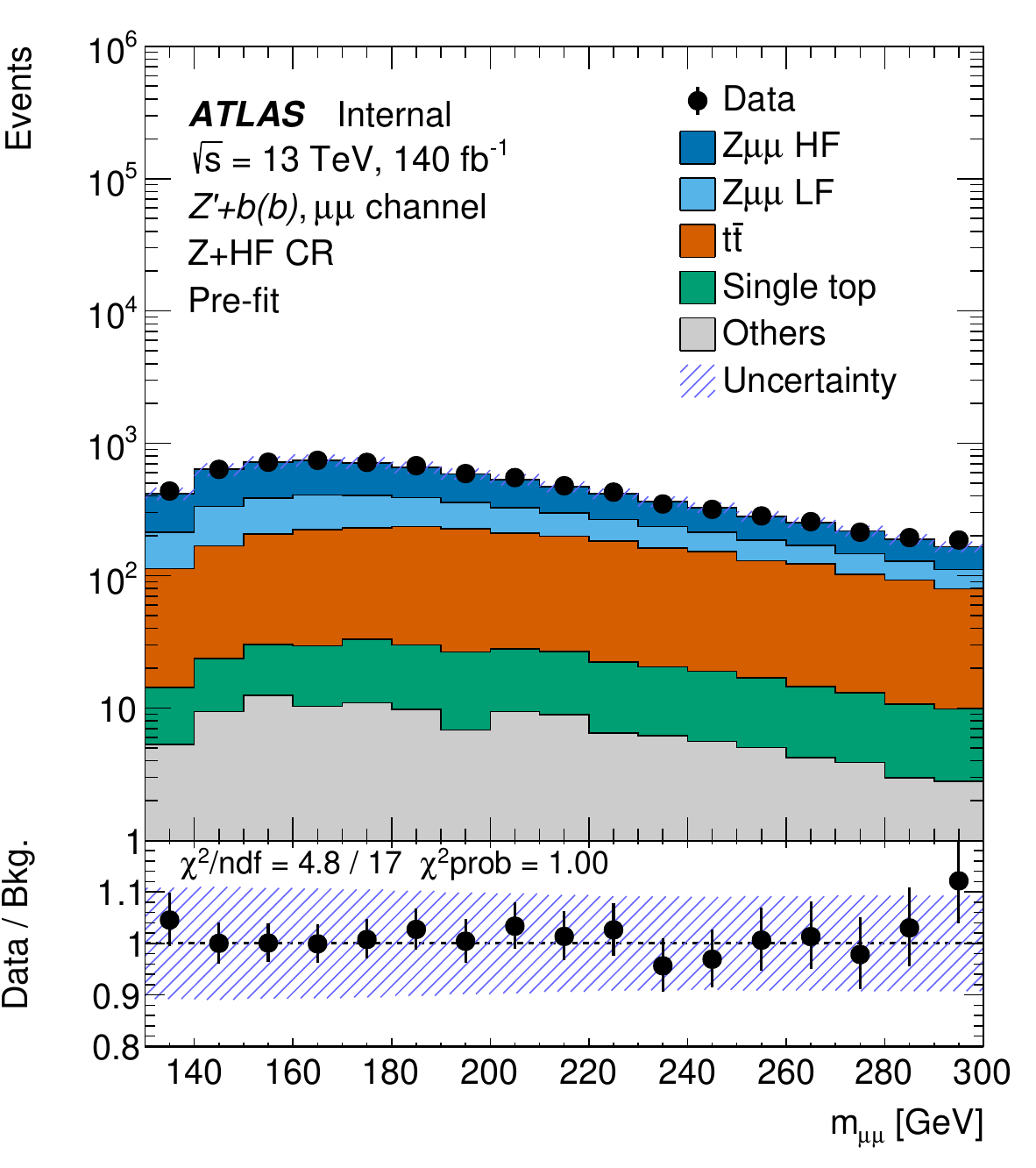}
		\label{fig:ZCR_atleast1b_datamc_inv_mass_mu}
	}

	\subfloat[(c)]{
		\includegraphics[width=0.33\textwidth]{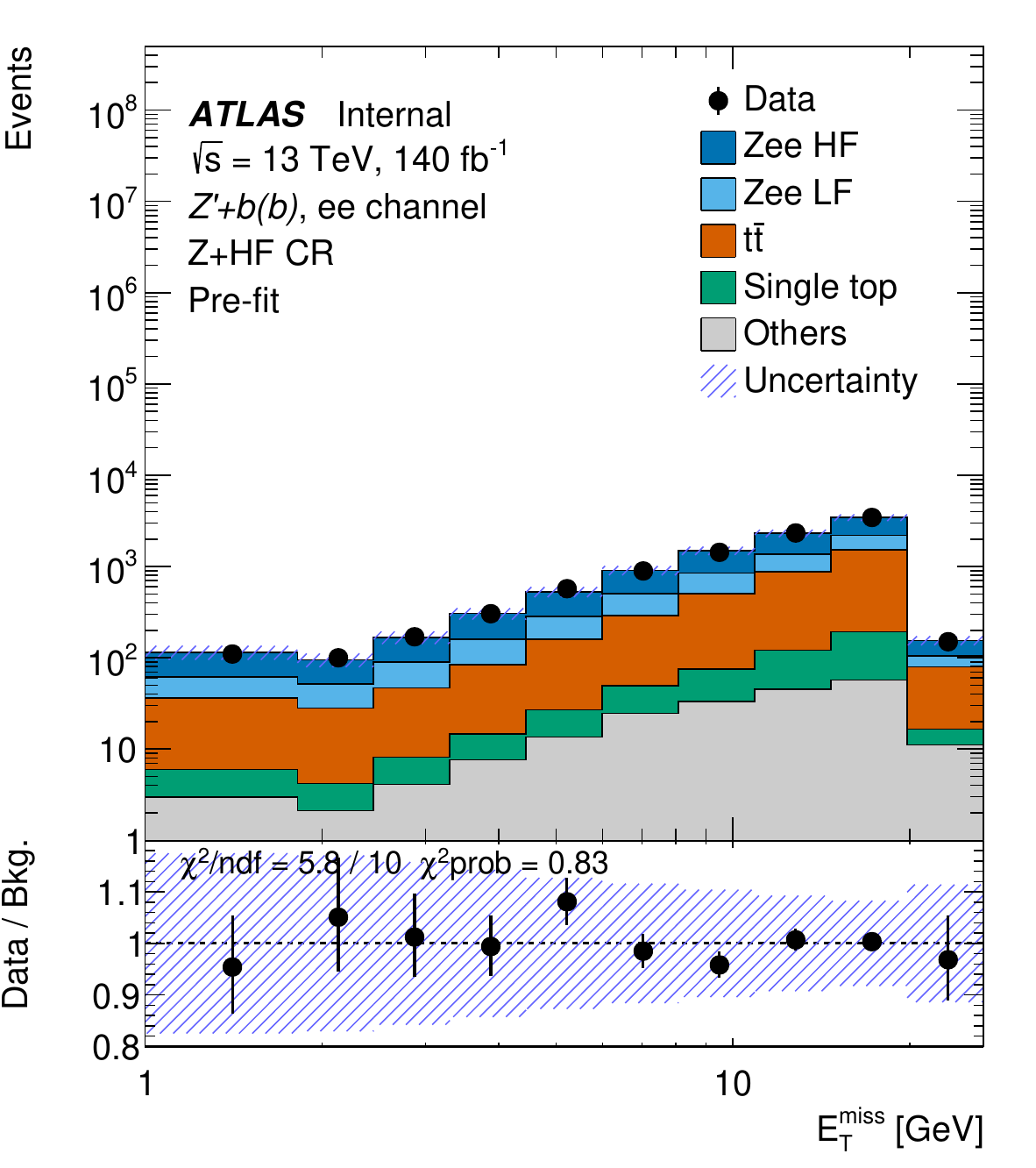}
		\label{fig:ZCR_atleast1b_datamc_met_ele}
	}
	\subfloat[(d)]{
		\includegraphics[width=0.33\textwidth]{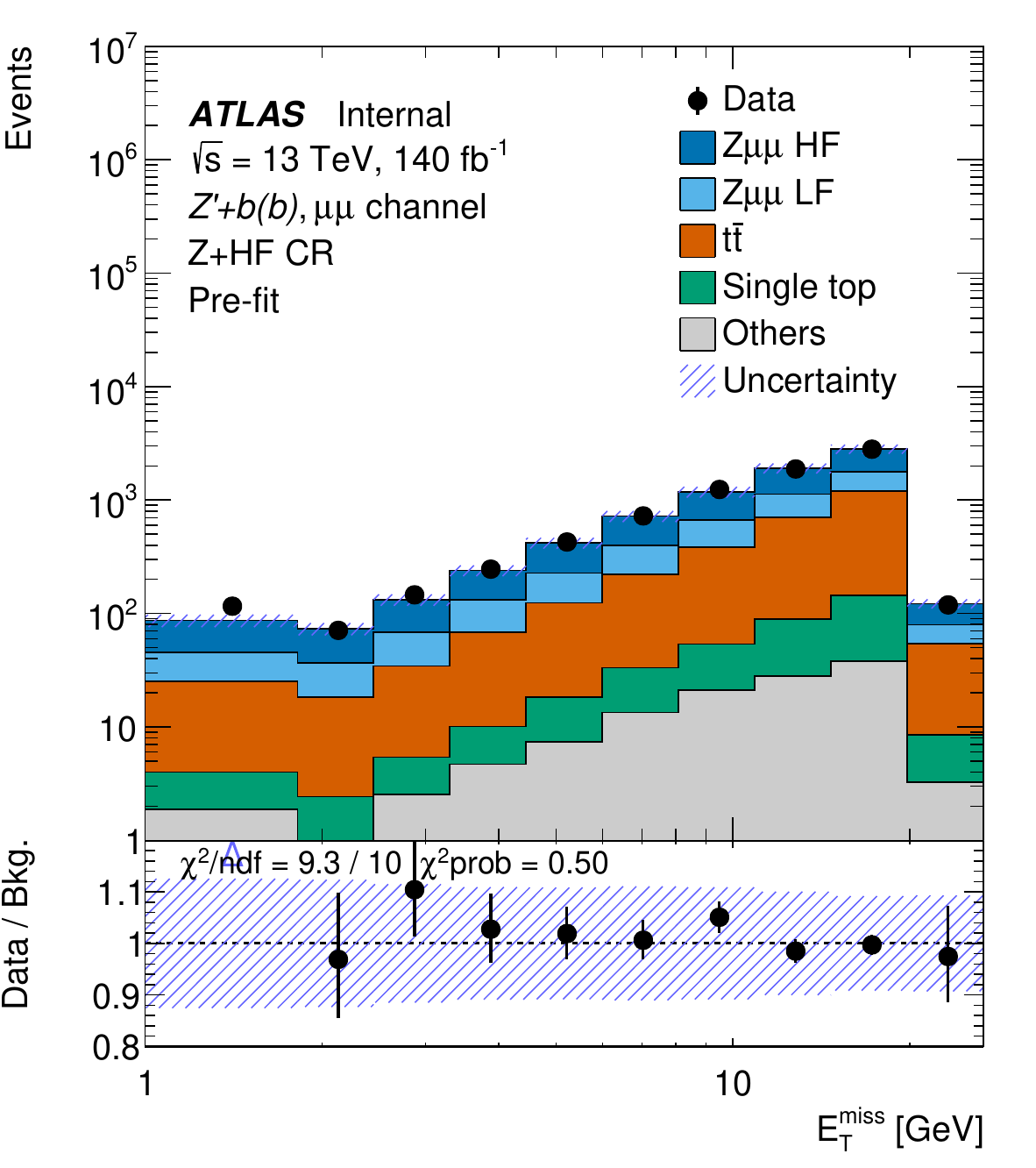}
		\label{fig:ZCR_atleast1b_datamc_met_mu}
	}

	\subfloat[(e)]{
		\includegraphics[width=0.33\textwidth]{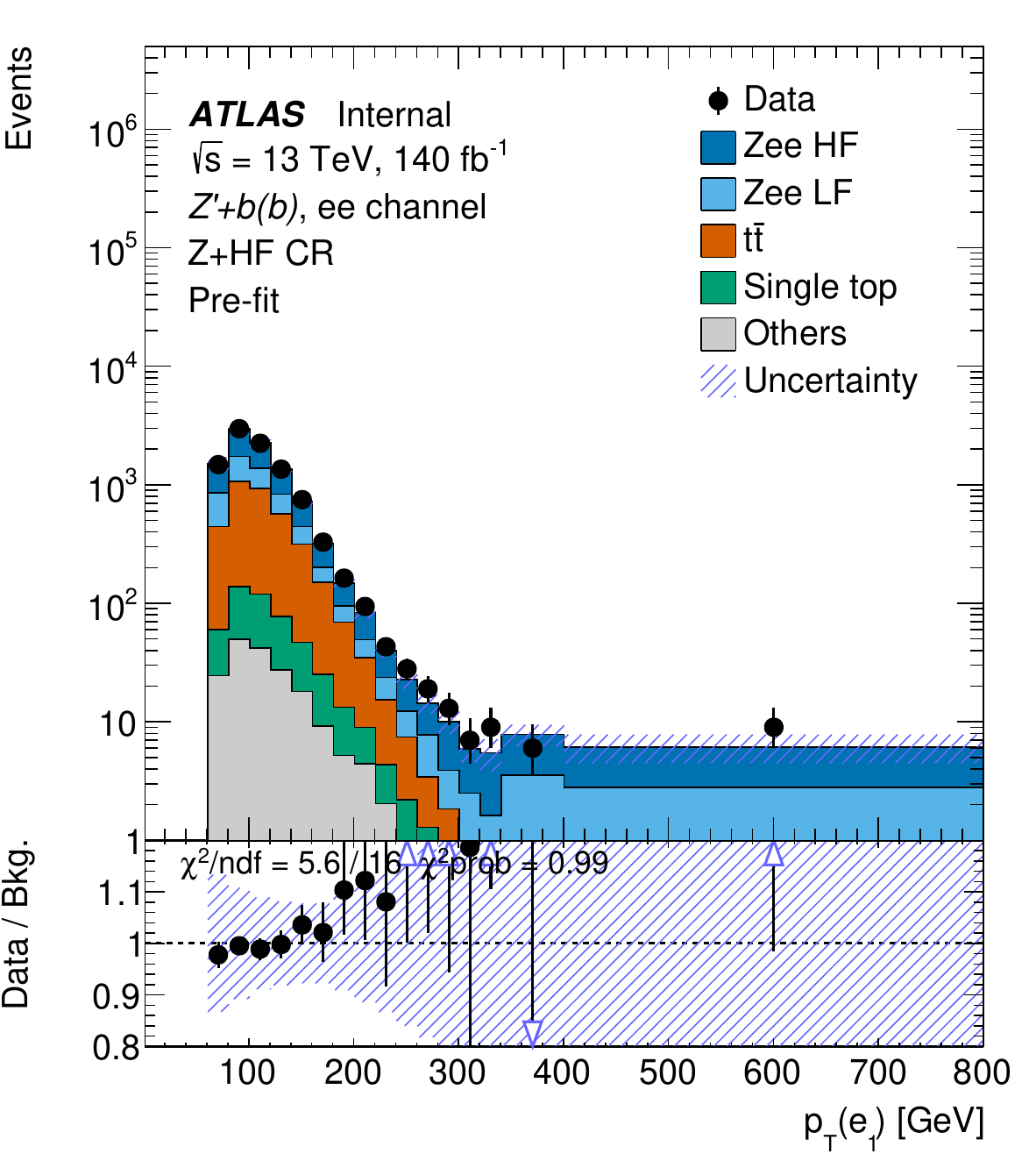}
		\label{fig:ZCR_atleast1b_datamc_pt_ele}
	}
	\subfloat[(f)]{
		\includegraphics[width=0.33\textwidth]{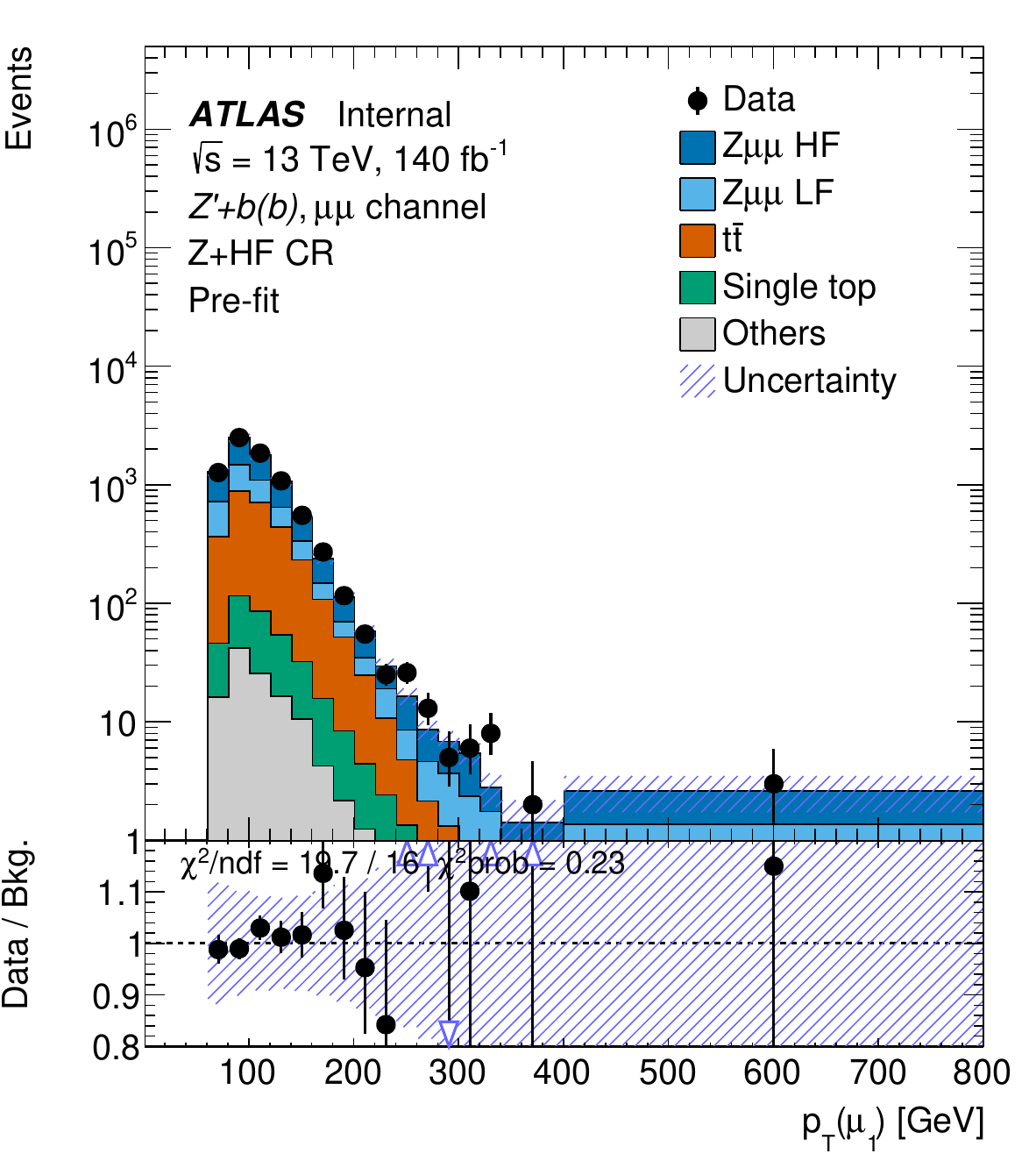}
		\label{fig:ZCR_atleast1b_datamc_pt_mu}
	}

	\caption{Data/MC comparison in the $Z$+HF Control Region: invariant dilepton mass (a--b), missing transverse momentum (c--d), and leading lepton $\pT$ (e--f) for the electron (left) and muon (right) channels. The uncertainty band includes statistical and background systematic uncertainties.}
	\label{fig:ZCR_atleast1b_datamc}
\end{figure}

FIG.~\ref{fig:ZCR_atleast1b_datamc} shows the data/MC comparison plots for the invariant dilepton mass distribution, the missing transverse momentum and the transverse momentum of the $p_T$-leading lepton. 

The data/MC comparison for the invariant dilepton mass distribution looks reasonable and the ratio is overall compatible with one. 

The distribution of the missing transverse momentum looks also reasonable apart from a deviation in one bin at low missing transverse momentum in the muon channel. 

For low values of the transverse momentum of the $p_T$-leading lepton, the data/MC-ratio is compatible with a value of one. 

\FloatBarrier

\subsection{Validation region}

The background estimates for the main backgrounds, namely the $\ttbar$ background and the Z background, split into a light and heavy flavour component, are tested in dedicated regions. Two VRs are defined based on the event pre-selection with an additional requirement of $300<m_{\ell\ell}<500\,\GeV$ for the invariant dilepton mass. For the $Z$+HF VRs, an additional requirement of $\met<20\,\GeV$ is applied, similar to the $Z$ CR with at least one $b$-jet.
The VR test the background normalisation factors extracted from the CR fits.

\begin{figure}[h]
	\centering
	\captionsetup[subfigure]{labelformat=empty}
	\subfloat[(a)]{
		\includegraphics[width=0.33\textwidth]{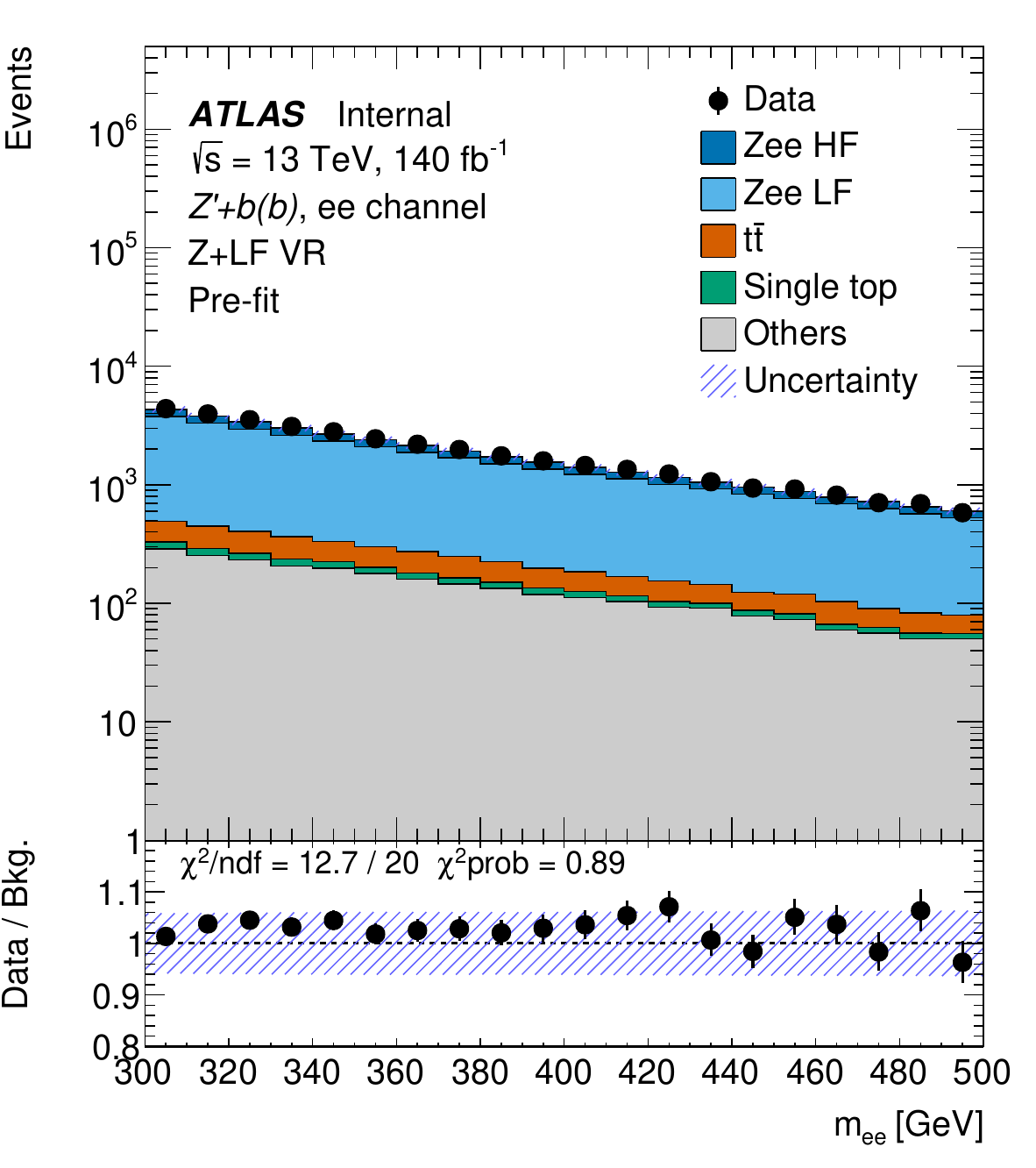}
		\label{fig:VR_0b_invmass_ele}
	}
	\subfloat[(b)]{
		\includegraphics[width=0.33\textwidth]{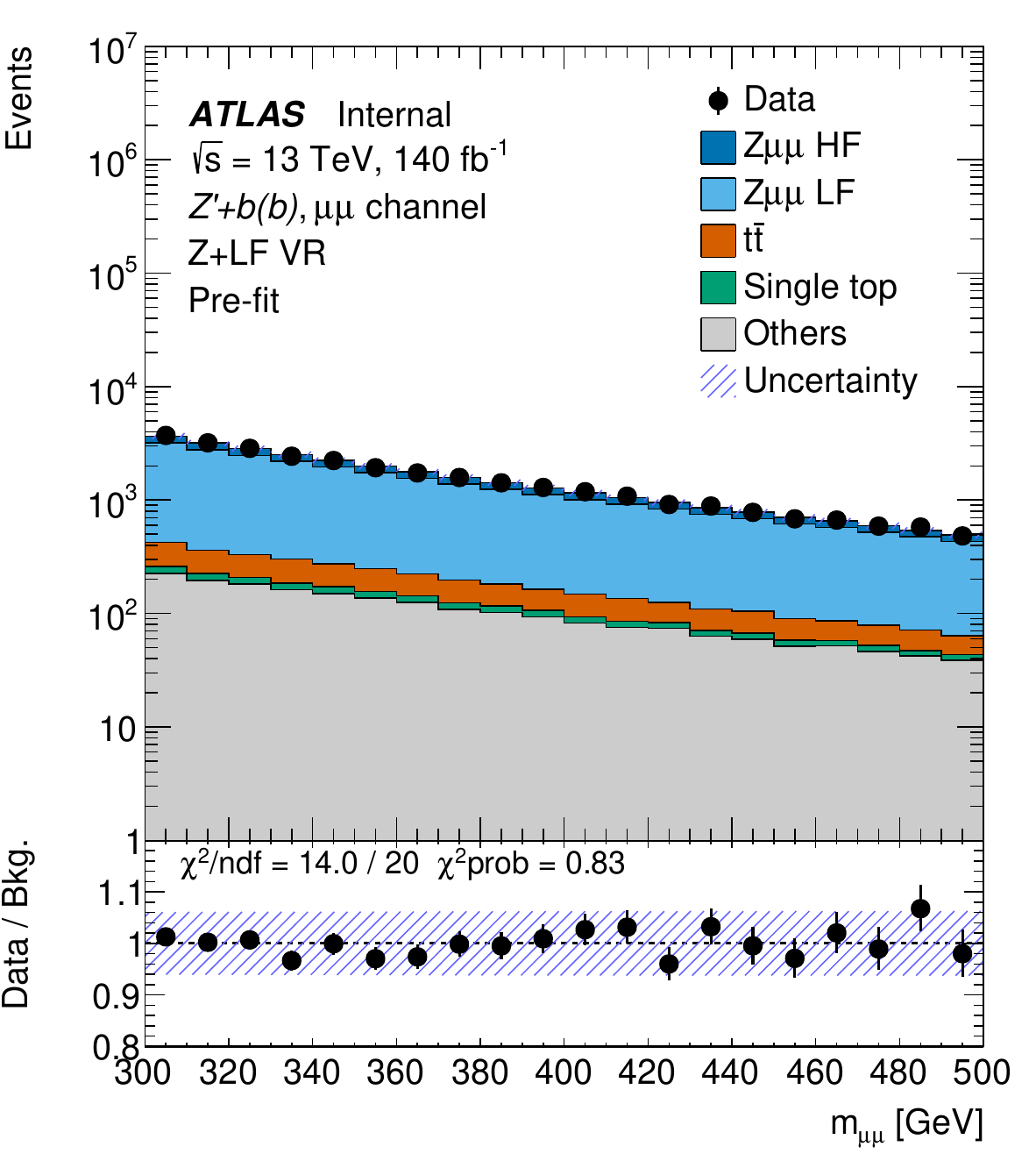}
		\label{fig:VR_0b_invmass_mu}
	}

	\subfloat[(c)]{
		\includegraphics[width=0.33\textwidth]{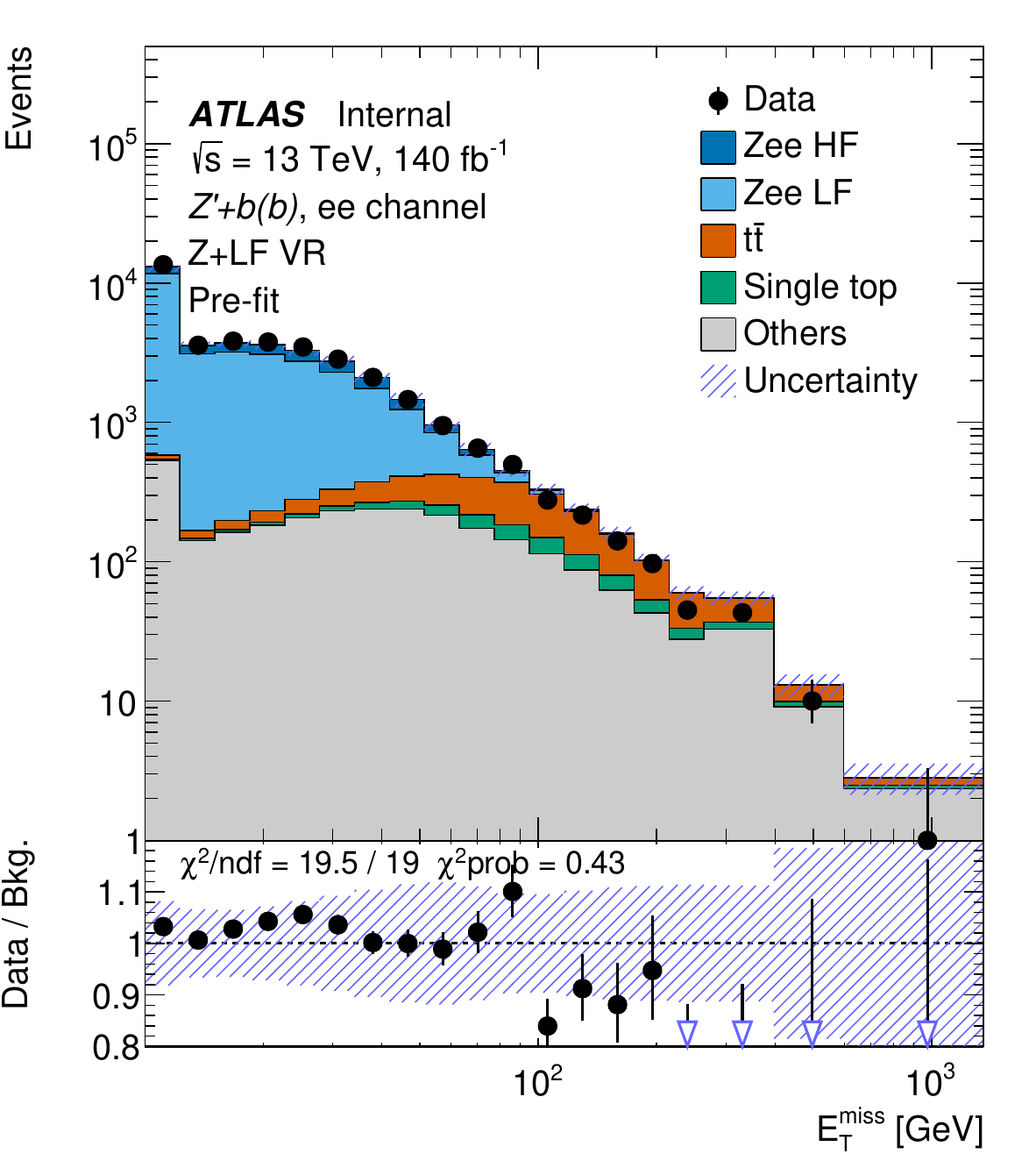}
		\label{fig:VR_0b_met_ele}
	}
	\subfloat[(d)]{
		\includegraphics[width=0.33\textwidth]{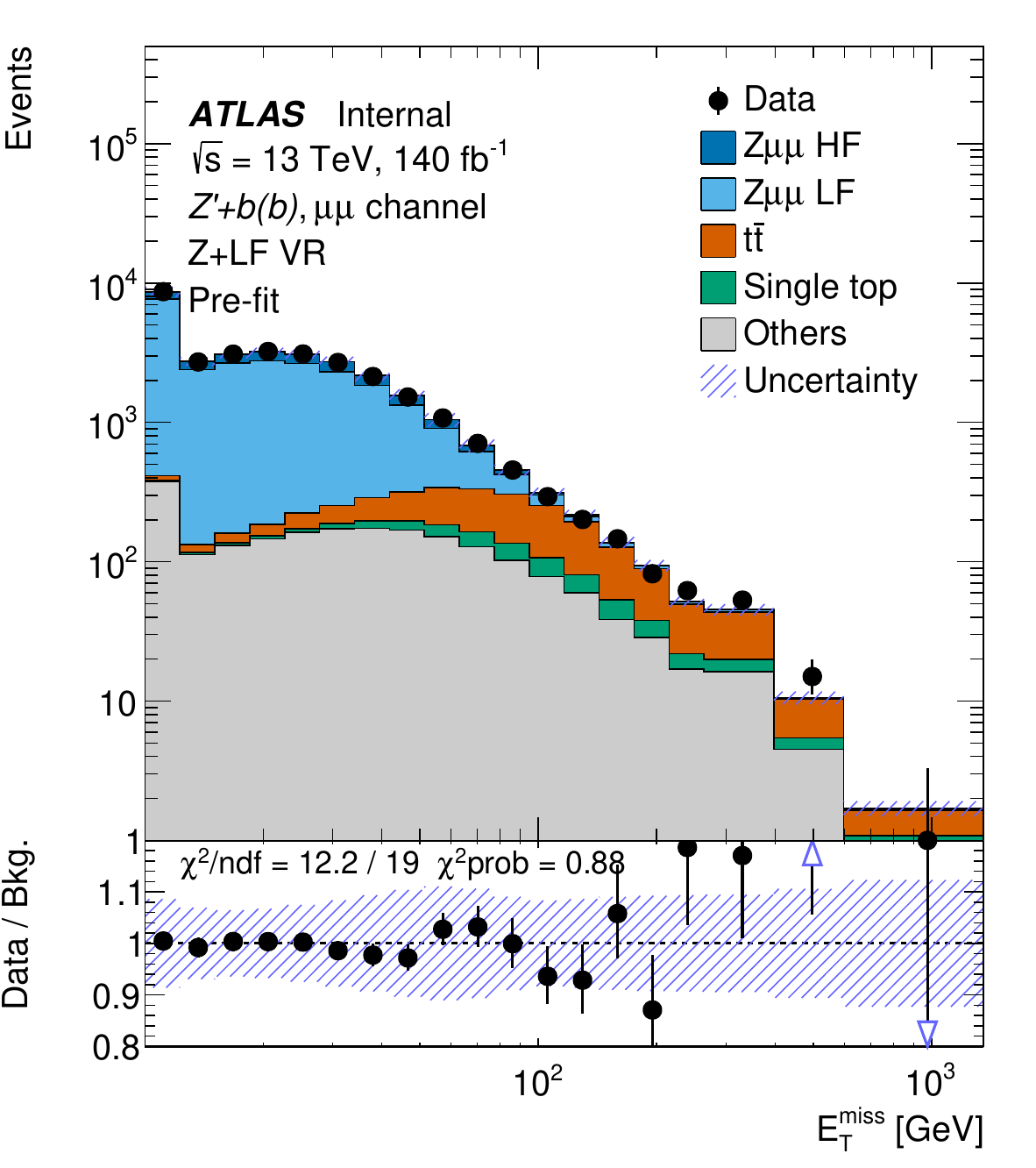}
		\label{fig:VR_0b_met_mu}
	}

	\subfloat[(e)]{
		\includegraphics[width=0.33\textwidth]{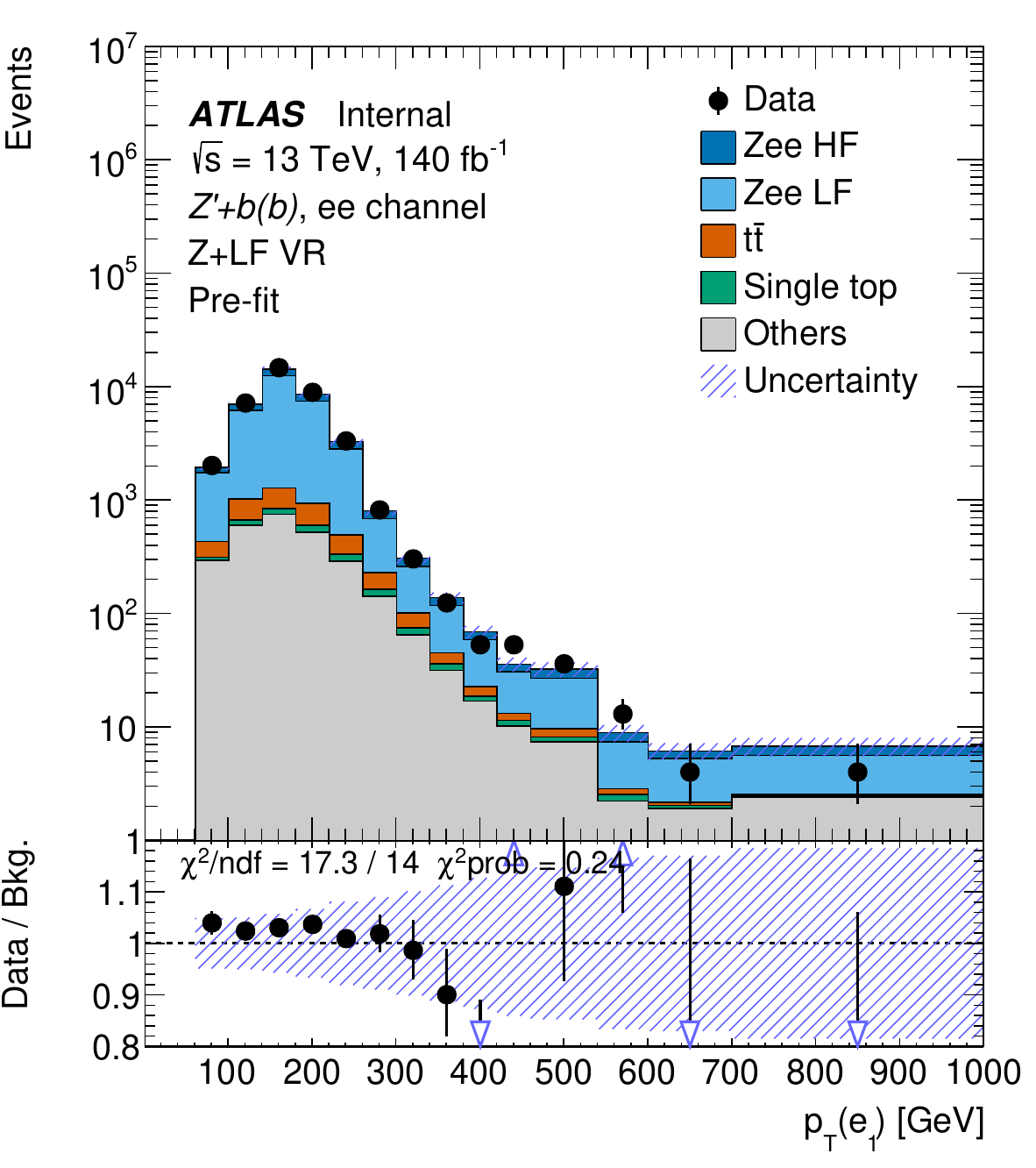}
		\label{fig:VR_0b_pt_ele}
	}
	\subfloat[(f)]{
		\includegraphics[width=0.33\textwidth]{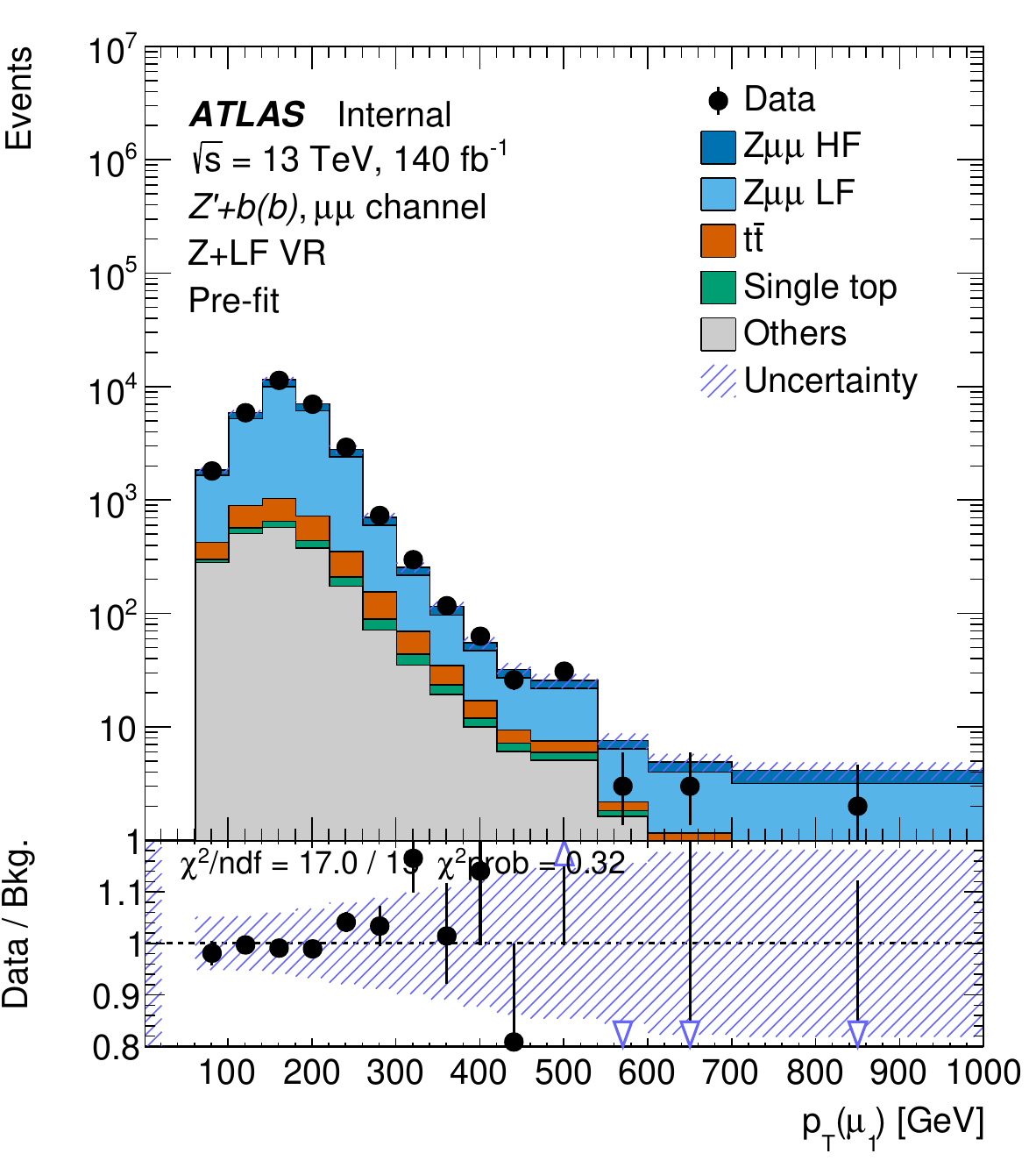}
		\label{fig:VR_0b_pt_mu}
	}

	\caption{Data/MC comparison in the 0 $b$-jet Validation Region: invariant dilepton mass (a--b), missing transverse momentum (c--d), and leading lepton $\pT$ (e--f) for the electron (left) and muon (right) channels. The uncertainty band includes statistical and background systematic uncertainties.}
	\label{fig:VR_0b}
\end{figure}

\begin{figure}[h]
	\centering
	\captionsetup[subfigure]{labelformat=empty}
	\subfloat[(a)]{
		\includegraphics[width=0.33\textwidth]{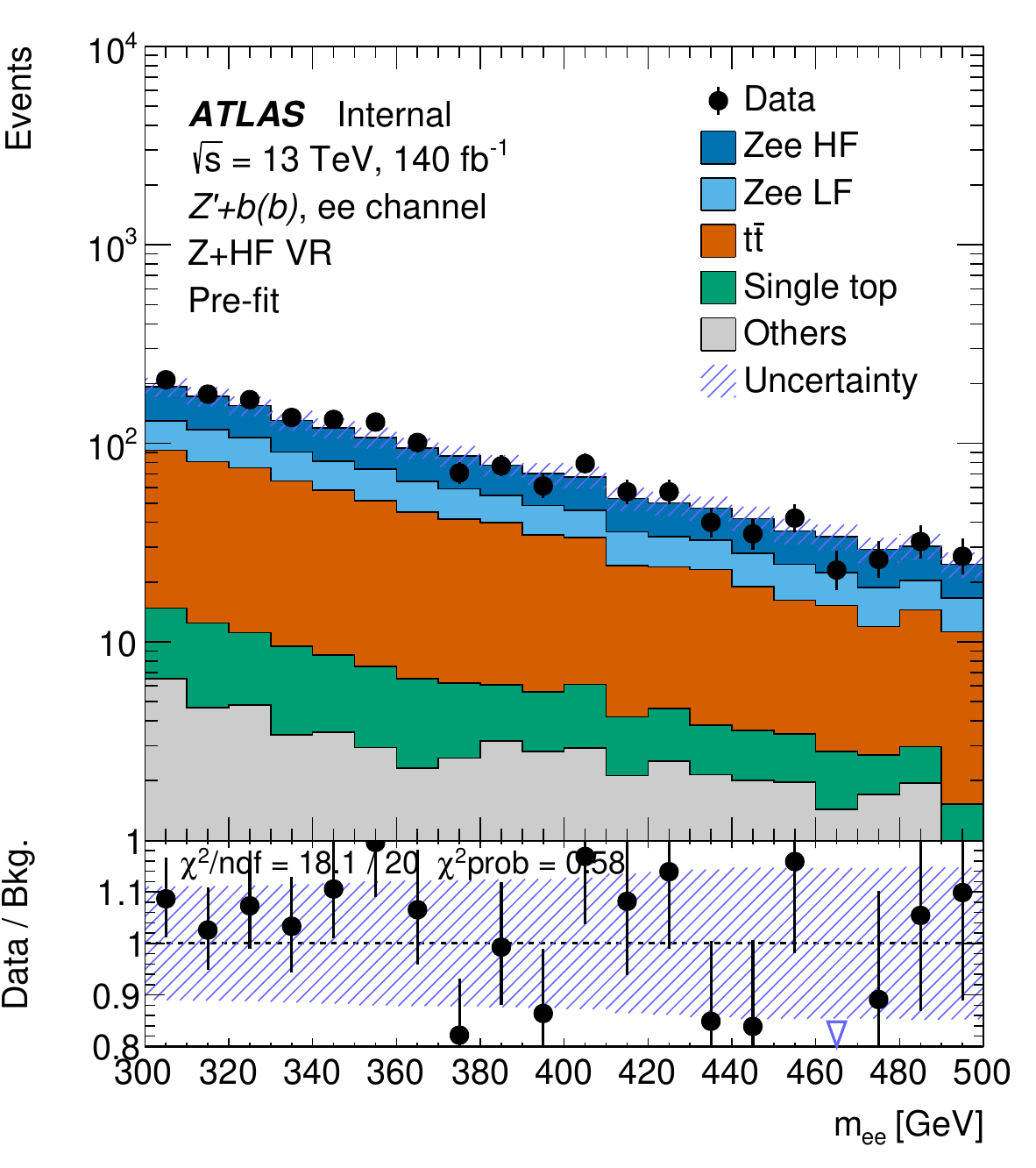}
		\label{fig:VR_atleast1b_invmass_ele}
	}
	\subfloat[(b)]{
		\includegraphics[width=0.33\textwidth]{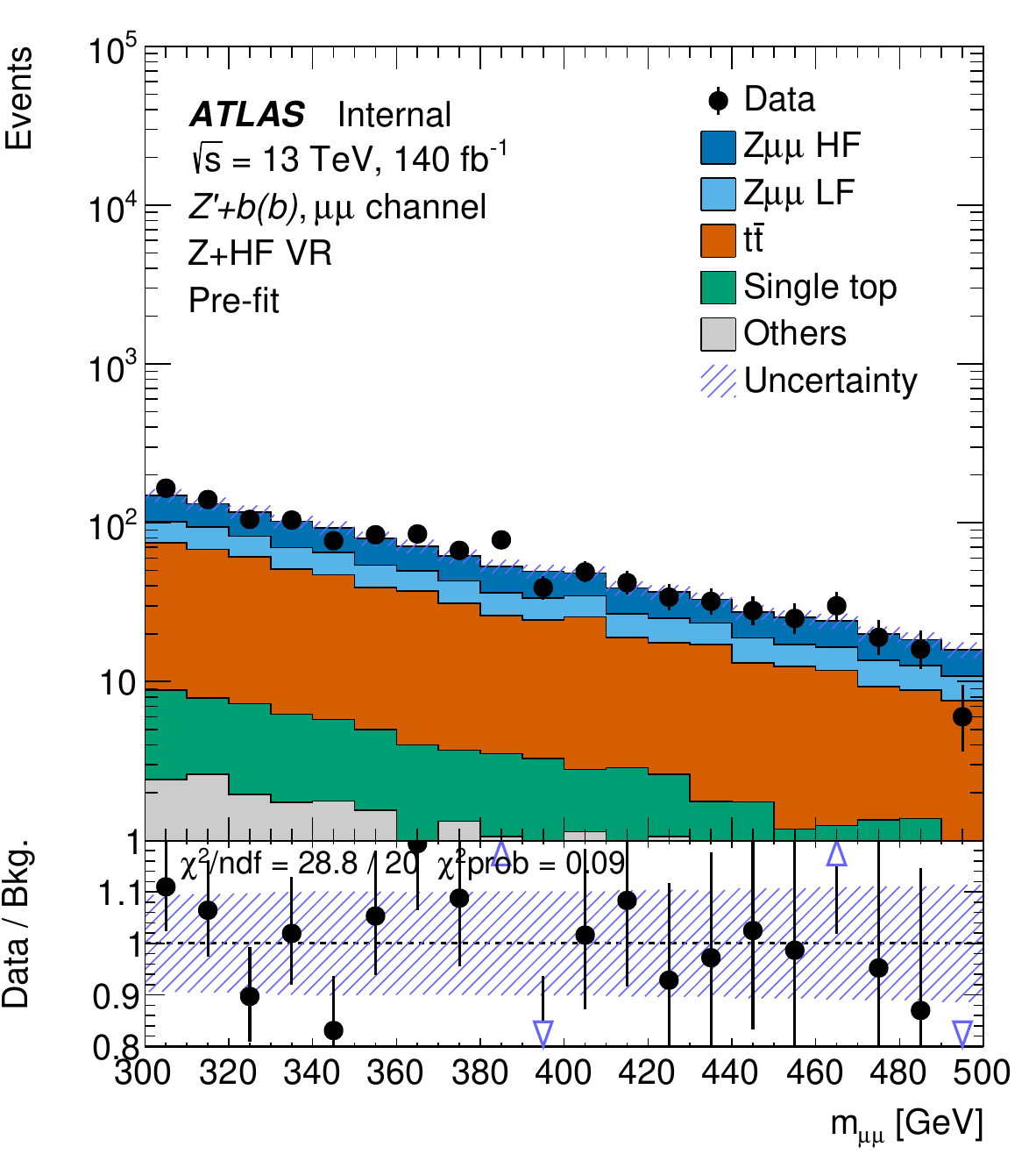}
		\label{fig:VR_atleast1b_invmass_mu}
	}

	\subfloat[(c)]{
		\includegraphics[width=0.33\textwidth]{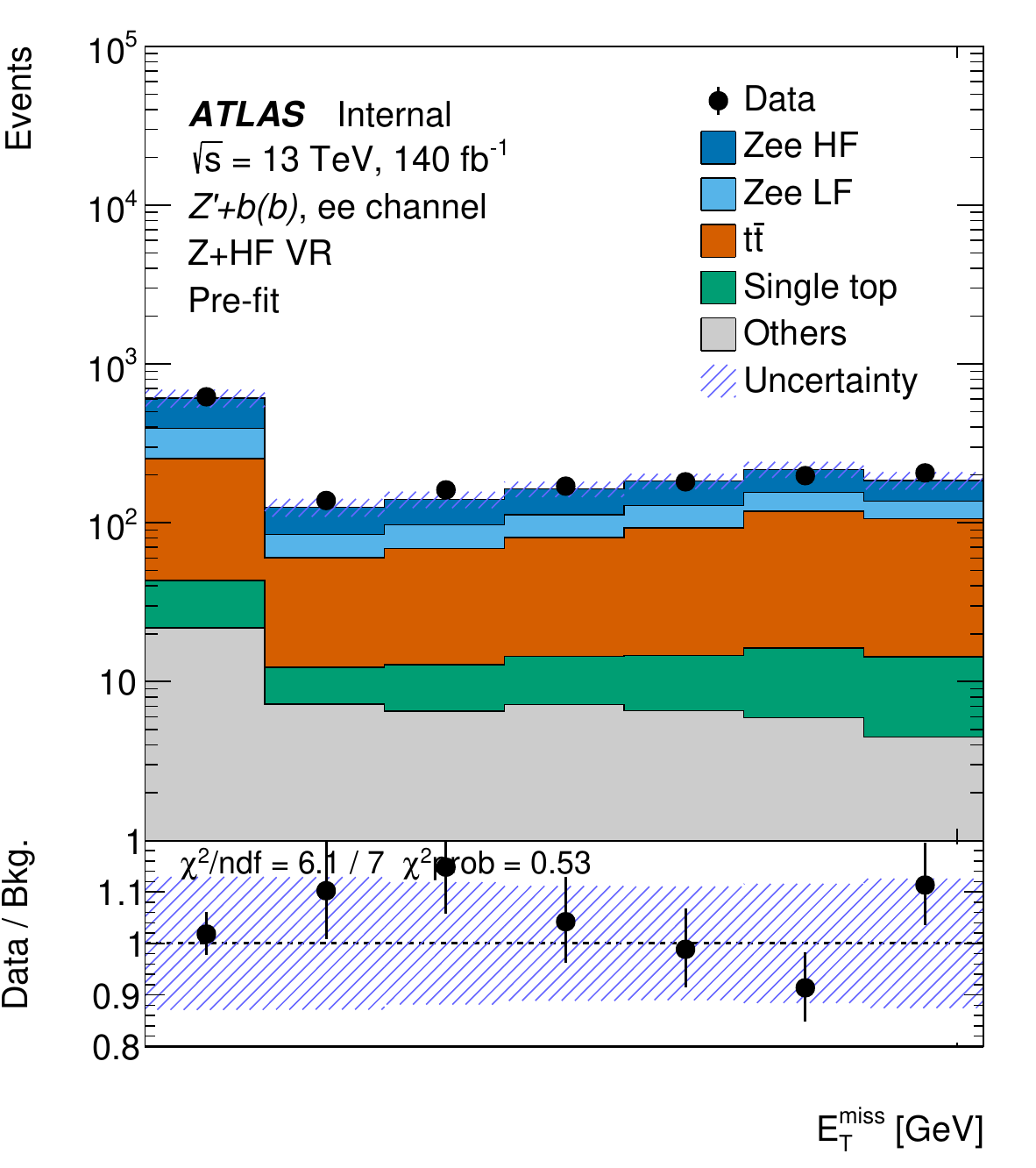}
		\label{fig:VR_atleast1b_met_ele}
	}
	\subfloat[(d)]{
		\includegraphics[width=0.33\textwidth]{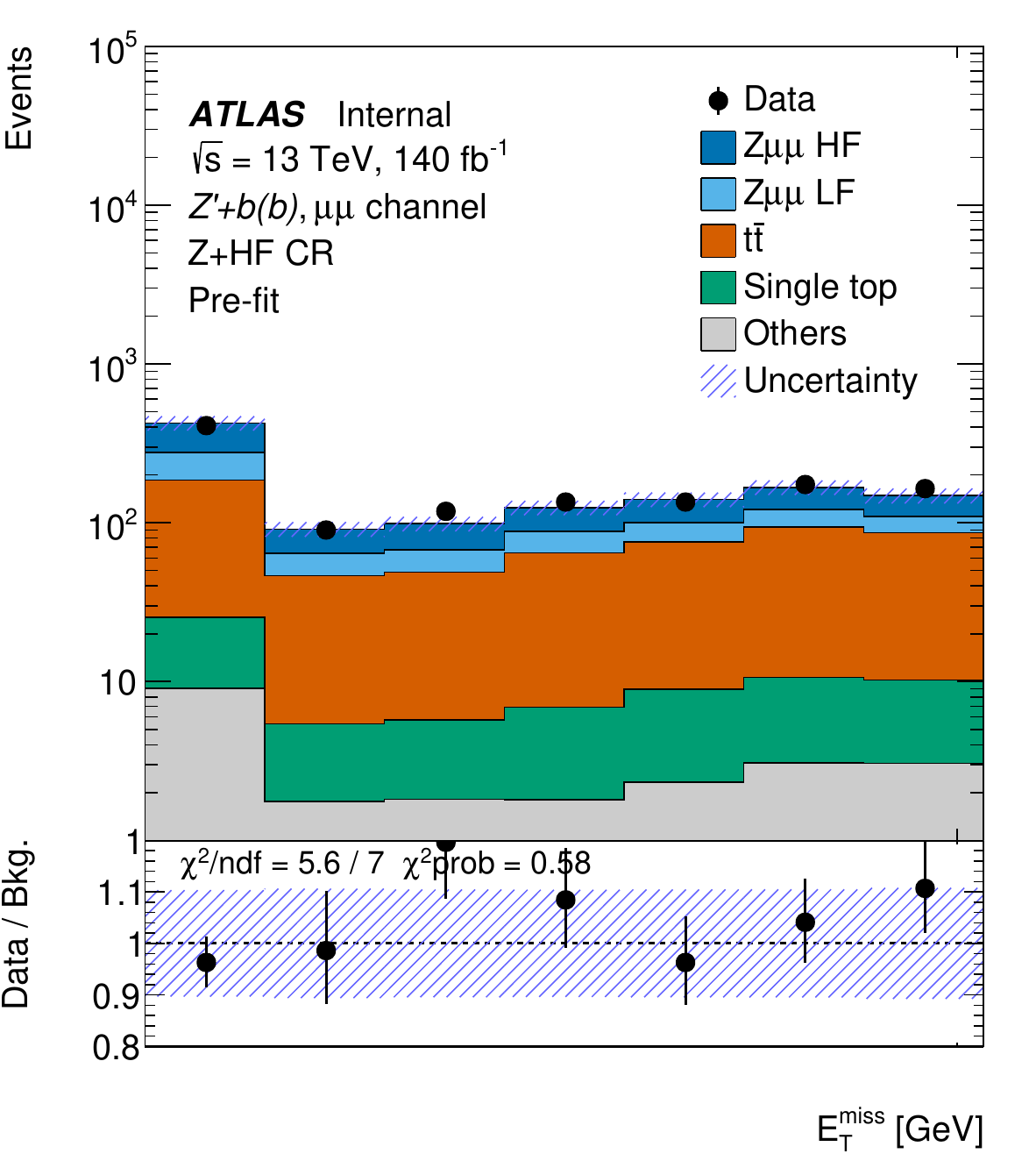}
		\label{fig:VR_atleast1b_met_mu}
	}

	\subfloat[(e)]{
		\includegraphics[width=0.33\textwidth]{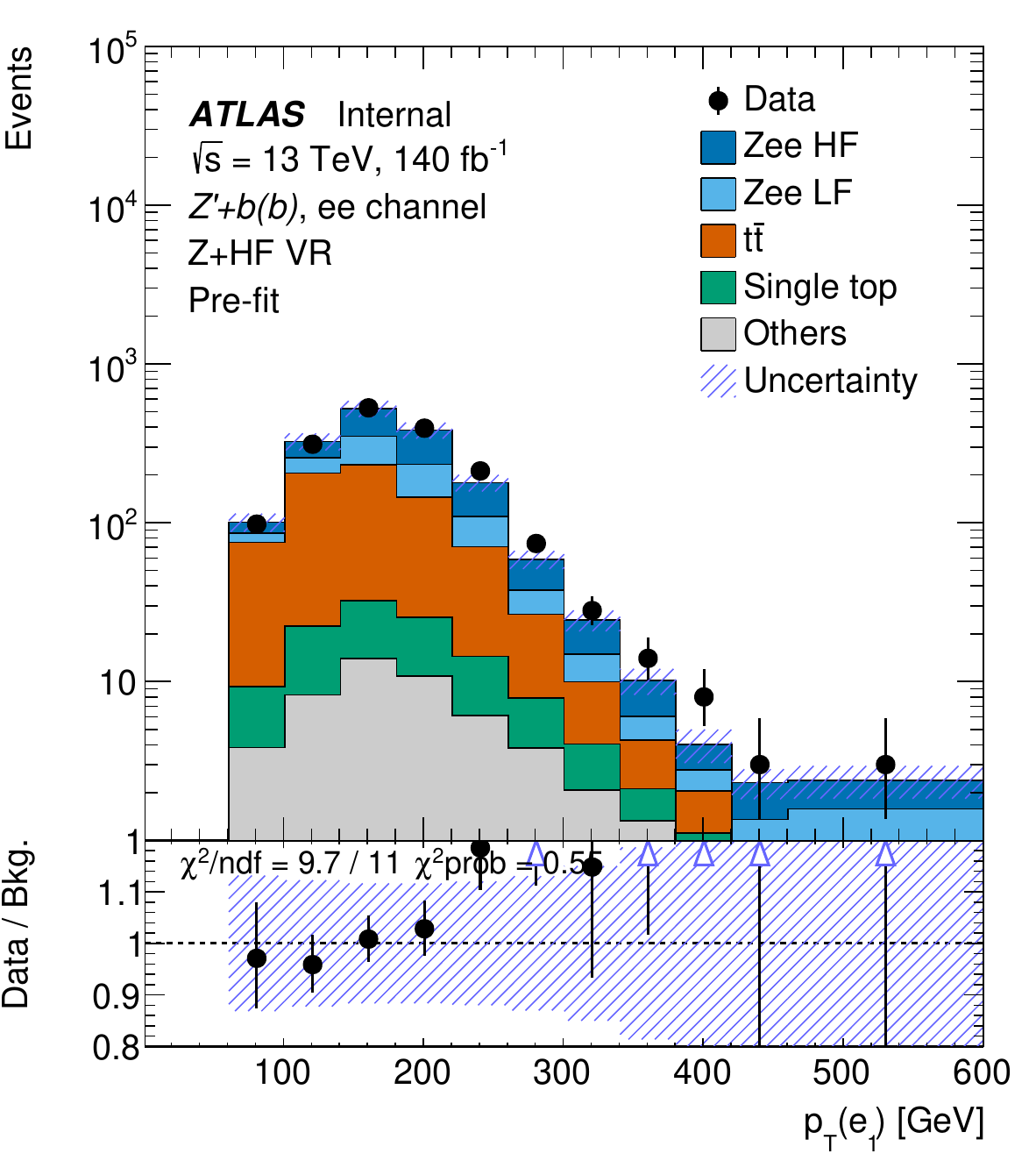}
		\label{fig:VR_atleast1b_pt_ele}
	}
	\subfloat[(f)]{
		\includegraphics[width=0.33\textwidth]{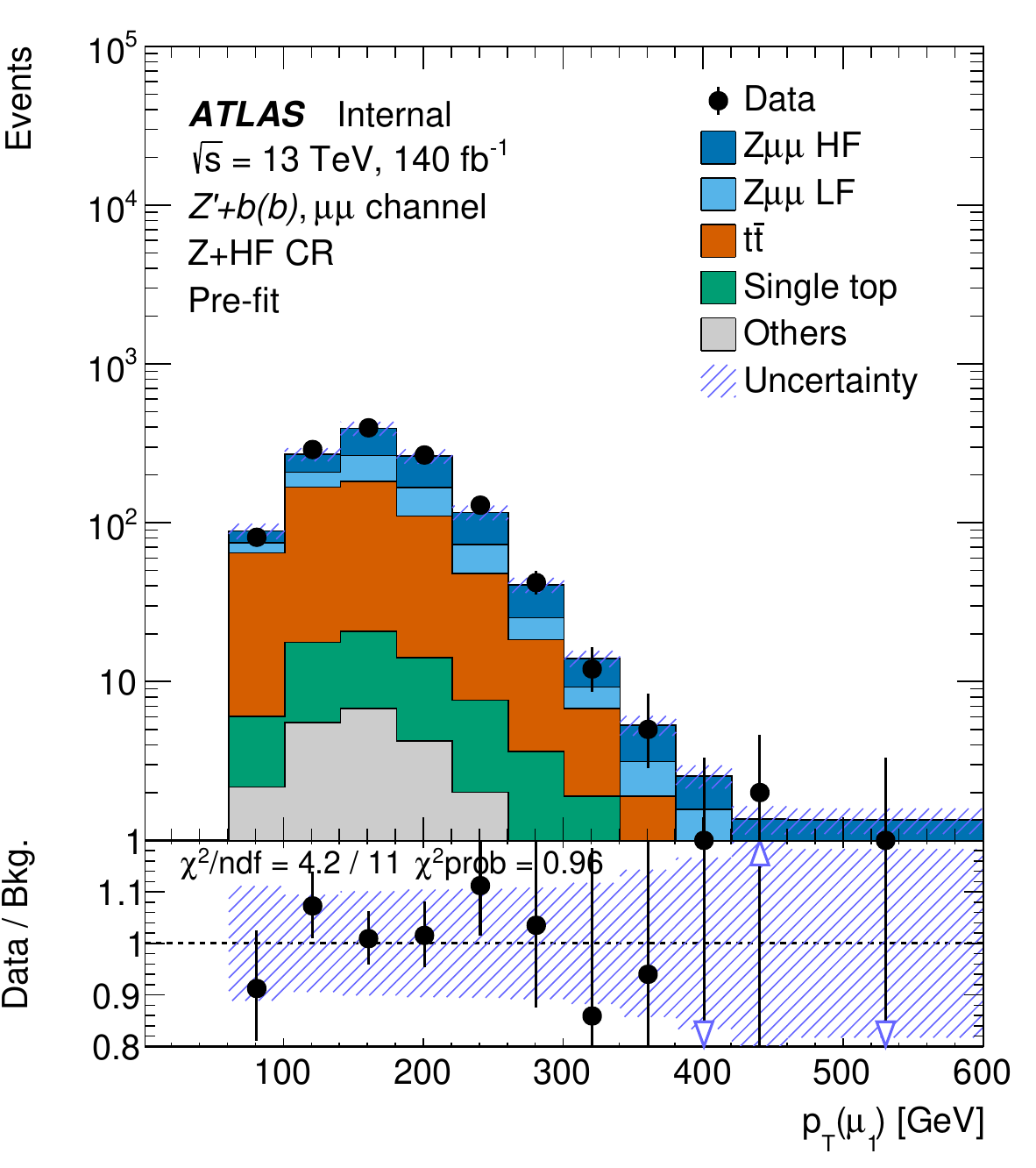}
		\label{fig:VR_atleast1b_pt_mu}
	}

	\caption{Data/MC comparison in the $\geq 1$ $b$-jet Validation Region: invariant dilepton mass (a--b), missing transverse momentum (c--d), and leading lepton $\pT$ (e--f) for the electron (left) and muon (right) channels. The uncertainty band includes statistical and background systematic uncertainties.}
	\label{fig:VR_atleast1b}
\end{figure}

A few data/MC comparison plots for the two VRs are shown in FIG.~\ref{fig:VR_0b} and \ref{fig:VR_atleast1b}. Since principally the same observations as for the CRs can be made, the VRs are not discussed in detail here. However, results and post-fit distributions are shown in Section~\ref{sec:zprime_results}.

\FloatBarrier

\subsection{Background normalisation}

The background normalisation factors are derived from a profile-likelihood fit to data simultaneously in the control and SRs. The fit is performed assuming a $Z'$ signal with $m_{Z'} = 1\,\TeV$ and coupling $g_{Z'} = 0.5$; results obtained with $g_{Z'} = 1.0$ are virtually identical, since the signal shape has negligible impact on the CRs and VRs, and are not shown separately.

\begin{figure}[h]
	\centering
	\captionsetup[subfigure]{labelformat=empty}
	\subfloat[(a)]{
		\includegraphics[width=0.45\textwidth]{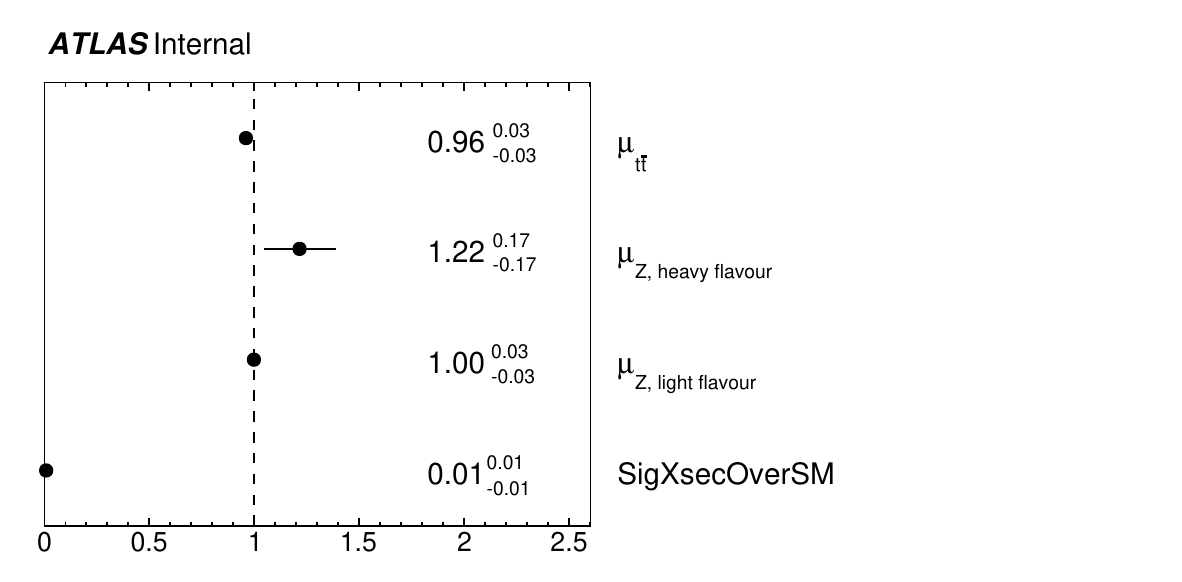}
		\label{fig:NP_CRs_mu_1000_g05_unblinded}
	}
	\hfill
	\subfloat[(b)]{
		\includegraphics[width=0.45\textwidth]{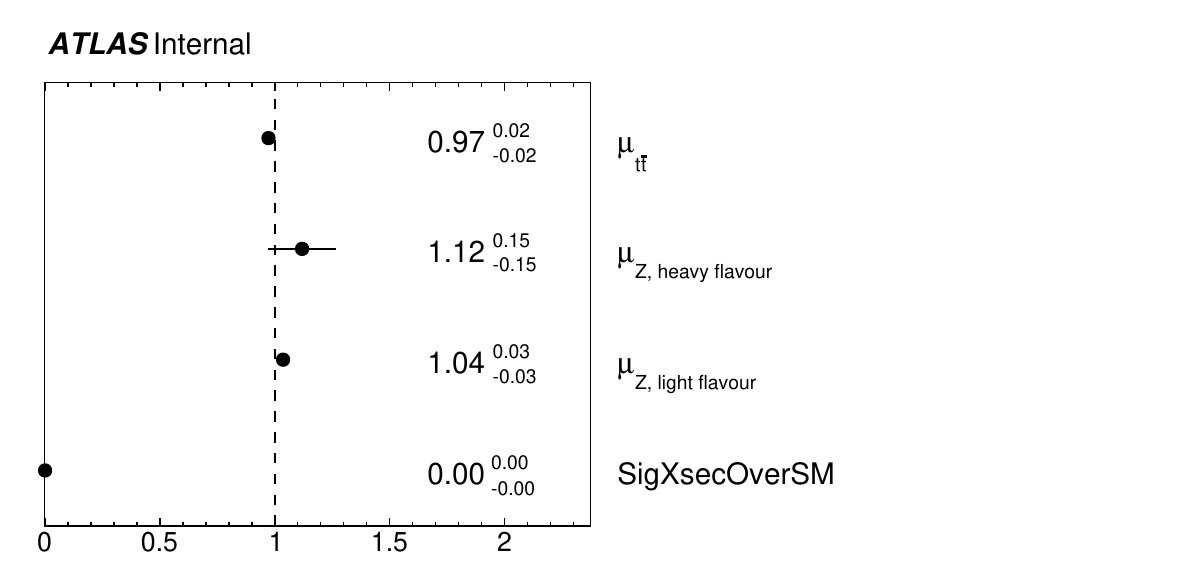}
		\label{fig:NP_CRs_ele_1000_g05_unblinded}
	}
	\caption{Normalisation factors obtained from the single-bin fits of the Control Regions in the (a) muon and (b) electron channel for the fit with $m_{\Zp}=1\,\TeV$ and $g_{\Zp}=0.5$. Statistical and systematic uncertainties are considered.}
	\label{fig:NP_CRs_1000_g05_unblinded}
\end{figure}

The normalisation factors for the $\ttbar$, $Z$+LF, and $Z$+HF backgrounds are shown in Figure~\ref{fig:NP_CRs_1000_g05_unblinded} for both channels.
In the muon channel, $\mu_{t\bar{t}} = 0.96\pm0.03$, $\mu_{\text{Z+LF}} = 1.00\pm0.03$, and $\mu_{Z+\text{HF}} = 1.22\pm0.17$.
In the electron channel, $\mu_{t\bar{t}} = 0.97\pm0.02$, $\mu_{\text{Z+LF}} = 1.04\pm0.03$, and $\mu_{Z+\text{HF}} = 1.12\pm0.15$.
All normalisation factors are compatible with unity within uncertainties, with the exception of $Z$+HF, which is pulled above one in both channels; this is attributed to correlations between background normalisation factors and systematic uncertainties. The normalisation factors are consistent between the two channels.

\begin{figure}[h]
	\centering
	\captionsetup[subfigure]{labelformat=empty}
	\subfloat[(a)]{
		\includegraphics[width=0.33\textwidth]{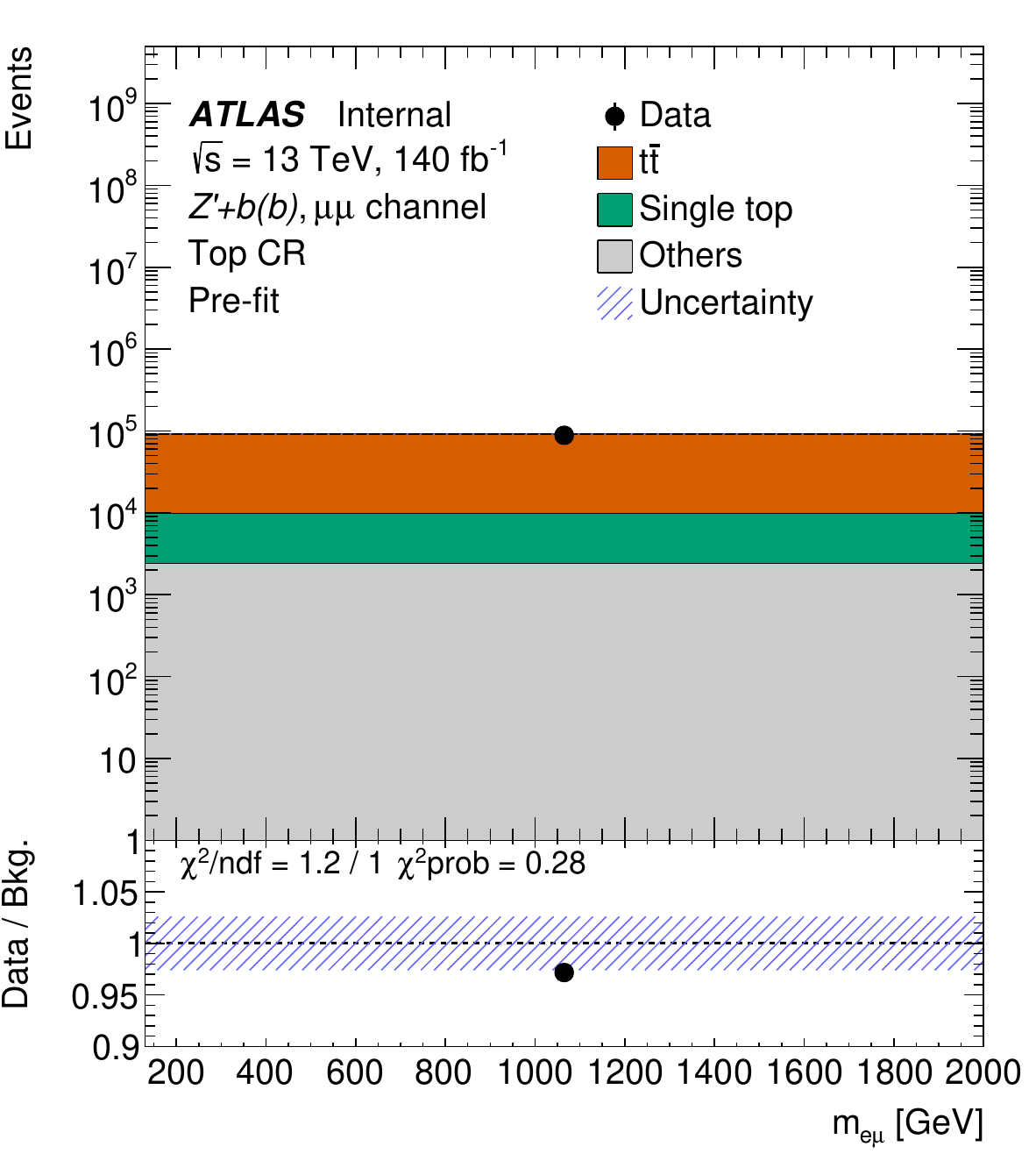}
		\label{fig:CR_fit_mu_TopCR_pre_1000_g05_unblinded}
	}
	\subfloat[(b)]{
		\includegraphics[width=0.33\textwidth]{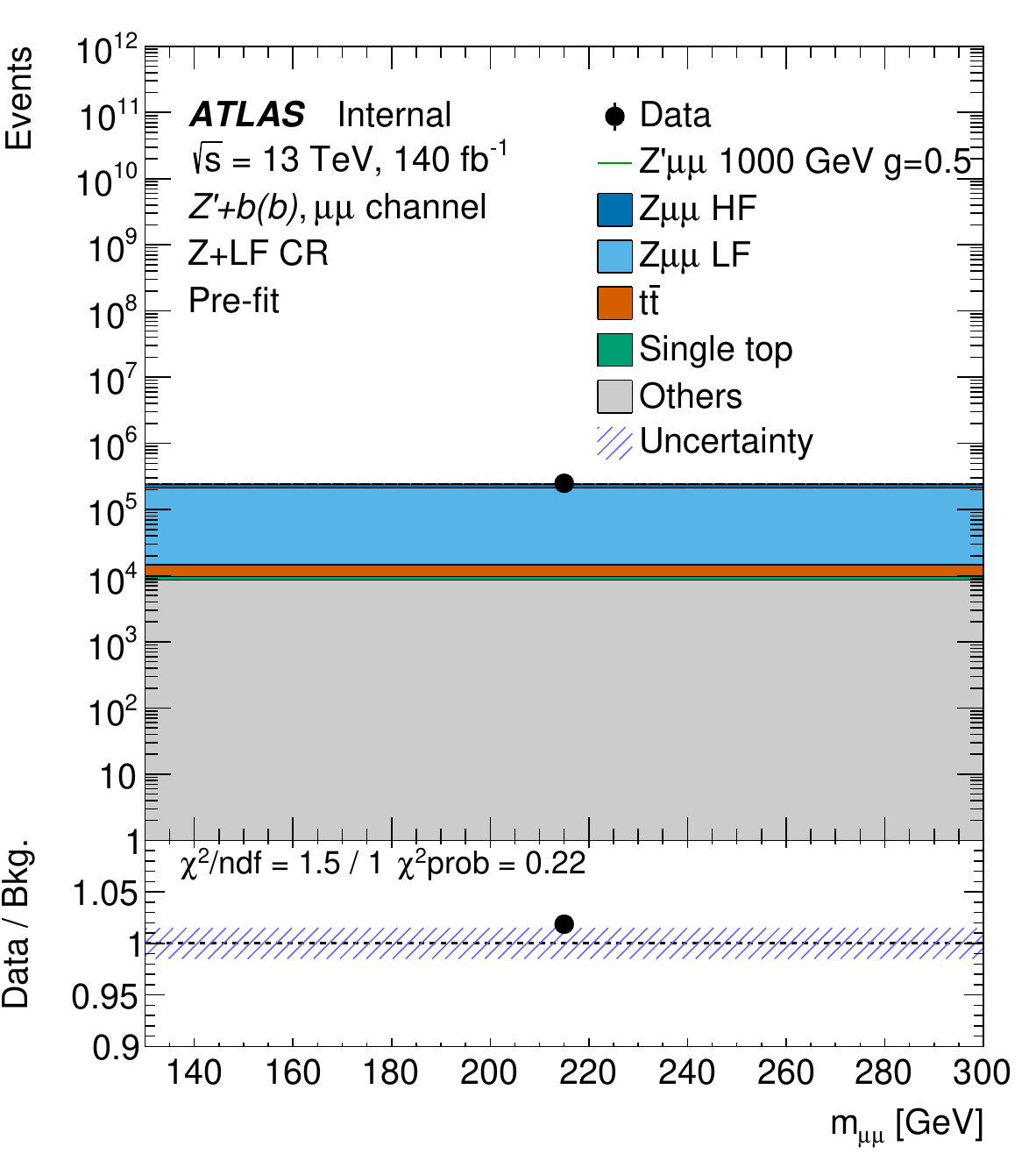}
		\label{fig:CR_fit_mu_ZCR_LF_pre_1000_g05_unblinded}
	}
	\subfloat[(c)]{
		\includegraphics[width=0.33\textwidth]{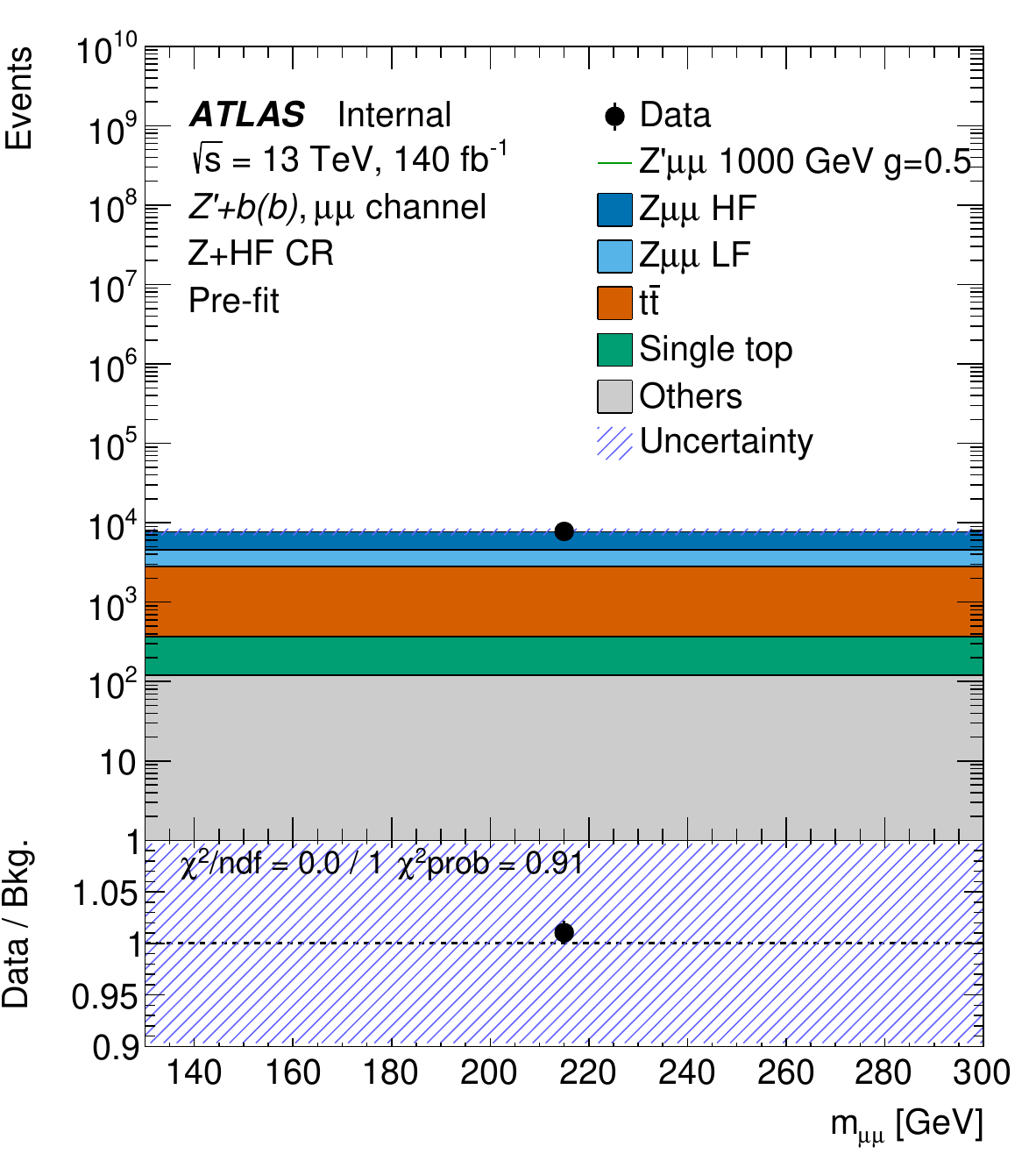}
		\label{fig:CR_fit_mu_ZCR_HF_pre_1000_g05_unblinded}
	}
	\hfill
	\subfloat[(d)]{
		\includegraphics[width=0.33\textwidth]{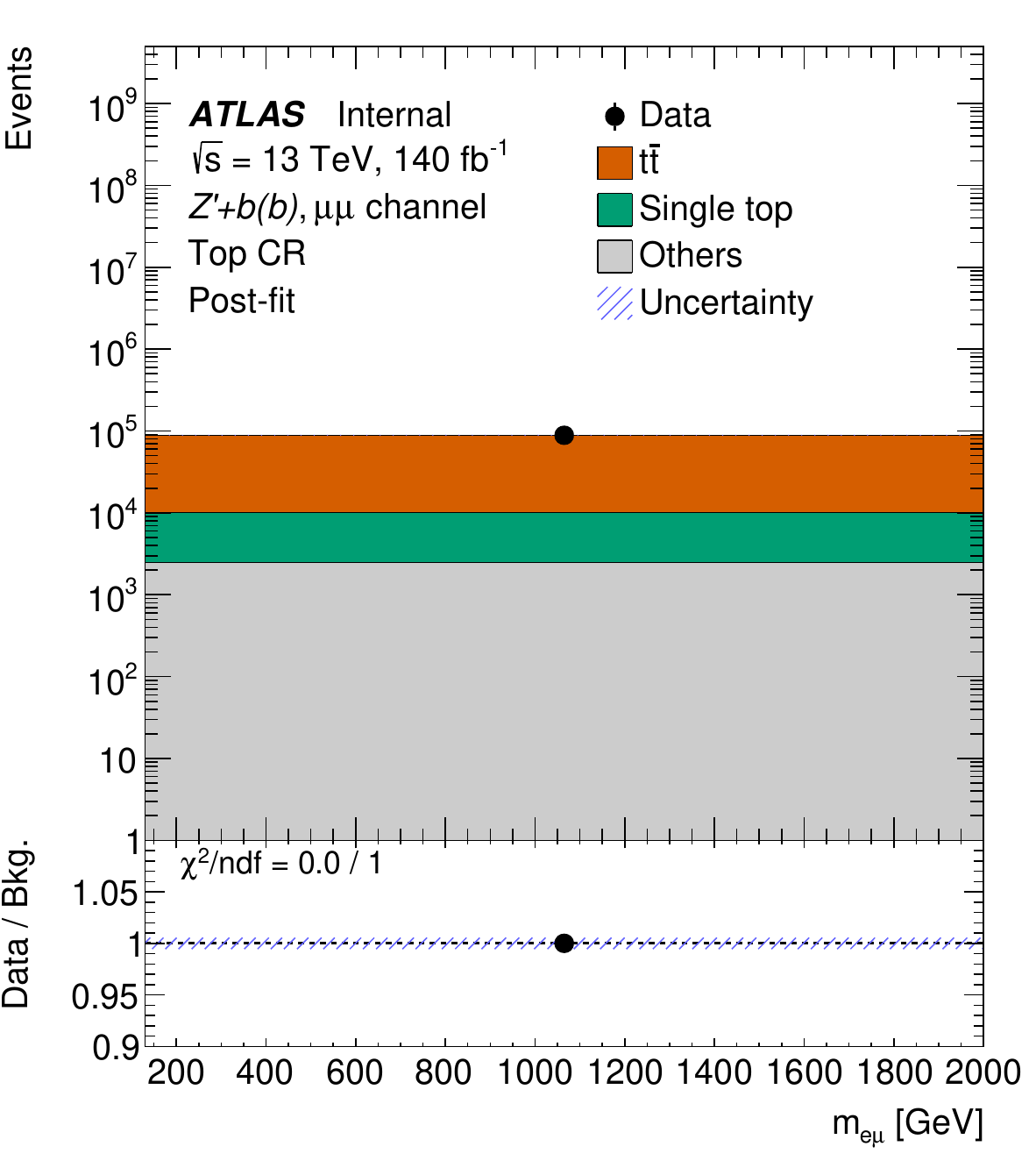}
		\label{fig:CR_fit_mu_TopCR_post_1000_g05_unblinded}
	}
	\subfloat[(e)]{
		\includegraphics[width=0.33\textwidth]{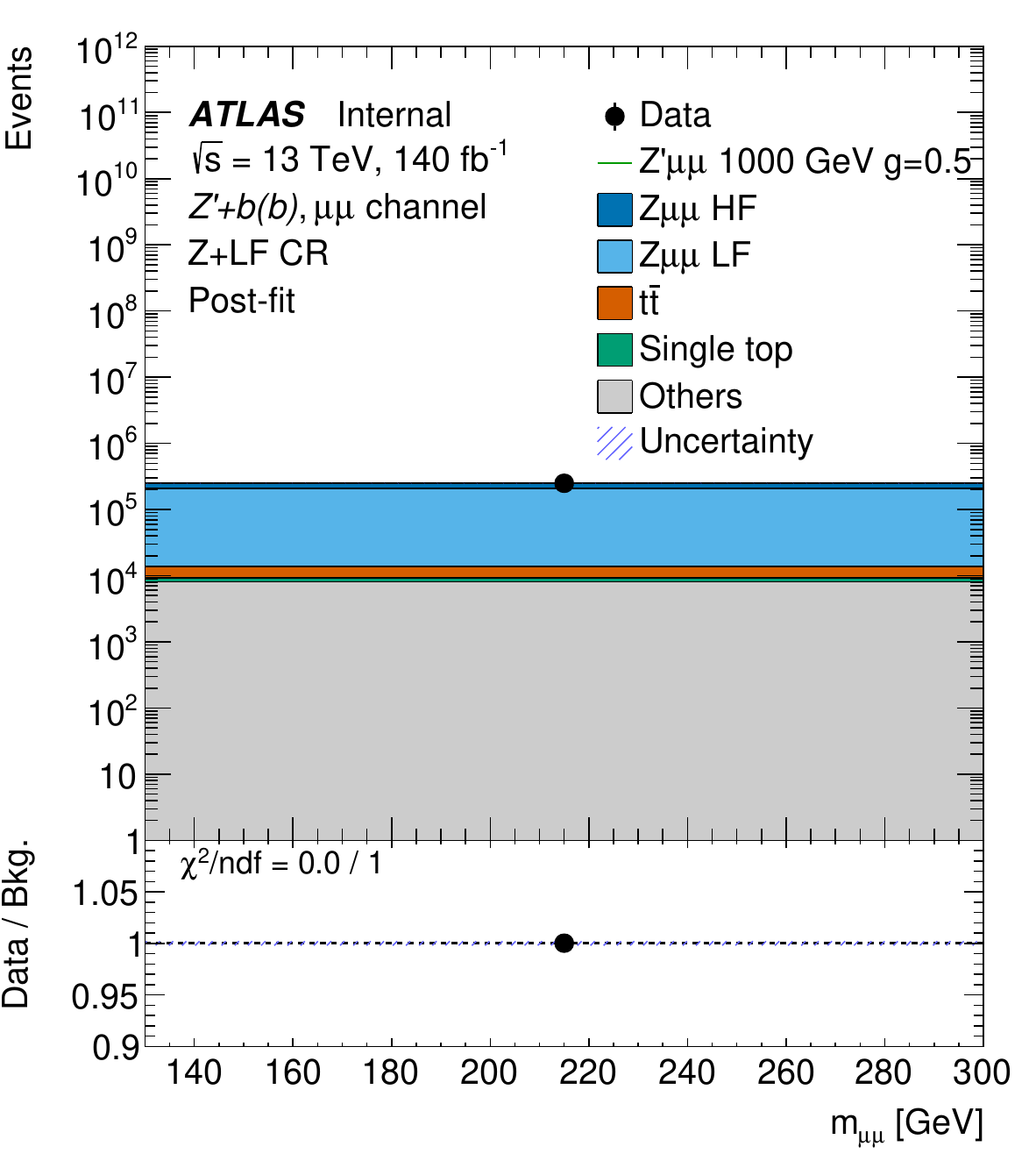}
		\label{fig:CR_fit_mu_ZCR_LF_post_1000_g05_unblinded}
	}
	\subfloat[(f)]{
		\includegraphics[width=0.33\textwidth]{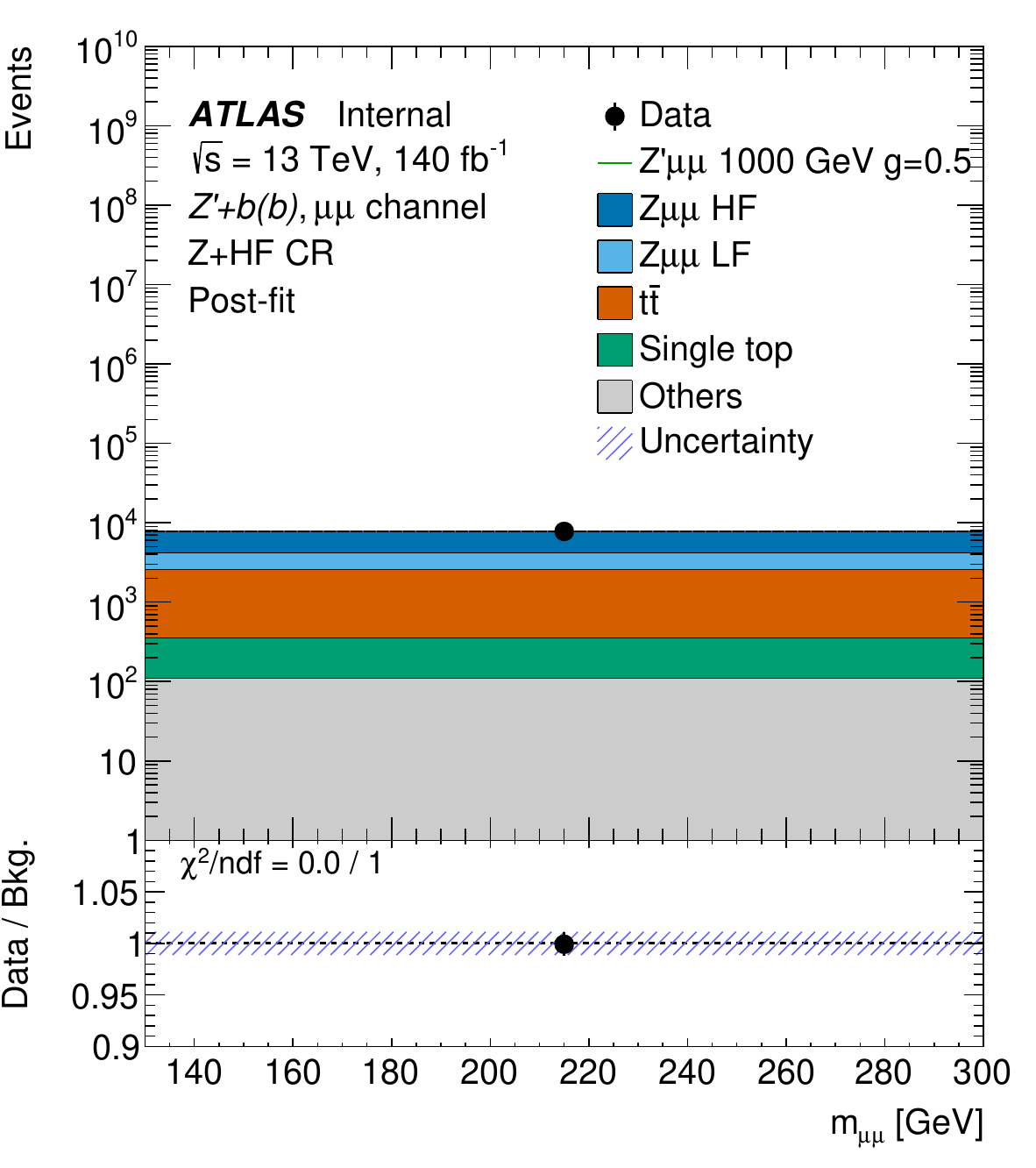}
		\label{fig:CR_fit_mu_ZCR_HF_post_1000_g05_unblinded}
	}

	\caption{Pre-fit (a)--(c) and post-fit (d)--(f) distributions in the muon channel Control Regions for $m_{\Zp}=1\,\TeV$, $g_{\Zp}=0.5$: top CR (a,d), $Z$+LF CR (b,e), and $Z$+HF CR (c,f). The uncertainty band includes statistical and systematic uncertainties.}
	\label{fig:CR_fit_mu_1000_g05_unblinded}
\end{figure}

\begin{figure}[h]
	\centering
	\captionsetup[subfigure]{labelformat=empty}
	\subfloat[(a)]{
		\includegraphics[width=0.33\textwidth]{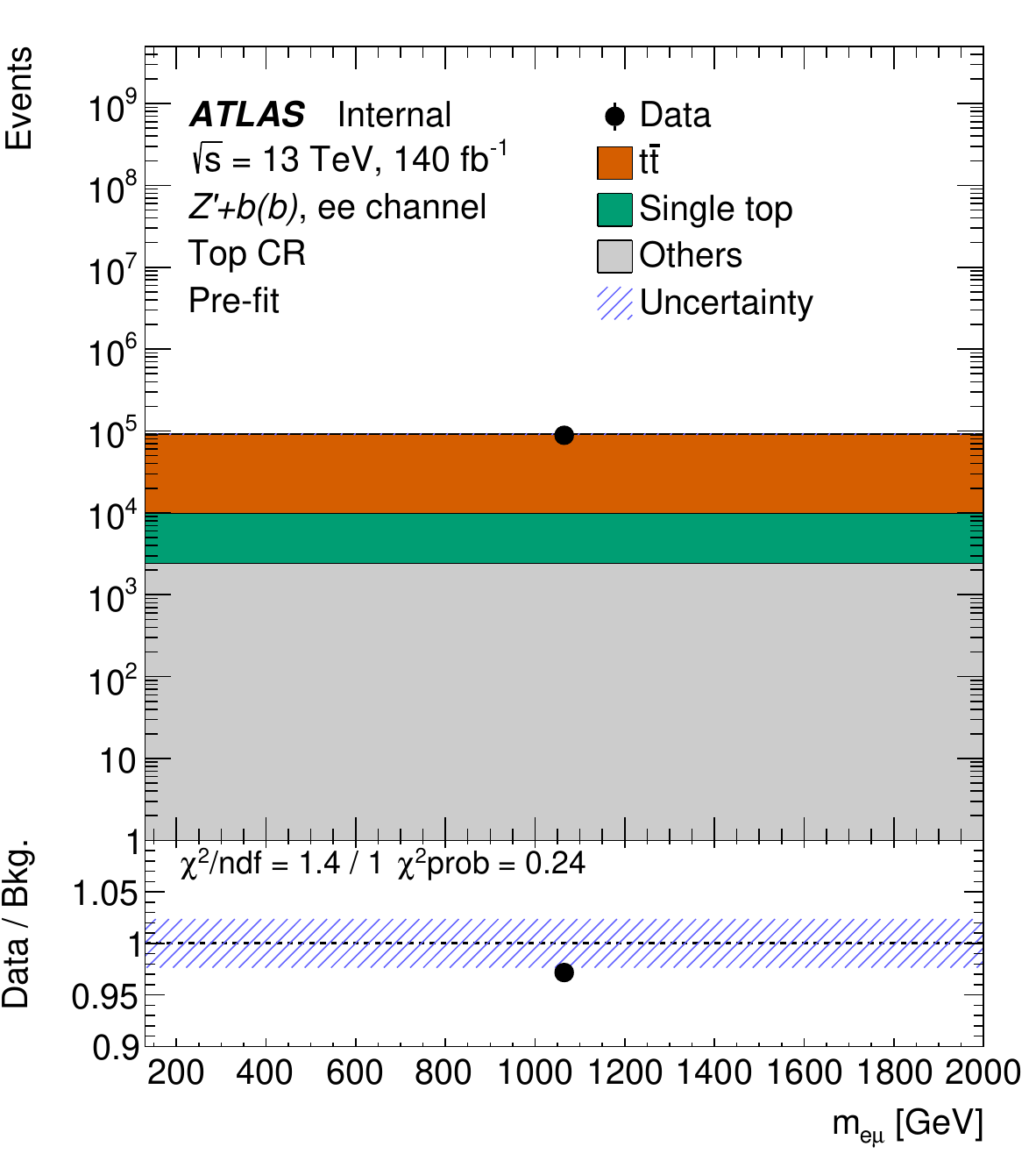}
		\label{fig:CR_fit_ele_TopCR_pre_1000_g05_unblinded}
	}
	\subfloat[(b)]{
		\includegraphics[width=0.33\textwidth]{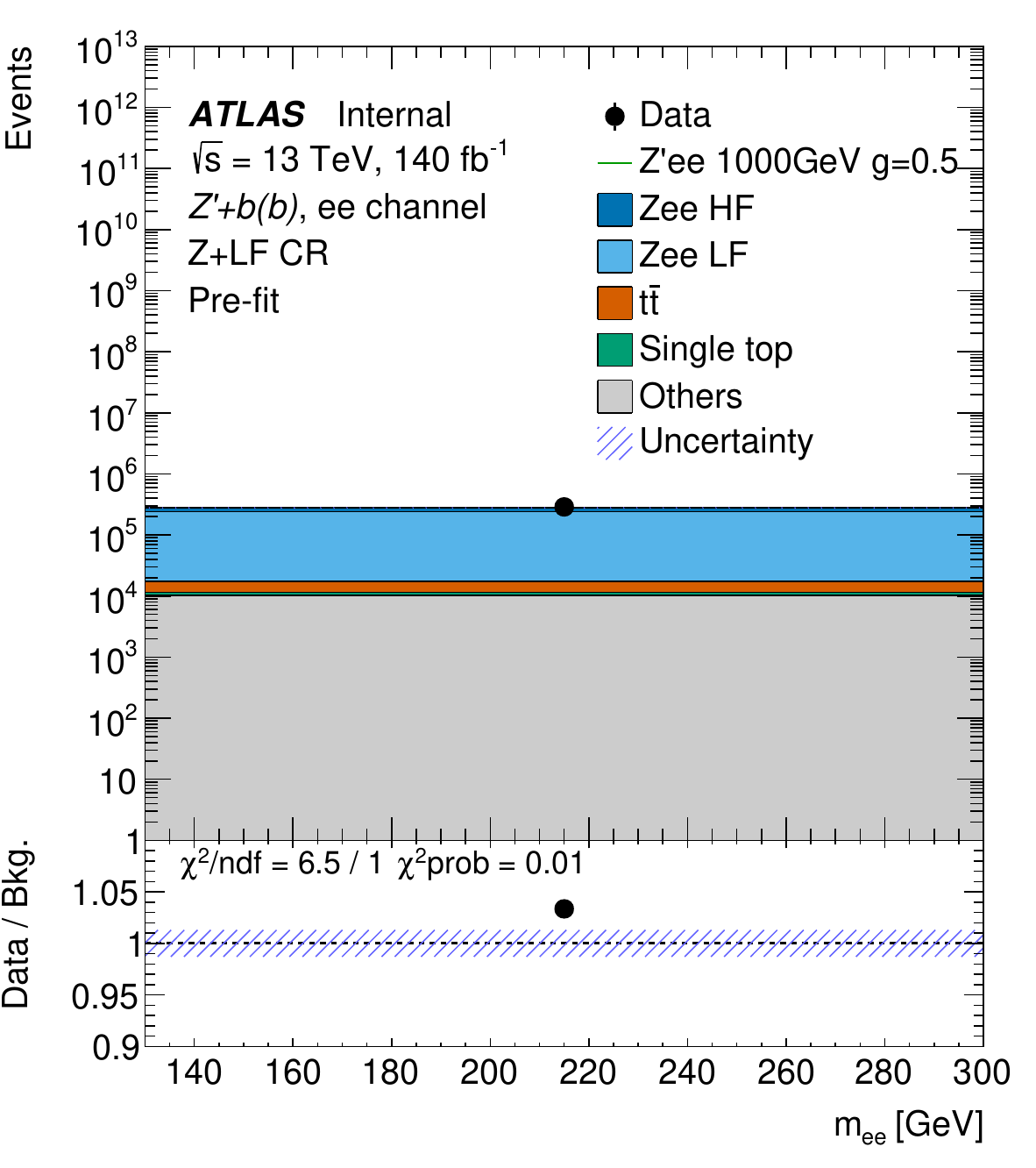}
		\label{fig:CR_fit_ele_ZCR_LF_pre_1000_g05_unblinded}
	}
	\subfloat[(c)]{
		\includegraphics[width=0.33\textwidth]{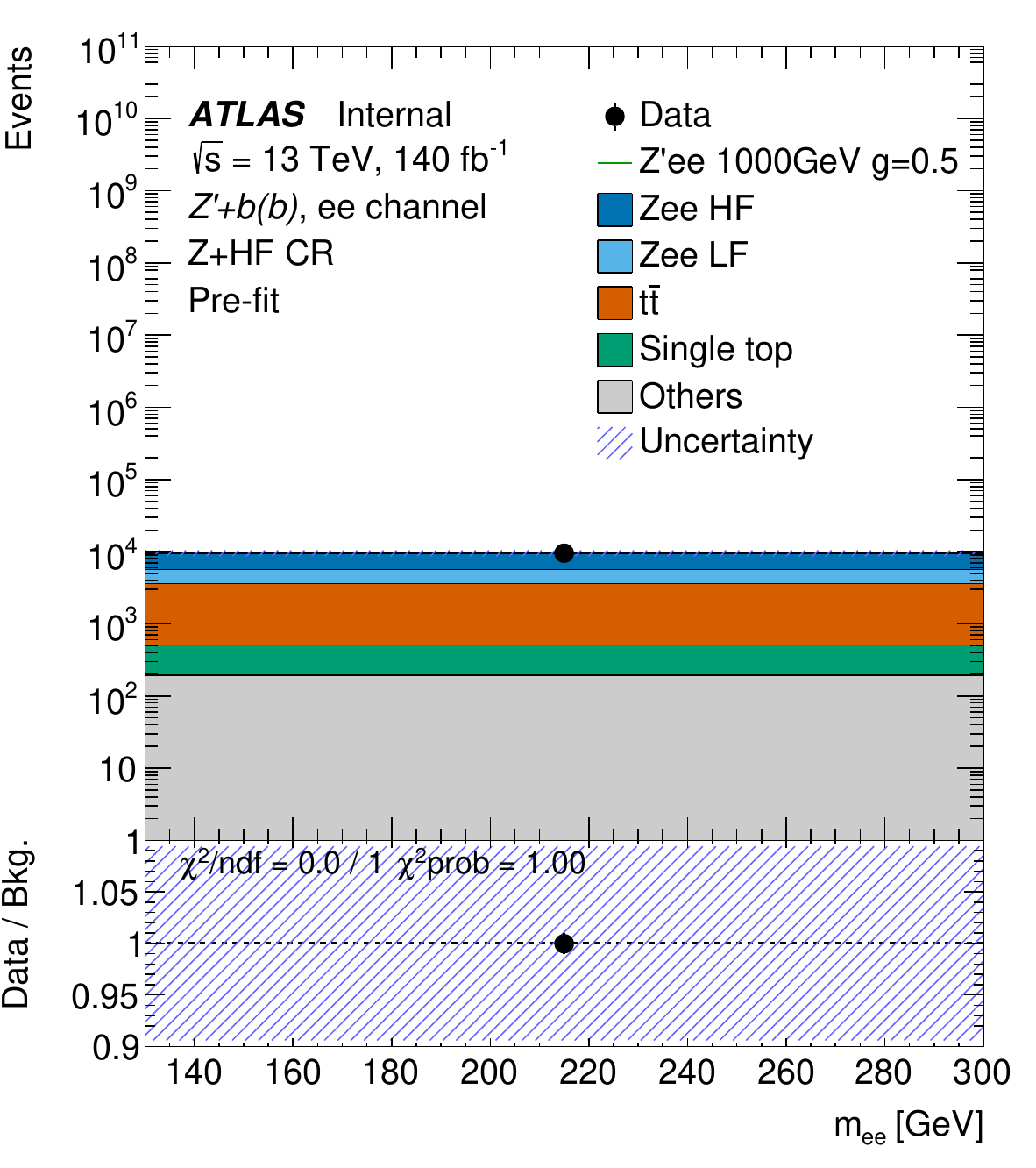}
		\label{fig:CR_fit_ele_ZCR_HF_pre_1000_g05_unblinded}
	}
	\hfill
	\subfloat[(d)]{
		\includegraphics[width=0.33\textwidth]{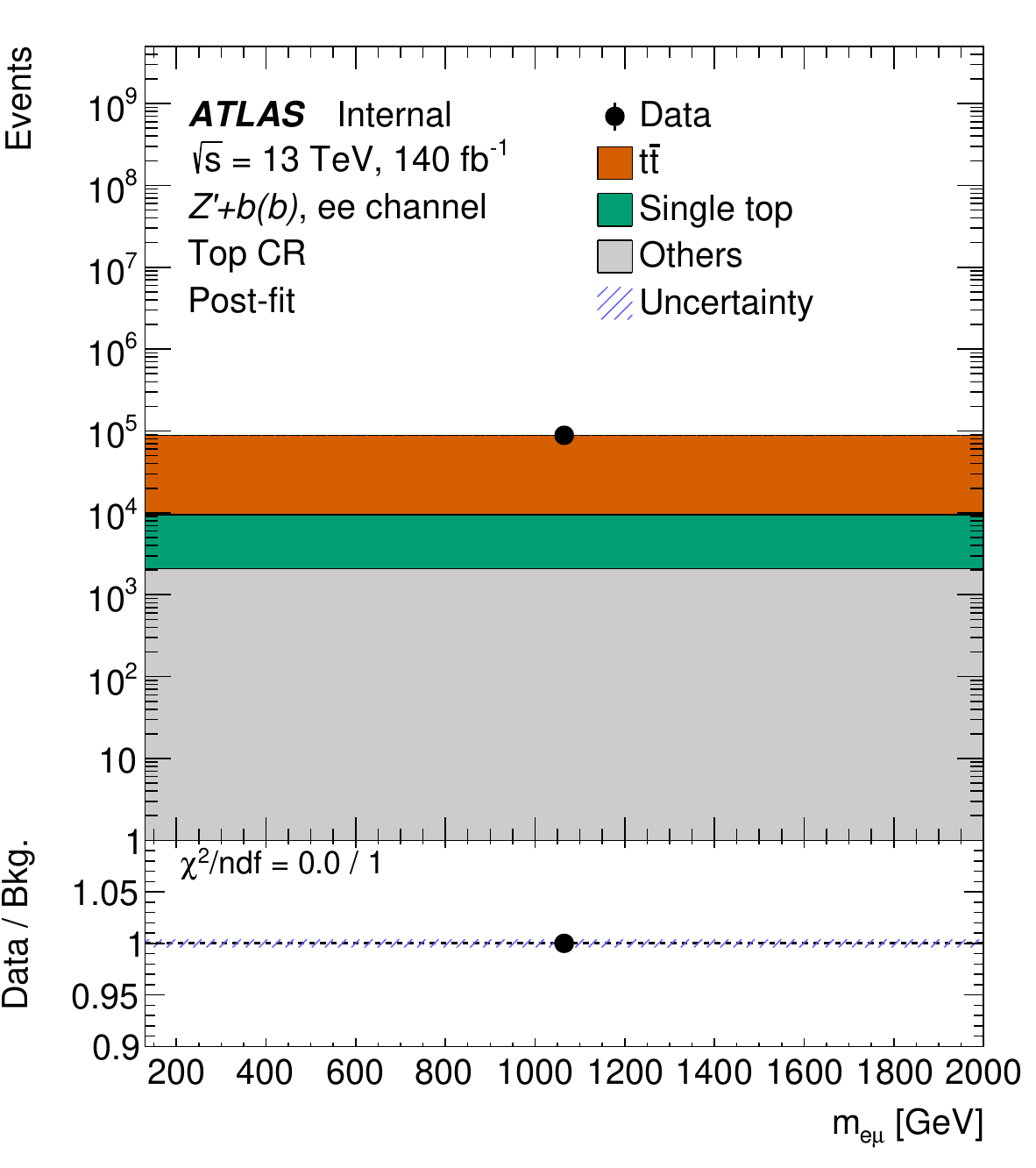}
		\label{fig:CR_fit_ele_TopCR_post_1000_g05_unblinded}
	}
	\subfloat[(e)]{
		\includegraphics[width=0.33\textwidth]{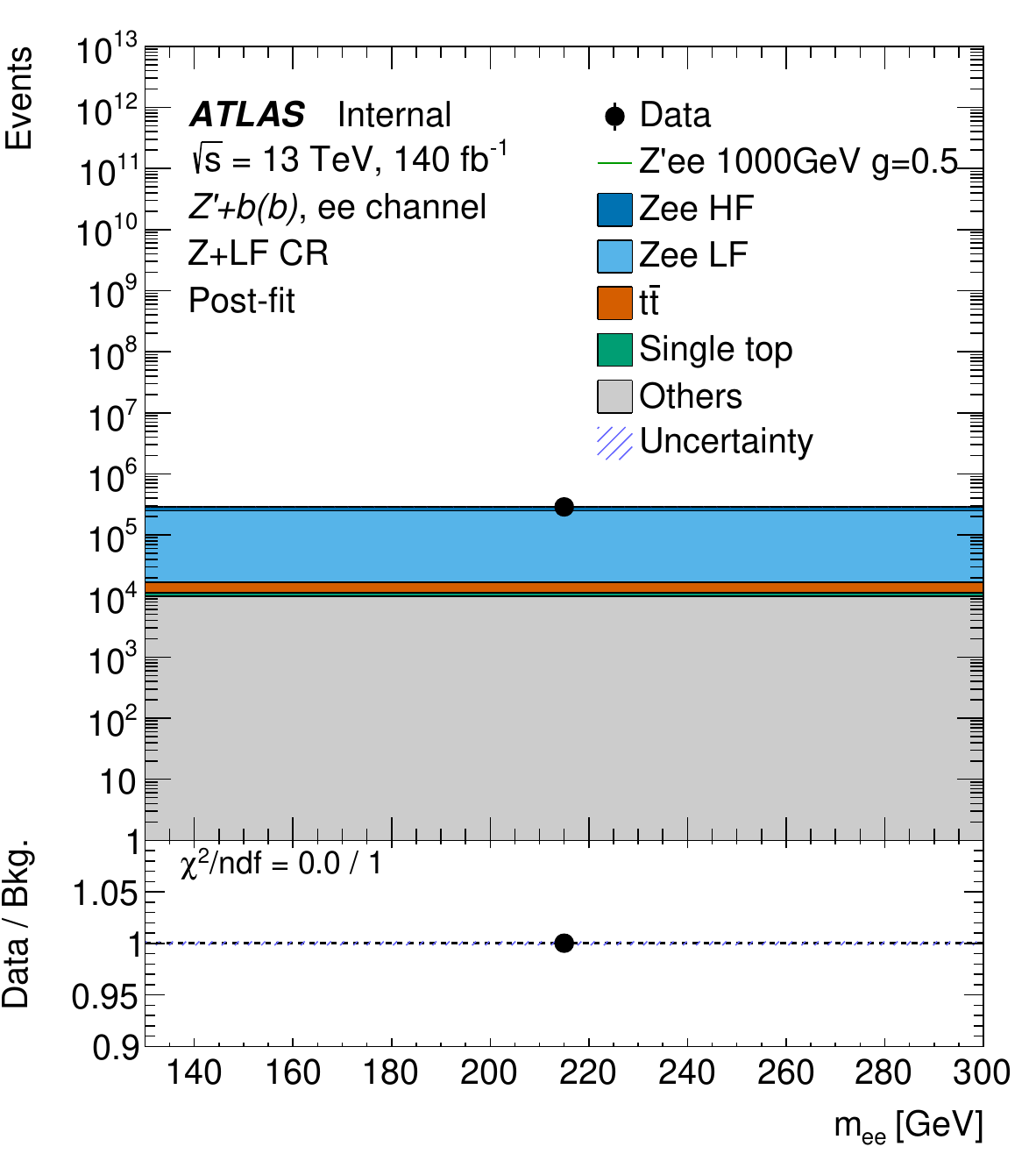}
		\label{fig:CR_fit_ele_ZCR_LF_post_1000_g05_unblinded}
	}
	\subfloat[(f)]{
		\includegraphics[width=0.33\textwidth]{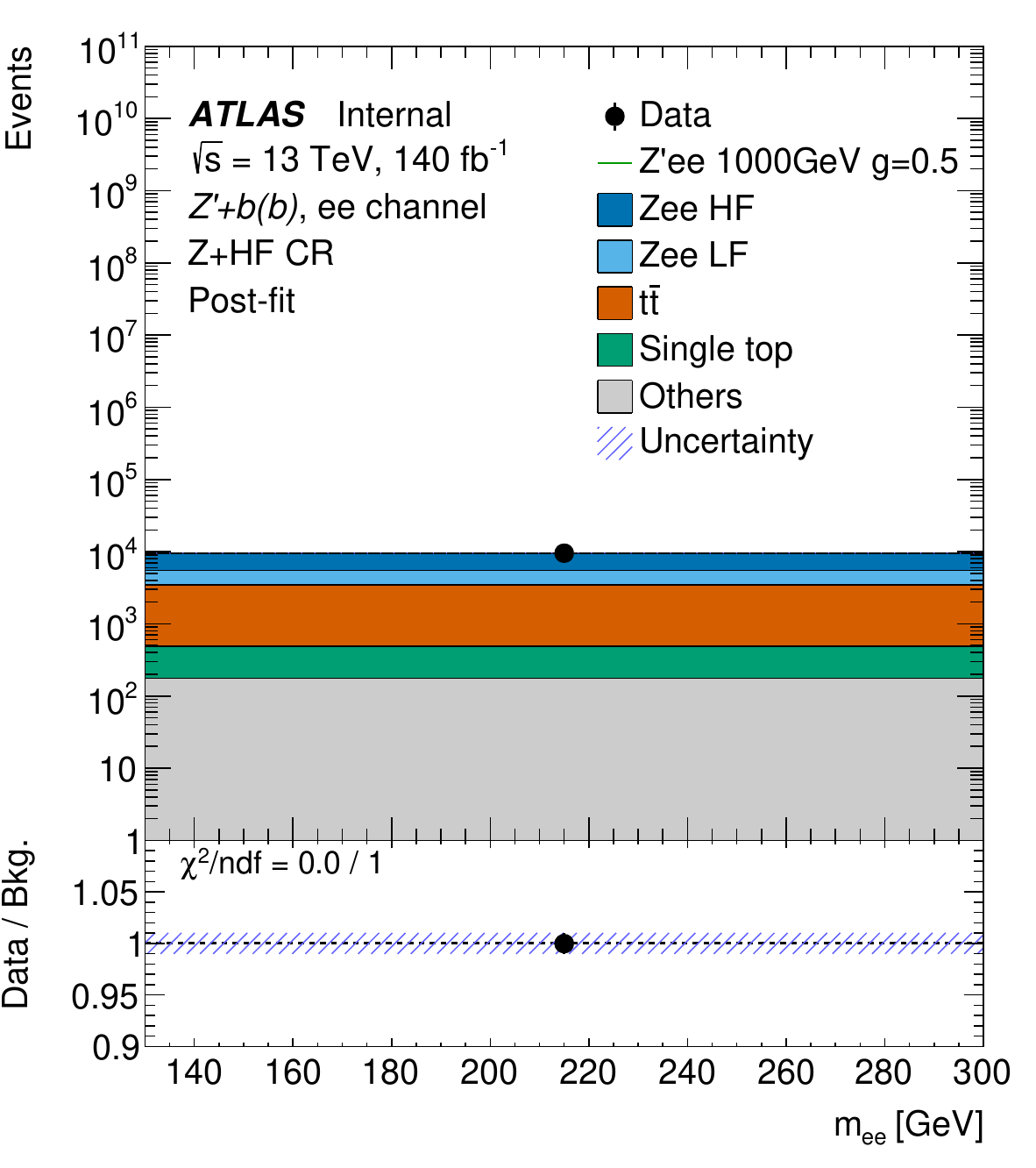}
		\label{fig:CR_fit_ele_ZCR_HF_post_1000_g05_unblinded}
	}

	\caption{Pre-fit (a)--(c) and post-fit (d)--(f) distributions in the electron channel Control Regions for $m_{\Zp}=1\,\TeV$, $g_{\Zp}=0.5$: top CR (a,d), $Z$+LF CR (b,e), and $Z$+HF CR (c,f). The uncertainty band includes statistical and systematic uncertainties.}
	\label{fig:CR_fit_ele_1000_g05_unblinded}
\end{figure}

FIG.~\ref{fig:CR_fit_mu_1000_g05_unblinded} and~\ref{fig:CR_fit_ele_1000_g05_unblinded} show the pre-fit and post-fit distributions in the top and $Z$ CRs for the muon and electron channels, respectively. Good agreement between data and the post-fit background prediction is observed in all CRs.

\section{Signal Regions}
\label{sec:zprime_SR}
Signal regions are defined by the $b$-jet multiplicity, consistent with Section~\ref{sec:zprime_strategy}. The lower bound on the dilepton invariant mass is $m_{\ell\ell}>300\;\mathrm{GeV}$. Additional requirements on $\met$ are applied. Missing transverse momentum has two sources: real $\met$ from neutrinos and from pile-up and detector acceptance, and fake $\met$ from mismeasurements of particle momenta. The $\met$ significance $\metsig$ (Section~\ref{sec:reco:met}, Equation~\ref{eq:metsig}) discriminates between the two.

\begin{figure}[h]
\centering
\subfloat[]{
  \includegraphics[width=0.45\textwidth]{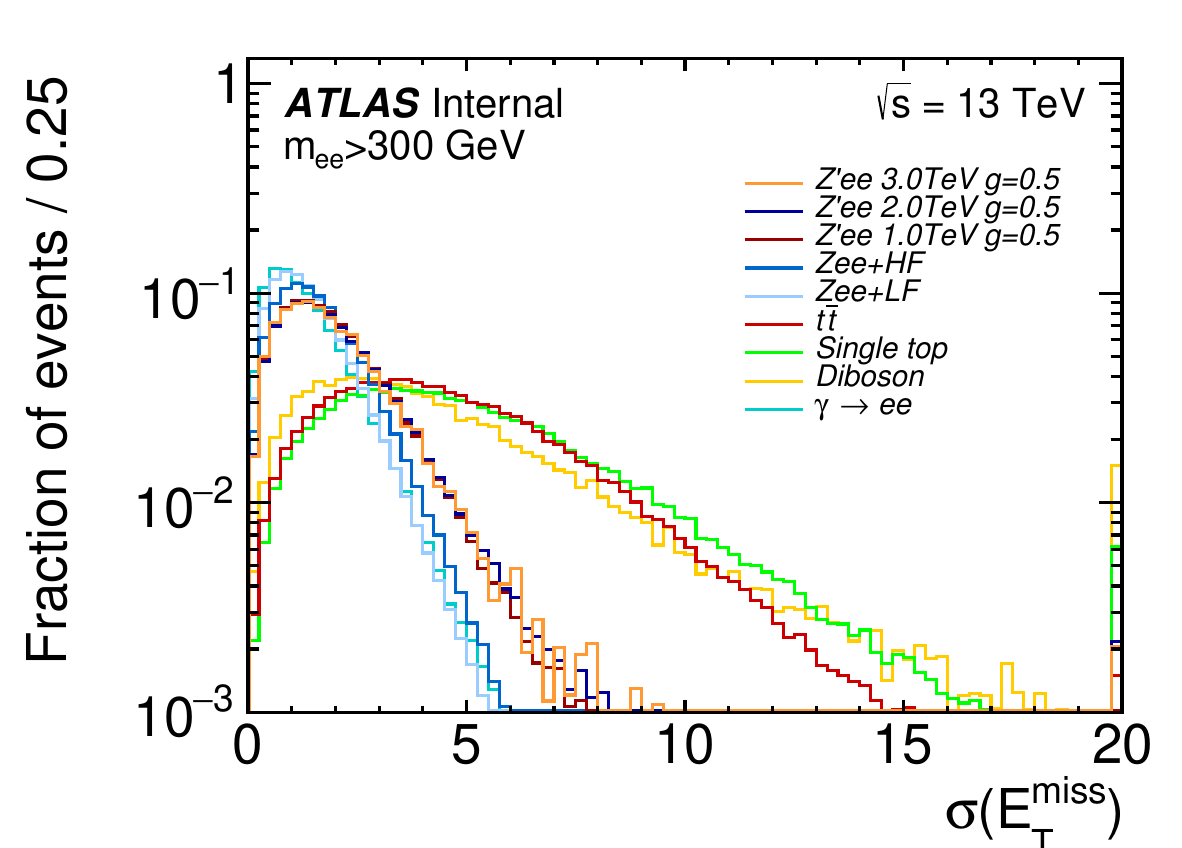}
  \label{fig:metsig_ele}
}
\hfill
\subfloat[]{
  \includegraphics[width=0.45\textwidth]{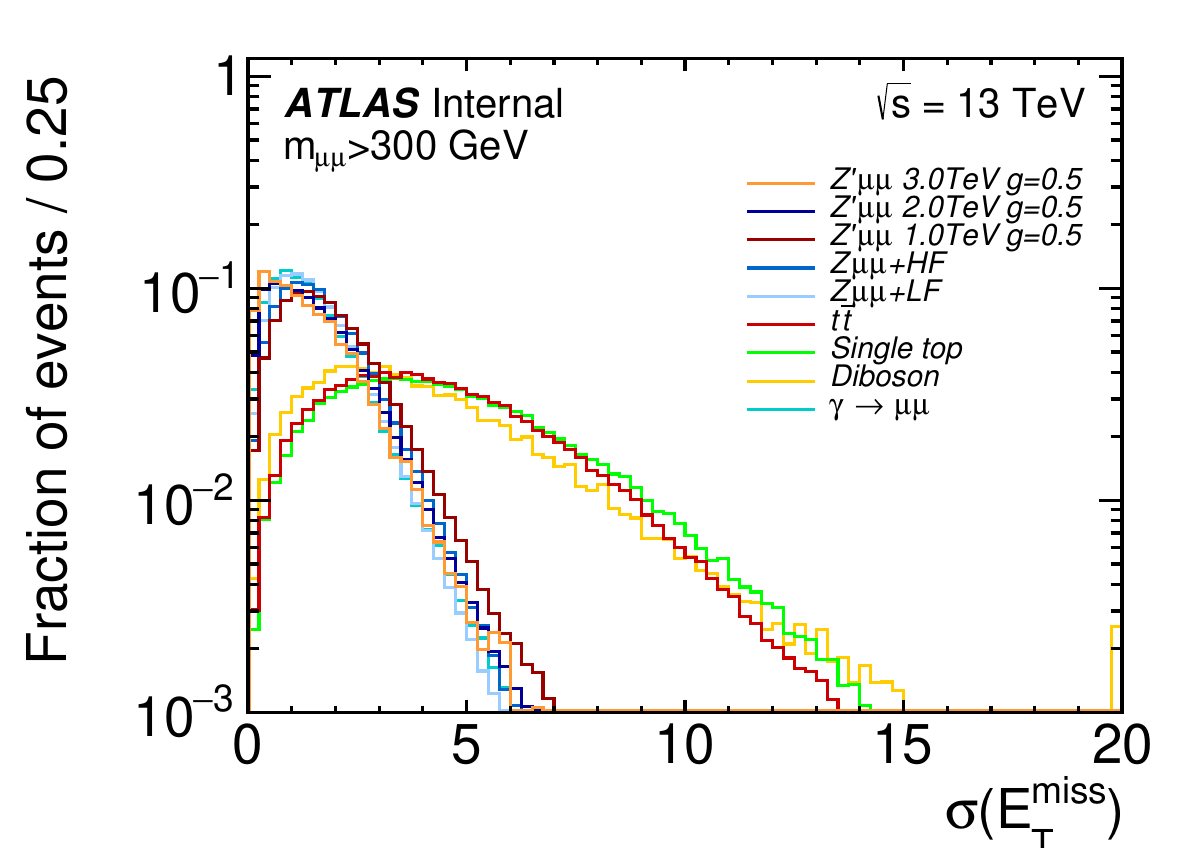}
  \label{fig:metsig_mu}
}
\caption{$\metsig$ of the signal and background processes for the electron (left) and muon (right) channels.}
\label{fig:metsig}
\end{figure}

FIG.~\ref{fig:metsig} shows the $\metsig$ distribution of the signal and background processes (normalised per process) for the muon channel and the electron channel, respectively. The $\ttbar$ background and also the 
Single Top and diboson background tend to have higher values of $\metsig$, which indicates the existence of neutrinos in the final state. Since the signal process has no neutrinos in the final state, a cut on $\metsig$
is a good possibility to reduce these backgrounds. Therefore, an event selection requirement of $\metsig < 5.0$ is introduced to reduce the background and to keep the loss of signal events small. 
The $\metsig$-cut was optimised in a dedicated study.

\begin{figure}
	\centering
	\includegraphics[width=0.5\linewidth]{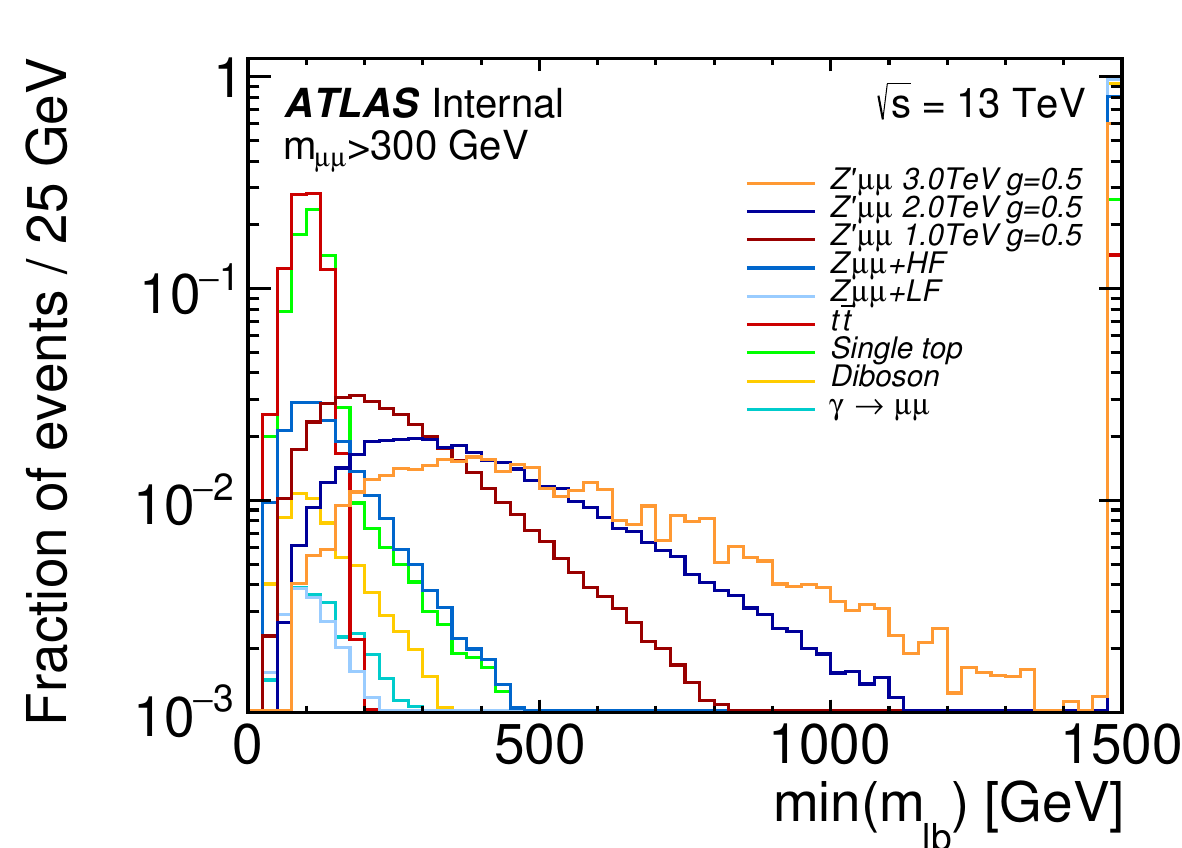}
	\caption{$\mathrm{min}(m_{\ell b})$ distribution of the signal and background processes for the muon channel.}
	\label{fig:minmlb_mu}
\end{figure}

A further discriminating variable is the minimum invariant mass of any lepton--$b$-jet pair in the event, $\mathrm{min}(m_{\ell b})$, shown for background and signal in FIG.~\ref{fig:minmlb_mu}. The distribution differs markedly between signal and background: the $\ttbar$ and single-top backgrounds peak near the top-quark mass, while the signal is concentrated at higher values. A requirement of $\mathrm{min}(m_{\ell b})>155\;\mathrm{GeV}$, optimised in a dedicated study, is applied in signal regions containing $b$-jets to suppress the top background.

\begin{figure}[h]
	\centering
	\subfloat[]{
		\includegraphics[width=0.4\textwidth]{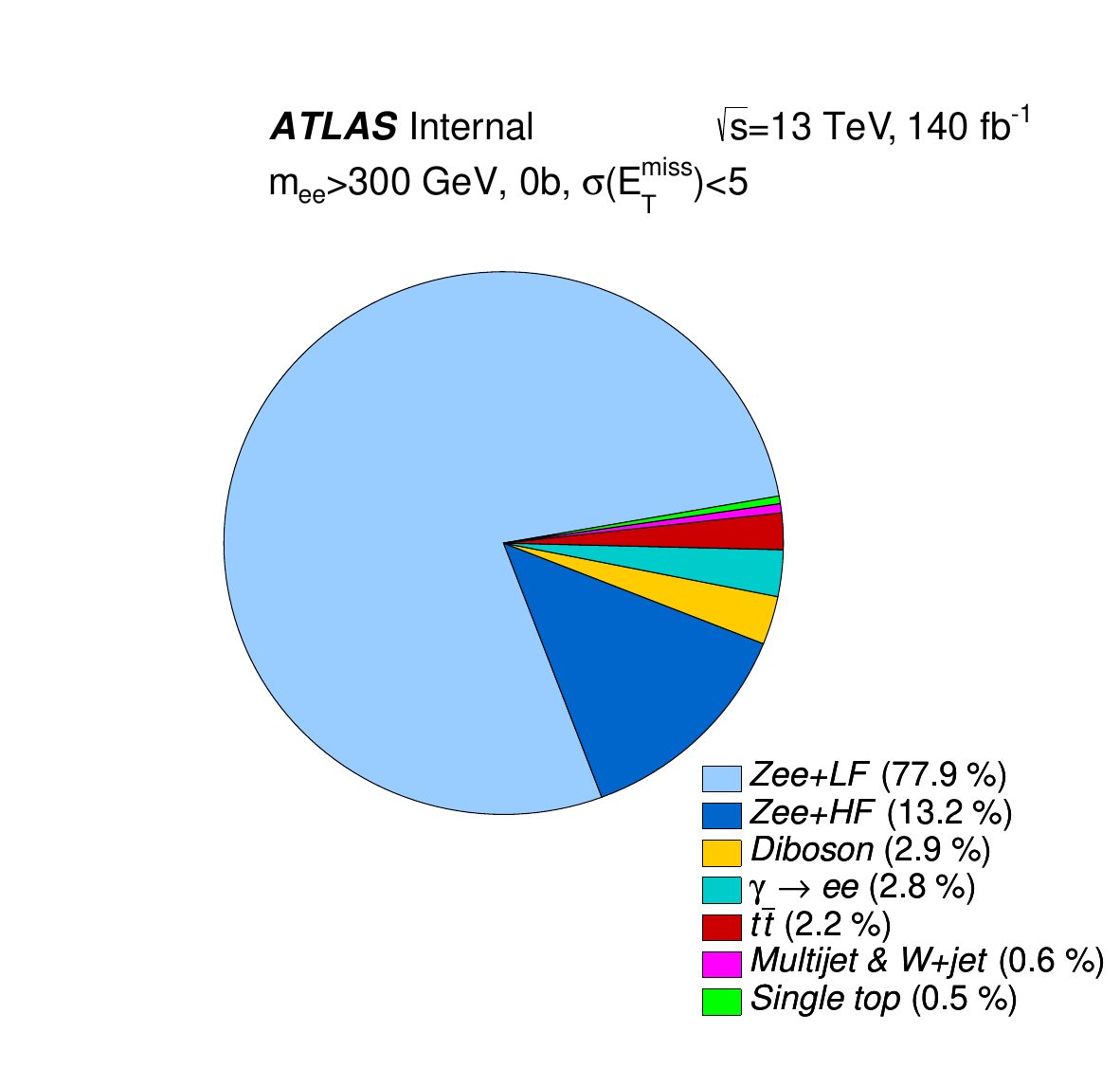}
		\label{fig:bkg_SR_0b_ele}
	}
	\subfloat[]{
		\includegraphics[width=0.4\textwidth]{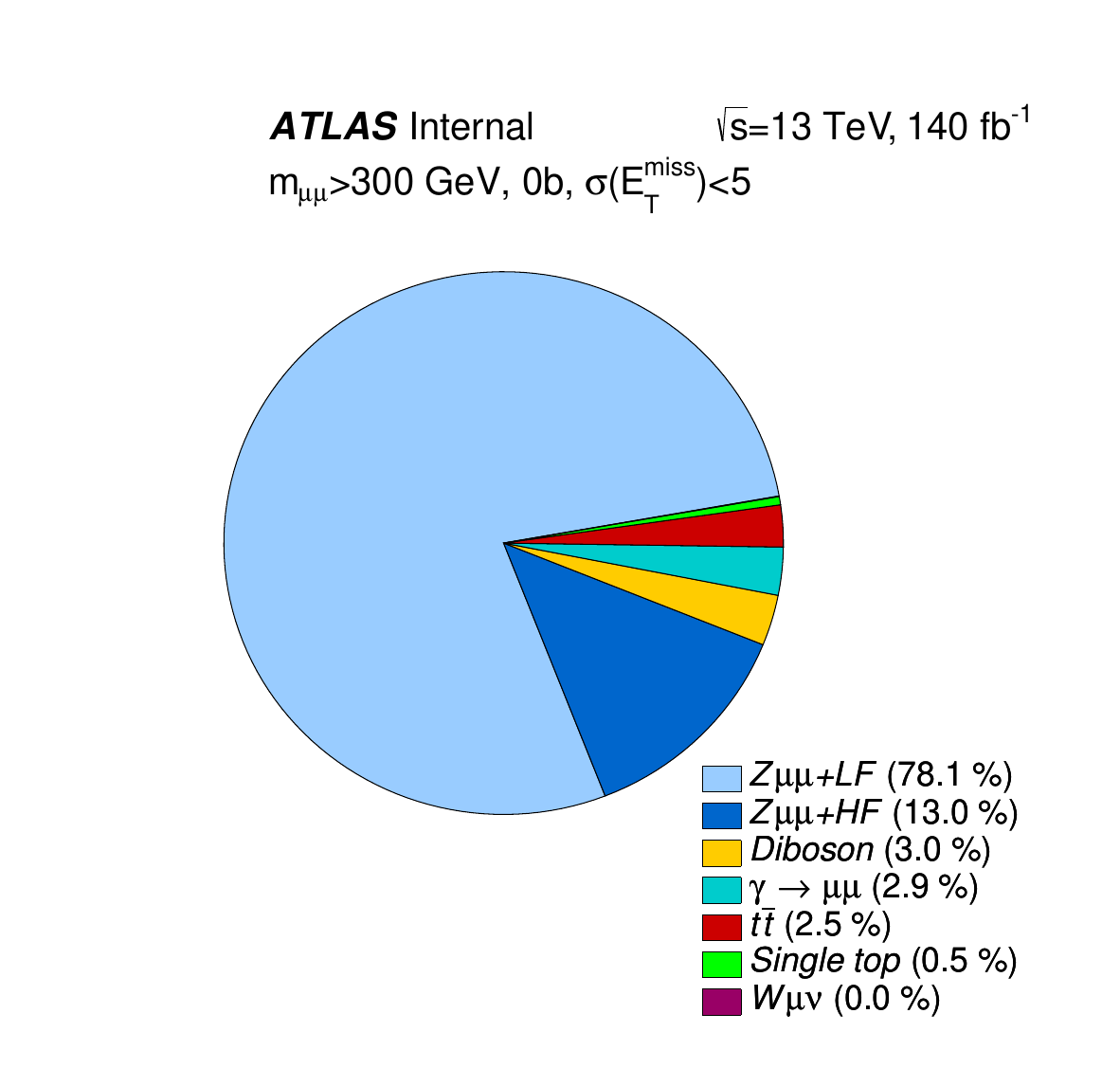}
		\label{fig:bkg_SR_0b_mu}
	}
	
	\subfloat[]{
		\includegraphics[width=0.4\textwidth]{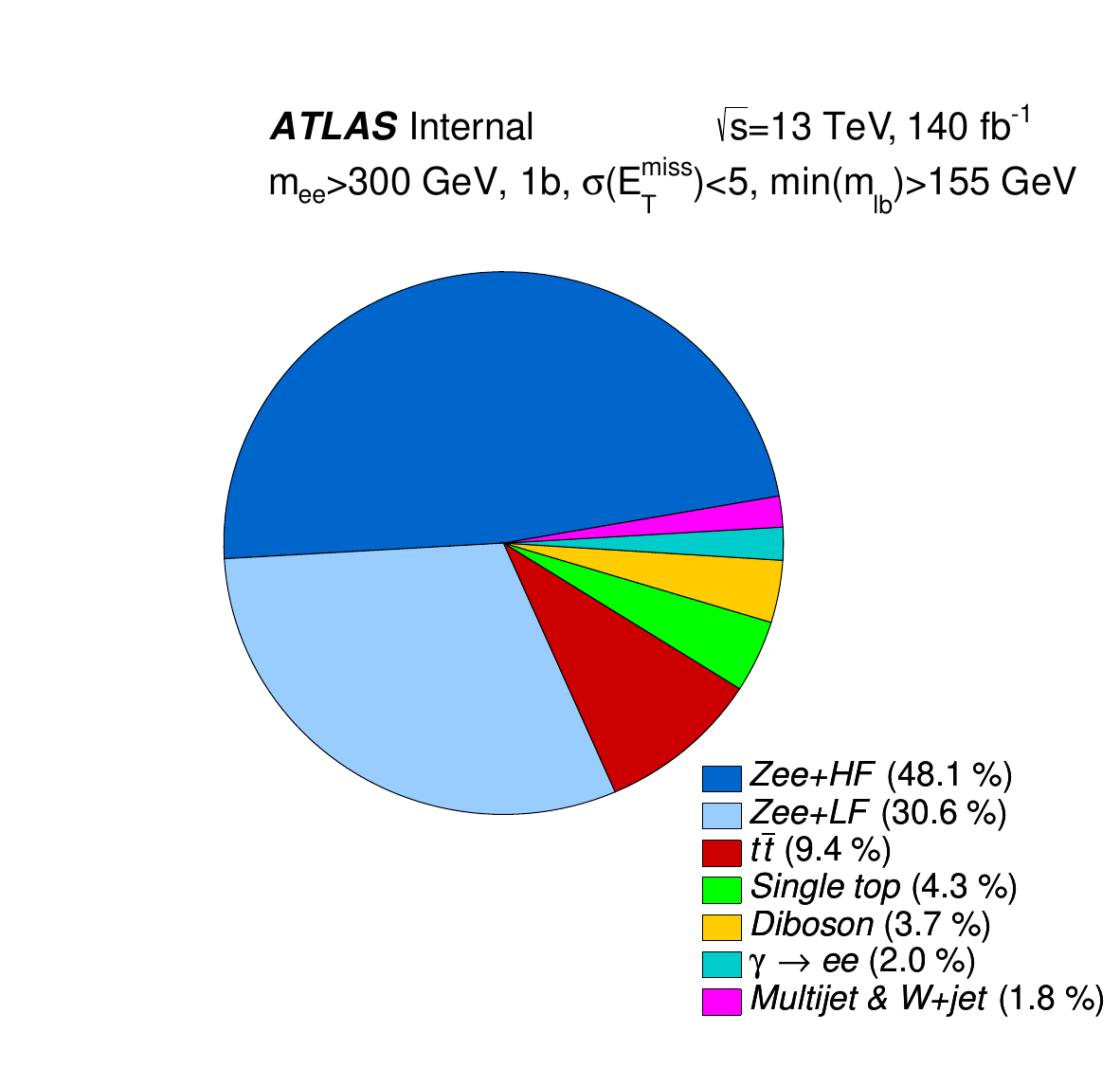}
		\label{fig:bkg_SR_1b_ele}
	}
	\subfloat[]{
		\includegraphics[width=0.4\textwidth]{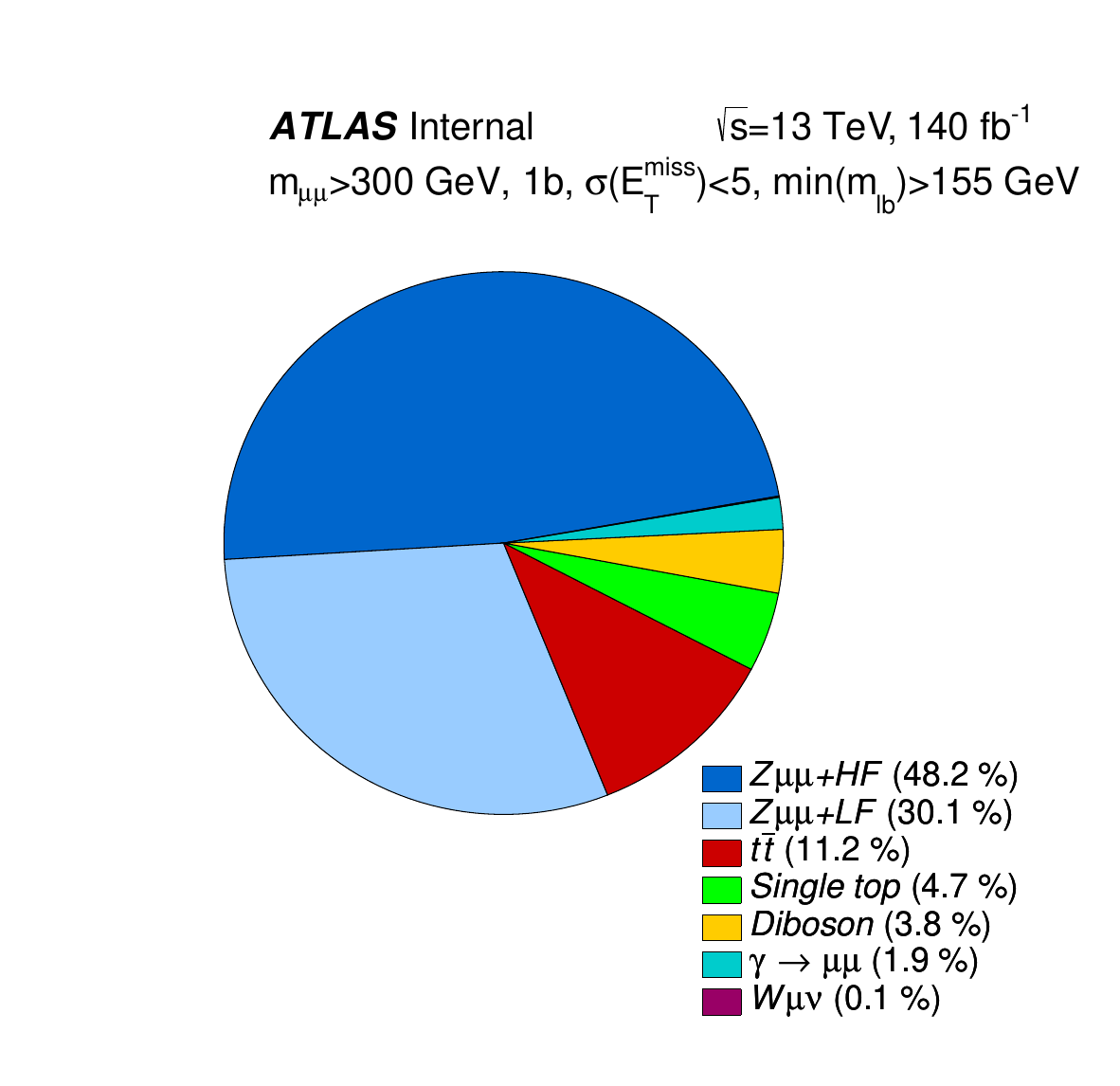}
		\label{fig:bkg_SR_1b_mu}
	}
	
	\subfloat[]{
		\includegraphics[width=0.4\textwidth]{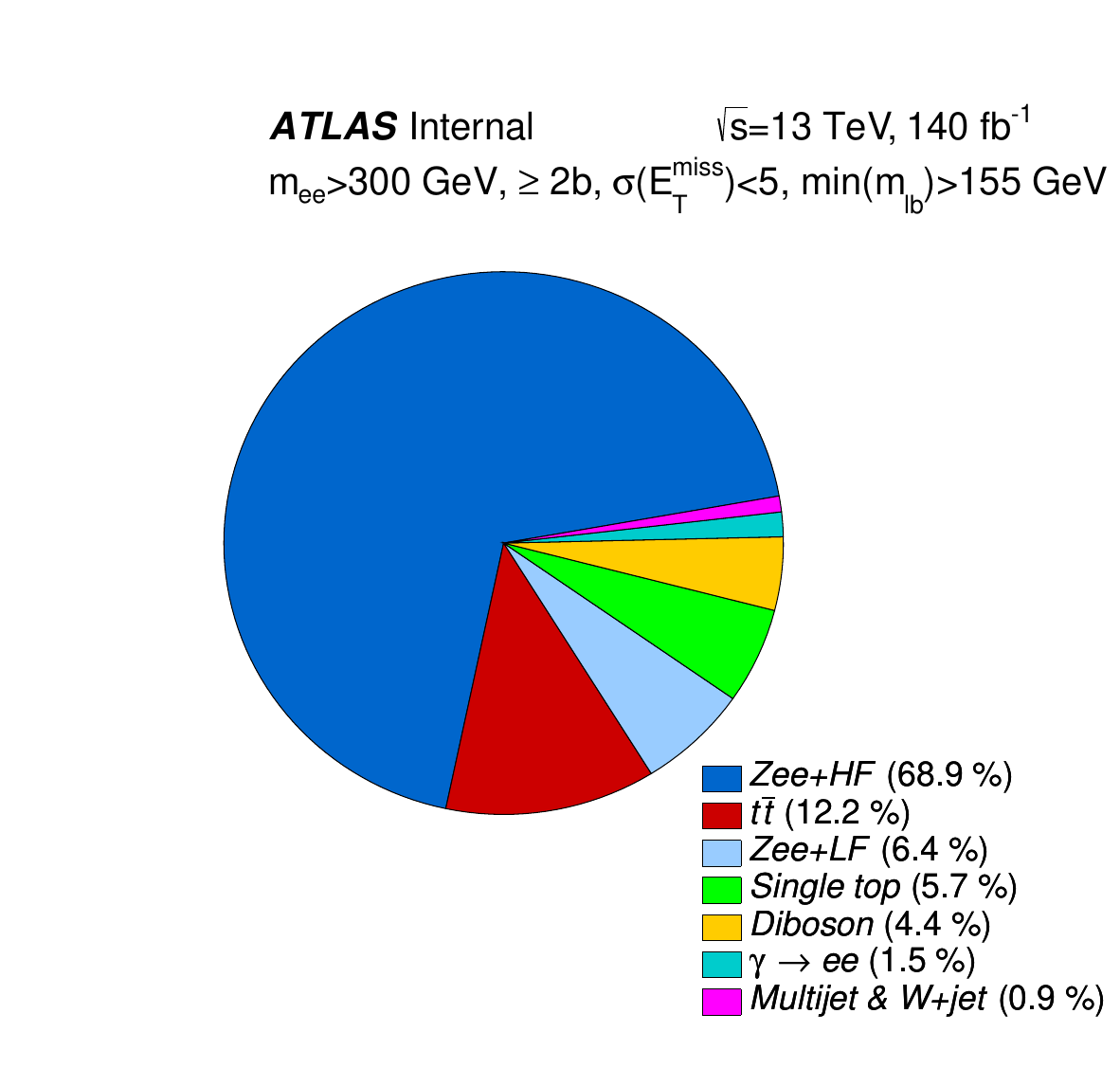}
		\label{fig:bkg_SR_atleast2b_ele}
	}
	\subfloat[]{
		\includegraphics[width=0.4\textwidth]{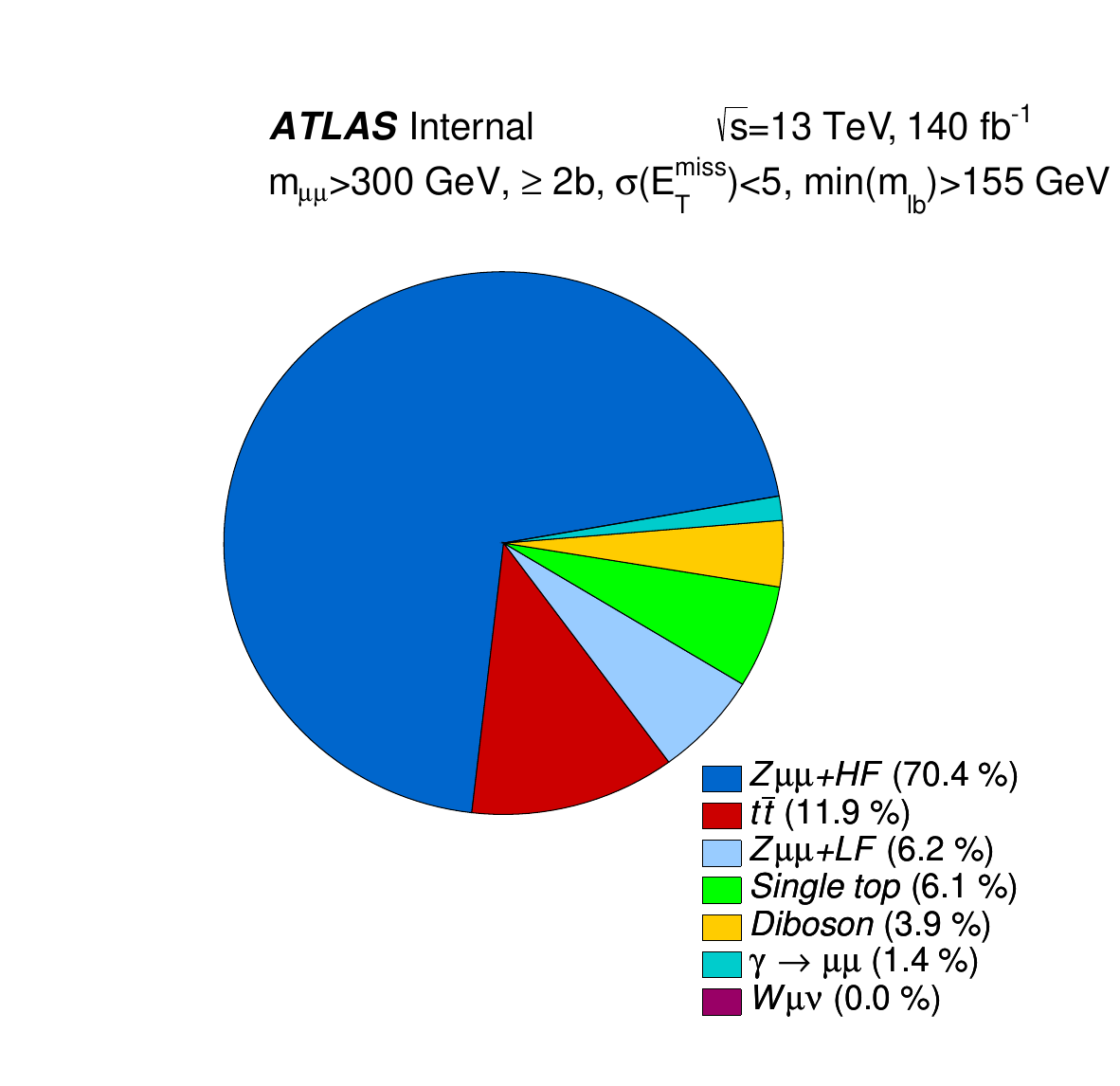}
		\label{fig:bkg_SR_atlaest2b_mu}
	}

	\caption{Background composition for the Signal Region with different $b$-jet multiplicities: electron (left) and muon (right) channels.}
	\label{fig:bkg_SR}
\end{figure}

The background composition of the signal regions after applying the aforementioned selection cuts is shown in FIG.~\ref{fig:bkg_SR}.

\begin{figure}[h]
	\centering
	\subfloat[]{
		\includegraphics[width=0.4\textwidth]{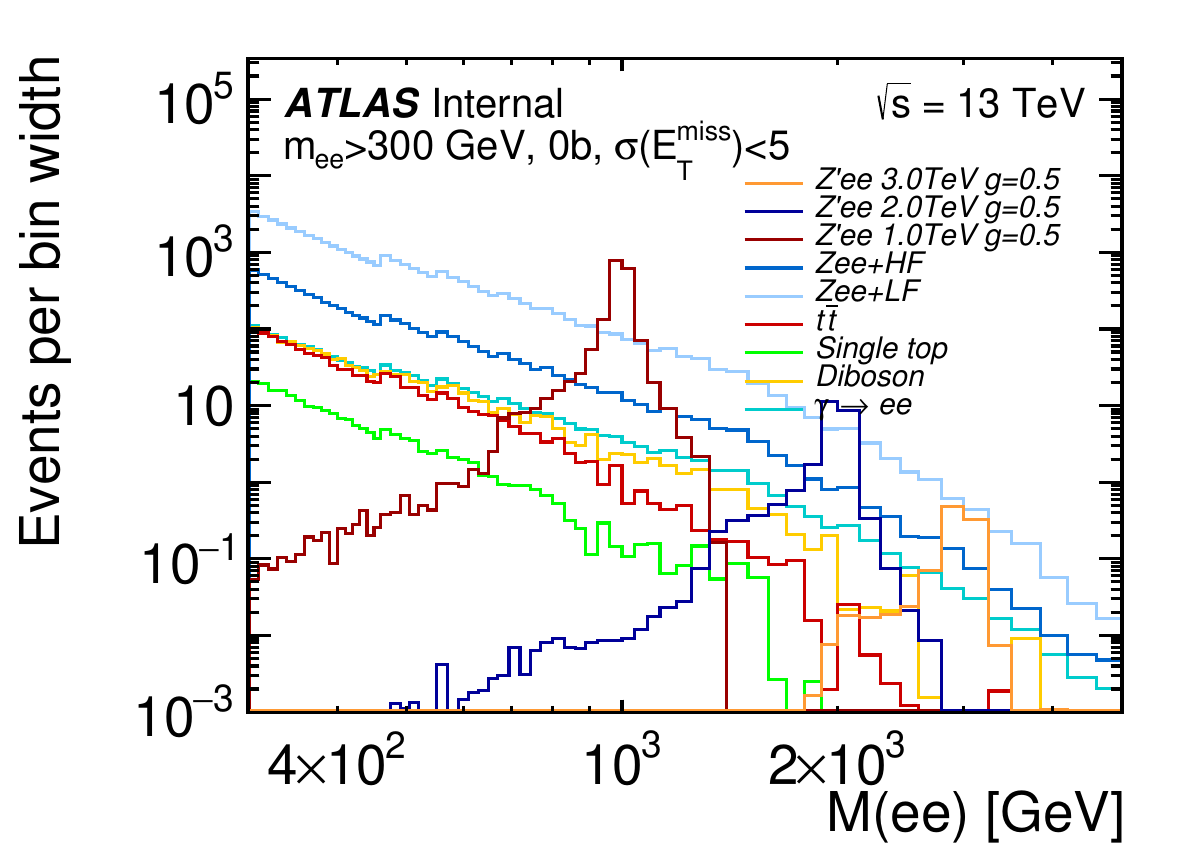}
		\label{fig:Mll_SR_0b_ele}
	}
	\subfloat[]{
		\includegraphics[width=0.4\textwidth]{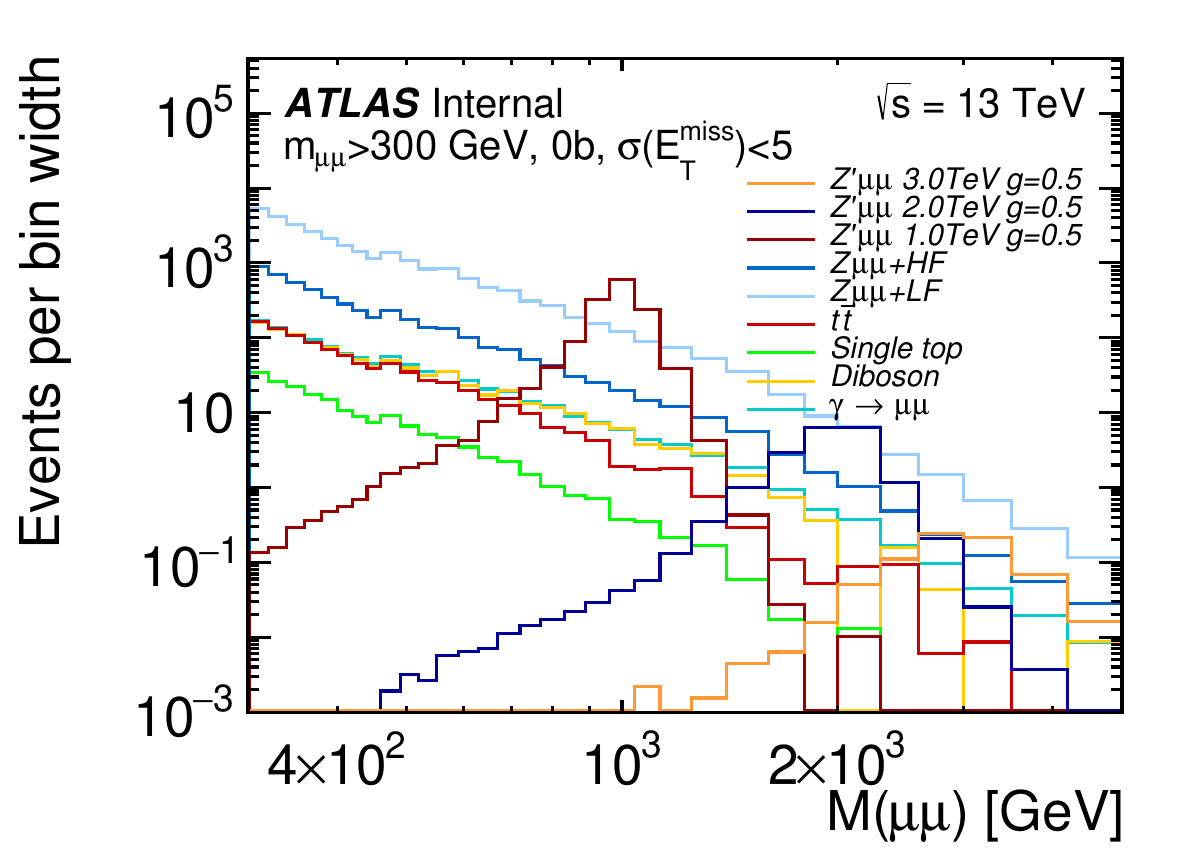}
		\label{fig:Mll_SR_0b_mu}
	}

	\subfloat[]{
		\includegraphics[width=0.4\textwidth]{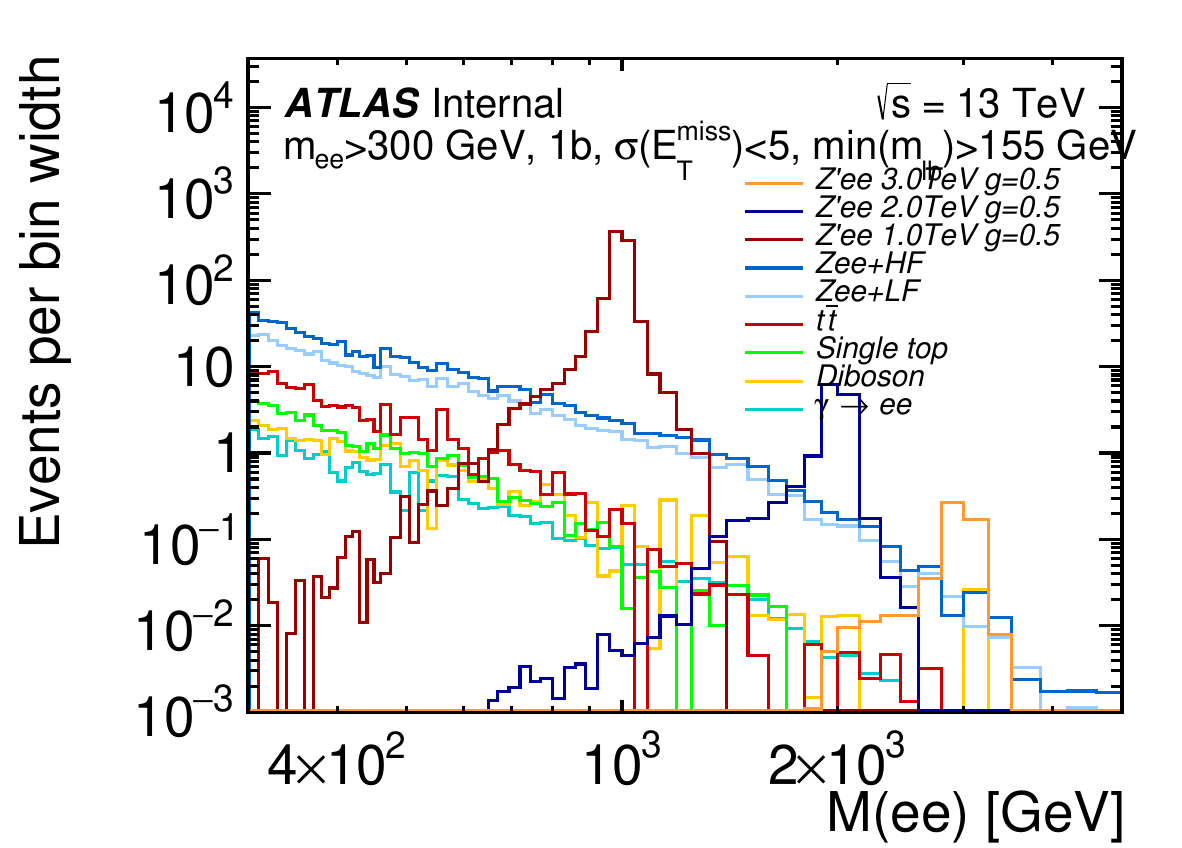}
		\label{fig:Mll_SR_1b_ele}
	}
	\subfloat[]{
		\includegraphics[width=0.4\textwidth]{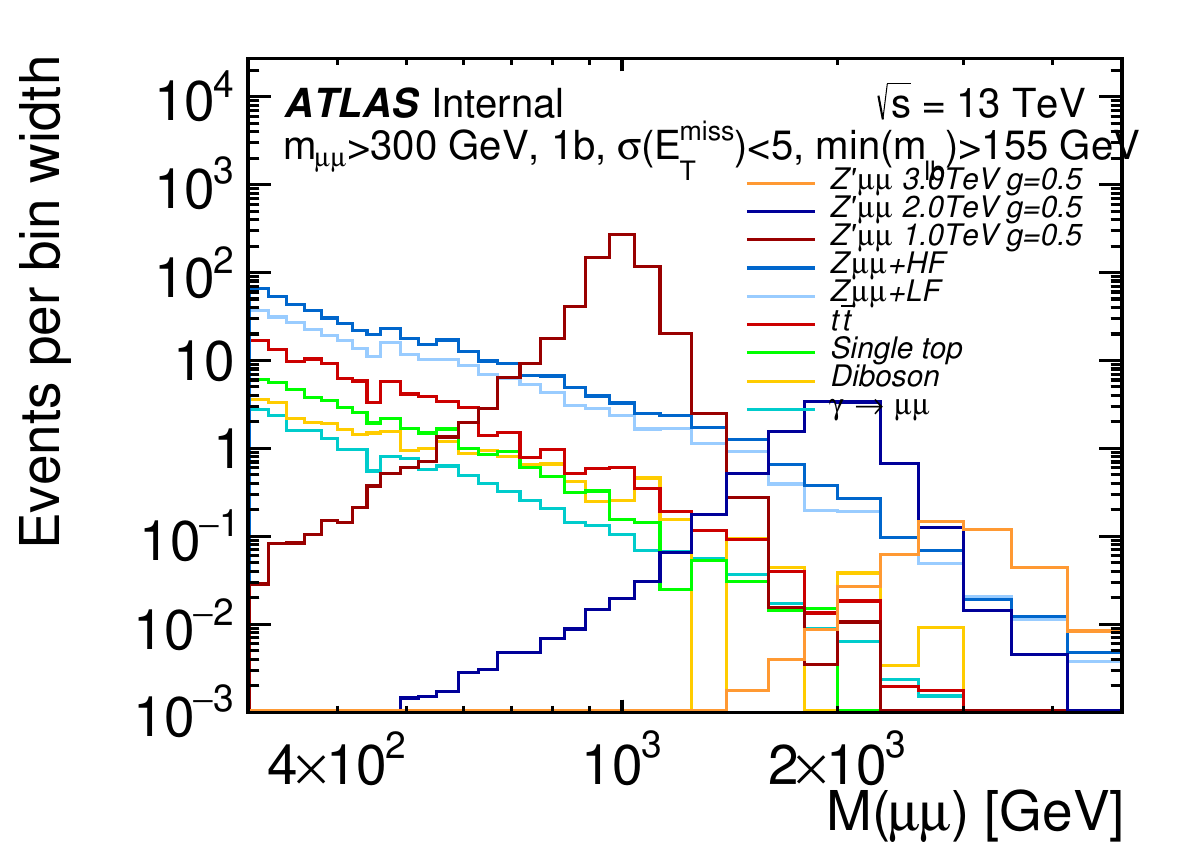}
		\label{fig:Mll_SR_1b_mu}
	}

	\subfloat[]{
		\includegraphics[width=0.4\textwidth]{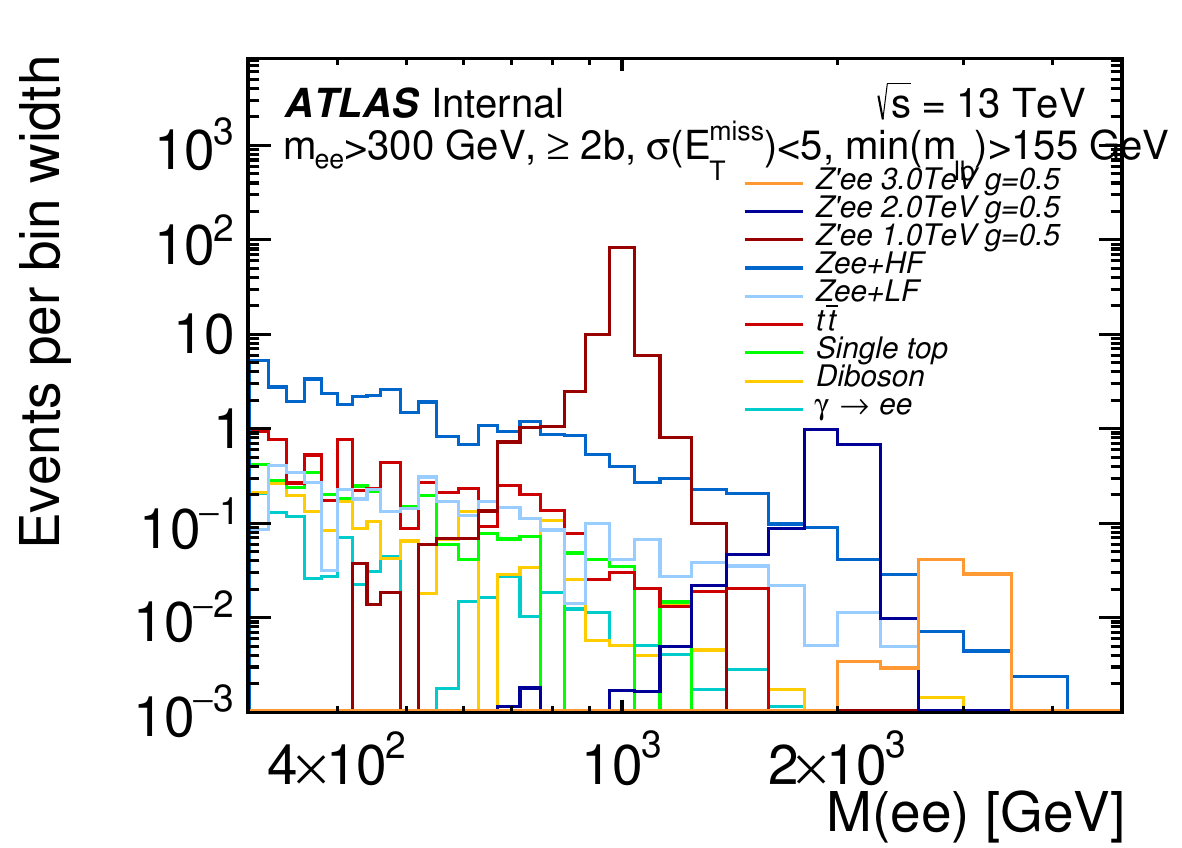}
		\label{fig:Mll_SR_atleast2b_ele}
	}
	\subfloat[]{
		\includegraphics[width=0.4\textwidth]{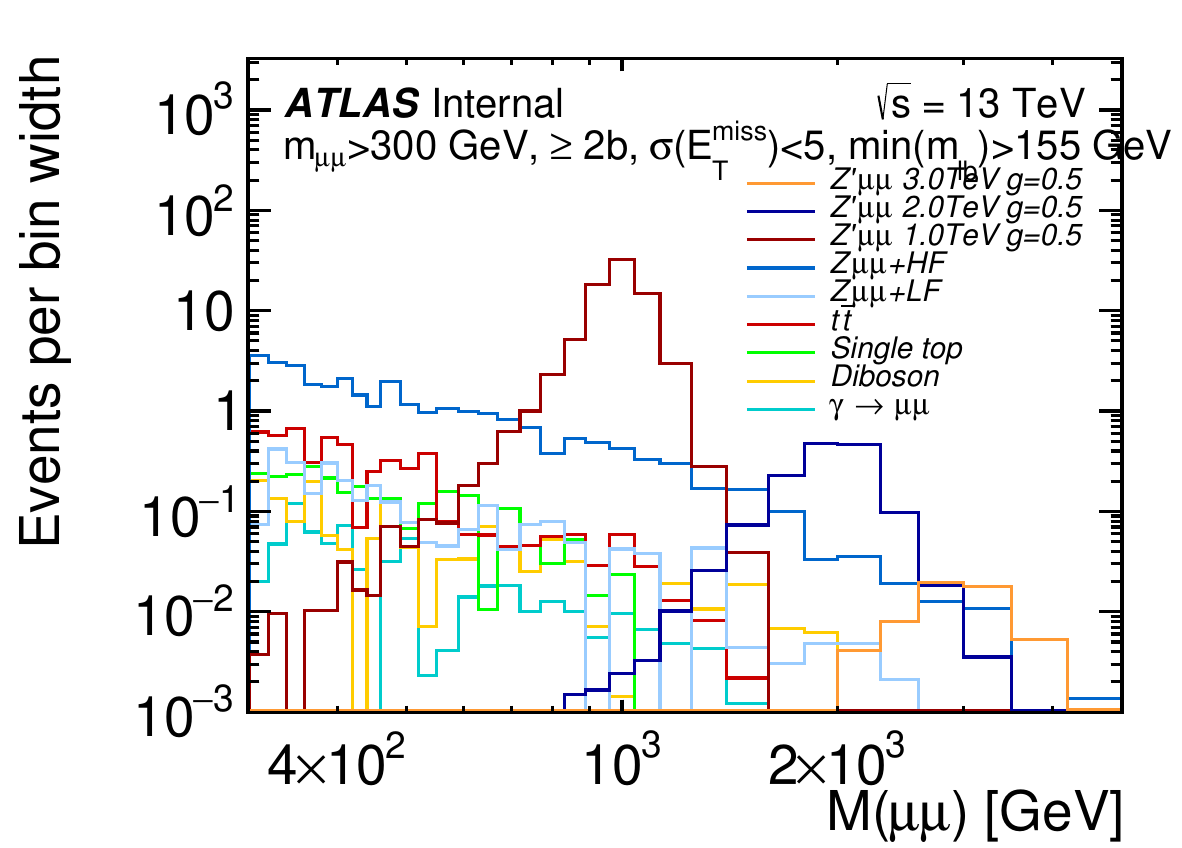}
		\label{fig:Mll_SR_atlaest2b_mu}
	}

	\caption{Invariant dilepton mass distribution in the signal regions for SM background processes and some benchmark signals for the electron (left) and muon (right) channels.}
	\label{fig:Mll_SR}
\end{figure}

The $m_{\ell\ell}$ distributions for background processes and benchmark signals are shown in FIG.~\ref{fig:Mll_SR}.

\section{Systematic Uncertainties}
\label{sec:zprime_systematics}
Systematic uncertainties are divided into experimental (detector performance) and theoretical (physics modelling) categories. Requiring $b$-jets in the final state shifts the background composition relative to an inclusive dilepton selection, making top-related uncertainties more prominent.

\subsection{Experimental systematic uncertainties}

Experimental systematic uncertainties arise from the reconstruction of physics objects using the ATLAS detector. The main sources are the finite precision of calibration and efficiency measurements. For each physics object, namely leptons, jets and $E^{\textrm{miss}}_{\textrm{T}}$, they may be subcategorised into two groups, calibration uncertainties and scale factor (efficiency) uncertainties. Such scale factors are used to correct for potential differences between data and MC. Several different scale factors are provided by the CP groups, e.g. for reconstruction, identification, isolation and triggers. Uncertainties on these scale factors are derived by varying various parameters in the calculation of the scale factors. Separate Calibration systematics refer to uncertainties on energy, momentum and spatial measurements of different ATLAS subdetectors. As such, they are calculated per kinematic variable, per object. Scale factor uncertainties are calculated per object, per event and applied as a variation on the event weight.

\subsubsection{Luminosity}

The total integrated luminosity collected by ATLAS is recorded using several detectors and algorithms. 
The uncertainty in the 2015--2018 total integrated luminosity is 0.83\%~\cite{DAPR-2021-01}, obtained using the LUCID-2 detector~\cite{LUCID2} for the primary luminosity measurements, complemented by measurements using the inner detector and calorimeters.

\subsubsection{Pile-up reweighting}

Another source of systematic uncertainty is induced by the pile-up reweighting procedure. Systematic variations of the pile-up weight are available from the pile-up reweighting tool. They are derived by varying the data scale factor from its nominal value $1.0/1.03$ to $1.0/0.99$ and $1.0/1.07$ for the up- and down-variations, respectively.

\subsubsection{Lepton uncertainties}

Lepton object definitions, working points, and the associated corrections applied to simulation are described in Chapter~\ref{chp:objects} (Sections~\ref{sec:reco:electrons} and~\ref{sec:reco:muons}).
Uncertainties in lepton calibration arise from reconstruction, identification, isolation and trigger efficiencies, referred to as ``RECO'', ``ID'', ``ISO'' and ``Trigger'' uncertainties. For these uncertainties, the so-called SIMPLIFIED model is used (as recommended by the ATLAS Electron Efficiency group). Further contributions arise from energy scale and resolution measurements. For muons, efficiency uncertainties are split into statistical and systematic components (``STAT'' and ``SYS''). Additional muon uncertainties arise from the bad-muon quality criterion, track-to-vertex association (TTVA), and the charge-dependent sagitta resolution bias.
The full list of lepton systematic uncertainties is given in Table~\ref{tab:lepton_systematics} in Appendix~\ref{app:zprime:systematics}.

\subsubsection{Jet uncertainties}

Jet reconstruction and calibration are described in Section~\ref{sec:reco:jets}.
Uncertainties on jet energy scale (JES) and jet energy resolution (JER) stem from the calibration of small-$R$ jets, using the JESGlobalReduction/SimpleJER systematic scenario. The JES uncertainty is split into several components for jet flavour composition, single-flavour response, pile-up modelling, and jets not fully contained in the calorimeter.
In the JER uncertainties, the non-closure in the dijet balance and alternative MC generators are taken into account; JER is decomposed into seven uncorrelated jet $p_{\textrm{T}}$- and $\eta$-dependent components.
Scale factor uncertainties relate to jet vertex tagging (JVT) and forward JVT (fJVT) requirements. The full list of jet systematic uncertainties is given in Table~\ref{tab:jet_systematics} in Appendix~\ref{app:zprime:systematics}.

\subsubsection{Flavour tagging uncertainties}

$b$-jet identification and the DL1r algorithm are described in Section~\ref{sec:reco:btag}.
Flavour tagging efficiency scale factors are corrected from fits to data, with uncertainties parameterised as eigenvector (EV) variations from an EV decomposition procedure~\cite{ATLAS:2022xlg}: nine for $b$-tagging, and four each for $c$- and light-flavour tagging. Two additional extrapolation uncertainties account for the extrapolation of tagging efficiencies beyond the calibration $p_{\textrm{T}}$ range~\cite{ATLAS:2021xox}. The full list is given in Table~\ref{tab:tagging_systematics} in Appendix~\ref{app:zprime:systematics}.

\subsubsection{$\met$ soft term uncertainties}

The $\met$ significance and its construction are described in Section~\ref{sec:reco:met}.
Systematic uncertainties on $\met$ arise from the track soft term (TST) and are derived via simulation-to-data matching. Projections of the soft-term momentum parallel and perpendicular to the hard-term direction yield scale and resolution uncertainties, respectively~\cite{ATLAS:2018ghb}. The three resulting uncertainties are listed in Table~\ref{tab:met_systematics} in Appendix~\ref{app:zprime:systematics}.

\subsubsection{Uncertainties on fake estimate}
\label{fake_systs}

The following sources of systematic uncertainty on the fake background are considered:

\begin{itemize}
	\item Statistical uncertainty of the real and fake efficiency
	\item Variation of the cross-section by $\pm 10\%$ in the prompt lepton subtraction  in the fake muon efficiency estimation
	\item Variation of the cross-section by $\pm 10\%$ in the prompt lepton subtraction  in the heavy flavour fake electron efficiency estimation
	\item Variation of the cross-section by $\pm 30\%$ in the prompt lepton subtraction  in the light flavour fake electron efficiency estimation
	\item Variation of the fraction of light and heavy flavour fakes by $\pm 20\%$
	\item The light flavour fake electron efficiency was estimated in a fake enriched region defined by three different cuts:
	\begin{enumerate}
		\item \textbf{noW: }Events with $E_T^{miss}>25\,\mathrm{GeV}$ are excluded to reduce the contribution from W decays
		\item \textbf{no DY: }Real electrons from DY processes are suppressed by rejecting events with at least 2 objects passing the \texttt{LHMedium} ID criterion
		\item \textbf{no Zmass:} Electrons from Z decays are eliminated by excluding events with a reconstructed mass close to the Z-peak
	\end{enumerate}
	We consider an uncertainty that accounts for the effect of the different cuts. For this purpose, the efficiencies are recalculated in the case where each of these cuts is not applied. By doing so, we get three systematic variations and we consider them by taking the envelope of those.
\end{itemize}

\subsection{Theoretical systematic uncertainties}

To estimate their theoretical systematic uncertainties, the four sets of processes contributing to the background, $\ttbar$, single-$t$, DY and $VV$, can be divided into two groups according to their generators. Top processes are modelled using Powheg+Pythia8, whereas DY and $VV$ are modelled using Sherpa. Processes in each group mostly share theoretical uncertainties that are hence evaluated in a similar fashion. For the two subdominant backgrounds, single-$t$ and $VV$, theoretical uncertainties are assigned as flat values based on ATLAS PMG recommendations: 4\% for single-$t$ (based on single-top cross-section uncertainties) and a conservative 10\% for $VV$.

Following ATLAS PMG recommendations, each set of theoretical variations is bundled into a single nuisance parameter (NP) for the uncertainty plots illustrating their effect on the dilepton invariant mass. For the profile likelihood fit, each individual variation is treated as a separate NP. The following text describes the theoretical uncertainties considered for the $t\bar{t}$ and DY backgrounds; a complete summary is given in Table~\ref{tab:theory_systematics} in Appendix~\ref{app:zprime:systematics}.

\subsubsection{$\ttbar$}

For the uncertainty plots in the next section, theoretical uncertainties on $\ttbar$ are evaluated following ATLAS PMG/top recipes. Most are estimated by internal reweighting of the nominal samples; others use alternative samples for 2-point uncertainties, compared to a dileptonic $\ttbar$ alternative, with relative differences propagated in the fit. Despite the bundling for presentation, each variation is treated as an individual NP in the fit. The following list summarises the theoretical uncertainties considered for the $\ttbar$ background:

\subsubsection*{Internal reweighting uncertainties}

\begin{itemize}

\item \textbf{Parton density function: } uncertainties due to the choice of the PDF are evaluated by applying the weights associated with each of the 30 functions consisting in the PDF4LHC15\_nlo set (PDFs 90901--90930). Since the PDF4LHC15\_nlo set constitutues a statistical combination of three independent PDF sets (CT14, MMHT2014, NNPDF3.0)~\cite{PDF4LHC}, the errors are evaluated relative to the nominal PDF4LHC15\_nlo sample (PDF 90900), each of which as an independent NP. The relative errors are then propagated onwards to the fit using the nominal sample produced with NNPDF3.0\_nlo.

\item \textbf{QCD scales:} 
Uncertainties related to the choice of QCD renormalisation and factorisation scales in the ME calculation are estimated by two NPs. There is each one NP for the $\mu_R$ scale uncertainty and the $\mu_F$ scale uncertainty. Each NP consists of two variations: one scale is kept at the nominal value of one, while the other scale is doubled and halfed.
\\For representational plots of the systematic uncertainties, these QCD scale uncertainties are estimated by independently taking each scale half and doubled: $\{ \mu_{\textrm{R}}, \mu_{\textrm{F}}\} \times \{1.0,0.5\} \times \{0.5,1.0\} \times \{2.0,1.0\} \times \{1.0,2.0\}$. The final uncertainty is taken from the envelope of these four variations.

\item \textbf{QCD FSR scales:} uncertainties attributed to the choice of FSR renormalisation scale, taking $\mu_{\textrm{R, FSR}}=0.5,2.0$ whilst keeping the ISR QCD scales constant (unity).

\item \textbf{Strong coupling constant:} a variation whereby $\alpha_{\textrm{S}}\pm 0.001$ around nominal, $\alpha_{\textrm{S}} = 0.118$. The average of the two is used as a single NP.

\item \textbf{ISR strong coupling constant:} a variation on  $\alpha_{\textrm{S}}$ in the ISR. 
In the fit, these two variations are considered as one NP (the average of the two variations is used).

\end{itemize}

\subsubsection*{2-point uncertainties}

\begin{itemize}

\item \textbf{Hard scatter matrix element:} a source of uncertainty which stems from matching between the hard-scattering generator where the matrix elements, $\mathcal{M}$, is computed and the parton shower (fragmentation and hadronisation). This uncertainty is determined by comparing the nominal MC sample where the \texttt{Pythia8} parameter \texttt{pThard} is set to its default value, 0, to an alternative sample where it is set to 1. This parameter defines the vetoed region of the showering, which is important to avoid holes/overlap in the phase space filled by \texttt{Powheg} and \texttt{Pythia8}. 

\item \textbf{Parton shower:} the uncertainty in the modelling of the parton shower (PS) is estimated by comparing the nominal sample to an alternative sample produced with \texttt{Herwig7} as PS generator. 

\item \textbf{Damping parameter} the \texttt{Powheg}-specific parameter $h_{\textrm{damp}}$ is used to control the damping of radiation with high transverse momenta. Its nominal value, where $h_{\textrm{damp}} = 1.5 m_t$, is compared to a sample with $h_{\textrm{damp}} = 3.0  m_t$ to determine the impact of the choice $h_{\textrm{damp}}$.

Note that the alternative samples use AFII simulation rather than full simulation; the difference is propagated as the uncertainty on the full-simulation result.

\end{itemize}

\subsubsection{Drell-Yan}

For the uncertainty plots in the next section, DY theoretical uncertainties are estimated following ATLAS PMG/weak boson recommendations. All are evaluated by internal reweighting. As for $\ttbar$, each variation is treated as an individual NP in the fit despite the bundled presentation. The following list summarises the theoretical uncertainties considered for the DY background:


\begin{itemize}

\item \textbf{Parton density function:} the uncertainty due to PDF choice is calculated by applying the weights associated with each of 100 variations in the NNPDF3.0NNLO PDF ensemble (PDF 303201--303300) and follows specifically from PMG PDF recommendations. All of these variations are used as individual NPs.
\\For representational systematic plots, this ensemble is made from fits to the ensemble test on the input data and therefore the uncertainty is the standard deviation. 

\item \textbf{QCD scales:} as above (under top).

\item \textbf{Strong coupling constant:} as above (under top). 


\end{itemize}

\subsubsection[\texorpdfstring{$Z'$}{Z'} Signal]{\texorpdfstring{$\bm{Z'}$}{Z'} Signal}

Theoretical uncertainties on the signal are estimated by internal reweighting. The following list summarises the theoretical uncertainties considered for the signal:

\begin{itemize}
	\item \textbf{Parton density function: }the uncertainty due to PDF choice is calculated by applying the weights associated of 100 replicas in the NNPDF3.0NLO PDF set. The different replicas are combined and the standard deviation of them is considered as PDF uncertainty in the fit. 
	\item \textbf{QCD scales: } as above (under top)
	\item \textbf{Strong coupling constant: } as above (under top)
\end{itemize}

The impact of the theoretical signal uncertainties on the expected cross-section limits was investigated.

A summary of all theoretical uncertainties and their treatment is given in Table~\ref{tab:theory_systematics} in Appendix~\ref{app:zprime:systematics}.

\subsection{Handling of systematic uncertainties}

Before systematic uncertainties are introduced to the profile likelihood fit, a pre-processing procedure is applied, which consists of several steps.

\textbf{Pruning:} In many cases, not all systematic uncertainties have an impact on the fit results. In order to avoid numerical instabilities and to speed up the fit procedure, systematic uncertainties with a negligible effect are discarded beforehand. The elimination can happen based either on their effect on the overall normalisation or on the contents of the different bins (shape). In the first case, all systematics are removed that do not vary the overall normalisation by more than 0.2\%. In the latter case, the same threshold is applied bin-by-bin, meaning that systematics changing the content of no bin more than 0.2\% are not considered further.

\textbf{Symmetrisation:} Systematic uncertainties can either be one-sided (only one variation is available) or two-sided (up and down variations are available). To increase the fit stability, one-sided uncertainties are symmetrised by mirroring the variation. Similarly, two-sided uncertainties are symmetrised by shifting the up- and down-variations such that the nominal value is in the centre of the two variations.

\textbf{Smoothening:} Systematics can be affected by statistical fluctuations. In order to avoid strange constraints caused by such fluctuations and to avoid double-counting the statistical uncertainties, a smoothing procedure is applied by rebinning the distribution until certain criteria are fulfilled.

\textbf{Rescaling of theoretical modelling uncertainties:} Theoretical modelling uncertainties are rescaled in a way that the sum of event yields in control regions and signal regions is the same as that of the nominal event yields in these regions. This is done to remove the normalisation component of such uncertainties and to only keep shape and acceptance effects of the theoretical modelling uncertainties (of signal and background). Due to this treatment, normalisation components of theoretical modelling uncertainties are not included in the uncertainty band in the pre-fit plots.

\section{Results}
\label{sec:zprime_results}

The background normalisation factors and control region fit results are presented in Section~\ref{sec:zprime_bkg_modelling}. Results of the profile-likelihood fit in the signal regions are presented below for coupling values $g_{Z'} = 0.5$ and $g_{Z'} = 1.0$. Nuisance parameter pulls, correlation matrices, and systematic ranking plots for both coupling values are provided in Appendix~\ref{app:zprime:fits}. Cross-section limits and their combination are presented at the end of this section.

\subsection{Signal regions: $g_{Z'} = 0.5$}

\begin{figure}[h]
	\centering
	\subfloat[]{
		\includegraphics[width=0.33\textwidth]{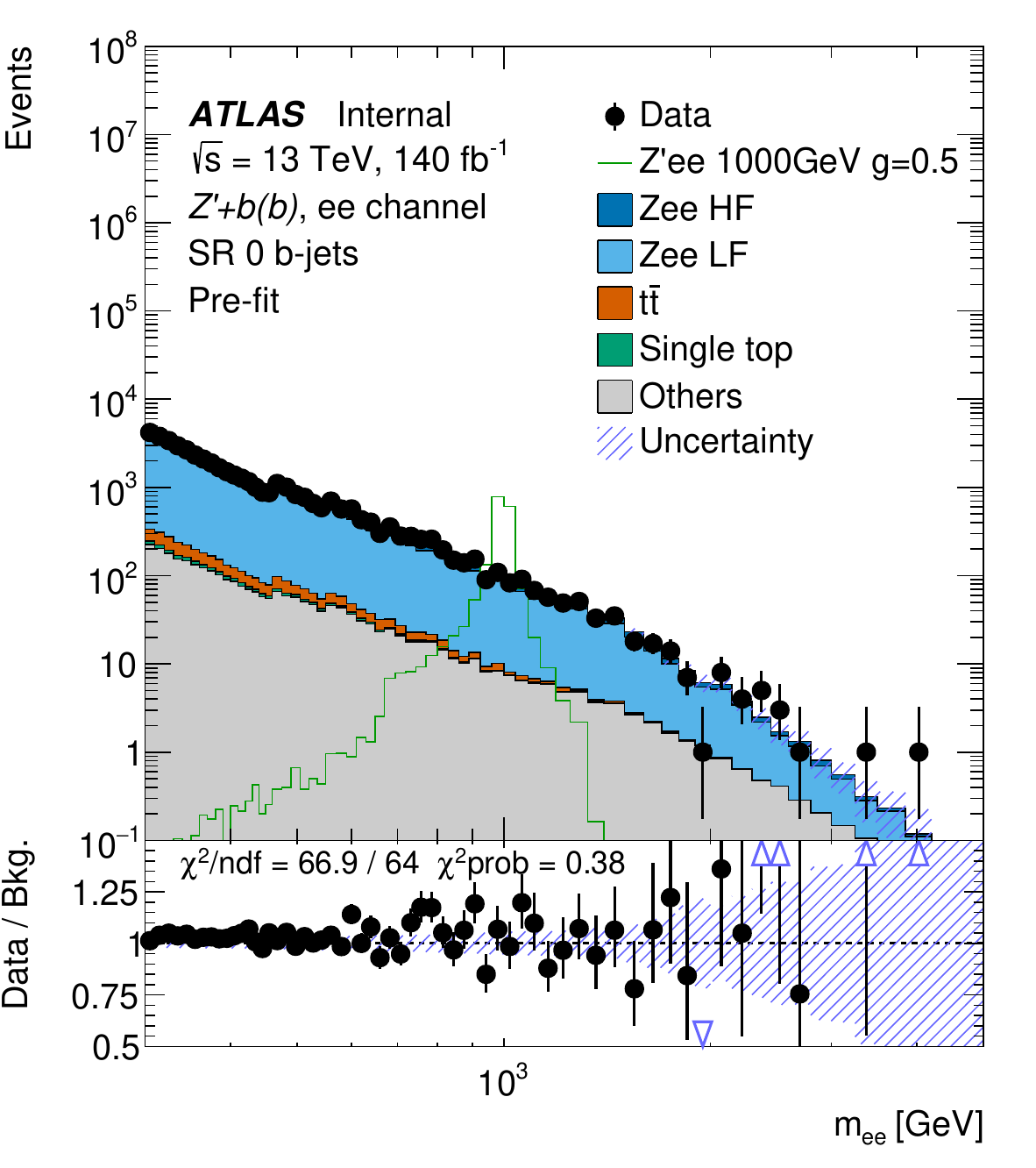}
		\label{fig:CR_fit_ele_SR_0b_pre_1000_g05_unblinded}
	}
	\subfloat[]{
		\includegraphics[width=0.33\textwidth]{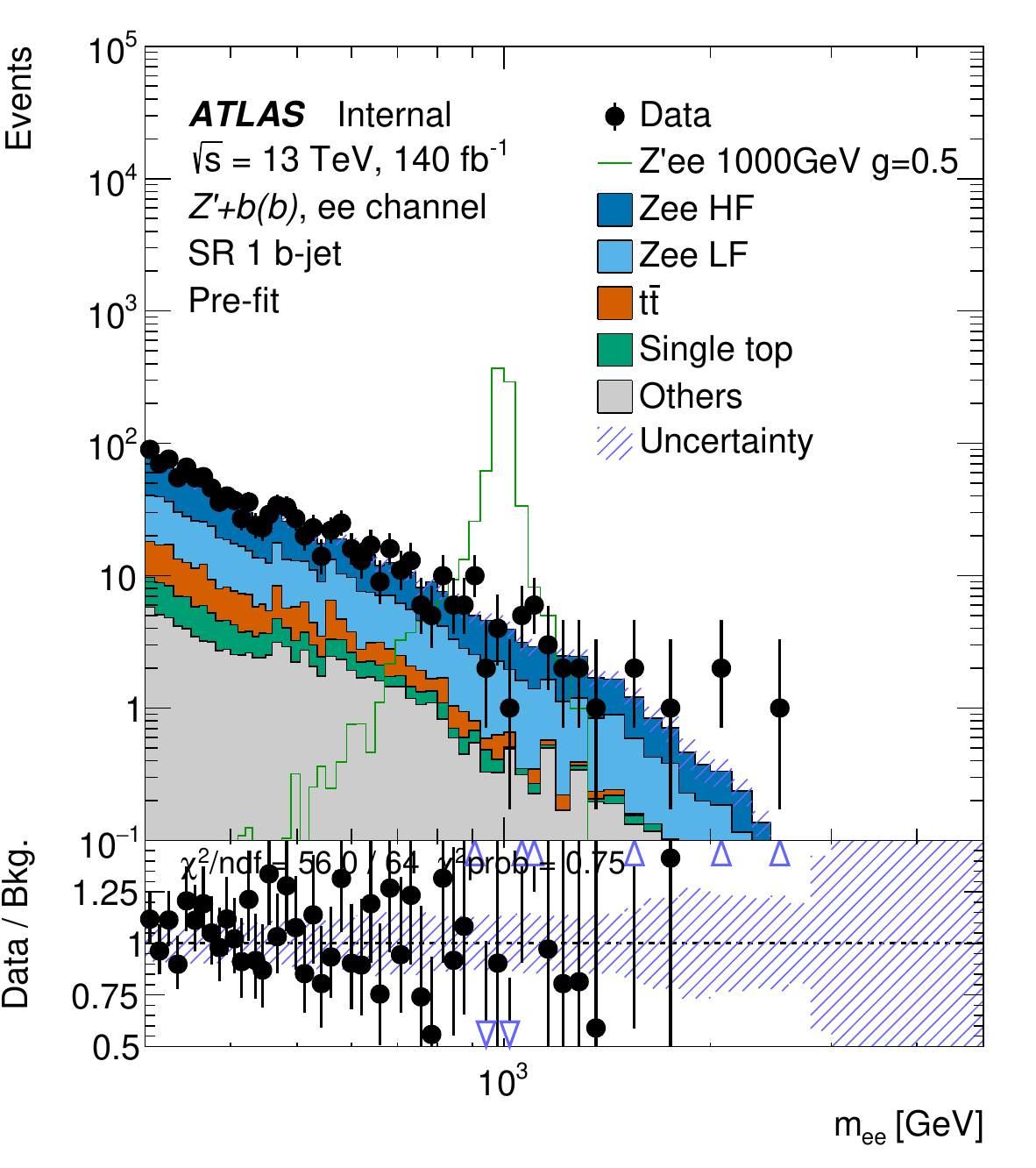}
		\label{fig:CR_fit_ele_SR_1b_pre_1000_g05_unblinded}
	}
	\subfloat[]{
		\includegraphics[width=0.33\textwidth]{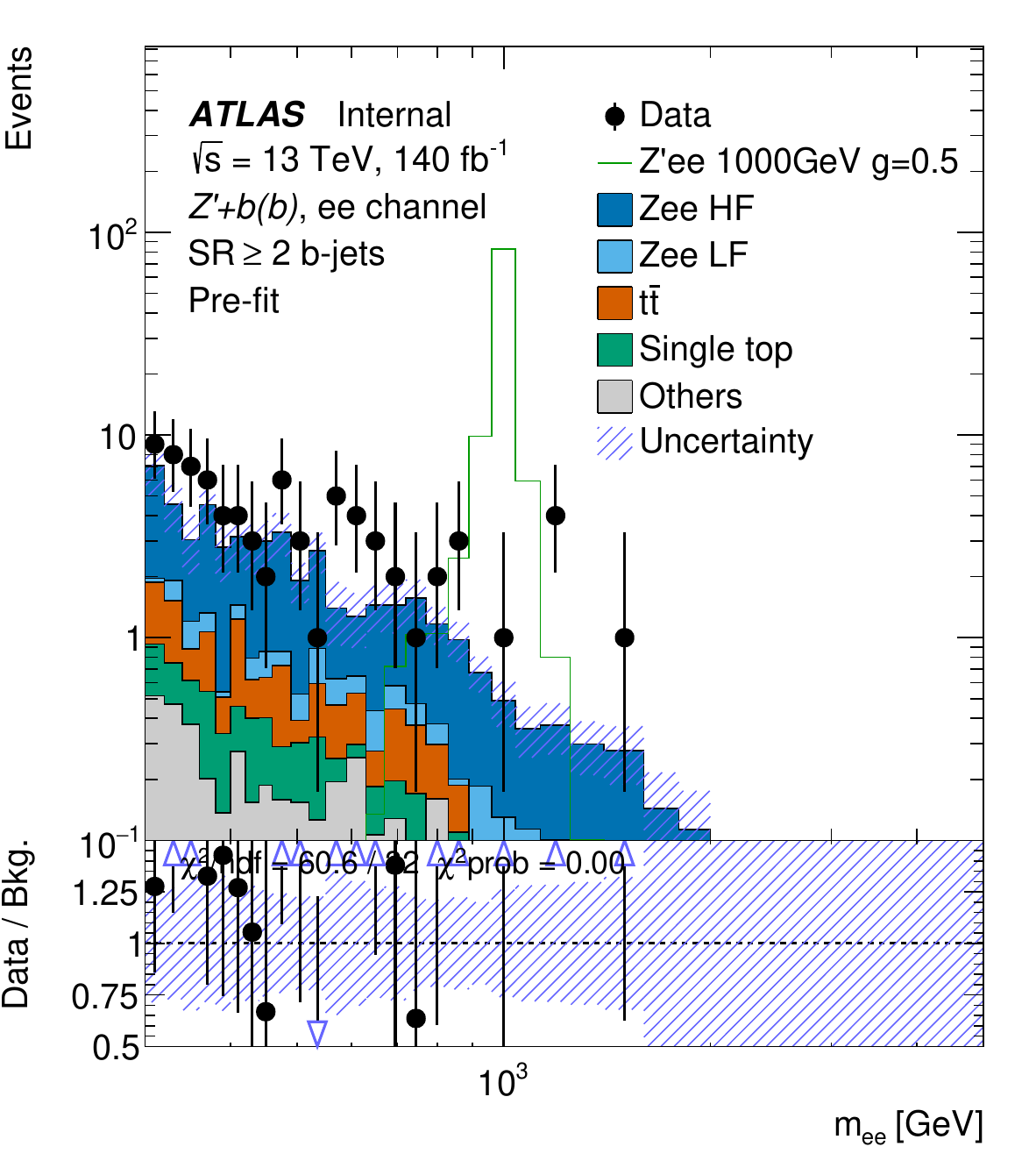}
		\label{fig:CR_fit_ele_SR_atleast2b_pre_1000_g05_unblinded}
	}
	\hfill
	\subfloat[]{
		\includegraphics[width=0.33\textwidth]{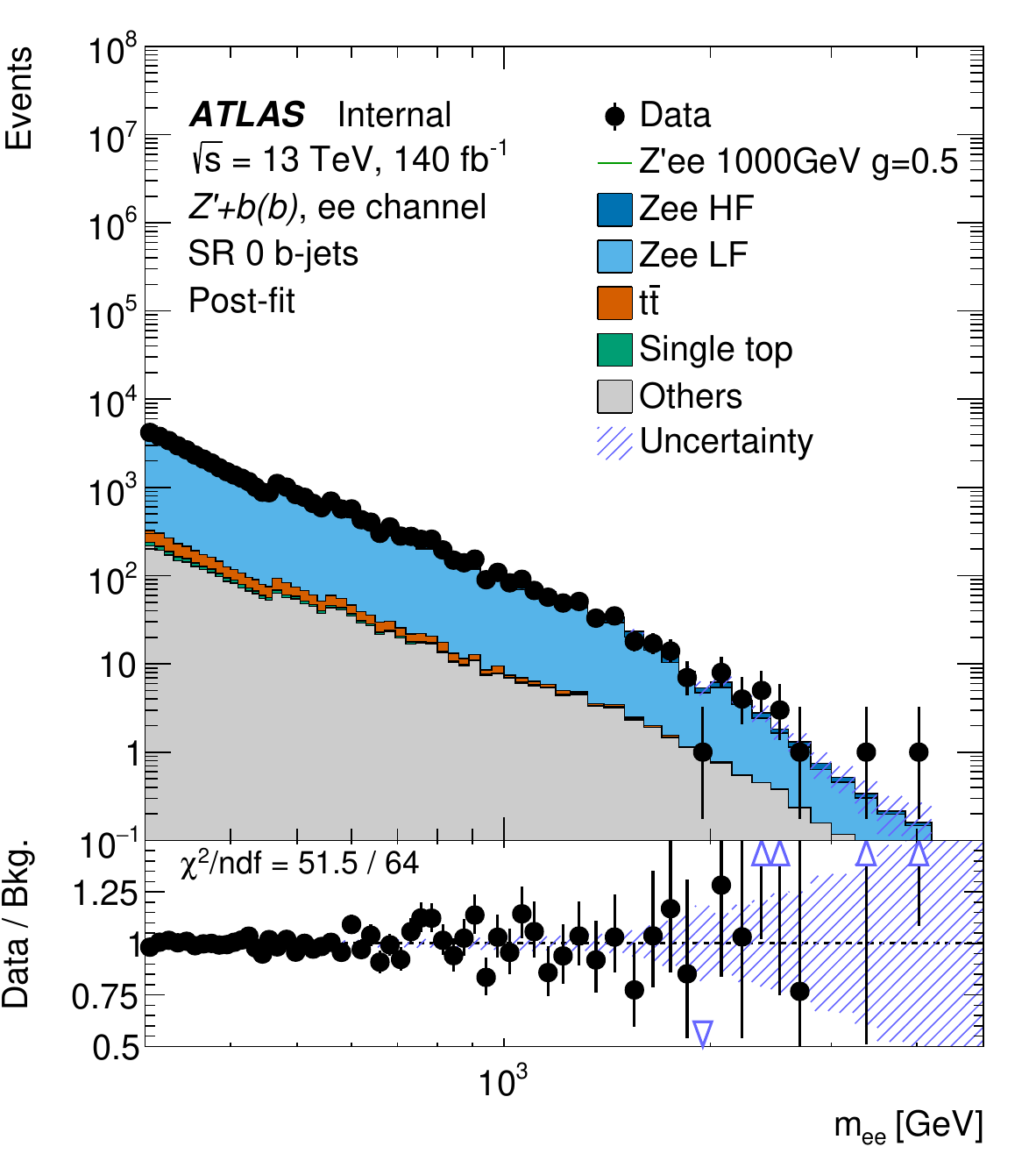}
		\label{fig:CR_fit_ele_SR_0b_post_1000_g05_unblinded}
	}
	\subfloat[]{
		\includegraphics[width=0.33\textwidth]{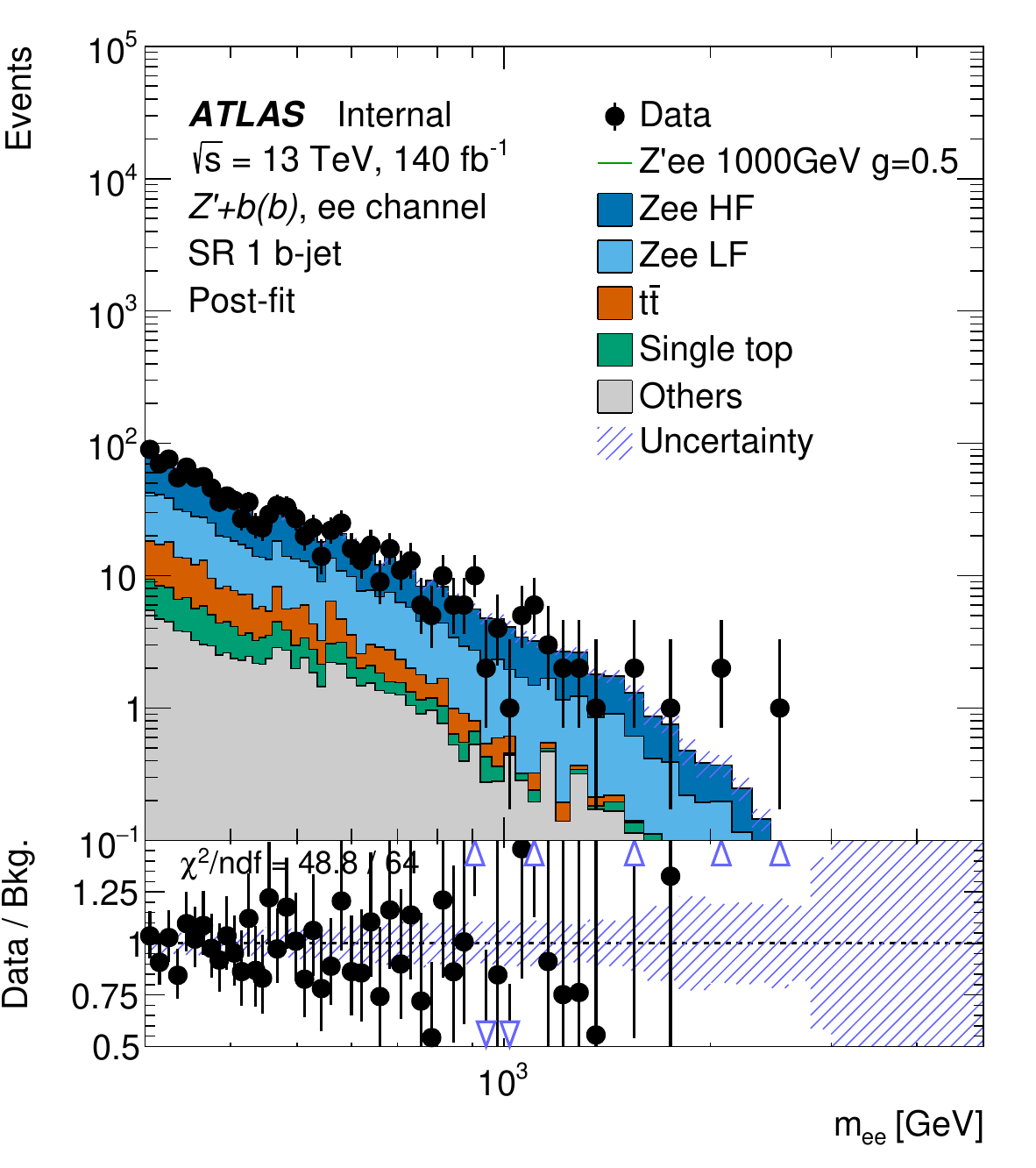}
		\label{fig:CR_fit_ele_SR_1b_post_1000_g05_unblinded}
	}
	\subfloat[]{
		\includegraphics[width=0.33\textwidth]{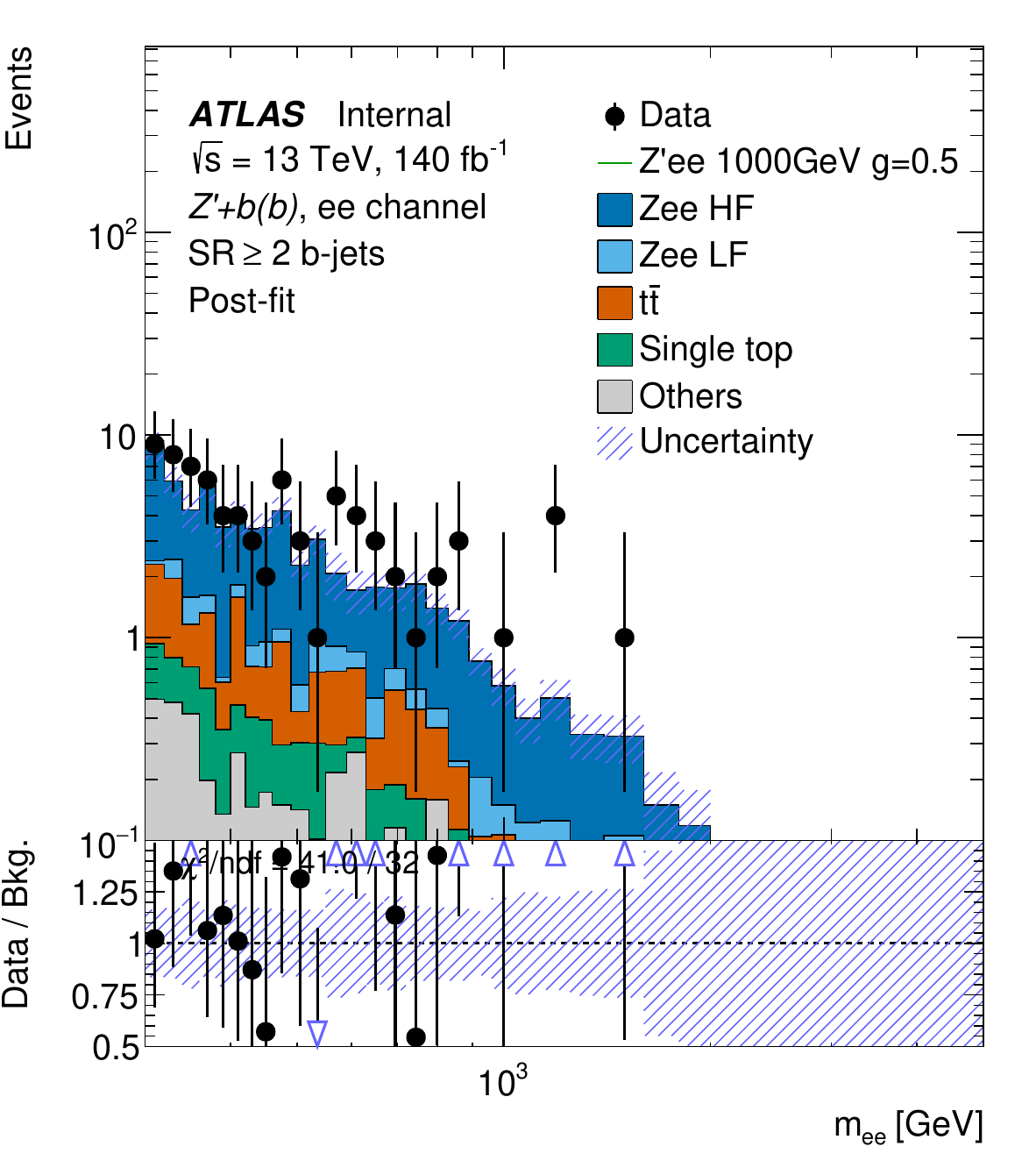}
		\label{fig:CR_fit_ele_SR_atleast2b_post_1000_g05_unblinded}
	}

	\caption{Pre-fit and post-fit plots for the single-bin fit of the $0b$ signal region (a, d), the $1b$ signal region (b, e) and the $\geq 2b$ signal region (c, f) in the electron channel. These plots are shown for $m_{Z'}=1\,\TeV$ and $g_{Z'}=0.5$. The uncertainty band includes statistical and systematic uncertainties.}
	\label{fig:SR_fit_ele_1000_g05_unblinded}
\end{figure}

\begin{figure}[h]
	\centering
	\subfloat[]{
		\includegraphics[width=0.33\textwidth]{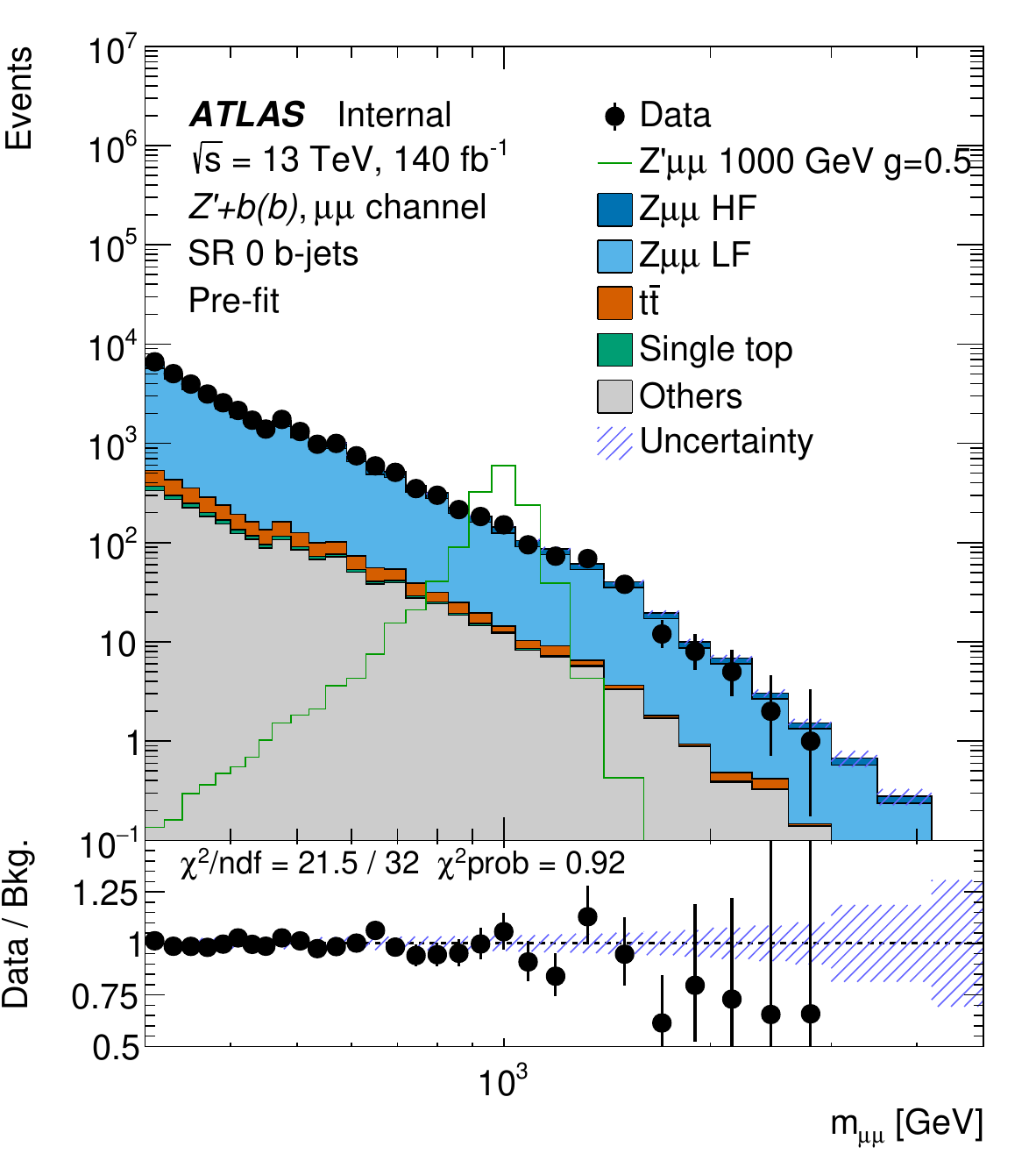}
		\label{fig:CR_fit_mu_SR_0b_pre_1000_g05_unblinded}
	}
	\subfloat[]{
		\includegraphics[width=0.33\textwidth]{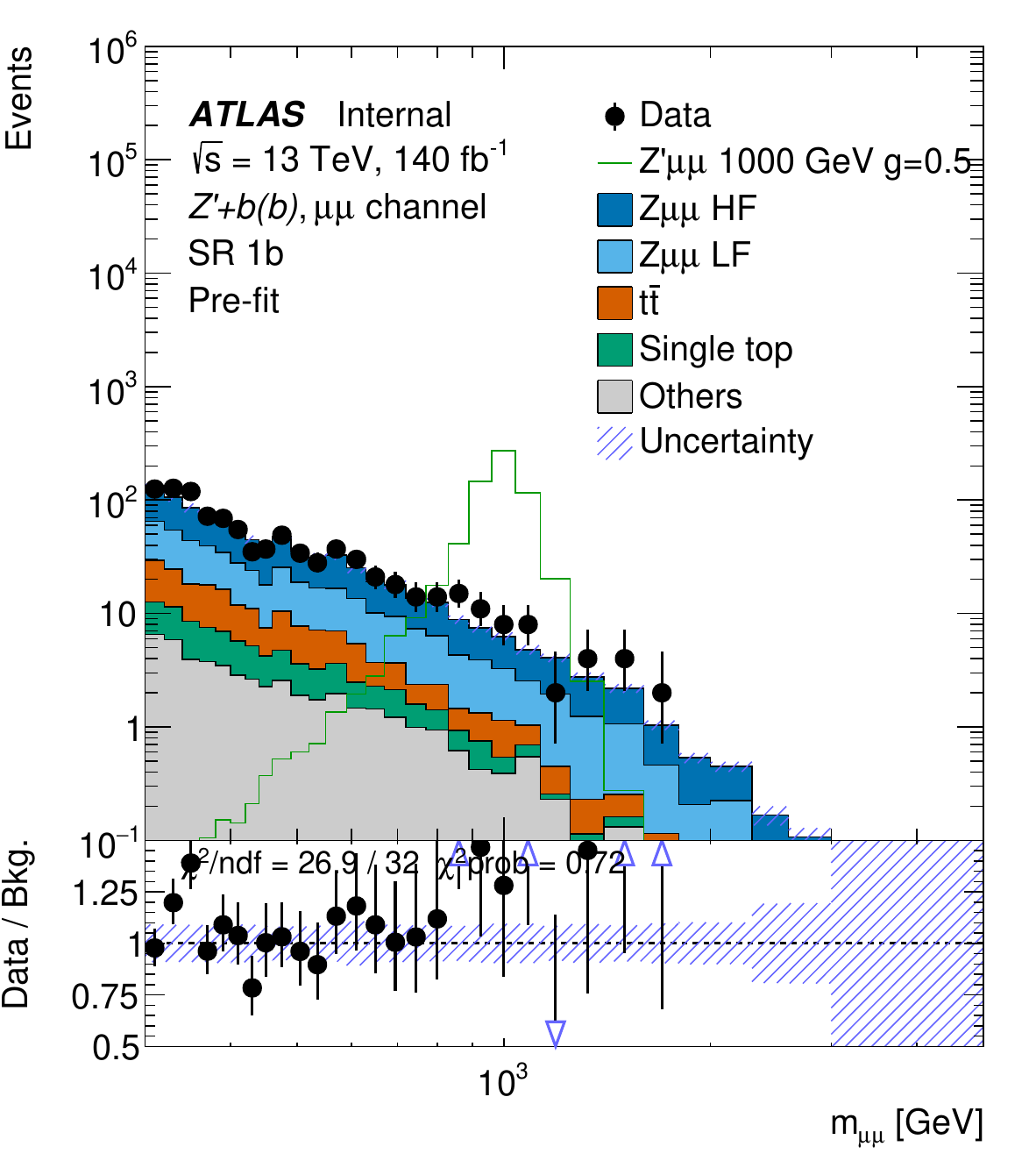}
		\label{fig:CR_fit_mu_SR_1b_pre_1000_g05_unblinded}
	}
	\subfloat[]{
		\includegraphics[width=0.33\textwidth]{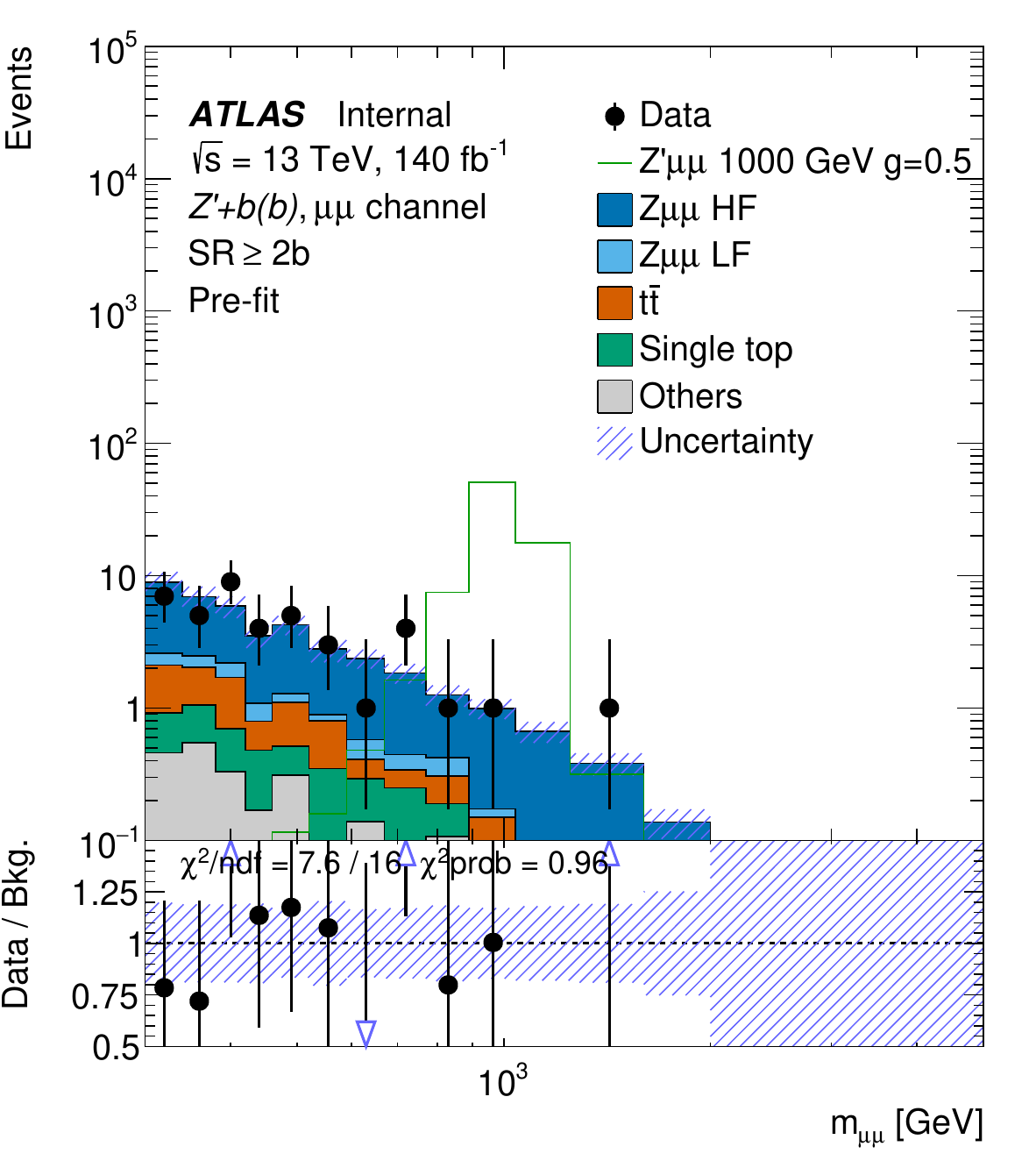}
		\label{fig:CR_fit_mu_SR_atleast2b_pre_1000_g05_unblinded}
	}
	\hfill
	\subfloat[]{
		\includegraphics[width=0.33\textwidth]{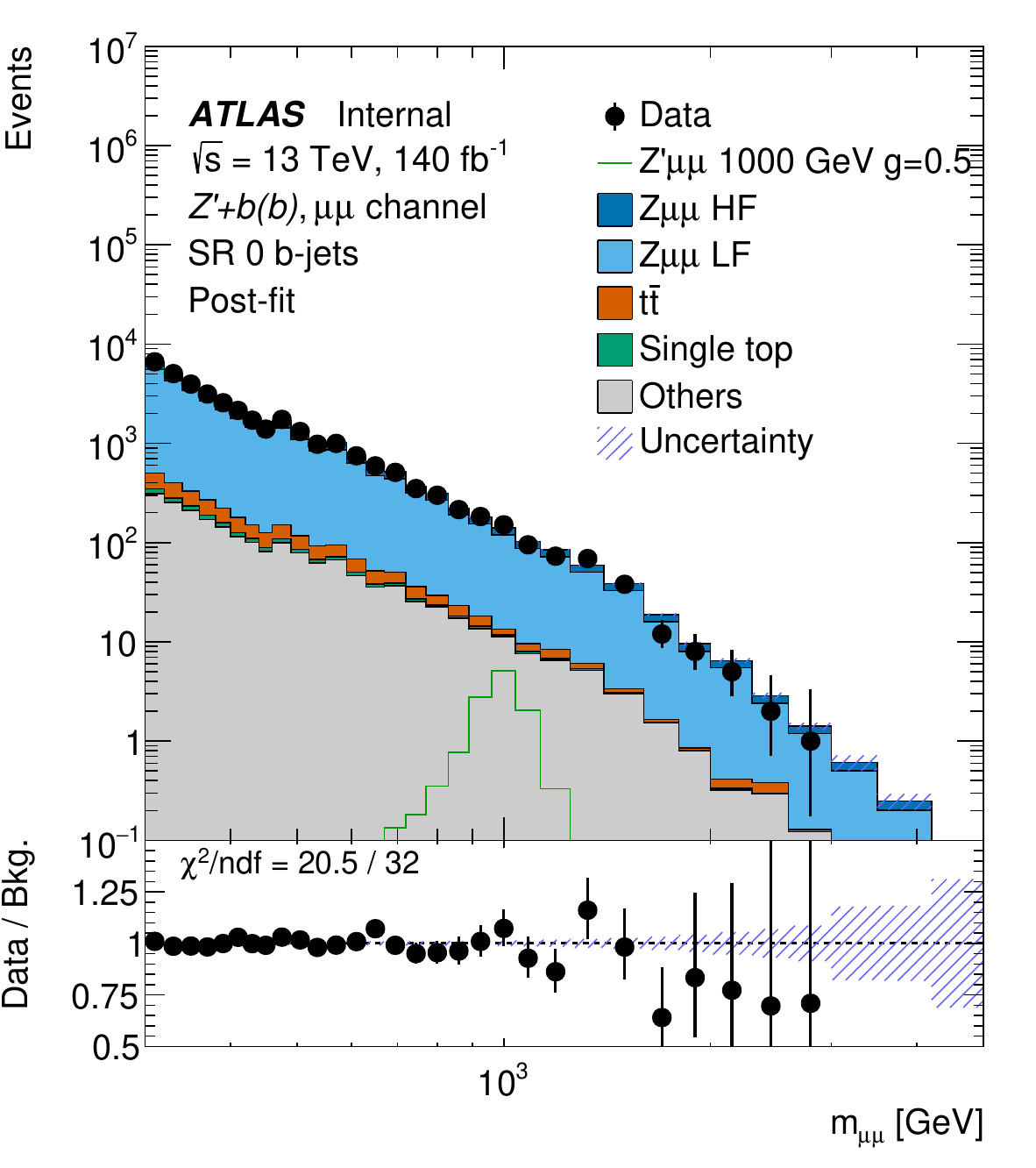}
		\label{fig:CR_fit_mu_SR_0b_post_1000_g05_unblinded}
	}
	\subfloat[]{
		\includegraphics[width=0.33\textwidth]{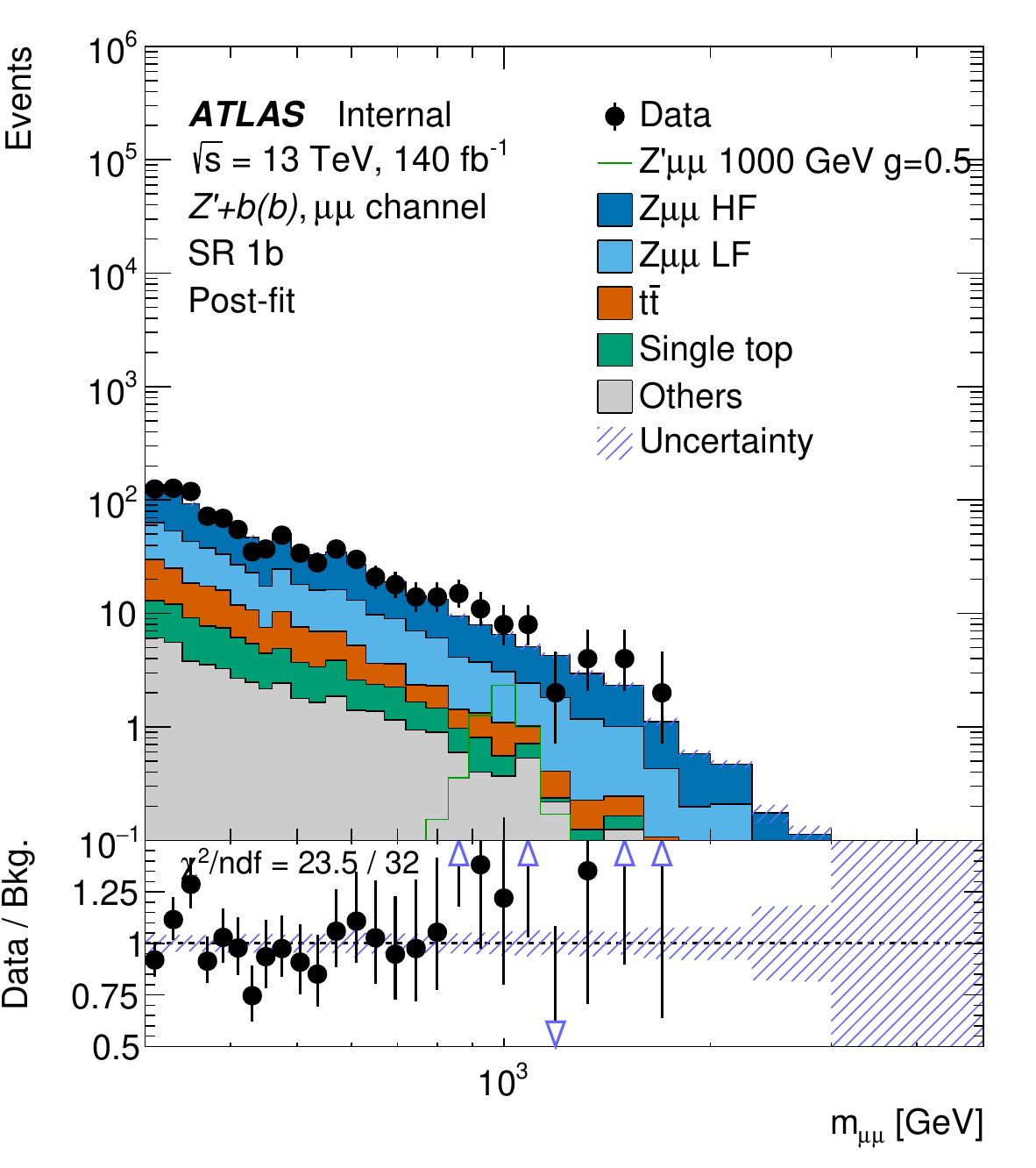}
		\label{fig:CR_fit_mu_SR_1b_post_1000_g05_unblinded}
	}
	\subfloat[]{
		\includegraphics[width=0.33\textwidth]{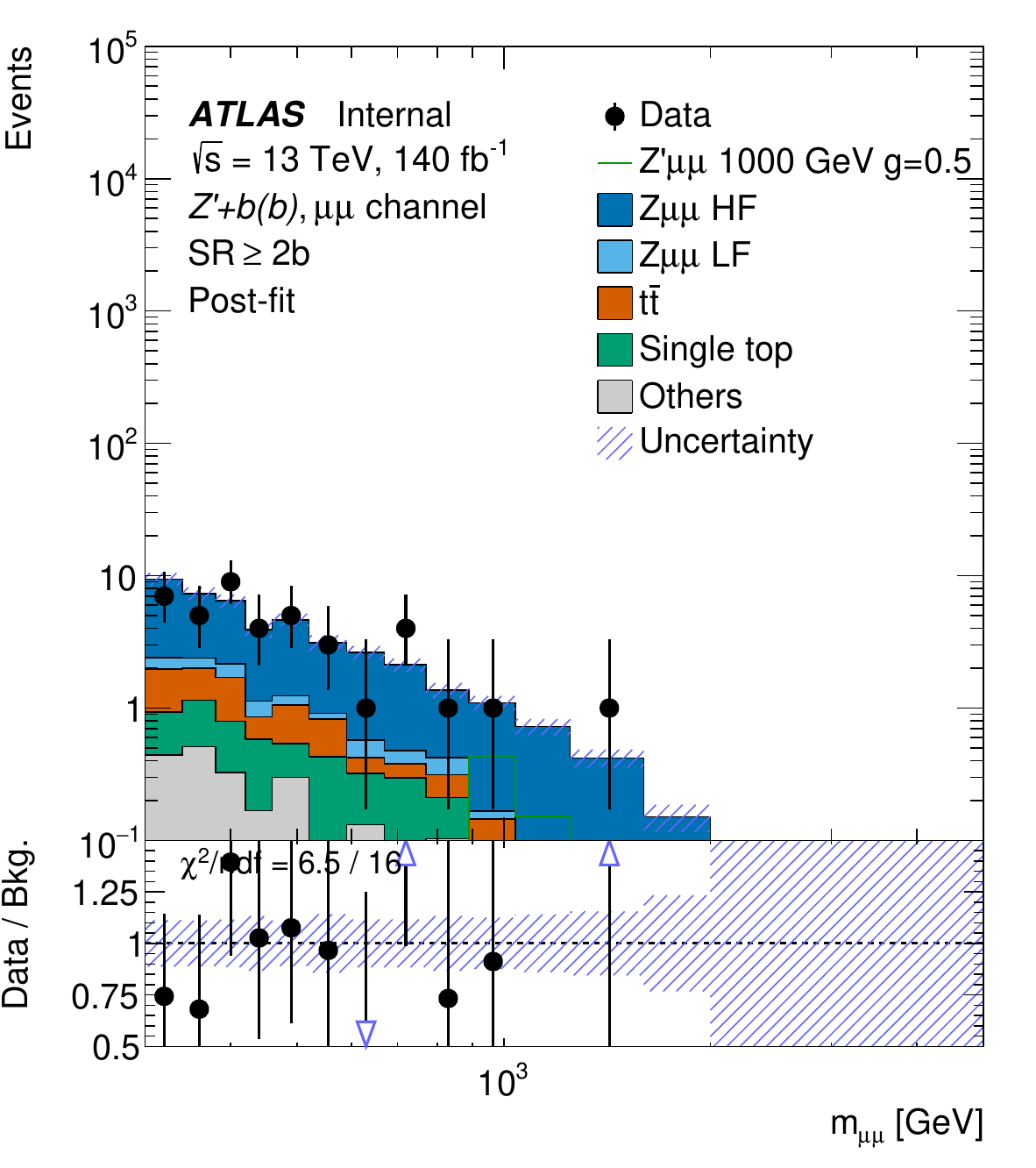}
		\label{fig:CR_fit_mu_SR_atleast2b_post_1000_g05_unblinded}
	}

	\caption{Pre-fit and post-fit plots for the single-bin fit of the $0b$ signal region (a, d), the $1b$ signal region (b, e) and the $\geq 2b$ signal region (c, f) in the muon channel. These plots are shown for $m_{Z'}=1\,\TeV$ and $g_{Z'}=0.5$. The uncertainty band includes statistical and systematic uncertainties.}
	\label{fig:SR_fit_mu_1000_g05_unblinded}
\end{figure}

The pre-fit and post-fit plots of the signal regions are shown in Figures~\ref{fig:SR_fit_ele_1000_g05_unblinded} and~\ref{fig:SR_fit_mu_1000_g05_unblinded} for the electron and muon channels. Overall data/MC agreement is good. In the $\geq2b$ signal region, statistical fluctuations are present, as the event count is strongly reduced compared to the other two signal regions. In the muon channel, a shape is visible in the data/MC ratio at high values of the invariant dilepton mass distribution; this is compensated by a pull of $\approx 1.7\sigma$ in the muon reconstruction efficiency uncertainty at high transverse momentum (referred to as \texttt{MUON\_EFF\_RECO\_PTDEPENDENCY}; see Section~\ref{sec:reco:muons}), the largest pull in the analysis.

\FloatBarrier

\subsection{Signal regions: $g_{Z'} = 1.0$}

\begin{figure}[h]
	\centering
	\subfloat[]{
		\includegraphics[width=0.33\textwidth]{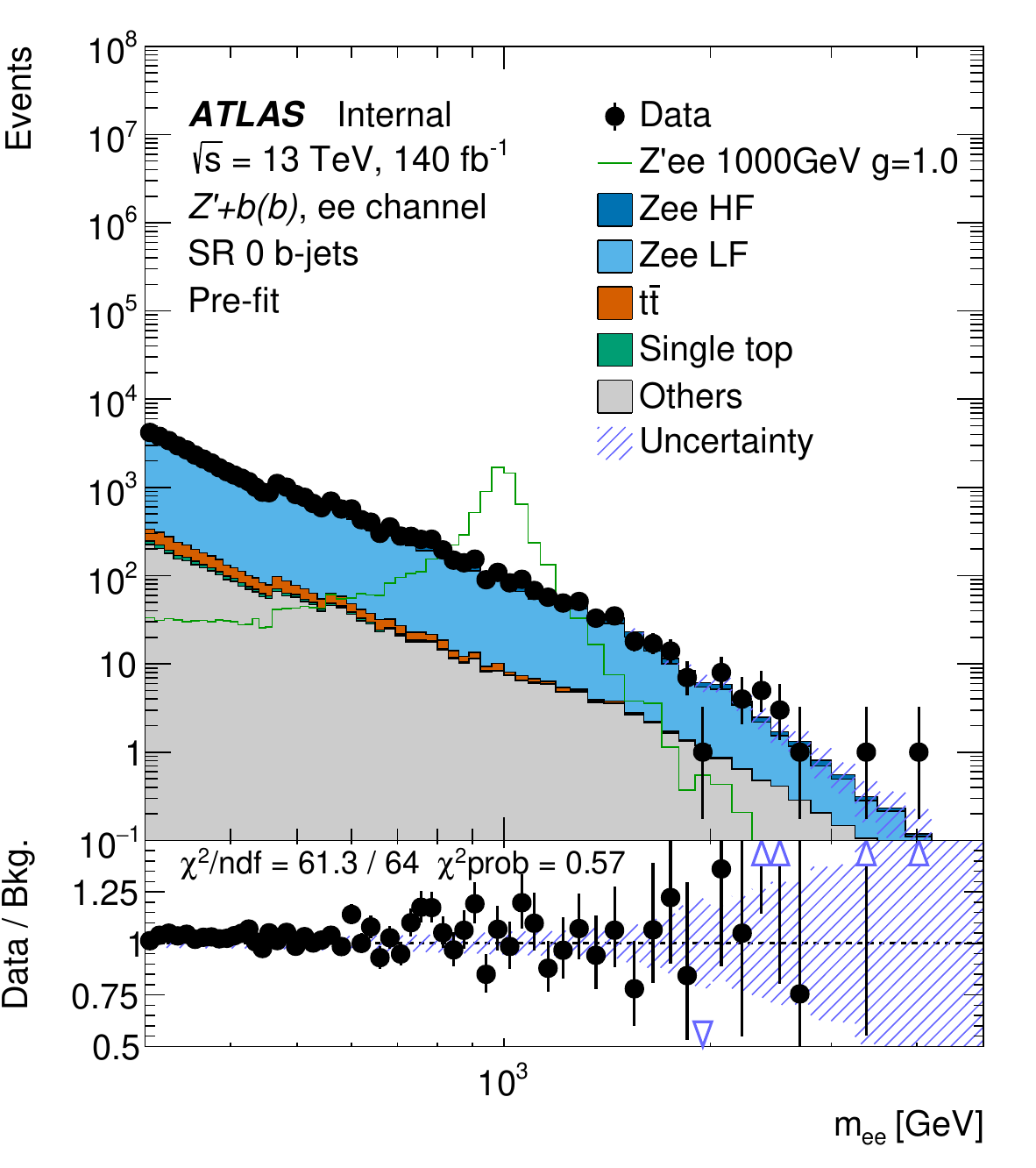}
		\label{fig:CR_fit_ele_SR_0b_pre_1000_g10_unblinded}
	}
	\subfloat[]{
		\includegraphics[width=0.33\textwidth]{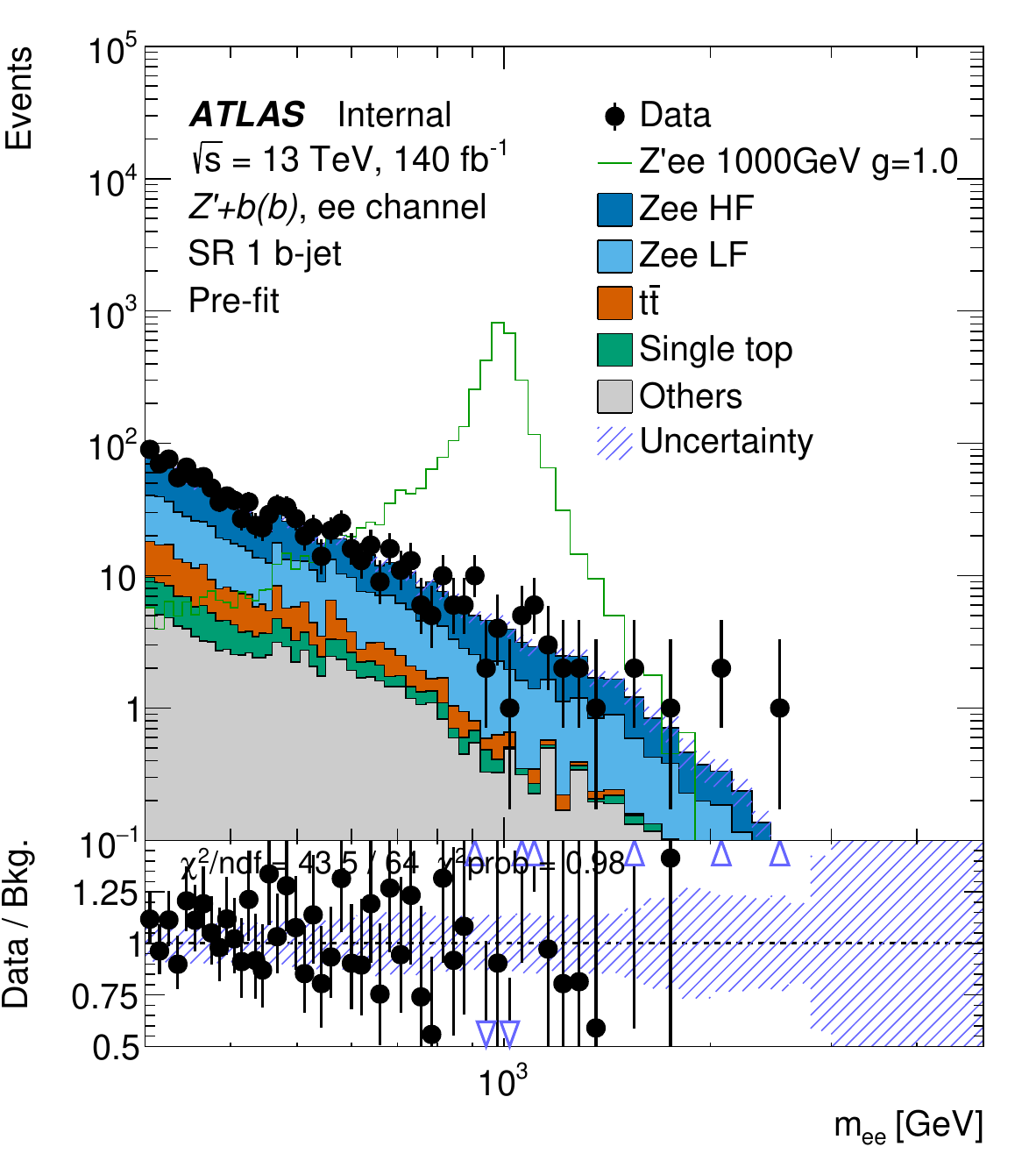}
		\label{fig:CR_fit_ele_SR_1b_pre_1000_g10_unblinded}
	}
	\subfloat[]{
		\includegraphics[width=0.33\textwidth]{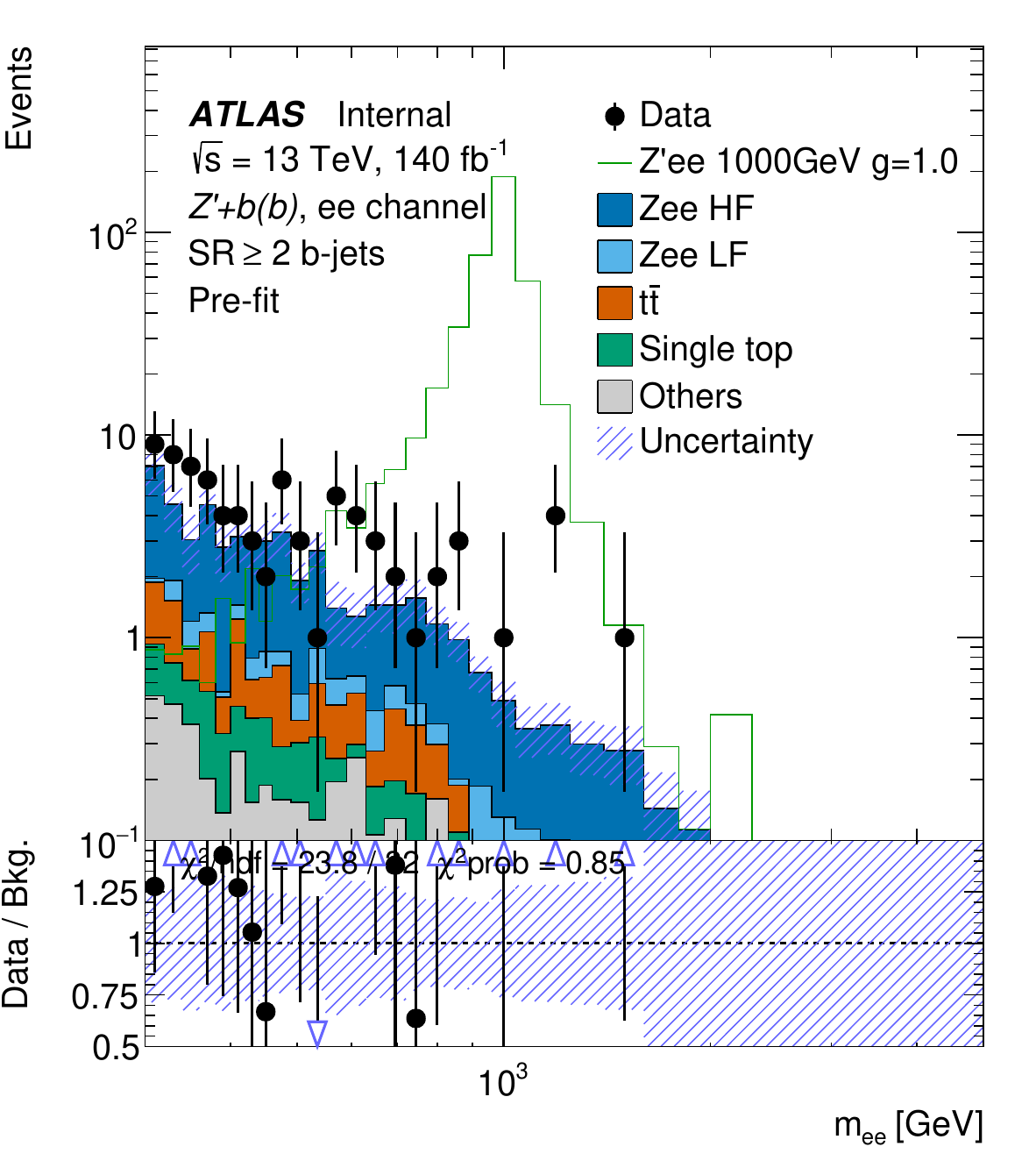}
		\label{fig:CR_fit_ele_SR_atleast2b_pre_1000_g10_unblinded}
	}
	\hfill
	\subfloat[]{
		\includegraphics[width=0.33\textwidth]{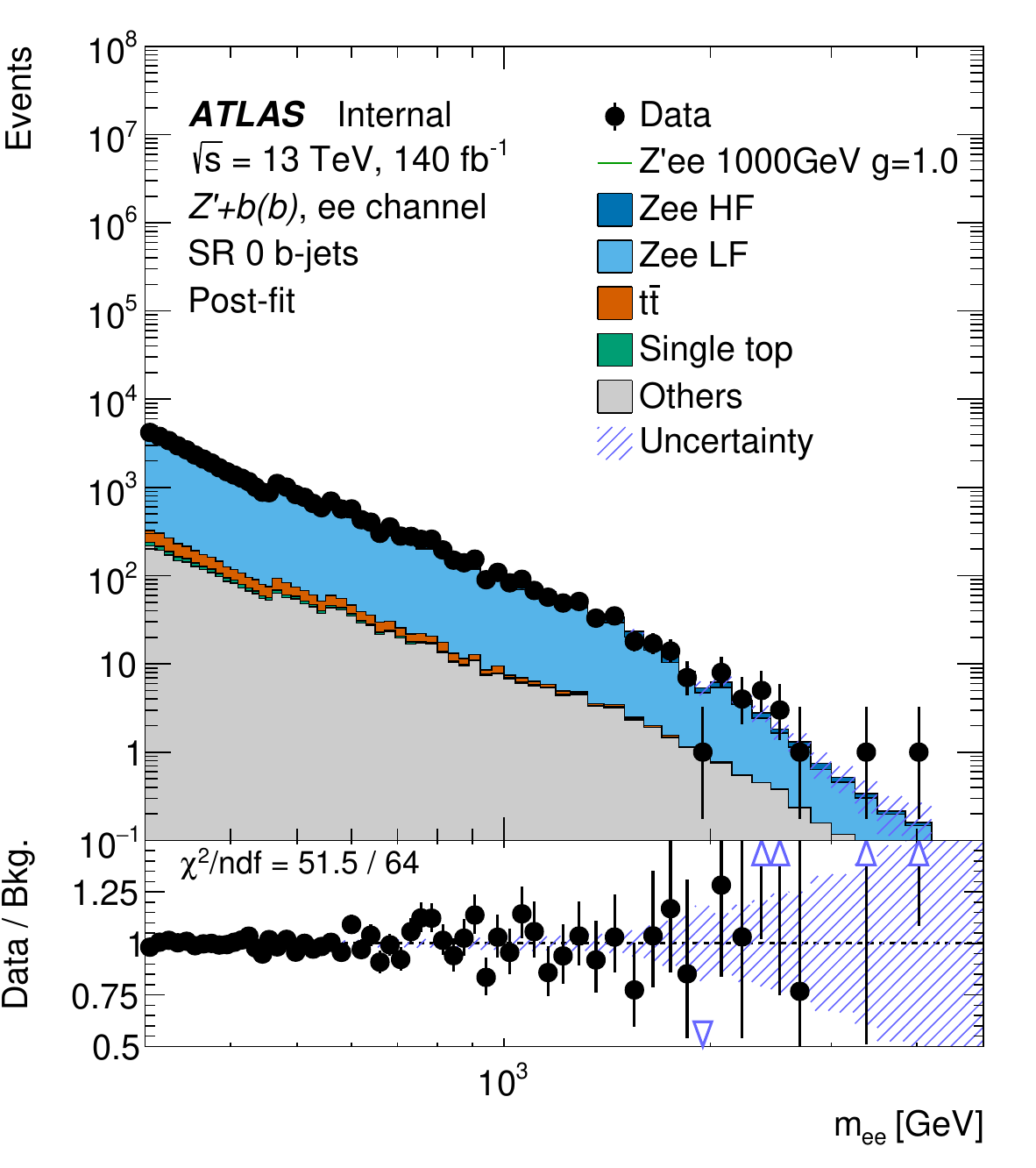}
		\label{fig:CR_fit_ele_SR_0b_post_1000_g10_unblinded}
	}
	\subfloat[]{
		\includegraphics[width=0.33\textwidth]{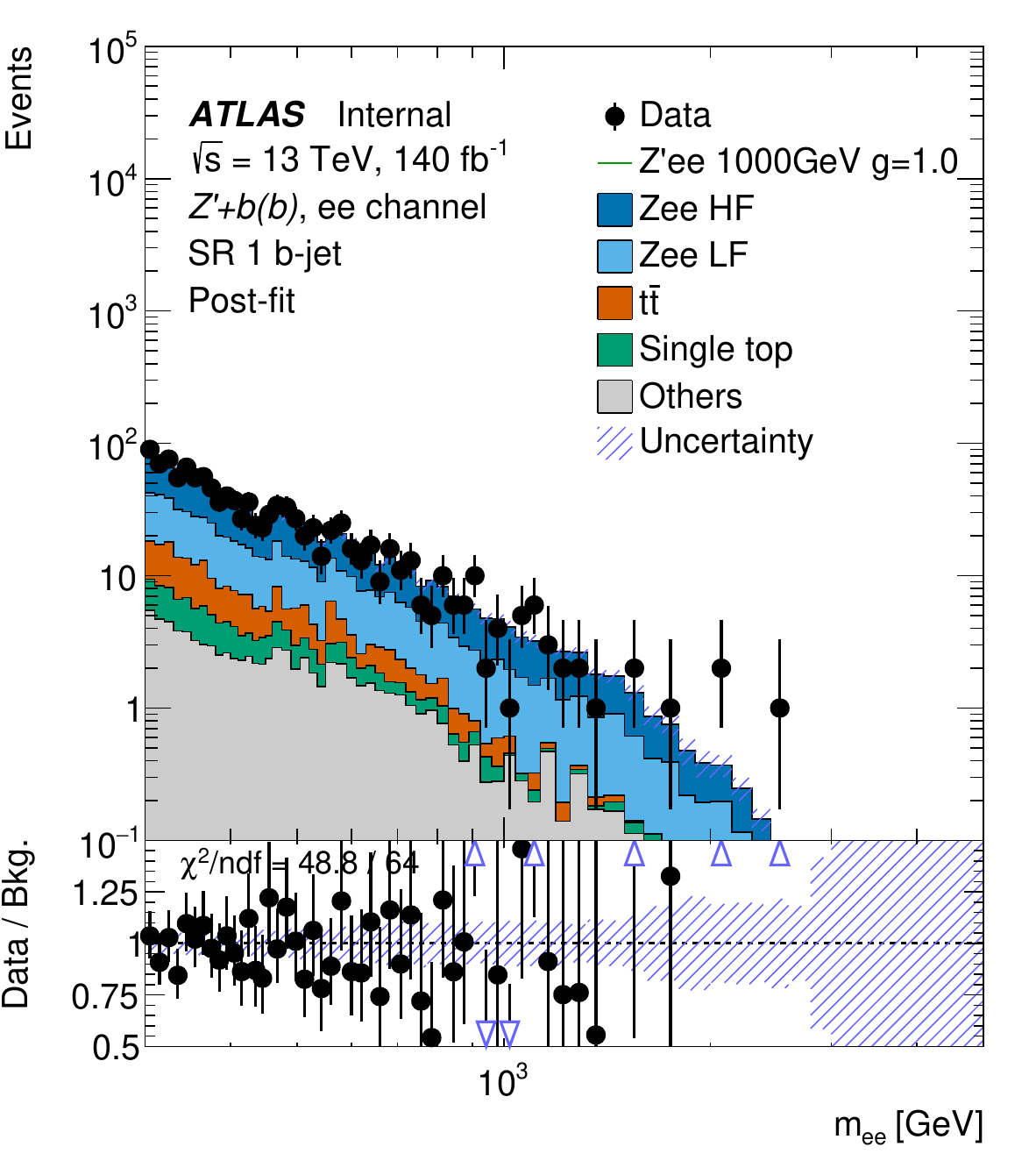}
		\label{fig:CR_fit_ele_SR_1b_post_1000_g10_unblinded}
	}
	\subfloat[]{
		\includegraphics[width=0.33\textwidth]{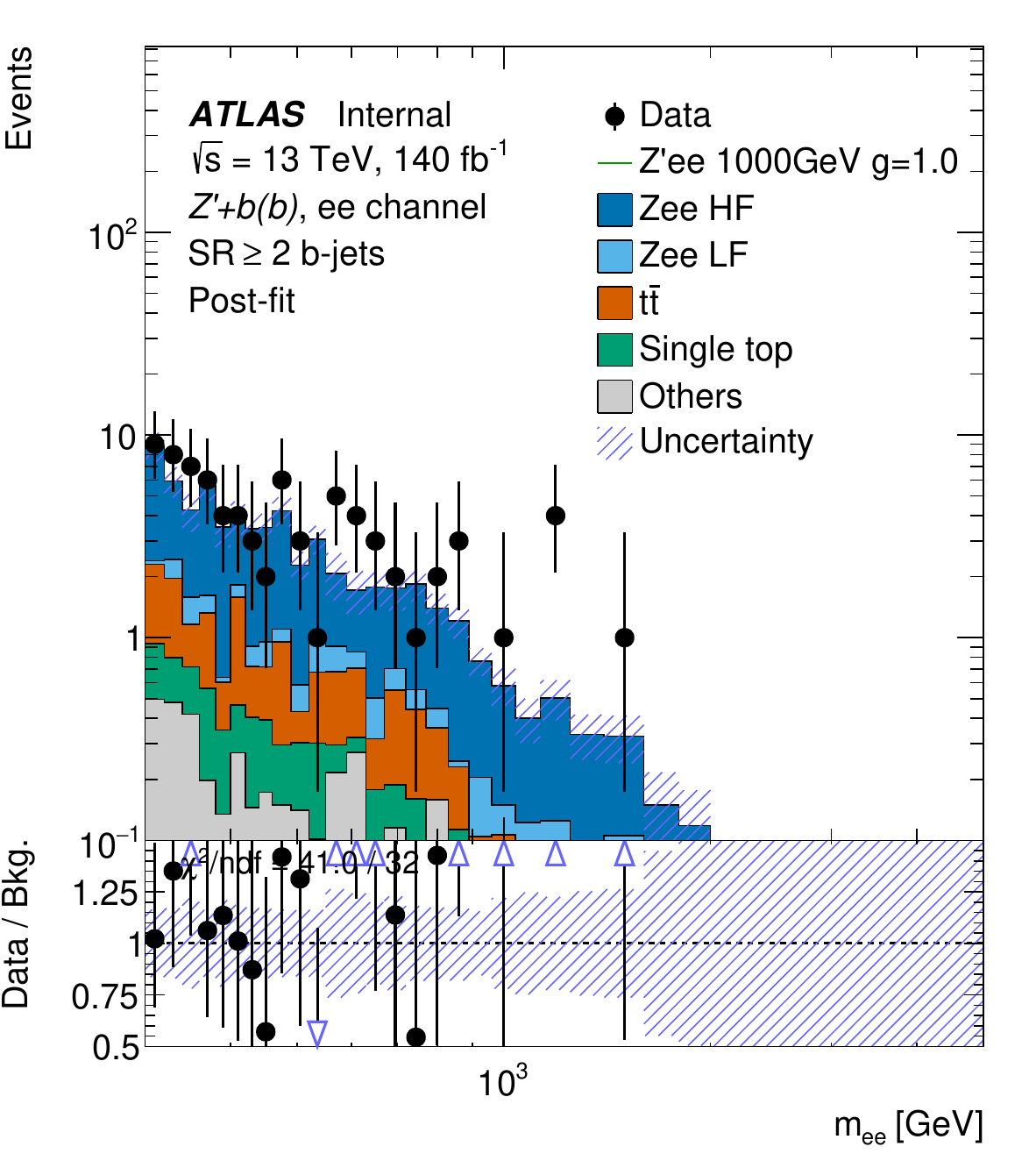}
		\label{fig:CR_fit_ele_SR_atleast2b_post_1000_g10_unblinded}
	}

	\caption{Pre-fit and post-fit plots for the single-bin fit of the $0b$ signal region (a, d), the $1b$ signal region (b, e) and the $\geq 2b$ signal region (c, f) in the electron channel. These plots are shown for $m_{Z'}=1\,\TeV$ and $g_{Z'}=1.0$. The uncertainty band includes statistical and systematic uncertainties.}
	\label{fig:SR_fit_ele_1000_g10_unblinded}
\end{figure}

\begin{figure}[h]
	\centering
	\subfloat[]{
		\includegraphics[width=0.33\textwidth]{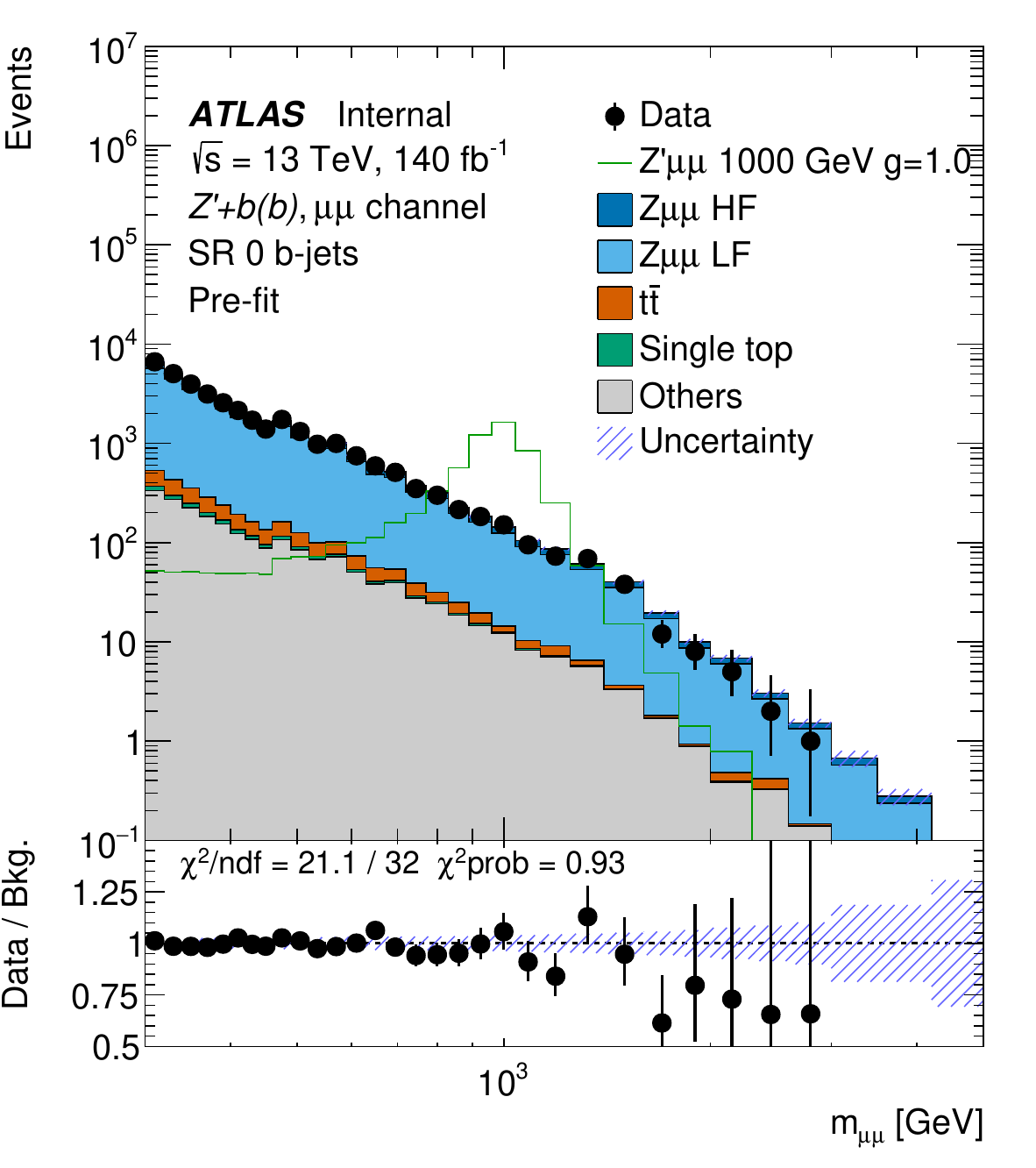}
		\label{fig:CR_fit_mu_SR_0b_pre_1000_g10_unblinded}
	}
	\subfloat[]{
		\includegraphics[width=0.33\textwidth]{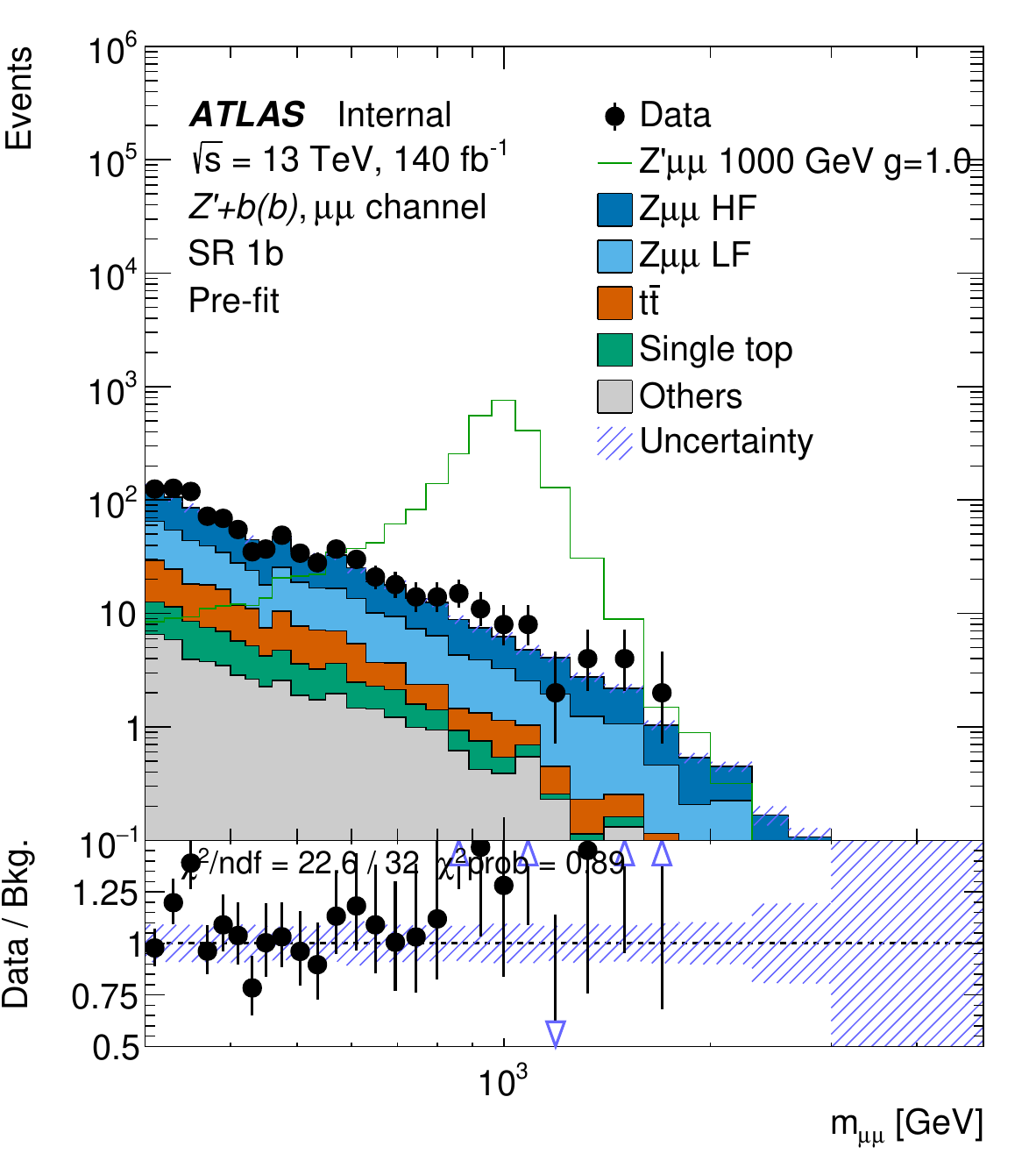}
		\label{fig:CR_fit_mu_SR_1b_pre_1000_g10_unblinded}
	}
	\subfloat[]{
		\includegraphics[width=0.33\textwidth]{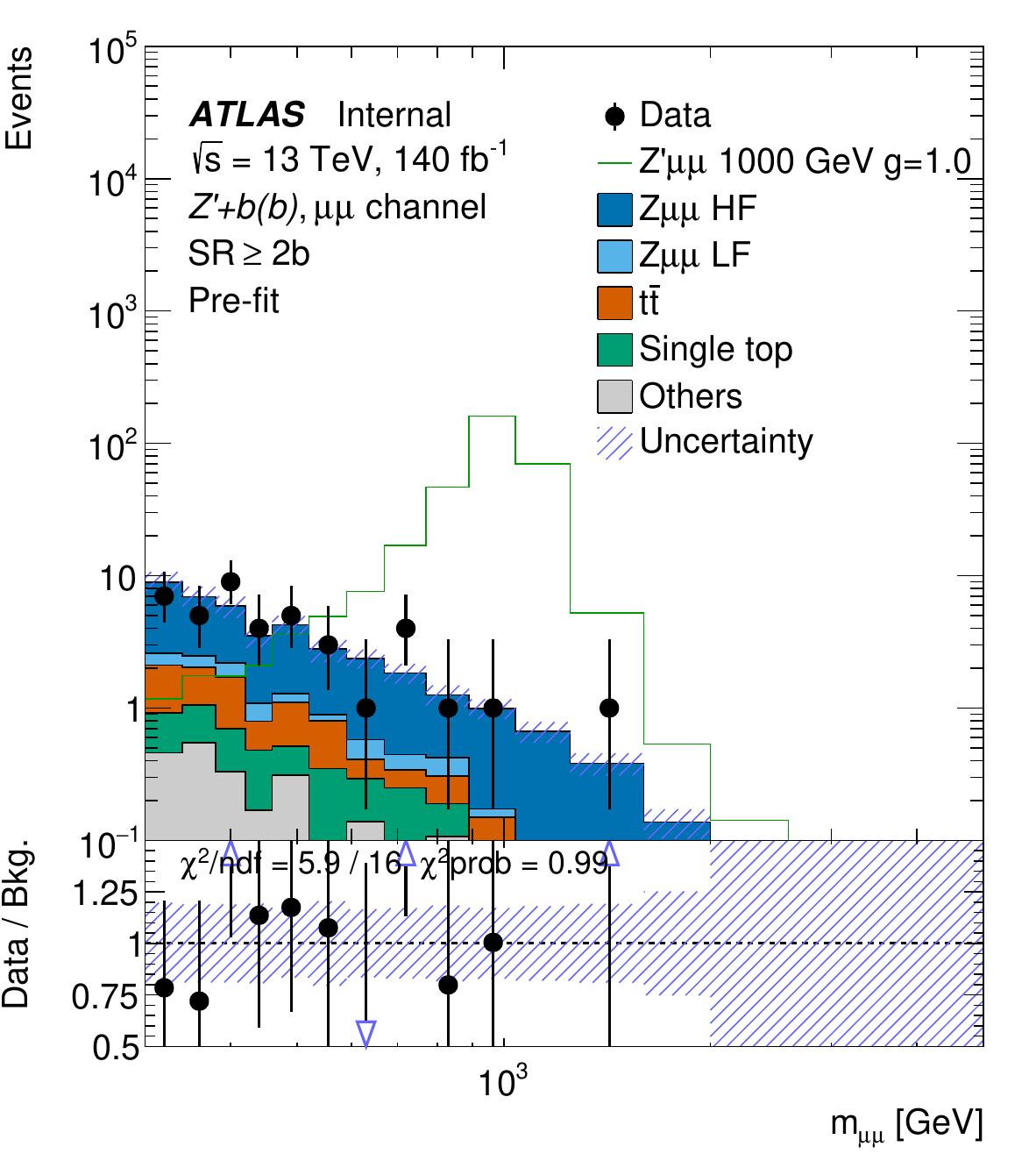}
		\label{fig:CR_fit_mu_SR_atleast2b_pre_1000_g10_unblinded}
	}
	\hfill
	\subfloat[]{
		\includegraphics[width=0.33\textwidth]{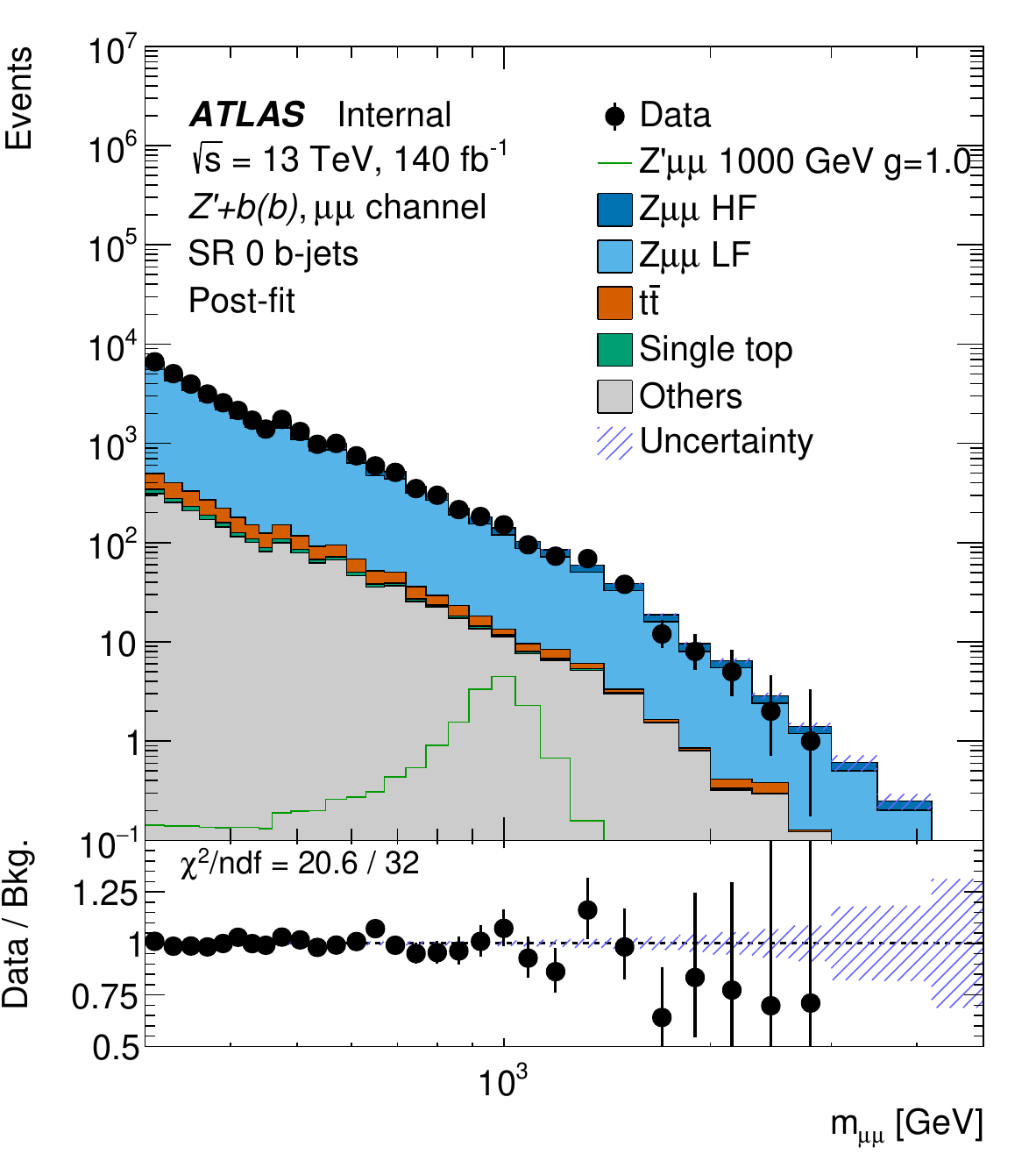}
		\label{fig:CR_fit_mu_SR_0b_post_1000_g10_unblinded}
	}
	\subfloat[]{
		\includegraphics[width=0.33\textwidth]{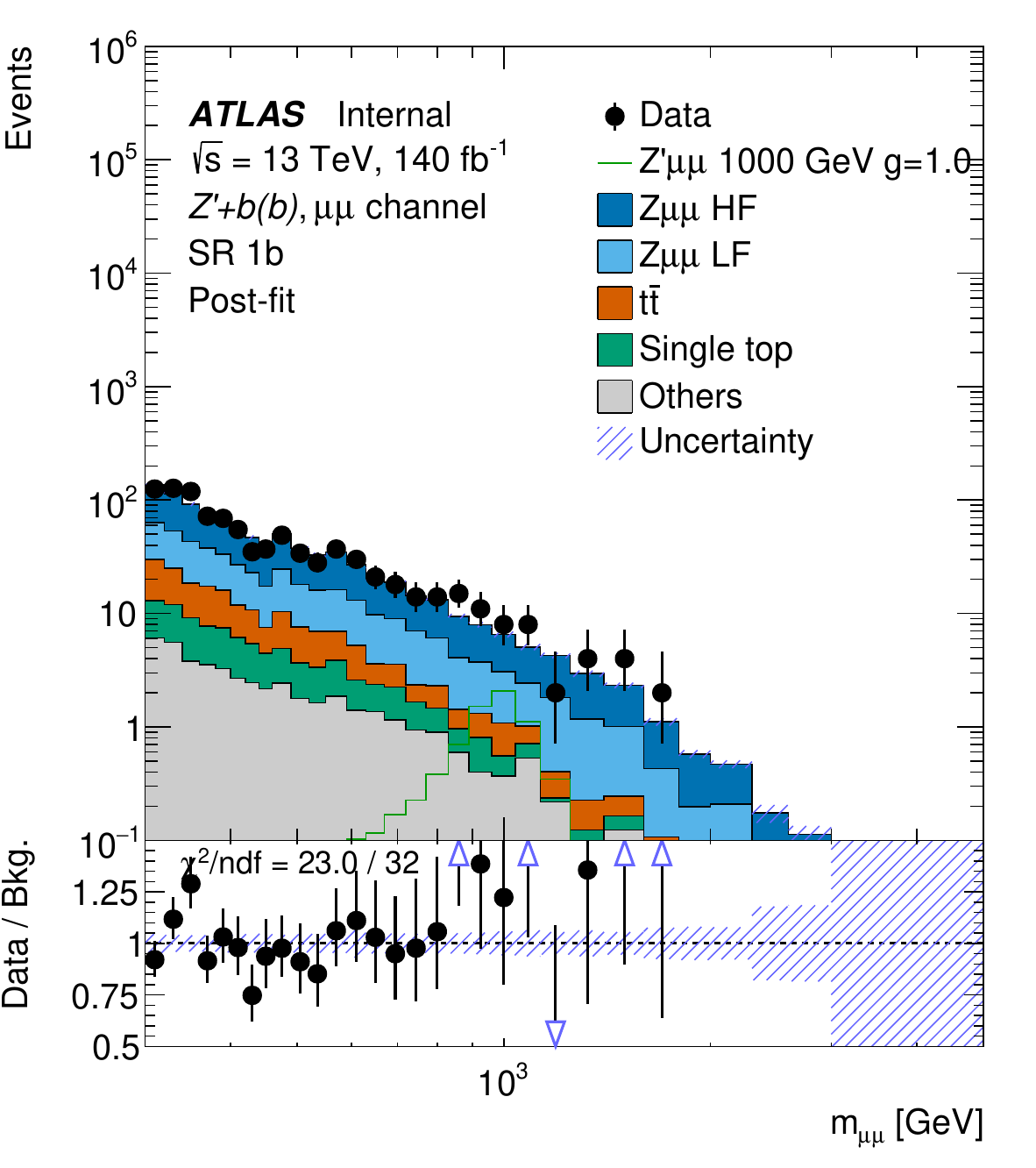}
		\label{fig:CR_fit_mu_SR_1b_post_1000_g10_unblinded}
	}
	\subfloat[]{
		\includegraphics[width=0.33\textwidth]{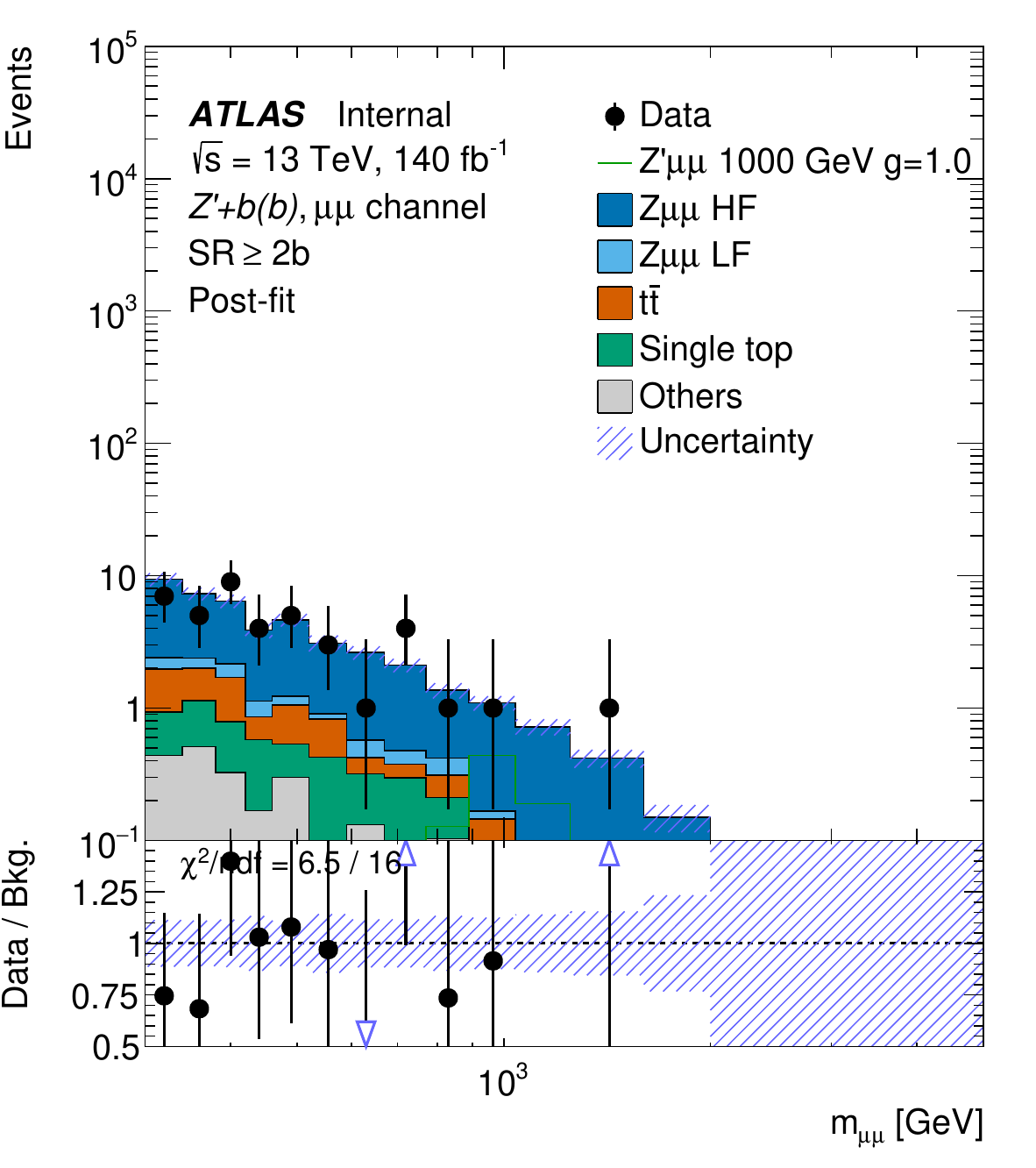}
		\label{fig:CR_fit_mu_SR_atleast2b_post_1000_g10_unblinded}
	}

	\caption{Pre-fit and post-fit plots for the single-bin fit of the $0b$ signal region (a, d), the $1b$ signal region (b, e) and the $\geq 2b$ signal region (c, f) in the muon channel. These plots are shown for $m_{Z'}=1\,\TeV$ and $g_{Z'}=1.0$. The uncertainty band includes statistical and systematic uncertainties.}
	\label{fig:SR_fit_mu_1000_g10_unblinded}
\end{figure}

The pre-fit and post-fit plots of the signal regions for $g_{Z'} = 1.0$ are shown in Figures~\ref{fig:SR_fit_ele_1000_g10_unblinded} and~\ref{fig:SR_fit_mu_1000_g10_unblinded}. The behaviour of the backgrounds and systematics is consistent with that observed for $g_{Z'} = 0.5$.

\FloatBarrier

\subsection{Cross-section limits}

\begin{figure}[h]
	\centering
	\subfloat[]{
		\includegraphics[width=0.47\textwidth]{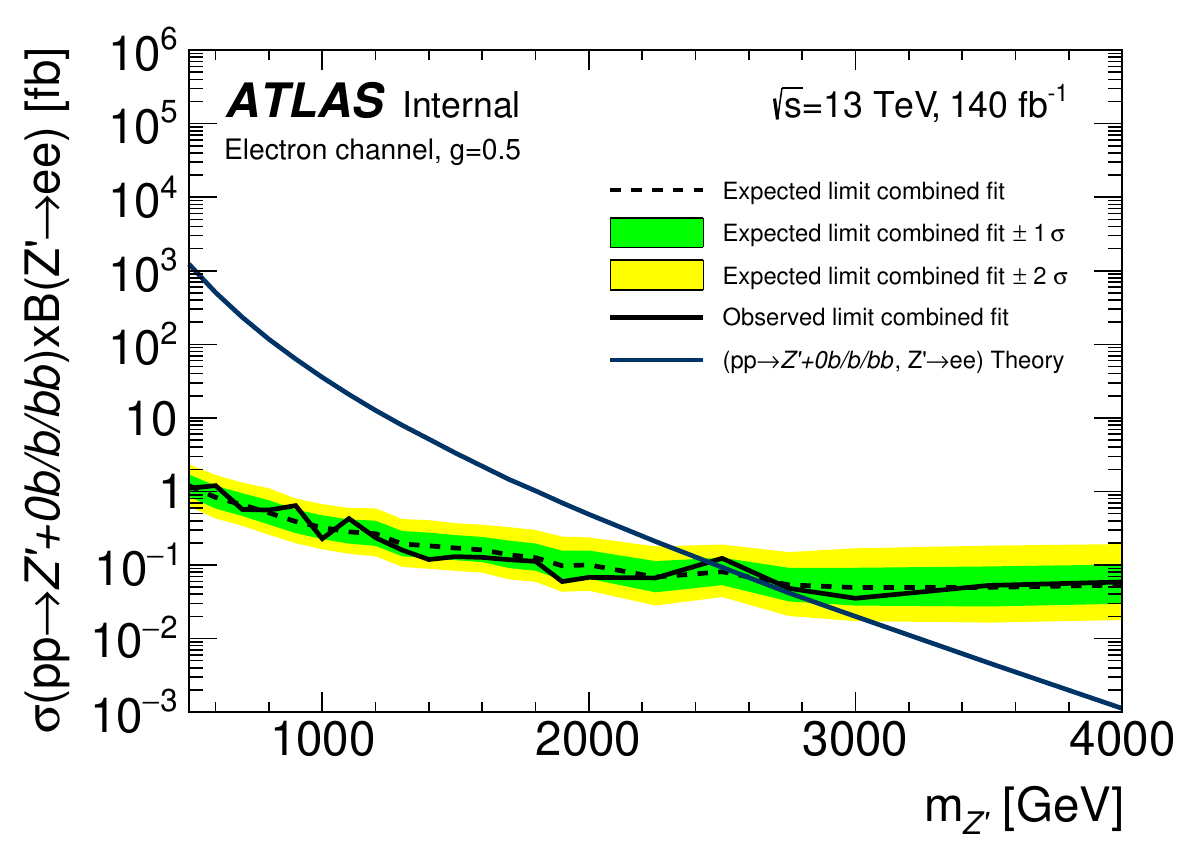}
		\label{fig:limits_ele_g05}
	}
	\subfloat[]{
		\includegraphics[width=0.47\textwidth]{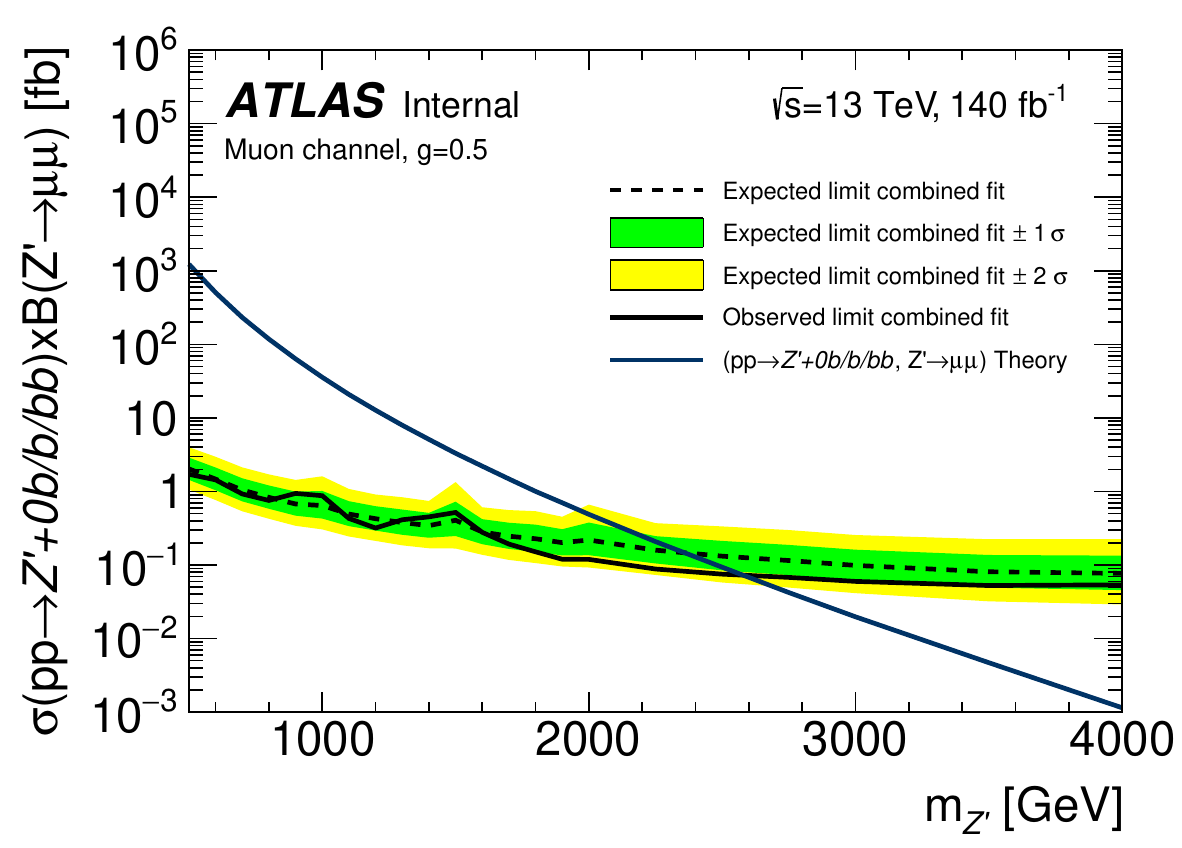}
		\label{fig:limits_mu_g05}
	}
	\hfill
	\subfloat[]{
		\includegraphics[width=0.47\textwidth]{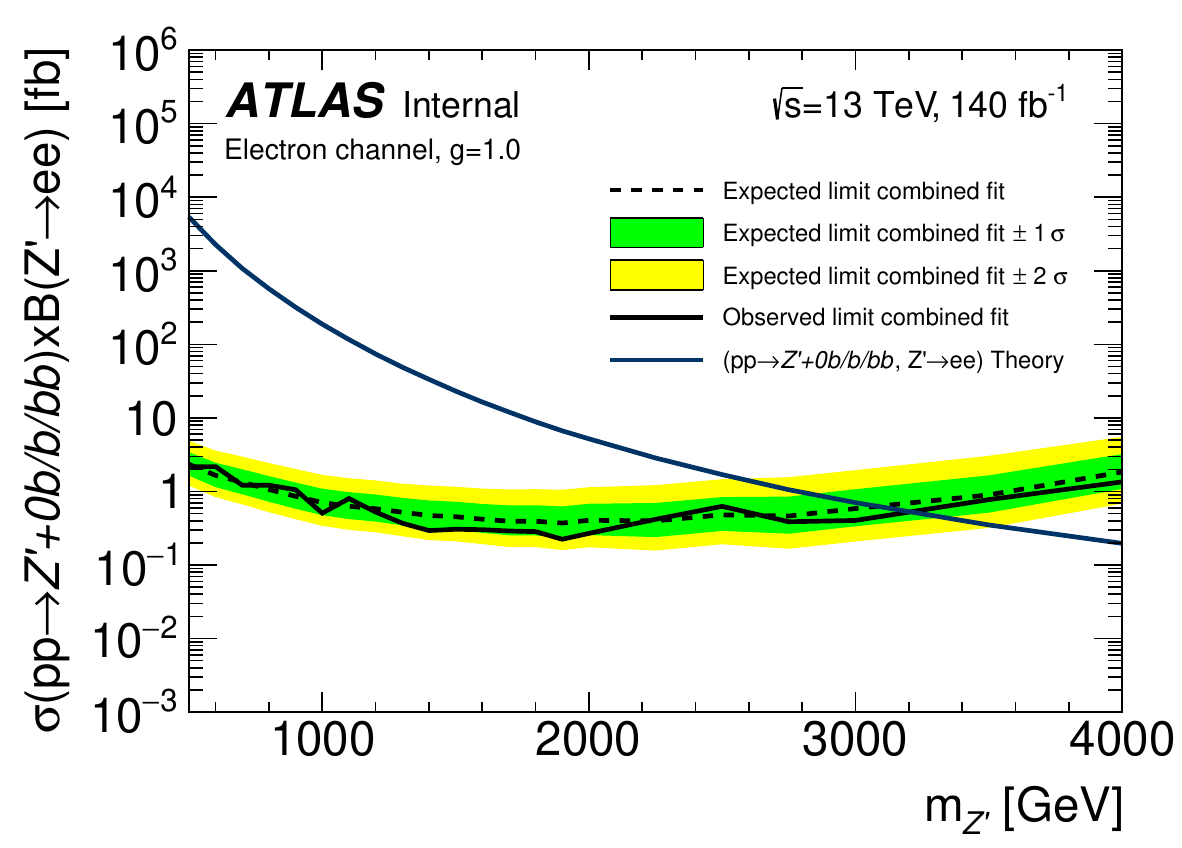}
		\label{fig:limits_ele_g10}
	}
	\subfloat[]{
		\includegraphics[width=0.47\textwidth]{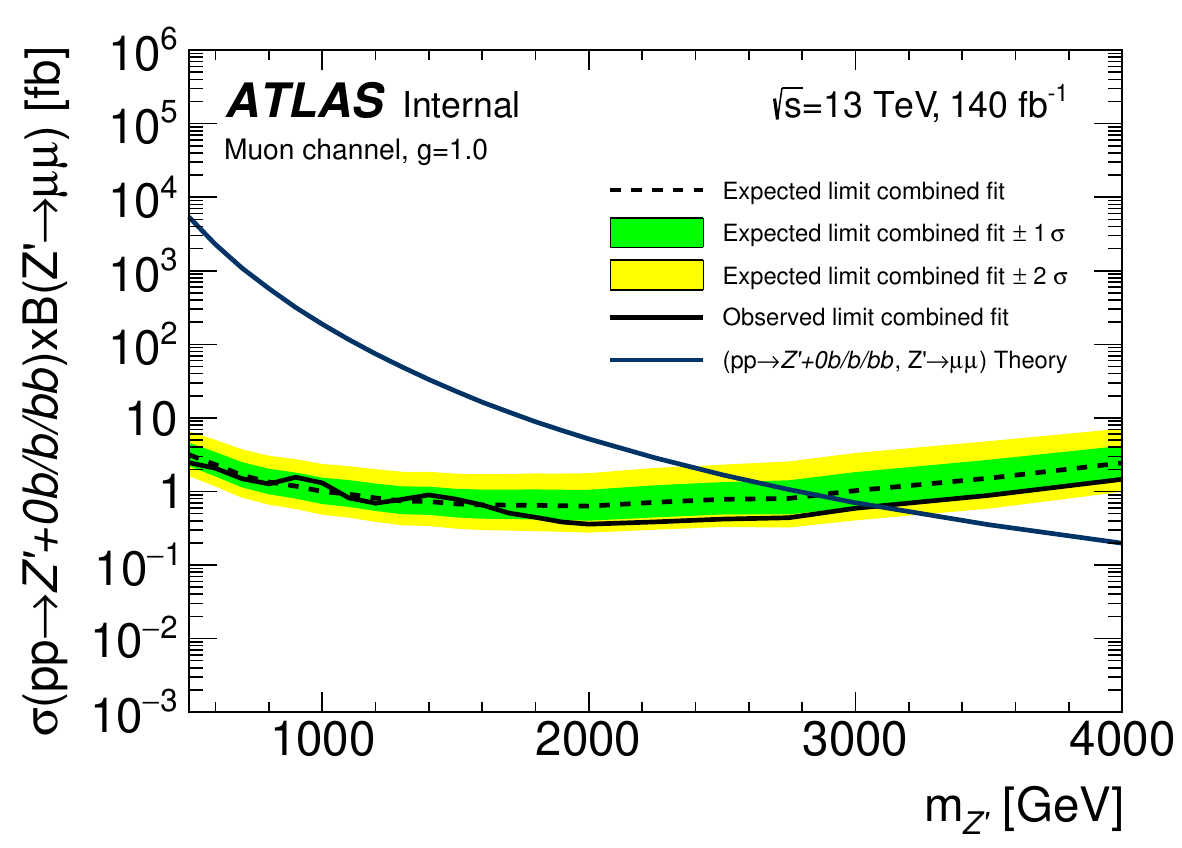}
		\label{fig:limits_mu_g10}
	}
	\caption{Expected and observed cross-section limits as a function of the $Z'$ boson mass for coupling parameters $g_{Z'}=0.5$ (a, b) and $g_{Z'}=1.0$ (c, d) for the electron and muon channels. Statistical and systematic uncertainties are considered when deriving the limits.}
	\label{fig:limits}
\end{figure}

\begin{figure}[h]
	\centering
	\subfloat[]{
		\includegraphics[width=0.47\textwidth]{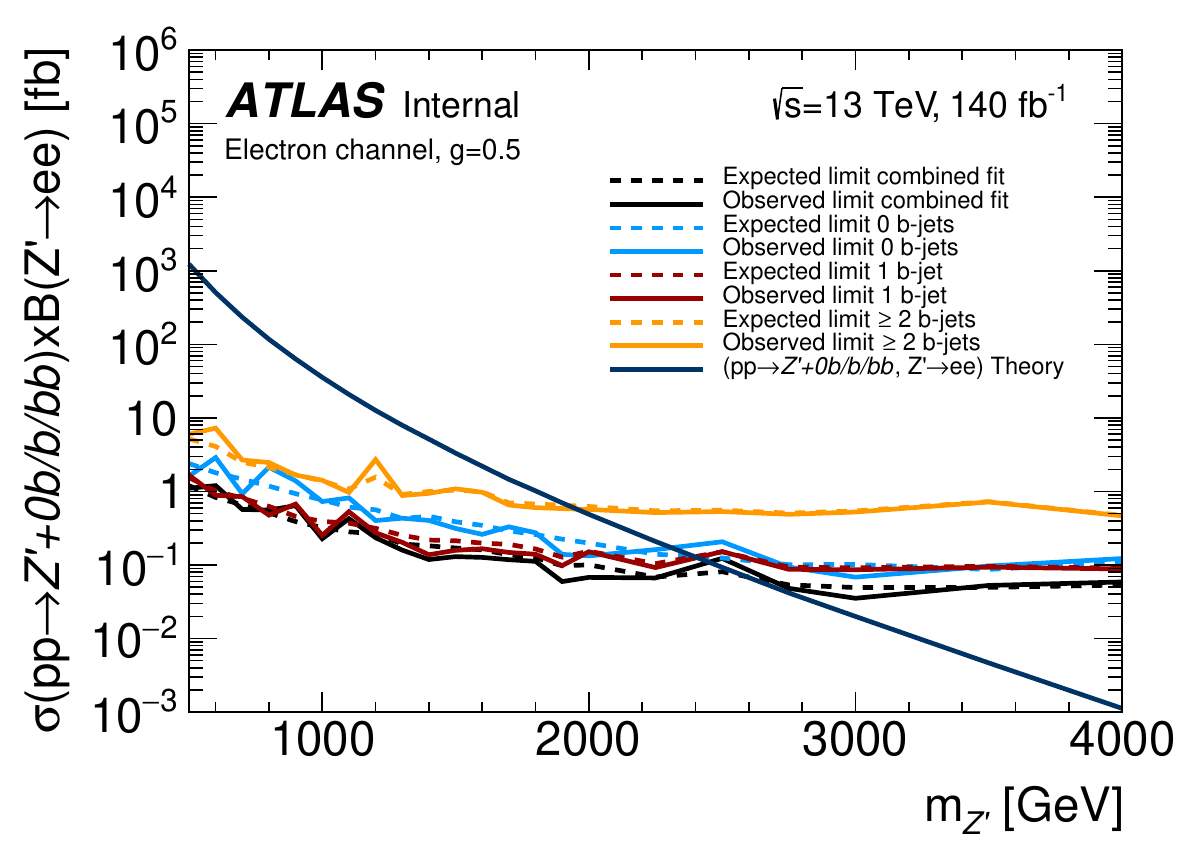}
		\label{fig:limits_ele_g05_comp}
	}
	\subfloat[]{
		\includegraphics[width=0.47\textwidth]{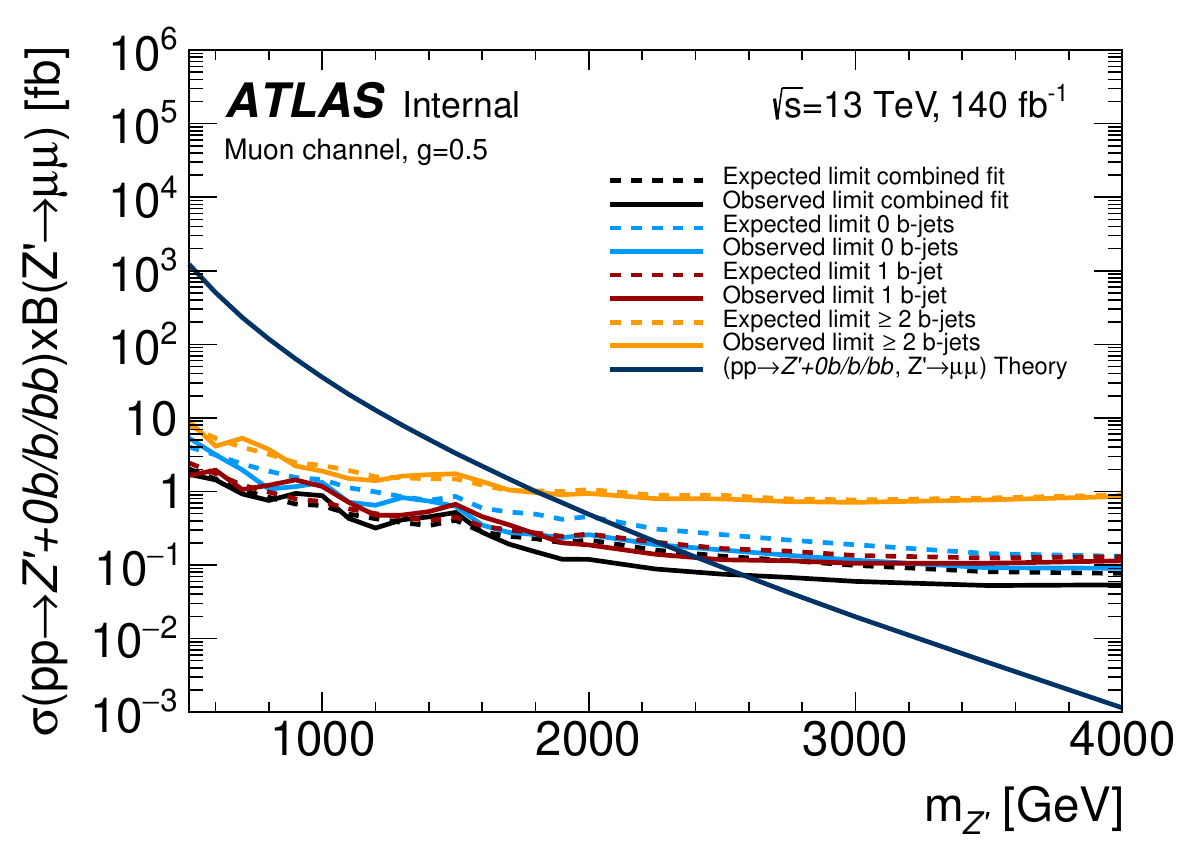}
		\label{fig:limits_mu_g05_comp}
	}
	\hfill
	\subfloat[]{
		\includegraphics[width=0.47\textwidth]{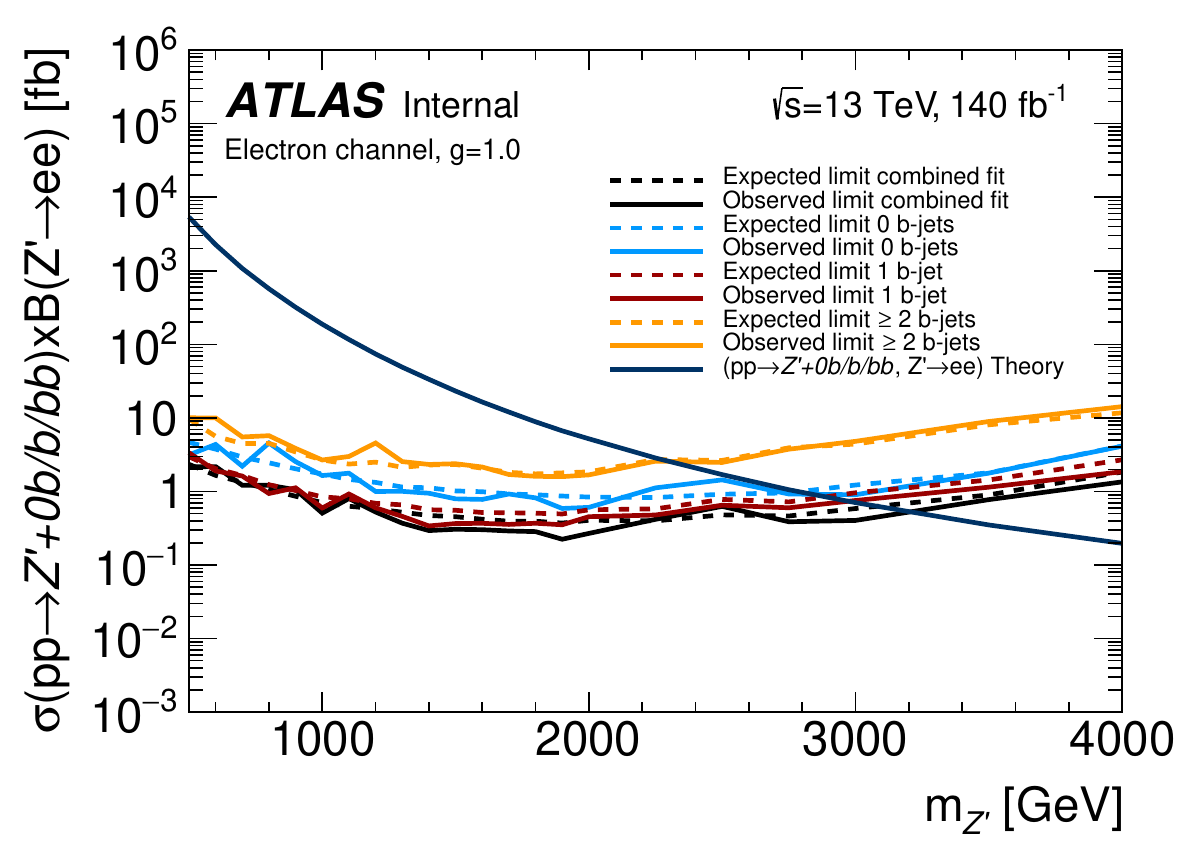}
		\label{fig:limits_ele_g10_comp}
	}
	\subfloat[]{
		\includegraphics[width=0.47\textwidth]{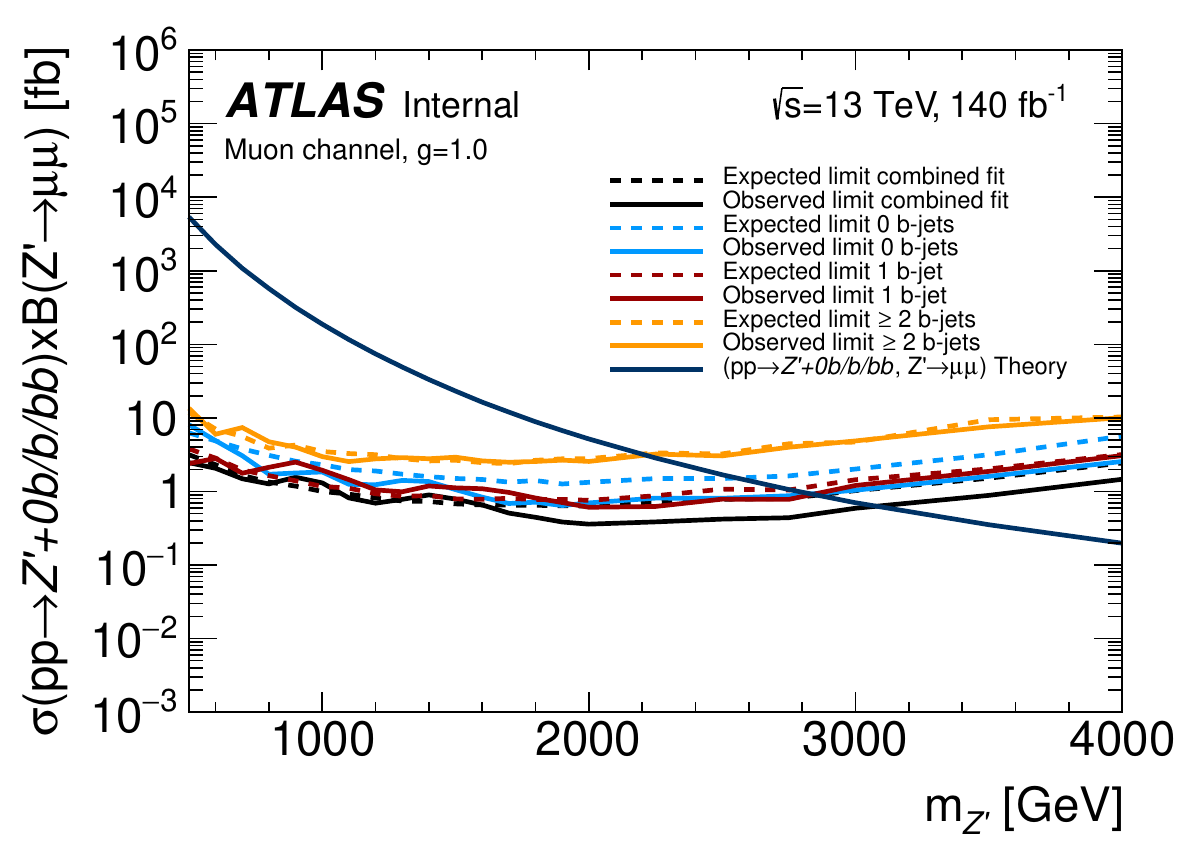}
		\label{fig:limits_mu_g10_comp}
	}
	\caption{Expected and observed cross-section limits as a function of the $Z'$ boson mass for coupling parameters $g_{Z'}=0.5$ (a, b) and $g_{Z'}=1.0$ (c, d) for the electron and muon channels. The limits are shown for the individual fits of the different signal regions as well as for the combined fit of all signal regions. Statistical and systematic uncertainties are considered when deriving the limits.}
	\label{fig:limits_comp}
\end{figure}

Figures \ref{fig:limits} and \ref{fig:limits_comp} show the cross-section limits for the two coupling parameters $g_{Z'}=0.5$ and $g_{Z'}=1.0$ for the muon and electron channel.
FIG.~\ref{fig:limits} shows the expected and observed limits together with the $\pm1\sigma$ and $\pm2\sigma$ bands for the combined fit of all signal regions. FIG.~\ref{fig:limits_comp} shows the expected and observed limits for the combined fit of all signal regions alongside the limits extracted from the individual signal region fits.

\FloatBarrier

\subsection{Combination of electron and muon channel}

In addition to the individual fits of the electron and muon channel, combined fits of both channels are performed.

\begin{figure}[h]
	\centering

	\includegraphics[width=0.85\textwidth]{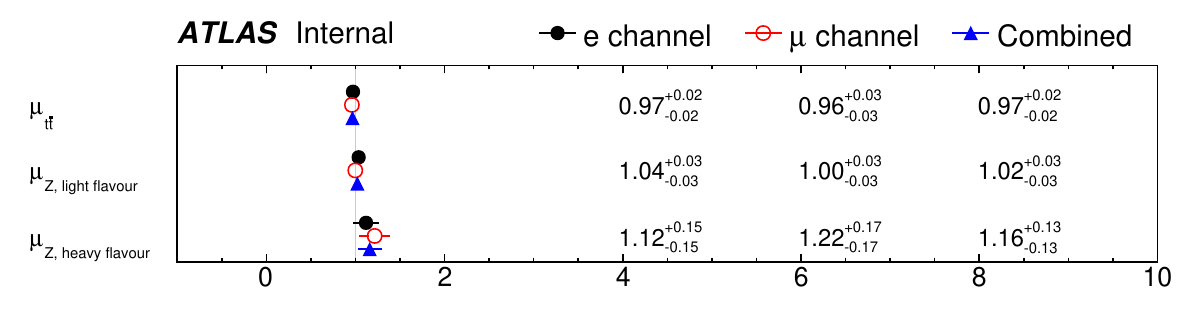}

	\caption{Background normalisation factors for the combined fit of the muon and electron channel for $m_{Z'}=1\,\mathrm{TeV}$ and $g_{Z'}=0.5$.}
	\label{fig:NormFactors_1000_g05_unblinded_comb}
\end{figure}

The background normalisation factors and nuisance parameter pull and ranking plots for the combined fit ($m_{Z'}=1\,\text{TeV}$, $g_{Z'}=0.5$) are shown in Figures~\ref{fig:NormFactors_1000_g05_unblinded_comb} and~\ref{fig:Ranking_1000_g05_unblinded_comb}; the corresponding nuisance parameter pull plot is provided in Appendix~\ref{app:zprime:fits} (FIG.~\ref{fig:NuisPar_1000_g05_unblinded_comb}).

\begin{figure}[h]
	\centering
	\subfloat[]{
		\includegraphics[width=0.47\textwidth]{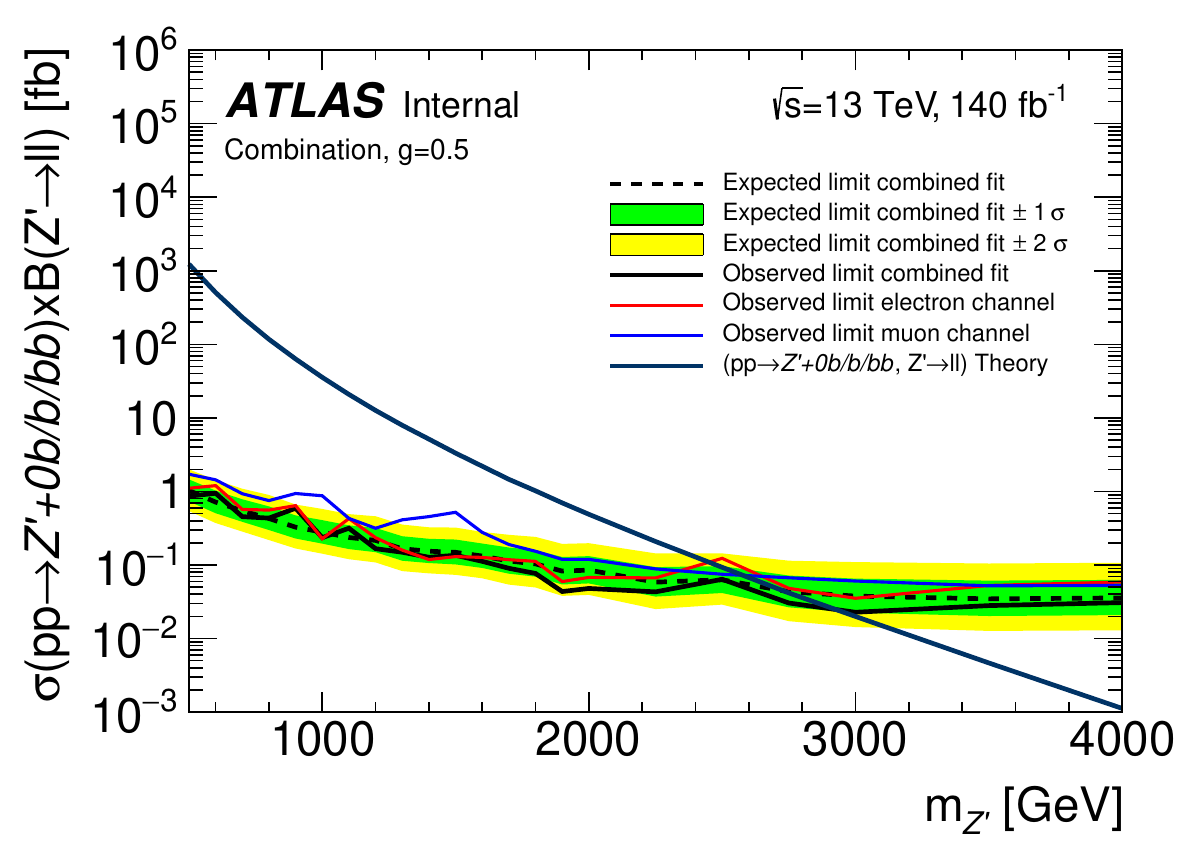}
		\label{fig:combined_limits_g05}
	}
	\subfloat[]{
		\includegraphics[width=0.47\textwidth]{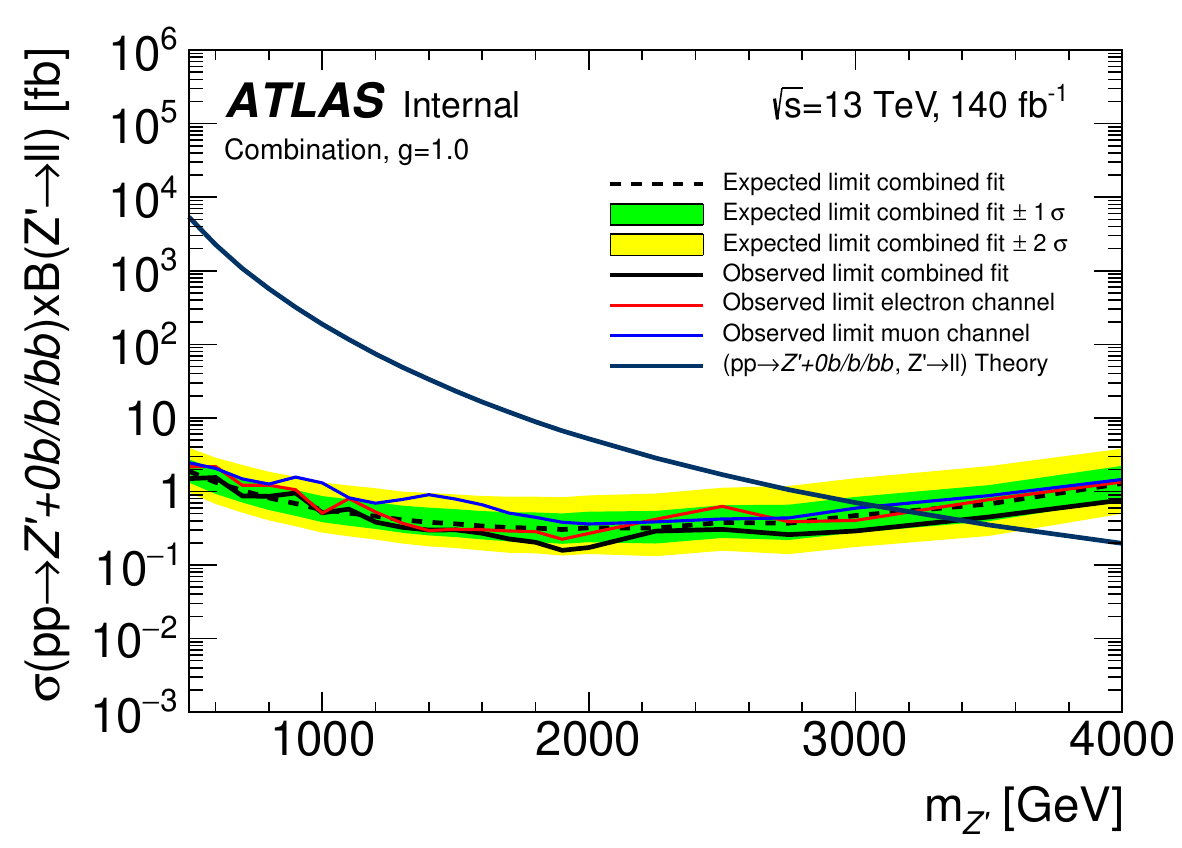}
		\label{fig:combined_limits_g10}
	}

	\caption{Expected and observed cross-section limits as a function of the $Z'$ boson mass for coupling parameters $g_{Z'}=0.5$ a and $g_{Z'}=1.0$ b for the combination of muon and electron channel. The observed limits are also shown for the individual fits of the two channels. Statistical and systematic uncertainties are considered when deriving the limits.}
	\label{fig:combined_limits}
\end{figure}

The resulting cross-section limits are shown in FIG.~\ref{fig:combined_limits} for the two coupling parameters $g_{Z'}=0.5$ and $g_{Z'}=1.0$. The observed limit of the combination is stronger than the limits in the individual channels, particularly at higher $Z'$ boson masses where the results are statistics-limited.

\begin{figure}[h]
	\centering

	\includegraphics[width=0.7\textwidth]{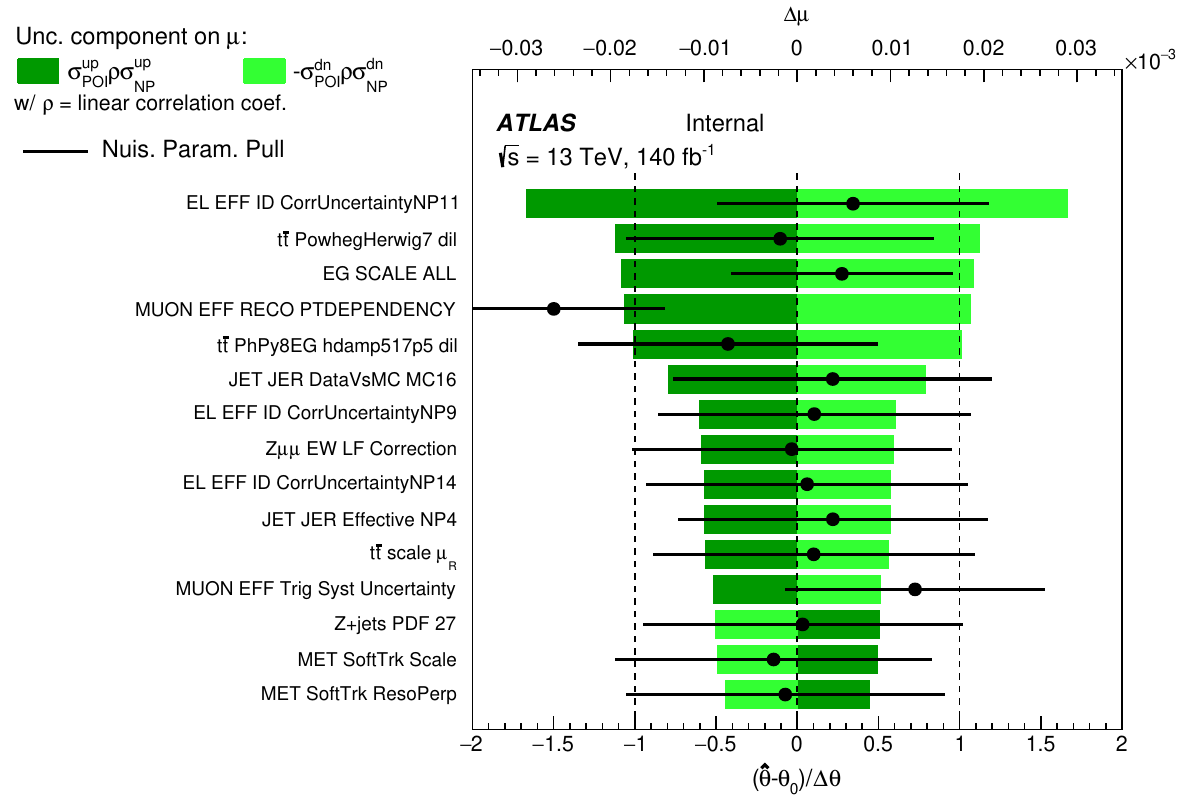}

	\caption{Nuisance parameter ranking plot for the combined fit of the muon and electron channel for $m_{Z'}=1\,\mathrm{TeV}$ and $g_{Z'}=0.5$.}
	\label{fig:Ranking_1000_g05_unblinded_comb}
\end{figure}

\FloatBarrier

\section{Summary and Status}
\label{sec:zprime_summary}
A search for $ee$ and $\mu\mu$ resonances in final states with associated $b$-jets has been presented, using the full Run~2 dataset recorded with the ATLAS detector.

The main backgrounds are estimated from MC simulation. The two dominant contributions, $\ttbar$ and DY, are corrected and validated in dedicated control and validation regions, which demonstrate good overall background modelling. Three signal regions are defined by $b$-jet multiplicity, with requirements of $\metsig<5$ and $\minmlb>155\;\text{GeV}$ to suppress the $\ttbar$ background.

Signals are modelled with MC simulation for $Z'$ masses from $500\;\mathrm{GeV}$ to $4\;\mathrm{TeV}$, for two coupling values $g_{Z'}$ of the $Z'$ to leptons and quarks.

A profile-likelihood fit~\cite{Cowan:2010js} is performed and $\mathrm{CL}_s$ exclusion limits~\cite{Read:2002hq,Junk:1999kv} are extracted (see Chapter~\ref{chp:stats}).
The strongest limits are obtained by combining the three signal regions. Exclusion limits are stronger for $g_{Z'}=1.0$ than $g_{Z'}=0.5$, and slightly better in the electron channel than in the muon channel. A combined fit of both channels yields better limits than either channel individually, particularly at high $m_{Z'}$.

This search is the primary effort within an umbrella of exclusive dilepton searches at ATLAS. The search was unblinded in May 2025 and the combined ATLAS publication, incorporating the $Z'\to\ell\ell+b$-jets search presented here alongside complementary analyses --- an updated search for a new neutral vector boson in $Z'\to\ell\ell+\Etmiss$ final states, building on the Run~2 CONF note~\cite{ATLAS-CONF-2023-045}, and a search targeting contact interactions in the $b\bar{b}\ell^+\ell^-$ final state~\cite{Afik:2019htr} --- is currently in preparation. The $Z'\to\ell\ell+b$-jets search presented here remains the primary focus of that effort.

\clearpage
\clearpage\chapter{Search for Periodic Signals in \texorpdfstring{$\bm{ee/\gamma\gamma}$}{ee/gg} Final States at \texorpdfstring{$\bm{\sqrt{s}=13}$}{sqrt(s)=13}~TeV}
\label{chp:clockwork}

\section{Introduction}
\label{sec:cw:intro}

The Clockwork/Linear Dilaton (CW/LD) model~\cite{Giudice:2016yja,Giudice:2017fmj}
predicts a dense, quasi-equally-spaced tower of KK graviton resonances
whose collective contribution to invariant mass spectra produces a distinctive
quasi-periodic signature --- qualitatively different from the isolated narrow
resonances or broad non-resonant deviations targeted by conventional BSM searches.
The model and its motivation as a solution to the hierarchy problem are described in
Section~\ref{sec:bsm:cwld}; the key phenomenological consequence for collider
searches is that the mass splittings between adjacent KK modes are typically a few
percent near the onset mass $k$ and fall below 1\% at high mode
numbers~\cite{Giudice:2017fmj}, so the resonances are not individually resolved but
instead superimpose into a long-range semi-periodic structure in the invariant mass
spectrum.
The squared mass of the $n$-th KK mode follows
\begin{equation}
  m_n^2 \approx k^2 + \frac{n^2}{R^2},
\end{equation}
where $k$ is the mass-gap parameter and $R$ is the size of the extra dimension.
The production cross section scales as $\sigma \propto M_5^{-3}$, where $M_5$ is
the five-dimensional reduced Planck mass, so an exclusion on the signal strength
translates directly into a lower limit on $M_5$ for a given $k$.
The dielectron and diphoton final states are chosen because the excellent energy
and momentum resolution of the ATLAS electromagnetic calorimeter and inner detector
makes them most sensitive to sub-percent-level mass splittings.

Conventional high-mass resonance searches in dilepton and diphoton final states ---
such as those performed by ATLAS~\cite{EXOT-2018-08,HIGG-2018-27} and
CMS~\cite{CMS_dilepton_2018} using Run~2 data --- are designed to detect isolated
peaks and are not optimised for quasi-periodic structures spanning a wide mass range.
No dedicated search for periodic signals in hadronic collision data existed prior to
this analysis.
The present search adapts the ATLAS high-mass dilepton~\cite{EXOT-2018-08} and
diphoton~\cite{HIGG-2018-27} event selections as its starting point and introduces
a novel analysis strategy based on the continuous wavelet transform (CWT) and
neural networks to exploit the periodic nature of the signal.

This chapter describes the search for such periodic structures in the dielectron
and diphoton invariant mass spectra using $139~\textrm{fb}^{-1}$ of $pp$ collision
data recorded by the ATLAS detector at $\sqrt{s} = 13~\TeV$ during Run~2 of the
LHC (2015--2018).
Two complementary neural network strategies are employed: a model-dependent
classifier trained to recognise CW/LD signal scalograms, and a model-independent
autoencoder that flags anomalous periodic structures without assuming a specific
signal model.
No significant deviation from the SM background is observed; 95\% CL exclusion
limits are set in the $k$--$M_5$ parameter plane, with the maximum excluded $M_5$
reaching approximately 10~\TeV\ in the diphoton channel.
These results represent the first direct search for periodic graviton resonances.

\section{Analysis Strategy}
\label{sec:cw:overview}

The search is performed on the dielectron ($ee$) and diphoton ($\gamma\gamma$)
invariant mass spectra in the ranges $225$--$6000$~\GeV\ and $150$--$5000$~\GeV,
respectively, adapting the ATLAS high-mass dilepton~\cite{EXOT-2018-08} and
diphoton~\cite{HIGG-2018-27} resonance searches.

Two complementary NN-based analysis strategies are employed:
\begin{itemize}
  \item \textbf{Model-dependent:} A convolutional neural network
    (CNN) trained to distinguish signal from background scalograms for specific
    CW/LD parameter values $(k, M_5)$.
    The classifier output score is used as a test statistic to set 95\% CL
    exclusion limits in the $k$--$M_5$ plane.
  \item \textbf{Model-independent:} An autoencoder (AE) NN trained
    on background-only scalograms.
    Anomalous (signal-like) events produce a large mean-squared-error (MSE) loss
    when reconstructed, providing a model-agnostic search for any periodic
    deviation.
\end{itemize}

The statistical inference procedure follows the framework described in
Chapter~\ref{chp:stats}.
For the model-dependent search, the classifier score is the test statistic and
CLs limits are derived from ensembles of pseudo-experiments incorporating
statistical and systematic uncertainties.
For the model-independent search, the AE loss $L_\text{MSE}$ plays the role of
the test statistic and local $p$-values are computed over the two-dimensional
scalogram.

\section{Event Selection}
\label{sec:cw:evsel}

Both search channels follow the object reconstruction and identification procedures
described in Chapter~\ref{chp:objects}, adapting the event selections from the ATLAS
high-mass dilepton~\cite{EXOT-2018-08} and diphoton~\cite{HIGG-2018-27} resonance searches.

\subsection*{Dielectron channel}

Events are required to contain exactly two signal electrons with $\ET > 65~\GeV$,
satisfying the \texttt{Medium} likelihood identification working point~\cite{EGAM-2018-01}
and the \texttt{FCTight} isolation requirement, within $|\eta| < 2.47$ (excluding the
calorimeter transition region $1.37 < |\eta| < 1.52$); see Section~\ref{sec:reco:electrons}.
The two electrons must form an opposite-sign pair.
The dielectron invariant mass spectrum is analysed in the range $m_{ee} > 225~\GeV$.

\subsection*{Diphoton channel}

Events are required to contain at least two photons satisfying the \texttt{Tight}
identification and \texttt{FixedCutTight} isolation working points~\cite{EGAM-2018-01},
within $|\eta| < 2.37$ (excluding $1.37 < |\eta| < 1.52$); see Section~\ref{sec:reco:photons}.
The leading (subleading) photon must satisfy $\ET/m_{\gamma\gamma} > 0.35$ ($>0.25$).
The diphoton invariant mass spectrum is analysed in the range $m_{\gamma\gamma} > 150~\GeV$.

\section{Continuous Wavelet Transform Technique}
\label{sec:cw:cwt}

The key analysis tool is the continuous wavelet transform (CWT), which converts
the one-dimensional invariant mass spectrum into a two-dimensional
mass--frequency representation called a \emph{scalogram}.
The CWT of a signal $f(\beta)$ is defined as
\begin{equation}
  W(\alpha, \beta) = \frac{1}{\sqrt{\alpha}}
    \int_{-\infty}^{\infty} f(\beta') \,
    \psi^*\!\left(\frac{\beta' - \beta}{\alpha}\right) d\beta',
  \label{eq:CWTdef}
\end{equation}
where $\psi$ is a mother wavelet, $\alpha > 0$ is the scale parameter
(inversely proportional to frequency), and $\beta$ plays the role of position
(here the invariant mass)~\cite{Beauchesne:2019tpx,Farina:2018fyg}.
The complex Morlet wavelet is chosen as the mother wavelet.

Periodic structures in the mass spectrum at a characteristic frequency manifest
as localised enhancements in $|W(\alpha, \beta)|$ at the corresponding scale $\alpha$.
A pure smooth background produces a nearly featureless scalogram,
whereas a CW/LD signal --- with its quasi-periodic KK graviton spectrum --- imprints a
localised ``island'' at the mass and scale corresponding to the onset and spacing of
the KK graviton tower.

\begin{figure}[htb!]
  \centering
  \subfloat[]{\includegraphics[width=0.48\textwidth]{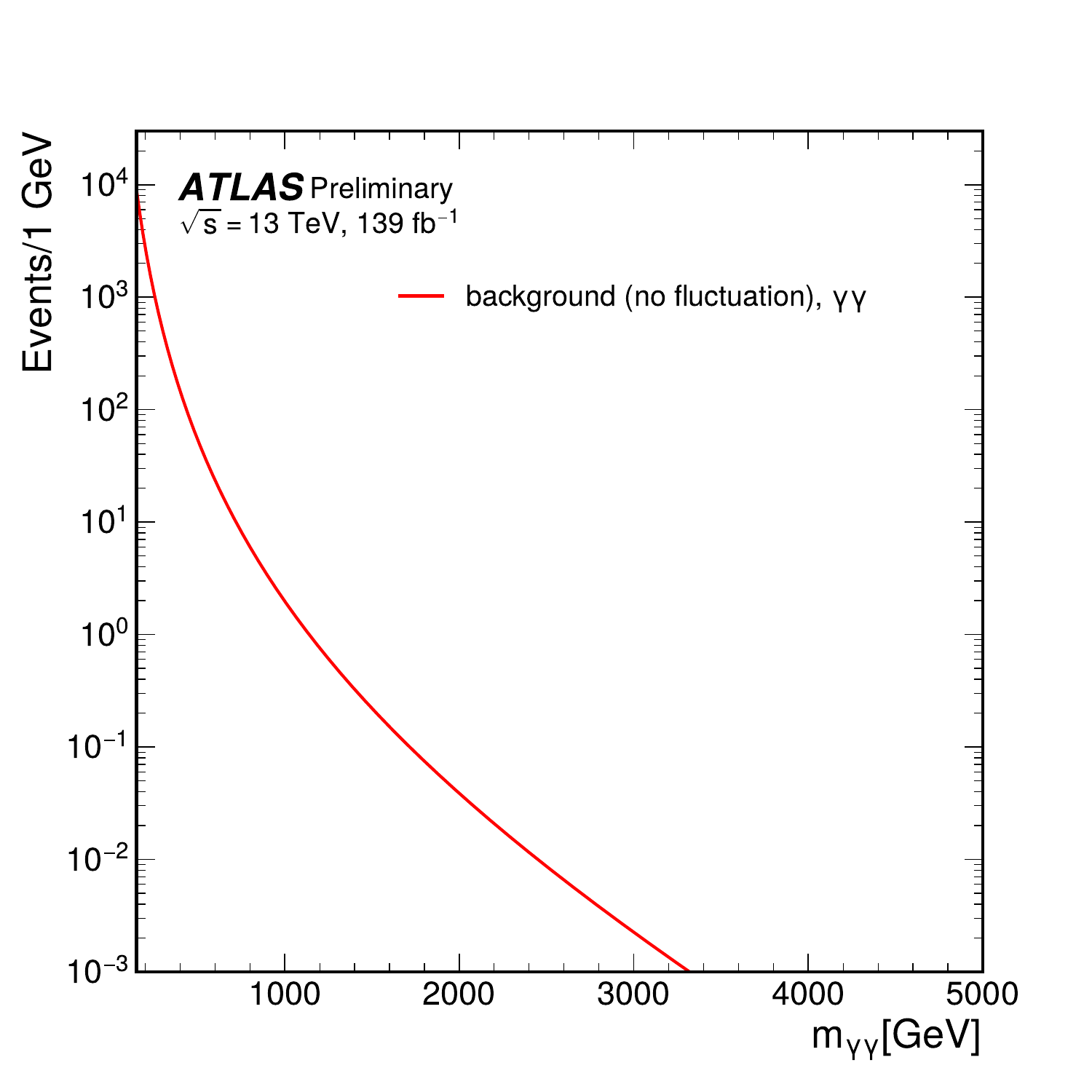}}
  \subfloat[]{\includegraphics[width=0.48\textwidth]{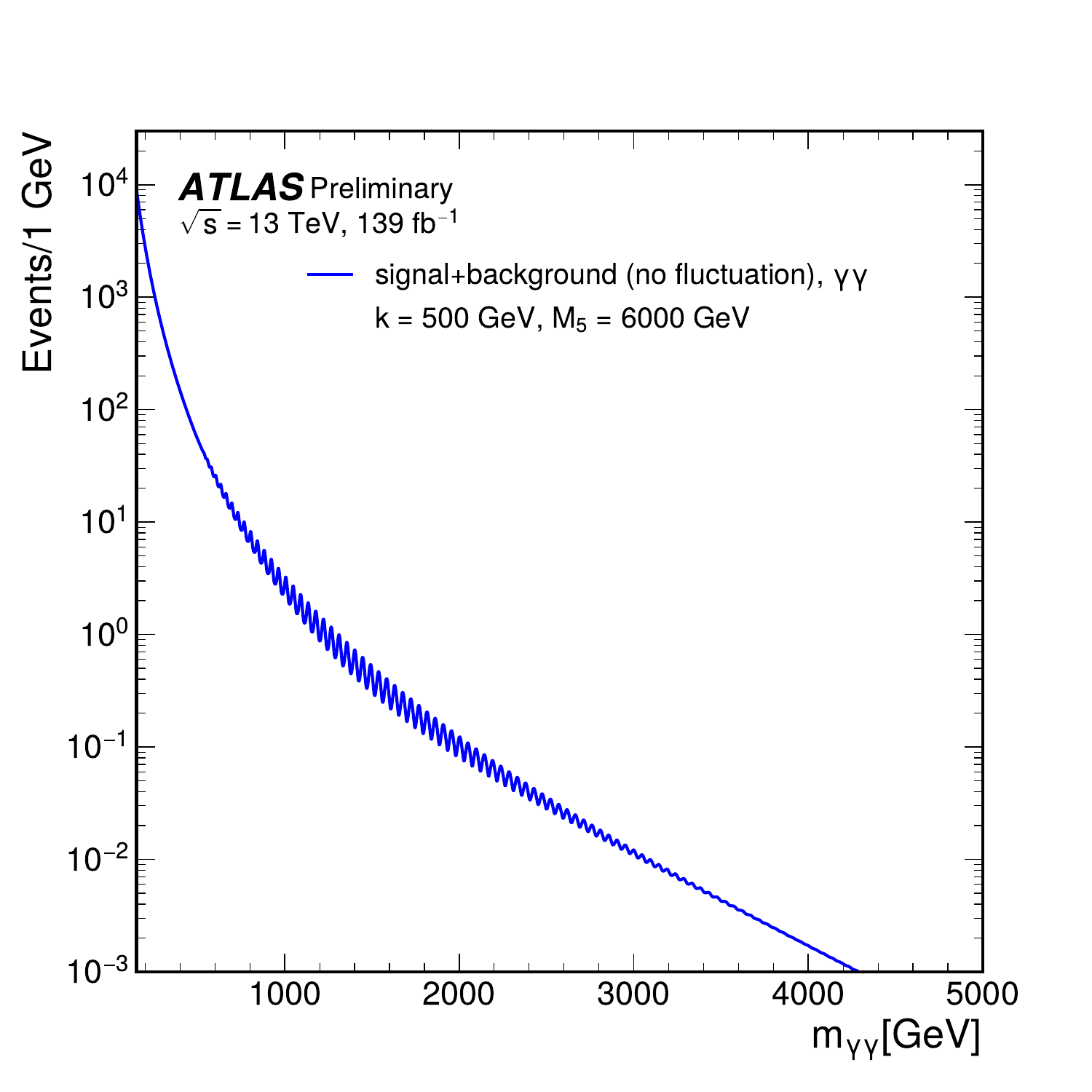}} \\
  \vspace{-0.3cm}
  \subfloat[]{\includegraphics[width=0.48\textwidth]{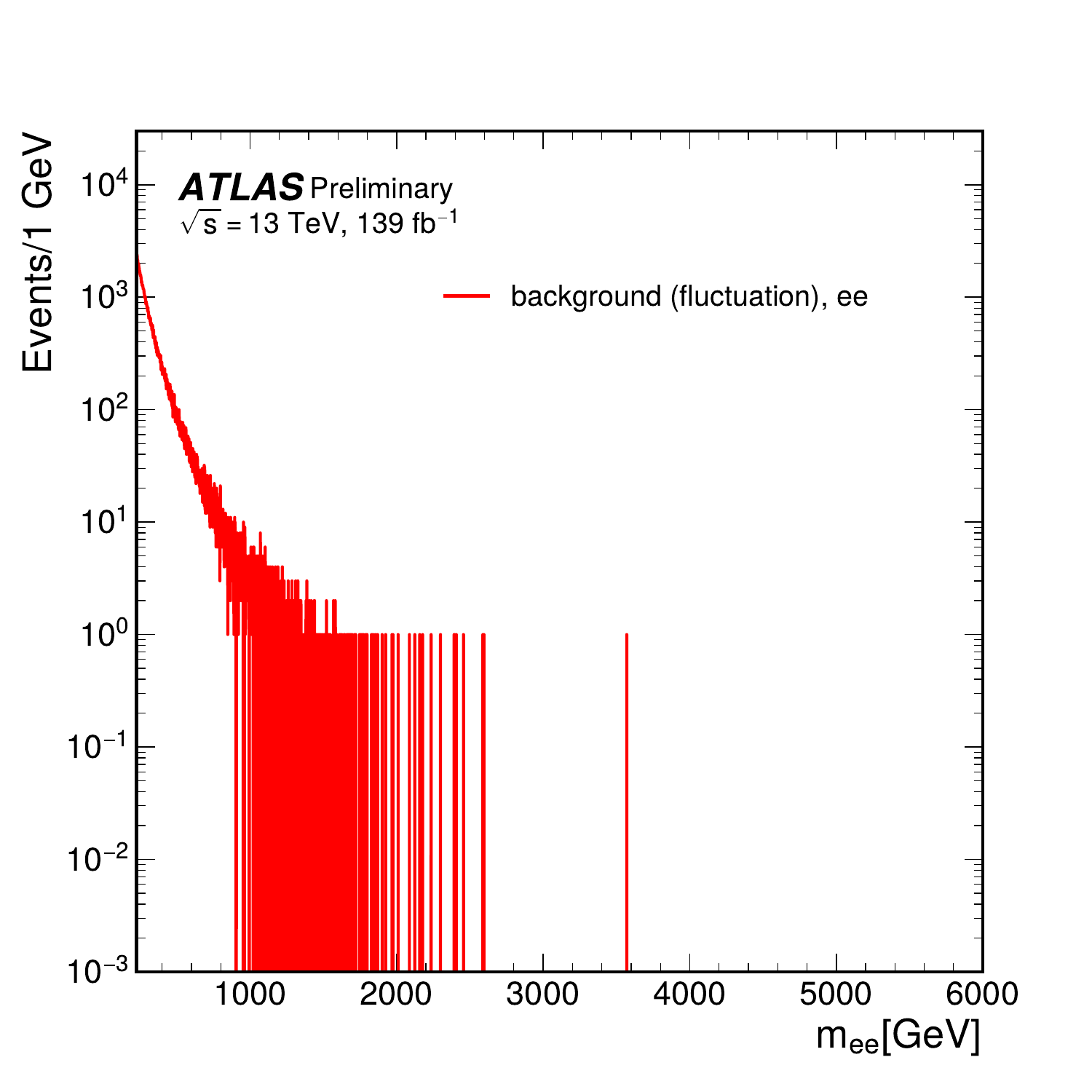}}
  \subfloat[]{\includegraphics[width=0.48\textwidth]{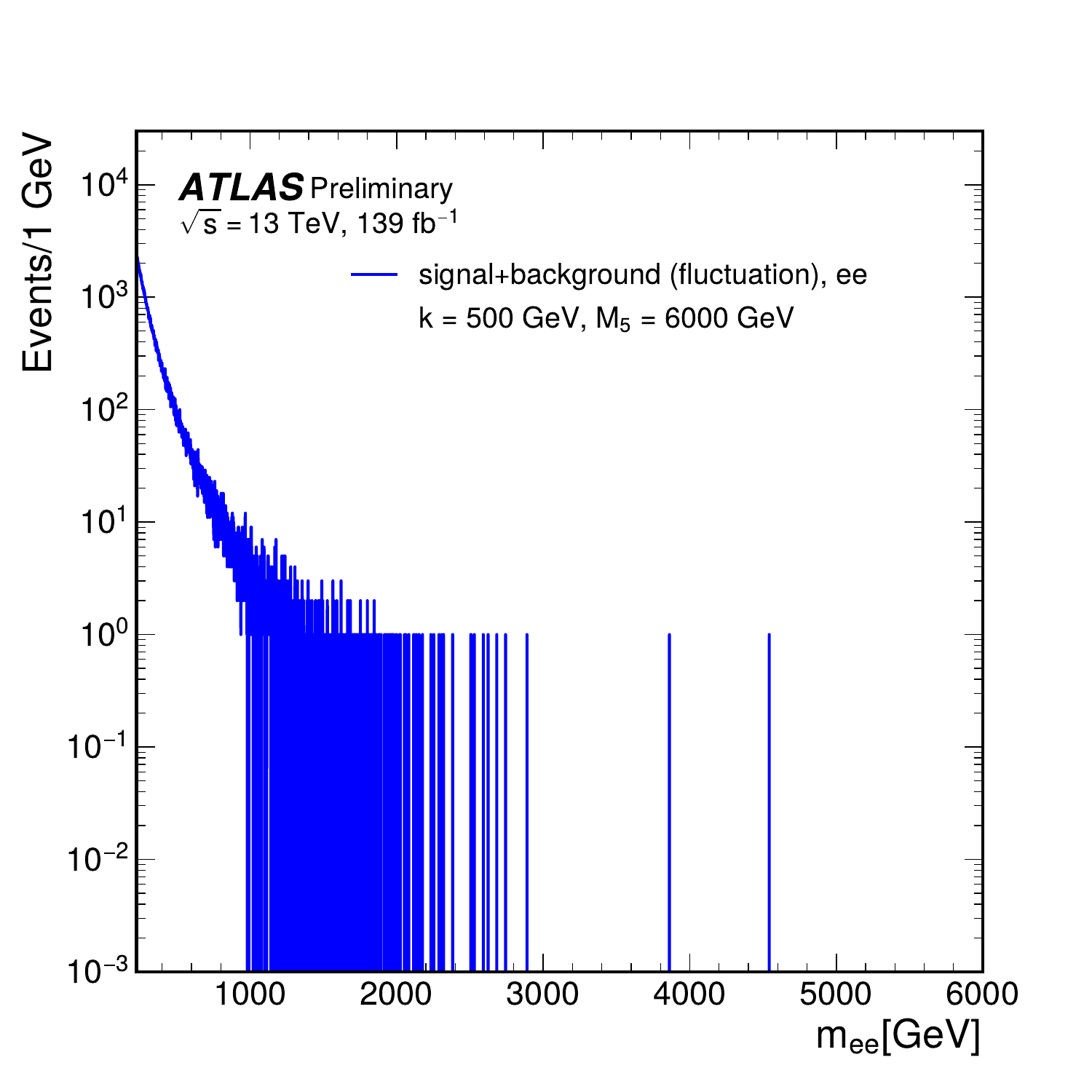}}
  \caption{One-dimensional dielectron invariant mass spectrum illustrating the signal
    topology before and after statistical smearing.
    Top row: analytical (pre-smearing) level, showing (a)~background-only and
    (b)~background plus an injected CW/LD signal ($k = 500$~\GeV, $M_5 = 6000$~\GeV).
    The quasi-periodic modulations from the dense KK graviton tower are clearly
    visible as a series of equally-spaced bumps on the falling continuum.
    Bottom row: the same spectra after Poisson fluctuations are applied, showing
    (c)~background-only and (d)~background plus signal.
    In (d) the periodic signal is present but largely buried under statistical noise,
    illustrating the challenge that the wavelet-based analysis is designed to overcome.}
  \label{fig:cw:1d_spectra}
\end{figure}

FIG.~\ref{fig:cw:1d_spectra} illustrates this directly in one dimension.
The top row shows the dielectron invariant mass spectrum at the analytical level,
before any statistical fluctuations are applied: panel~(a) shows the smooth,
steeply falling SM background, and panel~(b) overlays a CW/LD signal
($k = 500$~\GeV, $M_5 = 6000$~\GeV), revealing the characteristic quasi-periodic
oscillations --- a series of bumps with roughly equal fractional spacing in mass ---
superimposed on the continuum.
The bottom row shows the same spectra after Poisson statistical fluctuations are
applied, emulating the finite statistics of a real dataset.
The periodic signal structure in panel~(d) persists but is substantially obscured
by statistical noise, making a direct visual identification in the 1D spectrum
impractical and motivating the use of the CWT to extract the underlying frequency
content.

The invariant mass spectra are binned in 1~\GeV\ intervals and the CWT is applied;
the resulting scalograms are subsequently rebinned to a coarser grid before being
fed to the neural networks.
To suppress sensitivity to non-resonant low-frequency tails that could mimic a
periodic signal, a mass-threshold procedure is optionally applied: scalogram bins
are zeroed unless the local signal-to-background ratio $S_i/\sigma_{B_i}$ exceeds
50\% of the peak value across the mass range (\emph{$R < 50\%$} scenario) or the
peak scale amplitude (\emph{scale threshold}).

\begin{figure}[htb!]
  \centering
  \subfloat[]{\includegraphics[width=0.48\textwidth]{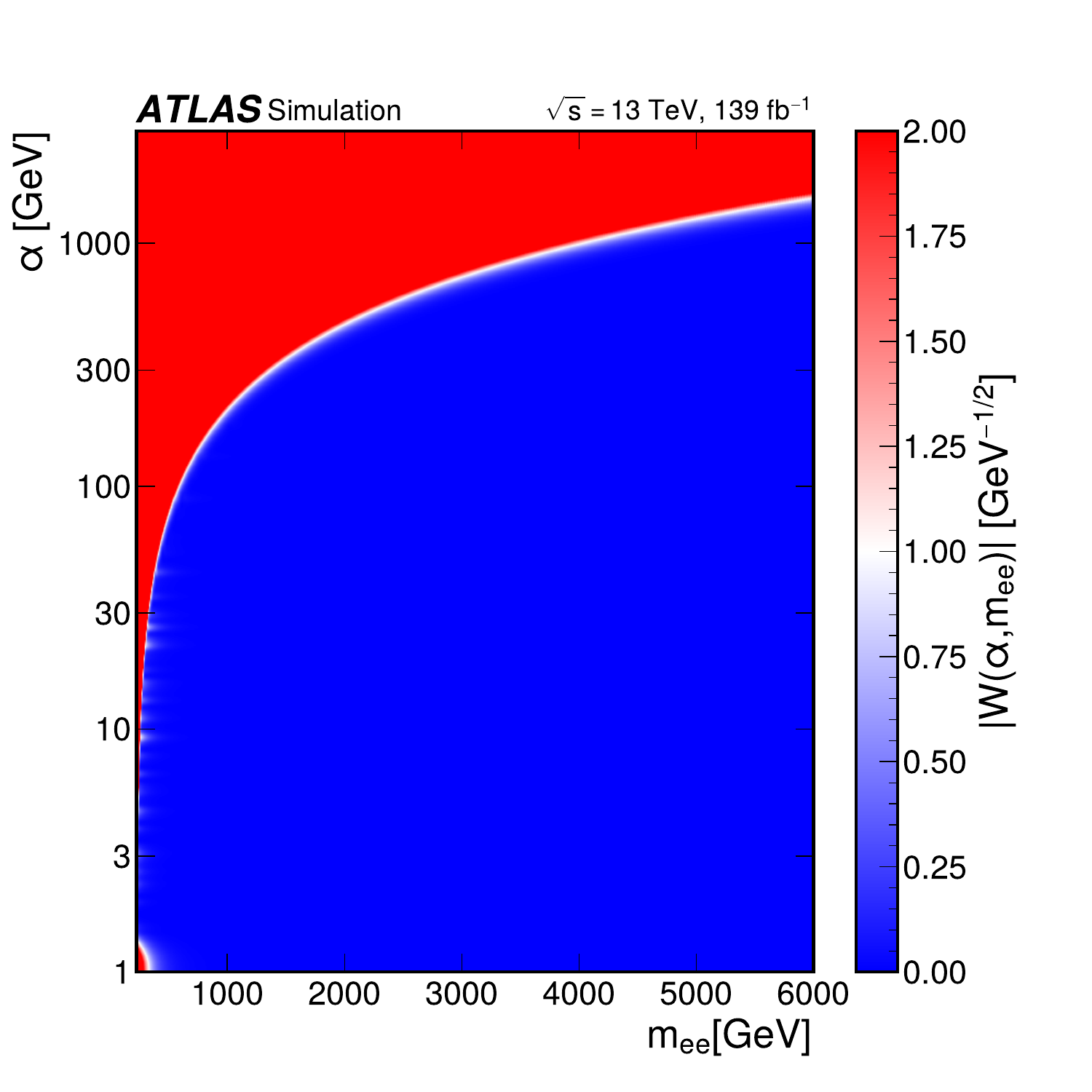}}
  \subfloat[]{\includegraphics[width=0.48\textwidth]{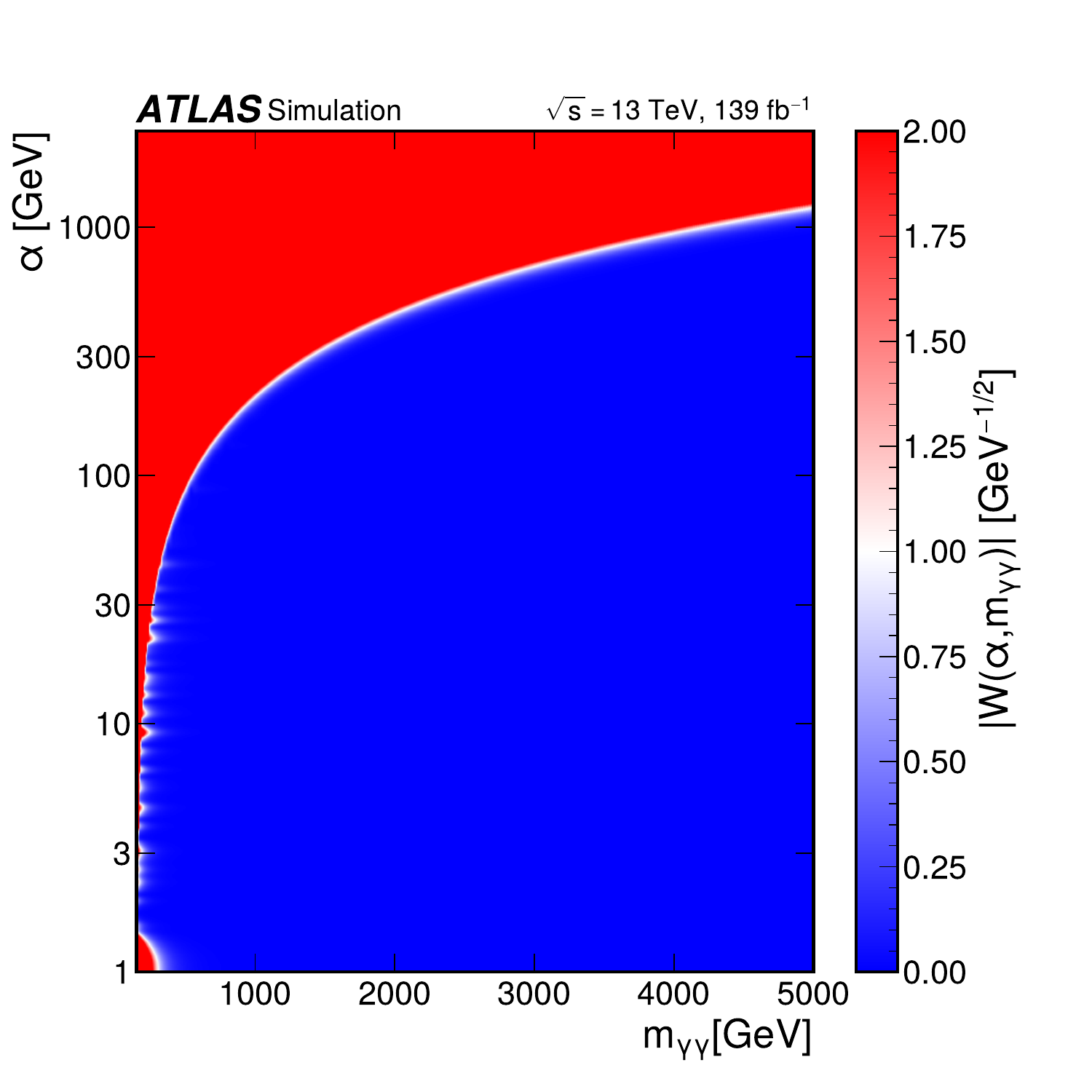}} \\
  \vspace{-0.3cm}
  \subfloat[]{\includegraphics[width=0.48\textwidth]{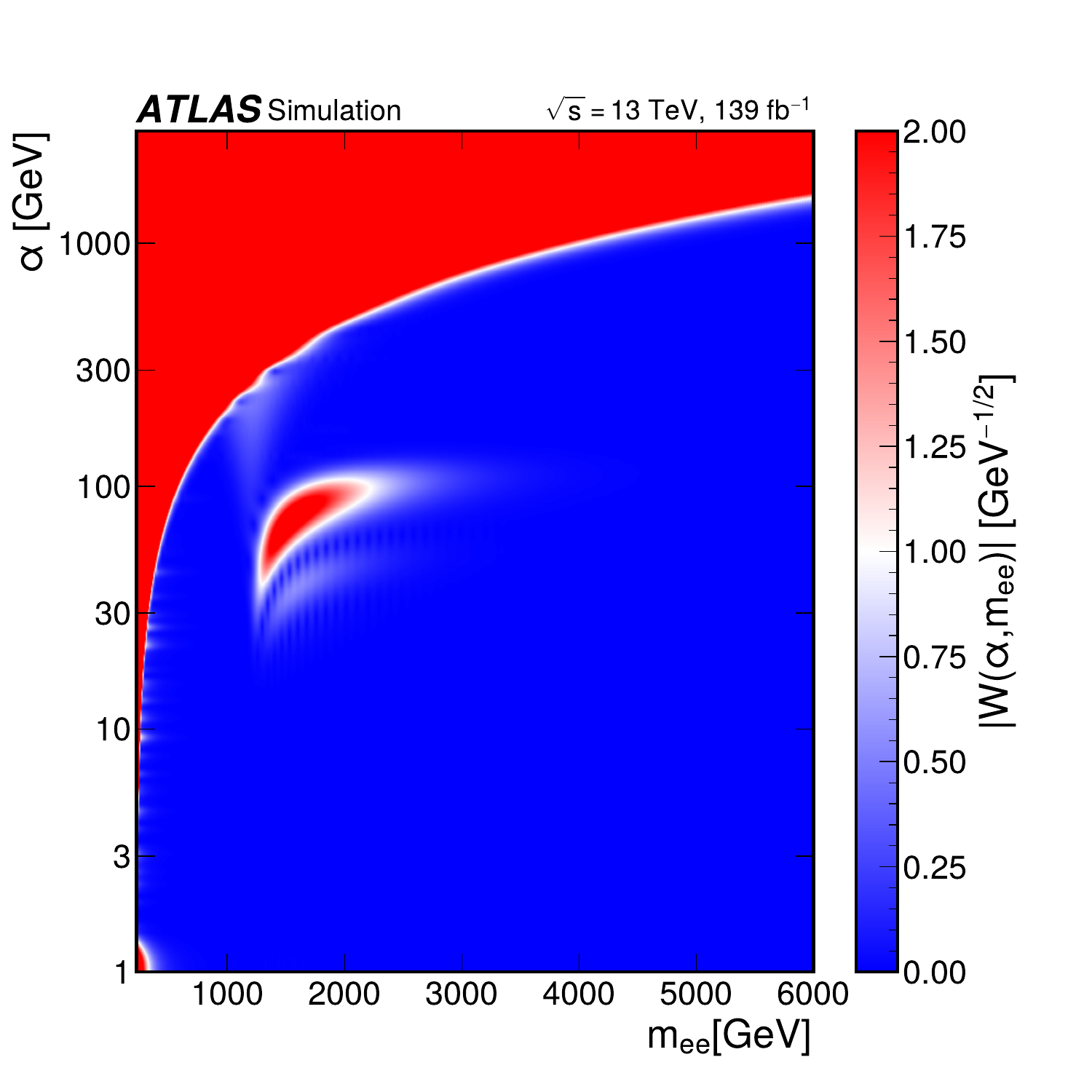}}
  \subfloat[]{\includegraphics[width=0.48\textwidth]{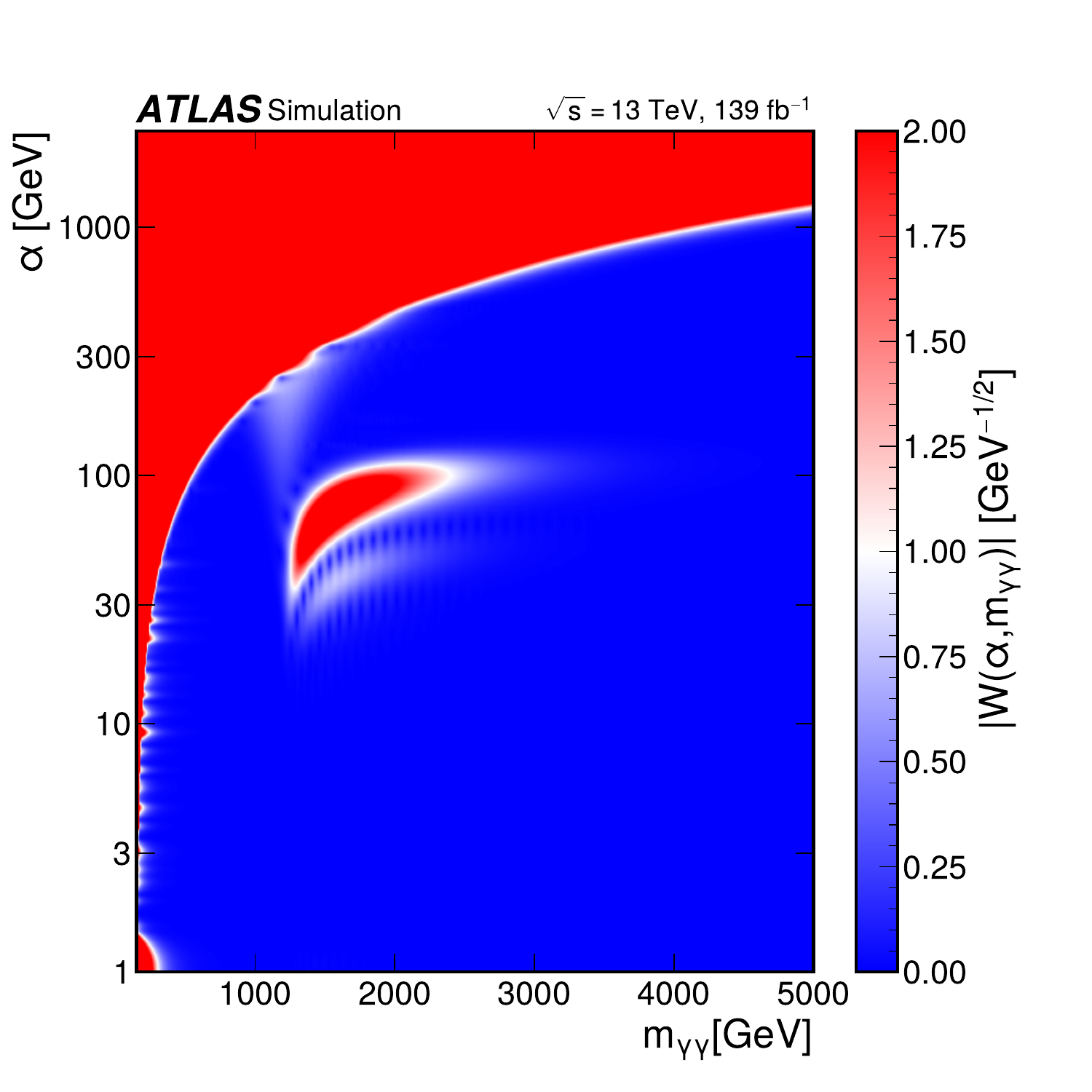}}
  \caption{Scalogram output of the CWT for analytical (pre-smearing) toy experiments
    in the dielectron and diphoton channels.
    Top row: background-only scalograms for the (a)~dielectron and (b)~diphoton
    channels.
    The smooth background produces no localised periodic feature; the saturation at
    large scales reflects the non-flat shape of the falling background distribution.
    Bottom row: signal-plus-background scalograms for the (c)~dielectron and
    (d)~diphoton channels, for $k = 1200$~\GeV and $M_5 = 3000$~\GeV.
    The CW/LD signal manifests as a localised island in mass and scale, clearly
    discernible from the featureless background.
    Here $\alpha$ is the CWT scale parameter and $W(\alpha,\beta)$ are the wavelet
    coefficients}
  \label{fig:cw:scalogram_analytic}
\end{figure}

The discriminating power of the CWT is most apparent in the two-dimensional
scalogram.
FIG.~\ref{fig:cw:scalogram_analytic} shows four scalograms computed from
analytical (pre-smearing) toy spectra for both search channels.
The top row shows background-only scalograms for the dielectron and diphoton
channels: the smooth falling distribution produces a saturated region at large
scales (low frequencies) due to the non-flat background, but no localised periodic
feature.
The bottom row adds a CW/LD signal ($k = 1200$~\GeV, $M_5 = 3000$~\GeV):
a clear localised island emerges at the mass where the KK tower turns on
($\beta \approx k$) and at the scale $\alpha$ corresponding to the graviton mass
splittings.
Fixing $k$ and decreasing $M_5$ increases the island prominence (larger cross
section); fixing $M_5$ and increasing $k$ shifts the island to higher masses and
larger scales.
This island is the discriminating feature exploited by the neural network
classifiers described in Section~\ref{sec:cw:nn_opt}.

\begin{figure}[htb!]
  \centering
  \subfloat[]{\includegraphics[width=0.49\textwidth]{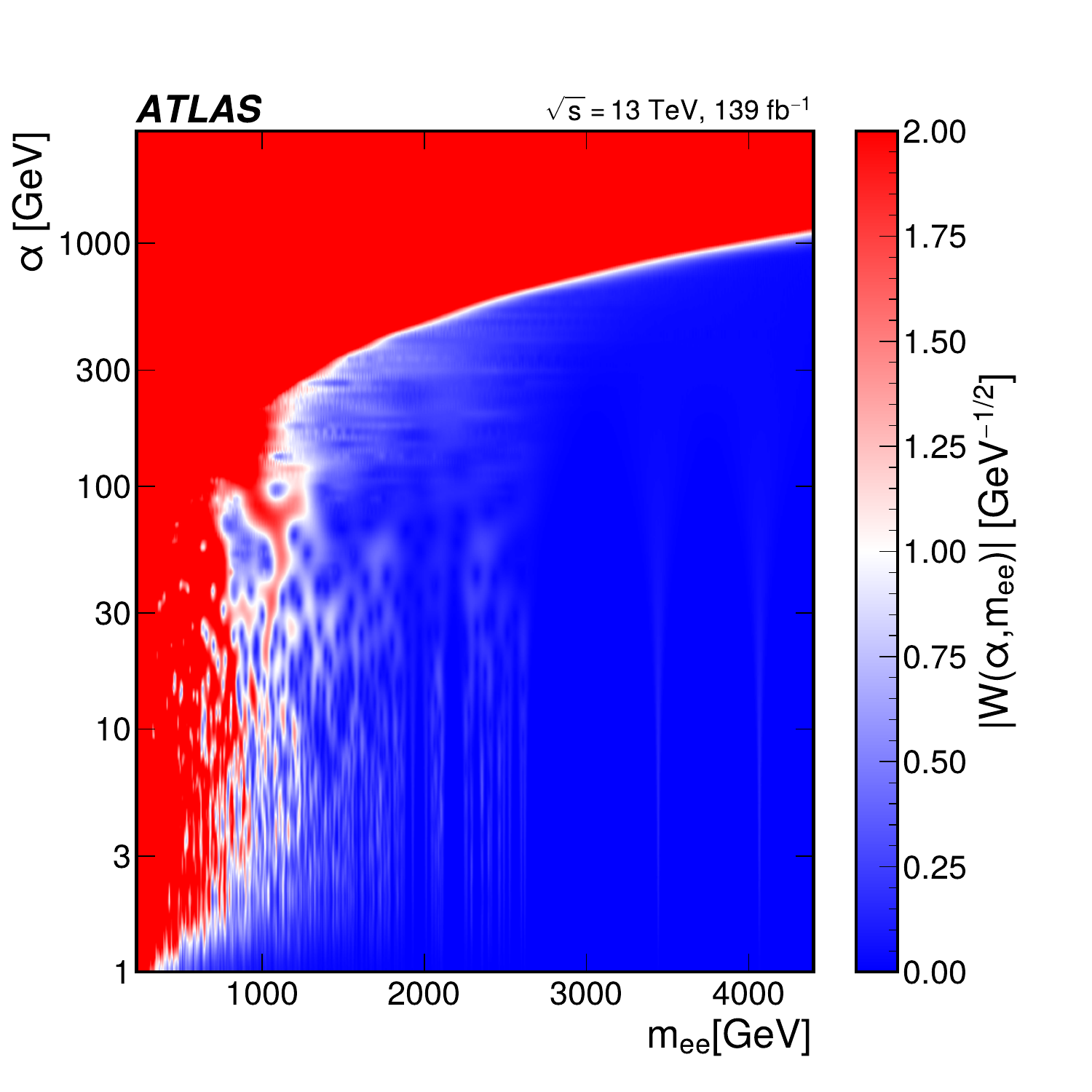}}
  \subfloat[]{\includegraphics[width=0.49\textwidth]{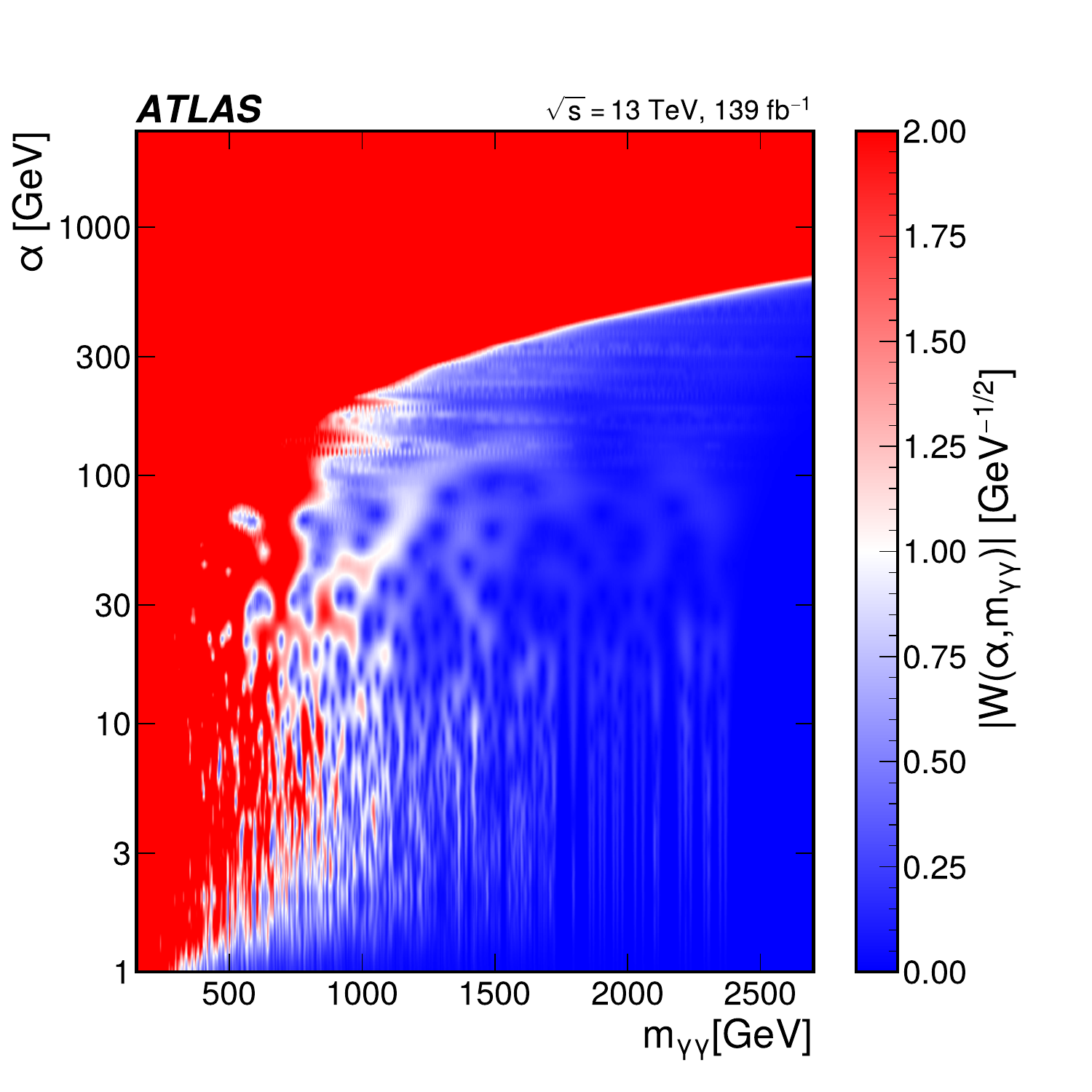}}
  \caption{Scalogram output of the CWT for the observed data in the
    (a)~dielectron and (b)~diphoton channels.
    Here $\alpha$ is the CWT scale parameter and $W(\alpha,\beta)$ are the wavelet
    coefficients of Eq.~(\ref{eq:CWTdef}).
    No localised island characteristic of a CW/LD signal is observed; the
    scalograms are consistent with the background-only expectation.}
  \label{fig:finalScalograms}
\end{figure}

FIG.~\ref{fig:finalScalograms} shows the scalograms of the observed data for both
search channels, after the full event selection and statistical smearing from
finite data.
No localised periodic island is observed in either channel; both scalograms are
consistent with the SM background expectation.

\section{Neural Network Optimisation}
\label{sec:cw:nn_opt}

Both the classifier NN and the autoencoder NN are optimised using a
systematic hyperparameter grid search performed with the \textsc{Talos}
framework~\cite{talos}.
For each configuration in the grid, the network is trained and the
configuration with the lowest validation loss is retained.

\subsection{Classifier NN}
\label{ssec:cw:classifier_opt}

The classifier NN is trained to discriminate signal-plus-background scalograms
from background-only scalograms using binary cross-entropy (BCE) loss.
Training uses 5000 dielectron and diphoton samples (80/20 train/validation
split).
Weak signal points are chosen for training -- specifically, signals where
$\sum_{i} S_i / \sum_{i} \sigma_{B_i} \sim \mathcal{O}(10^0)$ -- so that a
network able to separate a weak signal will naturally perform at least as well on
stronger signals.
This corresponds to the points
\begin{align}
  (k,\, M_5) &= (1479,\; 8172)~\text{GeV} \quad | \quad ee~\text{channel}, \notag\\
  (k,\, M_5) &= (1448,\; 9068)~\text{GeV} \quad | \quad \gamma\gamma~\text{channel}.
  \label{eq:talos_signals}
\end{align}

The hyperparameter search grid is summarised in Table~\ref{tab:cw:classifier_grid}.

\begin{table}[htb!]
  \centering
  \caption{Classifier NN hyperparameter optimisation grid.
    Values marked in \textbf{bold} indicate the nominal configuration.}
  \label{tab:cw:classifier_grid}
  \begin{tabular}{ll}
    \hhline{==}
    Hyperparameter & Values \\
    \hline
    Conv.\ neurons (layer A)   & \textbf{4}, 8, 16 \\
    Conv.\ neurons (layer B)   & \textbf{8}, 16, 32 \\
    Conv.\ neurons (layer C)   & \textbf{16}, 32, 64 \\
    Conv.\ activation          & \textbf{\texttt{elu}}, \texttt{relu} \\
    Output activation          & \textbf{\texttt{sigmoid}} \\
    Dense layers               & \textbf{2}, 3, 4 \\
    Dense neurons (outer)      & \textbf{100}, 200 \\
    Dense neurons (inner)      & \textbf{200}, 400 \\
    Batch size                 & \textbf{1000} \\
    Optimiser                  & \textbf{\texttt{adam}}, \texttt{nadam} \\
    Loss                       & \textbf{BCE} \\
    Epochs                     & \textbf{200} ($ee$) / \textbf{500} ($\gamma\gamma$) \\
    \hhline{==}
  \end{tabular}
\end{table}

\begin{figure}[htb!]
  \centering
  \subfloat[]{\includegraphics[width=0.49\textwidth]{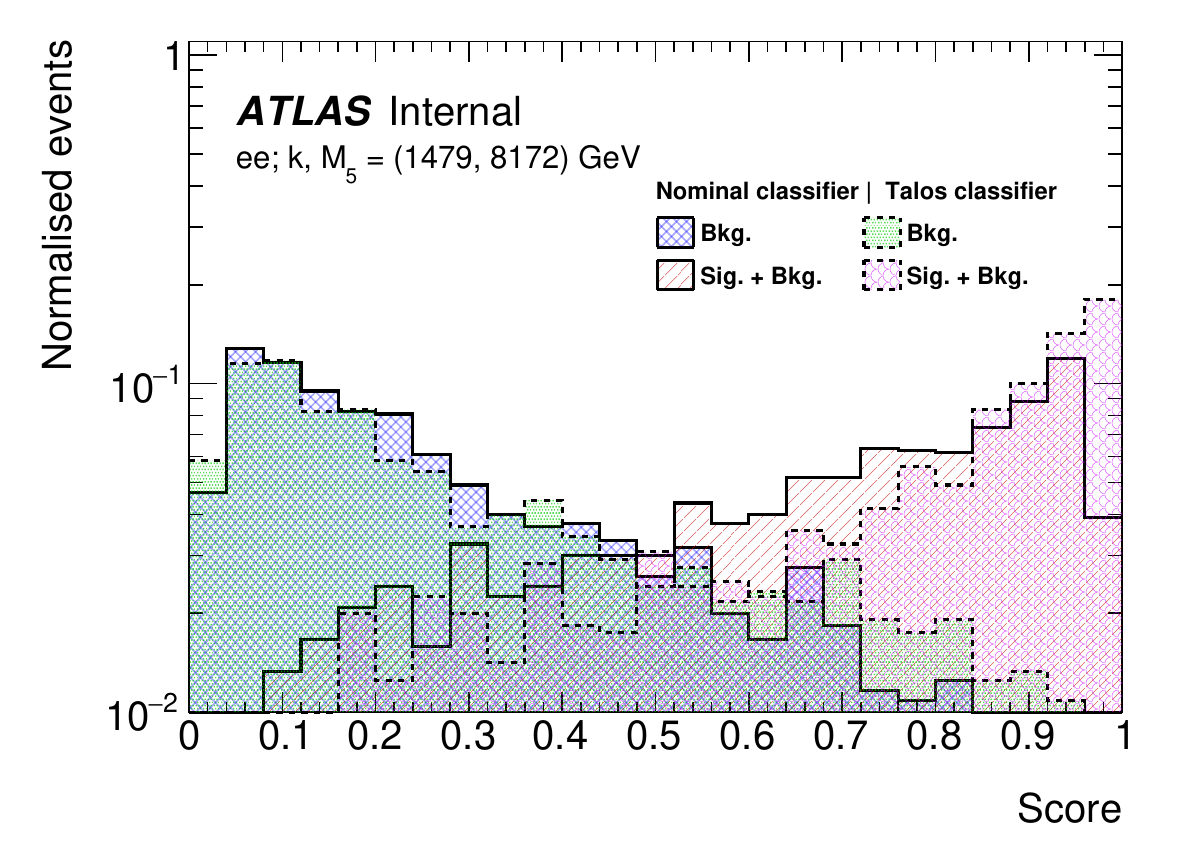}}
  \subfloat[]{\includegraphics[width=0.49\textwidth]{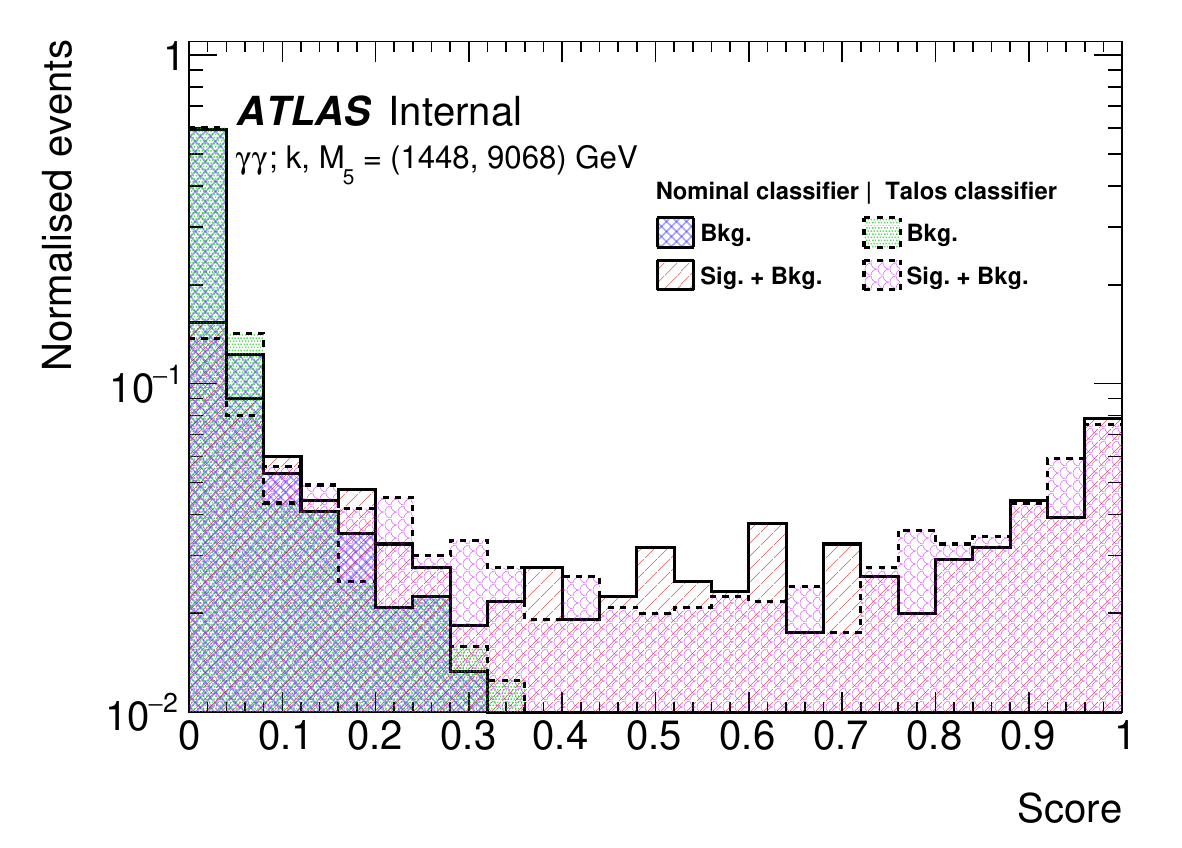}} \\
  \subfloat[]{\includegraphics[width=0.49\textwidth]{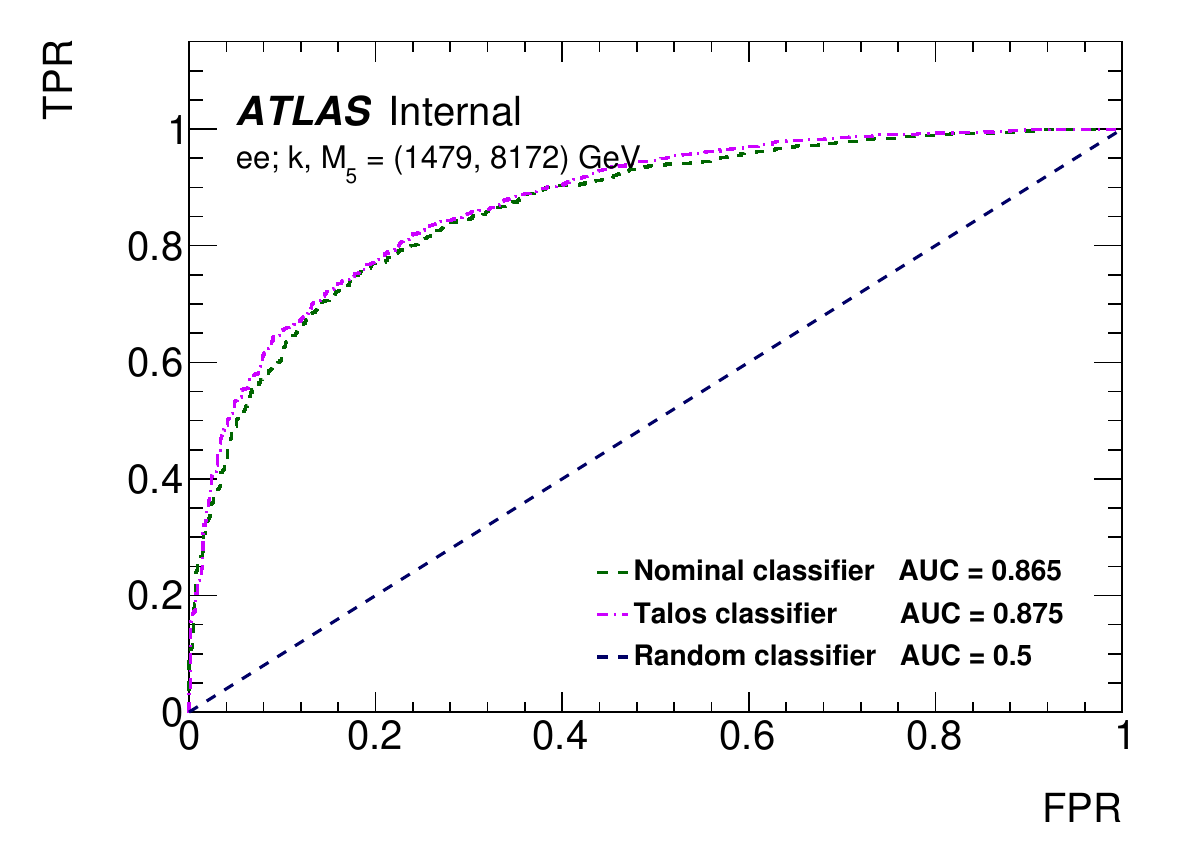}}
  \subfloat[]{\includegraphics[width=0.49\textwidth]{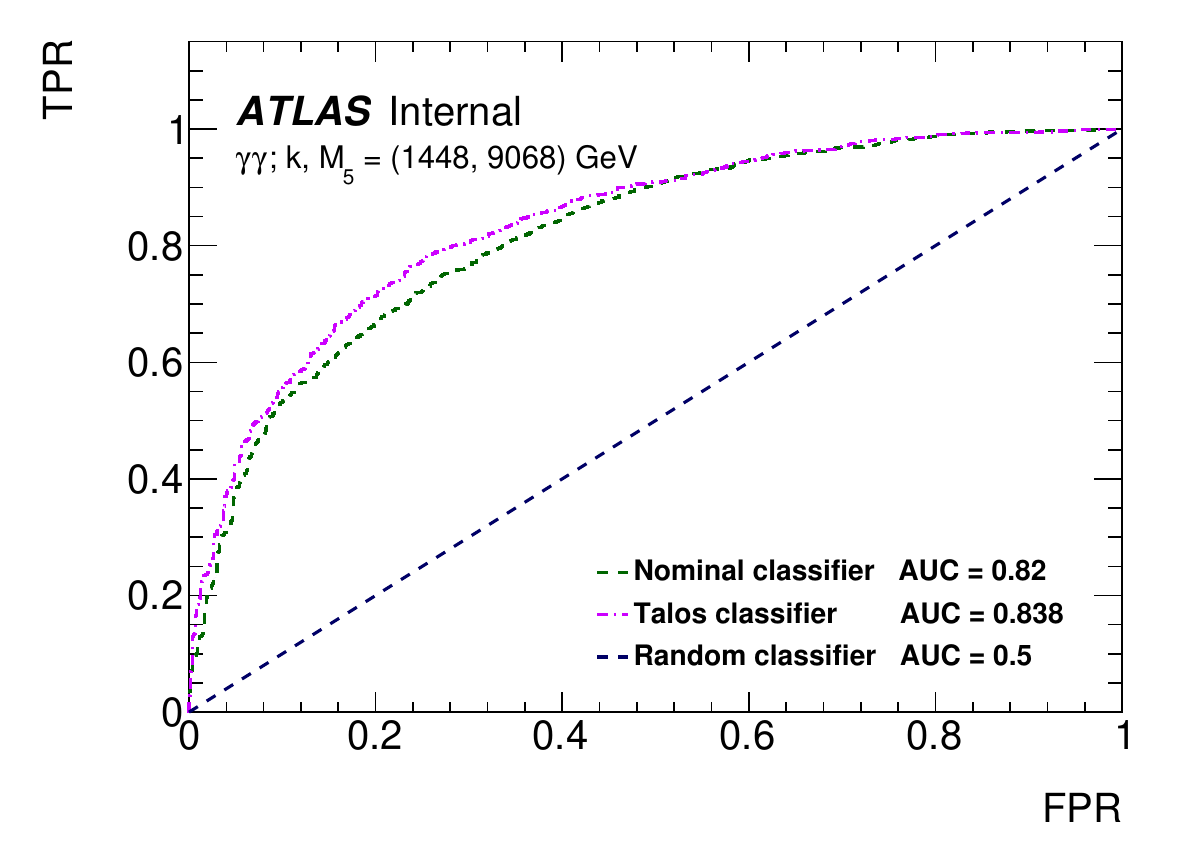}}
  \caption{Classifier NN performance for the (left)~dielectron and (right)~diphoton
    channels.
    (Top) score distributions for background-only and signal-plus-background
    samples for the nominal and grid-optimised networks.
    (Bottom) ROC curves with AUC values.}
  \label{fig:cw:roc_classifier}
\end{figure}

\begin{figure}[htb!]
  \centering
  \subfloat[]{\includegraphics[width=0.49\textwidth]{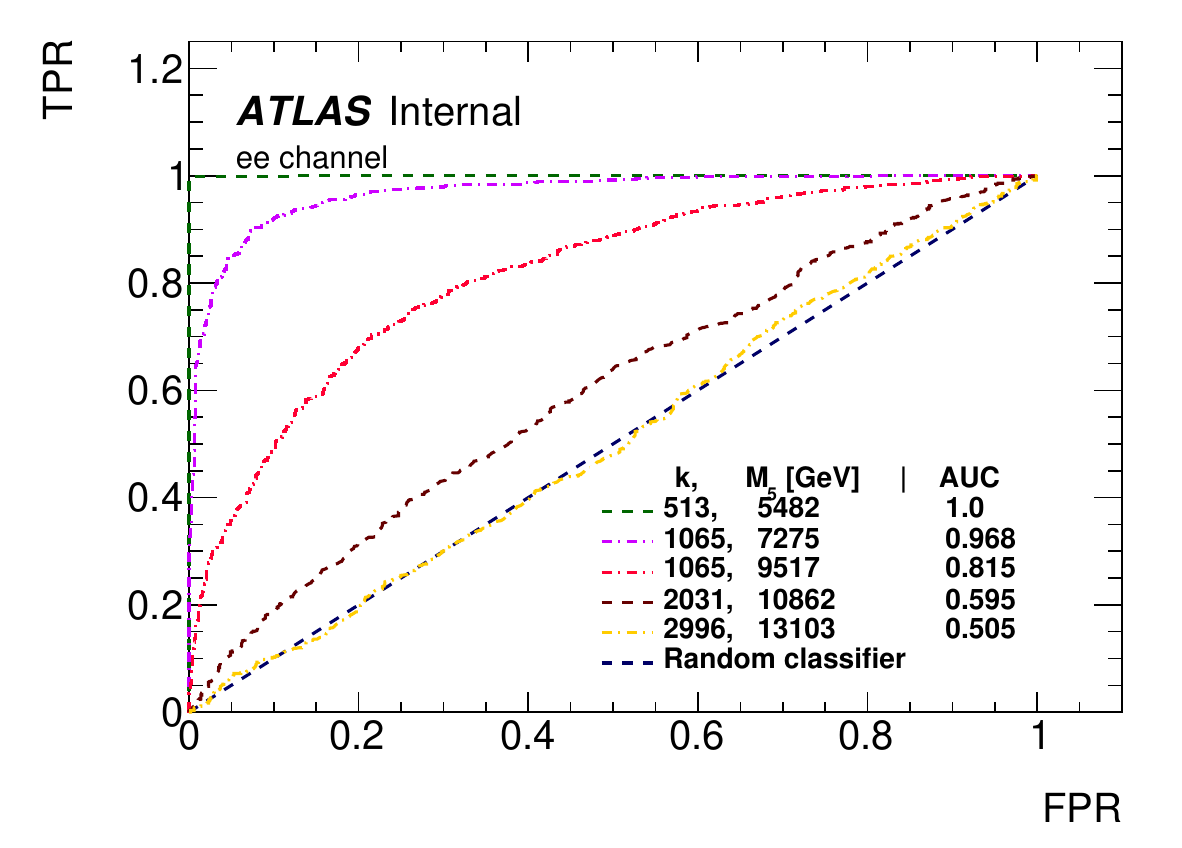}}
  \subfloat[]{\includegraphics[width=0.49\textwidth]{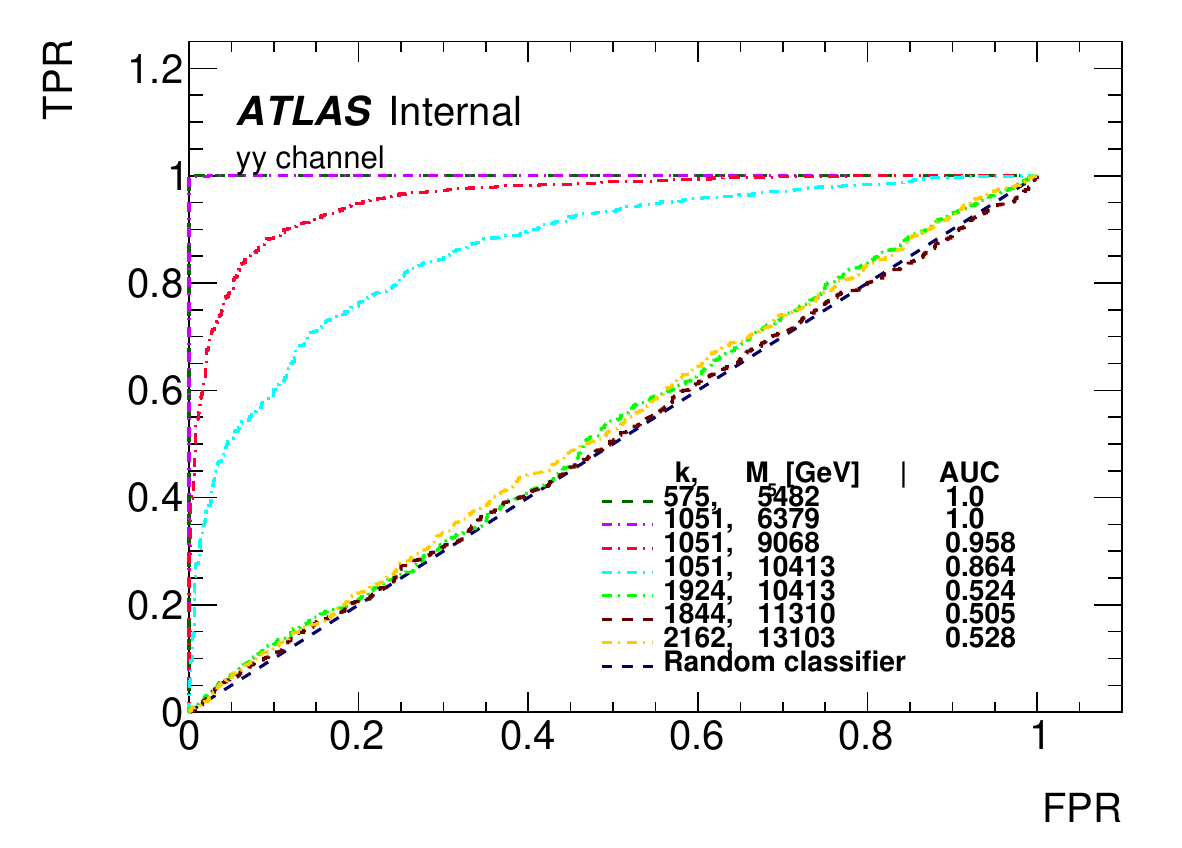}}
  \caption{ROC curves for the nominal classifier NN evaluated on a range of
    CW/LD signal points $(k, M_5)$ in the (a)~dielectron and
    (b)~diphoton channels.
    AUC values are shown for each signal model.}
  \label{fig:cw:roc_multi}
\end{figure}

The Receiver Operating Characteristic (ROC) curves and score distributions for the nominal and grid-optimised
classifier NNs are shown in FIG.~\ref{fig:cw:roc_classifier}.
The grid search yields only marginal improvements of $1.1\%$ and $2.2\%$ in Area Under Curve (AUC) for the dielectron and diphoton channels respectively, confirming that the
nominal configuration is near-optimal.
FIG.~\ref{fig:cw:roc_multi} demonstrates robust classification performance
across a range of $(k, M_5)$ values, not just at the training points.

\subsection{Autoencoder NN}
\label{ssec:cw:ae_opt}

The AE is trained exclusively on background-only scalograms, learning to
reconstruct the smooth SM background.
Anomalous scalograms (containing a periodic signal) incur a higher reconstruction
loss, providing an unsupervised anomaly-detection probe.
The MSE loss over all scalogram bins is used as the test statistic:
\begin{equation}
  L_\text{MSE} = \frac{1}{N_\text{bins}} \sum_{\text{bins}} \left(
    W_\text{input} - W_\text{predicted} \right)^2.
\end{equation}

Training uses 5000 background-only samples (80/20 split) and the loss is
evaluated as MSE over training and validation datasets.
The AE hyperparameter grid is given in Table~\ref{tab:cw:ae_grid}.

\begin{table}[htb!]
  \centering
  \caption{Autoencoder NN hyperparameter optimisation grid.
    Values marked in \textbf{bold} indicate the nominal configuration.}
  \label{tab:cw:ae_grid}
  \begin{tabular}{ll}
    \hhline{==}
    Hyperparameter & Values \\
    \hline
    Encoder neurons (layer A)  & \textbf{128}, 256 \\
    Encoder neurons (layer B)  & \textbf{128}, 256 \\
    Encoder neurons (layer C)  & \textbf{128}, 512 \\
    Latent space (inner)       & \textbf{20}, 30 \\
    Latent space (outer)       & \textbf{40}, 60 \\
    Activation                 & \textbf{\texttt{elu}}, \texttt{relu} \\
    Optimiser                  & \textbf{\texttt{adam}}, \texttt{nadam} \\
    Loss                       & \textbf{MSE} \\
    Batch size                 & \textbf{1000} \\
    Epochs                     & \textbf{25} \\
    \hhline{==}
  \end{tabular}
\end{table}

\begin{figure}[htb!]
  \centering
  \subfloat[]{\includegraphics[width=0.49\textwidth]{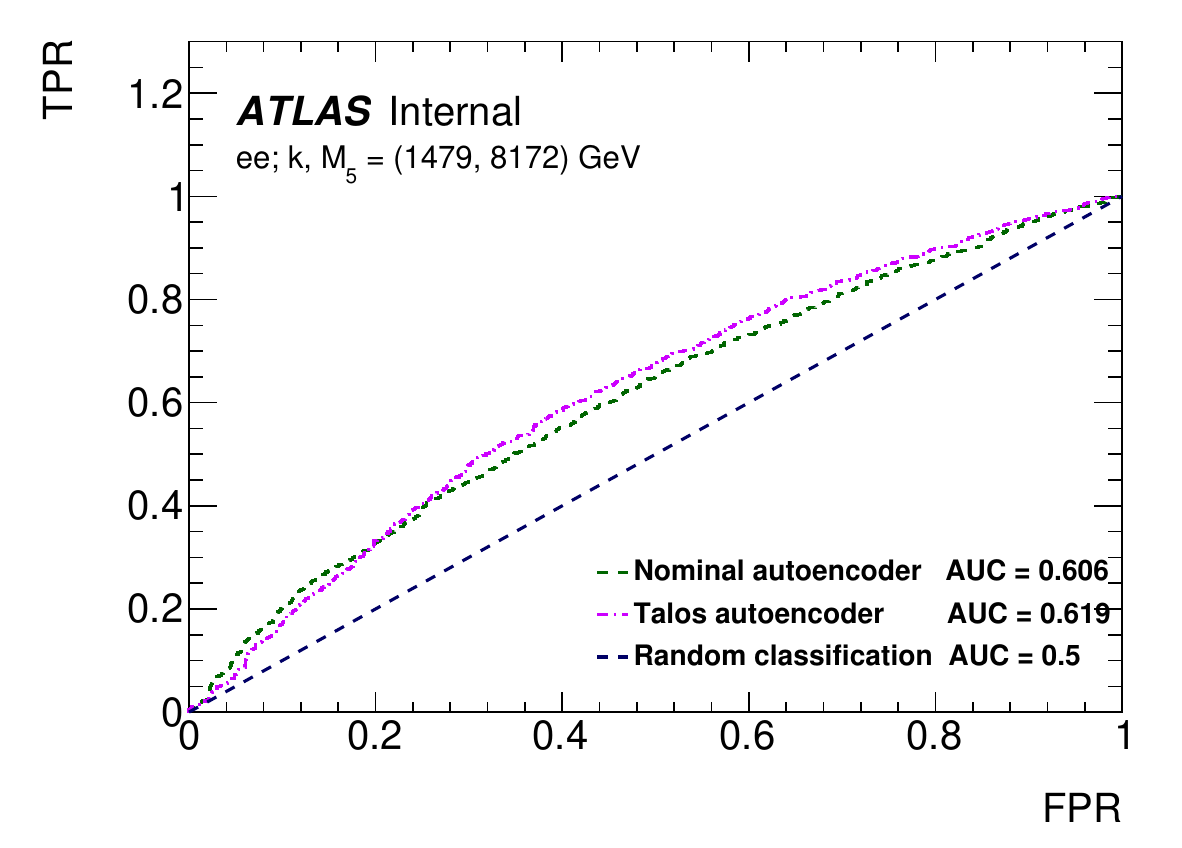}}
  \subfloat[]{\includegraphics[width=0.49\textwidth]{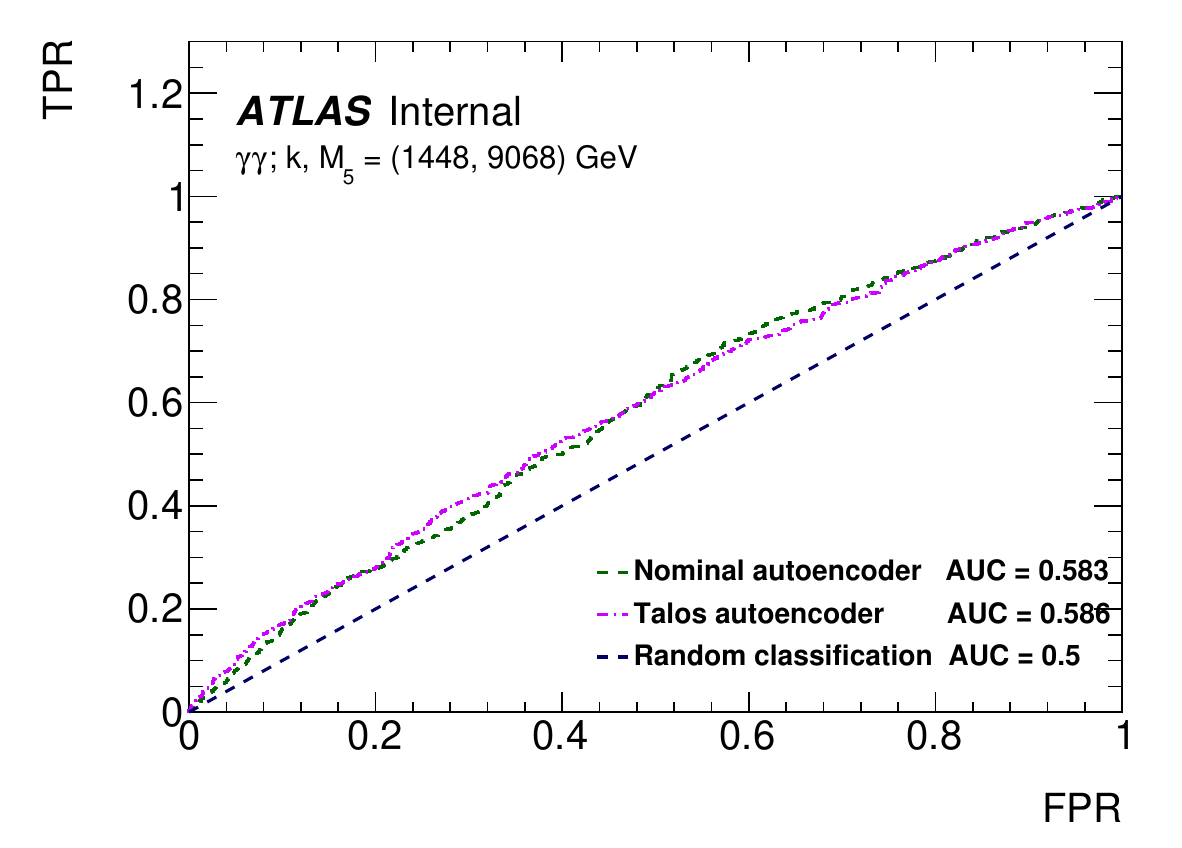}} \\
  \subfloat[]{\includegraphics[width=0.49\textwidth]{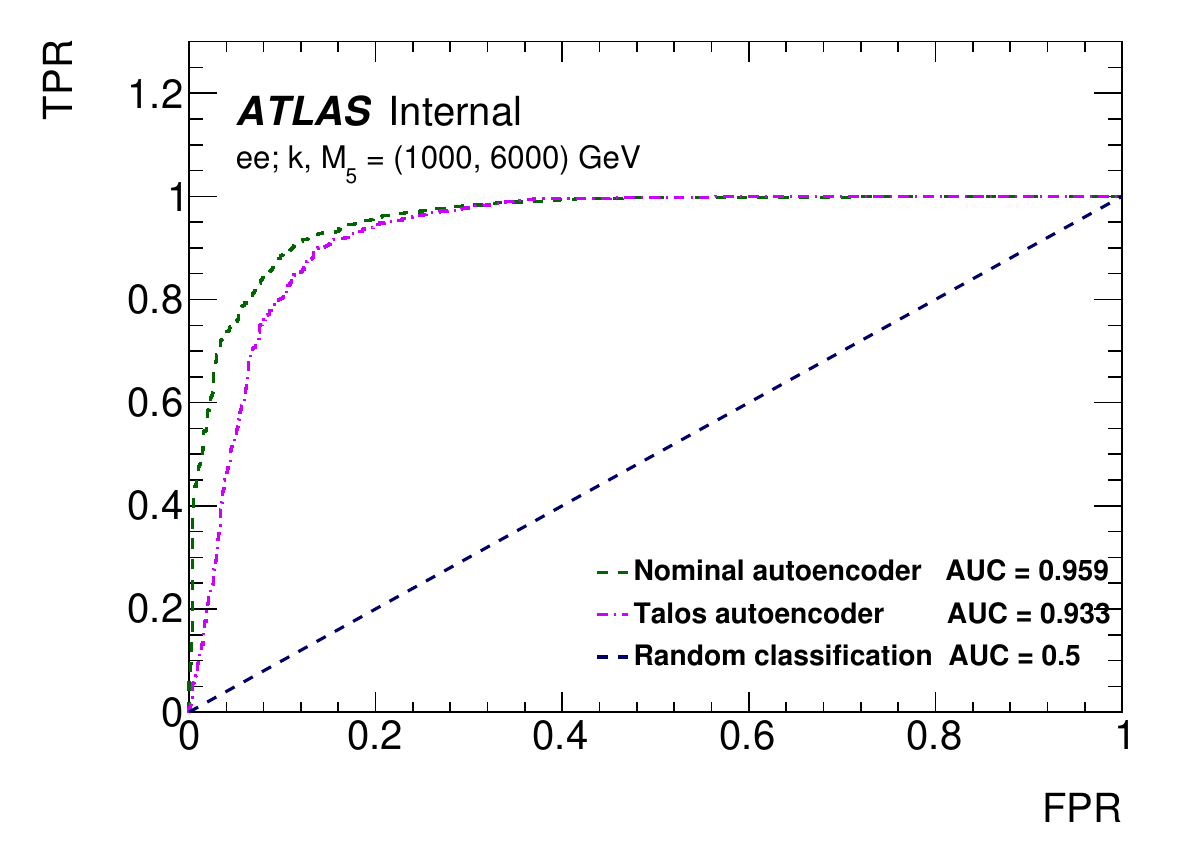}}
  \subfloat[]{\includegraphics[width=0.49\textwidth]{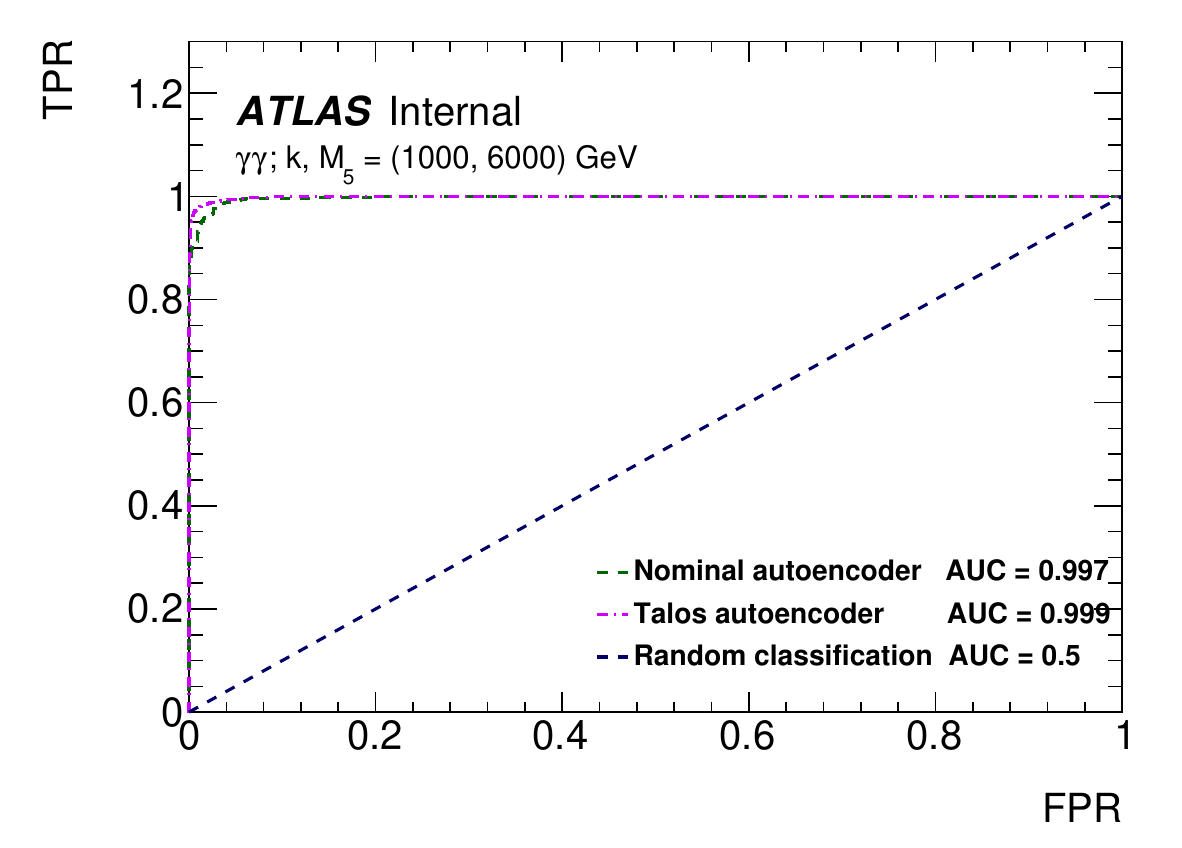}}
  \caption{AE ROC curves for the (left)~dielectron and (right)~diphoton channels.
    (Top) weak signal models. (Bottom) high-cross-section signal model at $(k, M_5) = (1000, 6000)$~\GeV.
    A random classifier curve (AUC~$= 0.5$) is shown for reference.}
  \label{fig:cw:roc_ae}
\end{figure}

The ROC curves of the nominal and grid-optimised AEs are shown in
FIG.~\ref{fig:cw:roc_ae} for both low-signal and high-signal scenarios.
As expected, the AE performs comparably to a random classifier for the weak
signal models used in Eq.~(\ref{eq:talos_signals}) (AUC $\approx 0.5$), since
those signals are barely distinguishable from background fluctuations.
For the high-cross-section point $(k, M_5) = (1000, 6000)$~\GeV, the AE
successfully separates signal from background.
The differences between the nominal and grid-optimised AE are negligible
($< 3\%$ in AUC), confirming that the nominal configuration is adequate.

\section{Results}
\label{sec:cw:results}

\subsection{Model-Independent Results}
\label{ssec:cw:mi_results}

\begin{figure}[htb!]
  \centering
  \subfloat[]{\includegraphics[width=0.49\textwidth]{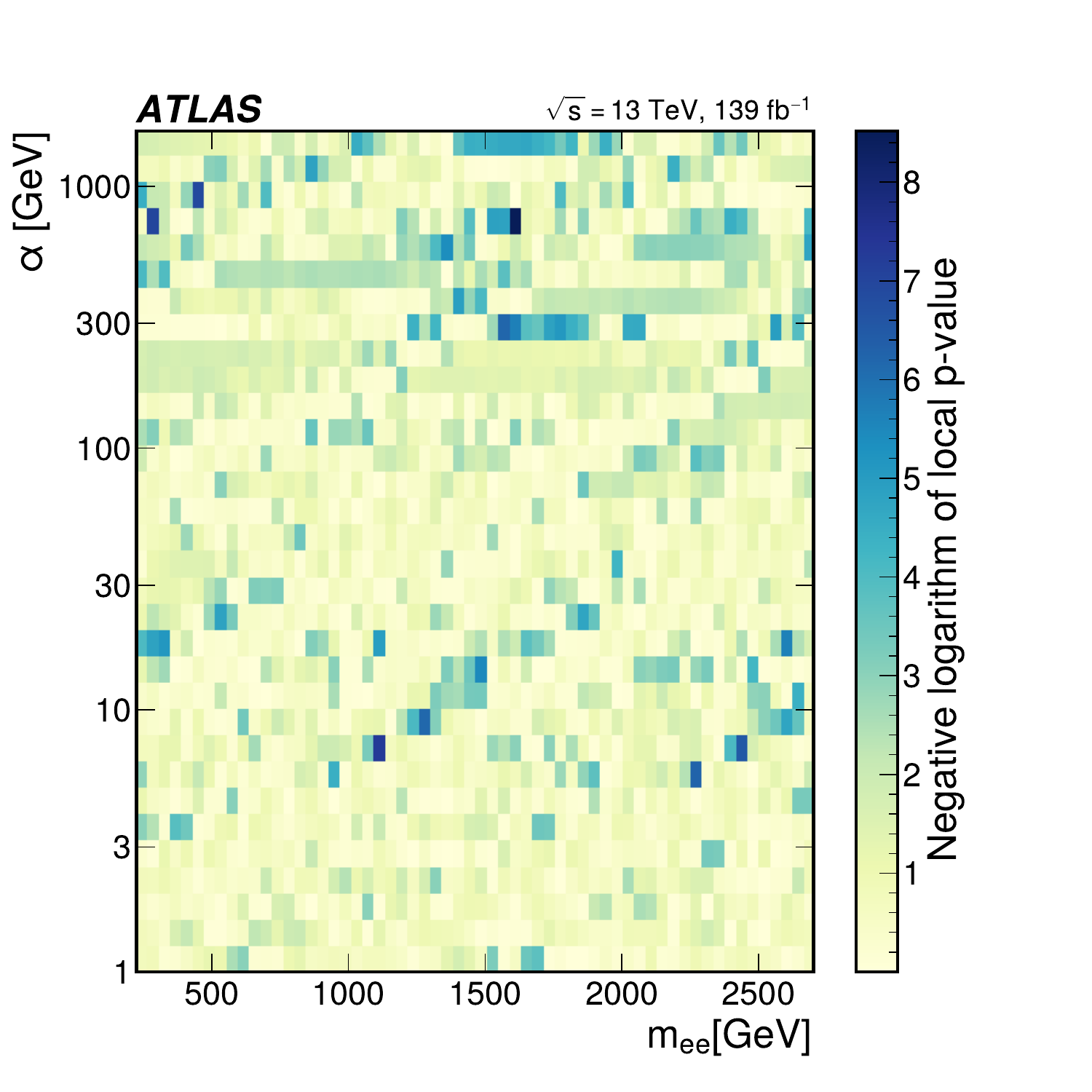}}
  \subfloat[]{\includegraphics[width=0.49\textwidth]{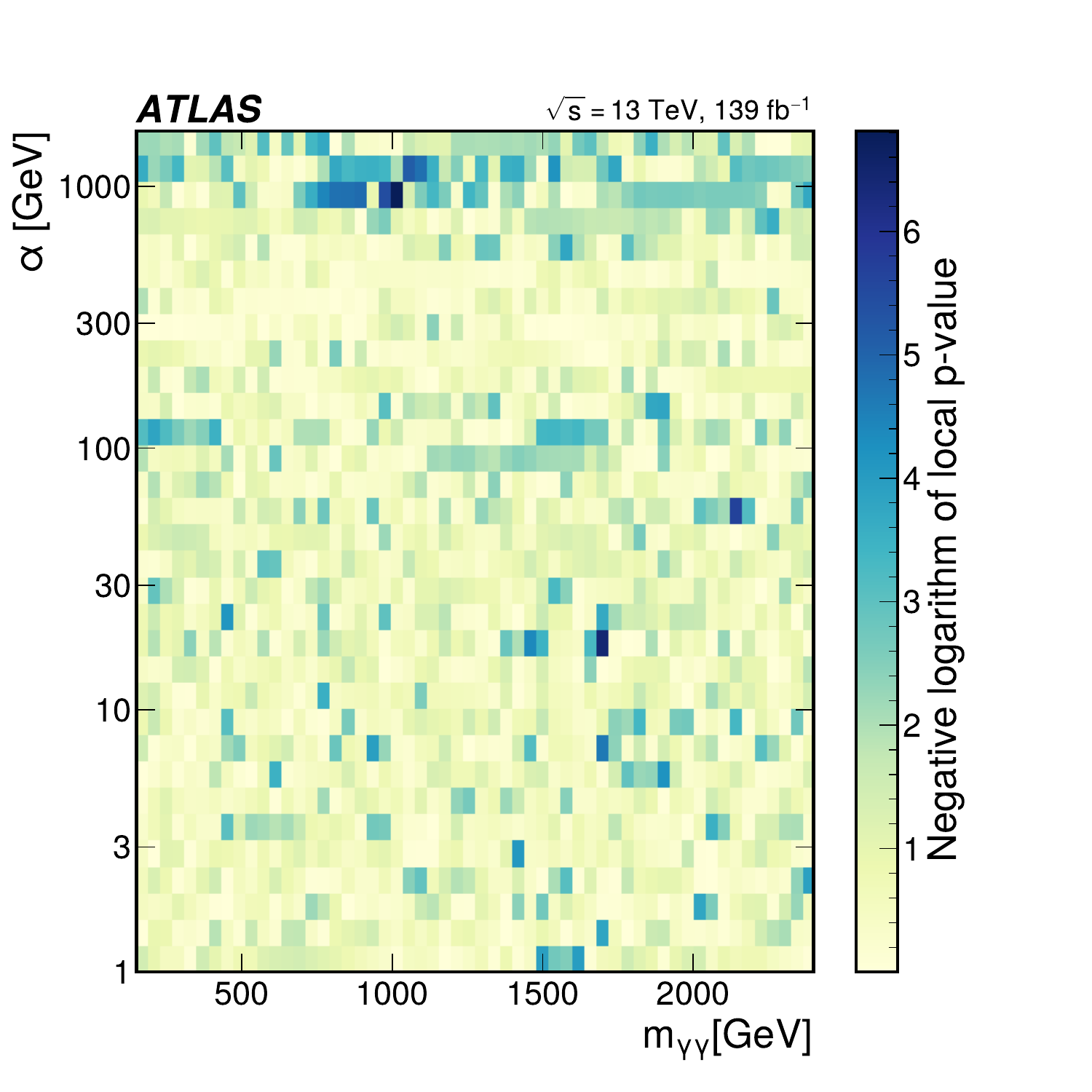}}
  \caption{Negative logarithm of the local $p$-values for the observed data in the
    (a)~dielectron and (b)~diphoton channels for the $R < 50\%$ threshold scenario.
    Here $\alpha$ is the CWT scale parameter.}
  \label{fig:localsignificance}
\end{figure}

The model-independent search uses the AE to compute local $p$-values across the
two-dimensional scalogram of the observed data.
The $-\log_{10}(p_\text{obs})$ maps for the dielectron and diphoton channels
are shown in FIG.~\ref{fig:localsignificance}.

The global significance is quantified by computing $L_\text{MSE}$ for the data
scalogram and comparing it against an ensemble of background-only pseudo-experiments.
The observed significances for all threshold scenarios considered are listed in
Table~\ref{table:significance_observed}.
No significant deviation from the background-only hypothesis is observed;
the largest excess is $1.5\sigma$ in the dielectron channel.

\begin{table}[htb!]
  \centering
  \caption{Observed significance in each analysis channel for the different
    thresholding scenarios.
    Positive (negative) values indicate that the data loss is higher (lower)
    than the median expected background loss.}
  \label{table:significance_observed}
  \begin{tabular}{ccc}
    \hhline{===}
    Threshold & Dielectron & Diphoton \\
    \hline
    $R < 10\%$     & $+0.4\sigma$ & $-1.8\sigma$ \\
    $R < 50\%$     & $+1.5\sigma$ & $-0.2\sigma$ \\
    No threshold   & $+0.7\sigma$ & $-0.7\sigma$ \\
    Scale threshold & $+1.5\sigma$ & $-0.6\sigma$ \\
    \hhline{===}
  \end{tabular}
\end{table}

\subsection{Model-Dependent Exclusion Limits}
\label{ssec:cw:md_results}

\begin{figure}[htb!]
  \centering
  \subfloat[]{\includegraphics[width=0.49\textwidth]{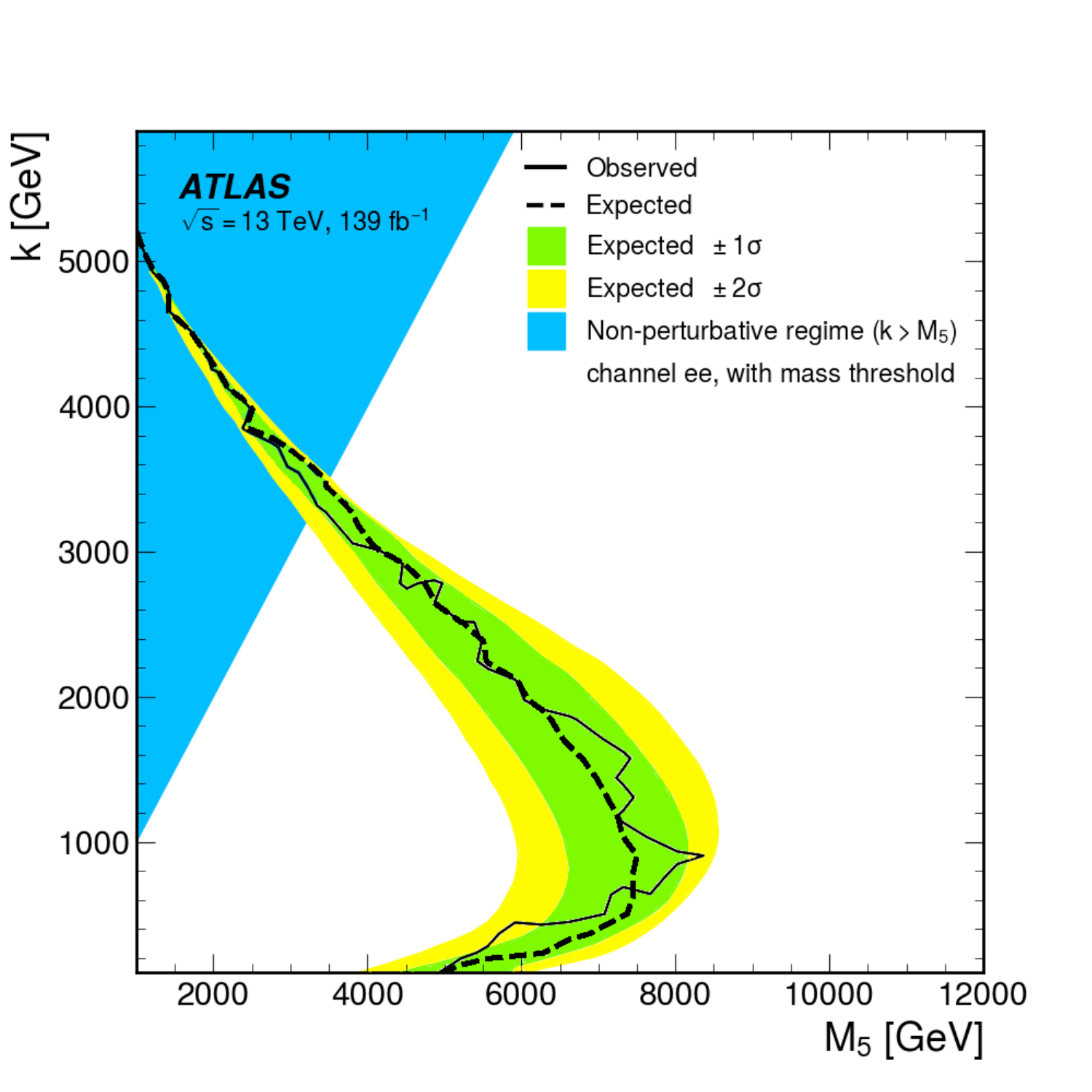}}
  \subfloat[]{\includegraphics[width=0.49\textwidth]{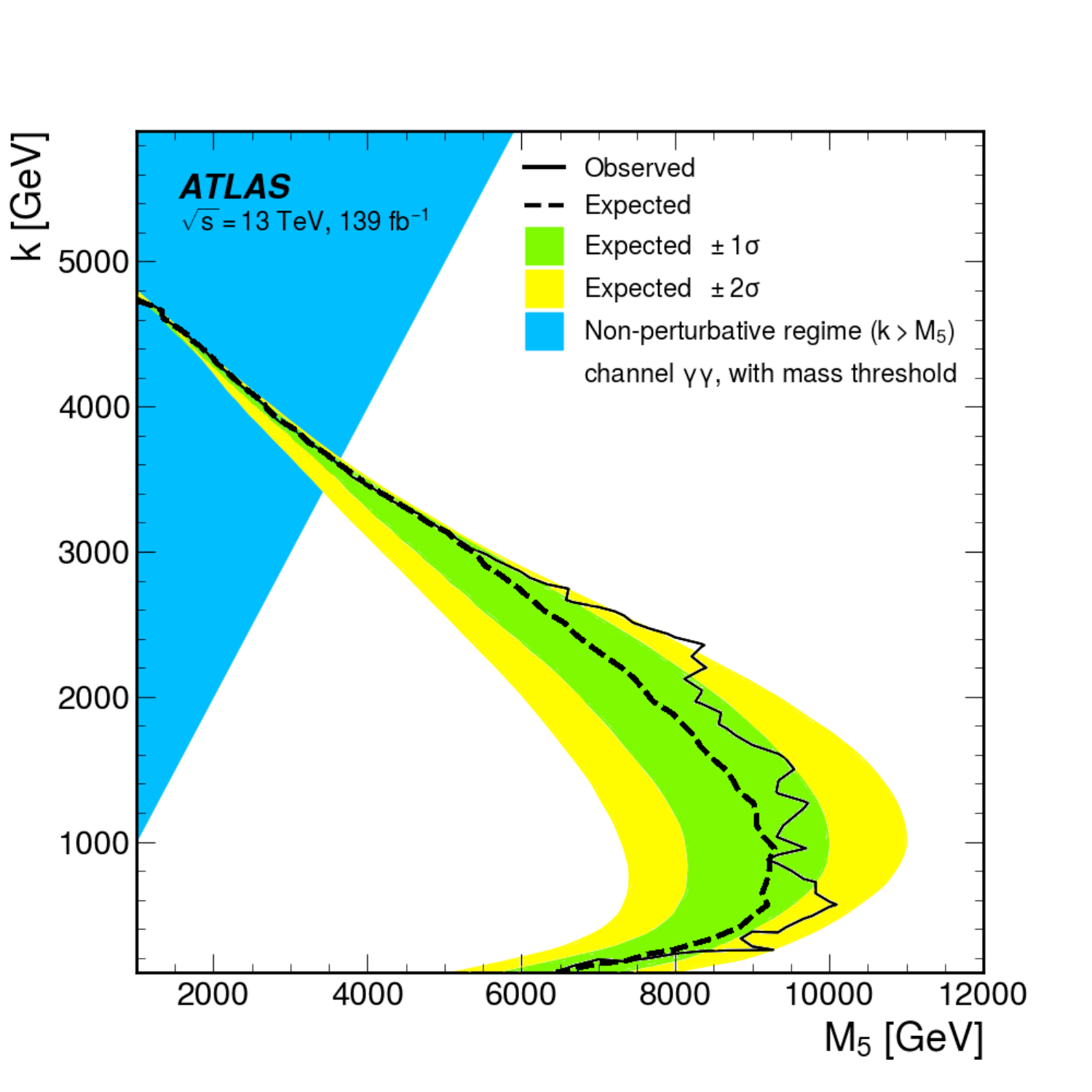}}
  \caption{Expected and observed exclusion limits at 95\% CL for the CW/LD model
    in the $k$--$M_5$ parameter space for the (a)~dielectron ($ee$) and
    (b)~diphoton ($\gamma\gamma$) channels with mass thresholds applied.
    Surrounding shaded bands represent the $\pm1\sigma$ and $\pm2\sigma$
    uncertainties on the expected limit.
    The shaded region with $k > M_5$ indicates where the theory becomes
    non-perturbative.}
  \label{fig:CWLDExclusion}
\end{figure}

In the absence of a significant signal, 95\% CL exclusion limits are set on the
CW/LD model in the $k$--$M_5$ plane using the CLs method.
FIG.~\ref{fig:CWLDExclusion} shows the expected and observed exclusion limits
for the case with mass thresholds applied.

With mass thresholds, the maximum excluded $M_5$ value is approximately 10~\TeV\
at $k \approx 600$~\GeV\ in the diphoton channel, and approximately 8.4~\TeV\
at $k \approx 900$~\GeV\ in the dielectron channel.
The dominant source of systematic uncertainty is the theoretical modelling of the
SM background; its effect weakens the $M_5$ exclusion by at most $\sim 1$~\TeV\
relative to the statistical-uncertainty-only case.
These results represent the first direct search for periodic graviton resonances
in invariant mass spectra at a hadron collider, and provide the most stringent
limits on the CW/LD model in the dielectron channel~\cite{ATLAS:2023hbp}.

\section{Summary and Status}
\label{sec:cw:summary}

The search for quasi-periodic graviton resonances from the Clockwork/Linear Dilaton model in dielectron and diphoton invariant mass spectra is complete. The analysis, using the full ATLAS Run~2 dataset of $139~\textrm{fb}^{-1}$ at $\sqrt{s}=13~\TeV$, has been published in JHEP~\cite{ATLAS:2023hbp}. It constitutes the first direct search for periodic KK graviton signals at a hadron collider and sets the most stringent limits on the CW/LD model in the dielectron channel. Maximum excluded five-dimensional Planck masses reach approximately $M_5 \approx 10~\TeV$ in the diphoton channel and $M_5 \approx 8.4~\TeV$ in the dielectron channel at 95\% CL. 
The results were published ~\cite{ATLAS:2023hbp} and featured in a CERN Courier article~\cite{CERN:Courier2023Clockwork}, highlighting the novel periodic-signal search methodology as a new direction for BSM searches at the LHC.\clearpage
\clearpage\chapter{Concluding Remarks}
\label{chp:conclusions}

The SM, for all its precision, leaves fundamental questions unanswered: the origin of the mass hierarchy between the weak and Planck scales, the persistent tensions in rare $B$-meson decays and the nature of DM.
This thesis presented three experimental searches for new phenomena bearing directly on these questions, all involving leptons in the final state. 
It also outlines detector performance work that contributes to the operation of ATLAS. 

The search for QBHs in lepton+jet final states exploits a unique feature of the model: the production rate of QBHs rises steeply with $\sqrt{s}$, so the step from Run~2 to Run~3 energies, despite only amounting to $0.6~\textrm{TeV}$, translates into an exceedingly-large gain in exclusion power at the highest accessible threshold masses.  
Using a partial Run~3 dataset --- the first QBH analysis in this final state at these energies — ADD threshold masses are excluded up to 9.4~\TeV\ ,  establishing the strongest limits to-date on QBHs. 
The results are public on arXiv and were submitted to Physics Letters B in April 2026~\cite{ATLAS:2026hzb}. 

The search for a $\Zp$ boson in dilepton final states with associated $b$-jets is motivated by residual tensions in $b\to s\ell^{+}\ell^{-}$ observables.  
This study is included in a broader efforts which also targets contact interactions, leptoquarks and a resonance produced in association with $E^{\textrm{miss}}_{\textrm{T}}$, with a preliminary result published for the latter as a CONF note~\cite{ATLAS-CONF-2023-045}. 
No significant excess is observed across masses from 500~\GeV\ to 4~\TeV. 
The search has been unblinded since May 2025 and the result forms the central component of a combined ATLAS publication, currently in preparation. 

The CW/LD search confronts the challenge that a tower of periodic heavy KK graviton resonances cannot be resolved individually against the fluctuating background. 
A two-dimensional treatment of the search is applied via performing wavelet transformations, resulting in spectograms fed to classifiers and autoencoders. 
This constitutes the first direct search for periodic graviton signals at a hadron collider and thus established a new method to tackle such signatures. 
The result was published in JHEP~\cite{ATLAS:2023hbp} and featured in a CERN Courier article~\cite{CERN:Courier2023Clockwork}, both in 2023. 

None of the searches finds evidence for BSM physics. 
However, they do contribute collectively to the advancement of our knowledge by pushing further the regions, where New Physics may exist. 
Each excluded region of parameter space narrows the landscape for what follows, and these results together set the baseline from which the High-Luminosity LHC programme will extend similar searches. 

Performance studies of the NSW sTGC strip detector characterise the detector response and identify a systematic offset between the pad-trigger firmware and the offline strip-to-band dictionary, explaining the absence of LUT matches in Run~3 data and providing the framework for commissioning the full strip-based trigger in subsequent data-taking. 

\clearpage

\clearpage
\phantomsection
\addcontentsline{toc}{chapter}{Bibliography}
\printbibliography[heading=bibliography,title={Bibliography}]

\captionsetup[figure]{list=no}
\captionsetup[table]{list=no}

\begin{appendices}
\clearpage\appendix

\chapter{QBH Analysis: Data and Monte Carlo Samples}
\label{app:qbh}

\addtocontents{toc}{\protect\setcounter{tocdepth}{0}}

\section{Collision Data}
\label{app:qbh:data}

The triggers used for each data-taking period in the QBH lepton+jet analysis are listed in
Table~\ref{tab:datasets::triggers}.
The Good Run List (GRL) files defining the accepted data quality intervals are listed in
Table~\ref{tab:GRL}, corresponding to a total integrated luminosity of $164~\textrm{fb}^{-1}$.

\begin{table}[h!]
    \centering
\caption{Summary of Run 3 single-lepton trigger chains used in the analysis.}
    \label{tab:datasets::triggers}
        \begin{tabular}{c c c}
        \hhline{===}
        Period      & Muon Triggers                                             & Electron Triggers                  \\ \hline
        \multirow{3}{*}{\centering 2022} & \multirow{3}{*}{\begin{tabular}{c} \texttt{HLT\_mu24\_ivarmedium\_L1MU14FCH} \\ \texttt{HLT\_mu50\_L1MU14FCH} \end{tabular}}      
             & \texttt{HLT\_e26\_lhtight\_ivarloose\_L1EM22VHI} \\ \Tstrut
             &  & \texttt{HLT\_e60\_lhmedium\_L1EM22VHI} \\ \Tstrut
             &  & \texttt{HLT\_e140\_lhloose\_L1EM22VHI} \\ \hline
        \multirow{3}{*}{\centering 2023} & \multirow{3}{*}{\begin{tabular}{c} \texttt{HLT\_mu24\_ivarmedium\_L1MU14FCH} \\ \texttt{HLT\_mu50\_L1MU14FCH} \end{tabular}}      
             & \texttt{HLT\_e26\_lhtight\_ivarloose\_L1eEM26M} \\ \Tstrut
             &  & \texttt{HLT\_e60\_lhmedium\_L1eEM26M} \\ \Tstrut
             &  & \texttt{HLT\_e140\_lhloose\_L1eEM26M} \\ \hline
        \multirow{3}{*}{\centering 2024} & \multirow{3}{*}{\begin{tabular}{c} \texttt{HLT\_mu24\_ivarmedium\_L1MU14FCH} \\ \texttt{HLT\_mu50\_L1MU14FCH}  \end{tabular}}      
              & \texttt{HLT\_e26\_lhtight\_ivarloose\_L1eEM26M} \\ \Tstrut
              &  & \texttt{HLT\_e60\_lhmedium\_L1eEM26M} \\ \Tstrut
              &  & \texttt{HLT\_e140\_lhloose\_L1eEM26M} \\ \hhline{===}
        \end{tabular}
\end{table}

\begin{table}
    \centering
\caption{List of Good Run Lists being used per year for this analysis.}
    \label{tab:GRL}
        \begin{tabular}{|c|c|c|}
    \hline
    \hline
    Year &  GRL &  Luminosity [$\textrm{pb}^{-1}$] \\
    \hline
    \hline

    2022 &   \makecell{\texttt{data22\_13p6TeV.periodAllYear\_DetStatus} \\ \texttt{-v134-pro28-10\_MERGED\_PHYS\_} \\ \texttt{StandardGRL\_All\_Good\_25ns\_ignore\_TRIGLAR.xml}} &  29294.4\\ \hline
    2023&   \makecell{\texttt{data23\_13p6TeV.periodAllYear\_DetStatus} \\ \texttt{-v133-pro31-11\_MERGED\_PHYS\_} \\ \texttt{StandardGRL\_All\_Good\_25ns\_ignoreTRIG\_JETCTPIN.xml}} &  26661.0\\ \hline
    2024 &   \texttt{physics\_25ns\_data24.xml} &  109400 \\ 
    \hline
    \hline
    \end{tabular}
\end{table}

\section{Background Monte Carlo Samples}
\label{app:qbh:bkg}

Tables~\ref{tab:wjets_MC}--\ref{tab:jj_MC} list the MC generator configurations for
each background process in the QBH analysis.

\begin{table}[h]
    \centering
\caption{$W$+Jets sample DSIDs, cross-section with branching ratios and efficiency of the generator filter.}
    \label{tab:wjets_MC}
        \begin{tabular}{|l|l|l|l|l|}
    \hline
    \hline
    Process             & DSID & $ \sigma \times BR [\textrm{pb}]$ & $ \epsilon $Filter gen & k-factor \\ 
    \hline
    \hline
    Sh\_2214\_Wenu\_maxHTpTV2\_BFilter        & 700777                    & 22937.000                 & 0.010  & 1.0                    \\
    Sh\_2214\_Wenu\_maxHTpTV2\_CFilterBVeto   & 700778                    & 22936.000                 & 0.148  & 1.0                    \\
    Sh\_2214\_Wenu\_maxHTpTV2\_CVetoBVeto     & 700779                    & 22936.000                 & 0.842  & 1.0                    \\
    Sh\_2214\_Wmunu\_maxHTpTV2\_BFilter       & 700780                    & 22938.000                 & 0.009  & 1.0                    \\
    Sh\_2214\_Wmunu\_maxHTpTV2\_CFilterBVeto  & 700781                    & 22939.000                 & 0.148  & 1.0                    \\
    Sh\_2214\_Wmunu\_maxHTpTV2\_CVetoBVeto    & 700782                    & 22939.000                 & 0.842  & 1.0                    \\
    Sh\_2214\_Wtaunu\_maxHTpTV2\_BFilter      & 700783                    & 22932.000                 & 0.010  & 1.0                    \\
    Sh\_2214\_Wtaunu\_maxHTpTV2\_CFilterBVeto & 700784                    & 22933.000                 & 0.148  & 1.0                    \\
    Sh\_2214\_Wtaunu\_maxHTpTV2\_CVetoBVeto   & 700785                    & 22933.000                 & 0.842  & 1.0                    \\
    \hline
    \hline   
    \end{tabular}
\end{table}

\begin{table}[h]
    \centering
\caption{$Z$+Jets sample DSIDs, cross-section with branching ratios and efficiency of the generator filter.}
    \label{tab:zjets_MC}
        \begin{tabular}{|l|l|l|l|l|}
    \hline
    \hline
    Process             & DSID & $ \sigma \times BR [\textrm{pb}]$ & $ \epsilon $Filter gen & k-factor \\ 
    \hline
    \hline
    Sh\_2214\_Zee\_maxHTpTV2\_BFilter          & 700786                    & 2336.100                  & 0.026     &     0.93            \\
    Sh\_2214\_Zee\_maxHTpTV2\_CFilterBVeto     & 700787                    & 2336.100                  & 0.130     &     0.93            \\
    Sh\_2214\_Zee\_maxHTpTV2\_CVetoBVeto       & 700788                    & 2336.100                  & 0.844     &     0.93            \\
    Sh\_2214\_Zmumu\_maxHTpTV2\_BFilter        & 700789                    & 2336.000                  & 0.025     &     0.93            \\
    Sh\_2214\_Zmumu\_maxHTpTV2\_CFilterBVeto   & 700790                    & 2335.900                  & 0.130     &     0.93            \\
    Sh\_2214\_Zmumu\_maxHTpTV2\_CVetoBVeto     & 700791                    & 2335.900                  & 0.844     &     0.93            \\
    Sh\_2214\_Ztautau\_maxHTpTV2\_BFilter      & 700792                    & 2337.200                  & 0.025     &     0.93            \\
    Sh\_2214\_Ztautau\_maxHTpTV2\_CFilterBVeto & 700793                    & 2337.100                  & 0.130     &     0.93            \\
    Sh\_2214\_Ztautau\_maxHTpTV2\_CVetoBVeto   & 700794                    & 2337.100                  & 0.844     &     0.93            \\
    \hline
    \hline   
    \end{tabular}
\end{table}

\begin{table}[h]
    \centering
\caption{Diboson $VV$ samples DSIDs, cross-section with branching ratios and efficiency of the generator filter.}
    \label{tab:vv_MC}
        \begin{tabular}{|l|l|l|l|l|}
    \hline
    \hline
    Process             & DSID & $ \sigma \times BR [\textrm{pb}]$ & $ \epsilon $Filter gen & k-factor  \\ 
    \hline
    \hline
    Sh\_2214\_lllljj           & 701000                       & 0.015                     & 1.000          &1.0             \\
    Sh\_2214\_lllvjj           & 701005                       & 0.053                     & 1.000              &1.0         \\
    Sh\_2214\_llvvjj\_os       & 701010                       & 0.208                     & 1.000              &1.0         \\
    Sh\_2214\_llvvjj\_ss       & 701015                       & 0.050                     & 1.000              &1.0         \\
    Sh\_2214\_lllljj\_Int      & 701020                       & 0.001                     & 1.000              &1.0         \\
    Sh\_2214\_lllvjj\_Int      & 701025                       & 0.003                     & 0.996              &1.0         \\
    Sh\_2214\_llvvjj\_os\_Int  & 701030                       & 0.004                     & 1.000              &1.0         \\
    Sh\_2214\_llvvjj\_ss\_Int  & 701035                       & 0.003                     & 1.000              &1.0         \\
    Sh\_2214\_llll             & 701040                       & 1.331                     & 1.000              &1.0         \\
    Sh\_2214\_lllv             & 701045                       & 4.772                     & 1.000              &1.0         \\
    Sh\_2214\_llvv\_os         & 701050                       & 12.652                    & 1.000              &1.0         \\
    Sh\_2214\_llvv\_ss         & 701055                       & 0.024                     & 1.000              &1.0         \\
    Sh\_2214\_lvvv             & 701060                       & 3.278                     & 1.000              &1.0         \\
    Sh\_2214\_vvvv             & 701065                       & 0.610                     & 1.000              &1.0         \\
    Sh\_2214\_ZqqZll           & 701085                       & 6.734                     & 0.264              &1.0         \\
    Sh\_2214\_ZbbZll           & 701090                       & 1.043                     & 0.479              &1.0         \\
    Sh\_2214\_ZqqZvv           & 701095                       & 8.963                     & 0.393              &1.0         \\
    Sh\_2214\_ZbbZvv           & 701100                       & 2.010                     & 0.491              &1.0         \\
    Sh\_2214\_WqqZll           & 701105                       & 3.550                     & 1.000              &1.0         \\
    Sh\_2214\_WqqZvv           & 701110                       & 7.031                     & 1.000              &1.0         \\
    Sh\_2214\_WlvZqq           & 701115                       & 9.219                     & 1.000              &1.0         \\
    Sh\_2214\_WlvZbb           & 701120                       & 2.593                     & 1.000              &1.0         \\
    Sh\_2214\_WlvWqq           & 701125                       & 116.820                   & 0.438              &1.0         \\
    \hline
    \hline   
    \end{tabular}
\end{table}

\begin{table}[h]
    \centering
\caption{Top sample DSIDs, cross-section with branching ratios and efficiency of the generator filter.}
    \label{tab:top_MC}
        \begin{tabular}{|l|l|l|l|l|}
    \hline
    \hline
    Process             & DSID & $ \sigma \times BR [\textrm{pb}]$ & $ \epsilon $ Filter gen & k-factor \\ 
    \hline
    \hline
    PhPy8EG\_A14\_ttbar\_hdamp258p75\_SingleLep & 601229                    & 811.290                   & 0.438        & 1.14              \\
    PhPy8EG\_A14\_ttbar\_hdamp258p75\_dil       & 601230                    & 85.482                    & 1.000       & 1.14              \\
    PhPy8EG\_A14\_ttbar\_hdamp258p75\_allhad    & 601237                    & 811.290                   & 0.456        & 1.14             \\
    PhPy8EG\_tb\_lep\_antitop                   & 601348                    & 1.350                     & 1.000      & 1.09              \\
    PhPy8EG\_tb\_lep\_top                       & 601349                    & 1.100                     & 1.000      &1.10                \\
    PhPy8EG\_tqb\_lep\_antitop                  & 601350                    & 24.200                    & 1.000       &1.09                \\
    PhPy8EG\_tqb\_lep\_top                      & 601351                    & 39.940                    & 1.000       &1.11                \\
    PhPy8EG\_tW\_dyn\_DR\_incl\_antitop         & 601352                    & 39.837                    & 1.000       &1.10                \\
    PhPy8EG\_tW\_dyn\_DR\_incl\_top             & 601355                    & 39.871                    & 1.000       &1.10                \\
    \hline
    \hline   
    \end{tabular}
\end{table}

\begin{table}[h]
    \centering
\caption{$t\bar{t}V$ sample DSIDs, cross-section with branching ratios and efficiency of the generator filter.}
    \label{tab:ttv_MC}
        \begin{tabular}{|l|l|l|l|l|}
    \hline
    \hline
    Process             & DSID & $ \sigma \times BR [\textrm{pb}]$ & $ \epsilon $Filter gen & k-factor \\ 
    \hline
    \hline
    aMCPy8EG\_NNPDF30NLO\_A14N23LO\_ttee\_run3     & 522024                       & 0.041                     & 1.000     & 1.0                 \\
    aMCPy8EG\_NNPDF30NLO\_A14N23LO\_ttmumu\_run3   & 522028                       & 0.041                     & 1.000     & 1.0                 \\
    aMCPy8EG\_NNPDF30NLO\_A14N23LO\_tttautau\_run3 & 522032                       & 0.041                     & 1.000     & 1.0                 \\
    aMCPy8EG\_NNPDF30NLO\_A14N23LO\_ttZqq\_run3    & 522036                       & 0.593                     & 1.000     & 1.0                 \\
    aMCPy8EG\_NNPDF30NLO\_A14N23LO\_ttZnunu\_run3  & 522040                       & 0.174                     & 1.000     & 1.0                 \\
    Sh\_2214\_ttW\_0Lfilter                        & 700995                       & 0.649                     & 0.200     & 1.0                 \\
    Sh\_2214\_ttW\_1Lfilter                        & 700996                       & 0.649                     & 0.372     & 1.0                 \\
    Sh\_2214\_ttW\_2Lfilter                        & 700997                       & 0.649                     & 0.428     & 1.0                 \\
    \hline
    \hline   
    \end{tabular}
\end{table}

\begin{table}[h]
    \centering
\caption{Dijet ($jj$) sample DSIDs, cross-section with branching ratios and efficiency of the generator filter.}
    \label{tab:jj_MC}
        \begin{tabular}{|l|l|l|l|l|}
    \hline
    \hline
    Process             & DSID & $ \sigma \times BR [\textrm{pb}]$ & $ \epsilon $Filter gen & k-factor \\ 
    \hline
    \hline
    Py8EG\_A14NNPDF23LO\_jj\_JZ0     & 801165                    & 78580000000.000           & 0.974        & 1.0             \\
    Py8EG\_A14NNPDF23LO\_jj\_JZ1     & 801166                    & 93901000000.000           & 0.035        & 1.0             \\
    Py8EG\_A14NNPDF23LO\_jj\_JZ2     & 801167                    & 2582600000.000            & 0.010       & 1.0             \\
    Py8EG\_A14NNPDF23LO\_jj\_JZ3     & 801168                    & 28528000.000              & 0.012     & 1.0             \\
    Py8EG\_A14NNPDF23LO\_jj\_JZ4     & 801169                    & 280140.000                & 0.014   & 1.0             \\
    Py8EG\_A14NNPDF23LO\_jj\_JZ5     & 801170                    & 5132.800                  & 0.015  & 1.0            \\
    Py8EG\_A14NNPDF23LO\_jj\_JZ6     & 801171                    & 297.650                   & 0.010  & 1.0           \\
    Py8EG\_A14NNPDF23LO\_jj\_JZ7     & 801172                    & 19.413                    & 0.012 & 1.0            \\
    Py8EG\_A14NNPDF23LO\_jj\_JZ8     & 801173                    & 0.798                     & 0.012 & 1.0           \\
    Py8EG\_A14NNPDF23LO\_jj\_JZ9incl & 801174                    & 0.028                     & 0.015 & 1.0            \\
    \hline
    \hline   
    \end{tabular}
\end{table}

\section{Signal Monte Carlo Samples}
\label{app:qbh:sig}

The full list of QBH signal samples produced for this analysis, including the dataset
identifiers (DSIDs), generator settings, and production campaigns, is given in
Table~\ref{tab:QBH_Sig_Production_Table}.

\begin{table}[h!]
    \centering
\caption{Summary of QBH signal samples produced for the electron and muon channels, process $qq \rightarrow \textrm{QBH}\rightarrow lq$. $n$ is the number of extra dimensions. The cross-section times branching fraction is summed over all QBH $Q$ states which are allowed to decay to $l q$. The number of events is split among the three periods in the \texttt{mc23} production campaign, namely \texttt{mc23a}, \texttt{mc23d}, and \texttt{mc23e}.}
    \label{tab:QBH_Sig_Production_Table}
        \begin{tabular}{c|c|c|c|c|c|c}
    \hline \hline
    Generator & Model & $M_{\textrm{th}}$ {[}TeV{]} & $n$ & DSID (e channel) & DSID ($\mu$ channel) & $\sum \sigma \times BR$ {[}pb{]} \\ \hline    
    
    \multirow{24}{*}{\begin{tabular}[c]{@{}c@{}}\texttt{QBHv3.02}+\\ \texttt{Pythia8.309}; \\ \texttt{CTEQ6L1} PDF\\ A14 tune\end{tabular}} 
    & \multirow{18}{*}{ADD} & 8.0  & \multirow{6}{*}{6} & 901972 & 901996 & 1.011e-03 \\
    & & 8.5  & & 901973 & 901997 & 3.097e-04 \\
    & & 9.0  & & 901974 & 901998 & 8.678e-05 \\
    & & 9.5  & & 901975 & 901999 & 2.172e-05 \\
    & & 10.0 & & 901976 & 902000 & 4.709e-06 \\
    & & 10.5 & & 901977 & 902001 & 8.452e-07 \\ \cline{3-7} 
    
    & & 8.0  & \multirow{6}{*}{4} & 901978 & 902002 & 4.647e-04 \\
    & & 8.5  & & 901979 & 902003 & 1.415e-04 \\ 
    & & 9.0  & & 901980 & 902004 & 3.939e-05 \\ 
    & & 9.5  & & 901981 & 902005 & 8.913e-06 \\ 
    & & 10.0 & & 901982 & 902006 & 2.108e-06 \\ 
    & & 10.5 & & 901983 & 902007 & 3.753e-07 \\ \cline{3-7} 
    
    & & 8.0  & \multirow{6}{*}{2} & 901984 & 902008 & 1.081e-04 \\
    & & 8.5  & & 901985 & 902009 & 3.255e-05 \\
    & & 9.0  & & 901986 & 902010 & 8.953e-06 \\
    & & 9.5  & & 901987 & 902011 & 2.221e-06 \\
    & & 10.0 & & 901988 & 902012 & 4.803e-07 \\
    & & 10.5 & & 901989 & 902013 & 8.596e-08 \\ \cline{2-7} 
    
    & \multirow{6}{*}{RS} & 6.0  & \multirow{6}{*}{1} & 901990 & 902014 & 2.796e-07 \\
    & & 6.5  & & 901991 & 902015 & 1.051e-07 \\
    & & 7.0  & & 901992 & 902016 & 3.814e-08 \\
    & & 7.5  & & 901993 & 902017 & 1.323e-08 \\
    & & 8.0  & & 901994 & 902018 & 4.335e-09 \\
    & & 8.5  & & 901995 & 902019 & 1.322e-09 \\ \hline \hline
    \end{tabular}
\end{table}

\begin{figure}[h]
    \subfloat[(a)]{
      \includegraphics[width=0.5\textwidth]{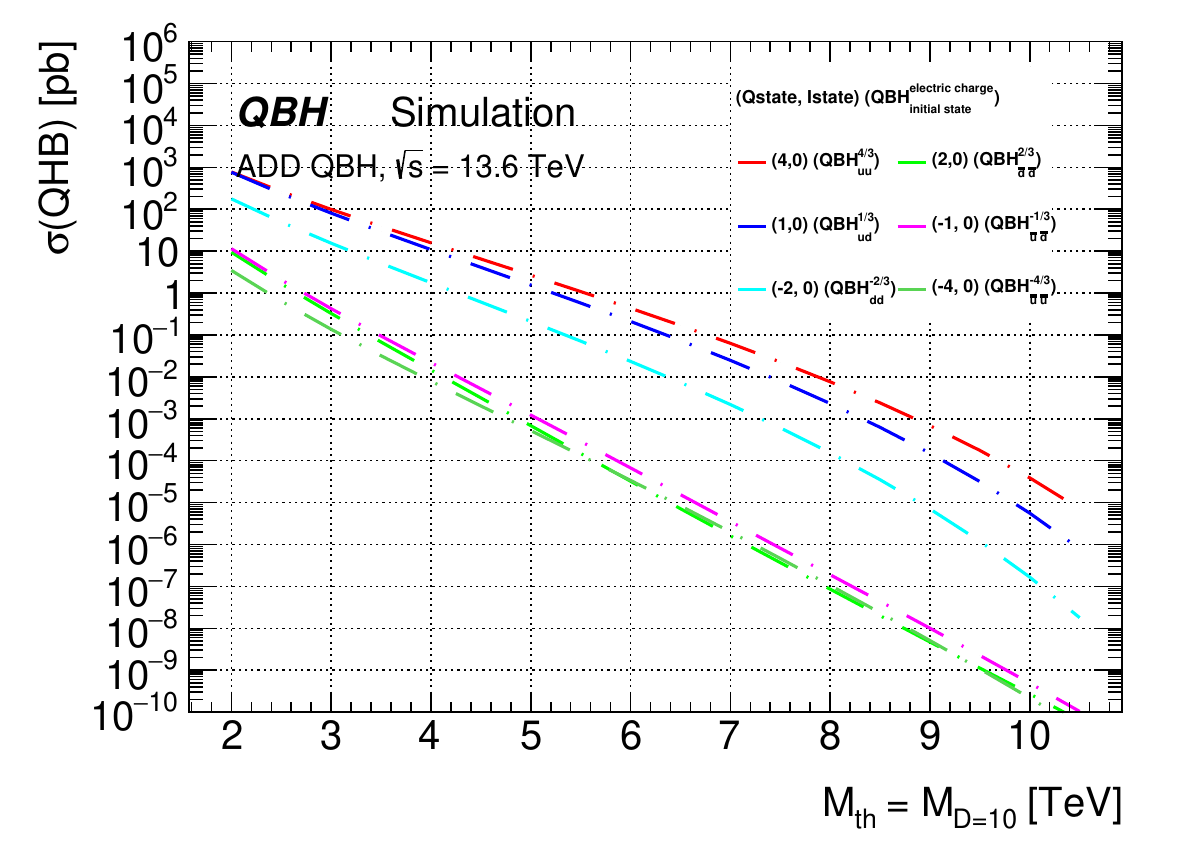}
    }
    \hfill
    \subfloat[(b)]{
      \includegraphics[width=0.5\textwidth]{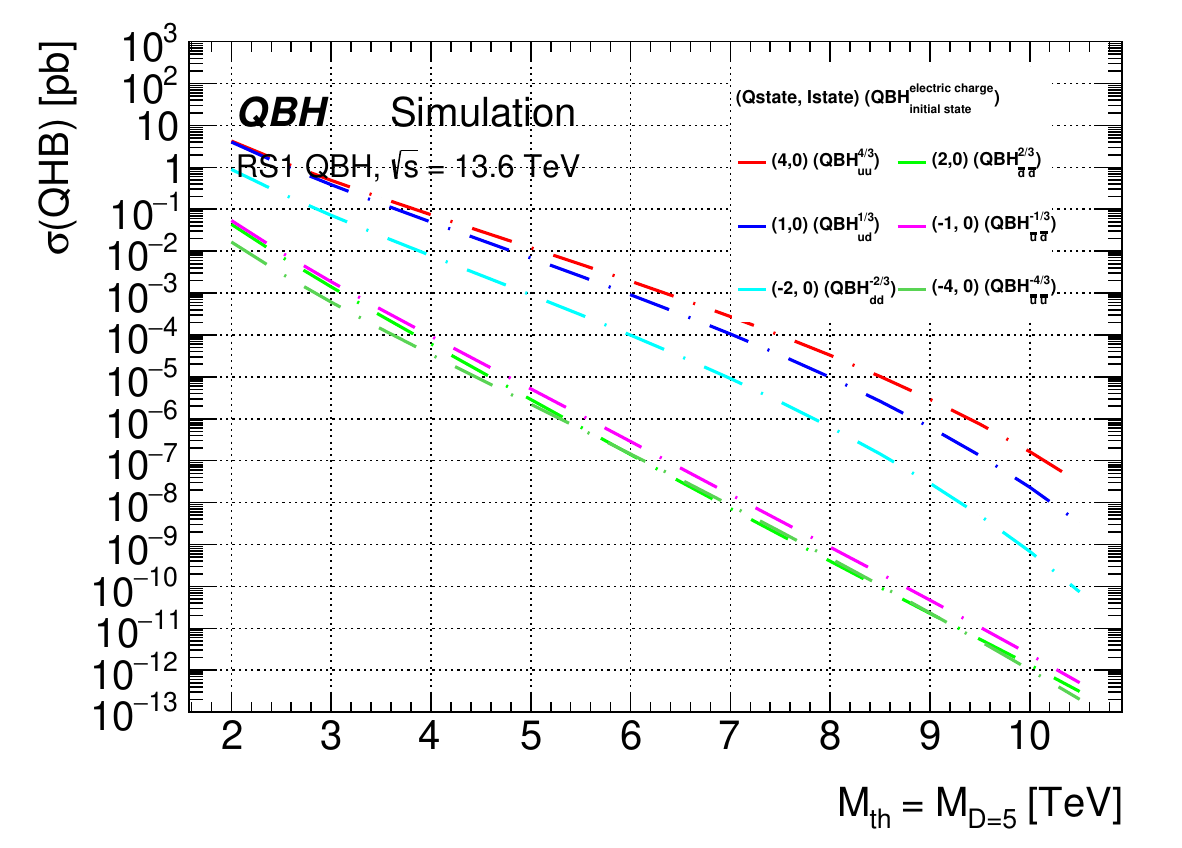}
    }
\caption{Production cross-section of quantum black holes at different threshold masses, $M_{\textrm{th}}$, for (a) the ADD model with $n=6$ ($D=10$) and (b) the RS model with $n=1$ ($D=5$) extra (total) dimensions, at centre-of-mass energy $\sqrt{s}=13.6~\TeV$. The colours indicate all possible quantum states of the quantum black hole that can decay to a $\ell q$ final state.}
\label{fig:QBH_XS_Run3_ADD_RS}
\end{figure}

The production cross-sections of quantum black holes for the ADD ($D=10$) and RS ($D=5$) models at $\sqrt{s}=13.6~\TeV$ are shown in FIG.~\ref{fig:QBH_XS_Run3_ADD_RS}.

\section{Systematic Uncertainties: Tables}
\label{app:qbh:exp_systs}

Tables~\ref{tab:lep_exp_syst} and~\ref{tab:jet_exp_syst} list the experimental systematic
uncertainties on the lepton and jet/$E_{\textrm{T}}^{\textrm{miss}}$ objects considered in the
QBH analysis, as discussed in Section~\ref{sec:Syst_exp}. Table~\ref{tab:syst:theory_overview}
lists the theoretical systematic variations applied to the $V$+jets backgrounds, as discussed
in Section~\ref{sec:Syst_th}.

\begin{table}[h]
\centering
\caption{Theory systematic variations applied to the $V$+jets backgrounds.}
\label{tab:syst:theory_overview}
\begin{tabular}{|l|l|l|}
\hline
\hline
Process & Variation & Evaluation \\
\hline
\hline
\multirow{3}{*}{$V$+jets}
    & $\mu_{\textrm{F}},\mu_{\textrm{R}}$ scales & 7-point reweighting \\
    & $\alpha_{S}$             & 2-point uncertainty \\
    & PDF                      & 100 PDF sets \\
\hline
\hline
\end{tabular}
\end{table}

\begin{table}[h]
    \centering
\caption{Experimental systematic uncertainties on the electron and muon objects.}
    \label{tab:lep_exp_syst}
        \begin{tabular}{|l|l|l|}
    \hline
    \hline
    Systematic & Type & Description \\
    \hline
    \hline
    \multicolumn{3}{|c|}{Electron}                                       \\
    \hline
    EL\_EFF\_ID\_TOTAL  & SF  & Uncertainty of ID efficiency                 \\
    EL\_EFF\_Trigger\_TOTAL  & SF  & Uncertainty of trigger efficiency                       \\
    EL\_EFF\_Reco\_TOTAL  & SF  & Uncertainty of reconstruction efficiency                     \\
    EL\_EFF\_Iso\_TOTAL  & SF  & Uncertainty of isolation efficiency                      \\
    EG\_RESOLUTION\_ALL  & Calib  & Uncertainty of the energy scale                    \\
    EG\_SCALE\_AF2  & Calib  & Uncertainty of the energy scale - FastSim                      \\
    EG\_SCALE\_ALL  & Calib  & Uncertainty of the energy resolution                      \\
    \hline
    \multicolumn{3}{|c|}{Muon}                                       \\
    \hline
    MUON\_EFF\_ISO\_STAT  & SF  & \multirow{2}{*}{Uncertainty of isolation efficiency}   \\
    MUON\_EFF\_ISO\_SYS  & SF  &    \\
    MUON\_EFF\_RECO\_STAT  & SF  & \multirow{2}{*}{Uncertainty of reconstruction efficiency}  \\
    MUON\_EFF\_RECO\_SYS  & SF  &   \\
    MUON\_EFF\_TrigStatUncertainty  & SF  & \multirow{2}{*}{Uncertainty of trigger efficiency}     \\
    MUON\_EFF\_TrigSystUncertainty  & SF  &                 \\
    MUON\_EFF\_TTVA\_STAT  & SF  & \multirow{2}{*}{Uncertainty of track to vertex association}     \\
    MUON\_EFF\_TTVA\_SYS  & SF  &                       \\
    MUON\_CB  & Calib  &   \makecell[l]{Uncertainty of energy resolution from \\ combined inner detector and muon systems}                    \\
    MUON\_SAGITTA\_DATASTAT  & Calib  &   \multirow{4}{*}{Momentum scale variations} \\
    MUON\_SAGITTA\_GLOBAL  & Calib  &                       \\
    MUON\_SAGITTA\_PTEXTRA  & Calib  &                       \\
    MUON\_SAGITTA\_RESBIAS  & Calib  &                       \\
    MUON\_SCALE  & Calib  & Uncertainty of the energy scale        \\
    \hline
    \hline
    \end{tabular}
\end{table}

\begin{table}[h!]
    \centering
\caption{Experimental systematic uncertainties on the jet and $E^{\textrm{miss}}_{\textrm{T}}$.}
    \label{tab:jet_exp_syst}
        \begin{tabular}{|l|l|l|}
    \hline
    \hline
    Systematic & Type & Description \\ 
    \hline
    \hline
    \multicolumn{3}{|c|}{Jet}                                       \\
    \hline
    JET\_NNJvtEfficiency        & SF  & Uncertainty of NNJvt efficiency         \\
    JET\_BJES\_Response           & Calib  &  b-quark-initiated jets                      \\
    JET\_EffectiveNP\_Detector(1,2)   & Calib  & \multirow{4}{*}{Effective nuisance parameters}                      \\
    JET\_EffectiveNP\_Mixed(1-3)   & Calib  &                       \\
    JET\_EffectiveNP\_Modelling(1-4)   & Calib  &                       \\
    JET\_EffectiveNP\_Statistical(1-6)   & Calib  &                       \\
    JET\_EtaIntercalibration\_Modelling   & Calib  & \multirow{3}{*}{\makecell[l]{$\eta$ intercalibration corrects energy \\ scale of forward jets}}              \\
    JET\_EtaIntercalibration\_NonClosure\_0p2\_PreRec   & Calib  &                       \\
    JET\_EtaIntercalibration\_TotalStat   & Calib  &                       \\
    JET\_Flavor\_Composition   & Calib  & \multirow{2}{*}{\makecell[l]{Difference of quark-initiated and \\ gluon-initiated jets}}         \\
    JET\_Flavor\_Response   & Calib  &                    \\    
    JET\_InSitu\_NonClosure\_PreRec   & Calib  & Description                      \\
    JET\_JER\_DataVsMC\_MC16   & Calib  & \multirow{4}{*}{Jet energy resolution}             \\
    JET\_JER\_EffectiveNP\_(1-11)   & Calib  &                       \\
    JET\_JER\_EffectiveNP\_12restTerm   & Calib  &                       \\
    JET\_JERUnc\_Noise\_PreRec   & Calib  &                       \\
    JET\_JESUnc\_Noise\_PreRec   & Calib  & \multirow{2}{*}{Jet energy scale}                       \\
    JET\_JESUnc\_VertexingAlg\_PreRec   & Calib  &                       \\
    JET\_Pileup\_OffsetMu   & Calib  & \multirow{4}{*}{Description}       \\
    JET\_Pileup\_OffsetNPV   & Calib  &                       \\
    JET\_Pileup\_PtTerm   & Calib  &                       \\
    JET\_Pileup\_RhoToppology   & Calib  &                       \\
    JET\_SingleParticle\_HighPt   & Calib  & \makecell[l]{Uncertainty from single-particle\\ and test-beam measurements}                       \\
    
    \hline
    \multicolumn{3}{|c|}{MET}                                       \\
    \hline
    MET\_SoftTrk\_ResoPara           & Calib  & \multirow{3}{*}{Soft term calculation} 
                      \\
    MET\_SoftTrk\_ResoPerp           & Calib  &                       \\
    MET\_SoftTrk\_Scale           & Calib  &                       \\

    \hline
    \hline
    \end{tabular}
\end{table}

\chapter{Sytematic Uncertainties}
\label{sec:app_systematics}

\section{Muon}

\begin{figure}[h]
  \subfloat{
    \includegraphics[width=0.5\textwidth]{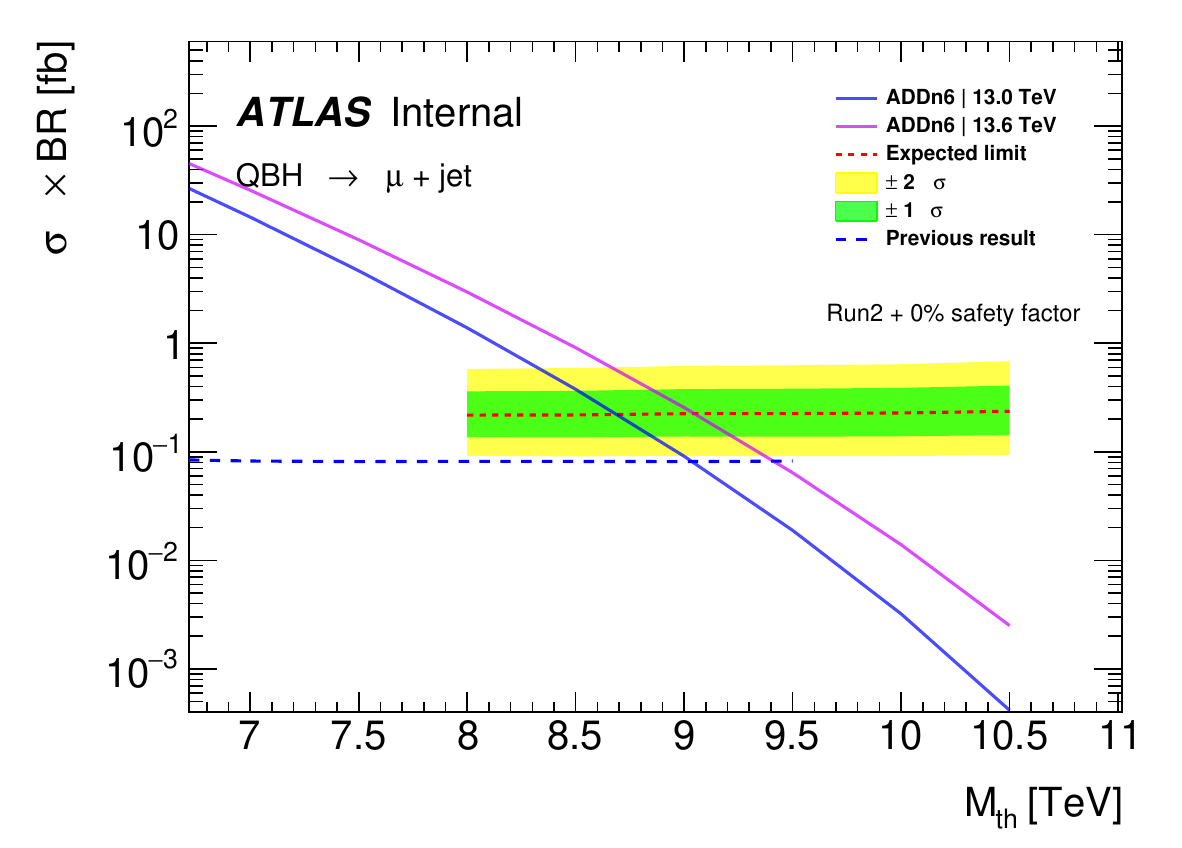}
  }
  \subfloat{
    \includegraphics[width=0.5\textwidth]{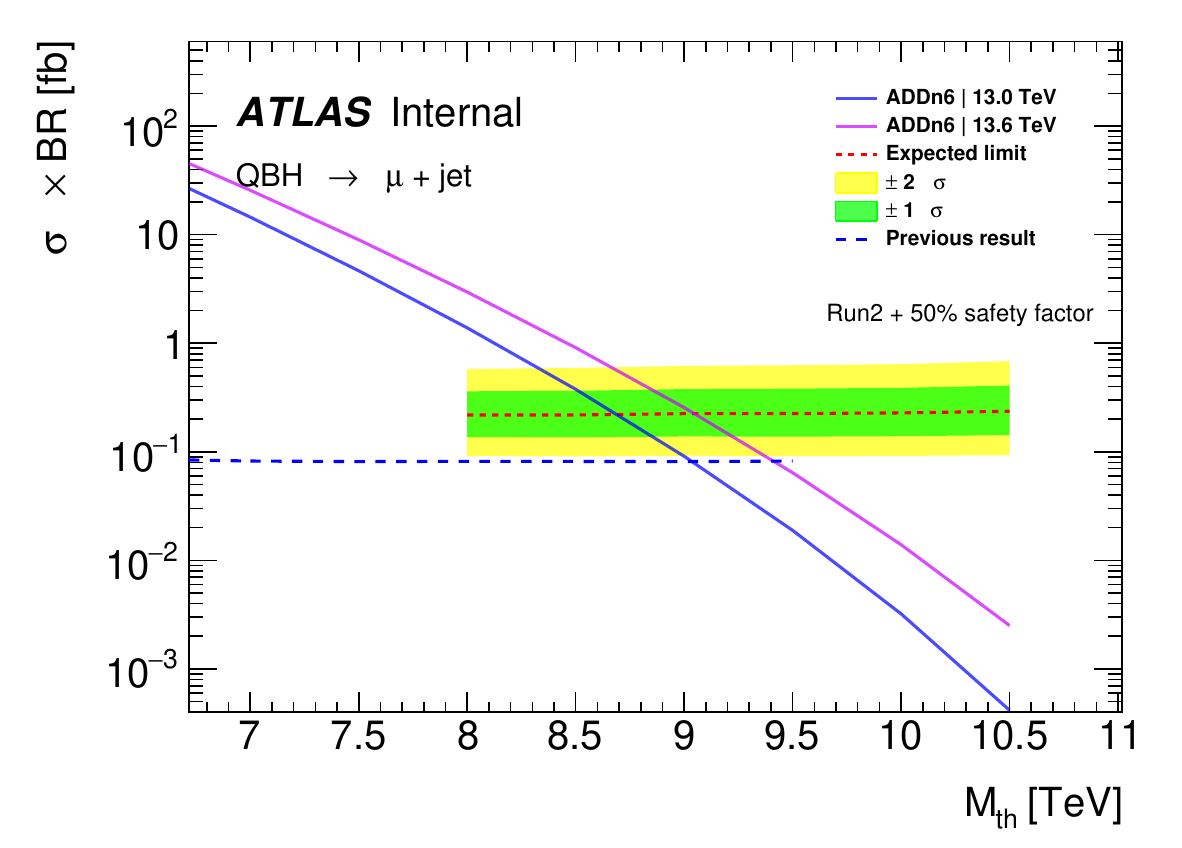}
  }
  \caption{Expected limits for the ADD n=6 model, calculated using Run2 values \texttt{MUON\_EFF\_RECO\_SYS} with an additional safety factor of (a) 0\% and (b) 50\%.}
  \label{fig:MUON_EFF_Limits}
\end{figure}

The systematic \texttt{MUON\_EFF\_RECO\_SYS} was found to be very large, as shown in figure~\ref{fig:MUON_EFF_before}. Following discussions with MCP this is believed to be due to an extremely conservative estimate of this systematic when extrapolating to high $m_{\ell j}$. 
Due to time constraints MCP approved the use of Run2 uncertainties plus an additional saftey factor. The Run2 uncertainties were 0.19\% in the WCRVR and SR, and 0.21\% in the ZCRVR. A study was conducted to see the impact of expected limits of an additional saftey factor. Two examples are shown in figure~\ref{fig:MUON_EFF_Limits} and all limit values summarised in table~\ref{tab:MUON_EFF_limits}. This shows that an effect is only begun to be seen at 100\% saftey factor, therefore approval to use Run2 values without a saftey factor was given. 

\begin{figure}[h]
  \subfloat{
    \includegraphics[width=0.9\textwidth]{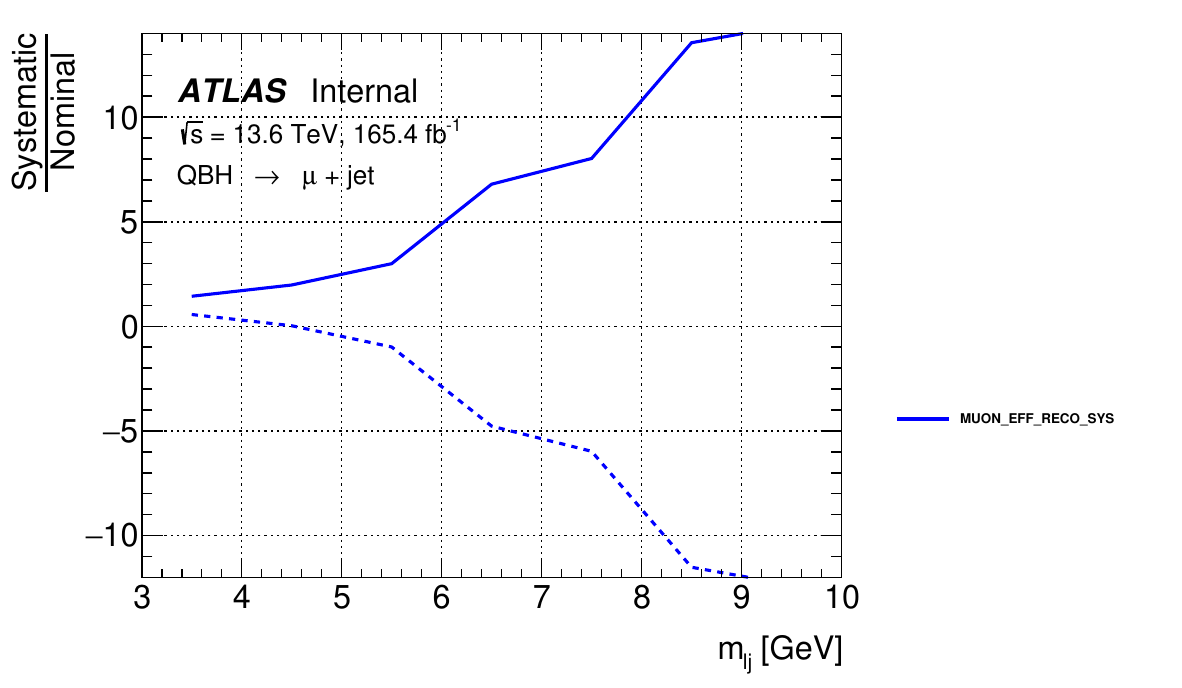}
  }
  \caption{\texttt{MUON\_EFF\_RECO\_SYS} for total background in the signal region for the $\mu+j$ channel, before solution is applied.}
  \label{fig:MUON_EFF_before}
\end{figure}

\begin{table}[h]
    \centering
    \begin{tabular}{c|ccccc}
        \hline
        Safety factor: & 0\% & 10\% & 20\% & 50\% & 100\% \\
        \hline
        M\_th & \multicolumn{5}{c}{$\sigma \times BR$ [fb]} \\
        \hline
        8.0  & 0.2179 & 0.2176 & 0.2178 & 0.2180 & 0.2166 \\
        8.5  & 0.2183 & 0.2181 & 0.2182 & 0.2185 & 0.2171 \\
        9.0  & 0.2244 & 0.2241 & 0.2243 & 0.2245 & 0.2231 \\
        9.5  & 0.2249 & 0.2250 & 0.2247 & 0.2250 & 0.2236 \\
        10.0 & 0.2279 & 0.2277 & 0.2278 & 0.2281 & 0.2267 \\
        10.5 & 0.2360 & 0.2362 & 0.2360 & 0.2362 & 0.2347 \\
        \hline
    \end{tabular}
    \caption{Expected limits for the ADD n=6 model, calculated using Run2 values \texttt{MUON\_EFF\_RECO\_SYS} with varying additional safety factors.}
    \label{tab:MUON_EFF_limits}
\end{table}

\section{Electron}

The EGamma CP group have found an issue with their calorimeter isolaiton, which affects the systematic \texttt{EL\_EFF\_Iso\_TOTAL}. This issue was found to have a large effect on the HighPtCaloOnly working point of up to a 10\% difference at high $p_T$. FIG.~\ref{fig:MUON_EFF_before} shows this systematic in our signal region, which is always less than ~10\%. Therefore a solution to use flat 20\% systematic was agreed as covers any possibe changes from the bug. Again the effect of this on the limits was checked, this is summarised in table~\ref{tab:EL_ISO_limits}, which shows there is no significant effect.  

\begin{figure}[h]
  \subfloat{
    \includegraphics[width=0.9\textwidth]{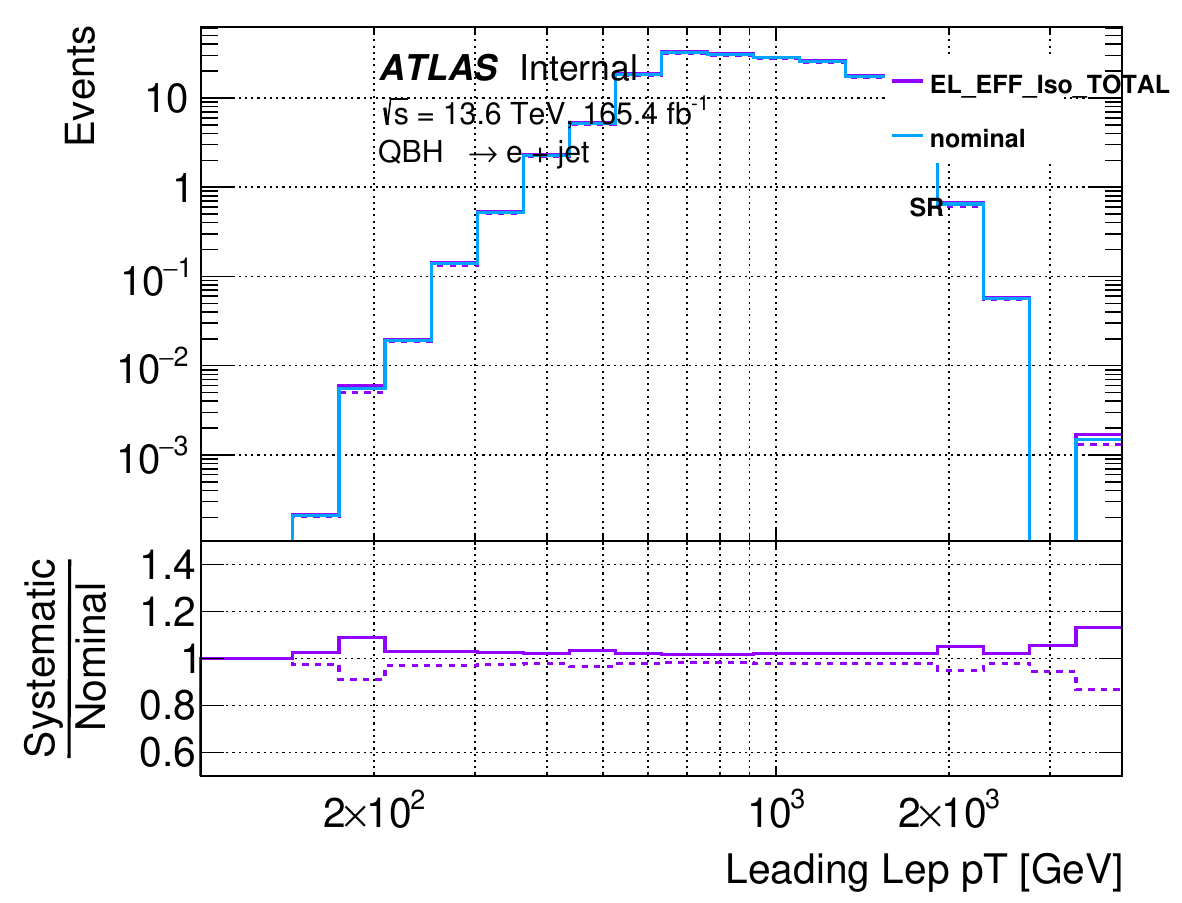}
  }
  \caption{\texttt{EL\_EFF\_Iso\_TOTAL} for total background in the signal region before solution is applied.}
  \label{fig:MUON_EFF_before}
\end{figure}

\begin{table}[h]
\centering
\begin{tabular}{c|cc}
\hline
 & Original & Fixed 20\% \\
\hline
M\_th & \multicolumn{2}{c}{$\sigma \times BR$ [fb]} \\
\hline
8.0  & 0.0189851 & 0.0188353 \\
8.5  & 0.0188295 & 0.0186812 \\
9.0  & 0.0187821 & 0.0186325 \\
9.5  & 0.0188229 & 0.0186728 \\
10.0 & 0.0186615 & 0.0185152 \\
10.5 & 0.0187193 & 0.0185798 \\
\hline
\end{tabular}
\caption{Expected limits for the ADD n=6 model, calculated using original \texttt{EL\_EFF\_Iso\_TOTAL} systematic vs a flat 20\% uncertainty.}
\label{tab:EL_ISO_limits}
\end{table}

\section{Per-object Systematic Uncertainty Breakdowns}
\label{app:qbh:exp_syst_plots}

The figures below show the per-object contributions to the experimental systematic uncertainty
for the total background in each analysis region, complementing the grouped summaries shown in
Section~\ref{sec:Syst_exp}.

\begin{figure}[h]
  \subfloat{
    \includegraphics[width=0.5\textwidth]{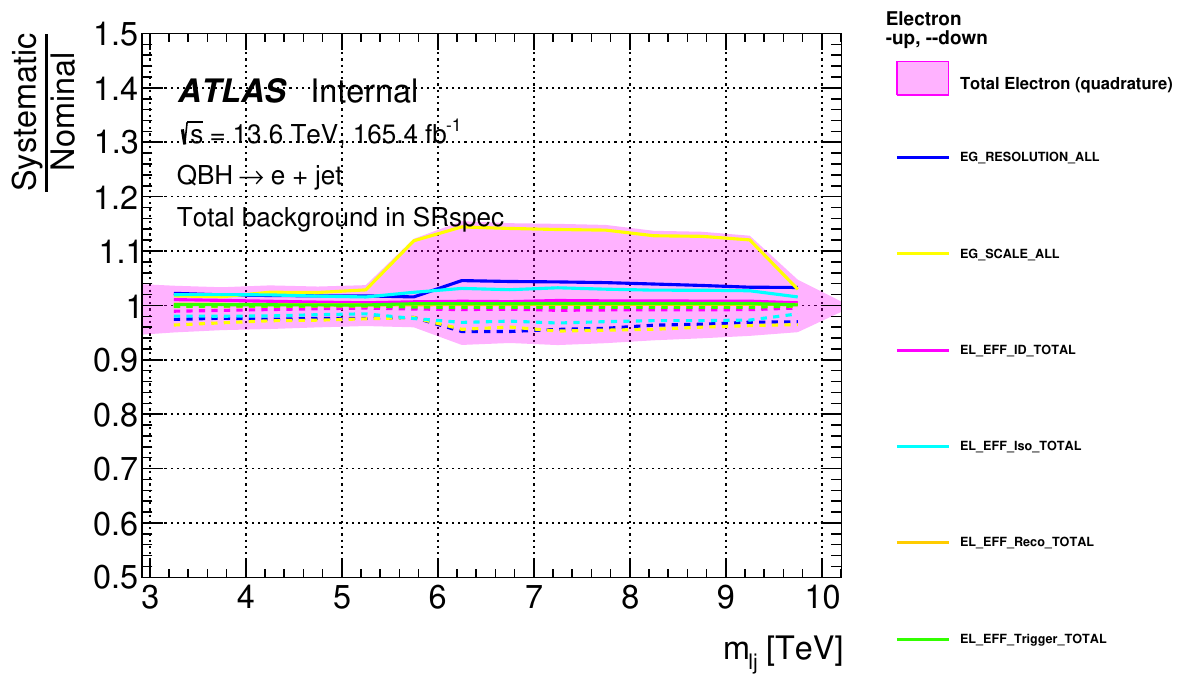}
  }
  \subfloat{
    \includegraphics[width=0.5\textwidth]{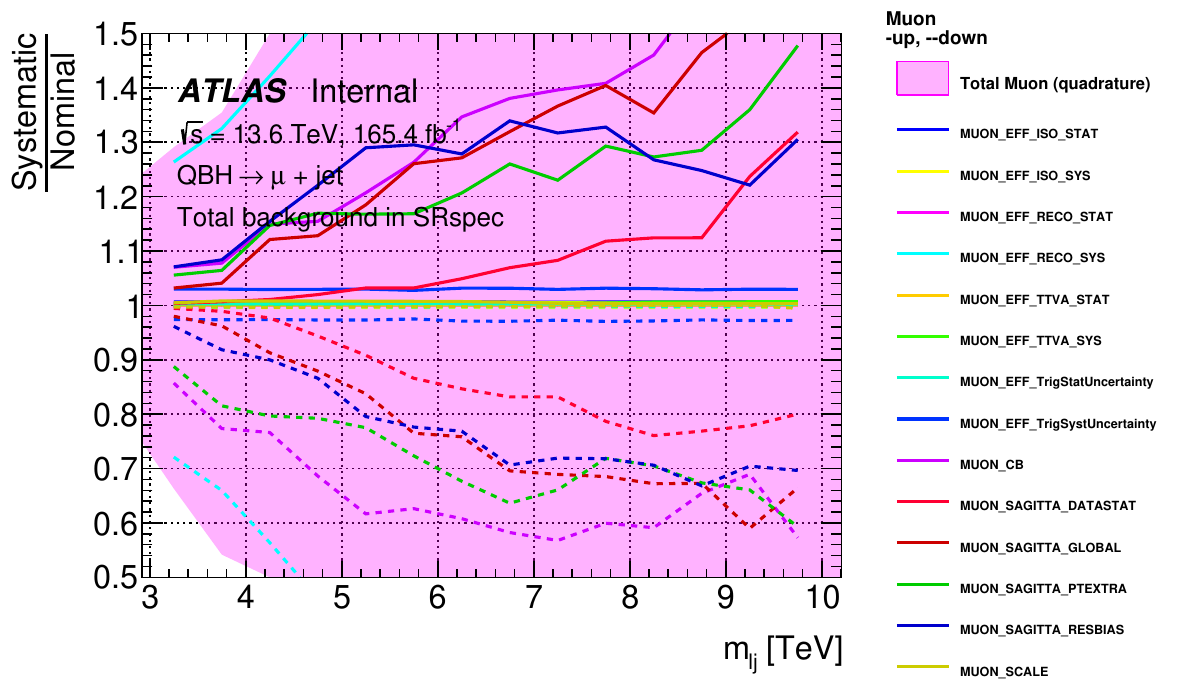}
  }
  \caption{Contributions to electron (left) and muon (right) systematics for the total background in the signal region.}
  \label{fig:SR_Lep_exp_syst}
\end{figure}

\begin{figure}[h]
  \subfloat{
    \includegraphics[width=0.5\textwidth]{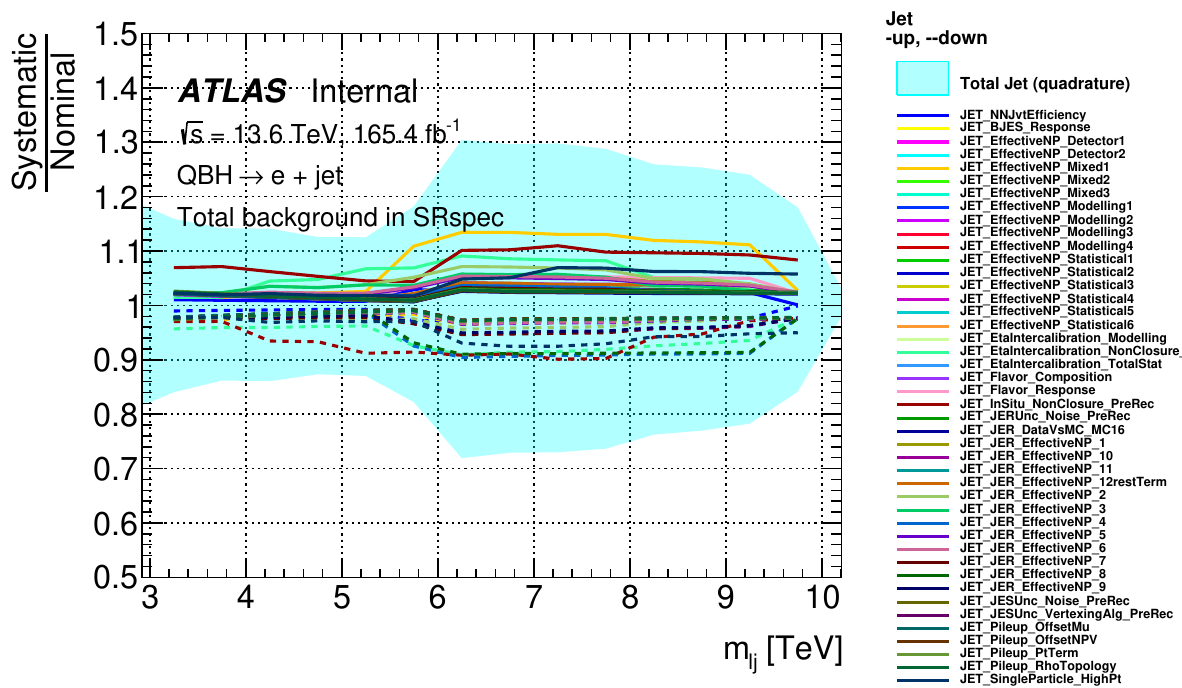}
  }
  \subfloat{
    \includegraphics[width=0.5\textwidth]{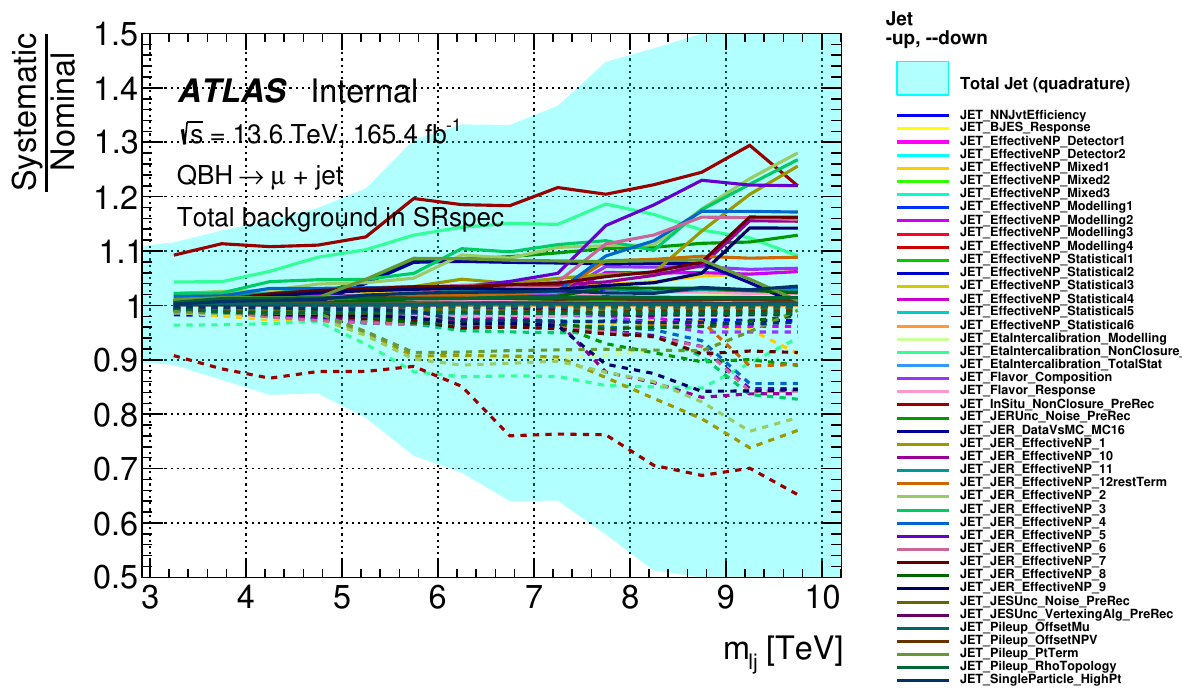}
  }
  \caption{Contributions to jet systematics for the electron (left) and muon (right) channel for the total background in the signal region.}
  \label{fig:SR_Jet_exp_syst}
\end{figure}

\begin{figure}[h]
  \subfloat{
    \includegraphics[width=0.5\textwidth]{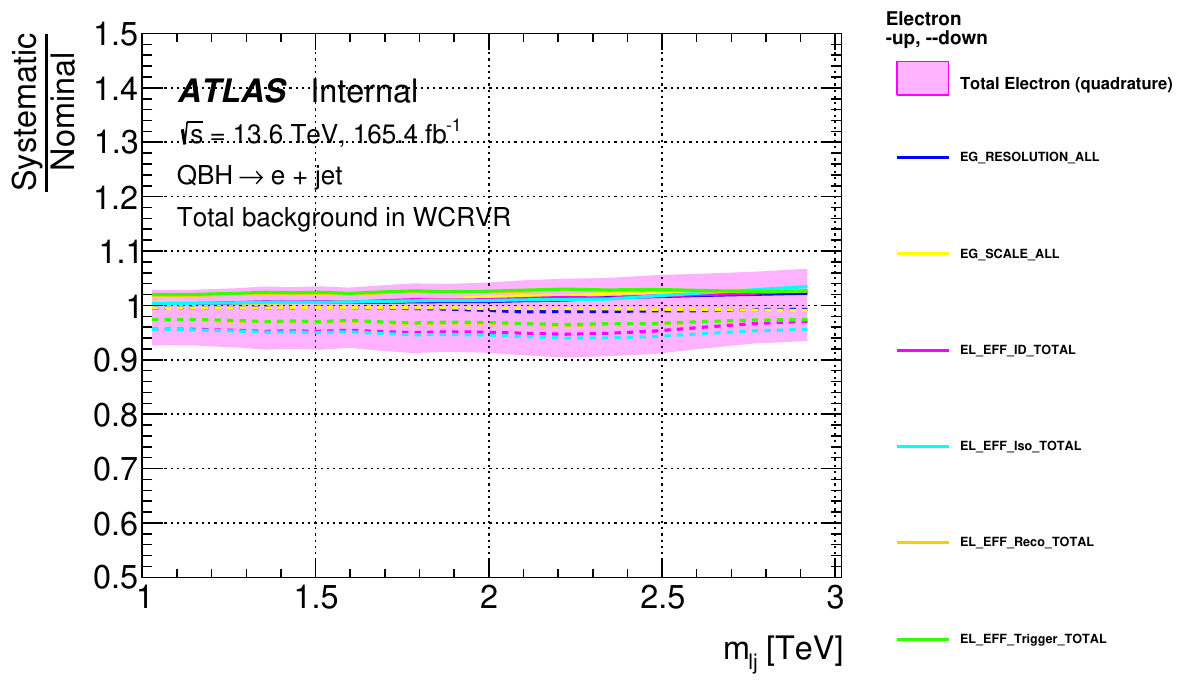}
  }
  \subfloat{
    \includegraphics[width=0.5\textwidth]{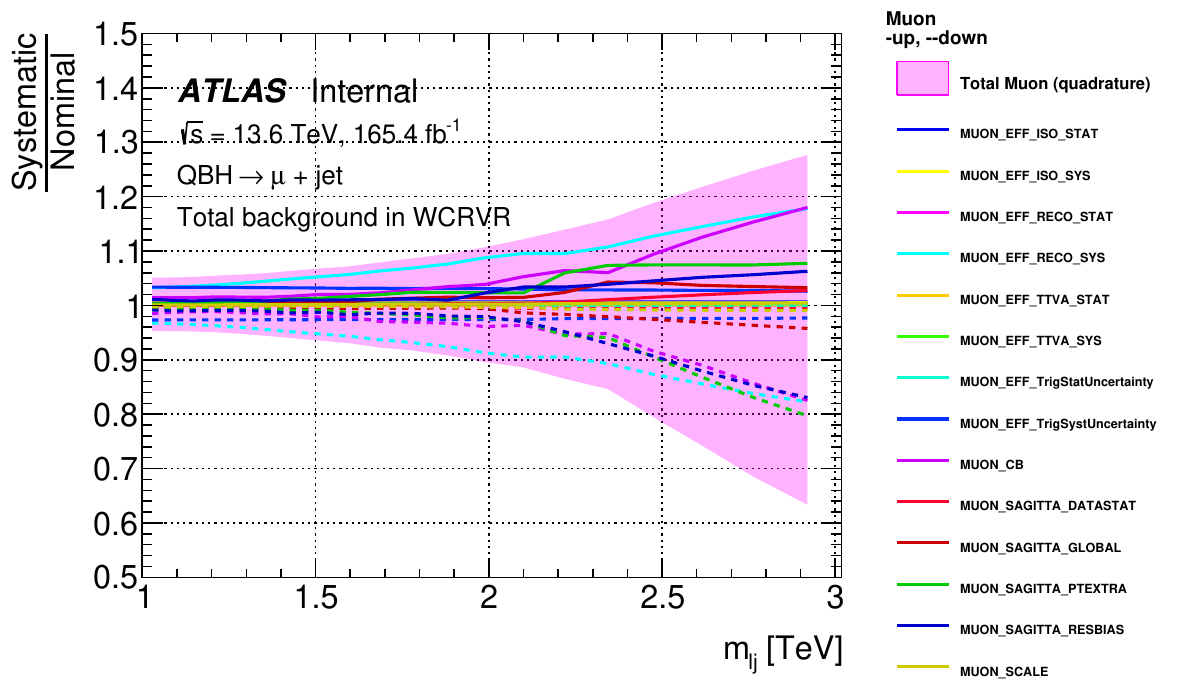}
  }
  \caption{Contributions to electron (left) and muon (right) scale factor systematics for the total background in the $W$+jets control and validation regions.}
  \label{fig:WCRVR_Lep_exp_syst}
\end{figure}

\begin{figure}[h]
  \subfloat{
    \includegraphics[width=0.5\textwidth]{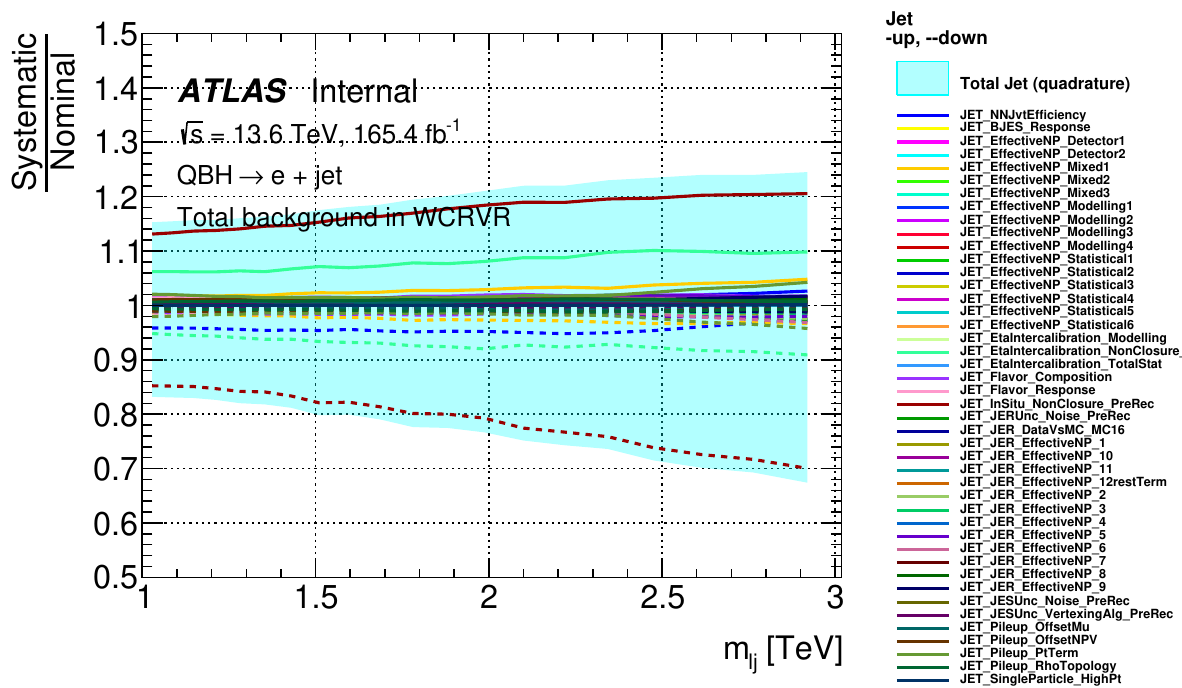}
  }
  \subfloat{
    \includegraphics[width=0.5\textwidth]{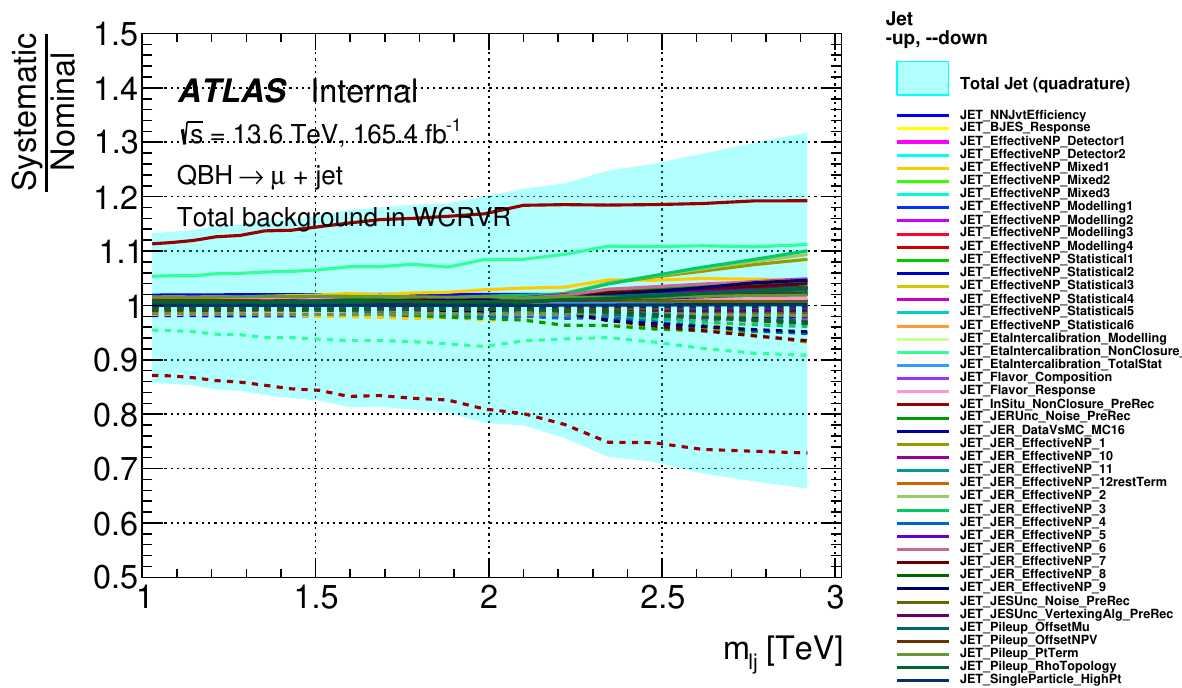}
  }
  \caption{Contributions to jet systematics for the electron (left) and muon (right) channel for the total background in the $W$+jets control and validation regions.}
  \label{fig:WCRVR_Jet_exp_syst}
\end{figure}

\begin{figure}[h]
  \subfloat{
    \includegraphics[width=0.5\textwidth]{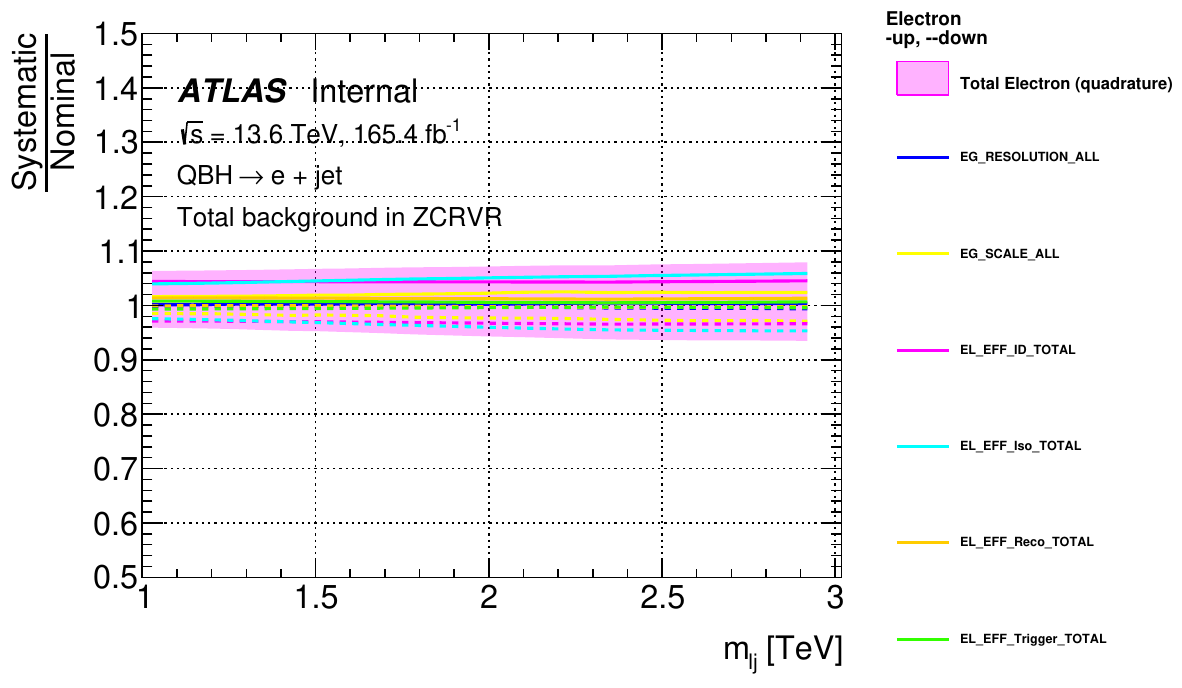}
  }
  \subfloat{
    \includegraphics[width=0.5\textwidth]{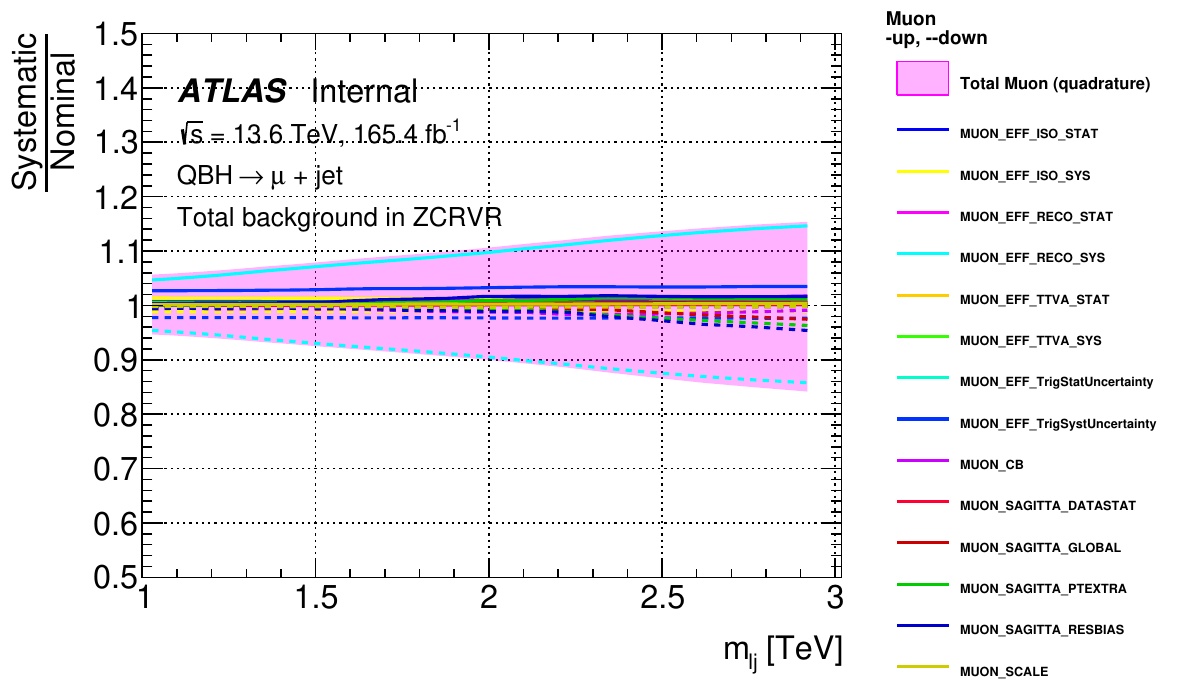}
  }
  \caption{Contributions to electron (left) and muon (right) scale factor systematics for the total background in the $Z$+jets control and validation regions.}
  \label{fig:ZCRVR_Lep_exp_syst}
\end{figure}

\begin{figure}[h]
  \subfloat{
    \includegraphics[width=0.5\textwidth]{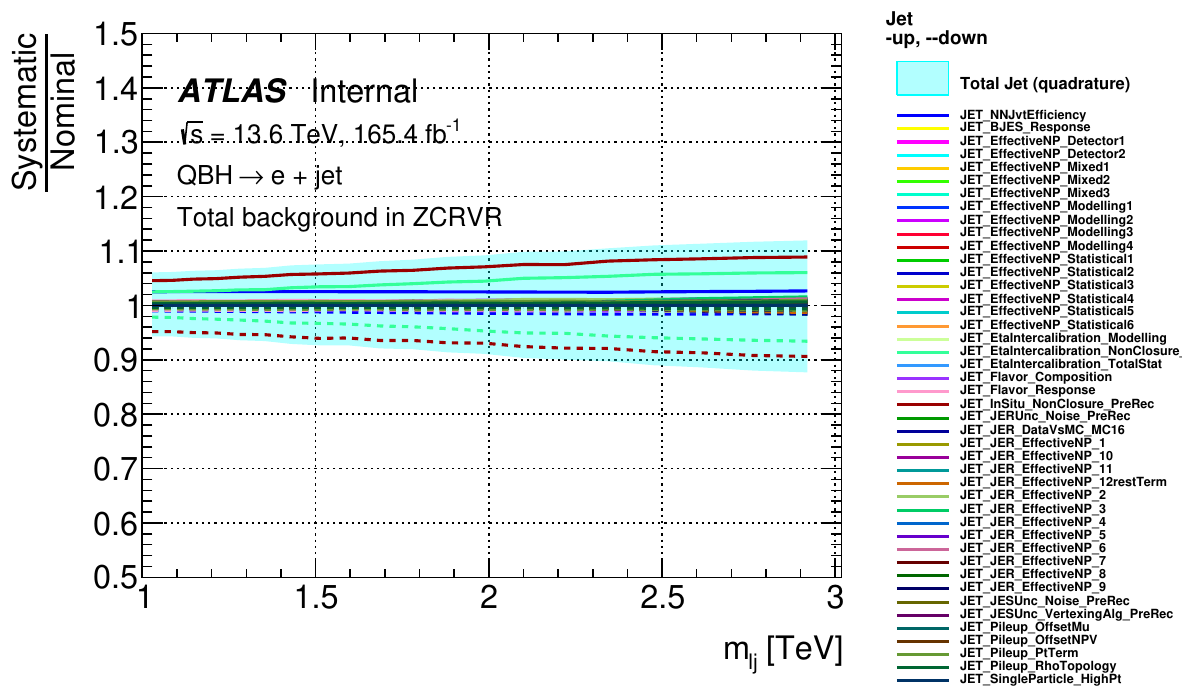}
  }
  \subfloat{
    \includegraphics[width=0.5\textwidth]{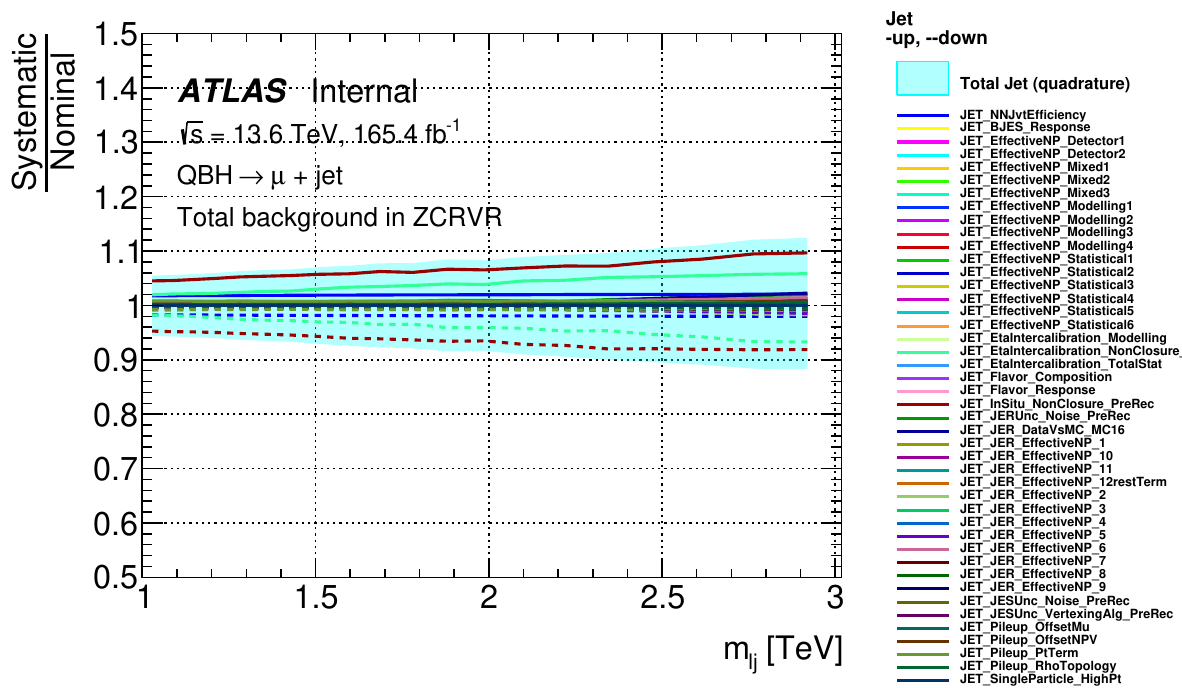}
  }
  \caption{Contributions to jet systematics for the electron (left) and muon (right) channel for the total background in the $Z$+jets control and validation regions.}
  \label{fig:ZCRVR_Jet_exp_syst}
\end{figure}

\FloatBarrier

\section{$Z$+jets Control and Validation Region Distributions}
\label{app:qbh:zcrvr}

\begin{figure}[h]
  \captionsetup[subfigure]{labelformat=empty}
  \subfloat{
    \includegraphics[width=0.5\textwidth]{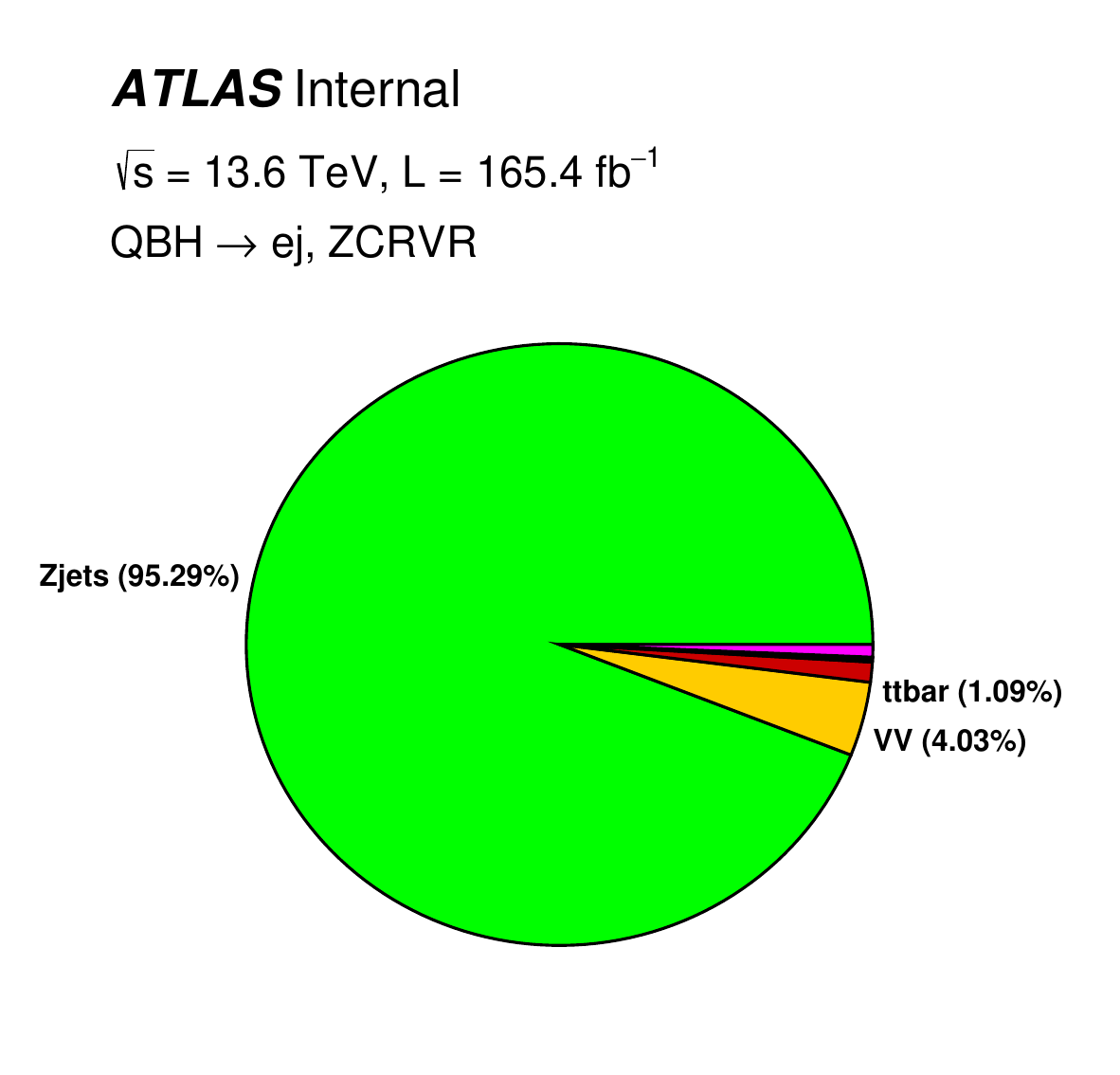}
  }
  \subfloat{
    \includegraphics[width=0.5\textwidth]{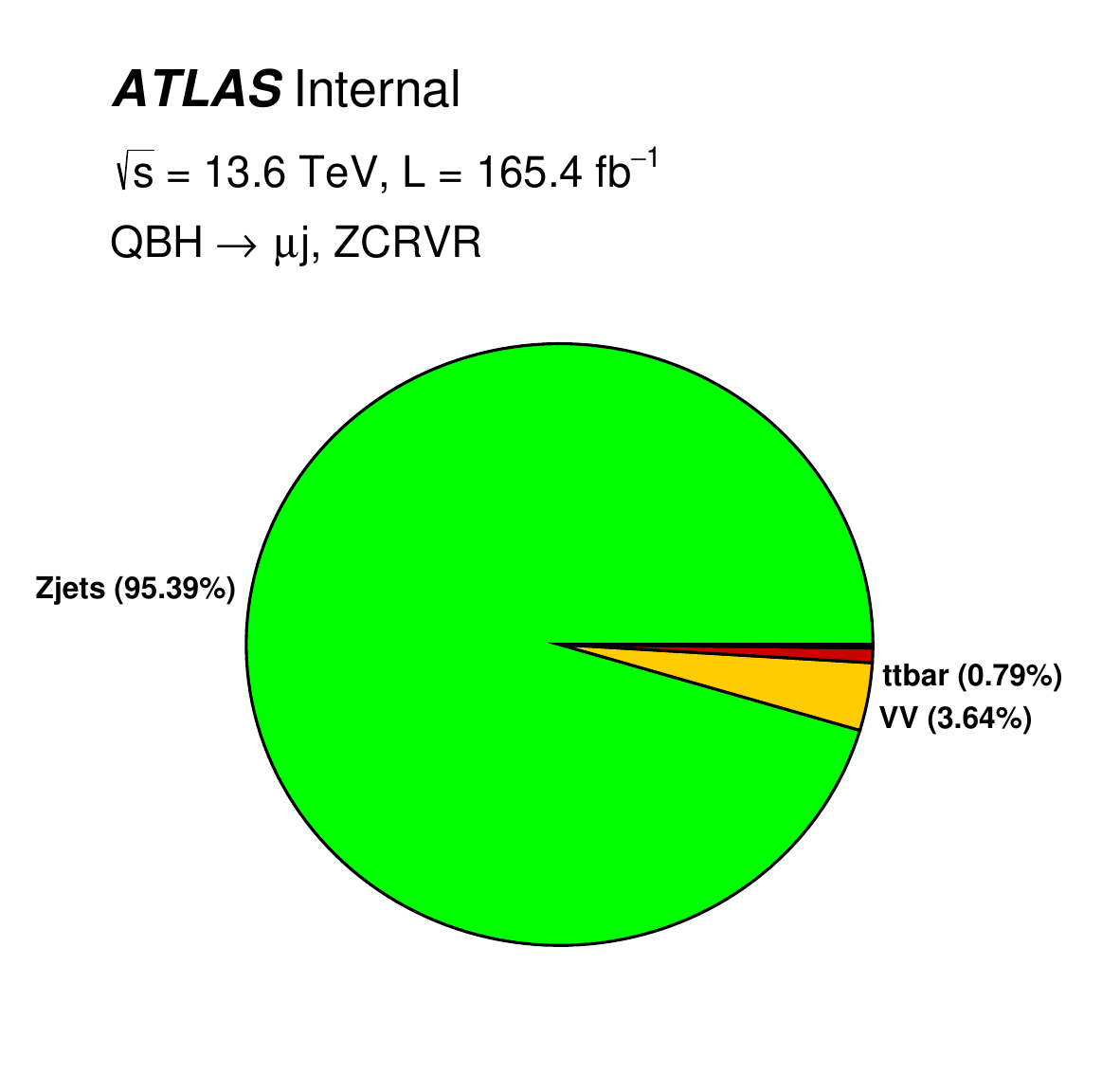}
  }
  \caption{Background composition pie charts in the $Z$+jets control and validation regions for the $e+j$ channel (left) and $\mu+j$ channel (right). The $Z$+jets process constitutes above 95\% of the total background in both channels.}
  \label{fig:Pie_Charts_ZCRVR}
\end{figure}

\begin{figure}[h]
  \captionsetup[subfigure]{labelformat=empty}
  \subfloat{
    \includegraphics[width=0.5\textwidth]{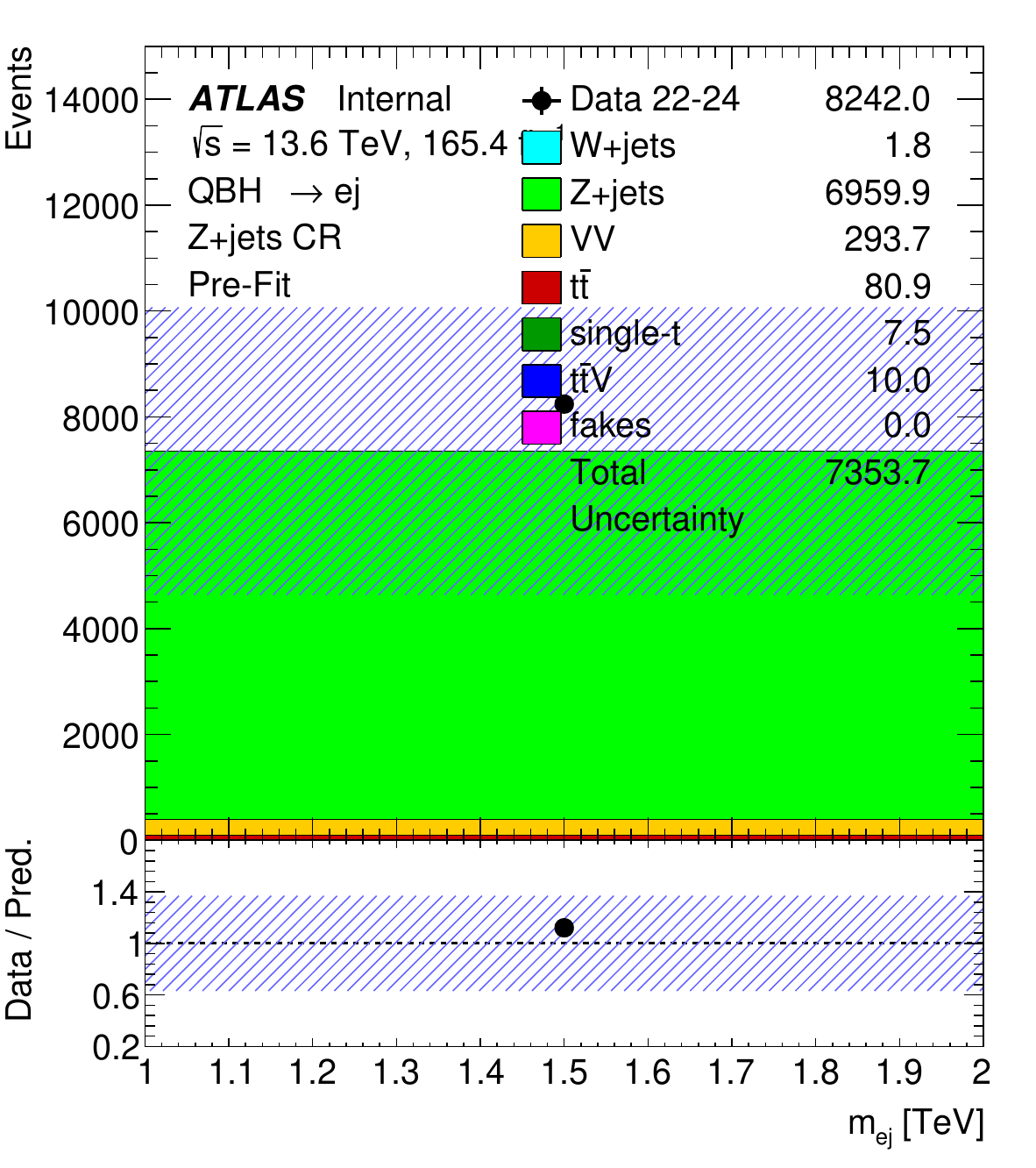}
  }
  \hfill
  \subfloat{
    \includegraphics[width=0.5\textwidth]{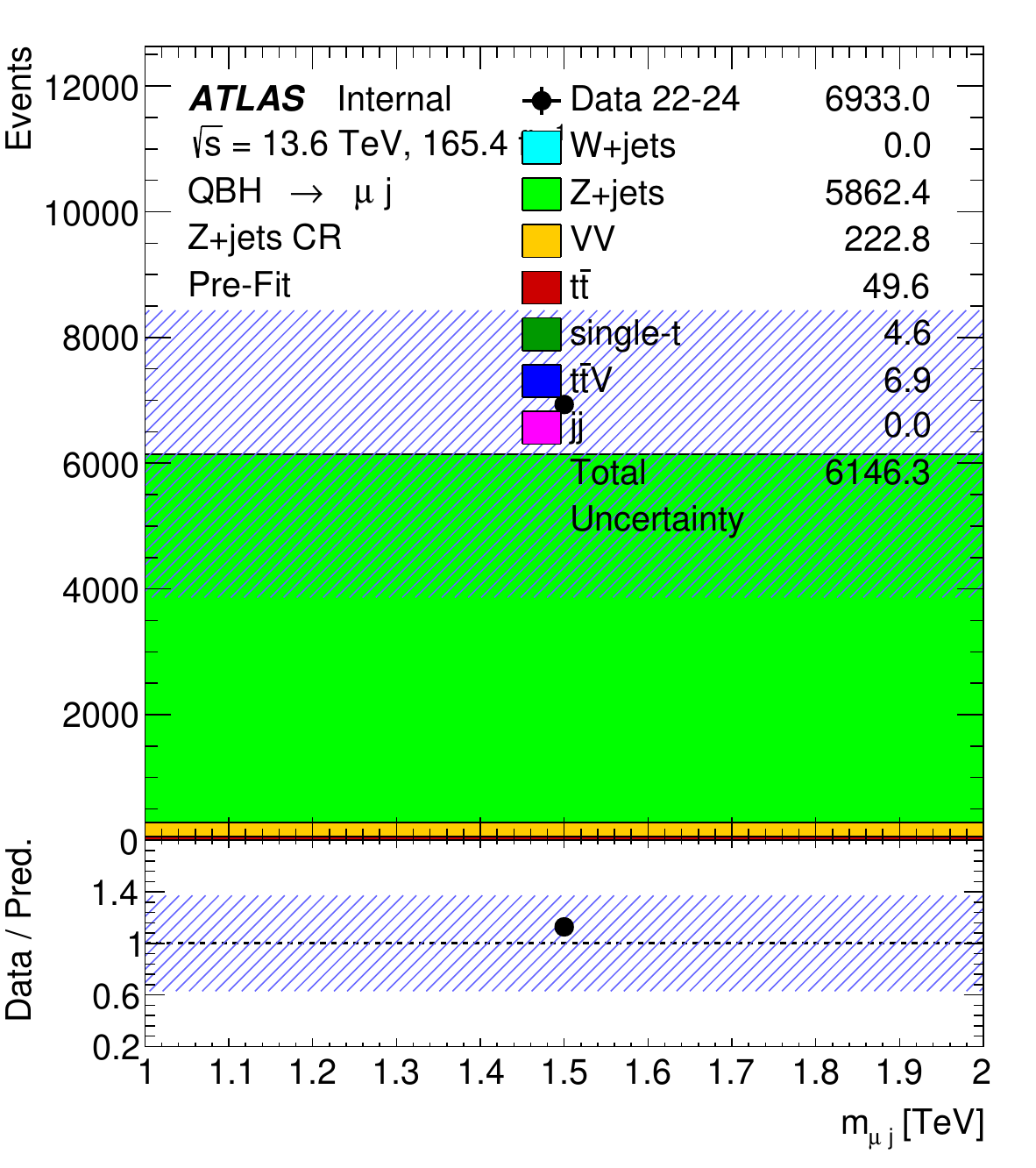}
  }
  \hfill
  \subfloat{
    \includegraphics[width=0.5\textwidth]{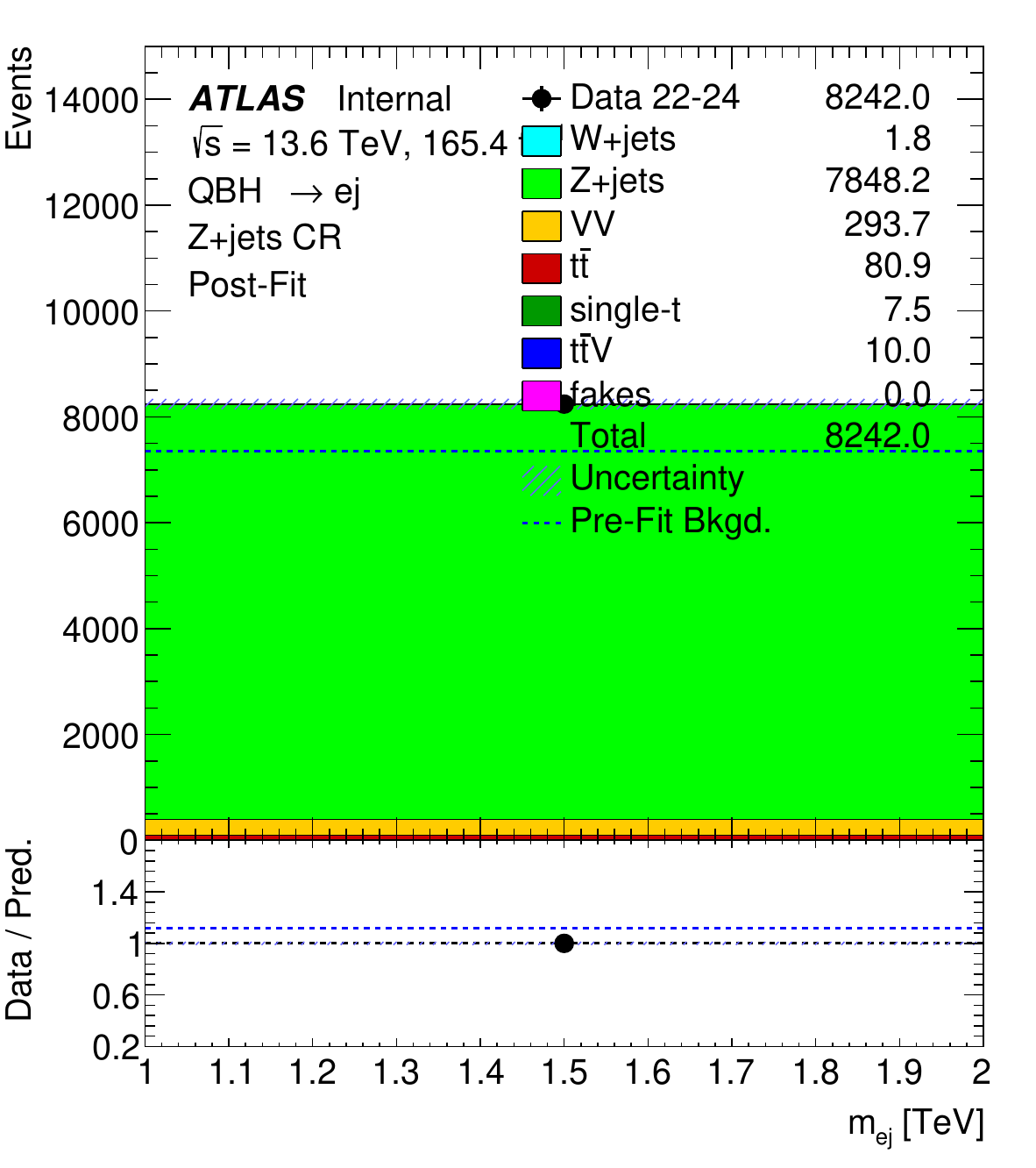}
  }
  \hfill
  \subfloat{
    \includegraphics[width=0.5\textwidth]{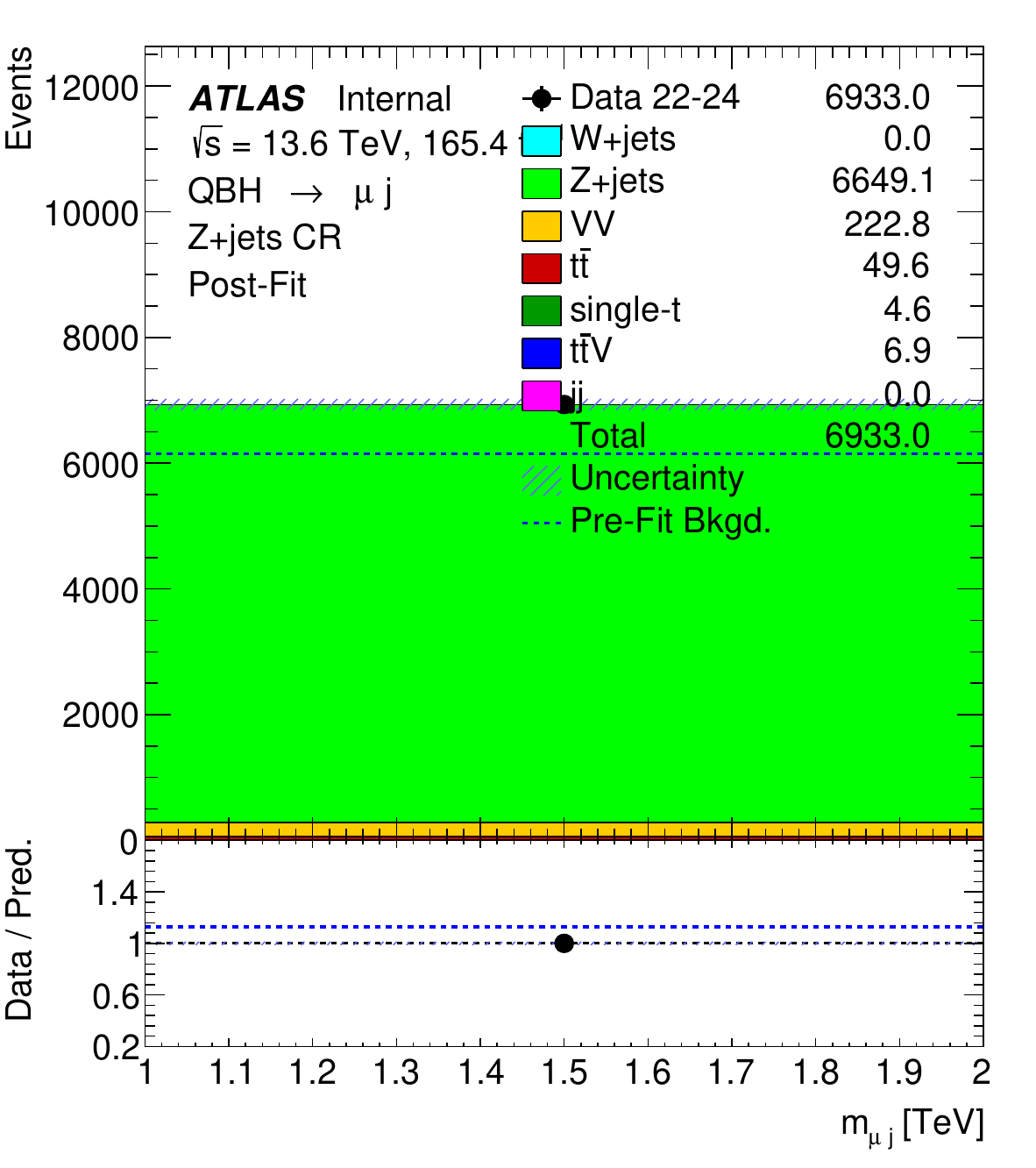}
  }
  \caption{Invariant mass of the lepton-jet pair in the single-binned $Z$+jets Control Region for the $e+j$ (left) and $\mu+j$ (right) channels, pre-fit (top) and post-fit (bottom).}
  \label{fig:Zjets_CR}
\end{figure}

\begin{figure}[h]
  \captionsetup[subfigure]{labelformat=empty}
  \subfloat{
    \includegraphics[width=0.5\textwidth]{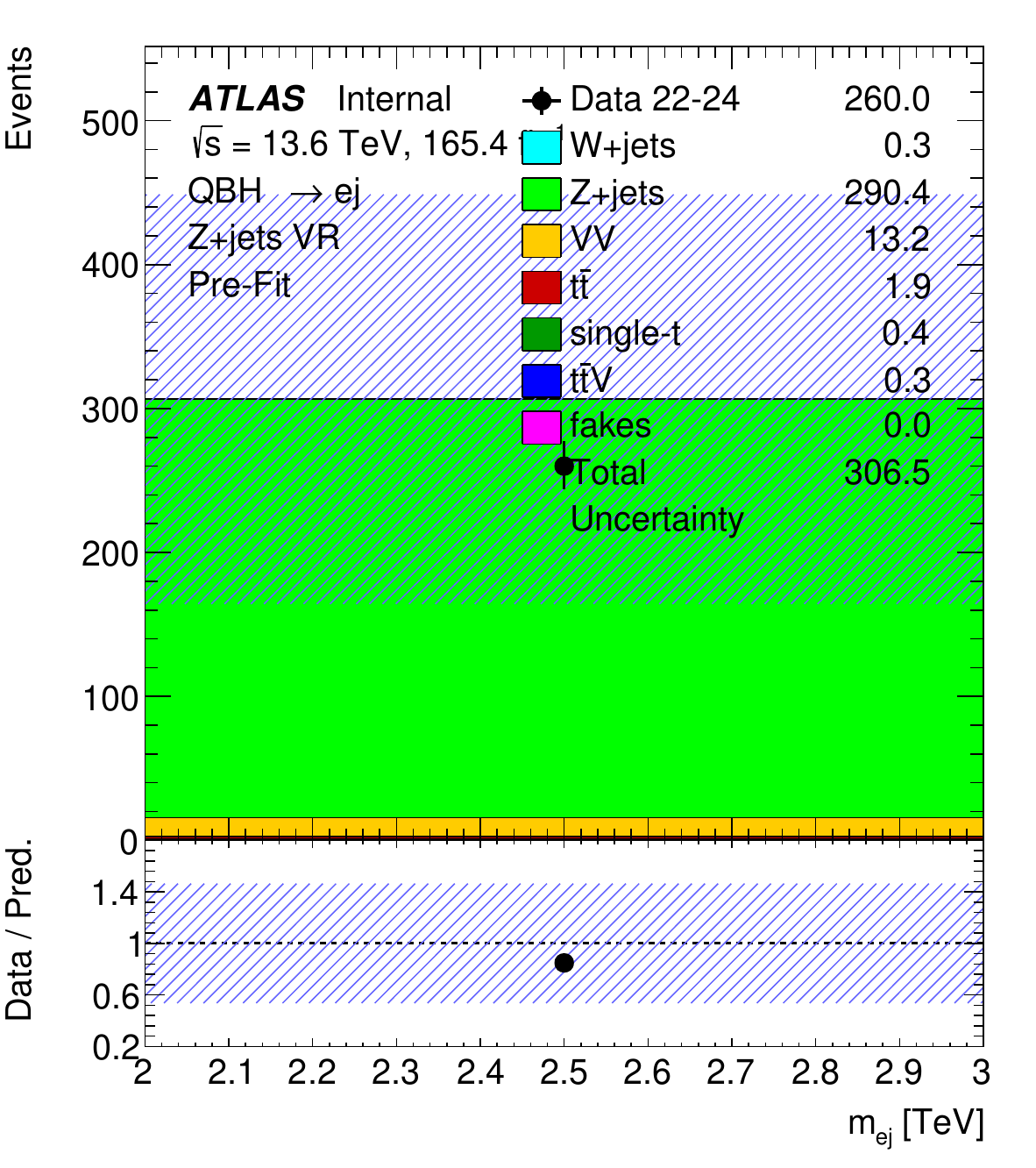}
  }
  \hfill
  \subfloat{
    \includegraphics[width=0.5\textwidth]{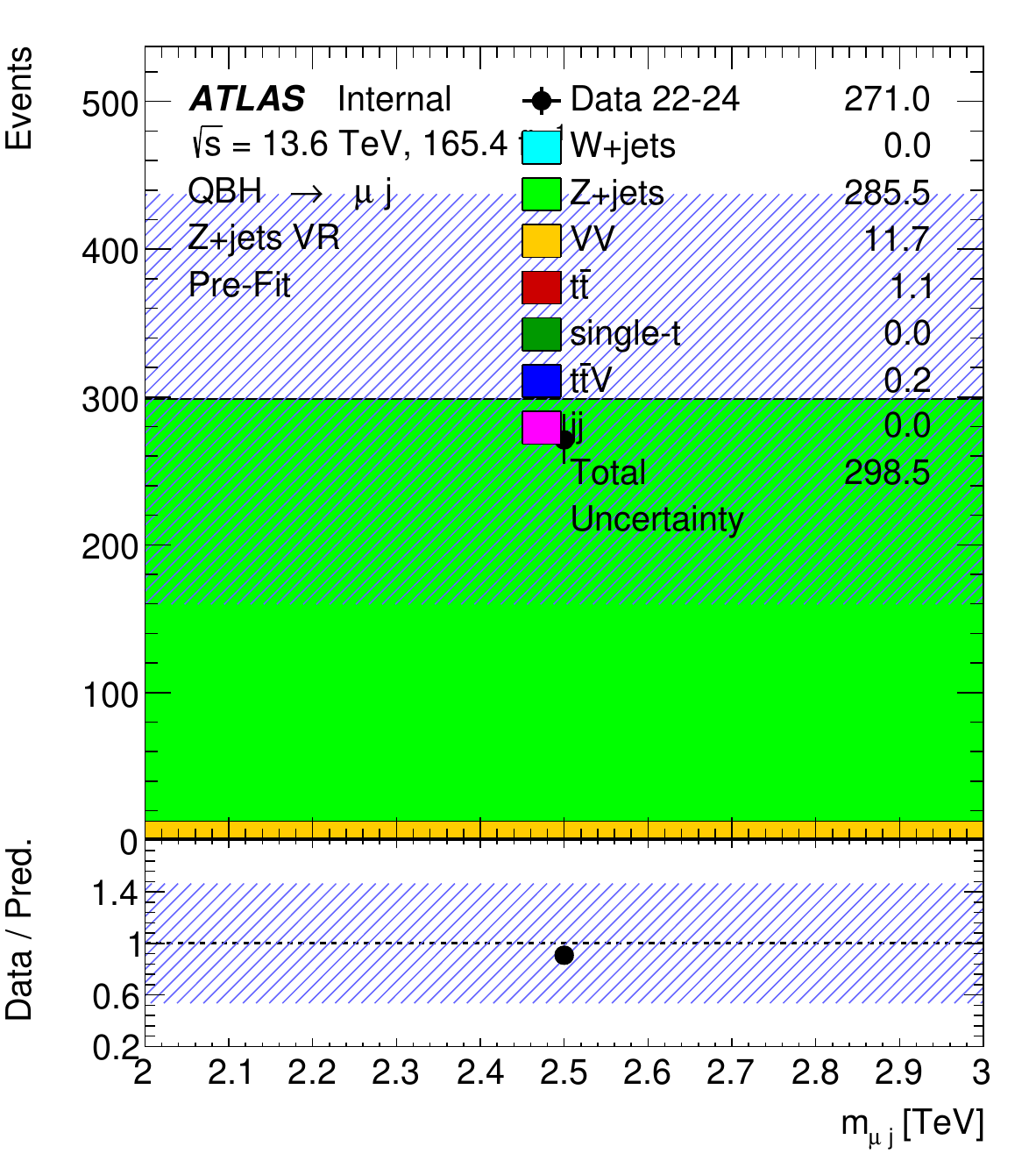}
  }
  \hfill
  \subfloat{
    \includegraphics[width=0.5\textwidth]{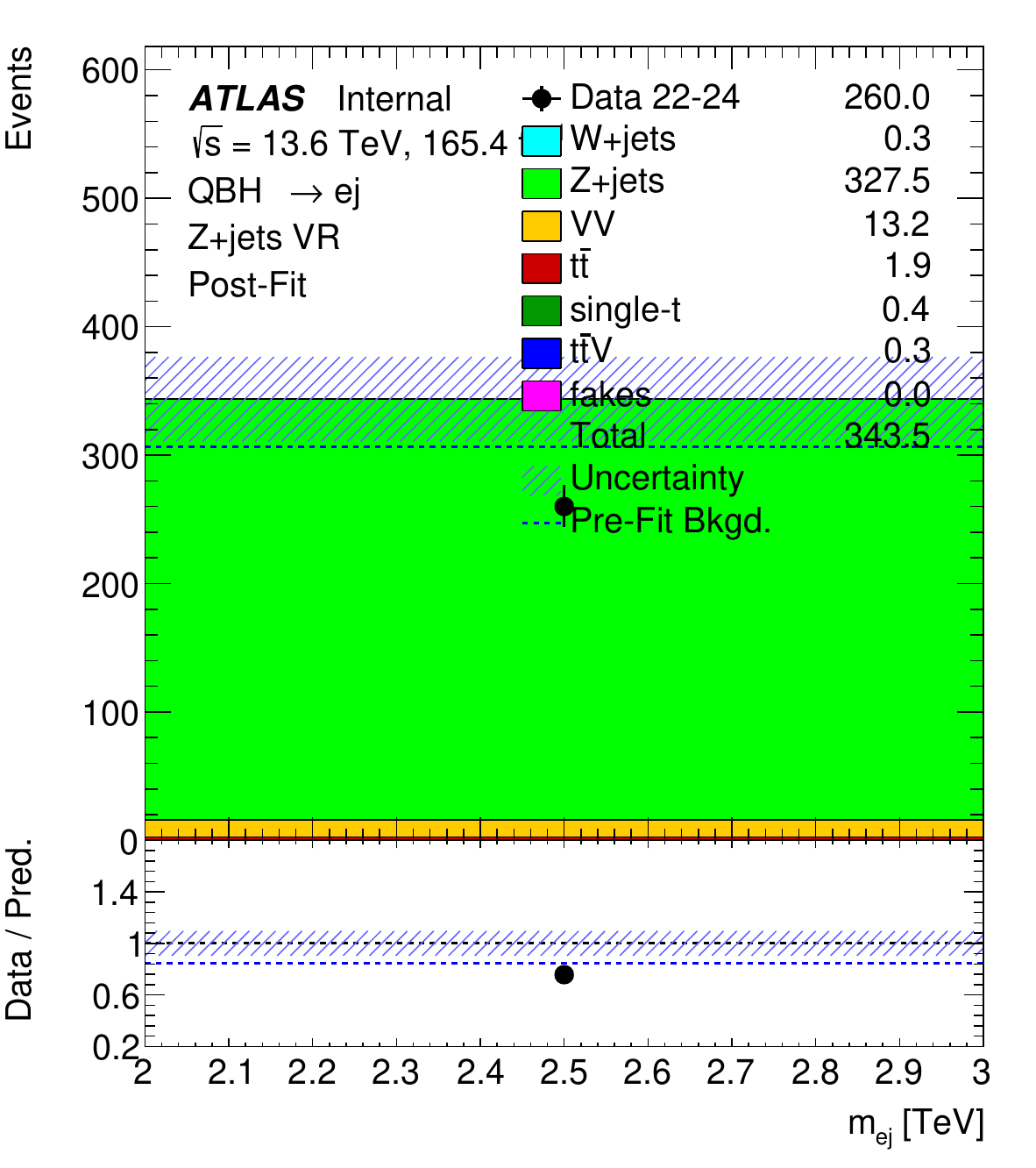}
  }
  \hfill
  \subfloat{
    \includegraphics[width=0.5\textwidth]{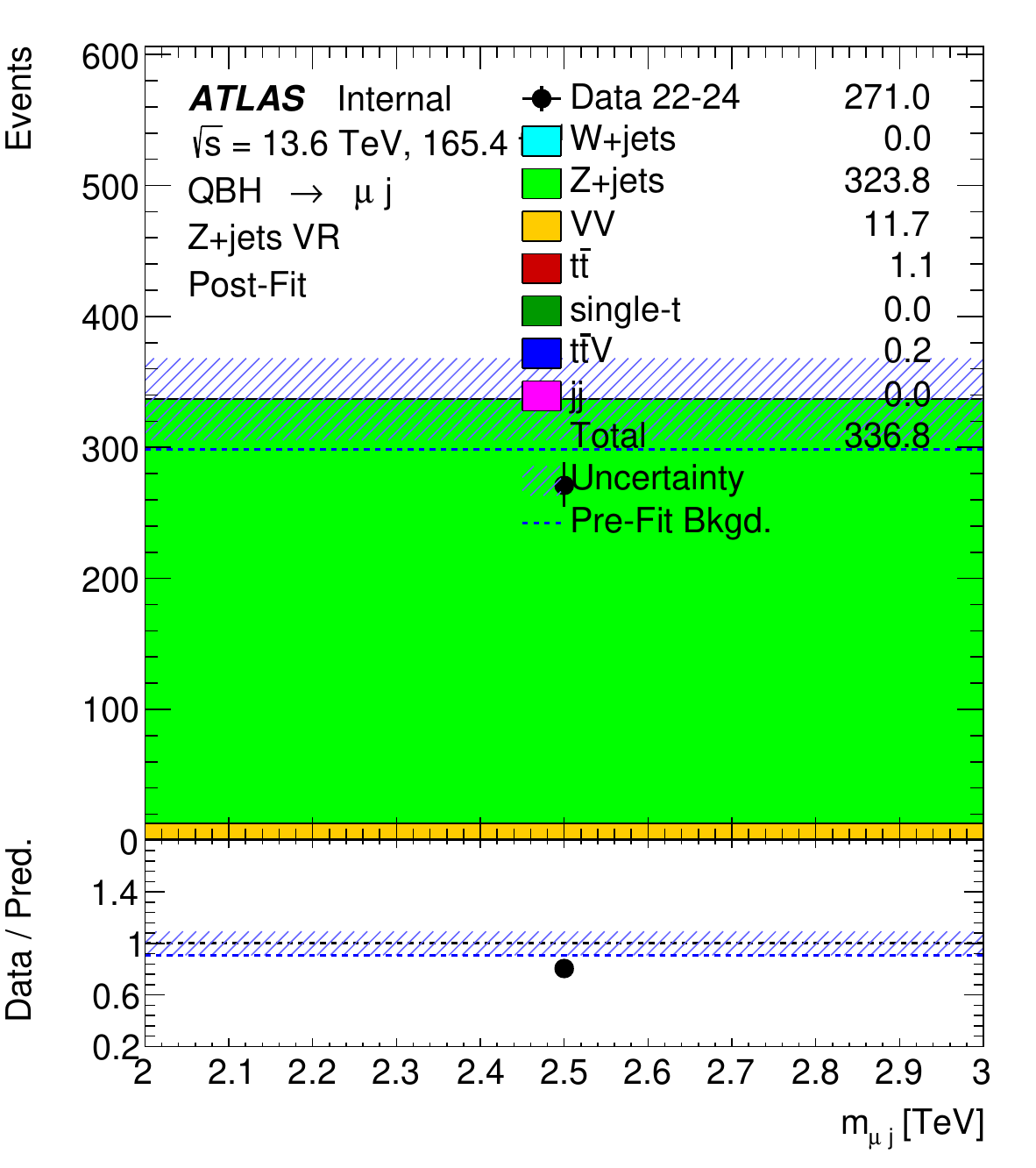}
  }
  \caption{Invariant mass of the lepton-jet pair in the single-binned $Z$+jets Validation Region for the $e+j$ (left) and $\mu+j$ (right) channels, pre-fit (top) and post-fit (bottom).}
  \label{fig:Zjets_VR}
\end{figure}

\FloatBarrier

\section{Fake Electron Systematic Uncertainty Distributions}
\label{app:qbh:fakes_systs}

The figures below show the effect of each class of systematic uncertainty on the total electroweak MC background as a function of $m_{ej}$ in the fake Control Region, used as a diagnostic to gauge the general size of each contribution. The actual propagation to the fake rate is done bin-by-bin in $(p_{\textrm{T}},|\eta|)$, as described in Section~\ref{sec:fakes}.

\begin{figure}[h!]
  \captionsetup[subfigure]{labelformat=empty}
  \subfloat[]{
    \includegraphics[width=0.5\textwidth]{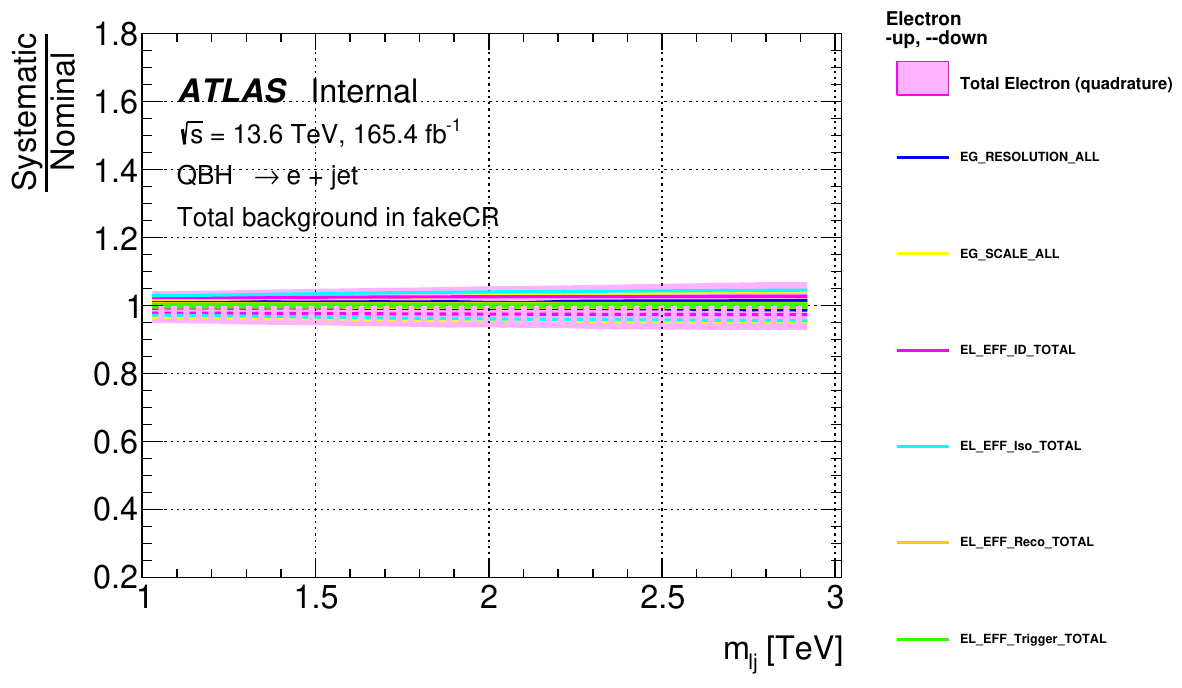}
  }
  \hfill
  \subfloat[]{
    \includegraphics[width=0.5\textwidth]{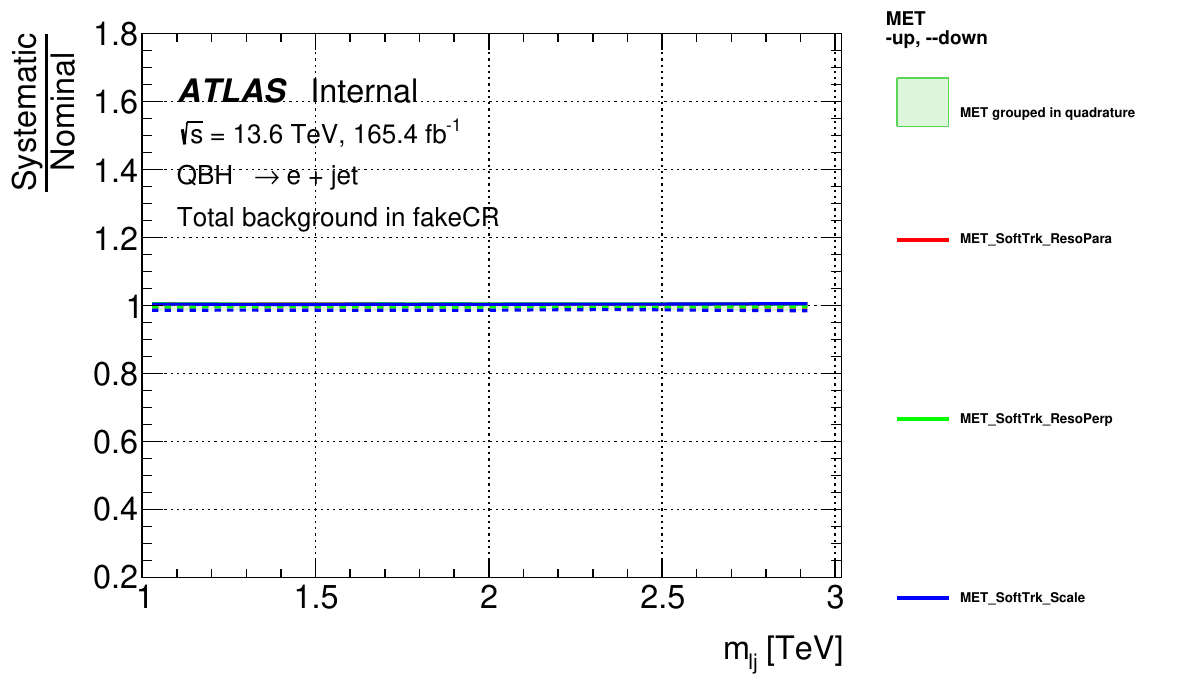}
  }
  \hfill
  \subfloat[]{
    \includegraphics[width=0.5\textwidth]{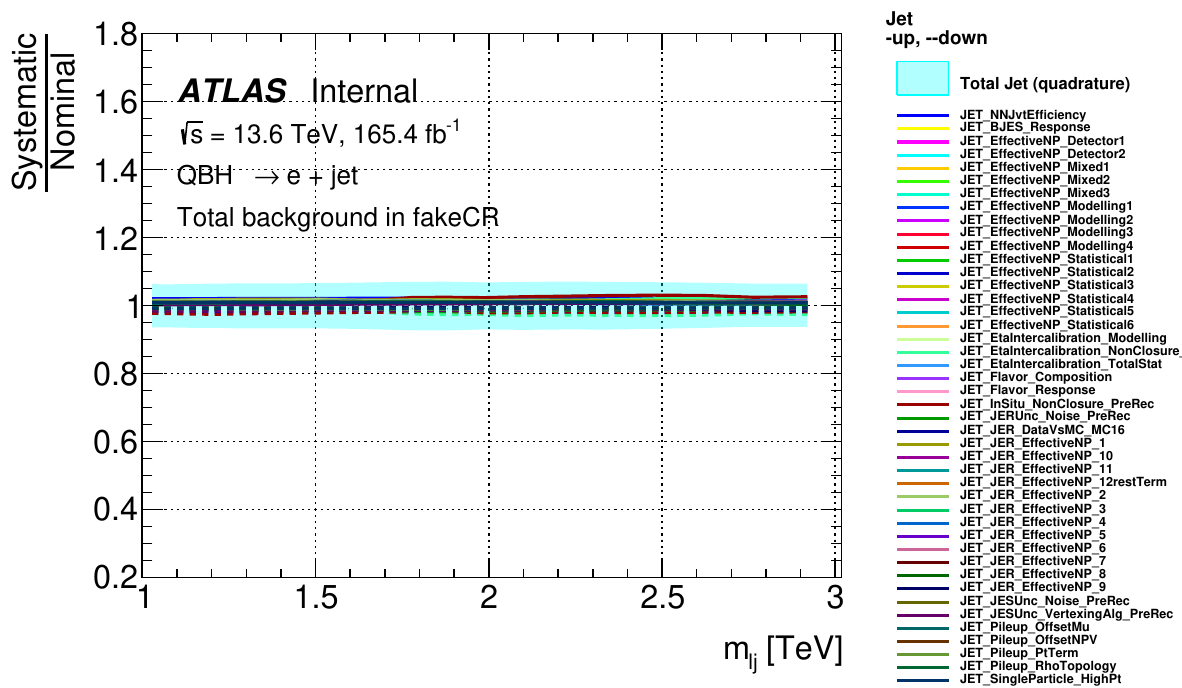}
  }
  \hfill
  \subfloat[]{
    \includegraphics[width=0.45\textwidth]{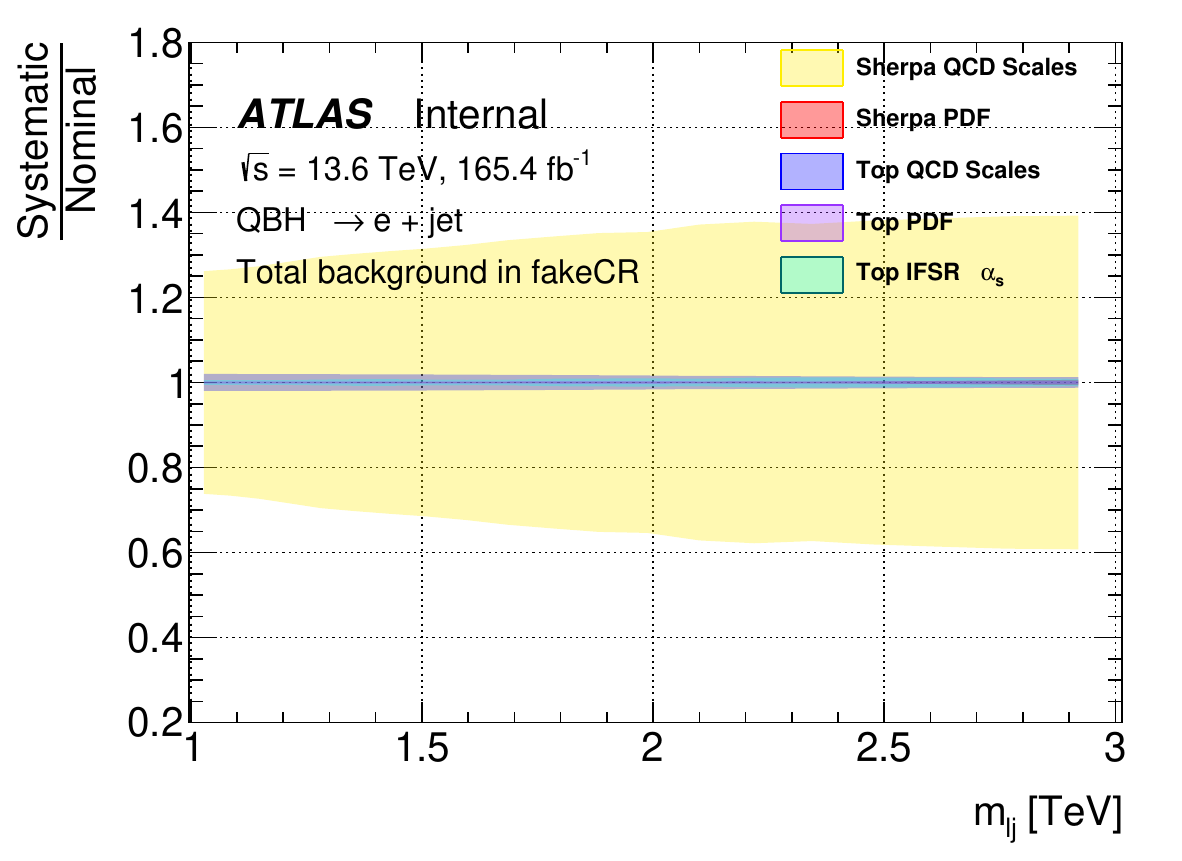}
  }
  \caption{Effect of systematic uncertainties on the total electroweak background in the fake Control Region, shown as a function of the invariant electron-jet mass $m_{ej}$. (Top left) electron systematics; (top right) $E^{\textrm{miss}}_{\textrm{T}}$ systematics; (bottom left) jet systematics; (bottom right) theory systematics.}
  \label{fig:Fake_systematics_mlj_CR}
\end{figure}

\FloatBarrier

\section{Fit Diagnostics}
\label{app:qbh:fit_diagnostics}

\begin{figure}[h]
    \subfloat{
      \includegraphics[width=0.45\textwidth]{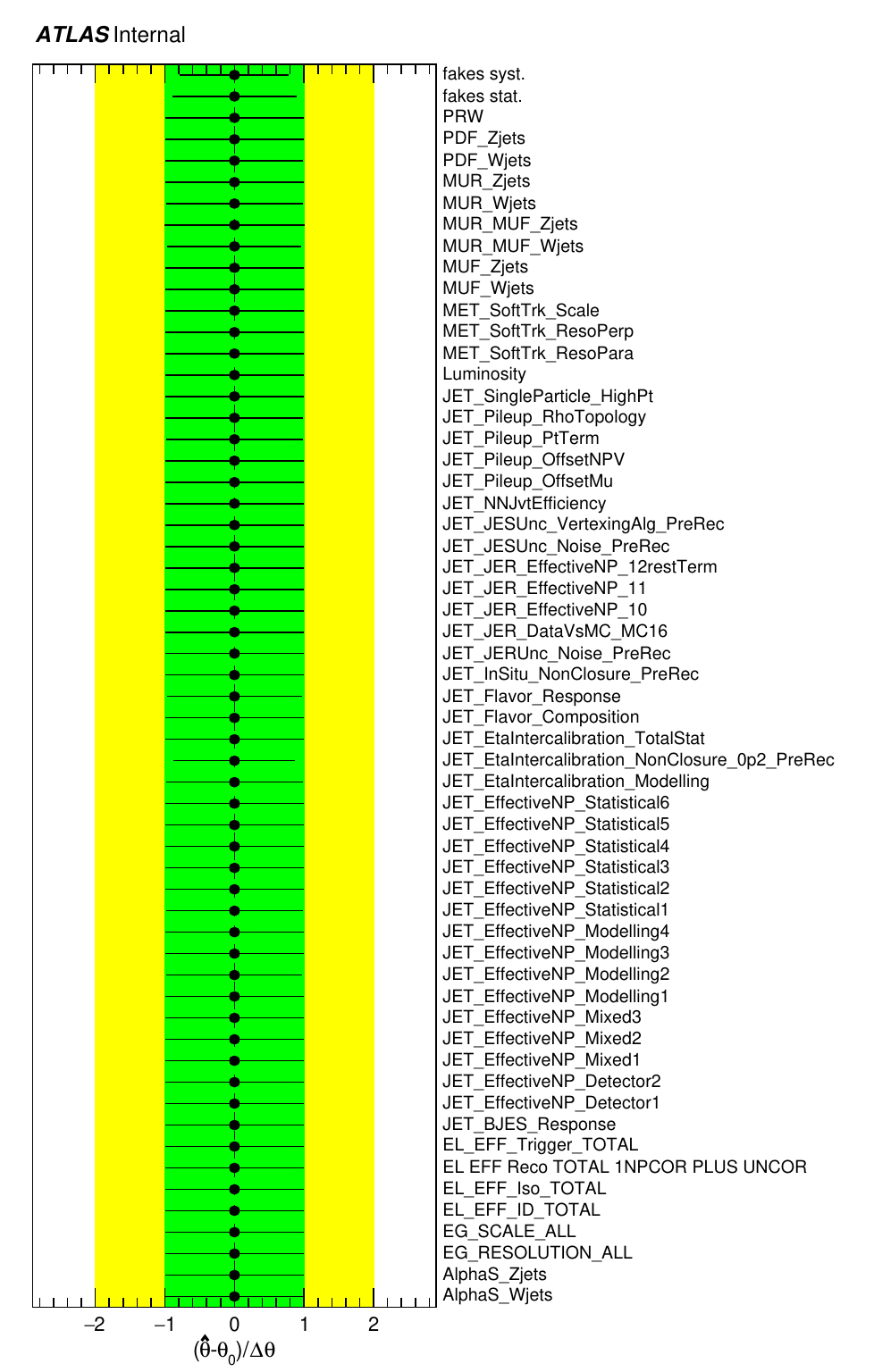}
    }
    \hspace{1cm}
    \subfloat{
      \includegraphics[width=0.45\textwidth]{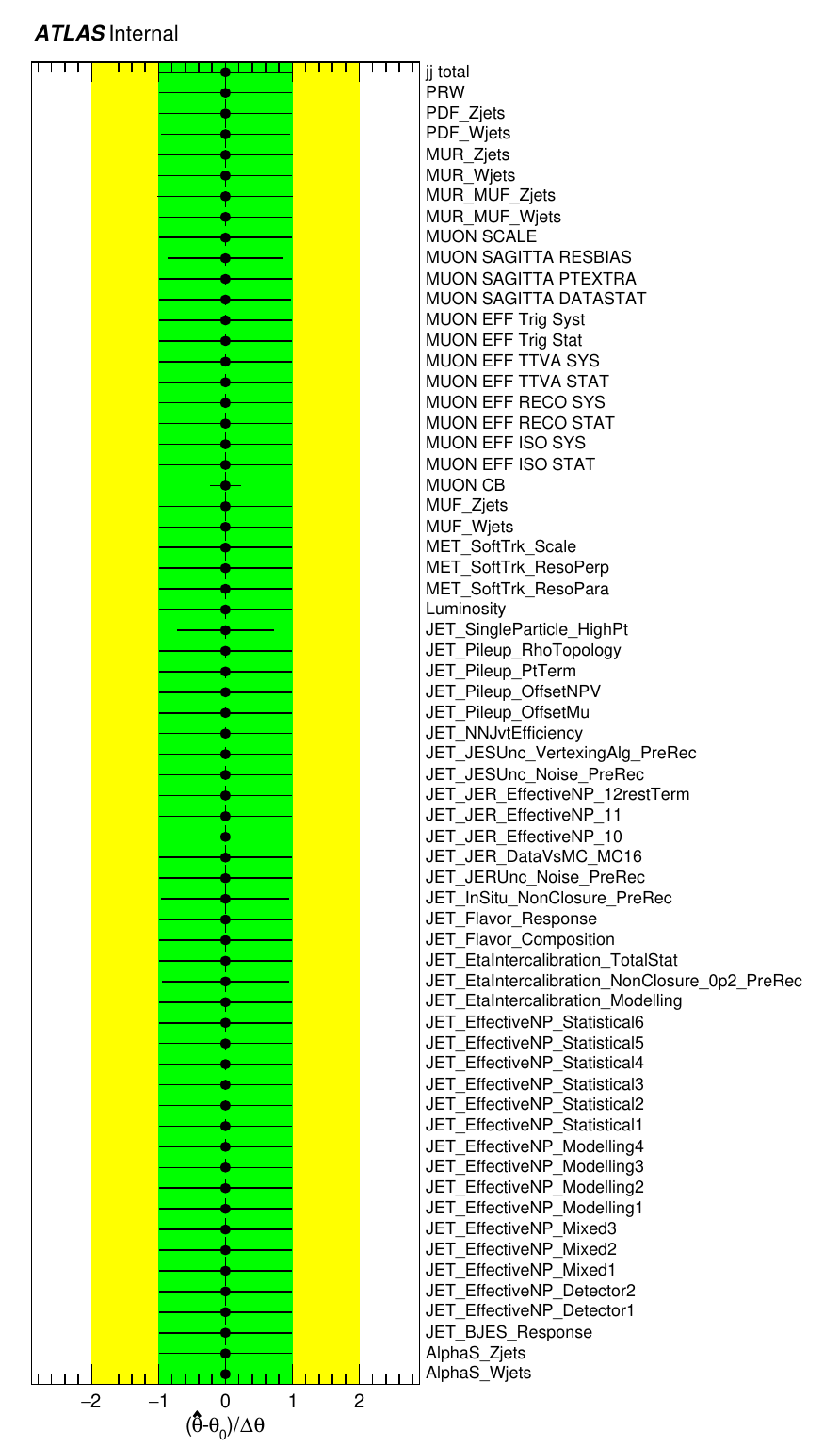}
    }
  \caption{Nuisance parameter pulls and constraints for the $e+j$ channel (left) and $\mu+j$ channel (right). Each entry shows the post-fit central value in units of its pre-fit uncertainty (pull) and the post-fit uncertainty relative to the prior (constraint). Most parameters are unconstrained; the muon momentum resolution (\texttt{MUON\_CB}) is the most tightly constrained, consistent with the high-\pT\ combined measurement precision.}
  \label{fig:Pulls_Constraints}
\end{figure}

\begin{figure}[h]
    \subfloat{
      \includegraphics[width=0.5\textwidth]{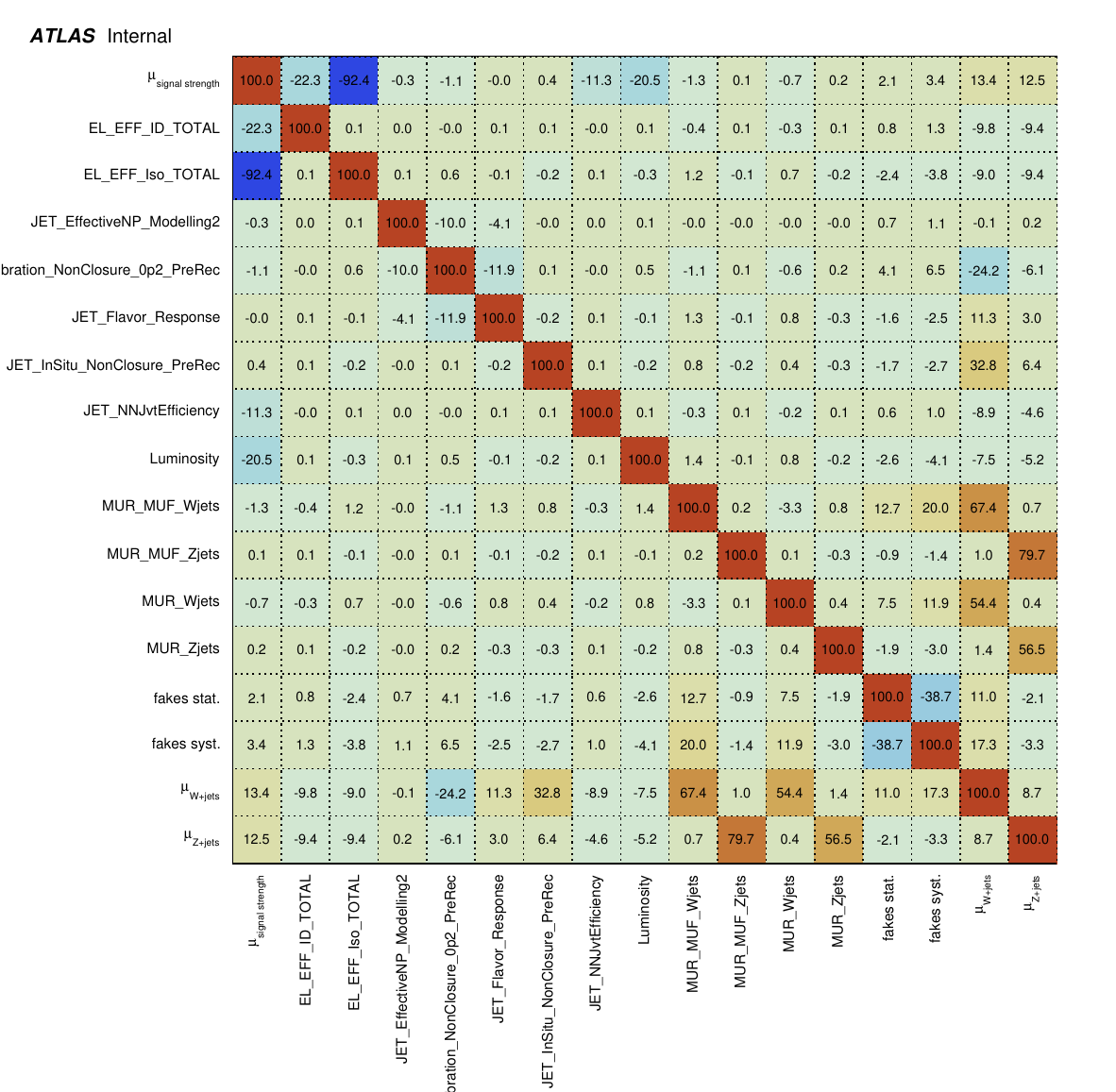}
    }
    \subfloat{
      \includegraphics[width=0.5\textwidth]{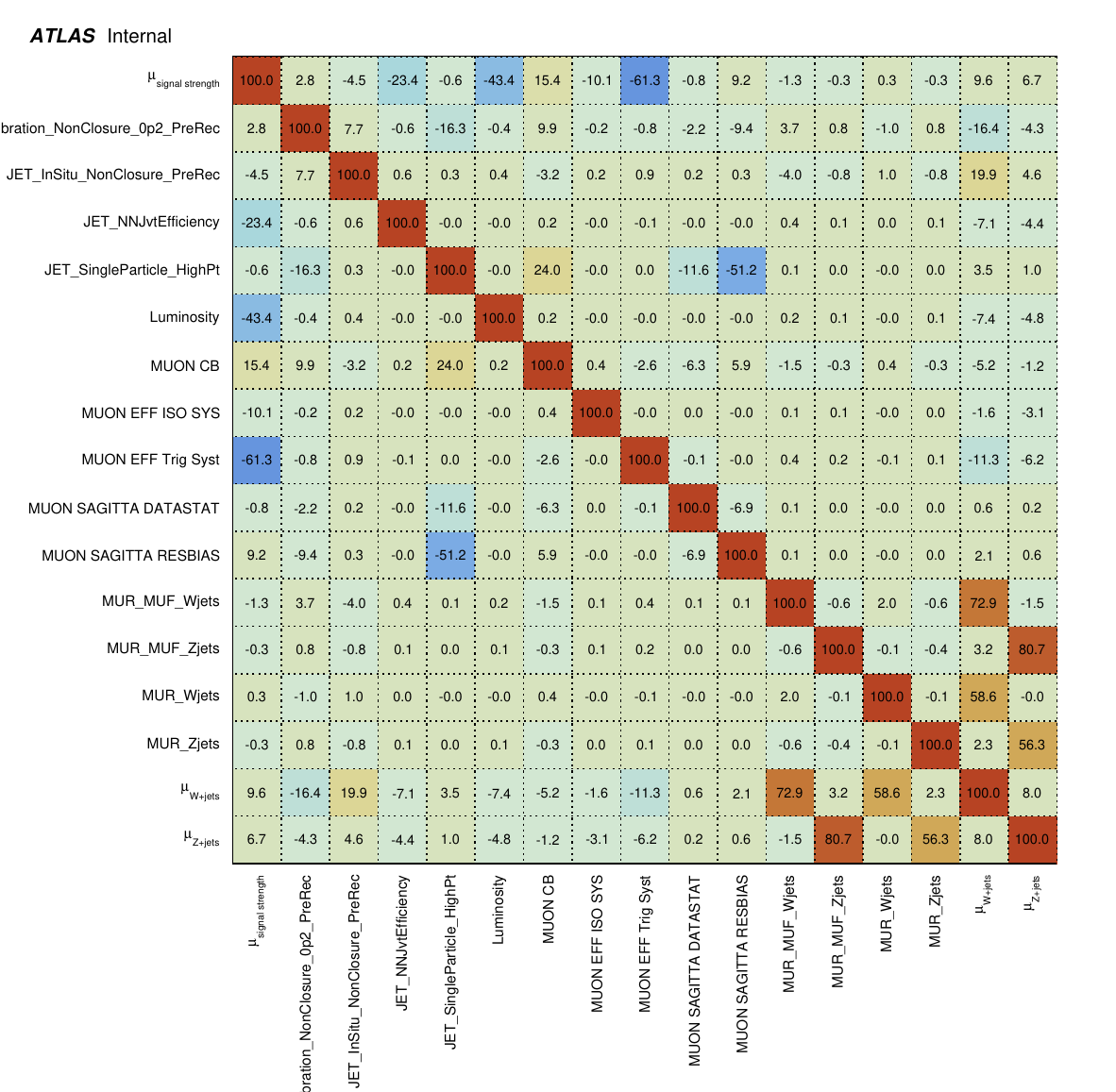}
    }
  \caption{Nuisance parameter correlation matrices for the $e+j$ channel (left) and $\mu+j$ channel (right). Only correlations above a threshold of 0.1 are shown. The matrices are dominated by diagonal entries, confirming that the systematic uncertainties are largely independent.}
  \label{fig:Correlations}
\end{figure}

\FloatBarrier


The angular cuts $\Delta\phi_{\ell j} > 2.8$ and $\Delta\eta_{\ell j} < 3.25$ applied in the signal region (Section~\ref{sec:EvSel}) were optimised on $56~\textrm{fb}^{-1}$ of partial Run~3 data. Signal efficiency above 90\% was required, and the combination of $\Delta\eta$ and $\Delta\phi$ cuts was found to improve background rejection by approximately 20\% (10\%) in the electron (muon) channel relative to a single $\Delta R$ cut, with negligible change in signal efficiency. The final values are tighter than those used in the Run~1 QBH search and were not applied in Run~2.

\addtocontents{toc}{\protect\setcounter{tocdepth}{2}}

\chapter{\texorpdfstring{$\bm{Z'}$}{Z'} Analysis: Data and Monte Carlo Samples}
\label{app:zprime}

\addtocontents{toc}{\protect\setcounter{tocdepth}{0}}

\section{Collision Data and Triggers}
\label{app:zprime:data}

Tables~\ref{tab:triggersEl}--\ref{tab:triggersElMu} list the trigger chains used in the
electron, muon, and electron--muon channels of the $Z'$ analysis for each Run~2
data-taking period.

\begin{table}[htbp]
    \centering
    \begin{tabular}{c|c}
      \hhline{==}
      Run period & Triggers \\ \hline
       276262--284484 (2015 data)  & \texttt{2e12\_lhloose\_L12EM10VH}   \\ \hline
       297730--311481 (2016 data)  & \texttt{2e17\_lhvloose\_nod0}            \\ \hline
       325713--326833 (2017 data)  & \texttt{2e24\_lhvloose\_nod0} OR \texttt{2e17\_lhvloose\_nod0\_L12EM15VHI}            \\ \hline
       326834--328393 (2017 data)  & \texttt{2e24\_lhvloose\_nod0}            \\ \hline
       328394--340453 (2017 data)  & \texttt{2e24\_lhvloose\_nod0} OR \texttt{2e17\_lhvloose\_nod0\_L12EM15VHI}            \\ \hline
       348885--364292 (2018 data)  & \texttt{2e24\_lhvloose\_nod0} OR \texttt{2e17\_lhvloose\_nod0\_L12EM15VHI}        \\
       \hhline{==}
    \end{tabular}
    \caption{Triggers used in the dielectron channel of the $Z'$ analysis. The special
      treatment of run period 326834--328393 for 2017 data reflects the prescaling of
      \texttt{2e17\_lhvloose\_nod0\_L12EM15VHI} in periods B5--B8.}
    \label{tab:triggersEl}
\end{table}

\begin{table}[htbp]
    \centering
    \begin{tabular}{c|c}
     \hhline{==}
      Run period & Triggers \\ \hline
      \multicolumn{2}{c}{Single-muon triggers} \\ \hline
        276262--284484 (2015 data)  & \texttt{mu26\_imedium} OR \texttt{mu50}    \\ \hline
        297730--311481 (2016 data)  & \texttt{mu26\_ivarmedium} OR \texttt{mu50} \\ \hline
        325713--340453 (2017 data)  & \texttt{mu26\_ivarmedium} OR \texttt{mu50} \\ \hline
        348885--364292 (2018 data)  & \texttt{mu26\_ivarmedium} OR \texttt{mu50} \\ \hline
      \multicolumn{2}{c}{Di-muon triggers} \\ \hline
        276262--284484 (2015 data)  & \texttt{mu18\_mu8noL1} OR \texttt{2mu10} \\ \hline
        297730--311481 (2016 data)  & \texttt{mu22\_mu8noL1} OR \texttt{2mu14} \\ \hline
        325713--340453 (2017 data)  & \texttt{mu22\_mu8noL1} OR \texttt{2mu14} \\ \hline
        348885--364292 (2018 data)  & \texttt{mu22\_mu8noL1} OR \texttt{2mu14} \\
        \hhline{==}
    \end{tabular}
    \caption{Triggers used in the dimuon channel of the $Z'$ analysis.}
    \label{tab:triggersMu}
\end{table}

\begin{table}[htbp]
    \centering \footnotesize
    \begin{tabular}{c|c}
    \hhline{==}
      Run period & Triggers \\ \hline
        276262--284484 (2015 data)  & \texttt{e17\_lhloose\_mu14} OR \texttt{e7\_lhmedium\_mu24}  \\ \hline
        297730--311481 (2016 data)  & \texttt{e17\_lhloose\_nod0\_mu14} OR \texttt{e26\_lhmedium\_nod0\_mu8noL1} OR \texttt{e7\_lhmedium\_nod0\_mu24} \\ \hline
        325713--340453 (2017 data)  & \texttt{e17\_lhloose\_nod0\_mu14} OR \texttt{e26\_lhmedium\_nod0\_mu8noL1} OR \texttt{e7\_lhmedium\_nod0\_mu24} \\ \hline
        348885--364292 (2018 data)  & \texttt{e17\_lhloose\_nod0\_mu14} OR \texttt{e26\_lhmedium\_nod0\_mu8noL1} OR \texttt{e7\_lhmedium\_nod0\_mu24}  \\
        \hhline{==}
    \end{tabular}
    \caption{Triggers used in the electron--muon channel of the $Z'$ analysis.}
    \label{tab:triggersElMu}
\end{table}

\section{Fit Diagnostics}
\label{app:zprime:fits}

Nuisance parameter pull plots, correlation matrices, and systematic ranking plots from the $Z'$ signal-plus-background fits are collected here for both coupling values $g_{Z'}=0.5$ and $g_{Z'}=1.0$. All plots correspond to the signal hypothesis $m_{Z'}=1\,\TeV$; results for $g_{Z'}=1.0$ are consistent with those for $g_{Z'}=0.5$.

\begin{figure}[h]
	\centering
	\subfloat[]{
		\includegraphics[width=0.23\textwidth]{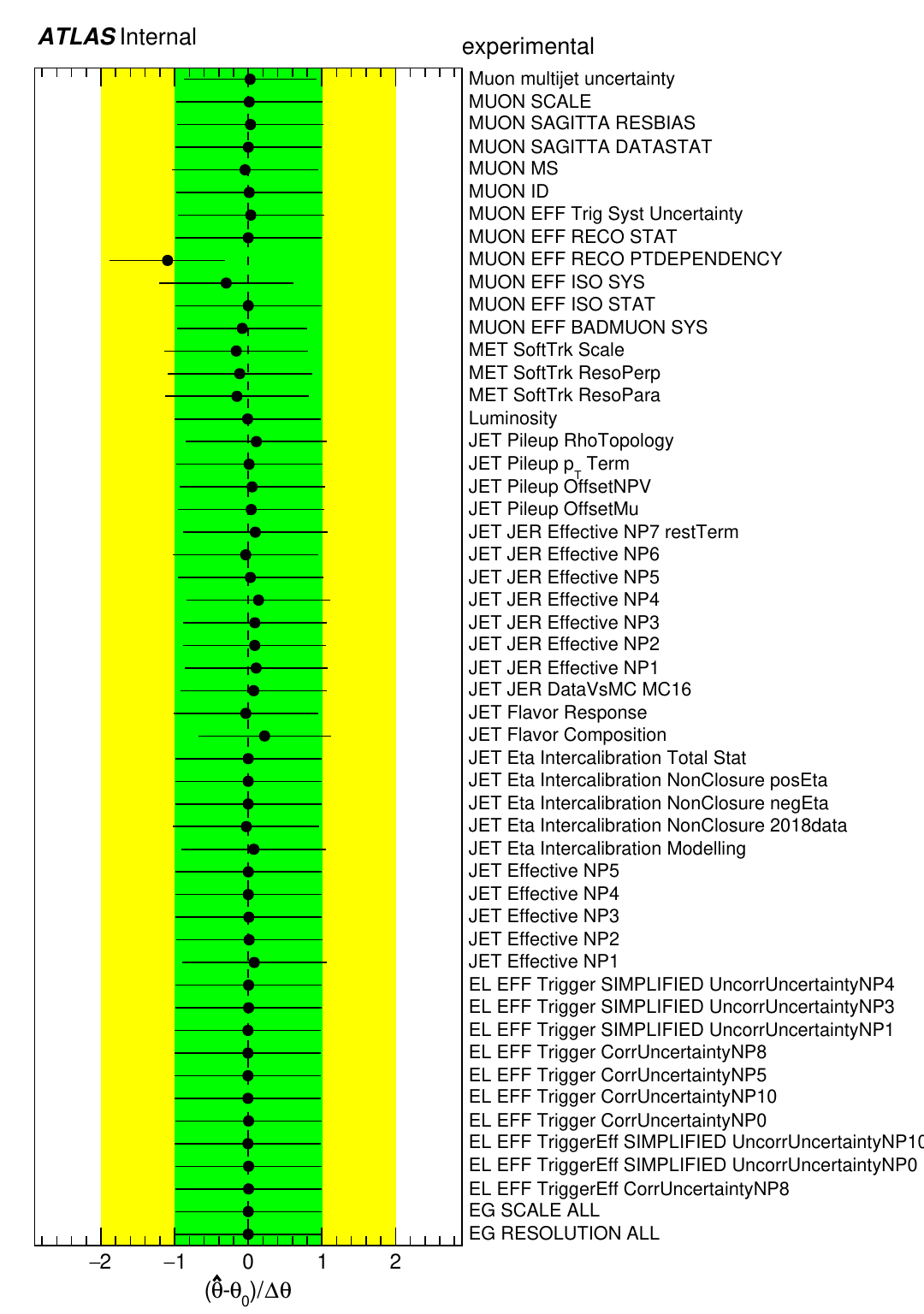}
		\label{fig:NuisPar_1000_g05_mu_experimental_unblinded}
	}
	\hfill
	\subfloat[]{
		\includegraphics[width=0.23\textwidth]{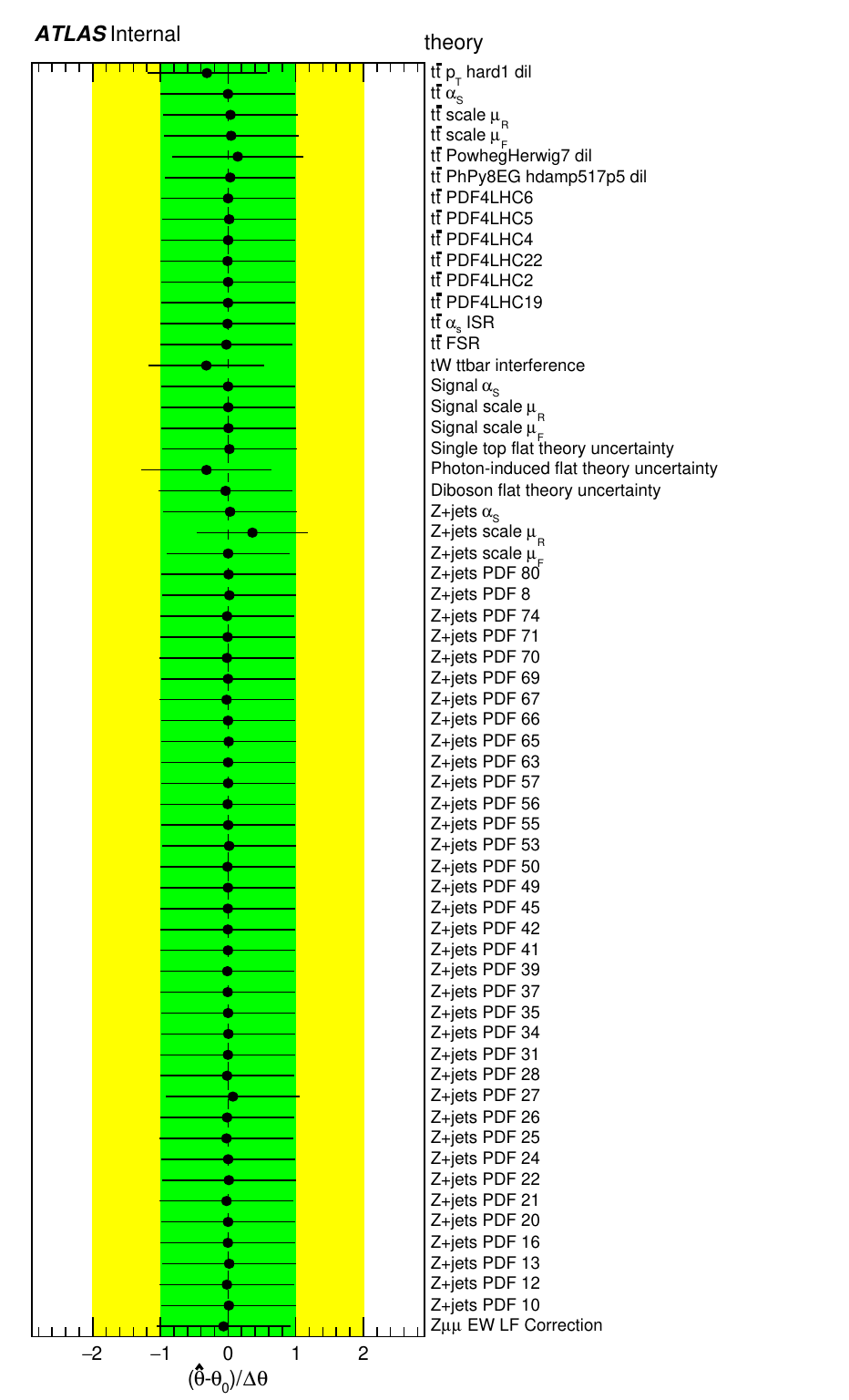}
		\label{fig:NuisPar_1000_g05_mu_theory_unblinded}
	}
	\hfill
	\subfloat[]{
		\includegraphics[width=0.23\textwidth]{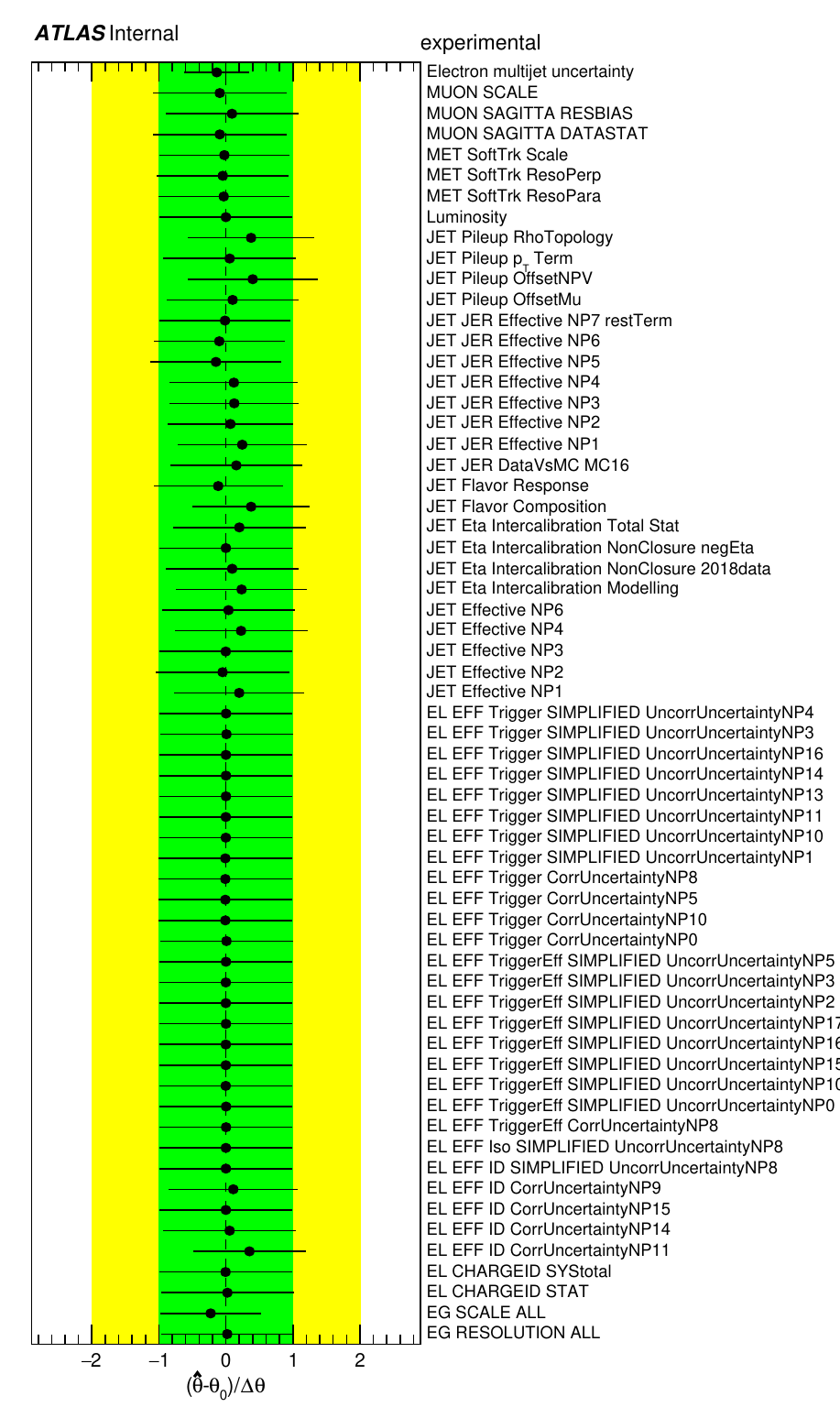}
		\label{fig:NuisPar_1000_g05_ele_experimental_unblinded}
	}
	\hfill
	\subfloat[]{
		\includegraphics[width=0.23\textwidth]{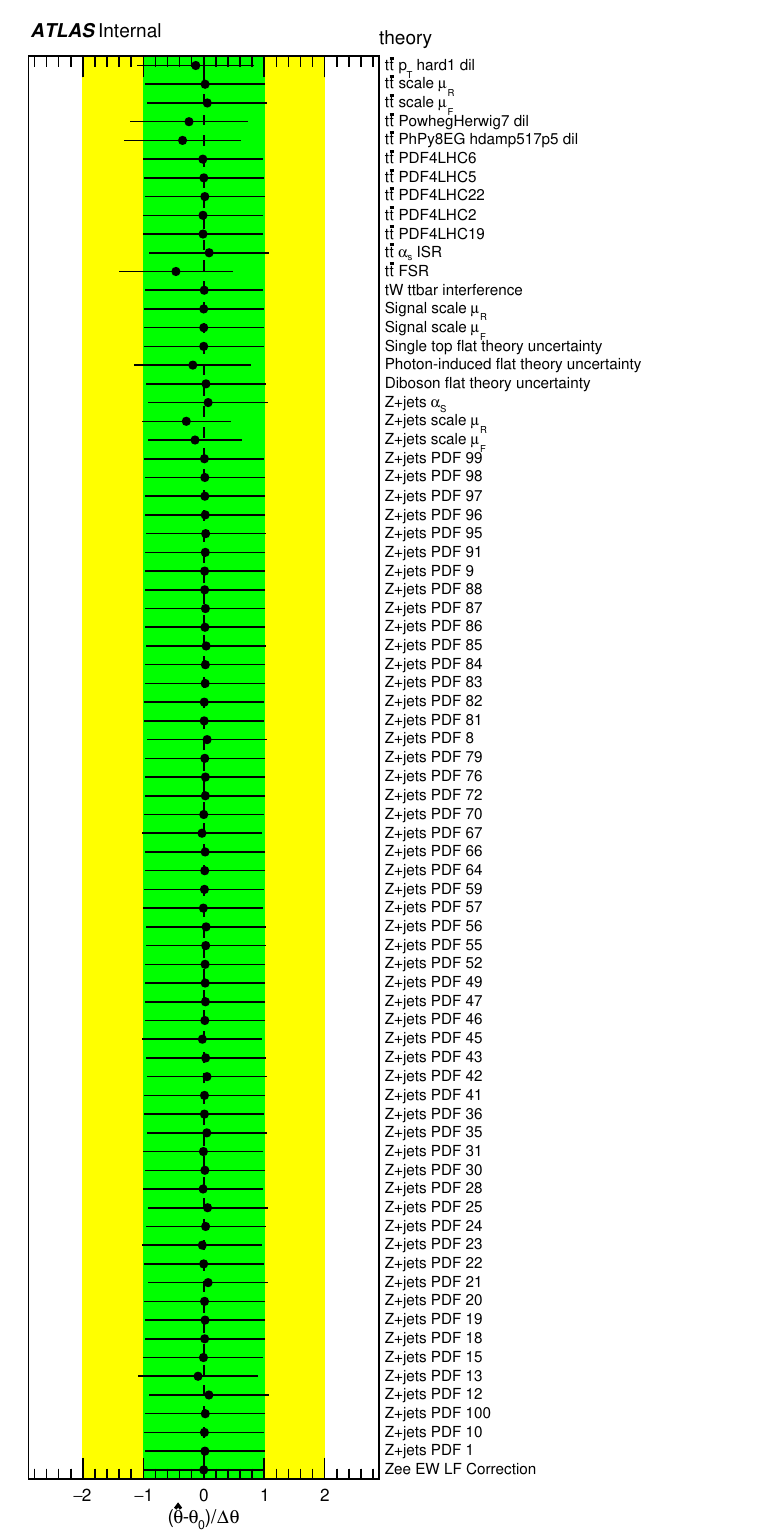}
		\label{fig:NuisPar_1000_g05_ele_theory_unblinded}
	}
	\caption{Nuisance parameter pull plots (experimental and theory) for the muon (a, b) and electron (c, d) channel for $m_{Z'}=1\,\TeV$ and $g_{Z'}=0.5$.}
	\label{fig:NuisPar_1000_g05_unblinded}
\end{figure}

\begin{figure}[h]
	\centering
	\subfloat[]{
		\includegraphics[width=0.45\textwidth]{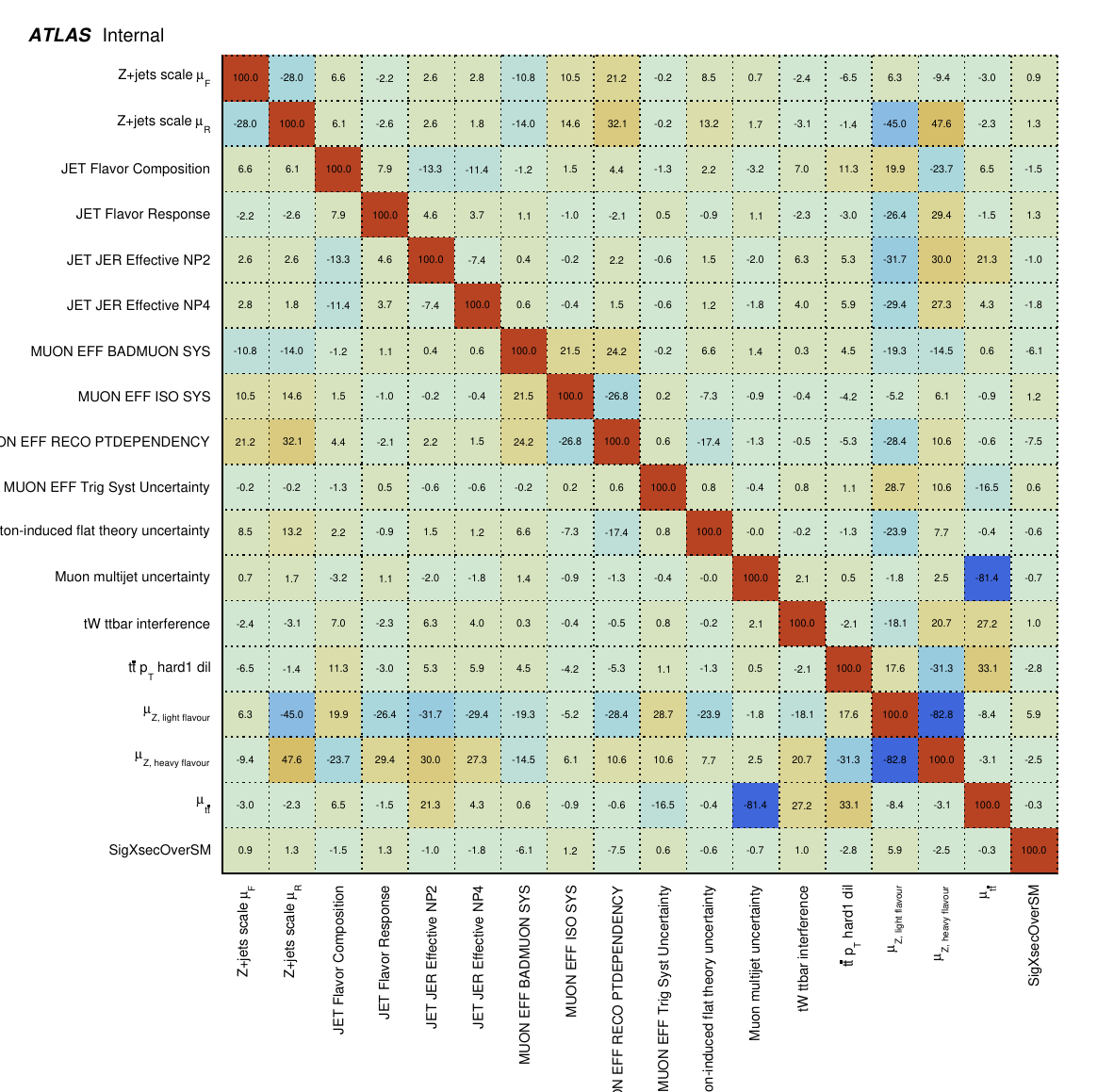}
		\label{fig:correlations_1000_g05_mu_unblinded}
	}
	\hfill
	\subfloat[]{
		\includegraphics[width=0.45\textwidth]{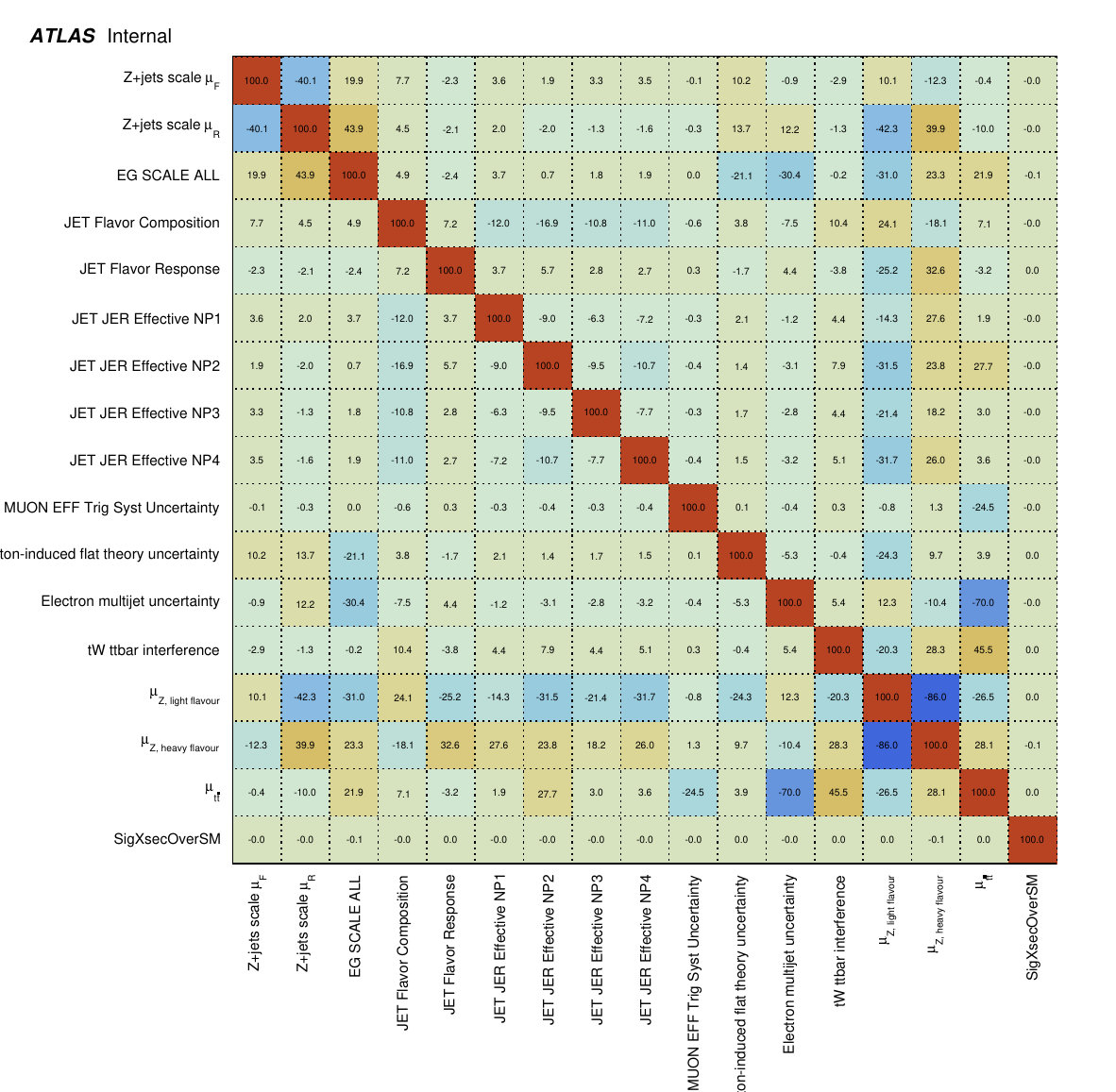}
		\label{fig:correlations_1000_g05_ele_unblinded}
	}
	\caption{Nuisance parameter correlation matrices for the muon a and electron b channel for $m_{Z'}=1\,\TeV$ and $g_{Z'}=0.5$. Only nuisance parameters with a correlation of at least 20\% with at least one other parameter are shown.}
	\label{fig:correlations_1000_g05_unblinded}
\end{figure}

\begin{figure}[h]
	\centering
	\subfloat[]{
		\includegraphics[width=0.47\textwidth]{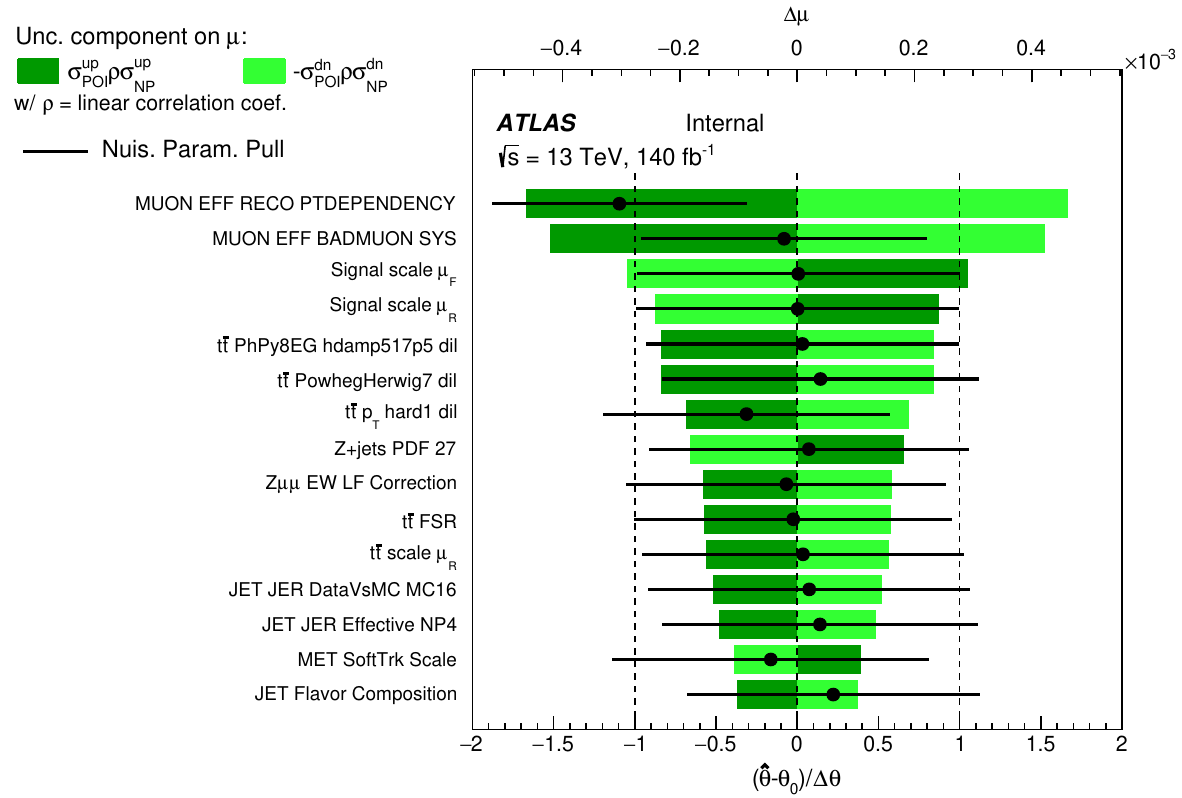}
		\label{fig:Ranking_1000_g05_mu_unblinded}
	}
	\hfill
	\subfloat[]{
		\includegraphics[width=0.47\textwidth]{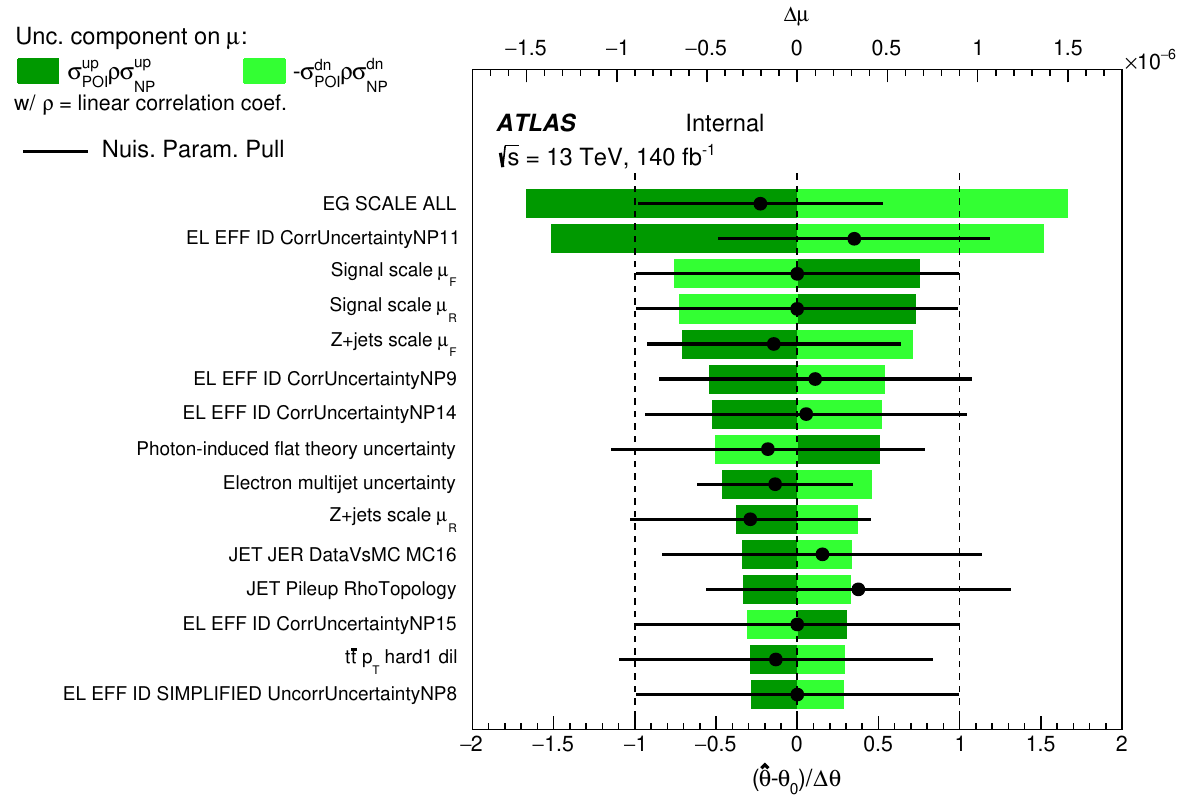}
		\label{fig:Ranking_1000_g05_ele_unblinded}
	}
	\caption{Nuisance parameter ranking plots for the muon a and electron b channel for $m_{Z'}=1\,\TeV$ and $g_{Z'}=0.5$.}
	\label{fig:Ranking_1000_g05_unblinded}
\end{figure}

\FloatBarrier

\begin{figure}[h]
	\centering
	\subfloat[]{
		\includegraphics[width=0.23\textwidth]{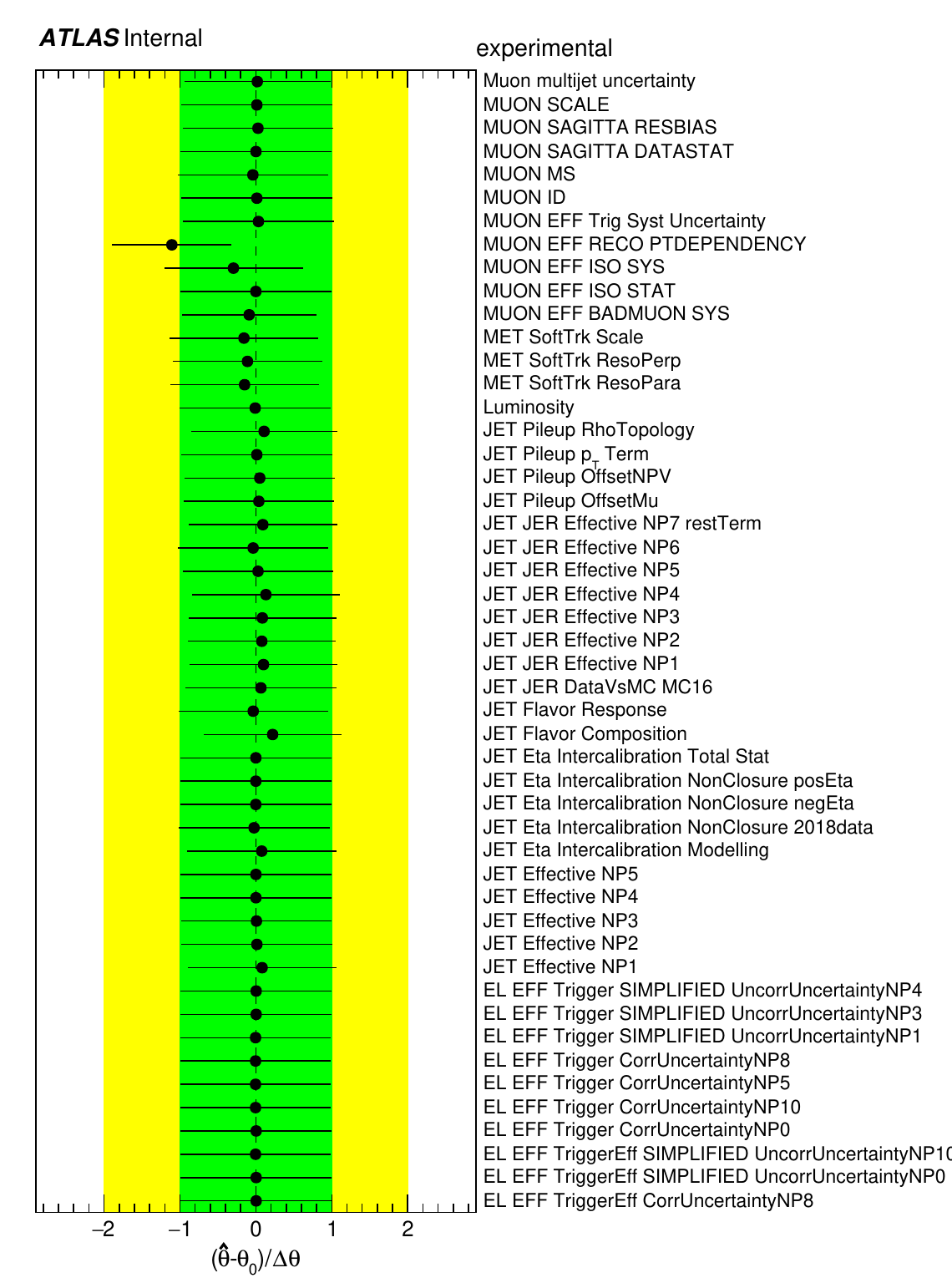}
		\label{fig:NuisPar_1000_g10_mu_experimental_unblinded}
	}
	\hfill
	\subfloat[]{
		\includegraphics[width=0.23\textwidth]{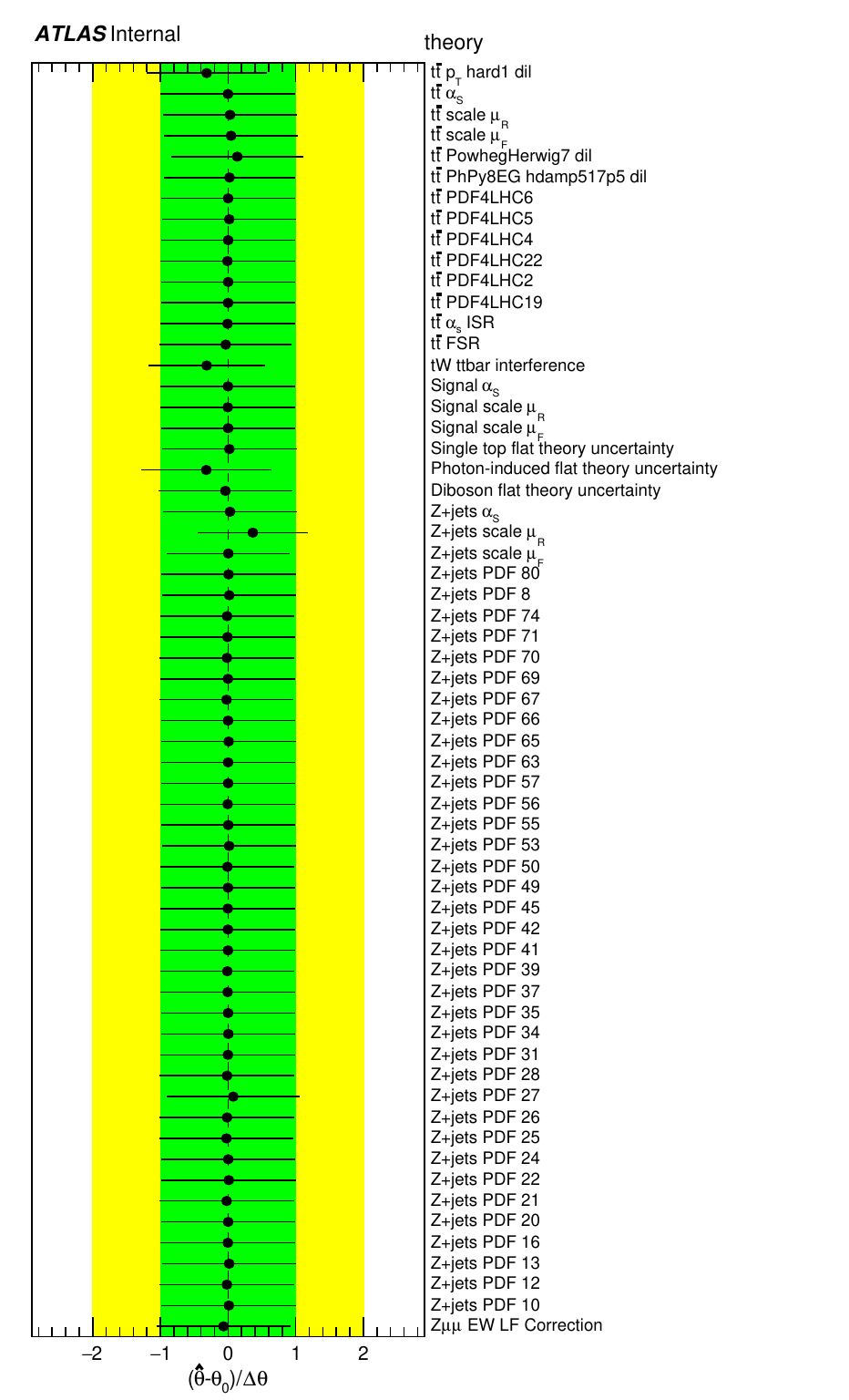}
		\label{fig:NuisPar_1000_g10_mu_theory_unblinded}
	}
	\hfill
	\subfloat[]{
		\includegraphics[width=0.23\textwidth]{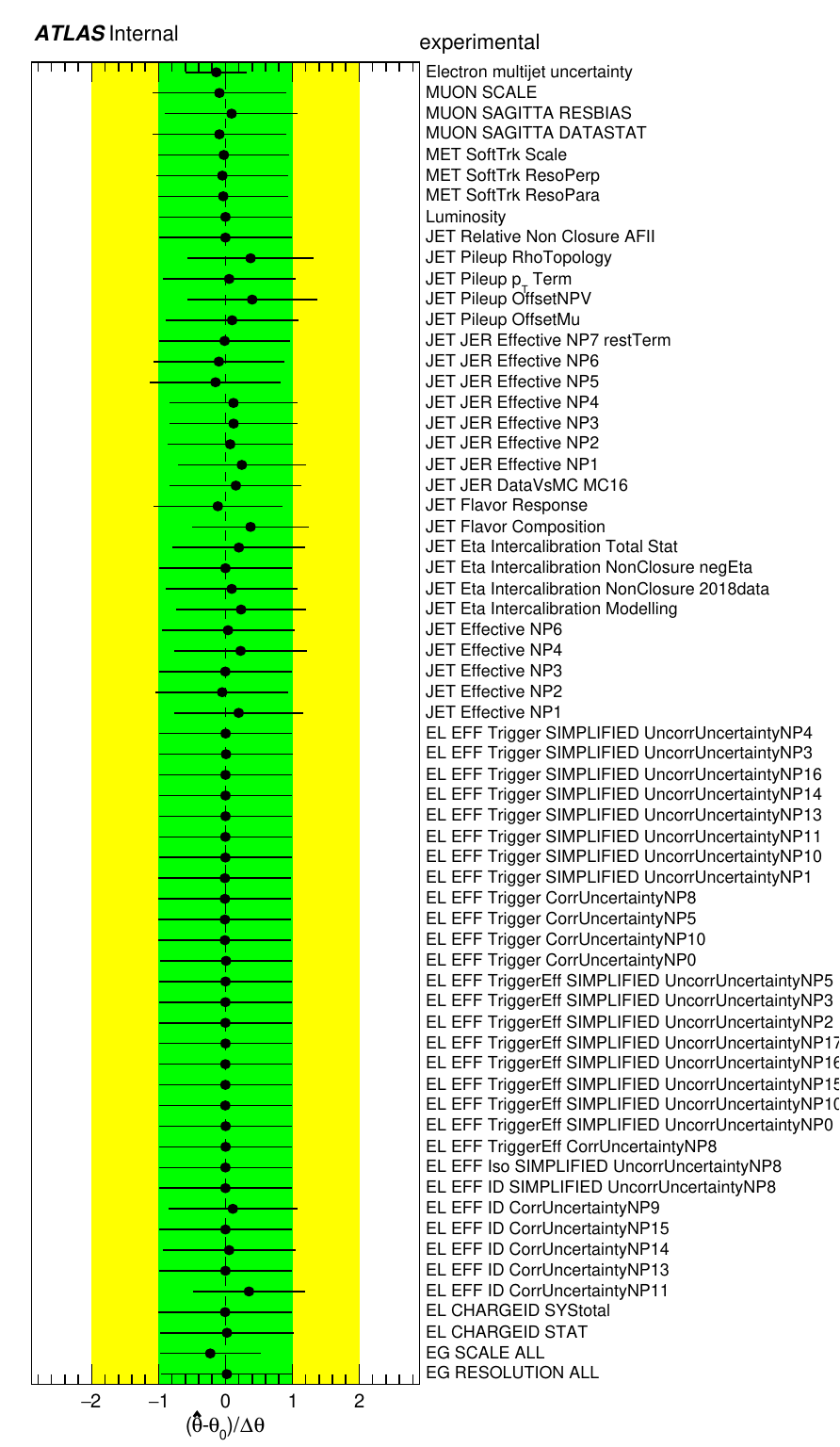}
		\label{fig:NuisPar_1000_g10_ele_experimental_unblinded}
	}
	\hfill
	\subfloat[]{
		\includegraphics[width=0.23\textwidth]{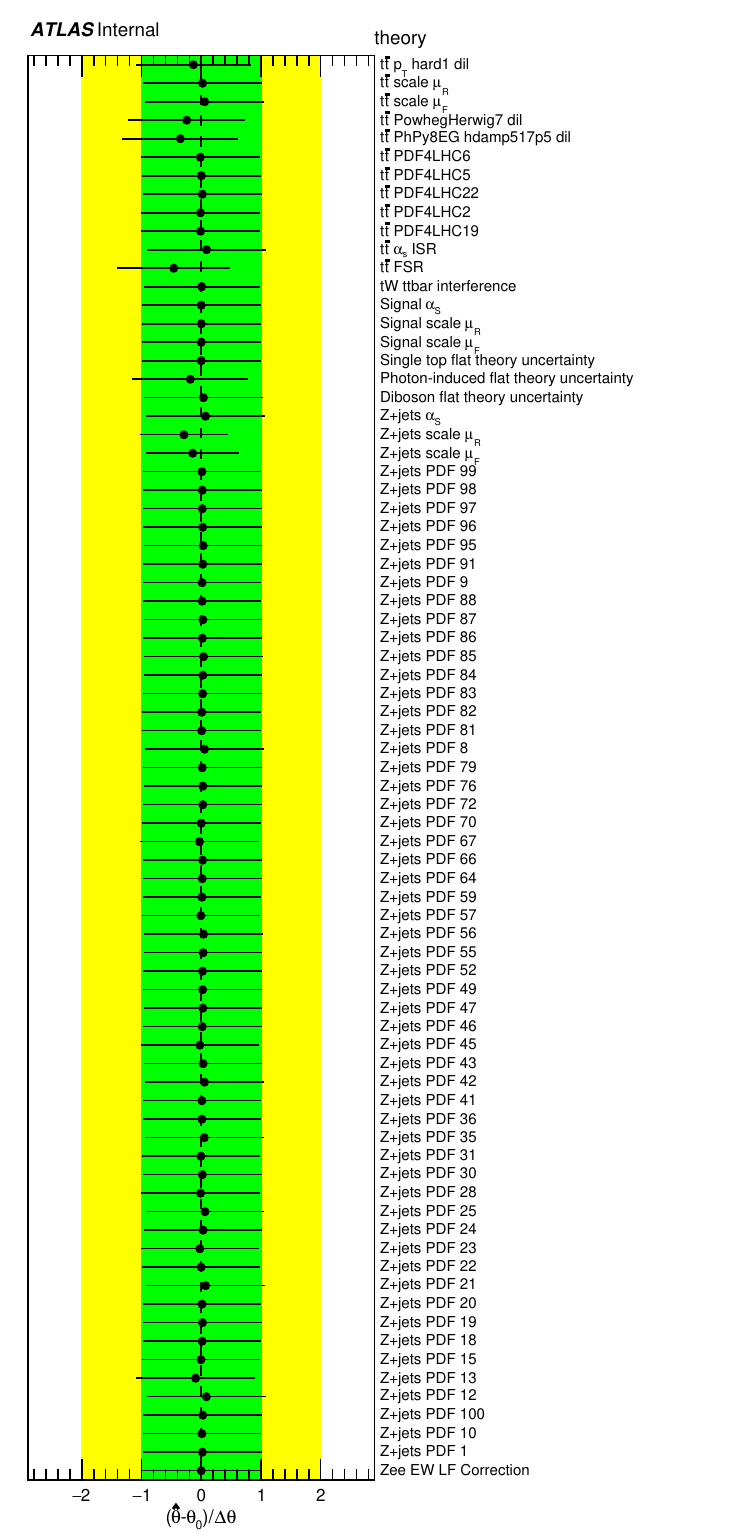}
		\label{fig:NuisPar_1000_g10_ele_theory_unblinded}
	}
	\caption{Nuisance parameter pull plots (experimental and theory) for the muon (a, b) and electron (c, d) channel for $m_{Z'}=1\,\TeV$ and $g_{Z'}=1.0$.}
	\label{fig:NuisPar_1000_g10_unblinded}
\end{figure}

\begin{figure}[h]
	\centering
	\subfloat[]{
		\includegraphics[width=0.45\textwidth]{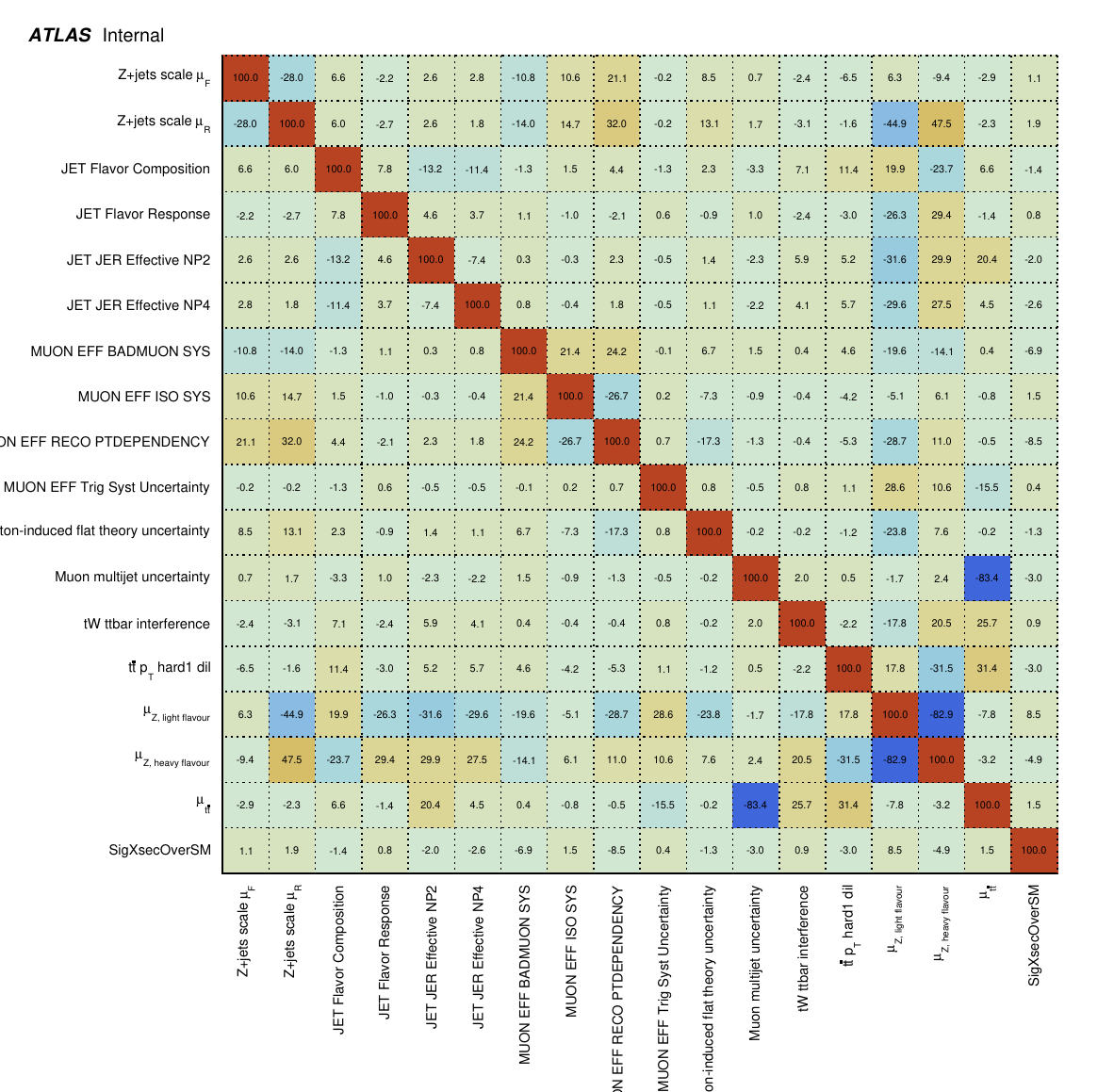}
		\label{fig:correlations_1000_g10_mu_unblinded}
	}
	\hfill
	\subfloat[]{
		\includegraphics[width=0.45\textwidth]{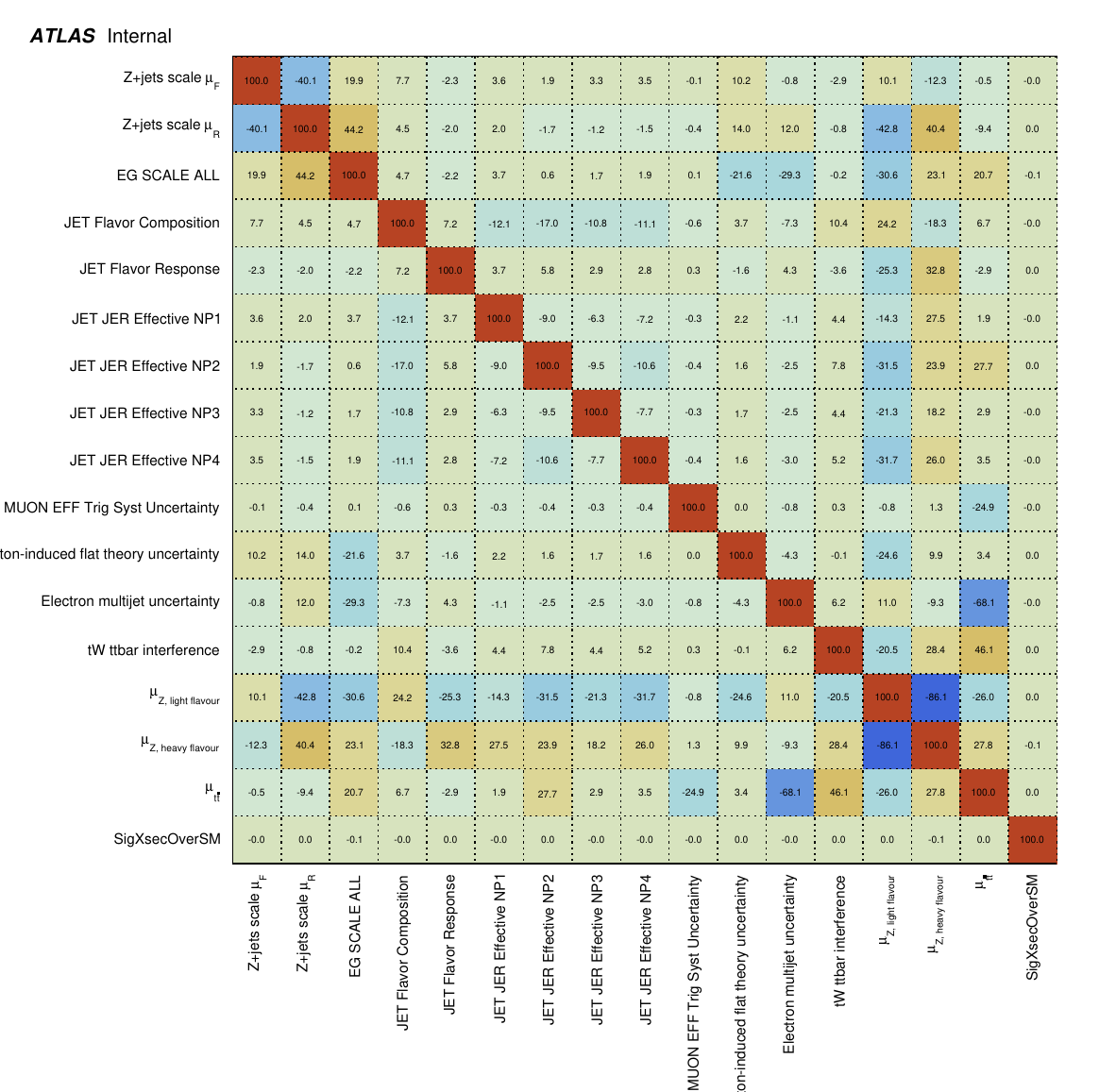}
		\label{fig:correlations_1000_g10_ele_unblinded}
	}
	\caption{Nuisance parameter correlation matrices for the muon a and electron b channel for $m_{Z'}=1\,\TeV$ and $g_{Z'}=1.0$.}
	\label{fig:correlations_1000_g10_unblinded}
\end{figure}

\begin{figure}[h]
	\centering
	\subfloat[]{
		\includegraphics[width=0.47\textwidth]{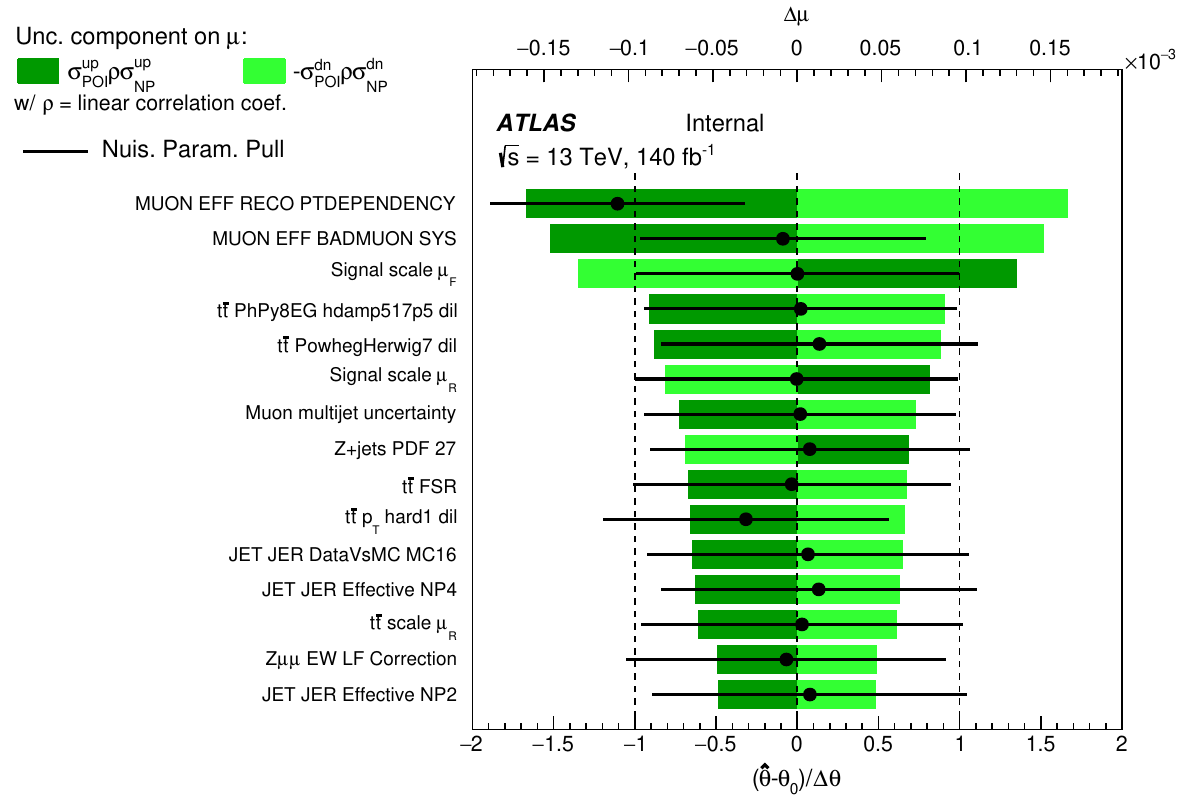}
		\label{fig:Ranking_1000_g10_mu_unblinded}
	}
	\hfill
	\subfloat[]{
		\includegraphics[width=0.47\textwidth]{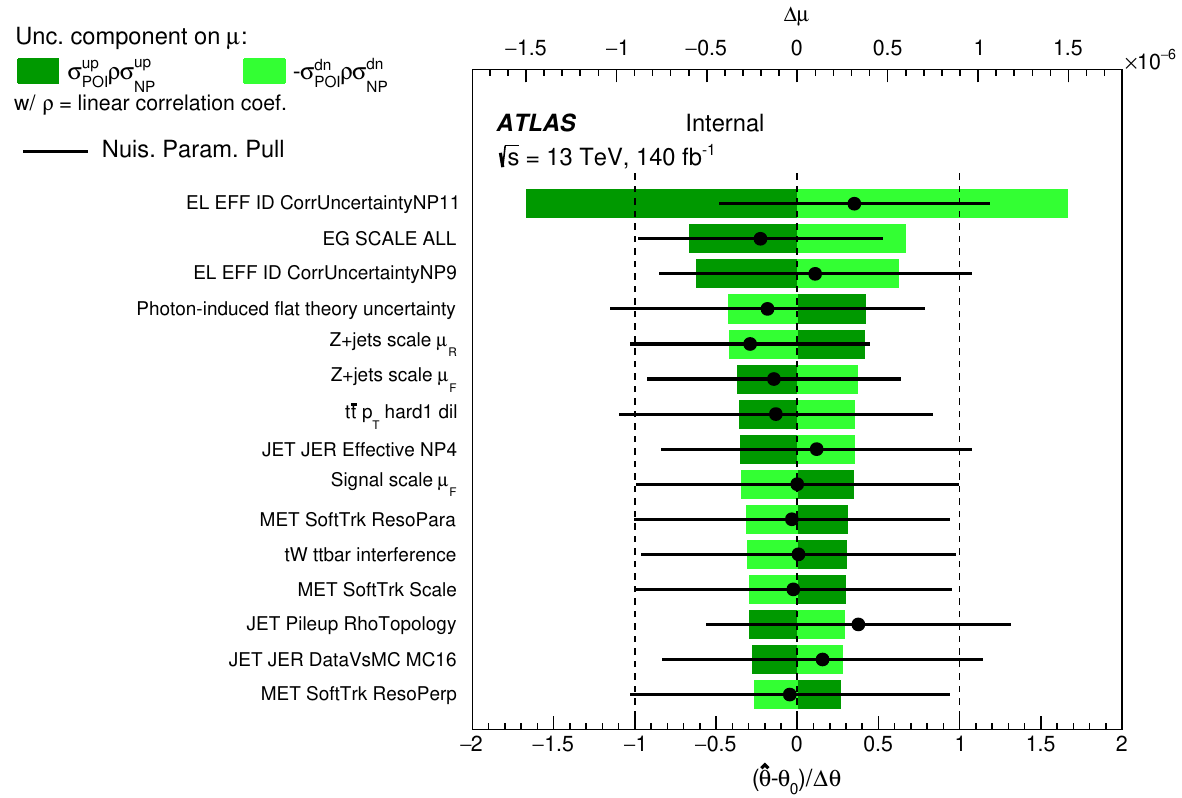}
		\label{fig:Ranking_1000_g10_ele_unblinded}
	}
	\caption{Nuisance parameter ranking plots for the muon a and electron b channel for $m_{Z'}=1\,\TeV$ and $g_{Z'}=1.0$.}
	\label{fig:Ranking_1000_g10_unblinded}
\end{figure}

\begin{figure}[h]
	\centering

		\includegraphics[width=0.23\textwidth]{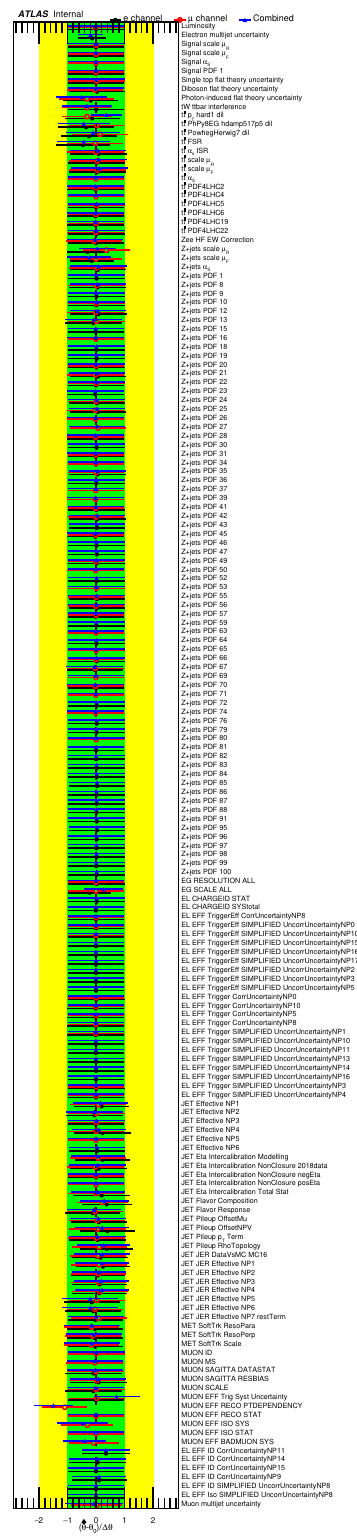}

	\caption{Nuisance parameter pull plot for the combined fit of the muon and electron channel for $m_{Z'}=1\,\mathrm{TeV}$ and $g_{Z'}=0.5$.}
	\label{fig:NuisPar_1000_g05_unblinded_comb}
\end{figure}

\FloatBarrier

\section{Monte Carlo Generator Summary}
\label{app:zprime:mc}
\label{sec:zprime_appendix}

Table~\ref{tab:MCsamples} summarises the generator chain used for each background and
signal process in the $Z'$ analysis.
Non-all-hadronic $\ttbar$ samples with $H_T$-slices are used for the top background instead
of purely dileptonic samples. This avoids a sharp drop in MC statistics at high dilepton
invariant masses, where the dileptonic $\ttbar$ cross-section falls steeply with mass and
the inclusive sample becomes statistically limited.

\begin{table}[htbp]
    \centering \footnotesize
    \resizebox{\textwidth}{!}{
    \begin{tabular}{c|c|c|c|c}
        \hline\hline
        Process & ME generator & ME PDF & Parton shower & Detector simulation \\ \hline
        $Z/\gamma^*\to\ell\ell$ & Sherpa 2.2.11 & NNPDF3.0nnlo & Sherpa & FullSim \\
        $t\bar{t}$ & Powheg-Box v2 & NNPDF3.0nlo & Pythia 8 (v8.230) & FullSim \\
        Single top & Powheg-Box v2 & NNPDF3.0nlo & Pythia 8 (v8.230) & FullSim \\
        Diboson & Sherpa 2.2.11/2.2.12 & NNPDF3.0nnlo & Sherpa & FullSim \\
        Photon-induced & Pythia 8 & NNPDF3.1NLO LUX QED & Pythia 8 & FullSim \\
        $Z'\to\ell\ell$ & MadGraph 2.9.3 & NNPDF3.0nlo & Pythia 8 (v8.245) & Atlfast~II \\
        \hline\hline
    \end{tabular}
    }
    \caption{Generator chain for the MC simulated background and $Z'$ signal processes.}
    \label{tab:MCsamples}
\end{table}

\section{Systematic Uncertainties: Tables}
\label{app:zprime:systematics}

Tables~\ref{tab:lepton_systematics}--\ref{tab:met_systematics} list the experimental systematic uncertainties for leptons, jets, flavour tagging, and $\met$ in the $\Zp$ analysis, as discussed in Section~\ref{sec:zprime_systematics}. Table~\ref{tab:theory_systematics} summarises the theoretical systematic uncertainties and their treatment.

\begin{table}[h]
	\centering
	\caption{Experimental systematic uncertainties for leptons. For electrons, calibration uncertainties begin with ``EG\_'' and efficiency (SF) uncertainties with ``EL\_EFF\_''; for muons, calibration uncertainties begin with ``MUON\_'' and efficiency uncertainties with ``MUON\_EFF\_''.}
	\label{tab:lepton_systematics}
\begin{tabular}{ll}
	\hline \hline
	Category & Systematic \\
	\hline
Electrons	&  EG\_RESOLUTION\_ALL\\
	&  EG\_SCALE\_ALL\\
& EL\_EFF\_ChargeIDSel\_CorrUncertaintyNP0-9\\
& EL\_EFF\_ChargeIDSel\_SIMPLIFIED\_UncorrUncertaintyNP0-17\\
& EL\_EFF\_ID\_CorrUncertaintyNP0-15\\
& EL\_EFF\_ID\_SIMPLIFIED\_UncorrUncertaintyNP0-17\\
& EL\_EFF\_Iso\_CorrUncertaintyNP0-9\\
& EL\_EFF\_Iso\_SIMPLIFIED\_UncorrUncertaintyNP0-17\\
& EL\_EFF\_Reco\_CorrUncertaintyNP0-7\\
& EL\_EFF\_Reco\_SIMPLIFIED\_UncorrUncertaintyNP0-17\\
 & EL\_EFF\_TriggerEff\_CorrUncertaintyNP0-9 \\
 & EL\_EFF\_TriggerEff\_SIMPLIFIED\_UncorrUncertaintyNP0-17\\
 & EL\_EFF\_Trigger\_CorrUncertaintyNP0-9\\
 & EL\_EFF\_Trigger\_SIMPLIFIED\_UncorrUncertaintyNP0-17\\
\hline
Muons	&  MUON\_ID\\
	& MUON\_MS \\
	& MUON\_SAGITTA\_DATASTAT \\
	& MUON\_SAGITTA\_RESBIAS \\
	& MUON\_SCALE  \\
	& MUON\_EFF\_TrigSystUncertainty \\
	& MUON\_EFF\_TrigStatUncertainty \\
	& MUON\_EFF\_TTVA\_SYS \\
	& MUON\_EFF\_TTVA\_STAT \\
	& MUON\_EFF\_RECO\_SYS \\
	& MUON\_EFF\_RECO\_STAT \\
	& MUON\_EFF\_ISO\_SYS \\
	& MUON\_EFF\_ISO\_STAT \\
	& MUON\_EFF\_BADMUON\_SYS \\
	\hline \hline
\end{tabular}
\end{table}

\begin{table}[h]
	\centering
	\caption{Experimental systematic uncertainties for jets.}
	\label{tab:jet_systematics}
	\begin{tabular}{ll}
		\hline \hline
		Category & Systematic \\
		\hline
	Jet energy scale & JET\_BJES\_Response \\
		& JET\_EffectiveNP (split into eight components) \\
		& JET\_EtaIntercalibration\_Modelling \\
		& JET\_EtaIntercalibration\_NonClosure\_2018data \\
		& JET\_EtaIntercalibration\_NonClosure\_highE\\
		& JET\_EtaIntercalibration\_NonClosure\_negEta \\
		& JET\_EtaIntercalibration\_NonClosure\_posEta \\
		& JET\_EtaIntercalibration\_TotalStat \\
		& JET\_Flavor\_Composition \\
		& JET\_Flavor\_Response \\
		& JET\_Pileup\_OffsetMu\\
		& JET\_Pileup\_OffsetNPV\\
		& JET\_Pileup\_PtTerm\\
		& JET\_Pileup\_RhoTopology\\
		& JET\_PunchThrough\_MC16 \\
		& JET\_SingleParticle\_HighPt\\
		\hline
Jet energy resolution & JET\_JER\_DataVsMC\_MC16 \\
		& JET\_JER\_EffectiveNP (split into seven components)\\ \hline
Jet vertex tagging & JET\_JvtEfficiency\\
		&  JET\_fJvtEfficiency\\
		\hline \hline
	\end{tabular}
\end{table}

\begin{table}[h]
	\centering
	\caption{Experimental systematic uncertainties for flavour tagging. EV decomposition uncertainties are summarised by the number of components per flavour.}
	\label{tab:tagging_systematics}
	\begin{tabular}{l}
		\hline \hline
		 Systematic \\
		\hline
		  FT\_EFF\_Eigen\_B\_0--8 \\
		  FT\_EFF\_Eigen\_C\_0--3 \\
		  FT\_EFF\_Eigen\_Light\_0--3 \\
		  FT\_EFF\_extrapolation \\
		  FT\_EFF\_extrapolation\_from\_charm \\
		\hline \hline
	\end{tabular}
\end{table}

\begin{table}[h]
	\centering
	\caption{Experimental systematic uncertainties for $\met$.}
	\label{tab:met_systematics}
	\begin{tabular}{l}
		\hline \hline
		 Systematic \\
		\hline
		 MET\_SoftTrk\_Scale  \\
		 MET\_SoftTrk\_ResoPara \\
		 MET\_SoftTrk\_ResoPerp  \\
		\hline \hline
	\end{tabular}
\end{table}

\begin{table}[]
    \centering
    \caption{List of theoretical systematic uncertainties. The third column indicates the treatment of the uncertainty in the context of visualisation of the overall error in the dilepton mass distribution, whereas for the fit the error estimation follows treating each variation as a separate nuisance parameter and prunning where necessary. Wherever ``individual NPs'' is indicated, each variation within an uncertainty is treated as a separate nuisance parameter in the visualisation too. The abbreviations ADD., MULT. and EXP. refer to additive, multiplicative and exponentiated respectively, in the context of NLO EW virtual (loop) corrections to the Drell-Yan background.}
    \label{tab:theory_systematics}
    \resizebox{1.1\textwidth}{!}{
        \begin{tabular}{cccc}
            \hline \hline 
            \textbf{Uncertainty} & \textbf{Description} & \textbf{Illustrative treatment} & \textbf{Fit treatment} \\ \hline
            \multicolumn{4}{c}{top (\texttt{PhPy8EG})} \\ \hline
            PDF4LHC15\_nlo & 30 variations from PDF4LHC15\_nlo set (90901--90930) & Individual NPs & Individual NPs \\
            $\mu_{\textrm{R}}, \mu_{\textrm{F}}$ scales & QCD renormalisation, factorisation scales & Envelope & Individual NPs \\
            FSR $\mu_{\textrm{R}}$ scale & QCD renormalisation scale variation in FSR & Individual NPs & Individual NPs \\
            $\alpha_{\textrm{S}}$ & Variations on the strong coupling constant & Average & Individual NPs \\
            ISR $\alpha_{\textrm{S}}$ (\texttt{Var3c}) & As above in the initial state radiation & Individual NPs & Individual NPs \\
            $h_{\textrm{damp}}$ & Variation on the damping parameter (2-point) & Individual NPs against $t\bar{t}$ dilep. & Individual NPs against $t\bar{t}$ dilep.\\
            PS & Variation on parton shower (2-point) & Individual NPs against $t\bar{t}$ dilep. & Individual NPs against $t\bar{t}$ dilep. \\
            $|\mathcal{M}|$ & Variation on hard-scatter ME (2-point) & Individual NPs against $t\bar{t}$ dilep. & Individual NPs against $t\bar{t}$ dilep. \\ \hline \hline 
            \multicolumn{4}{c}{DY (\texttt{Sherpa})} \\ \hline 
            NNPDF30nnloas0118 & 100 variations from NNPDF30nnloas0118 set (303201--303300) & Standard deviation & Individual NPs \\
            $\mu_{\textrm{R}}, \mu_{\textrm{F}}$ scales & QCD renormalisation, factorisation scales & Envelope & Individual NPs \\
            $\alpha_{\textrm{S}}$ & Variations on the strong coupling constant & Average & Individual NPs \\
            & & & \\ 
            NLO $\textrm{EW}_{\textrm{virt.}}$ & Virtual corrections inclusion method (ADD., MULT., EXP.) & Correct nominal with smallest; envelope remaining two & Individual NPs \\ \hline \hline
        \end{tabular}
    }
    \end{table}

\section{Signal Kinematics}
\label{app:zprime:kinematics}

\begin{figure}[h!]
	\captionsetup[subfigure]{labelformat=empty}
	\centering
	\subfloat[]{
		\includegraphics[width=0.49\textwidth]{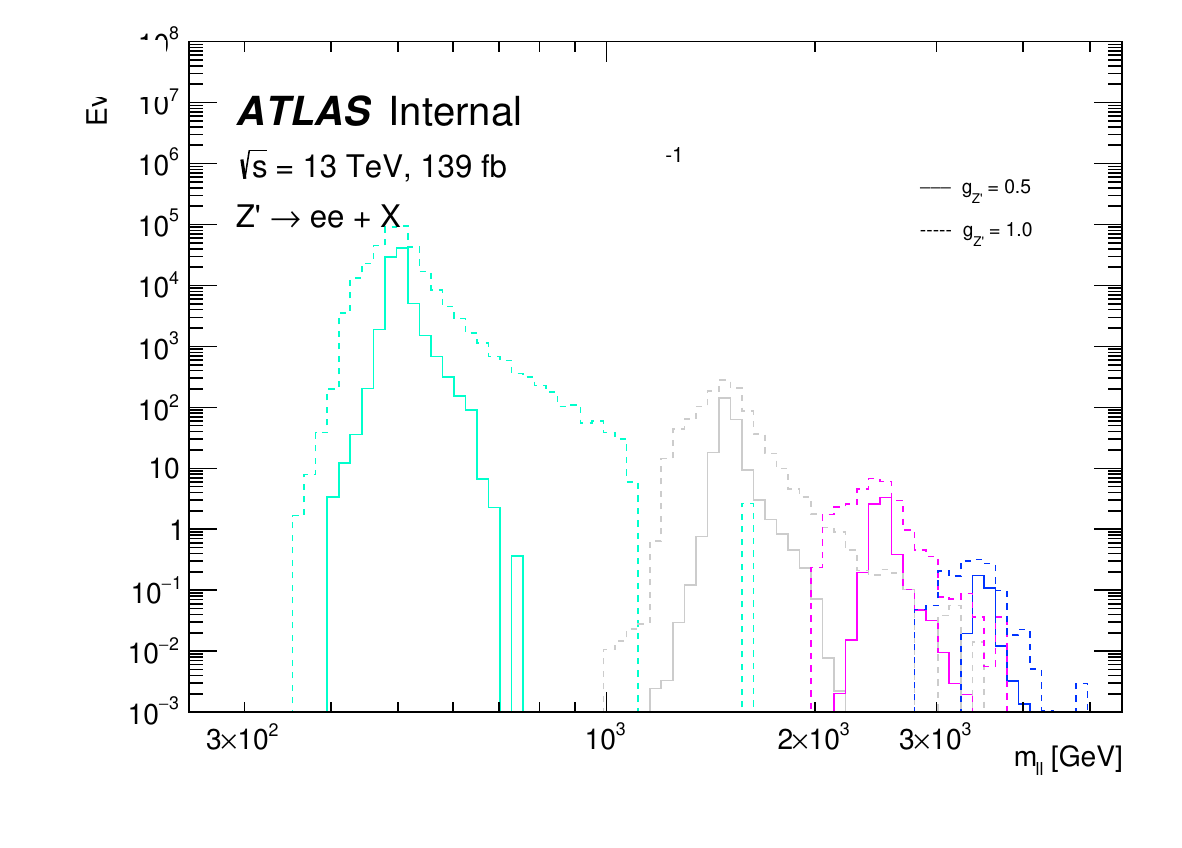}
		}
		\subfloat[]{
		\includegraphics[width=0.49\textwidth]{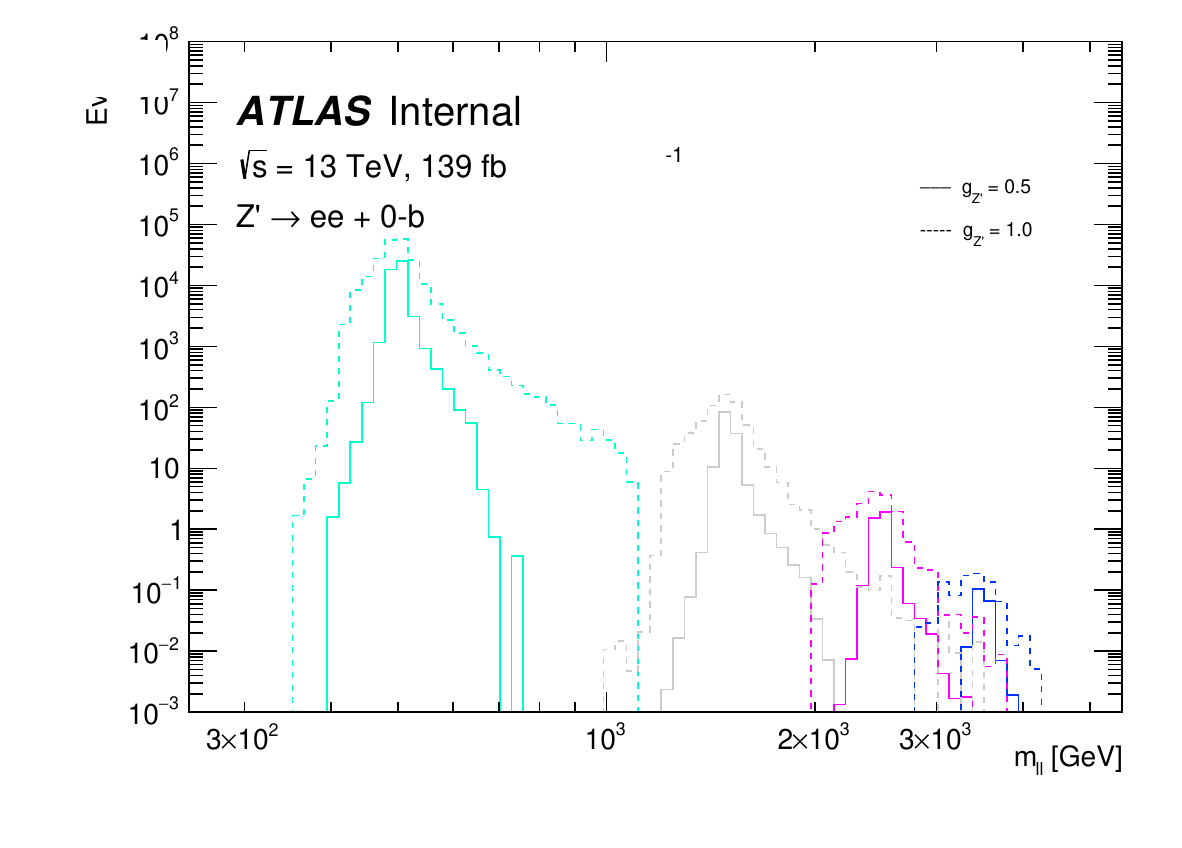}
		}
		\\ \vspace{-1cm}
		\subfloat[]{
		\includegraphics[width=0.49\textwidth]{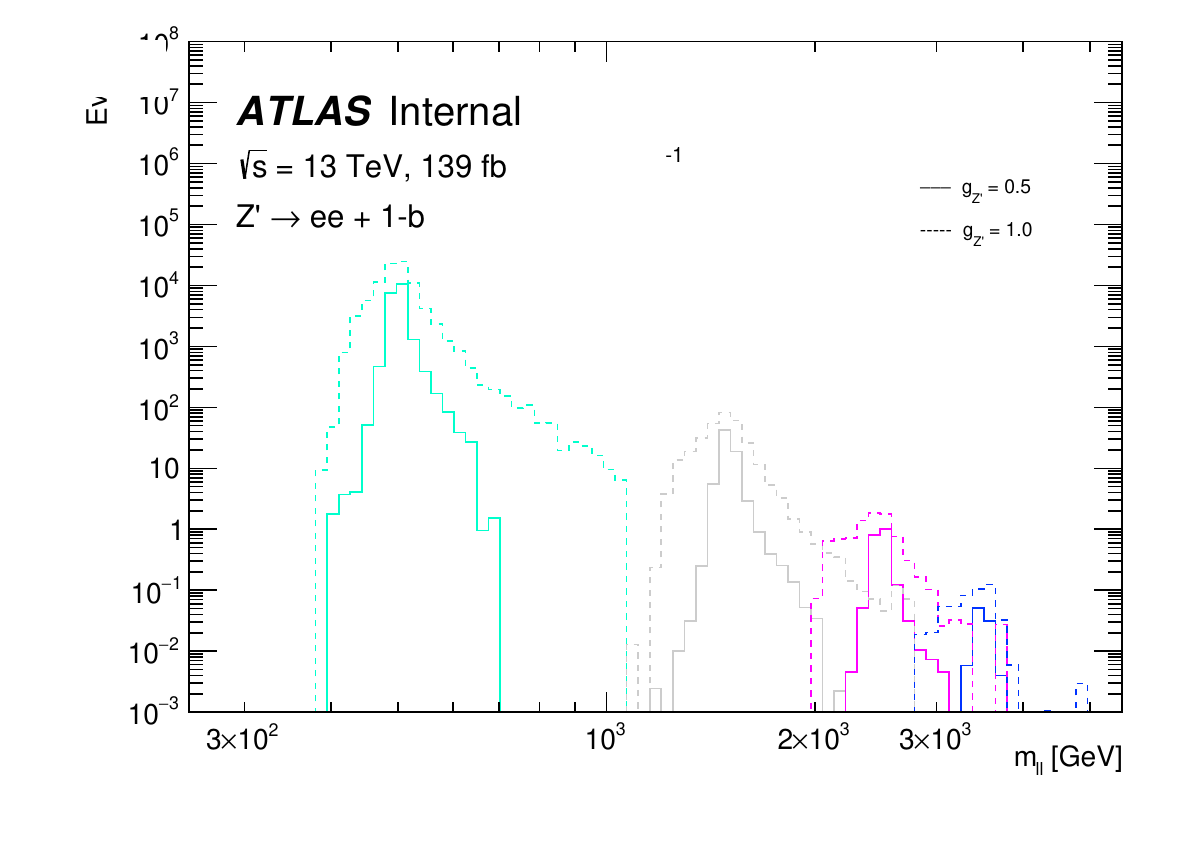}
		}
		\subfloat[]{
		\includegraphics[width=0.49\textwidth]{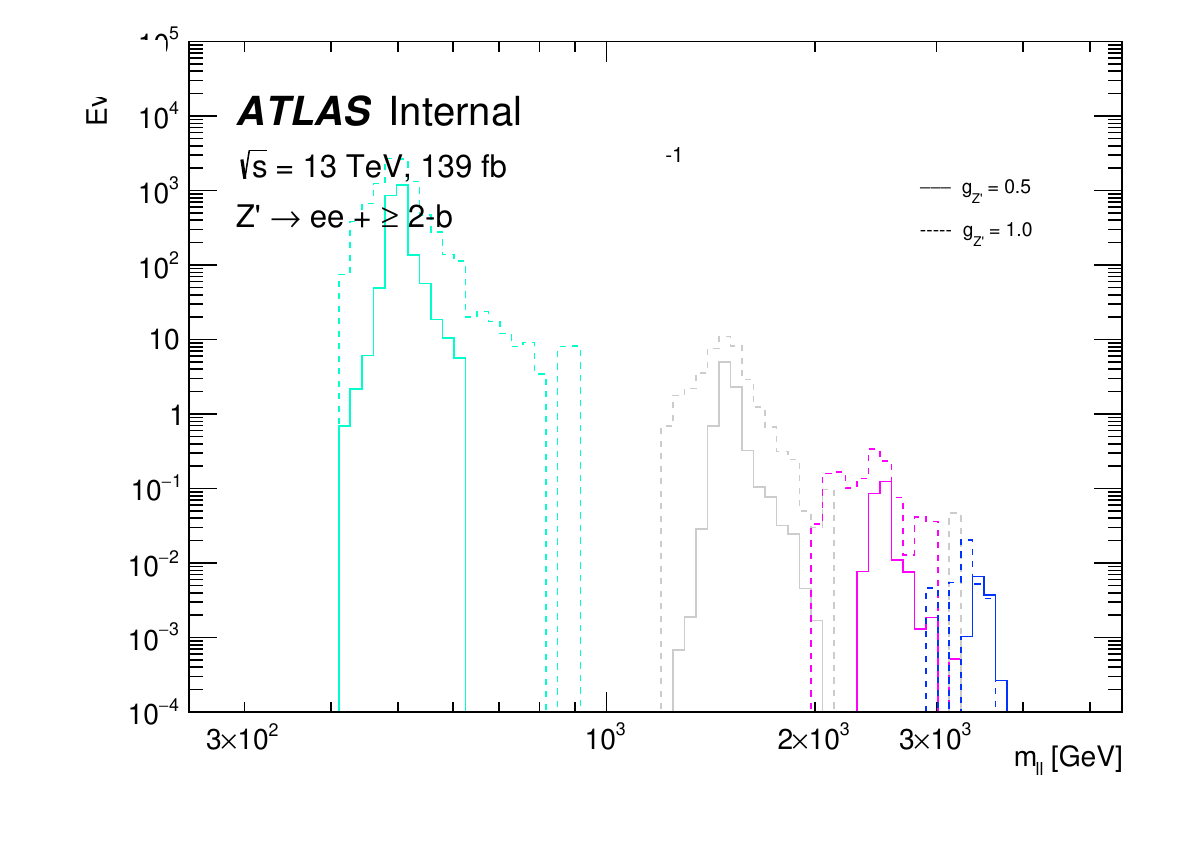}
		}
	\vspace{-1cm}
	\caption{Dielectron reconstructed invariant mass of $\Zp$ signals at various points in the 500--4000 GeV production range (cyan: $m_{Z'}=500\,\text{GeV}$, grey: $m_{Z'}=1500\,\text{GeV}$, pink: $m_{Z'}=2500\,\text{GeV}$, blue: $m_{Z'}=3500\,\text{GeV}$). (Top left) no requirements on $b$-jet multiplicity. (Top right) final state required to contain no $b$-jets. (Bottom left) final state required to contain exactly one $b$-jet. (Bottom right) final state required to contain at least two $b$-jets.}
	\label{fig:Dielectron_Signal_Mass}
\end{figure}

\begin{figure}[h!]
	\captionsetup[subfigure]{labelformat=empty}
	\centering
	\subfloat[]{
		\includegraphics[width=0.49\textwidth]{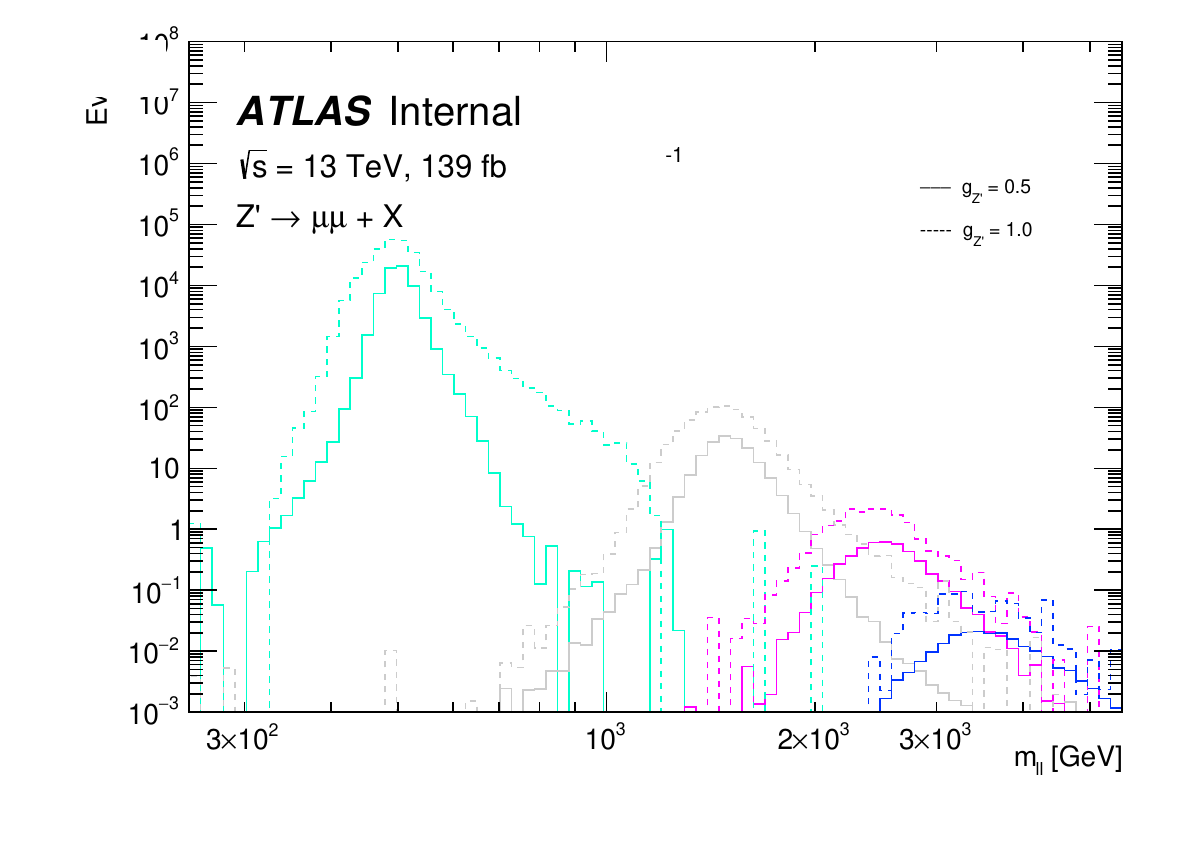}
		}
		\subfloat[]{
		\includegraphics[width=0.49\textwidth]{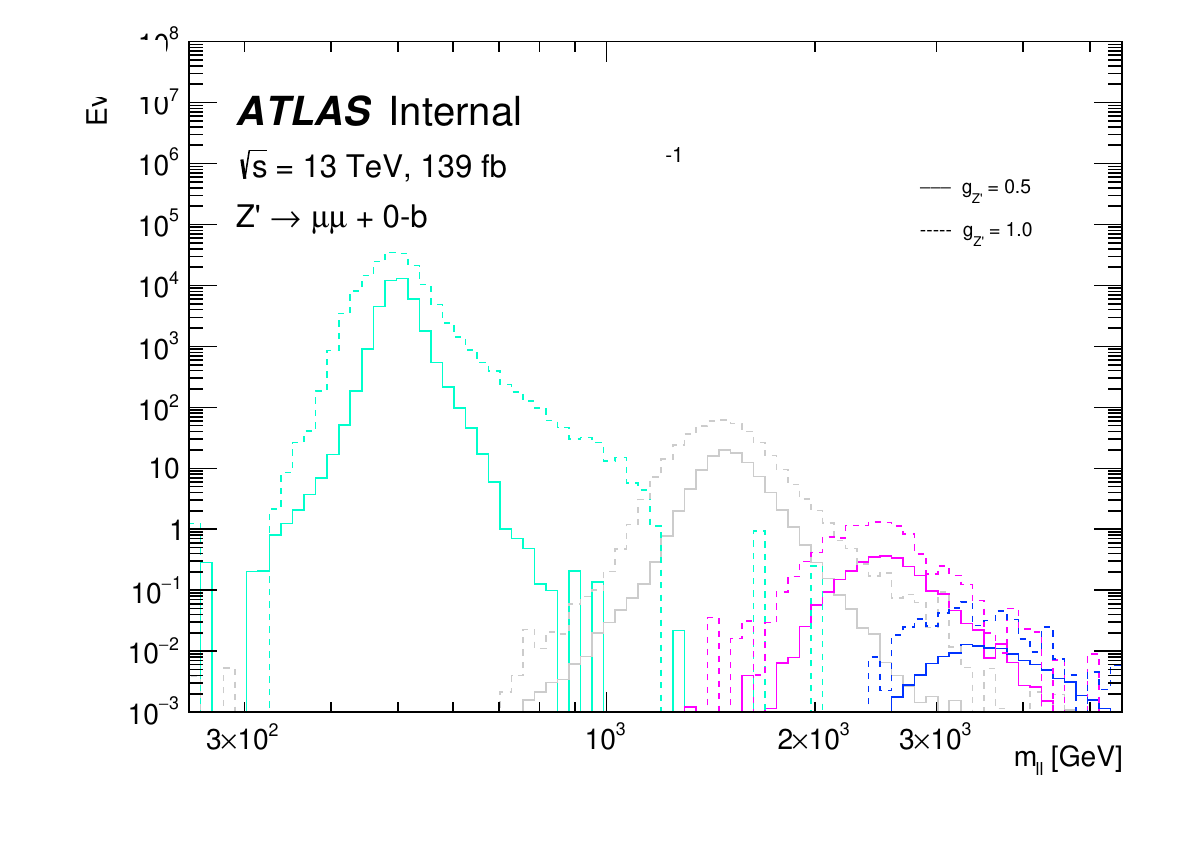}
		}
		\\ \vspace{-1cm}
		\subfloat[]{
		\includegraphics[width=0.49\textwidth]{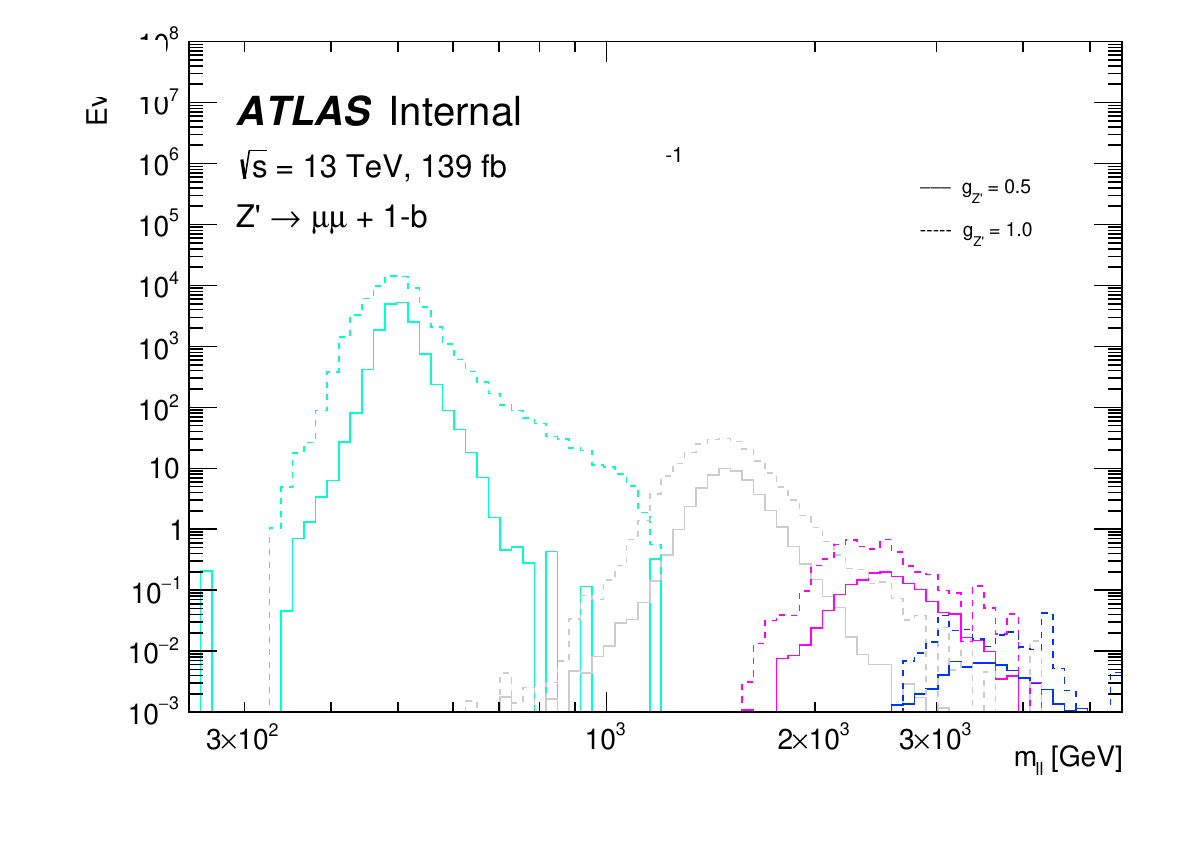}
		}
		\subfloat[]{
		\includegraphics[width=0.49\textwidth]{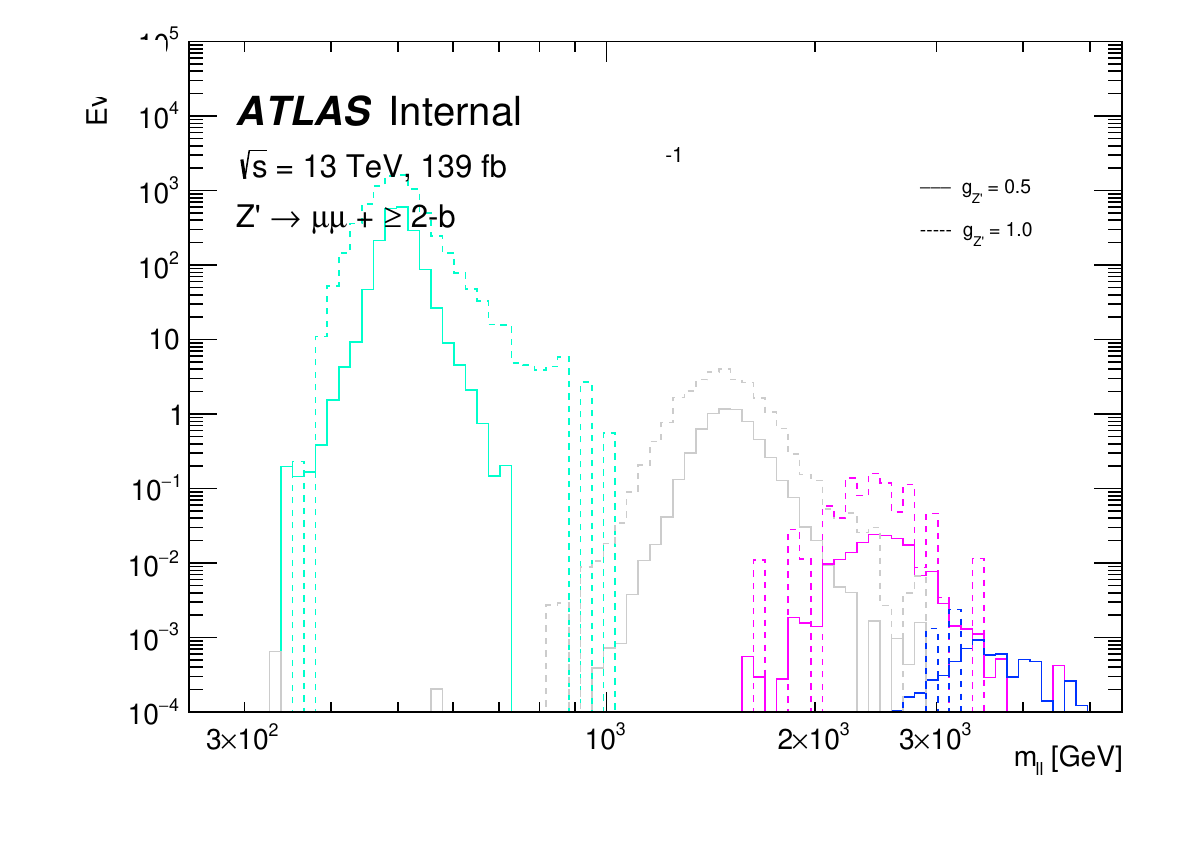}
		}
	\vspace{-1cm}
	\caption{Dimuon reconstructed invariant mass of $\Zp$ signals at various points in the 500--4000 GeV production range (cyan: $m_{Z'}=500\,\text{GeV}$, grey: $m_{Z'}=1500\,\text{GeV}$, pink: $m_{Z'}=2500\,\text{GeV}$, blue: $m_{Z'}=3500\,\text{GeV}$). (Top left) no requirements on $b$-jet multiplicity. (Top right) final state required to contain no $b$-jets. (Bottom left) final state required to contain exactly one $b$-jet. (Bottom right) final state required to contain at least two $b$-jets.}
	\label{fig:Dimuon_Signal_Mass}
\end{figure}

The resonance shape depends on the selection applied. The $b$-jet multiplicity in the final state determines both the signal efficiency and the peak height, while the coupling of the $\Zp$ to leptons and $b$-quarks controls the width, as dictated by Eq.~(\ref{eq:signal_Gamma}). In addition, the ATLAS momentum resolution for high-$p_{\textrm{T}}$ electrons is significantly better than for muons, resulting in narrower dielectron peaks. Figures~\ref{fig:Dielectron_Signal_Mass} and~\ref{fig:Dimuon_Signal_Mass} show the dielectron and dimuon invariant mass distributions for both coupling values and all $b$-jet multiplicity categories, after the full analysis selection. All distributions are shown at reconstruction level with the fiducial cuts of Section~\ref{sec:fiducial_signal_cuts} applied, visible as the lower-mass cutoffs.

As shown, the dielectron peaks are narrower than the dimuon peaks, contributing to the higher sensitivity expected in the electron channel. In the multi-TeV region, due to falling cross-section, there are very few events left.

\begin{figure}[h!]
	\captionsetup[subfigure]{labelformat=empty}
	\centering
	\subfloat[]{
		\includegraphics[width=0.49\textwidth]{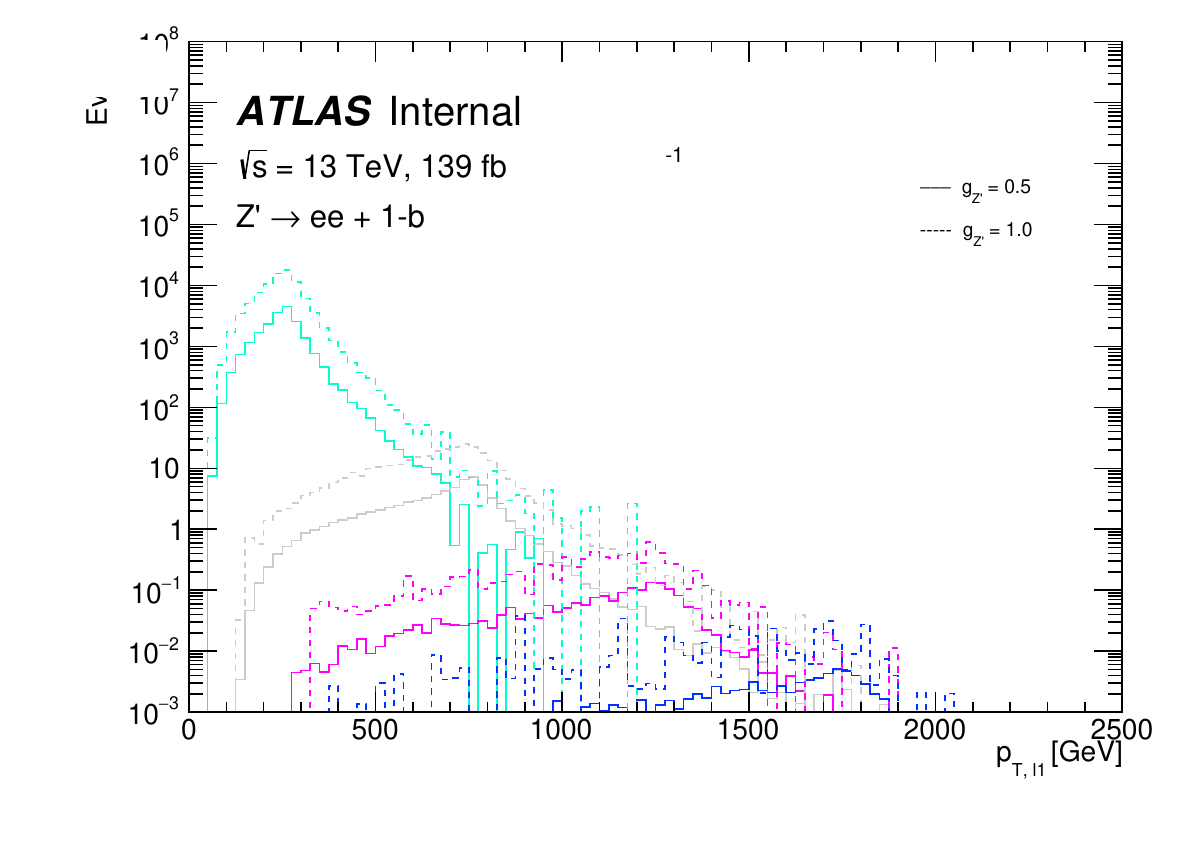}
	}
	\subfloat[]{
		\includegraphics[width=0.49\textwidth]{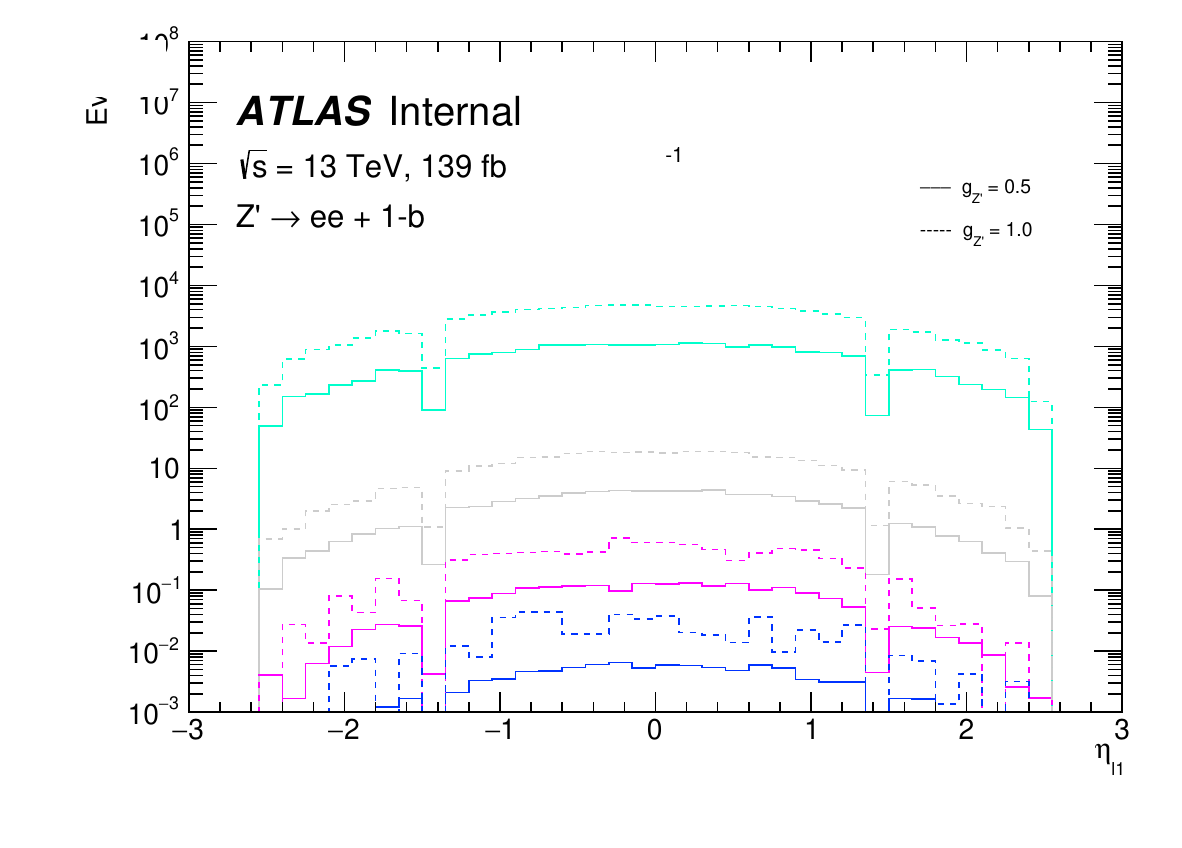}
	}
	\\ \vspace{-1cm}
	\subfloat[]{
		\includegraphics[width=0.49\textwidth]{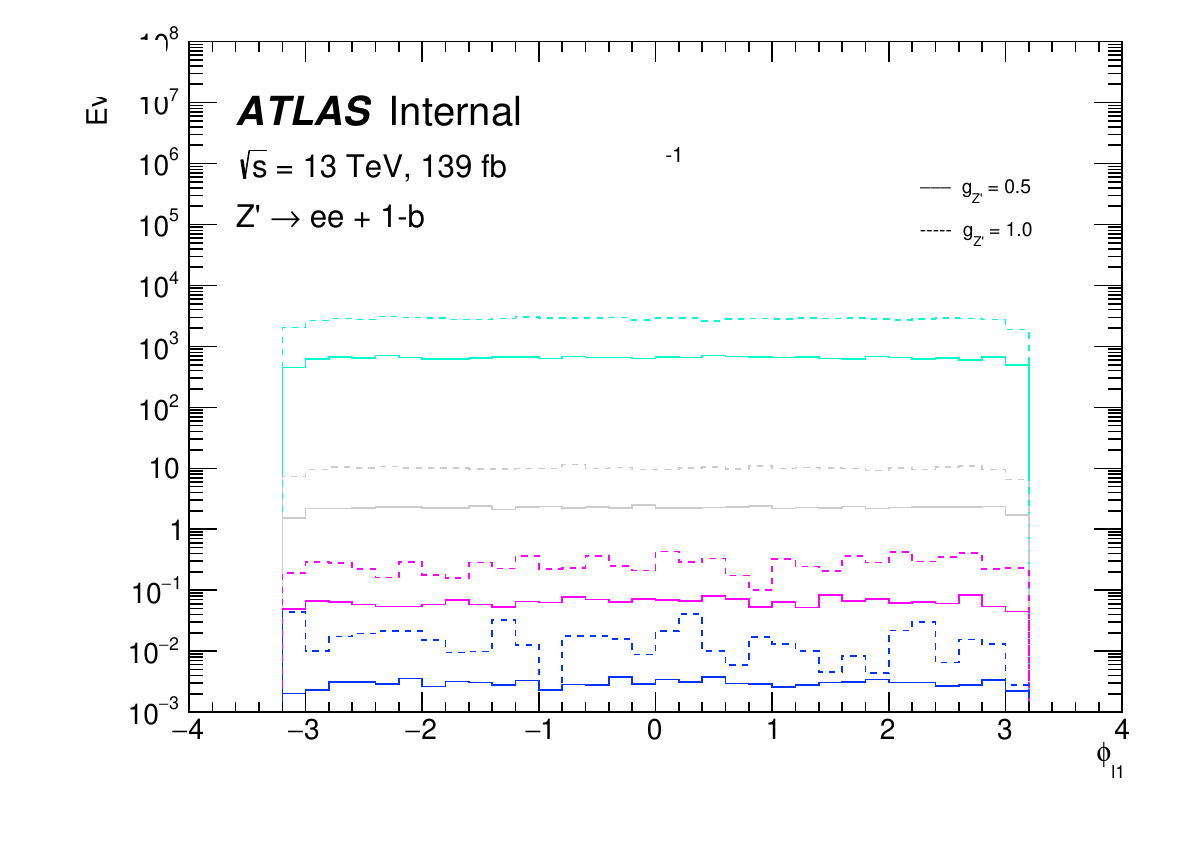}
	}
	\subfloat[]{
		\includegraphics[width=0.49\textwidth]{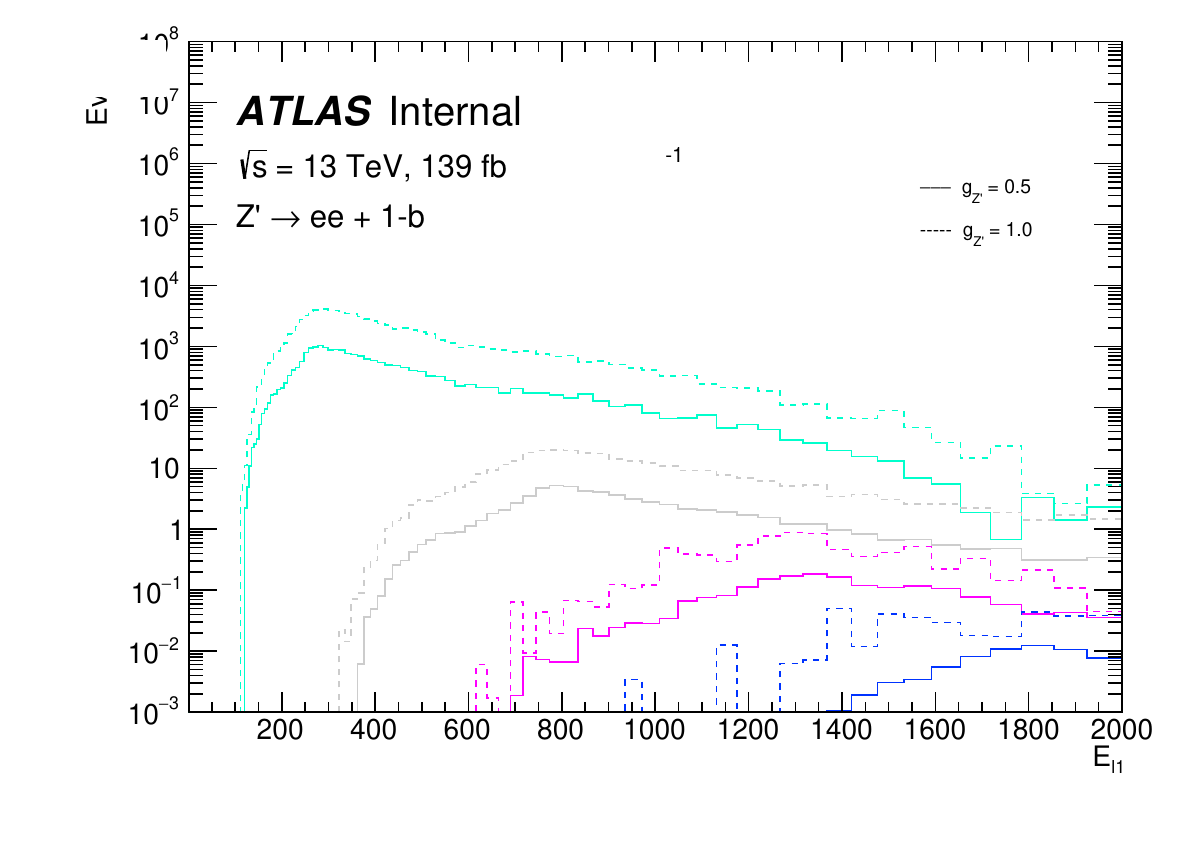}
	}
	\vspace{-1cm}
	\caption{Kinematic distributions of the leading electron for events with exactly one $b$-jet. (Top left) transverse momentum, (top right) pseudorapidity, (bottom left) azimuthal angle, (bottom right) energy (cyan: $m_{Z'}=500\,\text{GeV}$, grey: $m_{Z'}=1500\,\text{GeV}$, pink: $m_{Z'}=2500\,\text{GeV}$, blue: $m_{Z'}=3500\,\text{GeV}$).}
	\label{fig:leading_electron_kinematics_signal}
\end{figure}

\begin{figure}[h!]
	\captionsetup[subfigure]{labelformat=empty}
	\centering
	\subfloat[]{
		\includegraphics[width=0.49\textwidth]{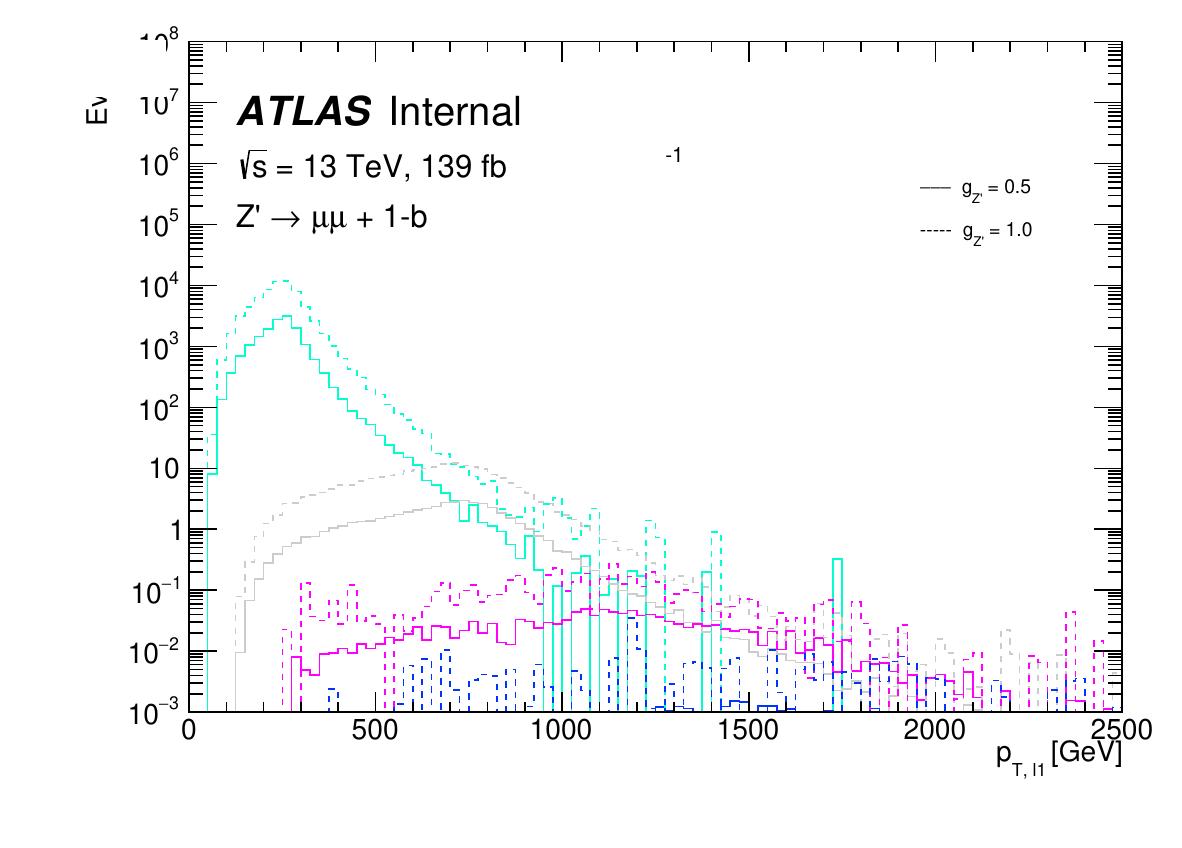}
		}
		\subfloat[]{
		\includegraphics[width=0.49\textwidth]{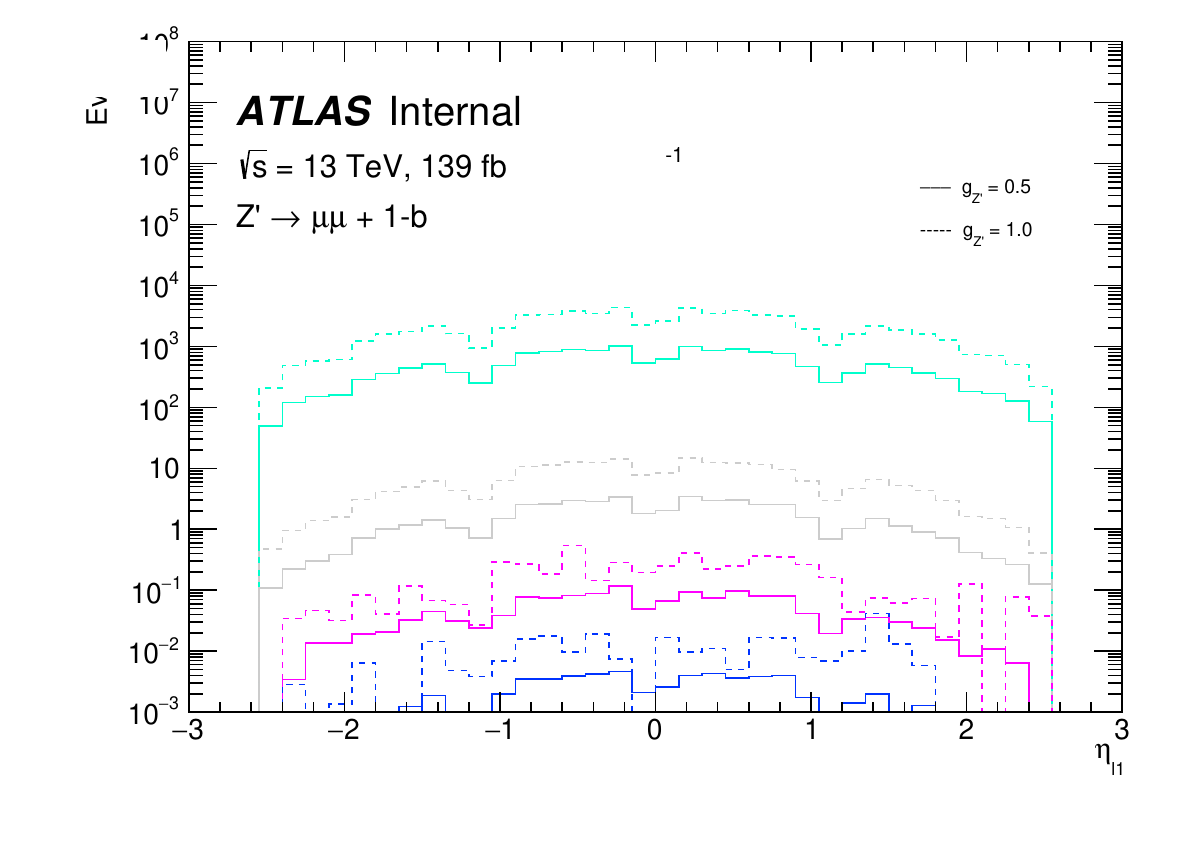}
		}
		\\ \vspace{-1cm}
		\subfloat[]{
		\includegraphics[width=0.49\textwidth]{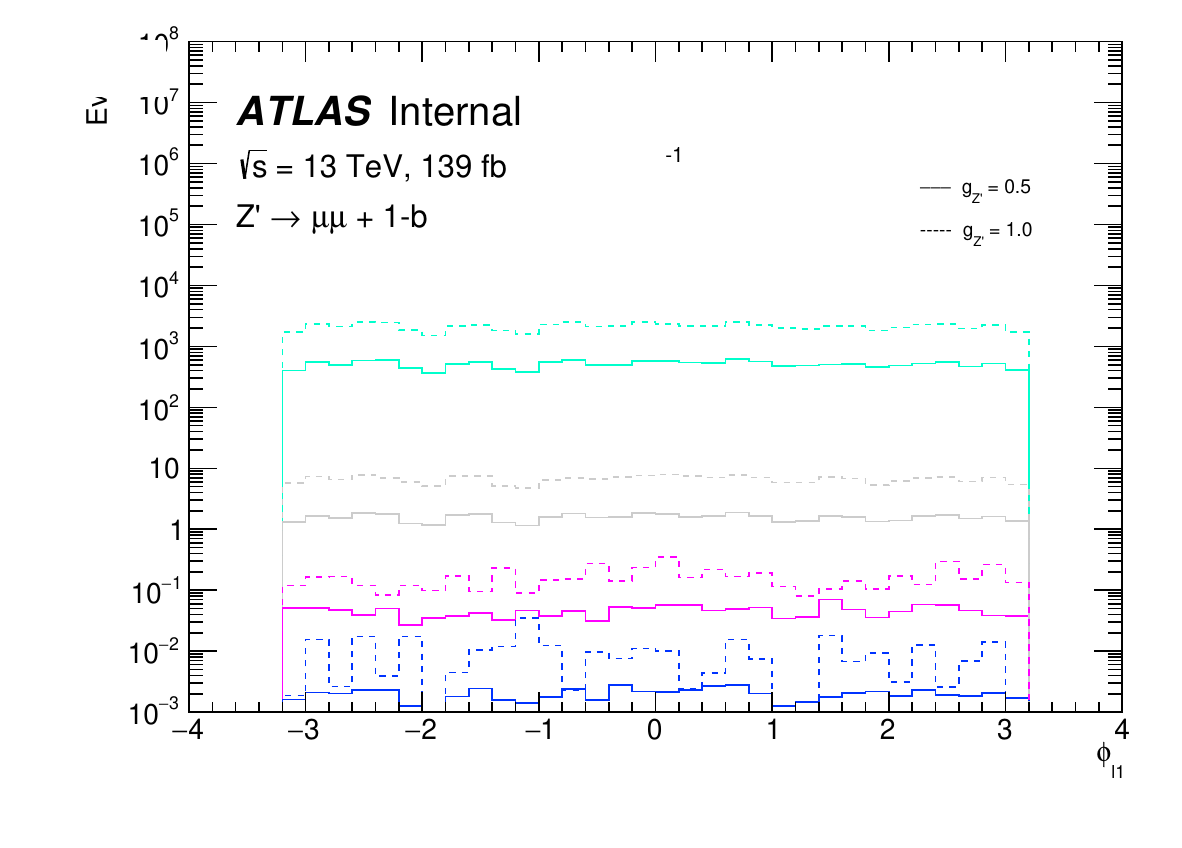}
		}
		\subfloat[]{
		\includegraphics[width=0.49\textwidth]{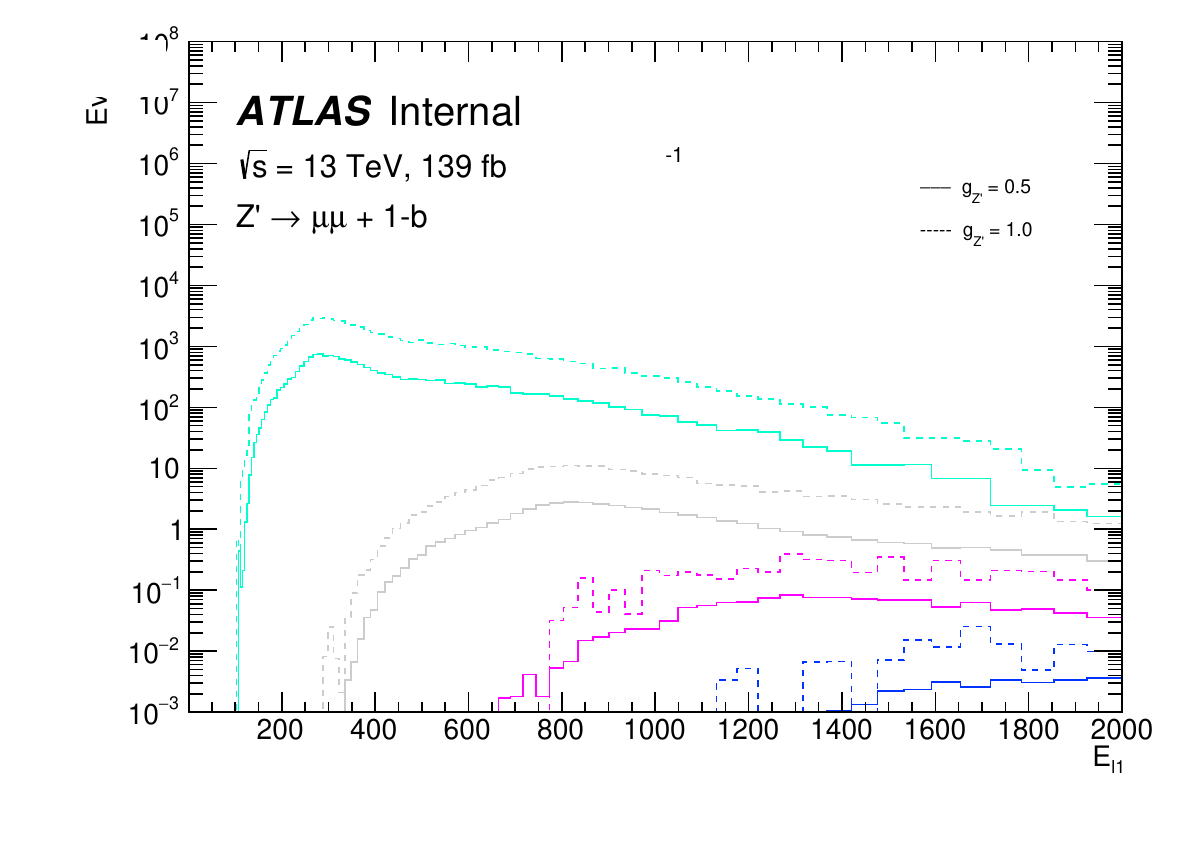}
		}
	\vspace{-1cm}
	\caption{Kinematic distributions of the leading muon for events with exactly one $b$-jet. (Top left) transverse momentum, (top right) pseudorapidity, (bottom left) azimuthal angle, (bottom right) energy (cyan: $m_{Z'}=500\,\text{GeV}$, grey: $m_{Z'}=1500\,\text{GeV}$, pink: $m_{Z'}=2500\,\text{GeV}$, blue: $m_{Z'}=3500\,\text{GeV}$).}
	\label{fig:leading_muon_kinematics_signal}
\end{figure}

Key observables for the analysis sensitivity include the leading and subleading leptons, leading and subleading $b$-jets, and missing transverse energy. Figures~\ref{fig:leading_electron_kinematics_signal} and~\ref{fig:leading_muon_kinematics_signal} show kinematic distributions of the leading lepton following the full event selection, for events with exactly one $b$-jet.

\FloatBarrier

\addtocontents{toc}{\protect\setcounter{tocdepth}{2}}

\addtocontents{toc}{\protect\setcounter{tocdepth}{0}}
\chapter{Integration of Strips into the New Small Wheel Trigger}
\label{app:nsw_trigger}

The New Small Wheel (NSW, Section~\ref{sec:atlas:nsw}) was installed during Long Shutdown~2
to maintain ATLAS muon trigger performance at the higher luminosities of Run~3.
Among its two detector technologies, the small-strip Thin Gap Chambers (sTGC) serve a dual role:
they provide precision muon tracking \emph{and} contribute directly to the first-level hardware trigger.
This appendix describes work carried out to integrate the sTGC strip readout
into the NSW trigger system using early Run~3 collision data.

\section{The sTGC Trigger Architecture}
\label{sec:nsw_trigger:architecture}

Each sTGC detector plane contains three types of readout elements sharing the same gas volume:
\begin{itemize}
  \item \textbf{Pads} --- large trapezoidal cathode electrodes providing a coarse $(\eta,\phi)$ coincidence;
  \item \textbf{Strips} --- narrow cathode strips measuring the precise
        $\eta$ (bending-plane) coordinate;
  \item \textbf{Wires} --- anode wire groups measuring the $\phi$ coordinate.
\end{itemize}
Each NSW endcap contains eight sTGC detector layers arranged in two \emph{multiplets},
each comprising four gas gaps: layers~1--4 form multiplet~1 (the inner quadruplet)
and layers~5--8 form multiplet~2 (the outer quadruplet).
Both A-side ($z > 0$) and C-side ($z < 0$) endcaps are instrumented identically.
Within each side the detector wheel is segmented azimuthally
into alternating \emph{large} and \emph{small} sectors,
with each sector further divided radially into three modules
(Q1, Q2, Q3, corresponding to $|\eta_\text{station}| = 1, 2, 3$).

\subsection{The Two-Stage Trigger Chain}
\label{sec:nsw_trigger:chain}

The sTGC trigger operates in two hardware stages.

\paragraph{Stage~1 --- Pad Trigger.}
When a muon crosses the detector, pads on the layers of a quadruplet fire.
The \emph{Pad Trigger} ASIC identifies a \emph{Band ID}:
an integer address (running from 2 to $\approx$\,94 per sector)
encoding the $\eta$ road of the hit.
Each Band ID corresponds to a narrow angular range in $\eta$ and
a specific set of strip channels expected to fire for a muon pointing straight
to the interaction vertex.
The Band ID is transmitted to the Trigger Processor within the same LHC bunch crossing.

\paragraph{Stage~2 --- Trigger Processor.}
The \emph{Trigger Processor} (TP) receives the Band ID and
uses the strip channels associated with that $\eta$ road to measure
the muon track angle $\Delta\theta$: the deviation of the muon trajectory
from a straight line pointing to the interaction vertex, projected onto the bending plane.
Because $\Delta\theta$ is proportional to the muon sagitta in the toroidal field,
it provides an online estimate of muon $p_T$,
which the Sector Logic uses to issue the L1 trigger decision.
The strip processing relies on a Look-Up Table (LUT) that maps observed
strip cluster patterns to valid muon cluster signatures.
Development and validation of this LUT against collision data is
an essential commissioning task for the sTGC trigger.

\section{The Strip-to-Band Mapping and Trigger Processor Emulator}
\label{sec:nsw_trigger:mapping}

\subsection{The Strip-to-Band Dictionary}

The geometrical correspondence between strip channels and Band IDs is encoded in a
mapping derived from the detector geometry:
for every combination of \{size (large/small), layer (1--8), module (Q1--Q3), sector\},
the mapping records the Band ID to which each strip channel belongs.
This information was stored in a lookup table (\texttt{Strips.csv})
covering the full strip inventory for all sectors on both sides.

Two complementary dictionaries were constructed from this table:
\begin{itemize}
  \item \textbf{Forward dictionary}: strip channel $\to$ Band ID ---
        given a fired strip, in which band does it fall?
  \item \textbf{Reverse dictionary}: Band ID $\to$ strip list ---
        given a pad-trigger Band ID, which strip channels are expected to fire?
\end{itemize}
The forward dictionary translates offline strip clusters into the TP's address space;
the reverse dictionary identifies the first strip of each band,
which serves as the reference for the 14-strip sliding window used in the LUT.

\subsection{The Trigger Processor Emulator}

The TP firmware classifies strip clusters using a LUT that maps 14-bit binary
patterns---encoding which of 14 consecutive strips fired within a band---to a validity flag.
A software emulator of this LUT (\texttt{TriggerProcessorEmulator.cpp})
was developed to enable direct comparison with data.

Valid cluster patterns recognised by the TP correspond to contiguous muon ionisation tracks:
groups of 3--6 strips with at most one missing strip (to tolerate single-strip inefficiency).
Table~\ref{tab:nsw_trigger:lut} lists the recognised patterns.
Single- and two-strip entries, as well as patterns with large gaps,
are rejected as noise or secondary hits.
The emulator generates the full LUT as a C++ header file (\texttt{lut\_output.h})
used at runtime by the analysis to test each observed cluster against the firmware acceptance.

\begin{table}[htbp]
  \centering
  \caption{Cluster templates recognised by the Trigger Processor LUT.
    Each pattern is a binary string of fired (1) and silent (0) strips
    within a 14-strip window, right-justified.}
  \label{tab:nsw_trigger:lut}
  \begin{tabular}{ccc}
    \hhline{===}
    Cluster ID & Strip pattern & Size \\
    \hline
    1  & \texttt{11111}  & 5 \\
    2  & \texttt{10111}  & 5 \\
    3  & \texttt{11011}  & 5 \\
    4  & \texttt{11101}  & 5 \\
    5  & \texttt{1111}   & 4 \\
    6  & \texttt{1011}   & 4 \\
    7  & \texttt{1101}   & 4 \\
    8  & \texttt{111}    & 3 \\
    9  & \texttt{101}    & 3 \\
    10 & \texttt{111111} & 6 \\
    11 & \texttt{101111} & 6 \\
    12 & \texttt{110111} & 6 \\
    13 & \texttt{111011} & 6 \\
    14 & \texttt{111101} & 6 \\
    \hhline{===}
  \end{tabular}
\end{table}

\section{Data Sample and Analysis Framework}
\label{sec:nsw_trigger:data}

The study uses Run~3 ATLAS data recorded at $\sqrt{s} = 13.6\,\TeV$ in 2023,
from the Muon Tester ntuple (MUTEST) format --- a dedicated commissioning stream
recording full NSW strip and pad readout alongside inner-detector and muon-spectrometer tracks.
Two runs are used:
\begin{itemize}
  \item Run~453319 (8\,GB): the primary dataset for cluster characterisation,
        band-ID multiplicity, and LUT matching studies.
  \item Run~455814 (1\,GB): used for the Band ID internal consistency study
        (Section~\ref{sec:nsw_trigger:bandid_consistency}),
        yielding 3476 on-track muon clusters.
\end{itemize}

The analysis framework (\texttt{TriggerAnalyser}) processes sTGC strip data
across all 32 readout categories
(8 layers $\times$ 2 sides $\times$ 2 sector sizes),
applying two preprocessing steps:
\begin{itemize}
  \item \textbf{Noise channel masking}: a first pass builds per-channel occupancy histograms;
        channels with occupancy more than $3.5\,\sigma$ above the mean are masked.
  \item \textbf{Muon selection}: for efficiency and matching studies, good muon candidates
        are required to satisfy $p_T > 15\,\GeV$, $1.3 < |\eta| < 2.4$,
        and muon author~1 (combined) or author~5 (MS standalone).
\end{itemize}

\section{Strip Cluster Properties}
\label{sec:nsw_trigger:clusters}

\subsection{Channel Occupancy}

\begin{figure}[htbp]
  \centering
  \includegraphics[width=0.72\textwidth]{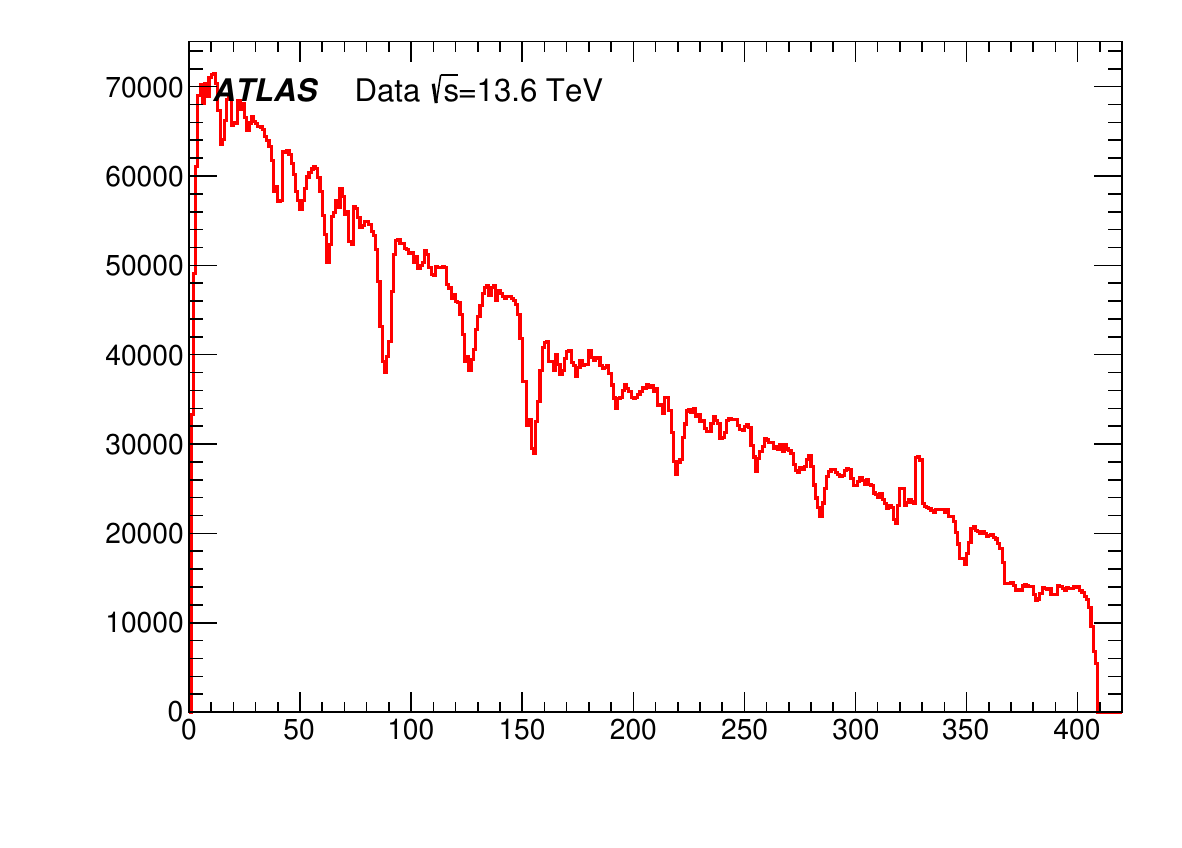}
  \caption{Strip channel occupancy for sTGC layer~1 in data at $\sqrt{s}=13.6\,\TeV$.
    The declining envelope from inner (low channel number) to outer (high channel number)
    strips reflects solid-angle coverage.
    The modulated multi-peak structure arises from the three radial modules (Q1, Q2, Q3)
    and their internal detector panel boundaries.}
  \label{fig:nsw_trigger:strip_occ}
\end{figure}

FIG.~\ref{fig:nsw_trigger:strip_occ} shows the raw per-channel occupancy of sTGC strips
in layer~1, summed over all events.
The overall envelope declines from the innermost channels (low channel numbers, inner radius)
towards higher channel numbers (outer radius),
reflecting solid-angle coverage:
strips closer to the beam axis subtend a larger $\eta$ range and accumulate more hits.
Superimposed on this global trend is a modulated multi-peaked structure
arising from the radial segmentation into three modules (Q1, Q2, Q3)
and the boundaries between adjacent detector panels within each module.
No extended dead regions are visible, confirming that all strip channels are active
and contributing to the readout; a small number of hot channels are subsequently
identified by the $3.5\,\sigma$ occupancy cut and masked before further analysis.

\begin{figure}[htbp]
  \centering
  \includegraphics[width=0.72\textwidth]{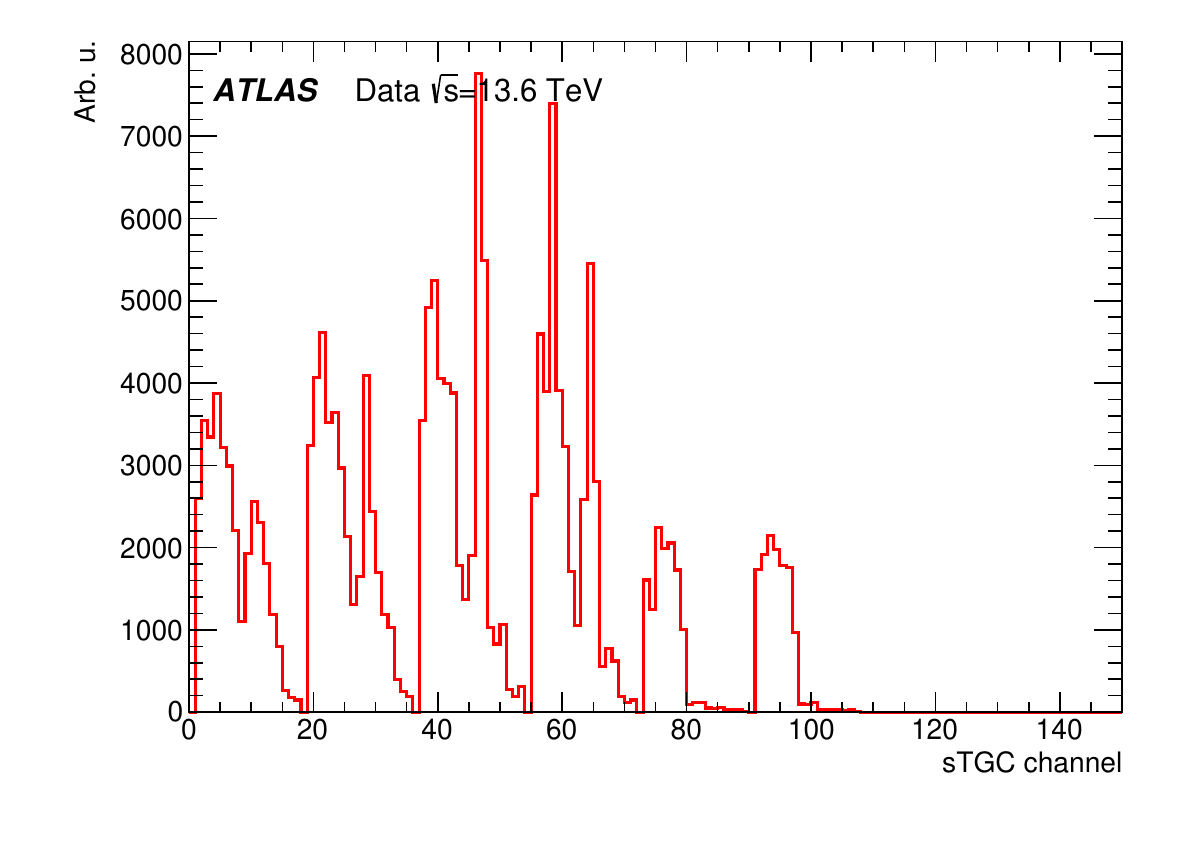}
  \caption{Pad-trigger hit channel distribution for sTGC layer~1 in data at $\sqrt{s}=13.6\,\TeV$.
    Each discrete peak corresponds to a group of pads forming a single pad-trigger readout unit.
    The structure spanning channels 5--95 reflects the complete $\eta$ segmentation of the pad readout.}
  \label{fig:nsw_trigger:pad_occ}
\end{figure}

FIG.~\ref{fig:nsw_trigger:pad_occ} shows the hit channel distribution of the sTGC pad trigger
for layer~1.
The distribution consists of discrete peaks separated by clear gaps,
each peak corresponding to a group of geometrically adjacent pads
belonging to a single pad-trigger readout unit.
The discrete structure spanning channels 5--95 reflects the complete
$\eta$ segmentation of the pad readout across the layer.

\subsection{Cluster Size}

\begin{figure}[htbp]
  \centering
  \includegraphics[width=0.72\textwidth]{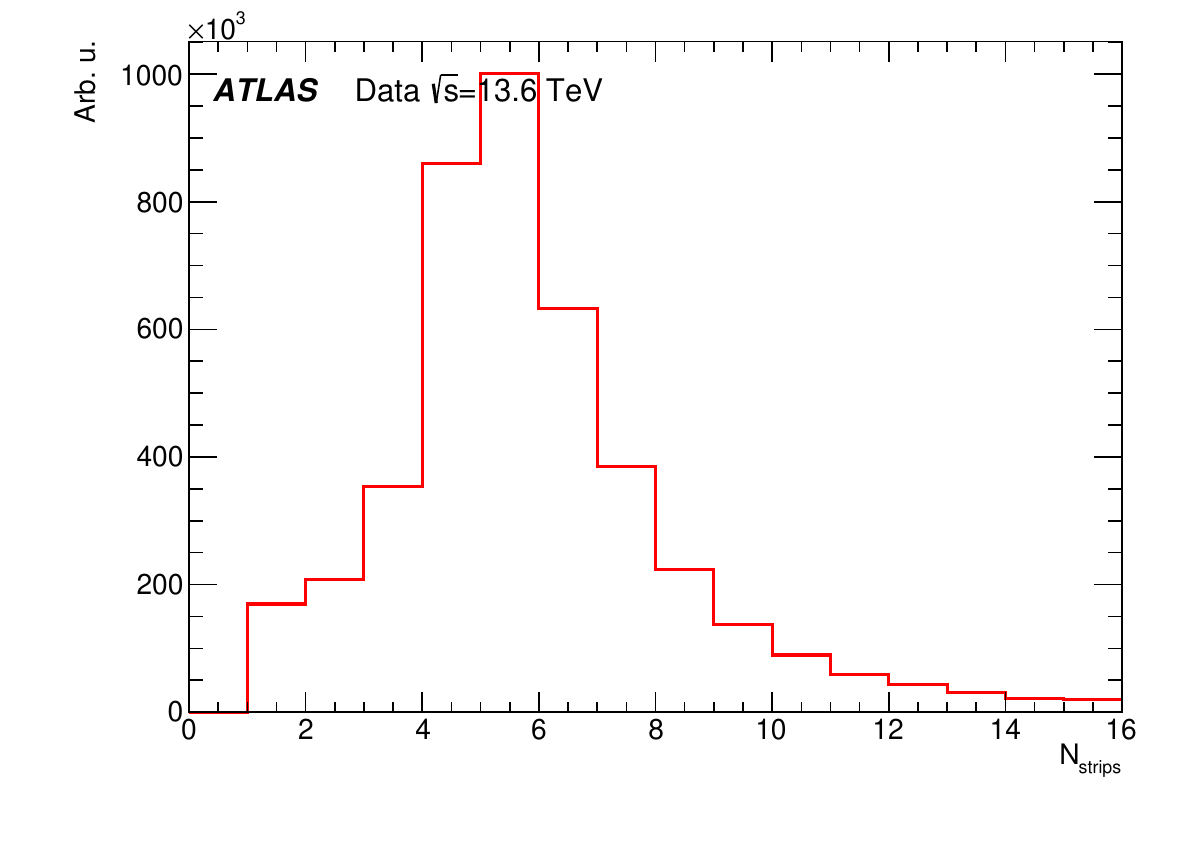}
  \caption{Distribution of the number of strips per cluster, $N_\text{strips}$,
    for sTGC layer~1, A-side large sector, in data at $\sqrt{s}=13.6\,\TeV$.
    The distribution peaks at $N_\text{strips}=5$ for all hits, consistent with
    the expected charge sharing in the sTGC gas.
    The LUT accepts clusters with 3--6 strips, covering the bulk of physical muon clusters.}
  \label{fig:nsw_trigger:nstrips}
\end{figure}

FIG.~\ref{fig:nsw_trigger:nstrips} shows the distribution of the number of strips
per cluster ($N_\text{strips}$) for layer~1 of the A-side large sector, for all hits.
The distribution peaks at $N_\text{strips} = 5$,
consistent with the expected charge sharing in the sTGC gas mixture:
a muon traversing the chamber at normal incidence ionises a column $\sim$2.7\,mm wide,
which at the strip pitch of $\sim$3.2\,mm projects to approximately 5 fired strips.
A secondary population at $N_\text{strips} = 1$--2 arises from noise hits and
edge-truncated clusters; the tail at $N_\text{strips} \geq 7$ from oblique tracks,
delta rays, and occasional cluster merging.
For good-muon selected clusters the peak shifts to $N_\text{strips} = 4$,
reflecting the cleaner isolation and tighter track-to-cluster association of that selection.
This range motivates the LUT acceptance window of 3--6 strips.

\subsection{Cluster Charge}

\begin{figure}[htbp]
  \centering
  \begin{subfigure}[b]{0.48\textwidth}
    \includegraphics[width=\textwidth]{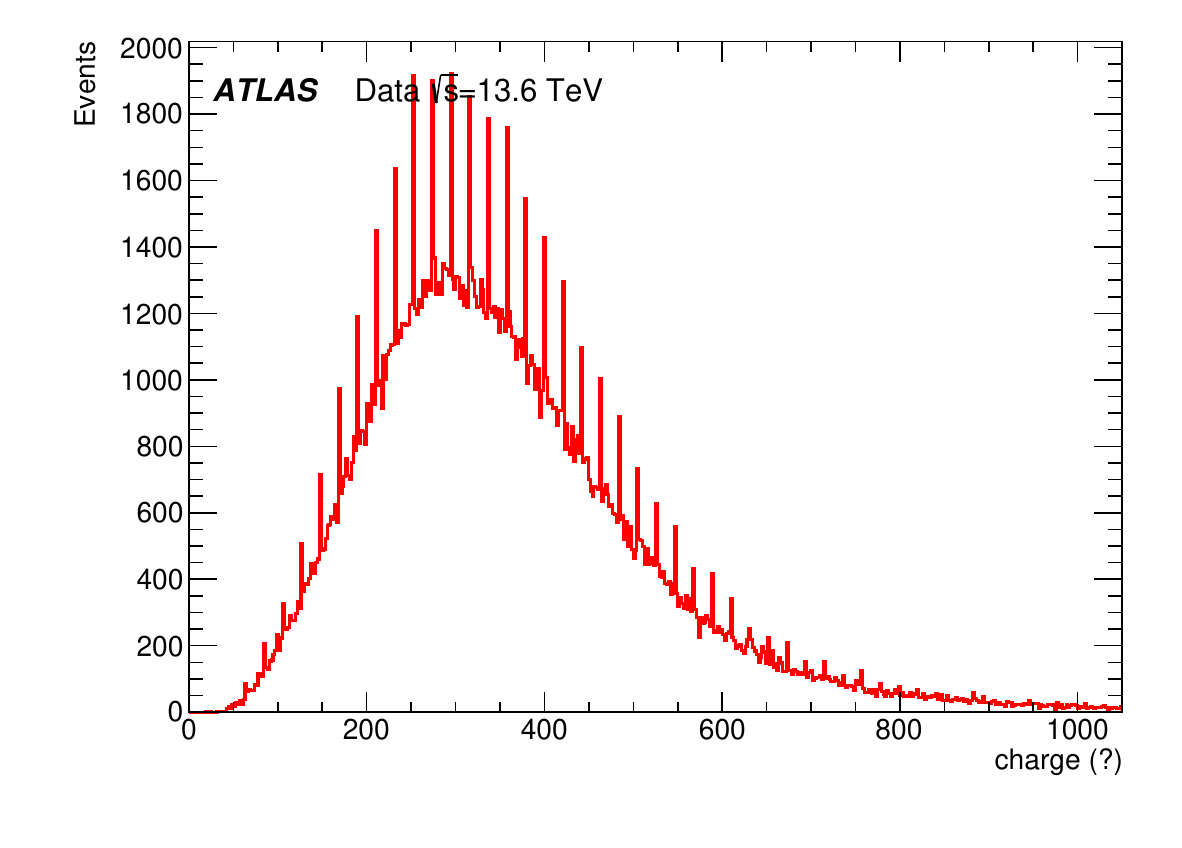}
    \caption{Normal clusters (3--5 strips).}
    \label{fig:nsw_trigger:charge_normal}
  \end{subfigure}
  \hfill
  \begin{subfigure}[b]{0.48\textwidth}
    \includegraphics[width=\textwidth]{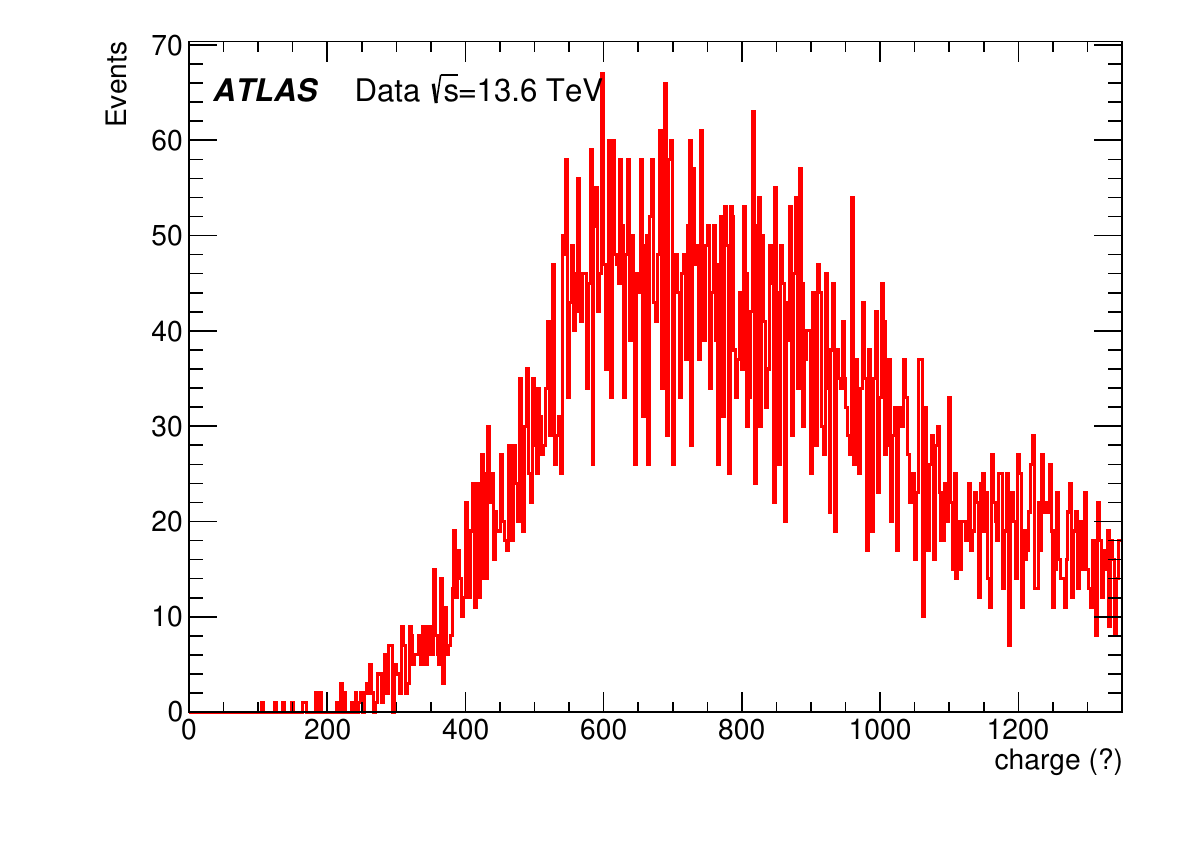}
    \caption{Large clusters ($\geq 6$ strips).}
    \label{fig:nsw_trigger:charge_large}
  \end{subfigure}
  \caption{Cluster charge distributions for good-muon matched sTGC clusters
    at $\sqrt{s}=13.6\,\TeV$.
    Normal clusters (left) show a Landau-like spectrum peaking at $\sim$300\,ADC counts,
    consistent with minimum-ionising particles.
    Large clusters (right) exhibit a broader distribution at higher values.
    The spike structure reflects ADC quantisation of individual strip charges.}
  \label{fig:nsw_trigger:charge}
\end{figure}

FIG.~\ref{fig:nsw_trigger:charge} shows the charge distribution for good-muon matched clusters
with 3--5 strips (``normal'' clusters, left) and for clusters with $\geq 6$ strips
(``large'' clusters, right).
Normal clusters exhibit a Landau-like spectrum peaking at $\sim$300\,ADC units
and extending to $\sim$1000, consistent with minimum-ionising particle energy loss
in the sTGC gas.
Large clusters have a charge distribution shifted to higher values (peak $\sim$600--700\,ADC),
as expected from the greater number of strips contributing to the total.
Both distributions exhibit a fine spike structure arising from the ADC quantisation
of individual strip charges.

\subsection{Charge Correlations}

\begin{figure}[htbp]
  \centering
  \begin{subfigure}[b]{0.48\textwidth}
    \includegraphics[width=\textwidth]{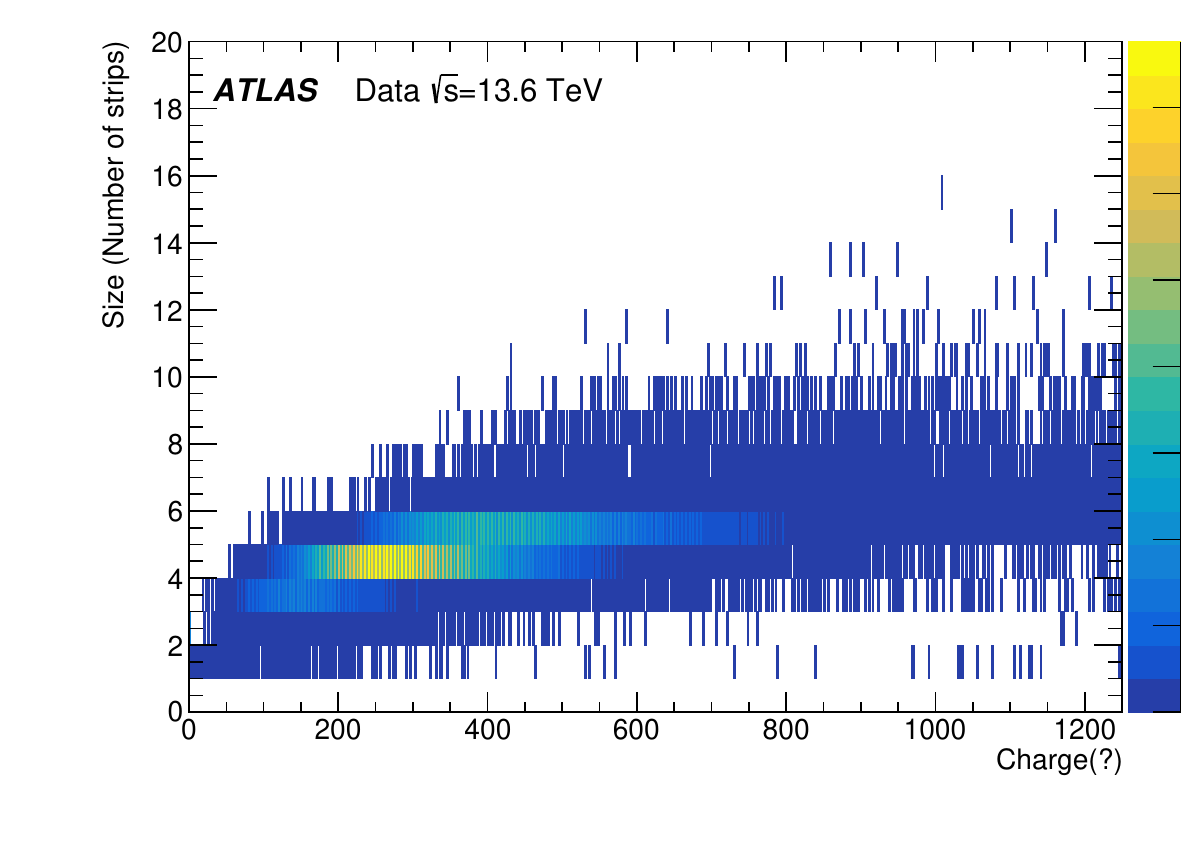}
    \caption{Charge vs.\ cluster size.}
    \label{fig:nsw_trigger:cvsize}
  \end{subfigure}
  \hfill
  \begin{subfigure}[b]{0.48\textwidth}
    \includegraphics[width=\textwidth]{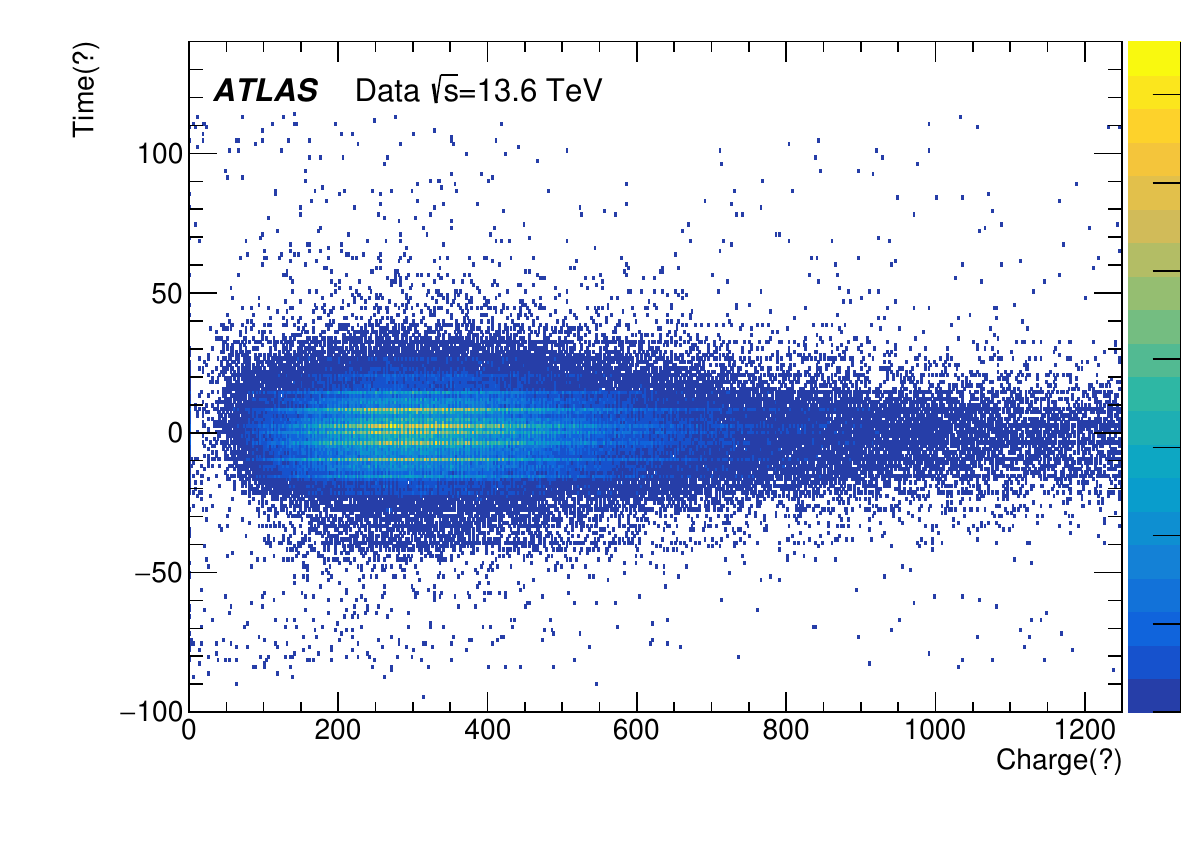}
    \caption{Charge vs.\ strip timing.}
    \label{fig:nsw_trigger:cvtime}
  \end{subfigure}
  \caption{Two-dimensional cluster charge distributions for good-muon LUT-matched clusters
    at $\sqrt{s}=13.6\,\TeV$.
    Left: charge versus $N_\text{strips}$. The highest density at $N_\text{strips}=5$ and
    200--400\,ADC represents typical muon clusters.
    Right: charge versus strip timing.
    In-time hits concentrate near $t=0$\,ns with $\lesssim\pm50$\,ns spread,
    and no time-walk correlation is observed.}
  \label{fig:nsw_trigger:charge2d}
\end{figure}

FIG.~\ref{fig:nsw_trigger:charge2d} shows two 2D distributions of cluster charge
for good-muon LUT-matched clusters.

The left panel shows charge versus cluster size $N_\text{strips}$.
The highest-density region (yellow) is concentrated at $N_\text{strips} = 5$
and charge $\sim$200--400\,ADC, with a clear positive correlation between the two:
larger clusters collect proportionally more total charge.
At a given cluster size the charge retains a Landau character;
at a given charge value the cluster size spans a range of a few strips,
reflecting shower fluctuations and strip granularity.

The right panel shows cluster charge versus strip timing in nanoseconds.
The distribution is centred near $t = 0$\,ns with charge concentrated
at 200--400\,ADC --- consistent with in-time muon hits aligned to the LHC clock.
The timing spread of $\lesssim\pm50$\,ns spans approximately two bunch-crossing windows.
The absence of any diagonal structure confirms that the charge measurement is not
affected by time-walk at these signal amplitudes.

\section{Per-Band-ID Multiplicity}
\label{sec:nsw_trigger:bandid}

A key diagnostic of the strip--pad trigger integration is whether the strip detector
responds uniformly across all Band IDs.
For each Band ID $b$ ($b = 2, \ldots, 92$ for small sectors),
the average number of strip clusters per event in band $b$
is computed using the read-out (RO) Band ID assignment:
the Band ID is derived offline from the strip cluster position
via the forward dictionary, independently of the pad trigger word.

\begin{figure}[htbp]
  \centering
  \includegraphics[width=0.72\textwidth]{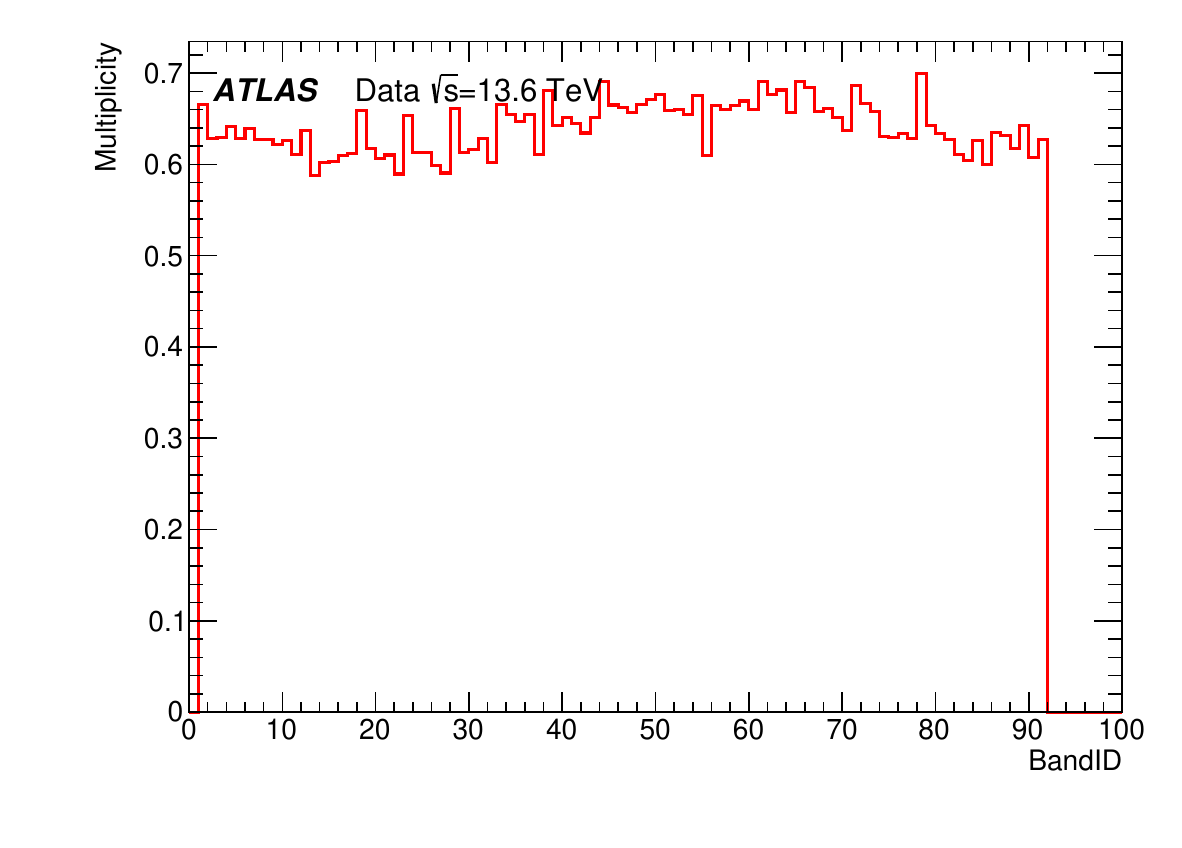}
  \caption{Average number of strip clusters per event as a function of Band ID,
    for sTGC layer~1, A-side large sector, at $\sqrt{s}=13.6\,\TeV$.
    The Band ID is assigned offline via the strip-to-band forward dictionary (RO convention).
    The flat profile at $\approx$\,0.65 across Band IDs $b = 2$--91
    demonstrates uniform strip detector response across the full $\eta$ acceptance.
    The drop at $b \approx 92$ reflects a known edge effect at the outermost module.}
  \label{fig:nsw_trigger:bandid_ro}
\end{figure}

FIG.~\ref{fig:nsw_trigger:bandid_ro} shows this multiplicity profile for layer~1
of the A-side large sector.
The distribution is flat at $\approx 0.62$--$0.70$ across all active Band IDs
from $b = 2$ to $b \approx 91$, with a sharp drop to zero at $b = 92$.
The flatness demonstrates that the sTGC strip detector responds uniformly
across the full $\eta$ acceptance of the NSW:
every angular band is equally likely to contain a strip cluster,
as expected for an isotropic muon flux from beam collisions.
The drop at the outermost Band ID is a known edge effect from the last detector module.
This result also validates the forward dictionary:
the strip-to-band mapping covers the full angular range without gaps or biases.

\begin{figure}[htbp]
  \centering
  \includegraphics[width=0.72\textwidth]{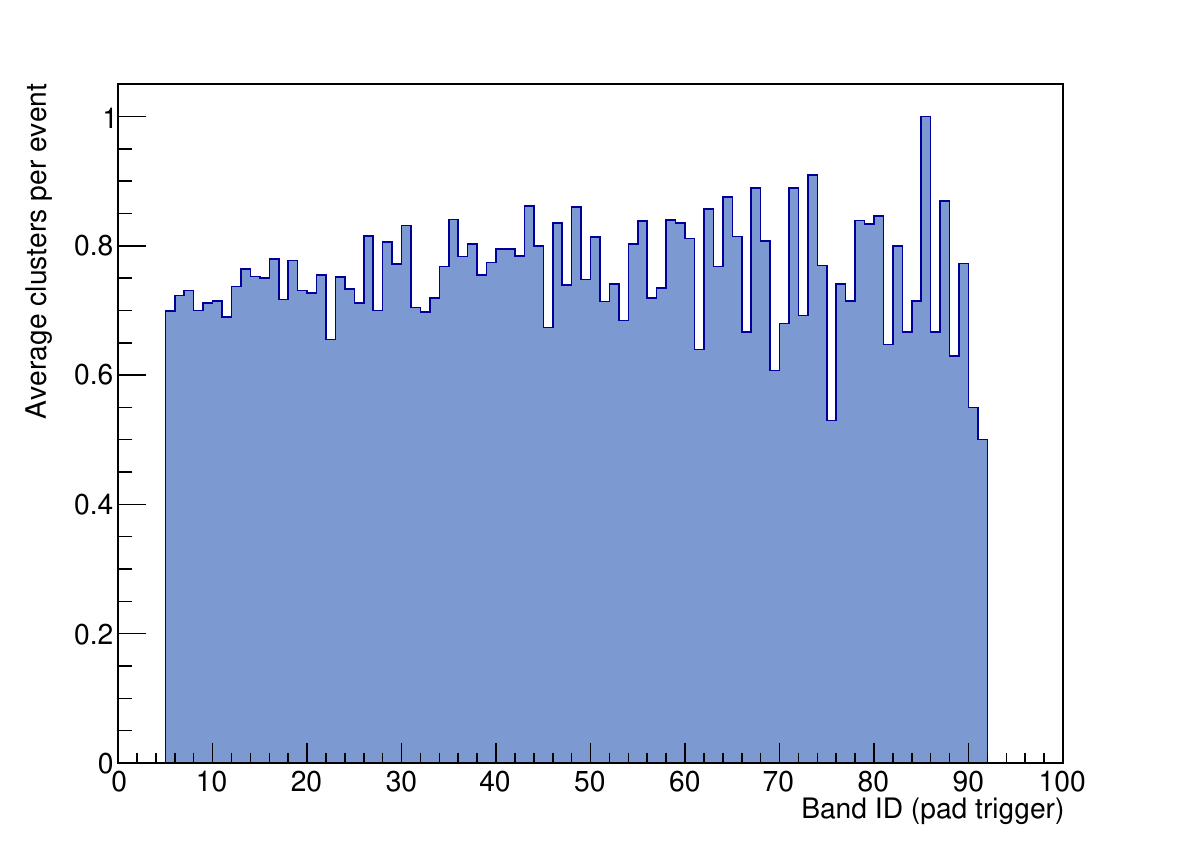}
  \caption{Average number of strip clusters per event as a function of Band ID,
    using the Trigger Processor (pad-trigger firmware) Band ID convention,
    for sTGC layer~1, A-side large sector, at $\sqrt{s}=13.6\,\TeV$.
    As with the offline (RO) assignment (FIG.~\ref{fig:nsw_trigger:bandid_ro}),
    the profile is flat across all active Band IDs,
    confirming uniform pad-trigger coverage of the full $\eta$ acceptance.
    The Band ID range populated differs from the RO convention,
    reflecting the systematic offset between the two Band ID numbering schemes.}
  \label{fig:nsw_trigger:bandid_tp}
\end{figure}

The same multiplicity profile computed using the pad-trigger Band ID
(the TP convention, derived from the firmware word rather than the offline dictionary)
is shown in FIG.~\ref{fig:nsw_trigger:bandid_tp}.
The distribution is similarly flat across the active Band IDs,
indicating that the pad trigger also observes uniform coverage.
Crucially, however, the two profiles are offset relative to each other:
the Band IDs assigned by the pad trigger firmware systematically differ from
those assigned by the offline strip-to-band dictionary for the same set of muon tracks.
This discrepancy directly underpins the LUT matching failure discussed in
Section~\ref{sec:nsw_trigger:matching}.

\section{Band ID Internal Consistency}
\label{sec:nsw_trigger:bandid_consistency}

An important cross-check of the strip-to-band mapping is whether all strips
within a single on-track cluster consistently map to the same Band ID.
For a physical single-muon cluster every fired strip should belong to the same
$\eta$ band; a cluster whose strips map to different Band IDs either
sits at a band boundary or contains contributions from more than one track.

\begin{figure}[htbp]
  \centering
  \includegraphics[width=0.72\textwidth]{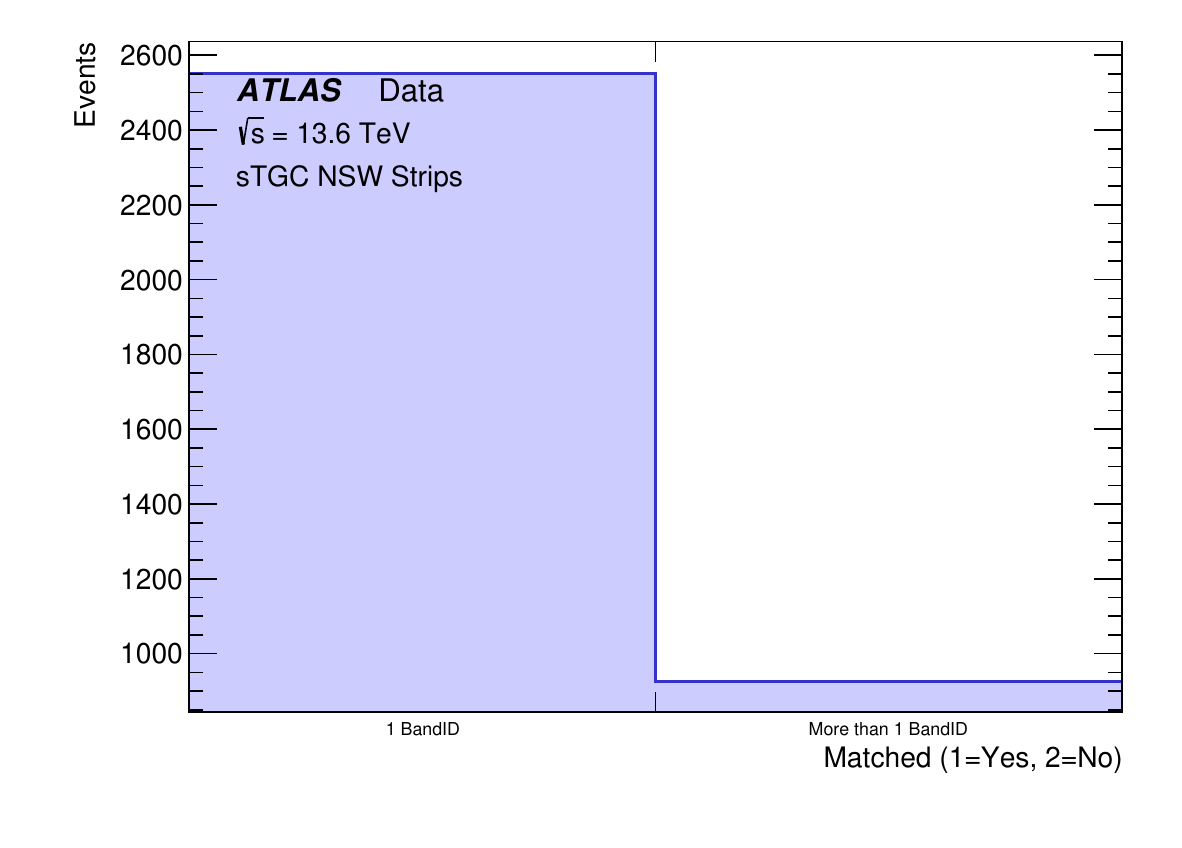}
  \caption{Fraction of on-track sTGC strip clusters whose strips all map to
    a single Band ID (left bin) versus clusters spanning more than one Band ID (right bin),
    in data at $\sqrt{s}=13.6\,\TeV$.
    Approximately 73\% of the 3476 on-track clusters are internally consistent
    with a single Band ID.}
  \label{fig:nsw_trigger:efficiency}
\end{figure}

FIG.~\ref{fig:nsw_trigger:efficiency} shows the result for the 3476 on-track clusters
from run~455814:
approximately 2550 clusters ($\approx 73\%$) map all their strips to a single Band ID,
while the remaining $\approx 27\%$ ($\approx 925$ clusters) span more than one Band ID.

\begin{figure}[htbp]
  \centering
  \begin{subfigure}[b]{0.48\textwidth}
    \includegraphics[width=\textwidth]{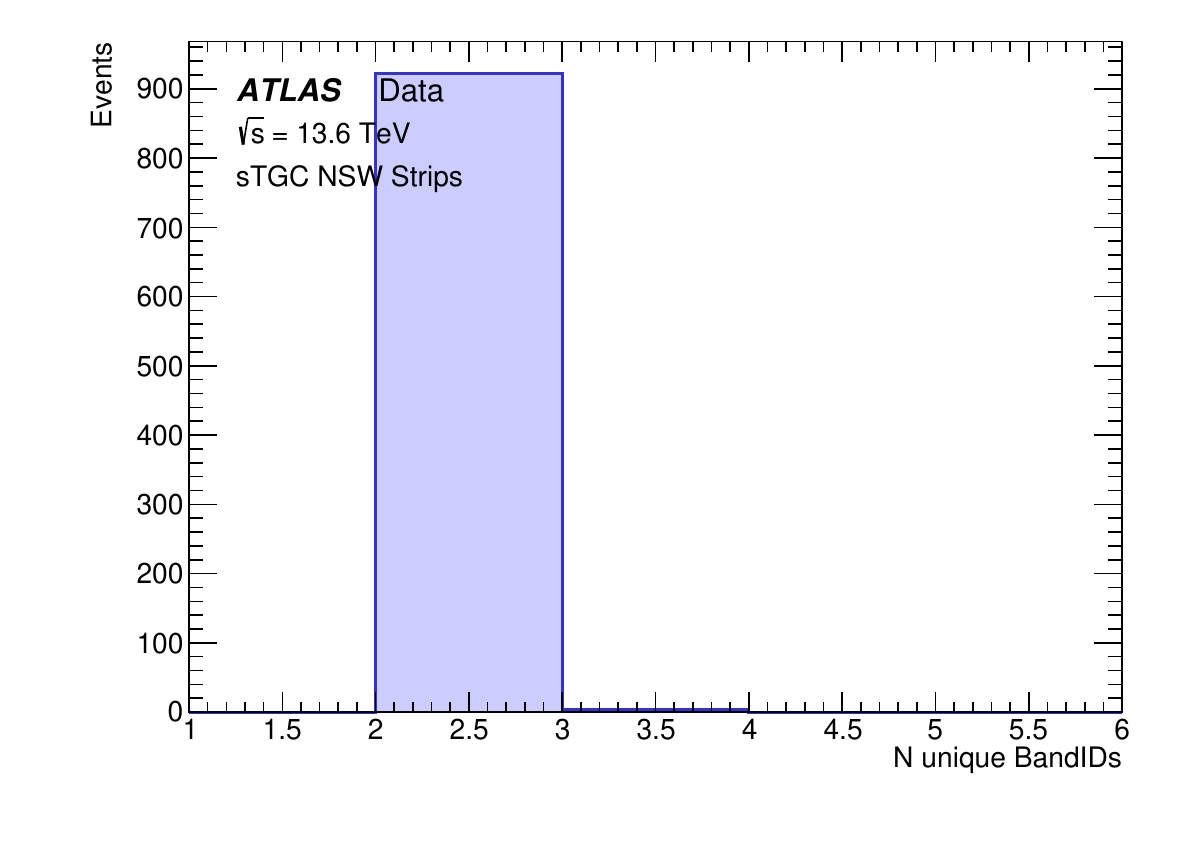}
    \caption{Number of distinct Band IDs per cluster.}
    \label{fig:nsw_trigger:n_unique_bandids}
  \end{subfigure}
  \hfill
  \begin{subfigure}[b]{0.48\textwidth}
    \includegraphics[width=\textwidth]{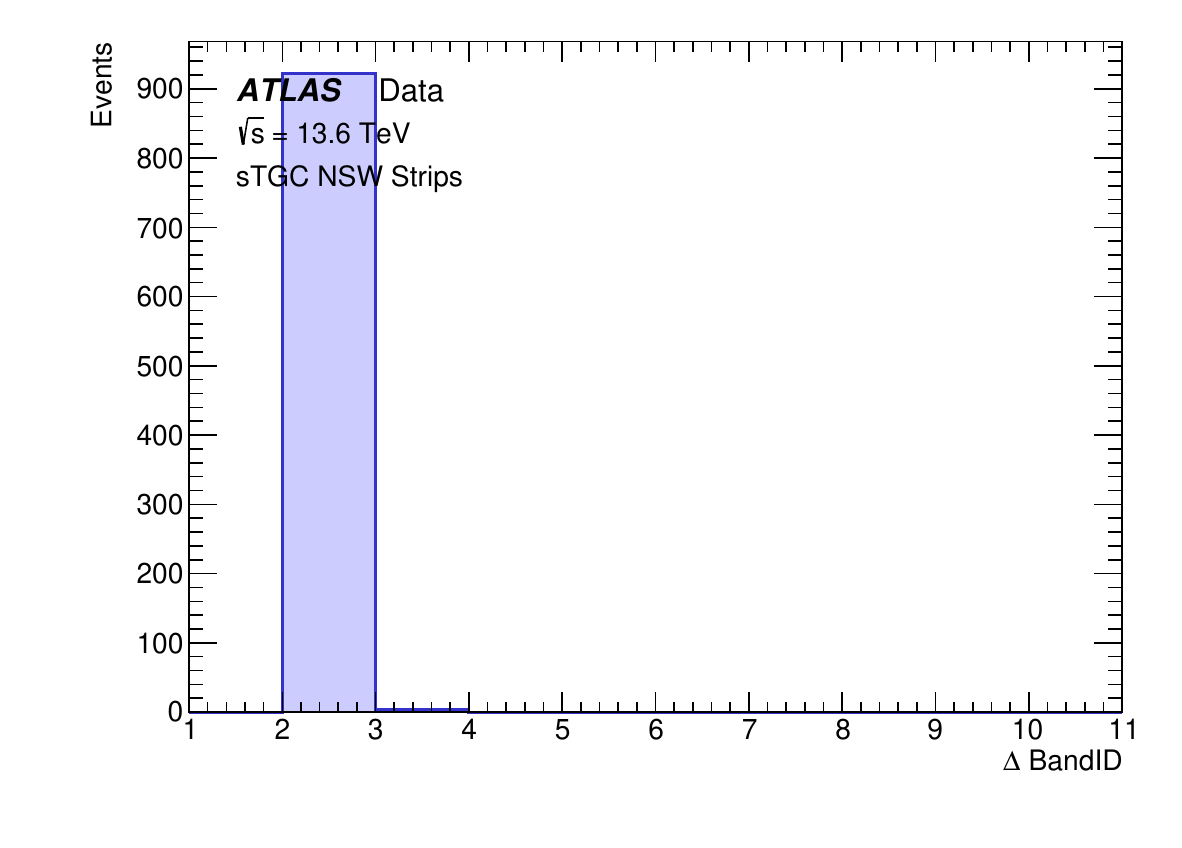}
    \caption{$|\Delta\text{Band\,ID}|$ for multi-Band-ID clusters.}
    \label{fig:nsw_trigger:bandid_diff}
  \end{subfigure}
  \caption{Properties of sTGC on-track clusters spanning more than one Band ID,
    in data at $\sqrt{s}=13.6\,\TeV$.
    Left: all multi-Band-ID clusters have exactly two distinct Band IDs.
    Right: the Band ID difference is always $|\Delta\text{Band\,ID}| = 2$,
    indicating that these clusters straddle a single band boundary.}
  \label{fig:nsw_trigger:multibandid}
\end{figure}

\begin{figure}[htbp]
  \centering
  \includegraphics[width=0.72\textwidth]{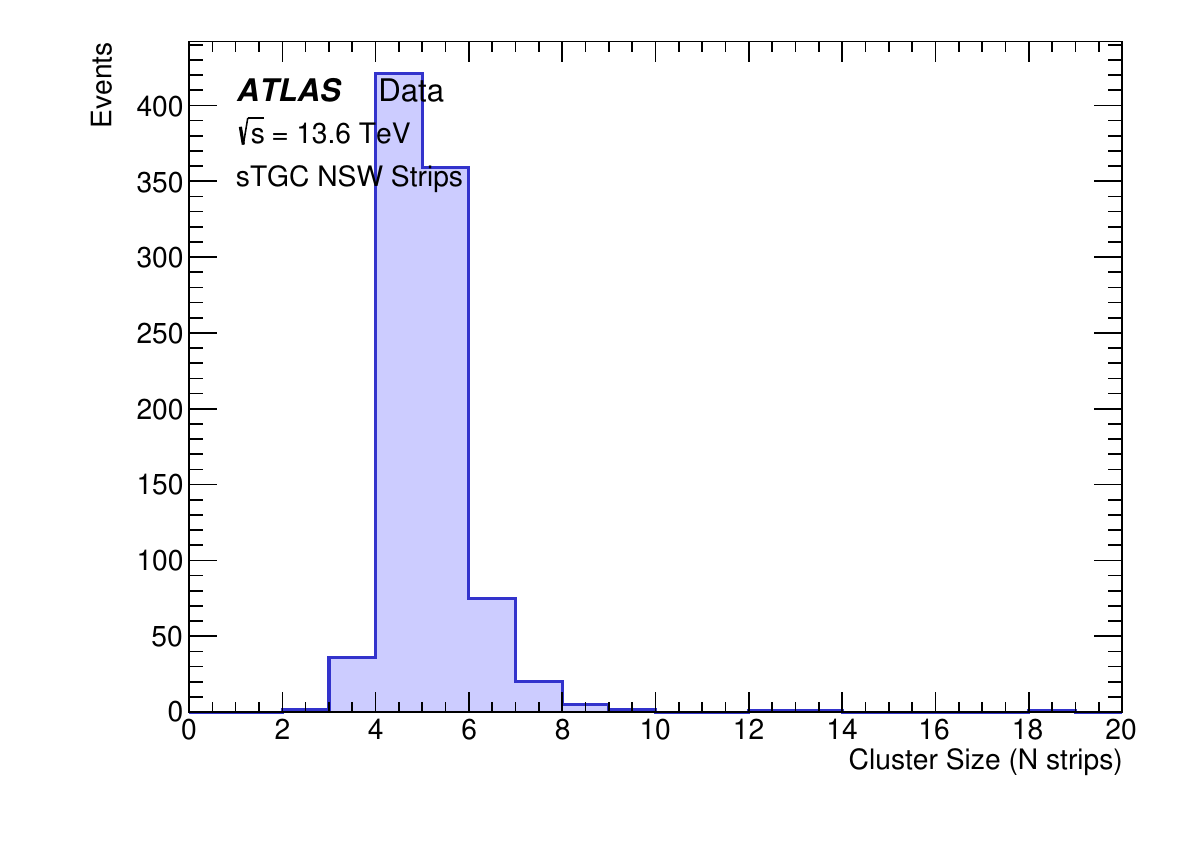}
  \caption{Cluster size $N_\text{strips}$ for multi-Band-ID clusters,
    in data at $\sqrt{s}=13.6\,\TeV$.
    The distribution peaks at $N_\text{strips} = 5$,
    consistent with larger clusters being more likely to straddle a Band ID boundary.}
  \label{fig:nsw_trigger:cluster_size_multi}
\end{figure}

The properties of the $\approx 27\%$ of multi-Band-ID clusters are examined in
FIG.~\ref{fig:nsw_trigger:multibandid}.
The left panel shows that \emph{all} multi-Band-ID clusters have exactly two distinct
Band IDs; no cluster spans three or more bands.
The right panel shows that the Band ID difference between the two assignments
is always $|\Delta\text{Band\,ID}| = 2$ for every one of the 925 multi-Band-ID clusters.
This highly specific value indicates that these clusters sit at a band boundary:
strips on one side of the cluster belong to Band ID $b$
while those on the other side belong to Band ID $b+2$.
FIG.~\ref{fig:nsw_trigger:cluster_size_multi} shows that these clusters
have cluster sizes peaking at $N_\text{strips} = 5$:
larger clusters are geometrically more likely to straddle a band boundary,
which is consistent with this interpretation.

\begin{figure}[htbp]
  \centering
  \includegraphics[width=0.72\textwidth]{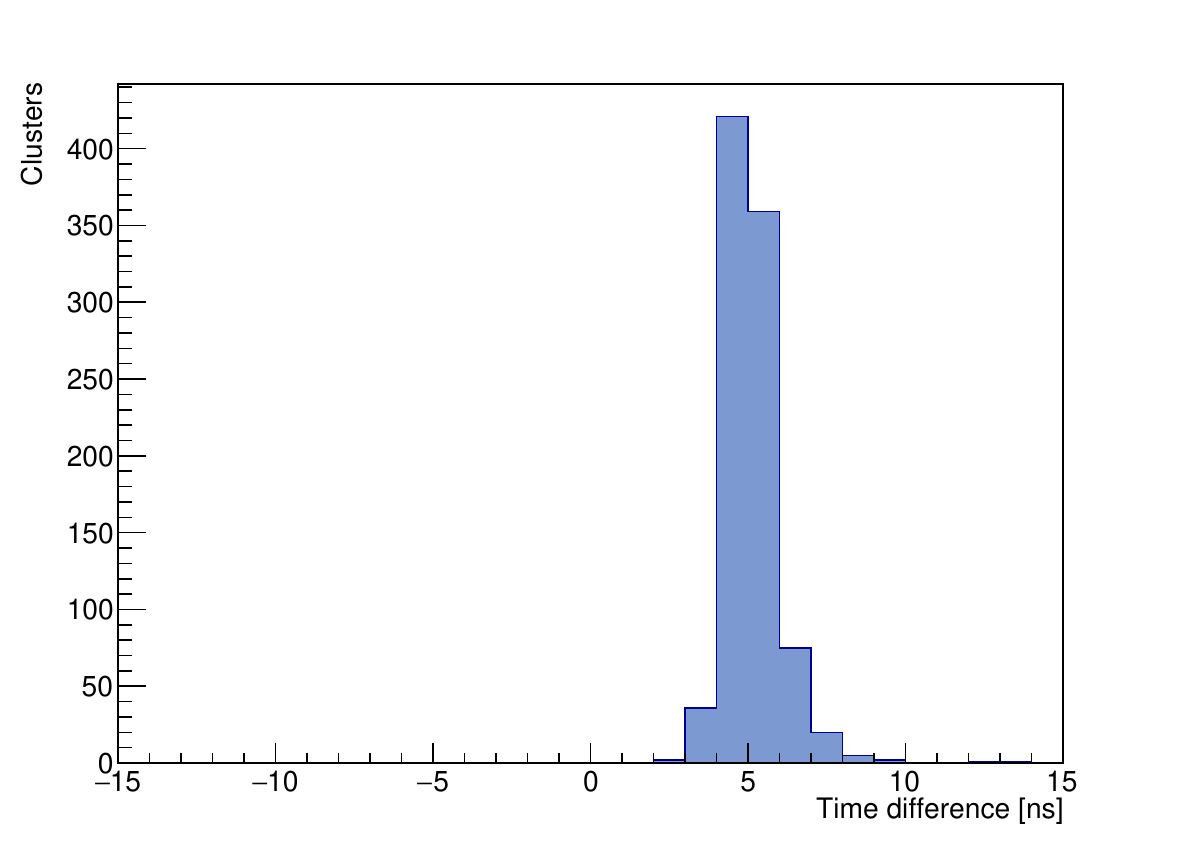}
  \caption{Distribution of the timing difference between strips of multi-Band-ID
    sTGC clusters, in data at $\sqrt{s}=13.6\,\TeV$.
    The distribution peaks at 4--6\,ns, well within one LHC bunch-crossing window (25\,ns),
    confirming that the strips belong to the same muon rather than to two separate tracks
    in adjacent bunch crossings.}
  \label{fig:nsw_trigger:bandid_timediff}
\end{figure}

FIG.~\ref{fig:nsw_trigger:bandid_timediff} shows the distribution of timing differences
between the strips of multi-Band-ID clusters.
The distribution is sharply peaked at $\approx 4$--$6\,\text{ns}$,
well within a single LHC bunch-crossing window of 25\,ns.
This confirms that the strips in a multi-Band-ID cluster are produced
by the same muon crossing rather than by two independent muons in adjacent bunch crossings.
The cluster therefore represents a single physical ionisation track straddling a band boundary,
consistent with the $|\Delta\text{Band\,ID}| = 2$ result.

\section{Strip--Pad Trigger Matching}
\label{sec:nsw_trigger:matching}

\subsection{Matching Procedure}

The central test of the strip integration is whether, for a given pad-trigger Band ID,
the strip clusters observed in data are consistent with the Trigger Processor's LUT.
The matching procedure for each strip cluster proceeds as follows:

\begin{enumerate}
  \item Retrieve the pad trigger's Band ID for the event.
  \item Using the reverse dictionary, identify the first strip channel belonging
        to that Band ID in the relevant layer/module/multiplet/sector.
  \item Compute the \emph{offset}: the position of the cluster's first strip
        relative to the band's first strip.
  \item Translate the cluster into a 14-bit binary string with each bit set to~1
        for a fired strip and 0 otherwise, positioned according to the offset.
  \item Query the TP LUT.
        A cluster is declared \emph{matched} if
        (a) the LUT returns a valid flag \emph{and}
        (b) the cluster does not overflow the 14-strip window
        ($\text{offset} + N_\text{strips} < 15$).
\end{enumerate}

\subsection{Results}

\begin{figure}[htbp]
  \centering
  \includegraphics[width=0.72\textwidth]{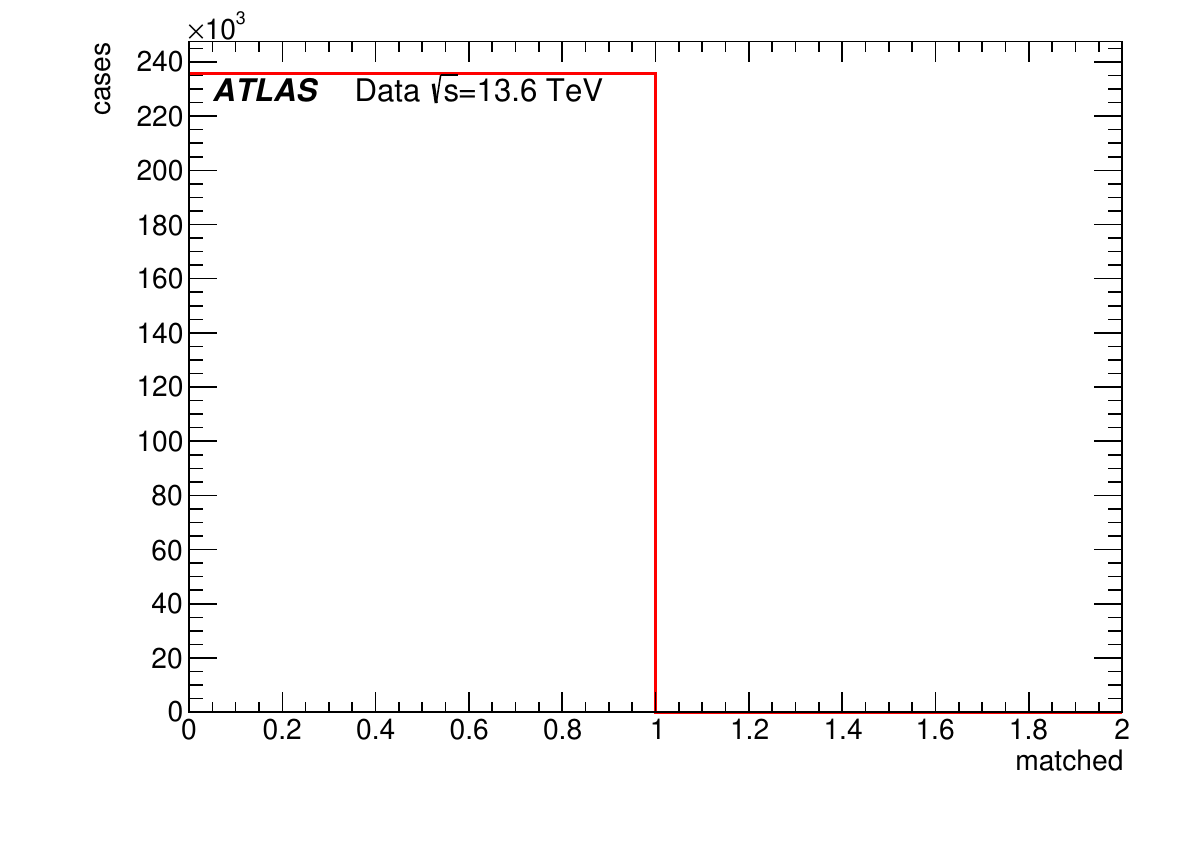}
  \caption{Result of the strip--pad-trigger LUT matching test:
    number of sTGC strip clusters not matched (0) or matched (1) to the pad-trigger
    Band ID via the Trigger Processor LUT, in data at $\sqrt{s}=13.6\,\TeV$.
    Essentially all $\sim$235\,000 clusters are unmatched,
    indicating a systematic offset between the offline strip-to-band dictionary
    and the pad trigger's Band ID convention.}
  \label{fig:nsw_trigger:matching}
\end{figure}

\begin{figure}[htbp]
  \centering
  \includegraphics[width=0.72\textwidth]{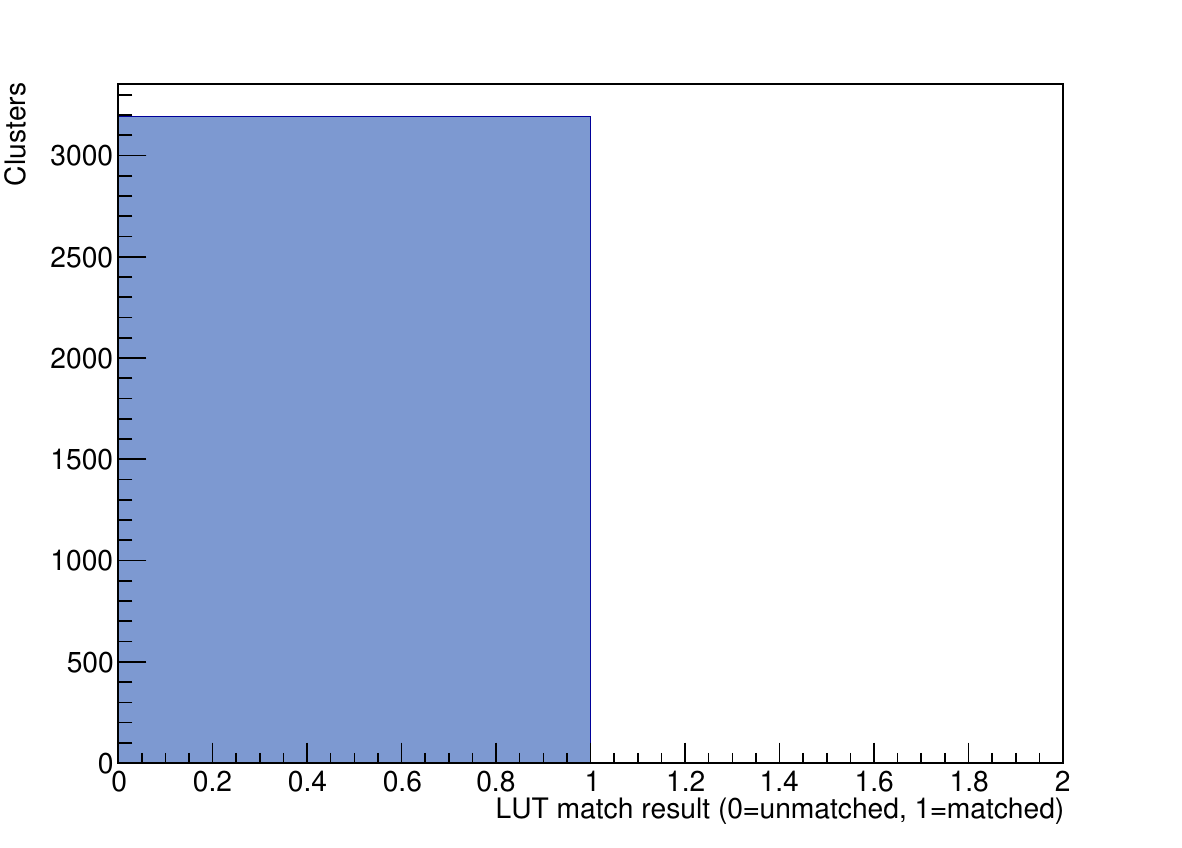}
  \caption{LUT matching result restricted to clusters with 3--5 strips
    (the LUT-accepted cluster size range), in data at $\sqrt{s}=13.6\,\TeV$.
    All $\sim$3\,200 clusters in the LUT-valid size window are unmatched (bin~0),
    confirming that the matching failure is independent of cluster size
    and reflects a systematic Band ID offset rather than an unphysical cluster topology.}
  \label{fig:nsw_trigger:matching_allowed}
\end{figure}

FIG.~\ref{fig:nsw_trigger:matching} shows the outcome of the matching test
applied to all sTGC strip clusters.
The histogram has two bins: bin~0 = not matched, bin~1 = matched.
Essentially all $\sim$235\,000 clusters fall in bin~0.
FIG.~\ref{fig:nsw_trigger:matching_allowed} shows the same test restricted to
clusters with 3--5 strips --- those most likely to produce a valid LUT entry ---
and the result is identical: all $\sim$3\,200 such clusters are unmatched.
The matching rate is consistent with zero across the full sample,
regardless of cluster size selection.

\begin{figure}[htbp]
  \centering
  \includegraphics[width=0.72\textwidth]{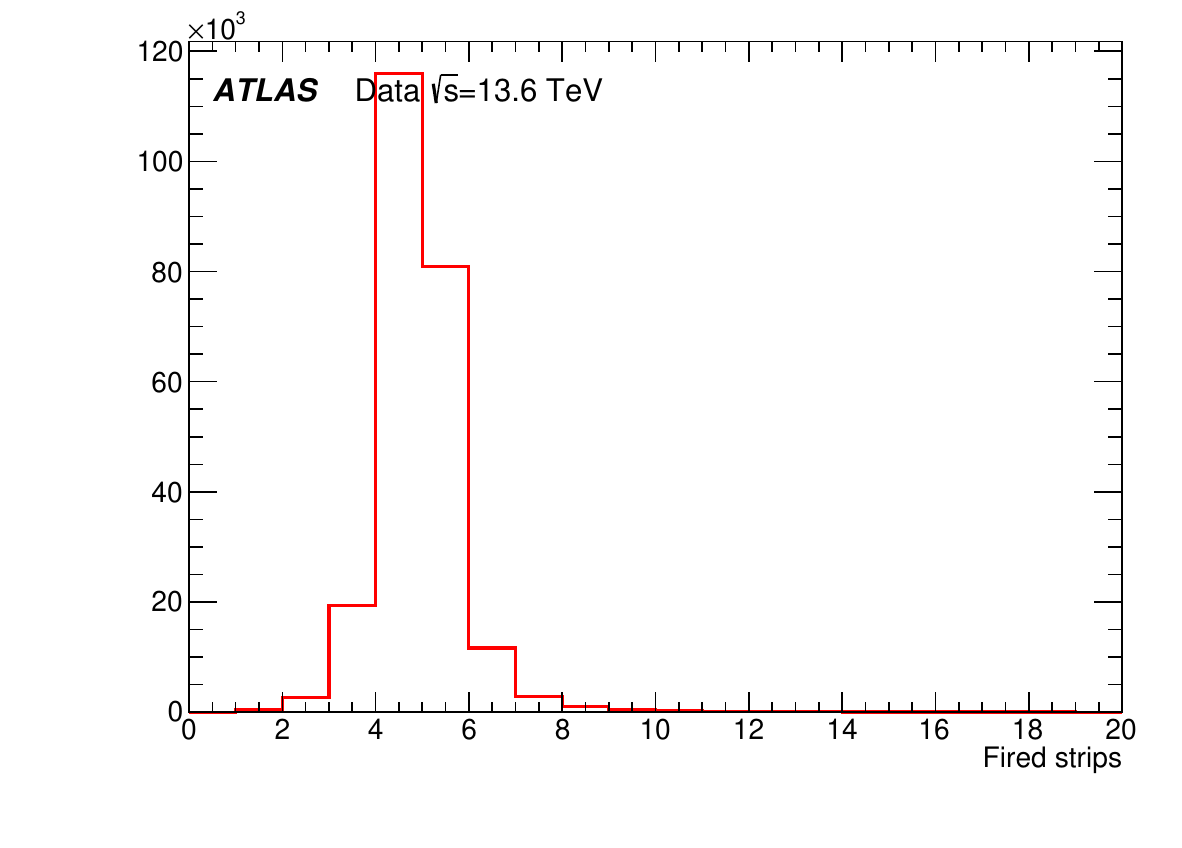}
  \caption{Cluster size $N_\text{strips}$ for sTGC strip clusters
    failing the pad-trigger matching test, in data at $\sqrt{s}=13.6\,\TeV$.
    The peak at 4--5 strips --- within the LUT-accepted range ---
    confirms that the matching failure is not due to unphysical cluster sizes.}
  \label{fig:nsw_trigger:unmatched}
\end{figure}

FIG.~\ref{fig:nsw_trigger:unmatched} shows the distribution of $N_\text{strips}$
for the unmatched clusters.
The distribution peaks at 4--5 strips --- within the LUT's accepted range of 3--6 ---
and is physically identical to the overall good-muon cluster size distribution.
This confirms that the matching failure is not due to anomalous cluster sizes:
the strips are forming normal physical clusters, but are systematically displaced
from the position predicted by the pad trigger's Band ID.

\subsection{Interpretation}

The universal matching failure, taken together with the uniform per-band multiplicity
(Section~\ref{sec:nsw_trigger:bandid}) and the high internal consistency
of the strip-to-band mapping (Section~\ref{sec:nsw_trigger:bandid_consistency}),
points to a specific root cause:
the Band ID convention used by the pad trigger firmware differs from
that used in the offline strip-to-band dictionary.
The two implementations use different reference frames, strip numbering conventions,
or band boundary definitions, causing a systematic offset between the Band ID
the TP assigns to a given $\eta$ position and the Band ID the offline dictionary
assigns to the same position.

This mismatch is not a hardware failure.
The strip detector is functioning correctly:
clusters form with the expected Landau charge distribution,
physical sizes, well-centred timing and uniform occupancy across all 32 readout categories.
Rather, it is a firmware--software alignment problem that must be resolved
by reconciling the Band ID definitions across the readout chain.
The diagnostic framework developed here --- the forward and reverse dictionaries,
the TP emulator, and the matching test --- provides the tools needed
to identify and correct this offset in subsequent commissioning.

\section{Anomalous Six-Strip Clusters}
\label{sec:nsw_trigger:sixstrip}

While the LUT accepts clusters of up to six strips to accommodate extended ionisation tracks,
anomalously large clusters with six or more strips can also arise from two nearby muon tracks
or out-of-time backgrounds producing overlapping clusters.
Such merged clusters may map to the wrong Band ID or fail the LUT entirely.

\begin{figure}[htbp]
  \centering
  \includegraphics[width=0.72\textwidth]{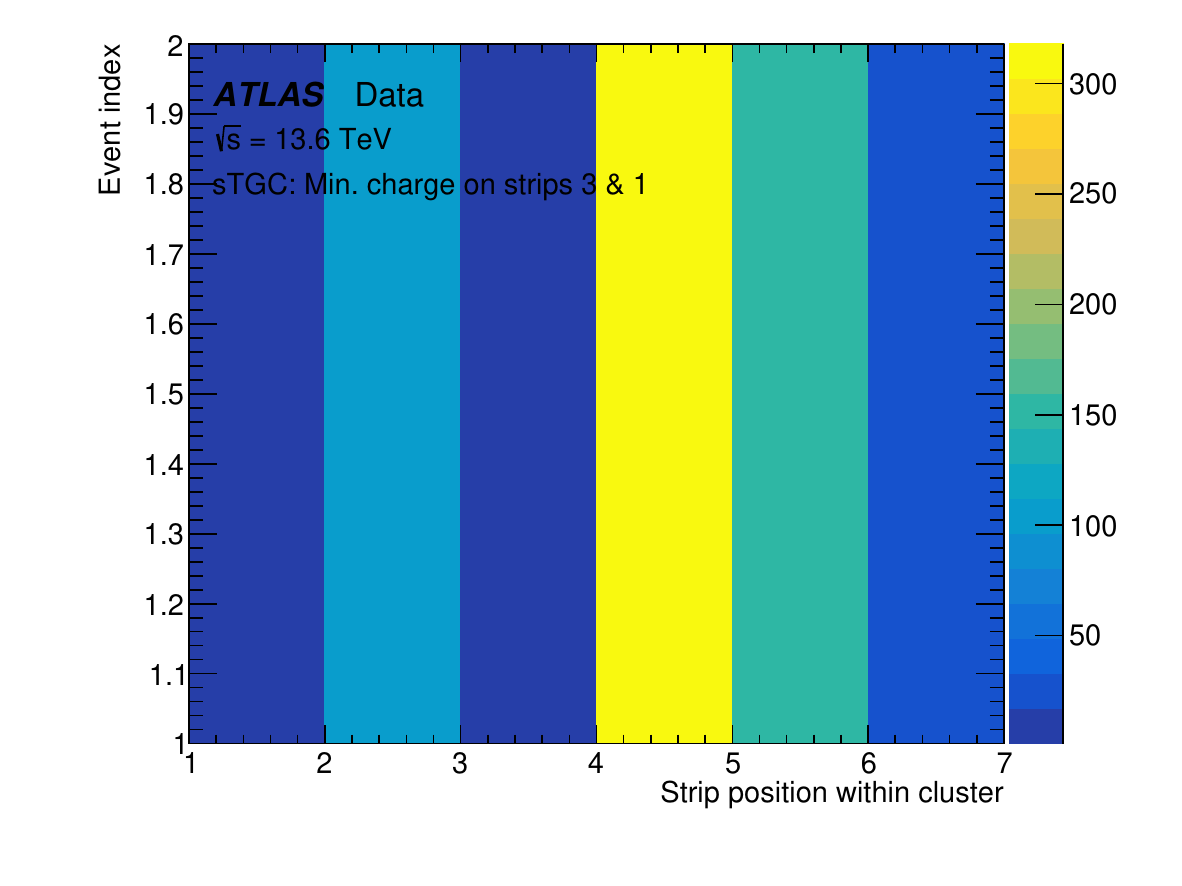}
  \caption{Strip charge profile for six-strip clusters in which the minimum charges
    occur at strip positions 1 and 3 within the cluster,
    in data at $\sqrt{s}=13.6\,\TeV$.
    The y-axis is the event index.
    The low charge at positions~1 and~3 (dark blue) and the high charge at position~4
    (yellow, $\sim$300\,ADC) reveal the double-cluster topology: two overlapping
    three-strip ionisation tracks sharing strips~2--4.}
  \label{fig:nsw_trigger:strip_charges}
\end{figure}

A dedicated study (\texttt{BasicTesterTree}) examined six-strip clusters whose
strip charge profile reveals a double-peak structure, suggesting two overlapping
three-strip clusters.
FIG.~\ref{fig:nsw_trigger:strip_charges} shows the strip-by-strip charge profile
for six-strip clusters where the two minimum-charge strips fall at positions~1 and~3
within the cluster.
The charge pattern exhibits low values at positions~1 and~3 (the minima),
a clear local maximum at position~4 (the dominant peak),
and a secondary plateau at positions~5--6.
This is the expected signature of two adjacent three-strip clusters:
the ionisation of the first track peaks around strips~2--3,
the ionisation of the second track peaks around strips~4--5,
and strip~3 sits between the two centres with suppressed charge.
Such merged clusters are rare ($\lesssim 0.1\%$ of all on-track clusters)
but represent a class of events that would not be matched by any single LUT entry
even after the Band ID offset described in the previous section is corrected.

\section{Summary}
\label{sec:nsw_trigger:summary}

This appendix described the commissioning of sTGC strip data in the NSW trigger chain
using early Run~3 data.
The work covered four components:
\begin{enumerate}
  \item Construction of the strip-to-band dictionary --- forward and reverse mappings
        between strip channels and pad-trigger Band IDs derived from the detector geometry.
  \item Implementation of a software Trigger Processor Emulator replicating the hardware
        LUT that classifies strip cluster patterns into valid ionisation-track signatures.
  \item Analysis of Run~3 ATLAS datasets to characterise strip cluster properties
        and test the strip--pad-trigger LUT matching.
  \item Studies of Band ID internal consistency and anomalous large clusters.
\end{enumerate}

The strip detector is performing well.
Clusters form with the expected Landau charge distribution,
sizes peaking at 5 strips for all hits (4 strips for good-muon selected clusters),
well-centred timing within $\lesssim\pm50$\,ns of the LHC clock,
and uniform occupancy across all 32 readout categories.
The per-band multiplicity ratio is flat at $\approx 0.65$ across all Band IDs 2--91
under both the offline (RO) and pad-trigger (TP) Band ID conventions,
confirming uniform strip and pad detector response across the full NSW $\eta$ acceptance.

The Band ID internal consistency study finds that $\approx 73\%$ of on-track clusters
map all their strips to a single Band ID.
The remaining $\approx 27\%$ are clusters straddling a band boundary,
each characterised by exactly two distinct Band IDs with $|\Delta\text{Band\,ID}| = 2$,
a cluster size peaking at 5 strips, and a strip--strip timing difference of 4--6\,ns
--- confirming they originate from a single muon crossing rather than two separate tracks.
A rare class of six-strip clusters ($\lesssim 0.1\%$ of all clusters)
shows charge profiles consistent with two overlapping three-strip ionisation tracks.

The pad-trigger LUT matching test, however, reveals that essentially no cluster
passes the full strip--pad matching requirement.
This is true regardless of cluster size: even restricting to 3--5 strip clusters
in the LUT-accepted range, the matching rate is zero.
The comparison of the offline (RO) and pad-trigger (TP) per-band multiplicity profiles
shows that both are flat but differ in their Band ID numbering,
providing direct evidence that the systematic offset lies between the firmware and offline
Band ID conventions rather than in the detector hardware itself.
The diagnostic framework developed here --- the forward and reverse dictionaries,
the TP emulator, and the matching test --- provides the tools needed to identify
and correct this offset, laying the groundwork for the full commissioning of strip-based
sTGC triggering in subsequent Run~3 data-taking.

\addtocontents{toc}{\protect\setcounter{tocdepth}{2}}

\clearpage
\end{appendices}

\end{document}